\documentclass[PhD]{sapthesis}

\usepackage[utf8]{inputenc}
\usepackage{csquotes}
\usepackage{amsfonts, amssymb}
\usepackage{float}
\usepackage{caption}
\usepackage{subcaption}
\usepackage{makeidx}
\usepackage{multicol} 
\usepackage[usenames,dvipsnames,svgnames,table]{xcolor}
\usepackage{latexsym}
\usepackage{tabularx}
\usepackage{epsfig}
\usepackage{setspace} 
\usepackage{empheq}
\usepackage{enumitem}

\usepackage{tikzsymbols}

\usepackage{stmaryrd} 

\usepackage[hyperfootnotes=true, linktocpage]{hyperref}
\hypersetup{colorlinks=true, linkcolor=Blue, citecolor=PineGreen,urlcolor=cyan}

\usepackage[skins,theorems]{tcolorbox}
\tcbset{enhanced, colframe=NavyBlue,colback=gray!9!white,arc=2pt,boxrule=1.5pt}

\usepackage{empheq}

{\endEmphEqMainEnv}

\usepackage[mathscr]{euscript}
\renewcommand{\H}{\mathscr{H}}

\usepackage{physics}

\renewcommand{\a}{\alpha}
\renewcommand{\th}{\theta}

\newcommand{\avg}[1]{\left\langle #1 \right\rangle} 

\newcommand{\uv}[1]{\underline{\vb*{#1}}}

\newtheorem{theorem}{Theorem}[section]

\usepackage{appendix}
\usepackage{dirtytalk}

\usepackage{array, makecell}
\usepackage{boldline}

\usepackage{pifont}

\usepackage{tocbibind}

\usepackage[backend=bibtex8, style=phys, sorting=none, backref=false]{biblatex}
\DefineBibliographyStrings{english}{%
	mathesis = {M.Sc. thesis},
}

\DeclareCiteCommand{\supercite}[\mkbibsuperscript]
{\usebibmacro{cite:init}%
	
	\iffieldundef{prenote}
	{}
	{\BibliographyWarning{Ignoring prenote argument}}%
	\iffieldundef{postnote}
	{}
	{\BibliographyWarning{Ignoring postnote argument}}%
	\bibopenbracket}%
{\usebibmacro{citeindex}%
	\usebibmacro{cite:comp}}
{}
{\usebibmacro{cite:dump}\bibclosebracket}

\usepackage[ruled,vlined,resetcount,algochapter]{algorithm2e}

\AtBeginEnvironment{subappendices}{%
\section*{Appendix}
\addcontentsline{toc}{section}{Appendices}
}
\graphicspath{{Figures/}{Figures/Phenomenology-glasses/}{Figures/SM-paper/}{Figures/FSS-paper/}{Figures/Mean-field-and-Gardner/}{Figures/Jamming/}{Figures/LP-algorithm/}}

\newcommand{\opt}[1]{{#1}^\star }

\newcommand{\ctc}[1]{\ensuremath{\qty[#1]}}

\let\vp\varphi
\newcommand{\tf}{\text{f}}
\newcommand{\tg}{\text{g}}
\renewcommand{\S}{\mathcal{S}}
\newcommand{\lpf}{\Gamma}
\newcommand{\ptar}{\ensuremath{p_{\text{tar}}} }

\author{Rafael Alberto Díaz Hernández Rojas}
\title{Numerical study of the microscopic structure of jammed systems: from inferring their dynamics to finite size scaling}
\IDnumber{1798865}
\course[Physics]{Fisica}
\courseorganizer{Scuola di dottorato Vito Volterra }
\AcademicYear{February 2021}
\copyyear{year 2021}

\advisor{Prof. Federico Ricci Tersenghi}
\advisor{Prof. Giorgio Parisi}

\authoremail{rafael.diazhernandezrojas@uniroma1.it}

\cycle{XXXIII}

\examdate{5 July 2021}
\examiner{Andrea Pelissetto}  \examiner[]{Andrea Gambassi} \examiner{Giuseppe Gonnella}

\begin{document}
\frontmatter
\maketitle

\begin{abstract}

In this thesis I present most of the results obtained during my PhD, where I worked on different subjects  regarding jamming in systems of frictionless spheres. In particular, I focused on microscopic properties of jammed packings, such as the distribution of contact forces and interparticle gaps, as well as the single particle dynamics that occur near the jamming point. 
Several of these results have already been presented in Refs.~\cite{paper-dynamics,paper-fss}, but here I include a more detailed analysis of some of them. Besides, all the results of the second chapter are new, even if related to such works.

The thesis is structured as follows. Chapter \ref{chp:intro} provides a quick introduction to the phenomenology of glassy systems, with an entire section dedicated to hard spheres (HS) as a minimalistic model of a glass former. I also include a brief survey of a recently developed mean-field theory, capable of providing an exact description of liquids, glasses and jammed systems in infinite dimensions. This theory serves as a framework for explaining many of the features of real glasses as “blurred” or imperfect analogies of the sharp transitions predicted in $d=\infty$. Moreover, its predictions about the distributions of contact forces and gaps seem to remain valid in low dimensions, \textit{i.e.} $d=2, 3$. Given that here I only consider tridimensional system, its relevance is obvious. But not only, the physical picture based on the meta-basin structure of the free energy landscape will be a guiding principle for many of the topics in later chapters. The final section of this first chapter is devoted to a general discussion about the jamming transition. Naturally, special attention is given to the microscopic features of jammed packings. Thus, I explain in detail some of the most important results derived in previous works. In particular, I give a careful description of the network of contact forces, showing that it is entirely determined by the particles' position, and that it contains important information about the packings stability. Similarly, I reproduce the proof that the exponents of the distribution of forces and gaps are connected through stability inequalities. Because such bounds are saturated, critical jammed packings are only marginally stable. This last feature is rationalized in terms of the constant density of states for vanishingly small frequencies. I hope 
readers will benefit from having all this material gathered in a single source.

Chapter \ref{chp:lp-algorithm} contains a detailed description of the iterative Linear Programming (iLP) algorithm we developed to generate jammed packings. This method has been previously introduced in Refs.~\cite{paper-dynamics,artiacoExploratoryStudyGlassy2020}, but here I present a complete explanation and proofs of several of its features. Moreover, I carry out a detailed characterization of it, although restricted to $3d$ systems. Additionally, in this chapter I review the Lubachevsky--Stillinger compression protocol, based on event-driven molecular dynamics simulations. This method is able to efficiently compress configurations of HS up to very high pressures, and at the same time allows to probe their glassy phase. Complementing our  iLP crunching algorithm with this procedure results in a reliable and reasonably fast method to generate jammed packings of HS. Even more, I show that we can apply the same technique to produce jammed configurations of the Mari--Kurchan (MK) model. 

The dynamics of particles near the jamming point is explored in Chapter \ref{chp:inferring-dynamics}. Given that few works have addressed this issue before us, the first step was to characterise the statistical properties of the trajectories of individual particles in this dynamical regime. Then, the idea was to investigate if the information of the network of contacts can be used to make statistical inferences of the particles' motion close to their jamming point. We found that by considering only the contact \emph{vectors} (\textit{i.e.} ignoring the magnitude of the forces) we were able to construct a couple of structural variables that correlate well with the dynamical features of the particles. More precisely, the vectorial sum of contact vectors is a good descriptor of the preferential directions in single-particle trajectories; while the sum of dot products between such vectors can be used as a predictor of a particle's mobility. The correlations thus obtained are significantly high, although rather short-lived. Importantly, our method proved to be superior than a normal modes approach, which fails to capture these dynamical features at the level of individual particles.
I should also mention that most of the results discussed in this chapter have appeared in Ref.~\cite{paper-dynamics}.

Finally, Chapter \ref{chp:fss} deals with a thorough analysis of the critical distributions of contact forces and interparticle gaps in configurations of HS, soft spheres, as well as in the MK model in $d=3$. The main results can also be found in Ref.~\cite{paper-fss}, where polydisperse disks and near crystals with an FCC structure were also considered. The purpose is to verify the expected power-laws using an analysis based on the finite size effects of such critical distributions. As mentioned above, current numerical estimations of the power-law exponents suggest that jammed packings saturate the stability criteria, so carefully validating their value is clearly important. 
%
Moreover, theoretical predictions of these exponents imply that jamming defines an universality class, to which a broad range of constraint-satisfaction problems belong. However, the calculations involved, being based on mean-field theory, are only exact in the $d\to\infty$ limit. And yet, available numerical results indicate that such universal criticality holds in low dimensions as well. Proving that the same scalings occur in finite dimensions is thus of major theoretical interest, as it would represent the most precise prediction of the so called replica method in realistic materials models. Our approach uses the scaling collapse of the distributions obtained from systems of different sizes and it amply confirms the predicted values in both models. However, it also reveals a striking difference in the size effects on the distribution of forces and gaps, namely, that such effects are negligible in the former but very pronounced in the latter. We rationalize this feature in terms of two correlation lengths that determine how fast the thermodynamic limit behaviour is reached.

I would be very glad if new graduate students entering the intriguing (and complex) world of jamming should find this thesis useful. With this in mind, I took advantage of the fact that after Chap.~\ref{chp:intro} all chapters are independent from each other, and thus I tried to make them as self-contained as possible. 
(Or, in any case, I refer to the relevant sections that provide the necessary context for the topics discussed.) For the same reasons, I decided to give independent conclusions in each chapter, rather than providing general ones at the end. However, I hope that it is sufficiently clear that the unifying thread of all the results presented here are the peculiar properties in the microscopic structure of jammed systems.

\end{abstract}

\begin{acknowledgments}

I've seen somewhere the advice that acknowledgements should be kept concise, or even be avoided all together, lest a work be considered unprofessional. I would hope that the last 15 months suffice to change the mind of anyone who still holds that view. At least for me, taking any opportunity at hand to express my deepest gratitude for the immense fortune I've had is now more important than ever. Or, in any case, let the following lines be a firm testimony that I rather be considered unprofessional than ungrateful.\\

First of all, I sincerely want to thank my supervisors, Prof. Federico Ricci Tersenghi and Prof. Giorgio Parisi, for guiding me through my PhD. It's been an honour to study with scientists of their stature. Working with them has been a most rewarding experience for several reasons, but mainly due to their genuine interest in discovering and understanding every possible aspect of a problem, and also because they transmit their (incredibly vast) knowledge in a very unassuming way. All of this combined with the friendly and cooperative environment of the Chimera group. And even if doing a PhD usually entails becoming overspecialized in a single topic, I am ready to admit at any given time that my major lesson has been that I know nothing but “$1/N^3$ contributions” of all the things they have to teach.

I also want to thank the Physics Department at Sapienza and the PhD coordination for the facilities and travel funds, as well as to the ERC for providing me with the grant that allowed me to pursue my PhD. Additionally, I want to express my gratitude to Profs. Walter Kob and Eric Corwin for reviewing this work as external referees. Their positive appraisal of my thesis has been truly encouraging.

The last chapter of this thesis presents the results obtained in a joint work with an outstanding group of colleagues: Patrick Charbonneau, Eric Corwin, Cameron Dennis and Harukuni Ikeda. I want to warmly thank them all for sharing their data with us and for all the things I learnt during this great collaboration.

I also benefited amply from many interactions with other PhD students: Angelo Cavaliere, Claudia Artiaco, Paolo Baldan; and other researchers: Patrick Charbonneau, Olivier Dauchot, Andrea Liu, Georgios Tsekenis, Franceso Zamponi. From informal talks to concrete suggestions and comments, all of them have undoubtedly helped me and influenced my understanding of the topics of this thesis. Thank you all for your good will in discussing physics with me!\\

The research I had the opportunity to carry out was in itself a terrific experience, but another very fulfilling aspect of my stay in Rome was the amazing persons I had pleasure to meet. In particular, the other PhD students and expats: Gui, Ilaria, Javi and Masry. But also the people from the Stanza 117/118 NEF: Angelo, Fabrizio, Giovanni, José, Silvia. Grazie mille, ragazzi! Tutti voi avete reso Roma una città indimenticabile per me.\\


Obviamente no puedo dejar de agradecer profundamente a toda la gente que siempre ha estado “ahí” para mí, cuando los he necesitado y cuando no. Primero, a mis padres, Antonieta y Rafael, mi abuela, mis tíos Roberto y Claudia, mis primos Emiliano y Constanza, además de mi tía Lina. Ustedes son la evidencia de que hay mucho más cariño que océano.
Por supuesto otra mención especial se merecen mis amigos: Andrea, Bofo, Motita, Karina, Bego, Tony, Eve, Ángela y Humberto. Y claro, los mexas que uno (re)encuentra de este lado del mundo y que hacen que el Atlántico se sienta menos ancho: Tania y Santi; Álvaro y Judith; David; Pato y Gabi.

Y a Viri, que ha estado \emph{aquí}, en los momentos más difíciles y en los más memorables. Además de tu paciencia para escucharme (usualmente a deshoras) cada vez que pienso en voz alta porque algo no cuadra en mis simulaciones, análisis de datos, etc.
Pensar que “todo fue casualidad; la misma beca, la misma universidad” me hace suponer que “debo caerle (muy) bien a alguien allá arriba”. Estos años a tu lado, viviendo con Megara en nuestra pseudo embajada mexicana y armados de mole, un molcajete y mayonesa McCormick (rigurosamente), me han transformado en un tipo capaz de cocinar algo más que quesadillas, en alguien más enamorado que {\normalfont{il pinguino innamorato}}, y sobre todo en una mejor persona. Y yo que pensé que el amor traía chueco el tino\dots \\

A todos ustedes, de verdad, infinitas gracias. De nada servirían los logros que he tenido si no pudiera compartirlos con ustedes.
\end{acknowledgments}

\tableofcontents

\mainmatter

\chapter{Introduction: Why do we care about jammed systems?}\label{chp:intro}

Broadly speaking, a system is said to be jammed when all its degrees of freedom are blocked (see Fig.~\ref{fig:jamming-examples} and Figs.~\ref{fig:jammed-config-2d}, \ref{fig:jammed configuration} below). In particular, in disordered systems such as glasses and amorphous solids, this is usually caused by geometric frustration. But in contrast to their lack of dynamics, research on jammed systems has been far from still. In this introductory chapter, I will give a general, and definitely non-exhaustive, overview of the myriad of fascinating phenomena that occur when a system approaches and \emph{reaches} its jamming point. Yet, an early warning is in place: the results presented in this thesis were obtained by working on configurations of frictionless spherical particles and, therefore, most of what will be discussed is restricted to this kind of systems. Despite such simplification, I will argue that these packings present features that are far from trivial. I will be mainly concerned with the \emph{microscopic} structural properties of jammed configurations, specifically (i) the small gaps formed between particles in near contact, and (ii) the network of contact forces. As I will explain below, it has been known for some time now that, among other things, their distributions exhibit non-trivial scalings at the jamming point. The main goal of this thesis is to investigate what other information they contain. It should be clear that such investigation is far from over, but I hope that the next chapters will contribute to expand our knowledge in that direction; or at least to better demarcate our (my) ignorance. 

\begin{figure}[!htb]
	\begin{subfigure}{0.30\linewidth}
		\includegraphics[width=\textwidth]{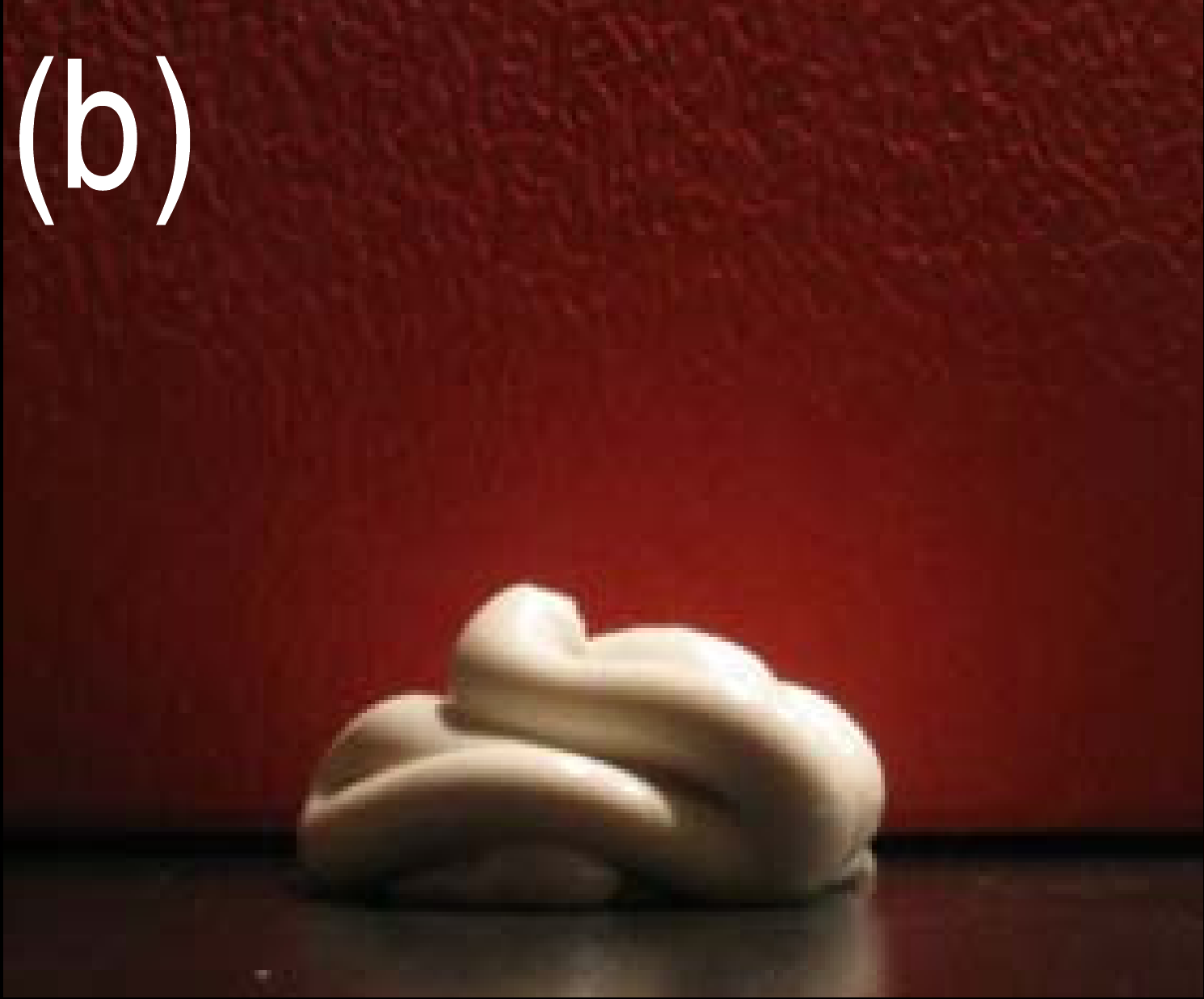}
		\caption{Colloids (toothpaste). Taken from \cite{vanheckeReviewJammingSoftParticles2010}.}
	\end{subfigure} \hfil
	\begin{subfigure}{0.30\linewidth}
		\includegraphics[width=\textwidth]{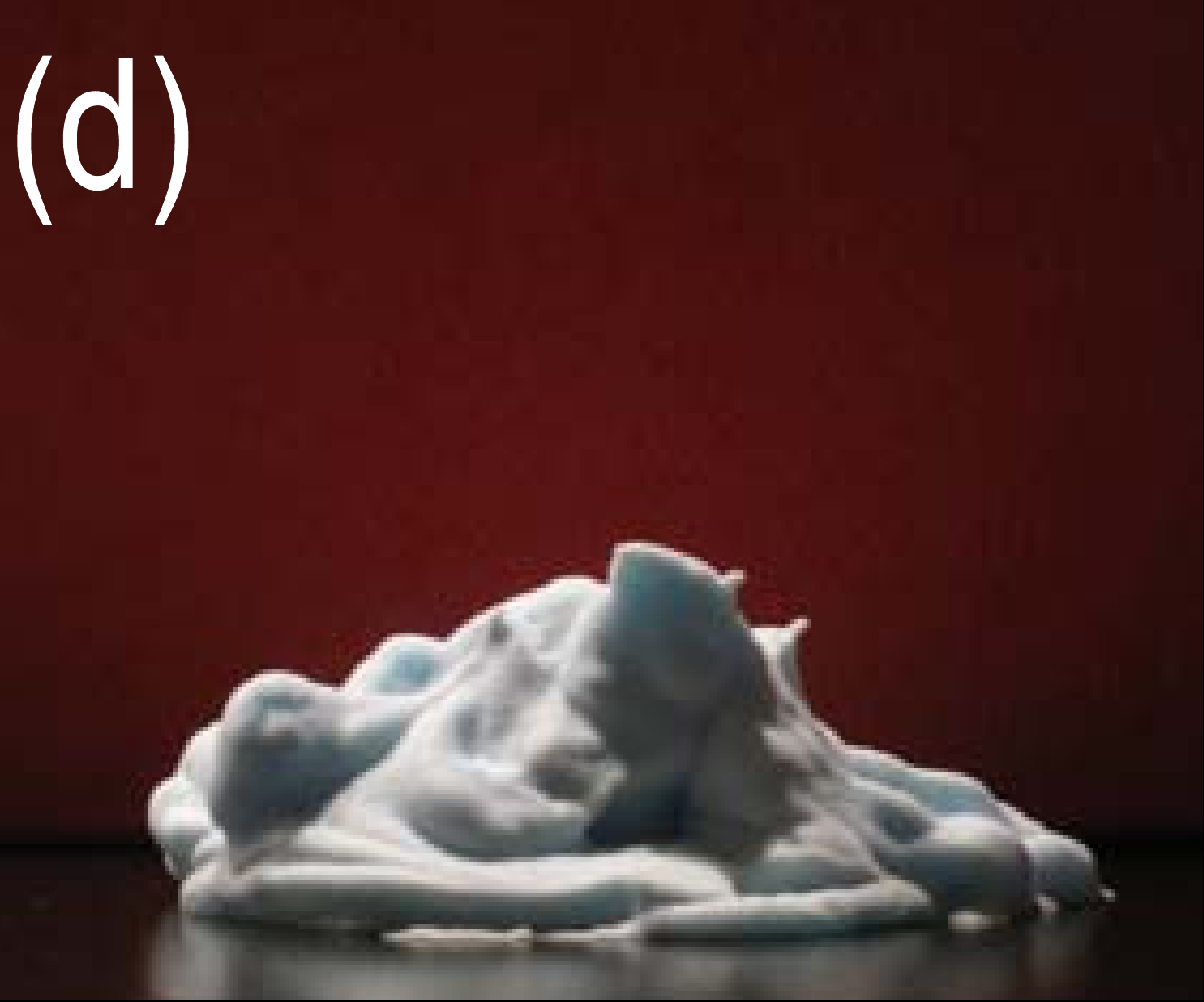}
		\caption{Foams (shaving foam). Taken from \cite{vanheckeReviewJammingSoftParticles2010}.}
	\end{subfigure} \hfil
	\begin{subfigure}{0.30\linewidth}
		\includegraphics[width=\textwidth]{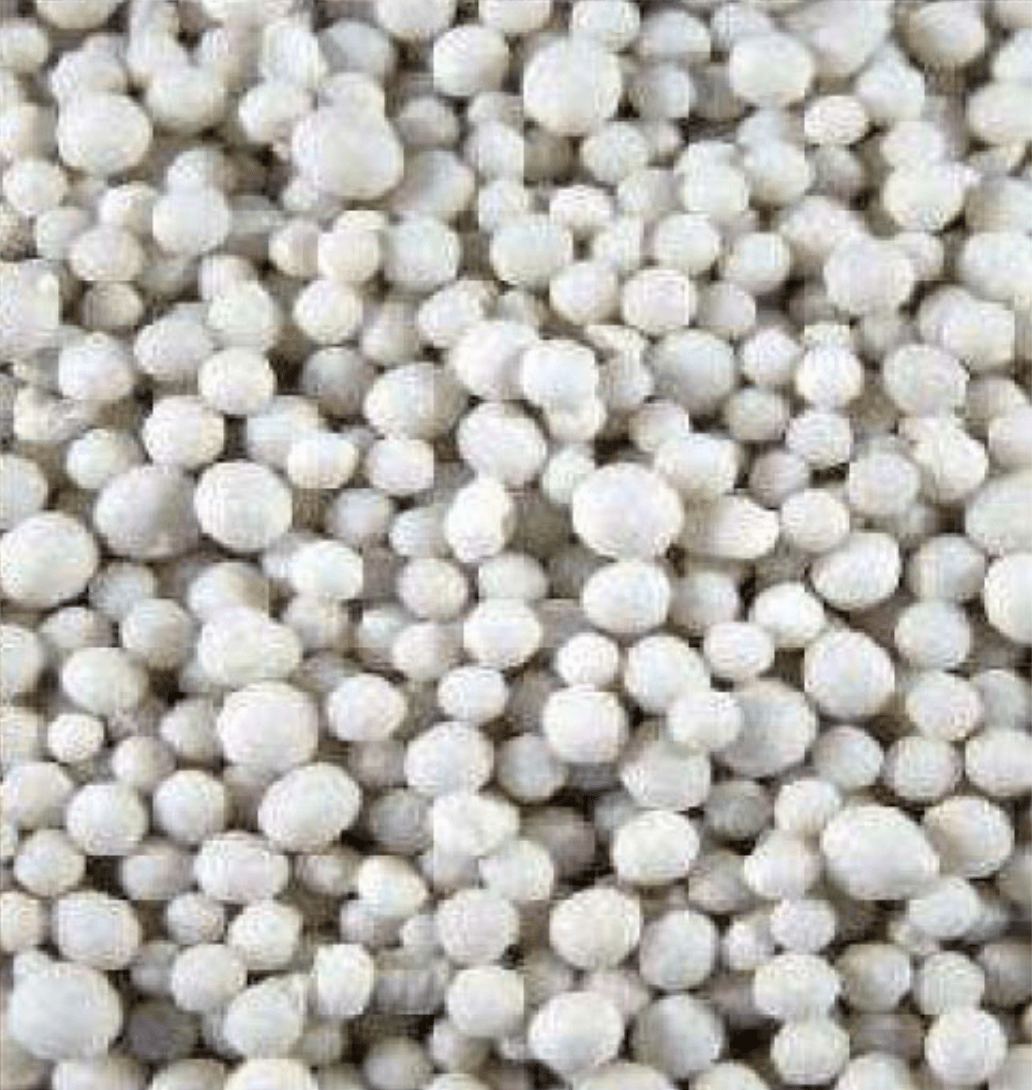}
		\caption{Granular matter (fertilizer). Taken from \cite{berthier_biroli_theoretical_2011}.}
	\end{subfigure} \\[5mm]
	\begin{subfigure}{0.30\linewidth}
		\includegraphics[width=\textwidth, height=3cm]{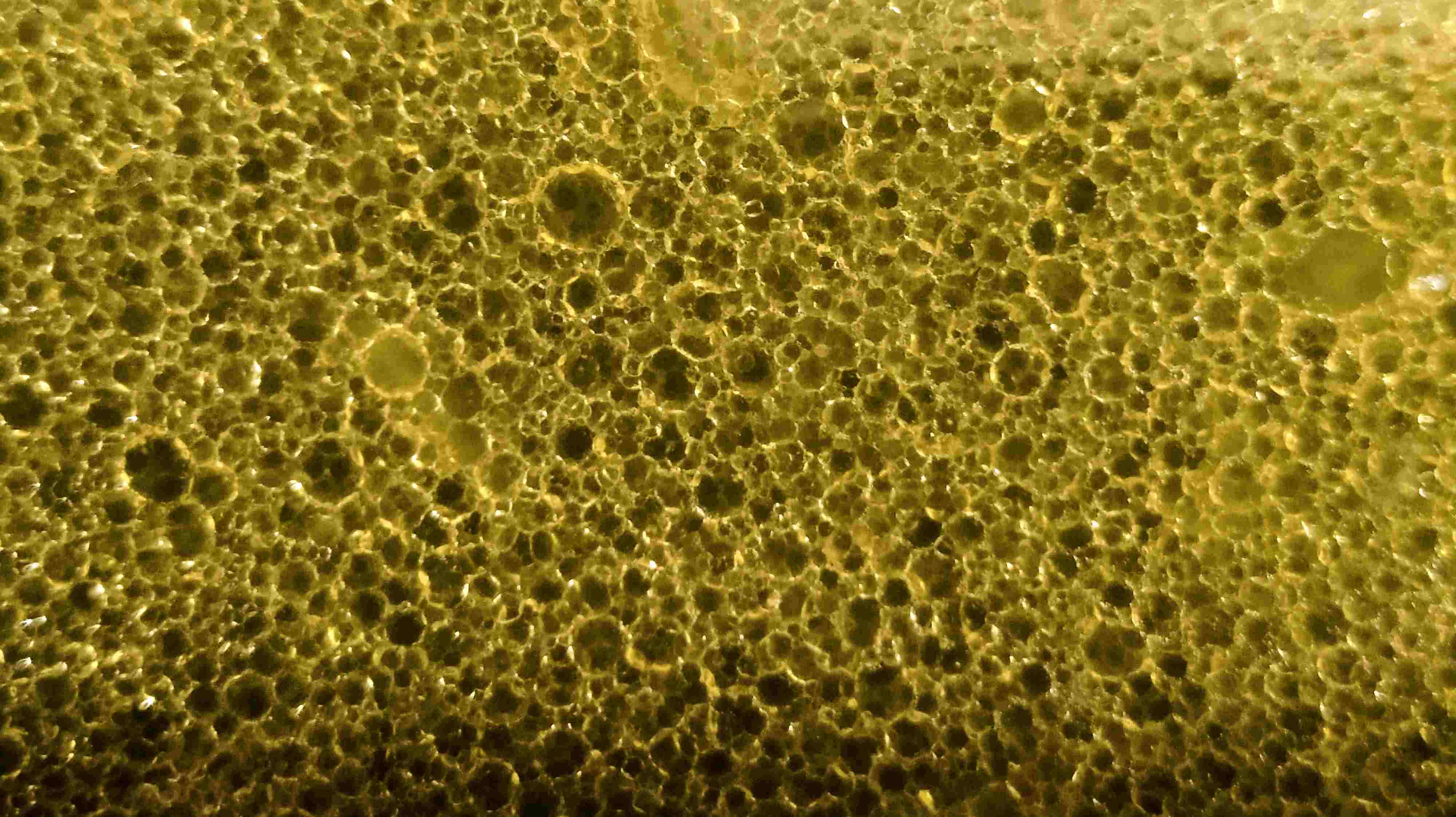}
		\caption{Emulsions (oil and vinegar)}
	\end{subfigure}\hfil
	\begin{subfigure}{0.30\linewidth}
		\includegraphics[width=\textwidth]{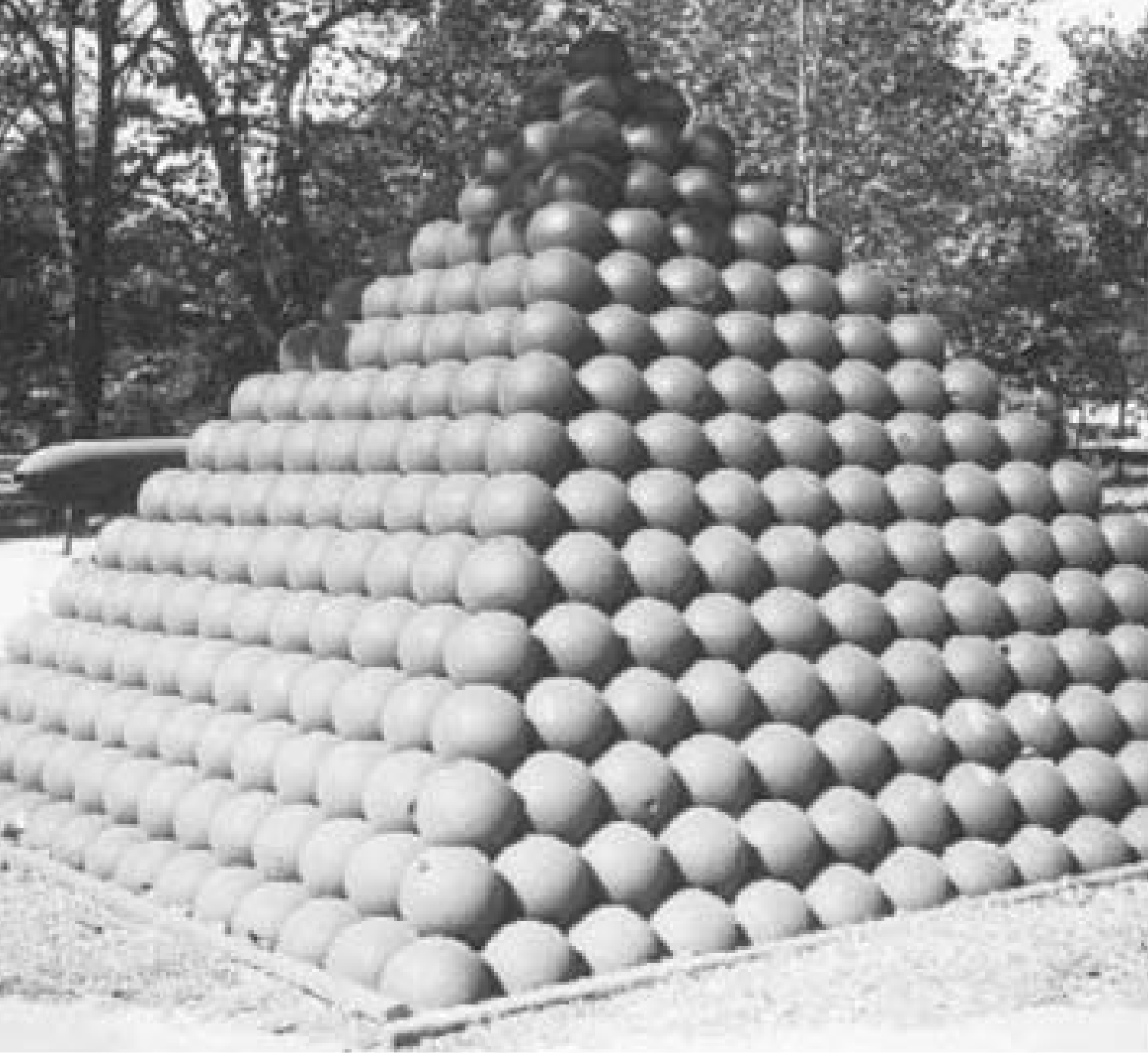}
		\caption{Ordered packings (FCC structure); usually \emph{hyper}-static. Taken from \cite{torquatoReviewJammedHardparticlePackings2010}.}
	\end{subfigure} \hfil
	\begin{subfigure}{0.30\linewidth}
	\includegraphics[width=\textwidth, height=4cm]{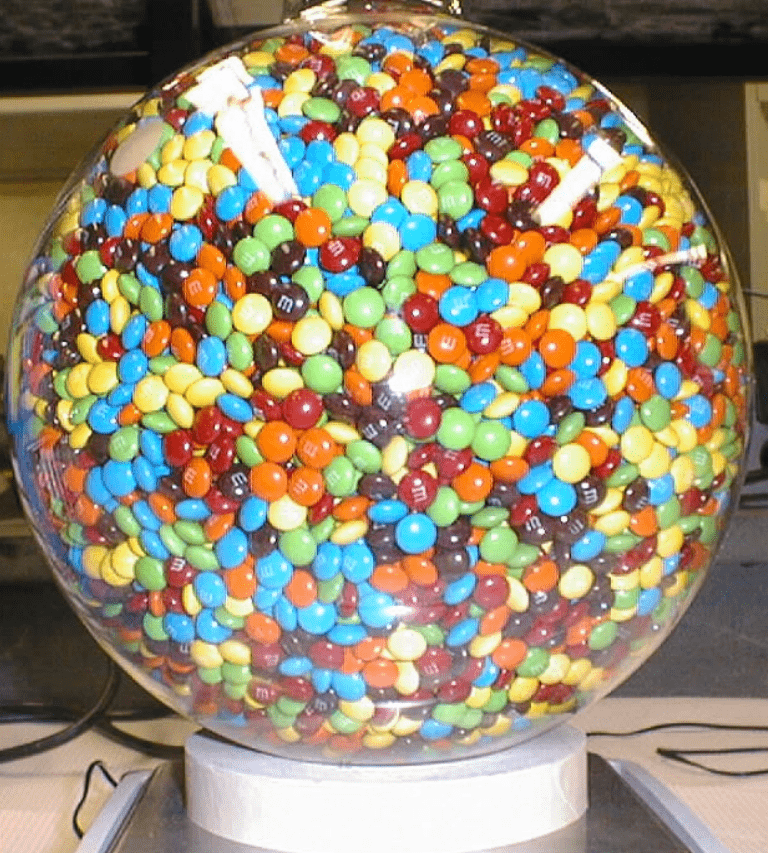}
	\caption{Ellipsoids (M\&M's); usually \emph{hypo}-static. Taken from \cite{donevPhdThesis}.}
	\end{subfigure} \hfil
\caption[Some examples of systems that exhibit a jammed regime.]{Some examples of systems that exhibit a jammed regime. However, the results in this thesis do not concern the last two, mainly due to their ordering and shape, but also because I will only focus on \emph{iso}static systems; see Sec.~\ref{sec:jamming-transition}.}
\label{fig:jamming-examples}
\end{figure}

Let me now sketch the path I will follow. In this first chapter I will discuss many of the ideas needed to better understand the main results of this work. Thus, Sec.~\ref{sec:glasses} is a very quick and superficial description of the phenomenology of supercooled liquids and glasses. This might seem unnecessary given that jamming is not exclusive of glasses. However, it will prove useful for appreciating that the glass and the jamming transitions are \emph{different} transitions. We now understand their differences clearly enough, but few years ago the situation was much blurrier. Furthermore, in  the algorithm described in Chp.~\ref{chp:lp-algorithm} we mimic the process of taking a liquid deep in its glass phase, and only then compress it further until the jamming point is reached. Having a clear idea of the physics that takes place in these regimes is thus important. Additionally, the discussion of Sec.~\ref{sec:dynamics-and-local-structure-glasses} contains many of the necessary background to the results of Chp.~\ref{chp:inferring-dynamics}. 
In Sec.~\ref{sec:hs-fluids} I will show that many of the peculiar features of glass formers can be reproduced using a very simple model: a hard-sphere (HS) fluid. The resemblance in their behaviour then justifies the usage of this minimal model for studying glassy and jammed systems. Hence, in this  second section, after a very basic introduction to the statistical mechanics of liquids, I analyse the equations of state for HS in liquid and glass phase, as well as other important phenomena, like the Gardner transition.
Moreover, a recently developed mean-field theory --exact, in the infinite dimensional limit-- uses both HS and soft spheres (SS) as archetypical models. The theory is powerful enough to consistently describe the behaviour of glass-forming systems from the point they fall out of equilibrium all the way down to jamming.
A brief outline of this theory and some of its many predictions are given in of Sec.~\ref{sec:MF theory}.
Finally, in Sec.~\ref{sec:jamming-transition} I discuss in detail the jamming transition. From the approach I follow, jamming is an (out of equilibrium) phase transition that marks the end of the glass in HS and SS systems. Surprisingly, this phase transition is shared by several other systems (Sec.~\ref{sec:jamming in many systems}) such as grains, foams, and colloids as depicted in Fig.~\ref{fig:jamming-examples}. All of them present common properties and can thus be studied using similar tools (Secs.~\ref{sec:jamming criticality} and \ref{sec:network of contacts}). The last parts of the chapter (Secs.~\ref{sec:forces-and-gaps} and \ref{sec:marginal stability}) are devoted to analyse the properties of the contact forces and interparticle gaps. These microscopic structural variables have a central role for the results of this work.


\section[Phenomenology of glassy systems: the departure point towards jamming]{Phenomenology of glassy systems: the departure point towards jamming%
	\sectionmark{Phenomenology of glassy systems}} \label{sec:glasses}
\sectionmark{Phenomenology of glassy systems}

Glasses are very common in our daily lives and, yet, incredibly hard to explain satisfactorily from a physical point of view. And every year that passes by without a complete theory of the \emph{glass transition} (defined soon below), helps to proof the now famous quote by Philip W. Anderson\supercite{anderson_glass-hard-problem}, \say{The deepest and most interesting unsolved problem in solid state theory is probably the theory of the nature of glass and the glass transition.} As with any other interesting problem, the lack of a complete theory of the glass phase is not caused by a shortage of efforts.
Indeed, the current literature on glassy systems is immense, as a (very personally biased) sample can show: \cite{berthier_biroli_theoretical_2011,cavagna_supercooled_2009,debenedetti_supercooled_2001,biroli_perspective_2013,binderGlassyMaterialsDisordered2011,gotzeComplexDynamicsGlassForming2008,les_houches_glasses,leuzziThermodynamicsGlassyState2007,puz_book,berthierDynamicalHeterogeneitiesGlasses2011,janssenModeCouplingTheoryGlass2018,biroliCrashCourseAgeing2005,reichmanModecouplingTheory2005,sciortinoPotentialEnergyLandscape2005,SpinglassTheoryPedestrians2005}. The contents of this section strongly follows Refs.~\cite{cavagna_supercooled_2009,berthier_biroli_theoretical_2011,debenedetti_supercooled_2001}, which provide very amenable reviews of the topic of glasses and supercooled liquids.

In very simple terms, what makes glasses so hard to understand is that they behave, in many aspects, just as (non-crystallized) solids, but structurally they are much more similar to liquids. In fact, it is “relatively simple” to produce a glass by cooling a liquid fast enough in order to avoid crystallization. (See however \cite[Sec.~2]{cavagna_supercooled_2009} for a detailed account of the obstacles that can occur in practice when cooling a liquid.) As temperature decreases below the liquid's melting point ($T_m$), the viscosity ($\eta$) increases drastically and, at some point, the liquid simply stops flowing and behaves, mechanically, as a solid. For instance, it develops a finite shear modulus that does not decay on time. Such point identifies the \emph{glass transition} of the liquid, and the temperature at which this happens is the so called \emph{glass transition temperature}, $T_g$. Pragmatically, a rule of thumb to define $T_g$ is the temperature such that\supercite{cavagna_supercooled_2009,biroli_perspective_2013}
\begin{equation}\label{def:Tg-viscosity}
\eta(T_g) \sim 10^{13}\ \text{Poise} = 10^{12} \text{ Pa s} \, .
\end{equation}
In comparison, the viscosity of water at room temperature is minute: $\eta_{\text{water}} \approx 10^{-3} \text{ Pa s}$.

\begin{figure}[h!]
	\centering
	\includegraphics[width=0.99\textwidth]{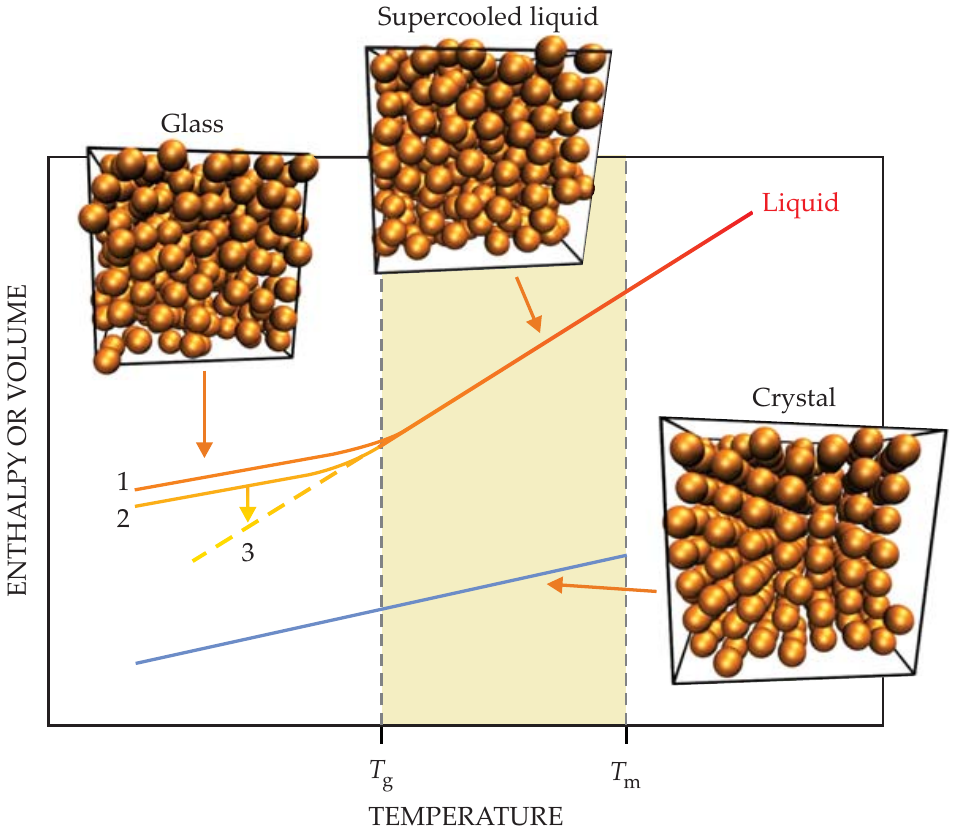}
	\caption[Glass -- supercooled liquid diagram.]{Main: Glass -- supercooled liquid phase diagram. For temperatures above (resp. below) the melting temperature ($T_m$), the liquid (resp. crystal) is the equilibrium state. When the temperature is not too small compared to $T_m$, crystallization can be avoided by fast cooling, allowing the liquid to remain in a metastable state: the supercooled liquid. In such metastable phase (shaded region) the relaxation time grows drastically.
	If $T$ is further decreased, the supercooled liquid is bound to become a glass, characterized by an “immense” relaxation time and viscosity, at the glass transition temperature ($T_g$). Interestingly, different cooling protocols locate $T_g$ at different values (curves labelled as “1” and “2”); see text for a discussion.
	Insets: Snapshots of the configurations obtained through molecular dynamics simulations of a crystal (right), supercooled liquid (centre), and glass (left). Notably, there is little structural difference between the supercooled liquid and glass configurations, but dynamically they are completely different as discussed throughout this section.
	Figure taken from Ref.~\cite{berthier_physics_today}.}
	\label{fig:glass-transition-phase-diagram}
\end{figure}

Yet, analysing the microscopic structure of the systems slightly above and slightly below $T_g$ reveals that little, if anything, has changed: glasses are, essentially, solidified non-crystalline liquids. This process is depicted in the volume-temperature plane of Fig.~\ref{fig:glass-transition-phase-diagram}. We can see that at for $T>T_m$ the liquid is the only stable phase, but for temperatures lower than $T_m$ the crystal becomes the equilibrium branch. Nevertheless, by applying a cooling rate fast enough to avoid crystal nucleation, but small enough to let the system relax, the liquid phase can be sustained in the so called supercooled phase, identified by the shaded region. In other words, we have to cool the system in such a way that it remains in \emph{metastable} equilibrium. As shown in the insets, there is a clear difference in the structure (\textit{i.e.} the arrangement of particles) between the crystalline and liquid phase. However, when reaching $T_g$ there does not seem to be any structural change in the configuration, but the physical properties become very different. This figure also shows a very peculiar feature of the glass transition: the exact location of $T_g$ actually depends on the protocol used to reach the glass state. That is, if a given cooling rate results in the glass state identified by the branch “1”, a slower protocol would result in the branch identified by “2”. Such behaviour immediately begs the question: what kind of transition does $T_g$ signals, if its value depends on the protocol employed? Even more, the definition of $T_g$ in the expression \eqref{def:Tg-viscosity} is arbitrary. Could we have chosen a different value for the threshold of $\eta$ instead of $10^{13}$ Poise?

\begin{figure}[htb!]
	\centering
	\begin{subfigure}{0.49\textwidth}
		\includegraphics[width=\textwidth]{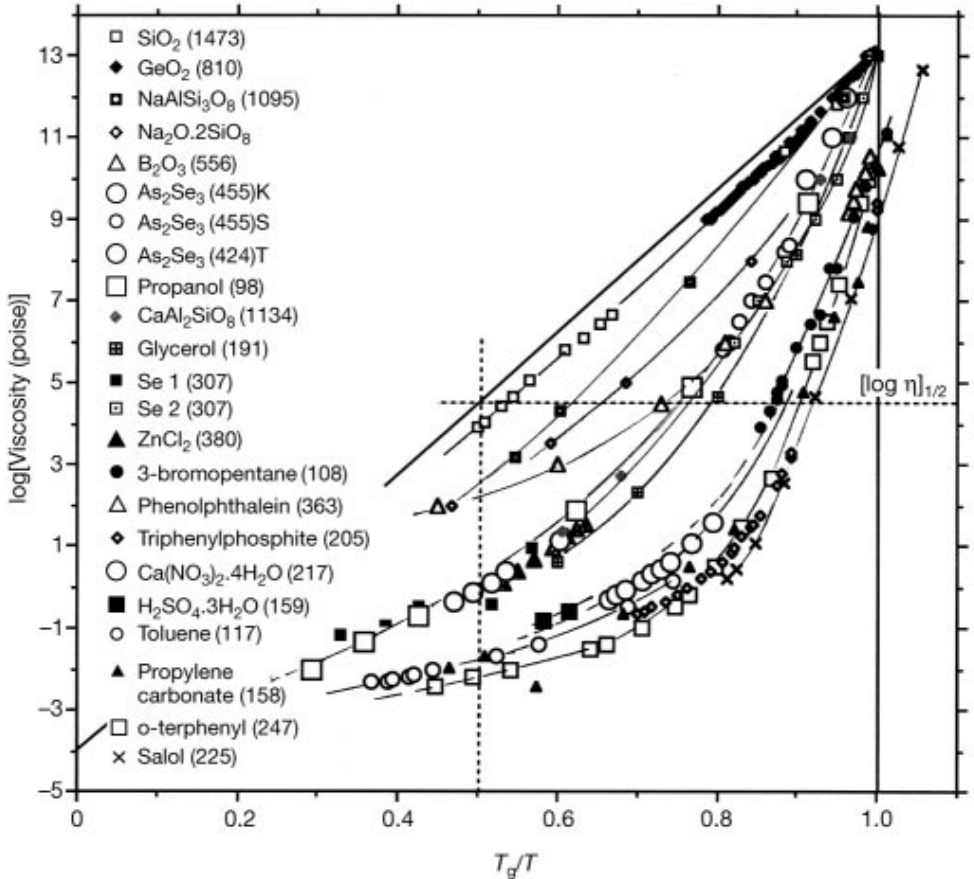}
		\caption{Angell plot of the logarithm of the viscosity as a function of (inverse) temperature, using experimental data from several materials. Taken from \cite{angellThermodynamicConnectionFragility2001}.}
		\label{fig:arrhenius-plot-angell}
	\end{subfigure}
	\begin{subfigure}{0.49\textwidth}
		\includegraphics[width=\textwidth, height=6.5cm]{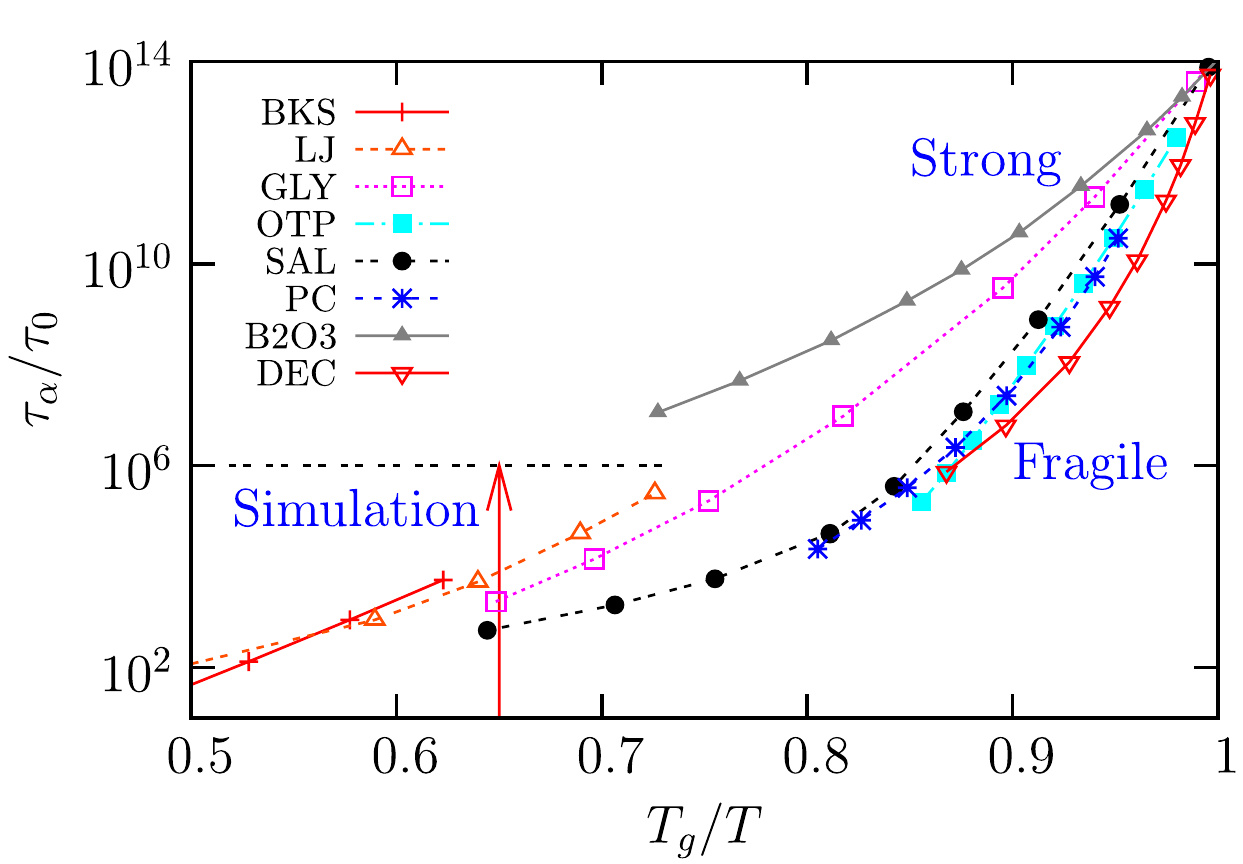}
		\caption{Angell plot of the relaxation time as a function of (inverse) temperature. Points show both experimental and numerical data. Taken from \cite{berthier_biroli_theoretical_2011}.}
		\label{fig:arrhenius-plot-2}
	\end{subfigure}
\caption[Angell plots obtained experimentally and from simulations.]{The so called \textit{Angell} (or sometimes \textit{Arrhenius}) plots obtained by analysing the dependence of $\log \eta$ or $\log \tau_\a$ as a function of $T_g/T$. They allow to determine whether a given material is a strong or fragile glass former if its corresponding curve is a straight line or not, respectively.}
\label{fig:arrhenius-plots}
\end{figure}

From these considerations it should be clear that the glass transition is not a usual thermodynamic transition,  while it is also true that the definition of $T_g$ entails a certain degree of arbitrariness. To better understand the physics of supercooled liquids and glasses, it is useful to recall the relation between the viscosity and relaxation time $\tau_\a$. For instance, from the Maxwell model of elasticity, we have that\supercite{cavagna_supercooled_2009,berthier_biroli_theoretical_2011} $\eta = G_{\infty} \tau_\a$, where $G_{\infty}$ is the instantaneous shear modulus.  $\tau_\a$ determines the typical time scale on which density fluctuations in a system relax. Hence, it is natural that as the system becomes more viscous, fluctuations take longer to relax in the system. In several systems it has been found empirically that as $T_g$ is approached, the increase in relaxation time follows an exponential behaviour, captured by the so call Arrhenius relation:
\begin{equation}\label{eq:arrhenius}
\tau_\a = \tau_0 \exp(\frac{E}{k_B T}) \qc
\end{equation}
where $E$ is an activation energy, and $k_B$ is the Boltzmann constant, which from now on I will assume equal to 1. Even more, there are various other materials whose viscosity increases at an even faster rate. Their sluggish behaviour is well captured by the Vogel--Fulcher--Tamman (VFT) law:
\begin{equation}\label{eq:vft}
\tau_\a = \tau_0 \exp( \frac{D}{T/T_0 -1}) \, .
\end{equation}
These two different types of growth are used to classify glass formers as \textit{strong} or \textit{fragile}, depending on whether they follow Eq.~\eqref{eq:arrhenius} or \eqref{eq:vft}, respectively. It is a rather unfortunate terminology, since it is not related to the rigidity of glasses themselves. In any case, plotting\supercite{angell_glass-forming_1994} the logarithm of relaxation time as a function of $T_g/T$ results in the curves shown in Fig.~\ref{fig:arrhenius-plots} for several materials.
Strong glass formers (such as SiO$_2$) follow straight lines, while fragile ones (\textit{e.g.} o-terphenyl) first exhibit a relatively slow increase of $\tau_\a$, followed by a sharp super-exponential growth as $T\to T_g$. Note that data in Fig.~\ref{fig:arrhenius-plot-angell} correspond to experiments where $\eta$ was measured, while Fig.~\ref{fig:arrhenius-plot-2} contains data of $\tau_\a$ both from experiments and from simulations. Thus, putting together these two figures validates the relation $\eta \propto \tau_\a$ mentioned above. In fact, an alternative definition for $T_g$ is such that $\tau_\a(T_g)\sim 10^2-10^3$ s.

This last statement seems to reinforce the idea that $T_g$ is a very arbitrary definition, because setting $100$ s as the limit for $\tau_\a$ apparently indicates that we are not patient enough to perform longer experiments. However, if considered carefully, the exponential and super-exponential growths shown in Fig.~\ref{fig:arrhenius-plots} indicate that any reasonable change in the threshold value of $\tau_\a$ or $\eta$ would have a very small effect in $T_g$\footnote{Indeed, it can be shown\supercite{cavagna_supercooled_2009} that if we increase the threshold by a certain factor, its effect in shifting $T_g$ to lower values is exponentially damped, at least.}. In specific, at the melting point typical values of the viscosity are\supercite{cavagna_supercooled_2009} $\eta(T_m) \sim 10^{-2}-10^{-3}$ Poise, which means that $\eta$ increases by 15 orders of magnitude in the range $T_m \gtrsim T \gtrsim T_g$. Importantly, usual values of the glass transition temperature are around\supercite{debenedetti_supercooled_2001} $T_g \approx 2T_m /3$. Therefore, it is clear that even if $T_g$ is not a unique nor precise temperature, it identifies the region where characteristic time scales change dramatically.

To continue, let me address the physical meaning of the different terms that appear in the VFT law \eqref{eq:vft}. In general, $D$ and $T_0$ can be used just as fitting parameters. So, for instance, the smaller the value of $D$ the more fragile the respective glass. Similarly, $T_0$ indicates a temperature at which a divergence in the relaxation time is expected. Despite the fact that it plays the role of a fitting parameter, several studies suggest that it might be a relevant thermodynamic quantity\supercite{angellThermodynamicConnectionFragility2001}. This connection stems from the fact that its value is very close\supercite{richertDynamicsGlassformingLiquids1998,berthier_biroli_theoretical_2011} to the Kauzmann temperature, $T_K$, which identifies a temperature at which the entropy of the glass (\textit{i.e.} a state with a liquid-like structure) becomes \emph{smaller} than the one of the crystal. A very counter-intuitive feature! Theoretically, the (supercooled) liquid's excess entropy with respect to the crystal, $S_{exc}(T) = S_{liq}(T)-S_{crys}(T)$ is considered and values of $T_K$ are obtained by extrapolating the excess entropy until it vanishes, \textit{i.e.} the condition $S_{exc}(T_K)=0$ identifies $T_K$. In practice however, this point can never be reached since $T_K < T_g$, implying that the glass transition always intervenes, making it impossible for the system to equilibrate and thus to measure $S_{exc}$. Nonetheless, the excess entropy is a very important quantity because it is closely related with the configurational entropy, which serves as a measure of the amount of local minima present in energy landscape of the system; see Fig.~\ref{fig:energy-landscape-glasses}. The so-called “Kauzmann paradox”\supercite{kauzmannNatureGlassyState1948} and the configurational entropy are very interesting topics that would deserve a section of their own, but length considerations dictate that I should omit them in this thesis. Fortunately, a good review is available in Ref.~\cite{berthierConfigurationalEntropyGlassforming2019}.

\begin{figure}[htb!]
	\centering
	\includegraphics[width=0.8\textwidth]{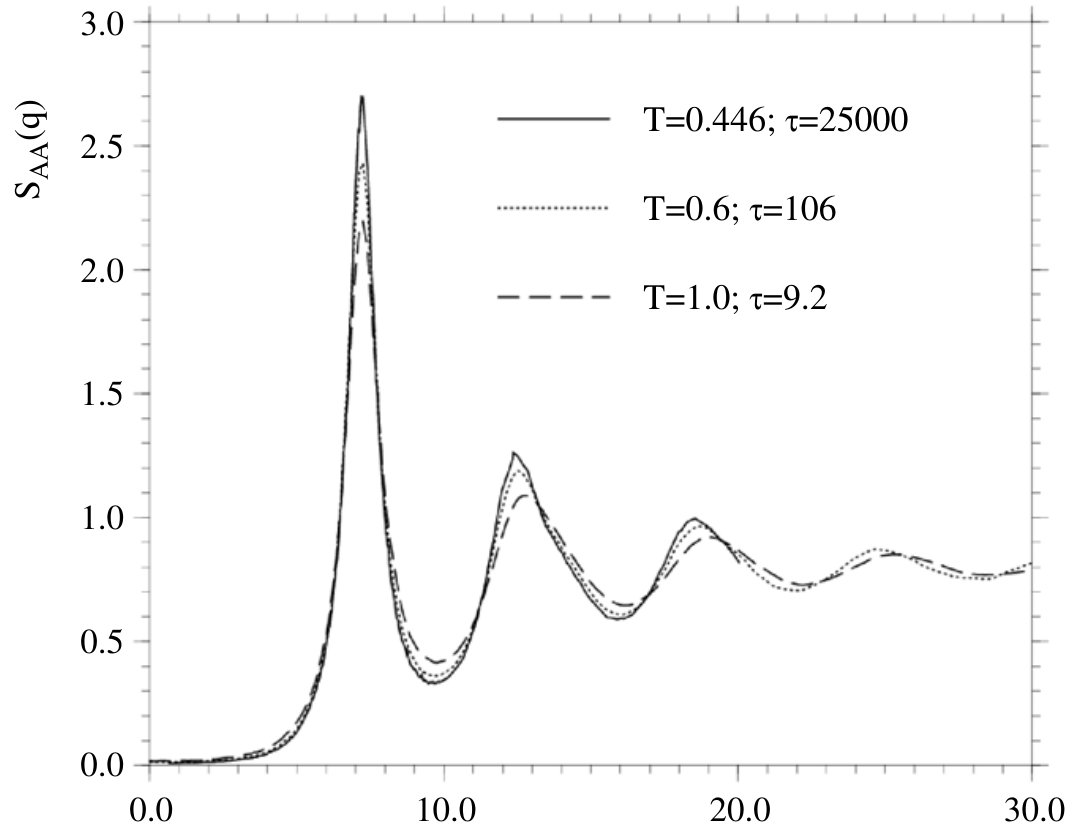}
	\caption[Structure factor in a LJ mixture at four different temperatures.]{(Partial) structure factor ($S$) of a Lennard-Jones mixture as a function of the  wave-vectors, $q$, and for different temperatures. $S(q)$ is defined in Eq.~\eqref{def:structure factor} below. The relaxation time at each temperature ($\tau$) is also indicated.
		Taken from \cite{kob_supercooled-and-glass-trans}.}
	\label{fig:structure-factor-diff-Ts}
\end{figure}

Now, given that the glass transition also marks an increase in the relaxation times of the system, it is only natural to investigate the effects on dynamical properties, instead of the static ones, as a system approaches $T_g$. For instance, Fig.~\ref{fig:structure-factor-diff-Ts} depicts the structure factor, $S(q)$, of a Lennard--Jones (LJ) fluid at different temperatures. $S(q)$ is defined in Eq.~\eqref{def:structure factor} below, but essentially it is the Fourier transform of the radial distribution function, $\tg(r)$ (see Eq.~\eqref{def:rdf}) and thus it provides information about spatial ordering present in the configuration. Data of the relaxation time (denoted in Fig.~\ref{fig:structure-factor-diff-Ts} by $\tau$) at each temperatures are also included. Note that $S_(q)$ is practically unchanged while $\tau$ increases by more than 3 orders of magnitude. This demonstrates quantitatively that, structurally, glasses are very similar to liquids.

\begin{figure}[h!]
	\begin{subfigure}{0.49\linewidth}
		\centering
		\includegraphics[width=\textwidth, height=5cm]{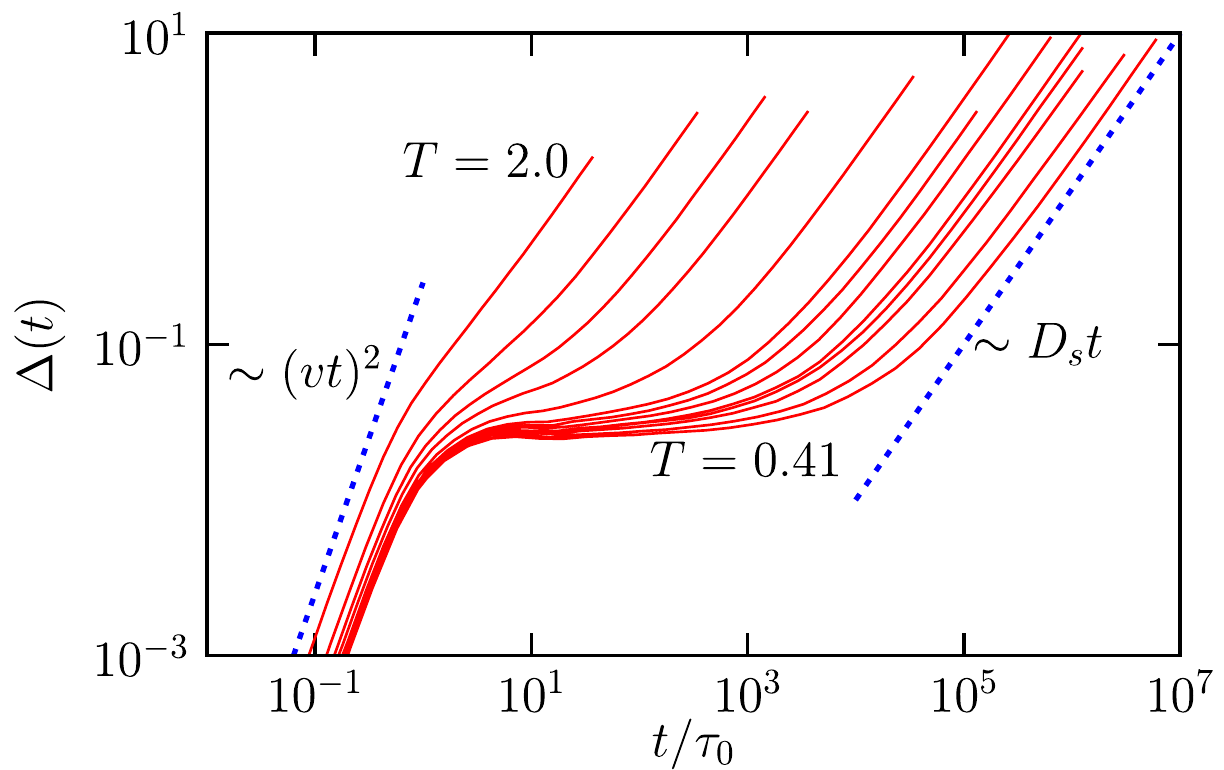}
		\caption{Time dependence of the MSD at different temperatures. Taken from \cite{berthier_biroli_theoretical_2011}.}
		\label{fig:two-steps-msd}
	\end{subfigure}
	\begin{subfigure}{0.49\linewidth}
		\centering
		\includegraphics[width=0.99\textwidth, height=5cm]{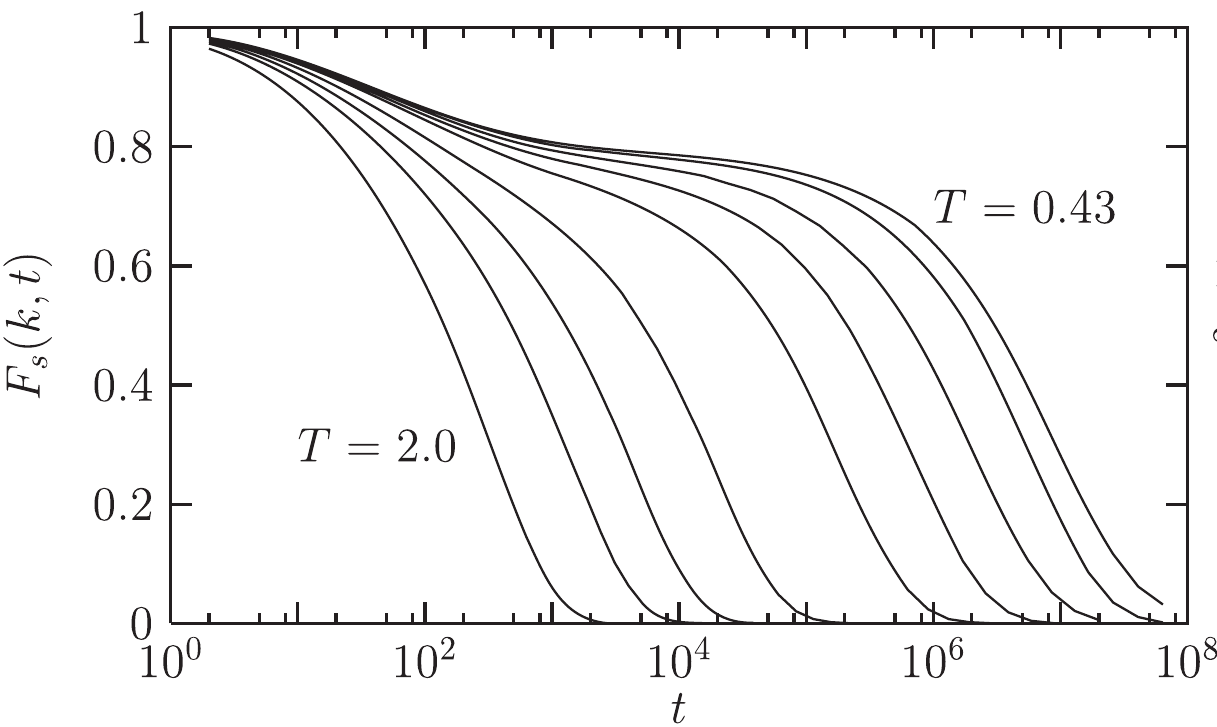}
		\caption{Self-intermediate scattering function at different temperatures. Taken from \cite{berthierMonteCarloDynamics2007}. Cf. with Fig.~\ref{fig:structure-factor-diff-Ts}.}
		\label{fig:two-steps-Fs}
	\end{subfigure}
\caption[Signatures of two-steps relaxations in the MSD and the self-intermediate scattering function.]{As $T\to T_g$, dynamical variables develop a plateau that separates their characteristic behaviours. Both plots show that the plateau lengthens as the temperature decreases. These curves are characteristic of the two-steps relaxation process, as discussed in the main text.}
\label{fig:two-step-relaxations}
\end{figure}

\begin{figure}[htb!]
	\centering
	\includegraphics[width=0.9\textwidth]{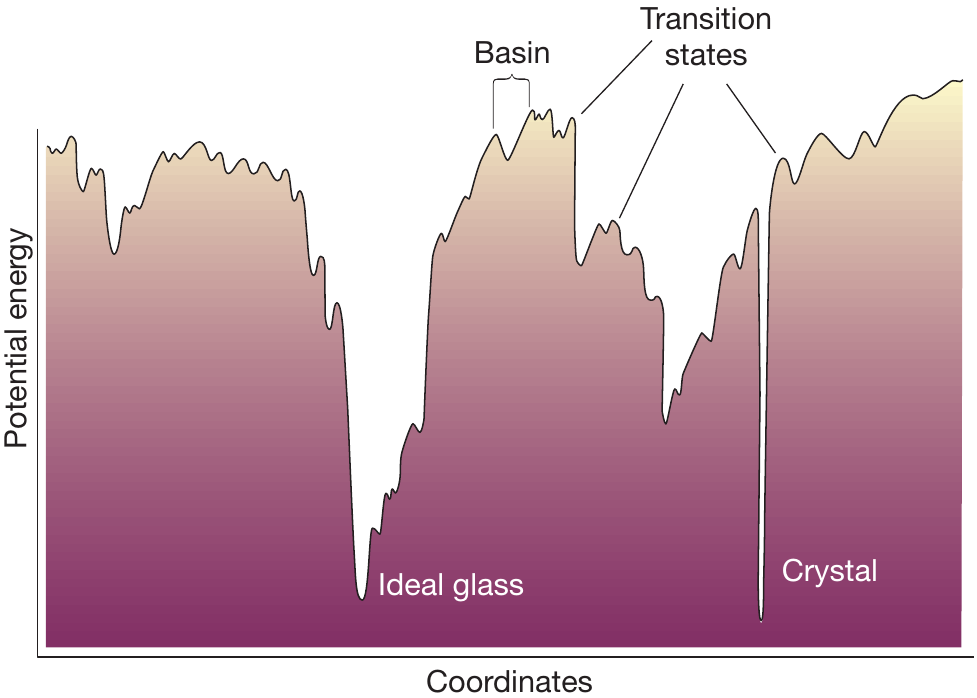}
	\caption[Sketch of the potential energy landscape of a glass former.]{Sketch of the potential energy landscape of a glass former. Different basins define metastable states in which the system can remain trapped for sufficiently low temperatures.
	Its very rough structure implies that a vast amount of (nearby) minima exist, separated by relatively small energy barriers. Of course, the global minimum corresponds to a crystalline state, and yet, the sampling is dominated by the local minima due to their abundance. The transition to an ideal glass state is not discussed in this work.
	 Taken from \cite{debenedetti_supercooled_2001}.}
	\label{fig:energy-landscape-glasses}
\end{figure}

In contrast, let us consider the mean square displacement (MSD),
\begin{equation}\label{def:msd}
\avg{\Delta (t) } = \avg{ \frac1N \sum_{i=1}^N \abs{ \vb{r}_i(t) - \vb{r}_i(0)}^2  } \qc
\end{equation}
where $\avg{\bullet}$ denotes thermal average. This variable will prove to be essential for many of the topics considered here, as shown later. In any case, in Fig.~\ref{fig:two-steps-msd} the behaviour of the MSD is reported, considering again a LJ mixture at different temperatures. We can see that at high temperatures (curves to the left), the initial ballistic behaviour --characterized by $\Delta(t) \simeq t^2$-- quickly leads to a diffusive one --where $\Delta(t) \simeq t$. This latter regime is characteristic of a liquid in equilibrium, or even metastable equilibrium, as in the supercooled phase. On the other hand, as $T$ decreases a plateau begins to form, separating these two different regimes of $\Delta$. Importantly this figure also shows that the lower the temperature the longer the plateau, but its height remains roughly constant. Other dynamical variables present very similar features. An important case are dynamical correlation functions, such as the intermediate scattering function $F(q,t)$. Just as $S(q)$ provides information about the \emph{static} density correlations at a length scale $\sim 1/q$, $F(q,t)$ measures the correlations of two points separated by the same length, but at different times. Its temperature dependence is shown in Fig.~\ref{fig:two-steps-Fs}, once again  for a LJ mixture. Note the marked slow down in the relaxation towards the (metastable) equilibrium state, identified by the lack of correlations. In fact, what these curves suggest is that as $T \to T_g$ relaxation bifurcates into two different processes, each of which has its own time scale.

The potential energy landscape\supercite{sciortinoPotentialEnergyLandscape2005} (PEL) picture, first introduced by Goldstein\supercite{goldsteinEnergyLandscape}, provides a fruitful scheme to understand these and several other properties of glassy systems. 
As sketched in Fig.~\ref{fig:energy-landscape-glasses}, the PEL is a very rough surface in the configuration space and its several local minima identify possible metastable states in which a liquid can be trapped forming a glass if the temperature decreases below $T_g$. 
From this perspective, when the system's temperature is high enough, it is able to explore the full landscape and therefore equilibrium can always be attained. As $T$ decreases, the system remains trapped for some relatively short time in a basin before it can escape and thus explore a larger region of the phase space. Then, as $T$ is further decreased the time spent trapped in a basin increases exponentially until,  at $T\lesssim T_g$, the system remains trapped in one of these local minima. At this point, ergodicity is broken and thus equilibration is impossible. Several clarifications are in place. First, in all this argument I blatantly ignored the crystalline state. So staying  in equilibrium should be understood with respect to the supercooled liquid state. It is equally important to stress that the mechanism by which a system escapes one basin and jumps into another, when $T>T_g$, is by \emph{very localized rearrangements of particles}. In other words, transitions between different local minima are accomplished by changing a sub-extensive number of coordinates\supercite{cavagna_supercooled_2009}. Hence, what in configuration space looks as jumping an energy barrier, in real space corresponds to small displacements of few nearby particles. Consequently, the vast majority of the system is virtually unaffected by the rearrangement of such cluster, and serves as a rather constant background. Therefore, changing $T$ does not impact the structure of the PEL itself, but instead influences how the system \emph{samples it}\supercite{debenedetti_supercooled_2001}.

With this in mind, let us see how the PEL helps to explain the plateaus of the MSD and $F(q,t)$ in Fig.~\ref{fig:two-step-relaxations}. The idea is that the initial and fast process (corresponding to the length of the plateaus) can be identified with relaxation \emph{within} a basin, as the system gets temporarily trapped, while the second and slower relaxation corresponds to the system escaping such basin and exploring the rest of the configuration space. These two different processes are customarily called $\beta$ and $\a$ relaxations, respectively. 
From this stand point we can better understand that the lower the $T$ the longer the plateau --since the system takes longer to escape-- but its height is roughly constant --because the process leading to a basin trapping is unaltered by $T$. Thus, the behaviour of $\Delta$ shown in Fig.~\ref{fig:two-steps-msd} can be explained in the following manner. After the initial ballistic regime has ended, the system has reached a minimum and remains trapped in a situation where the main source of relaxation are vibrations around such minimum, identified with the $\beta$ process. This coincides with the point where particles are \emph{caged} by their neighbours and remain mostly arrested, leading to the constant value of $\Delta$ in the plateau. And recall that because $T$ does not change the structure of the landscape, different minima are reached at roughly the same value of $\Delta$. 
Then, when enough time has passed, the system is able to leave the basin, with a concomitant relaxation process mostly driven by kinetic contributions, characteristic of the $\alpha$ branch. As expected, this also signals the point where $\Delta$ detaches from its plateau value and the diffusive regime takes over\footnote{It should be mention that the last part of the $\beta$ relaxation can coincide with the initial part of the $\alpha$ one, so there is no clear cut distinction between these two processes\supercite{kob_supercooled-and-glass-trans}.}. Naturally, when $T<T_g$ the $\alpha$ relaxation never takes place and $\Delta$ saturates at its constant plateau value even for very long times. See a similar description of the different regimes of $F(q,t)$ in \cite[Fig.~3]{kob_supercooled-and-glass-trans}. Therefore, the PEL offers a very intuitive and powerful perspective for the study of glassy systems: several of its predictions have been verified experimentally\supercite{debenedetti_supercooled_2001}, while it also provides crucial information of the decorrelation rate of $F(q,t)$. 

\begin{figure}[htb!]
	\centering
	\includegraphics[width=\linewidth]{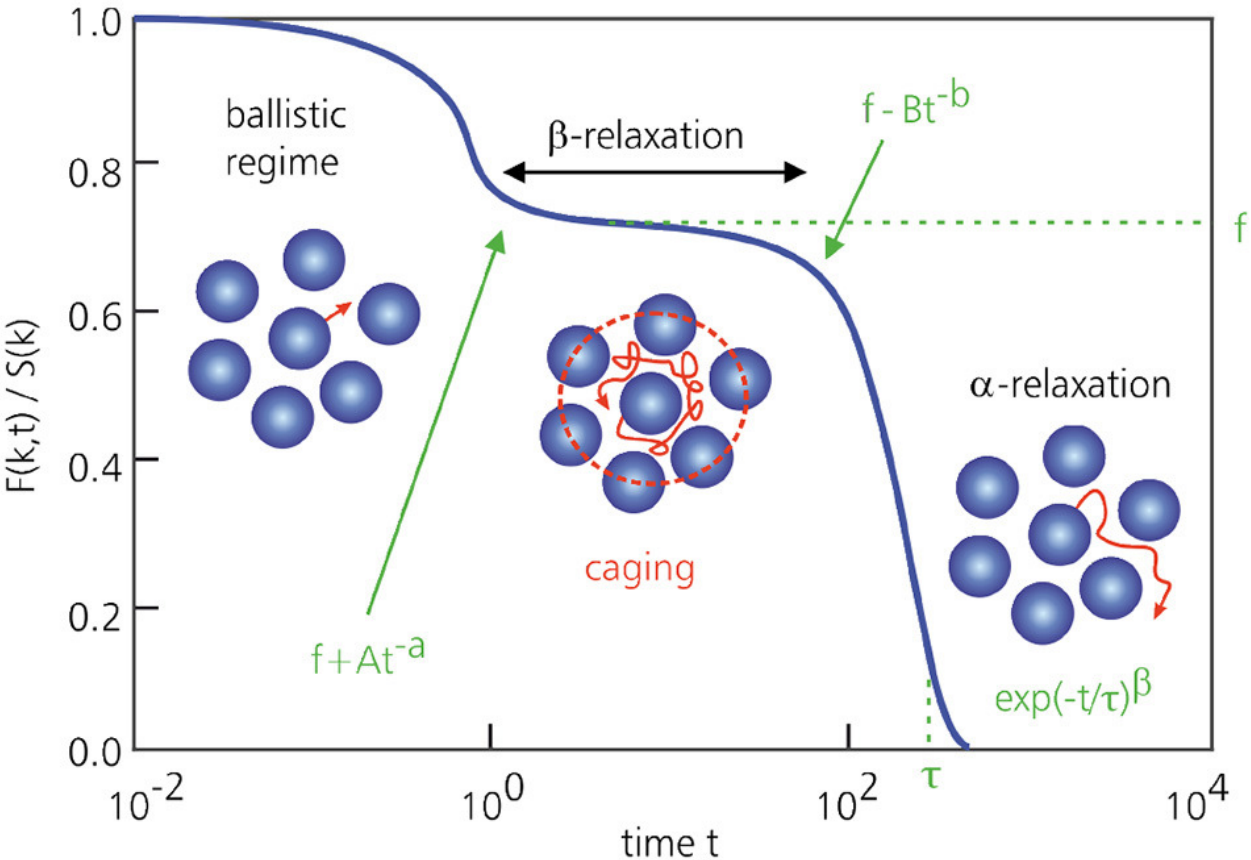}
	\caption[Prediction of typical behaviour of correlation functions according to MCT.]{Typical form of the intermediate scattering function, Eq.~\eqref{eq:Fq MCT}, as derived from MCT. Each regime is identified with a characteristic dynamical regime, as illustrated by the insets. See also Figs.~\ref{fig:two-step-relaxations} and  \ref{fig:4point-chi}.
	Taken from \cite{janssenModeCouplingTheoryGlass2018}.}
	\label{fig:F-MCT}
\end{figure}

Another important theoretical framework is the so called \textit{Mode Coupling Theory} (MCT). It is a very interesting and powerful scheme, but it certainly lies beyond the scope of this thesis. However, many good accounts are available, for instance: \cite{gotzeComplexDynamicsGlassForming2008,janssenModeCouplingTheoryGlass2018,kob_supercooled-and-glass-trans,reichmanModecouplingTheory2005}. Most of the phenomenology it predicts resembles closely the one obtained through the mean-field theory discussed in Sec.~\ref{sec:MF dynamics and glass transition}. Here, I will only mention that, a major success of MCT is being able to identify a “transition temperature”, $T_{MCT}$, at which the relaxation time diverges. Moreover, it successfully predicts the two steps relaxation of $F(q,t)$ by showing that its form is given by\supercite{janssenModeCouplingTheoryGlass2018}
\begin{equation}\label{eq:Fq MCT}
F(q,t) \sim f + At^{-a}  \, ; \qquad \text{and} \qquad F(q,t) \sim f - Bt^{-b} \, .
\end{equation}
Where the first relation holds as $F(q,t)$ approaches its plateau, and the second one as $F(q,t)$ decays from it. Sufficiently close to (but above) $T_{MCT}$, the two exponents $a$ and $b$ are related through
\begin{equation}\label{eq:critical dynamical exponents}
\frac{\Gamma(1-a)^2}{\Gamma(1-2a)} = \frac{\Gamma(1+b)^2}{\Gamma(1+2b)} \, . 
\end{equation}
This is an important relation to which I will return in a later section. The characteristic behaviour for $F(q,t)$ is sketched in Fig.~\ref{fig:F-MCT}. It also illustrates the connection of $F(q,t)$ with several important dynamical regimes discussed throughout this section. Notice the close resemblance with the intermediate scattering function of Fig.~\ref{fig:two-steps-Fs} obtained through simulations.

\begin{figure}[htb!]
	\centering
	\begin{subfigure}[c]{0.47\linewidth}
		\includegraphics[width=\textwidth]{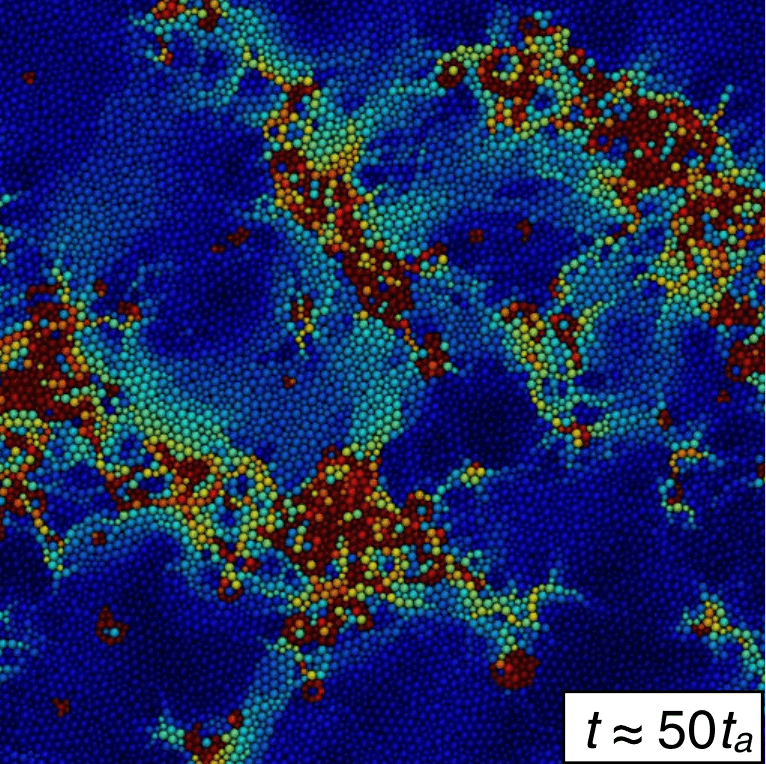}
		\caption{Spatial heterogeneity in the dynamics of particles. Colours differentiate mostly fixed particles (in blue) from the ones that have displaced at least one particle diameter (red). The time considered for computing the displacement is indicated, and is about a tenth of the relaxation time.
			Taken from \cite{heter-dyn-hierarchical-relaxation}; see this reference for a very illustrative video.}
		\label{fig:heter-dyn-coloured-particles}
	\end{subfigure} \hfill
	\begin{subfigure}[c]{0.52\linewidth}
		\includegraphics[width=\textwidth, height=5.5cm]{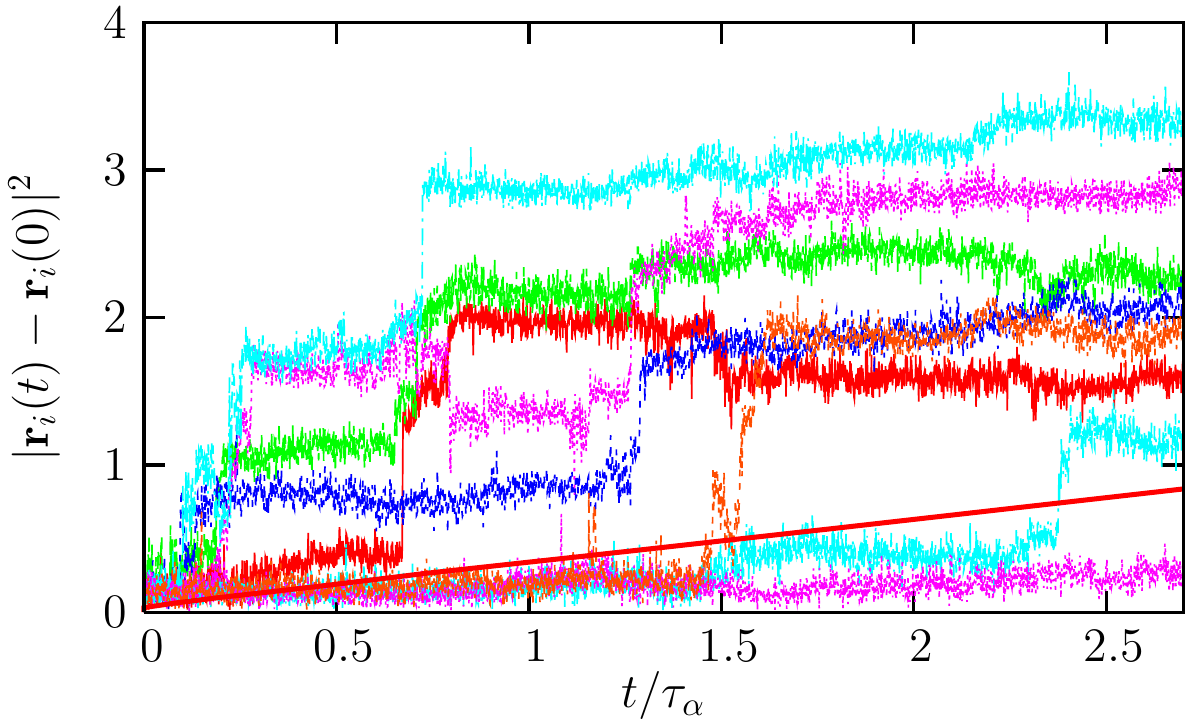}
		\caption{Square displacement of individual particles, and MSD (solid red straight line). Most of the time, particles vibrate around a fixed position, but eventually this behaviour is interrupted by long, quick jumps. Temporal heterogeneity is manifested by the fact that such jumps occur randomly and independently for each particle.
			Taken from \cite{berthierOverviewDifferentCharacterisations2010}.}
		\label{fig:heter-dyn-SD-individual-particles}
	\end{subfigure}
	\caption[Dynamical heterogeneity in glassy systems.]{Dynamical heterogeneity in glassy systems. Glass formers usually show both spatial (panel a) and temporal (panel b) heterogeneity. It is a salient feature of this type of systems, but its connection with local structure is far from understood. See Sec.~\ref{sec:dynamics-and-local-structure-glasses}.}
	\label{fig:heterogenous-dynamics}
\end{figure}

A final feature I would like to consider are the \emph{dynamical heterogeneities}\supercite{berthierDynamicalHeterogeneitiesGlasses2011} present in supercooled liquids and glasses alike. As the name suggests, it has been found that the dynamics in glassy systems are non-homogenous in space nor time. The first case is illustrated in Fig.~\ref{fig:heter-dyn-coloured-particles}, where particles are coloured according to their displacement at a given time, in such a way that highly mobile (mostly fixed) ones are shown in red (blue) tones. Heterogeneity is visually clear from the figure: particles able to travel a (relatively) large distance form clusters --wherein all particles have similar mobility-- that are surrounded by large portions of mostly immobile particles.  Nevertheless, at later times the picture can be very different because arrested particles can become mobile. Conversely, those whose motion was initially rather unimpeded can afterwards remain blocked by their neighbours. Intuitively, this is the main component of the temporal heterogeneity in particles' dynamics. Fig.~\ref{fig:heter-dyn-SD-individual-particles} illustrates this idea more clearly by plotting the square displacement of individual particles.
We can see that the particles' trajectories consist mostly of vibrational motion for long intervals, interrupted by large, spontaneous jumps. Moreover, not all the jumps occur at the same time, signalling the temporally heterogeneous character of the dynamics. This also means that the appearance and disappearance of domains of particles with similar mobility is not a synchronized process, but occurs randomly. 
It is worth noting that the abrupt jumps are not reflected in the MSD, as shown by the smooth straight red line in the same figure.

\begin{figure}[htb!]
	\centering
	\includegraphics[width=\linewidth]{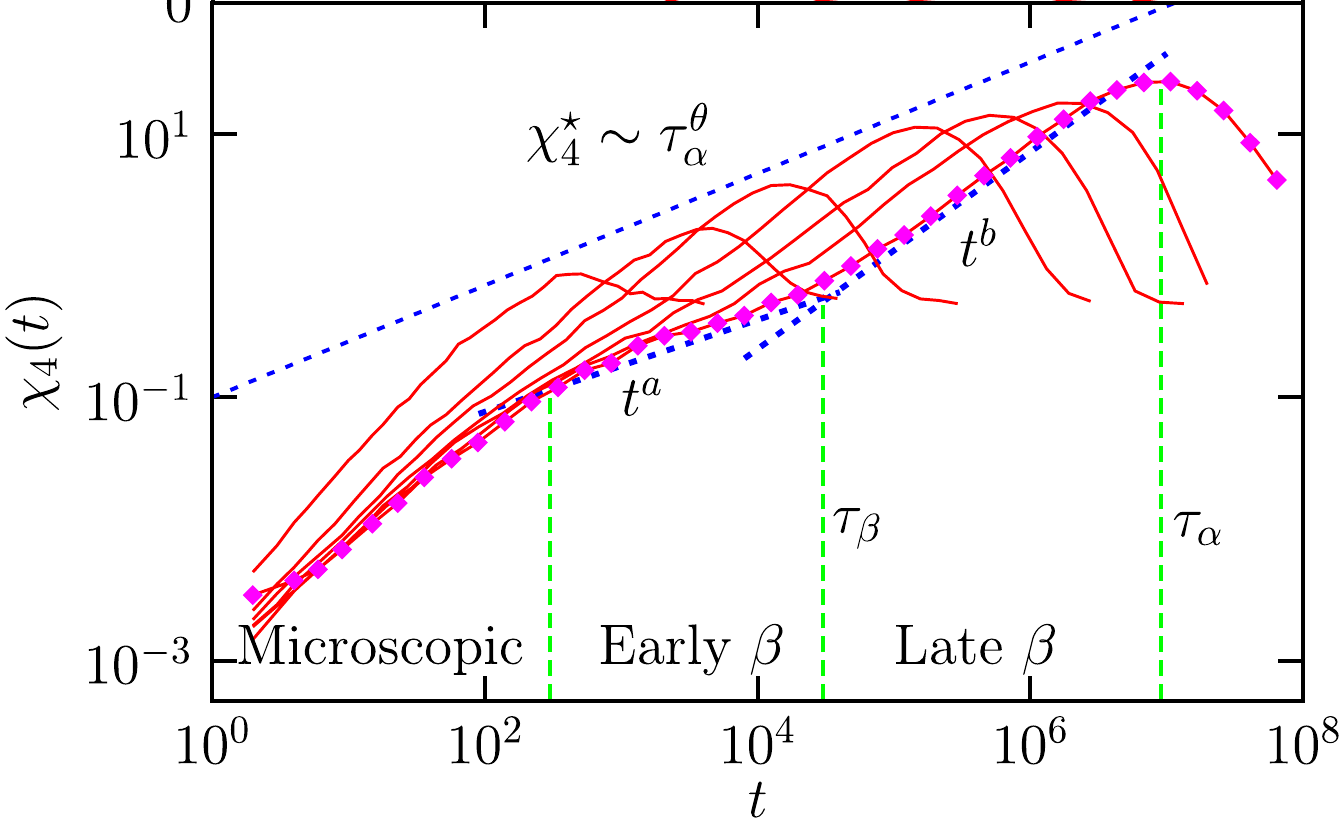}
	\caption[Criticality of the four point susceptibility.]{Four point susceptibility as a function of time in a LJ supercooled liquid. It is a very useful for studying dynamical heterogeneities in supercooled liquids. Different curves show that as temperature decreases (from left to right), mobility fluctuations grow in the system, as indicated by the larger values of $\chi_4$. In particular, its maximum, closely related to the relaxation time $\tau_\a$, is shifted to  larger times. The connection with the two steps relaxation, depicted in Fig.~\ref{fig:F-MCT}, is also indicated.
		 Taken from \cite{berthier_biroli_theoretical_2011}.}
	\label{fig:4point-chi}
\end{figure}

As peculiar as dynamical heterogeneity can be, its relevance is mainly due to its role in the study of dynamical correlation functions and dynamical susceptibilities. To understand the connection, consider a set of particles separated by a distance $r$ that are part of a high mobility cluster. If at later times the cluster transforms into a domain of mostly blocked particles, then the \emph{fluctuations} of their mobilities would be correlated in time, and on a scale $\sim r$. However, to properly identify this sort of correlations \emph{four point} correlation functions ($G_4(\vb{r},t)$) and susceptibilities ($\chi_4(t)$) are required. From the (very important) relation $\chi_4 (t) = \int \dd{\vb{r}} G_4(\vb{r},t)$, it is easy to see that a high value of the susceptibility signals mobility correlations on a long scale.
The typical behaviour of $\chi_4(t)$ at different temperatures (indicated by different curves) in a supercooled LJ mixture is shown in Fig.~\ref{fig:4point-chi}. For a fixed temperature, $\chi_4$ increases monotonically until it reaches a maximum $\chi_4^\star$ at $t=t^\star$, and then decays. 
This behaviour actually makes sense: considering that the initial regime of the dynamics is ballistic, the whole system lacks correlation, leading to small values of the susceptibility. As time passes, clusters of particles with similar mobility start to form and thus correlations begin to appear; the larger the clusters the higher the correlation. Yet, these clusters cannot be maintained forever, since the system, being in the supercooled phase, is bound to eventually relax to equilibrium. When this happens, the correlations are lost and consequently $\chi_4$ decreases. From this argument, it is natural to expect that $t^\star$ should be similar to the relaxation time, and this is indeed what happens\supercite{cavagna_supercooled_2009,berthier_biroli_theoretical_2011}. Moreover, as the temperature decreases, both $t^\star$ and $\chi_4^\star$ constantly increase, signalling the presence of higher, longer lived cluster correlations. 
Interestingly, Mode Coupling Theory predicts that $\chi_4$ should actually diverge at  $T_{MCT}$. Unfortunately, this temperature is rather high\supercite{cavagna_supercooled_2009,janssenModeCouplingTheoryGlass2018} when compared to $T_g$. 
At any rate, $\chi_4$ plays the role of the usual “two point“ susceptibility of critical phenomena, with its characteristic divergent behaviour at a phase transition. Within this framework, we are also able to identify the corresponding order parameter: the dynamical correlation function\footnote{Actually, a more precise statement is that the order parameter should be the \emph{long time limit} of such correlation function, see \cite[Sec.~7.8]{cavagna_supercooled_2009}}! What is more, a very similar scenario is found in a famous spin glass mean-field model: the $p$-spin model\supercite{kirkpatrickComparisonDynamicalTheories1988,franzRecipesMetastableStates1995,franzPhaseDiagramCoupled1997,franzNonlinearSusceptibilitySupercooled2000,donatiTheoryNonlinearSusceptibility2002}. There, the corresponding $\chi_4$ diverges as the system approaches its
\emph{dynamical} transition at $T_d$, the model's analogue of $T_{MCT}$.

These results support the hypothesis that by studying the behaviour of $\chi_4$ it is possible to detect the onset of “glassy behaviour”, characterized by the non-trivial dynamical correlations, whose length scales increase as the temperature is lowered. However, the analogies have to be taken with care since no phase transition occurs in real systems, as evinced by the fact that $\chi_4^\star$ just keeps on growing even \emph{below} $T_{MCT}$. Nevertheless, these theoretical considerations suggest that $T_{MCT}$ might be a temperature more physically meaningful than $T_g$, and that the putative divergence of $\chi_4$ is avoided possibly due to barrier crossing\supercite{cavagna_supercooled_2009}. Furthermore, just as in the usual case of critical phenomena, a correlation length\supercite{harrowellLengthScalesDynamic2010} ($\xi_4$) can be found that is expected to diverge at $T_{MCT}$, but that in supercooled liquids indicates the length scale of mobility correlations; thus it is essentially a measure of the size of the clusters of particles' with the same mobility. Moreover, as discussed in Sec.~\ref{sec:MF dynamics and glass transition}, mean-field theory provides an exact description that resembles closely the phenomenology of MCT and qualitatively agrees with the results observed empirically.

\subsection{Local structure and its connection with dynamics} \label{sec:dynamics-and-local-structure-glasses}

Now that we have a clearer picture of the intriguing dynamical phenomena that ensue near the glass transition, it is worth bringing back the main obstacle for a complete physical description of it: finding a connection between the complex dynamical features of supercooled liquids and glasses and their local structure. In other words, so far I have described the peculiar dynamical properties of supercooled liquids as well as some tools to unveil and analyse them. Yet, I have not mention how such properties can be explained in terms of structural (and hence \emph{static}) information of the system. The problem is notably intricate, so an entire sub-section seemed (just) enough to give an account of such pressing question. Nevertheless, I should anticipate that, maybe unsurprisingly, there is a lack of consensus on which are the correct variables to consider nor on their relative degree of success.

\begin{figure}[htb!]
	\begin{subfigure}[t]{0.4\linewidth}
	\includegraphics[width=\textwidth, height=5cm]{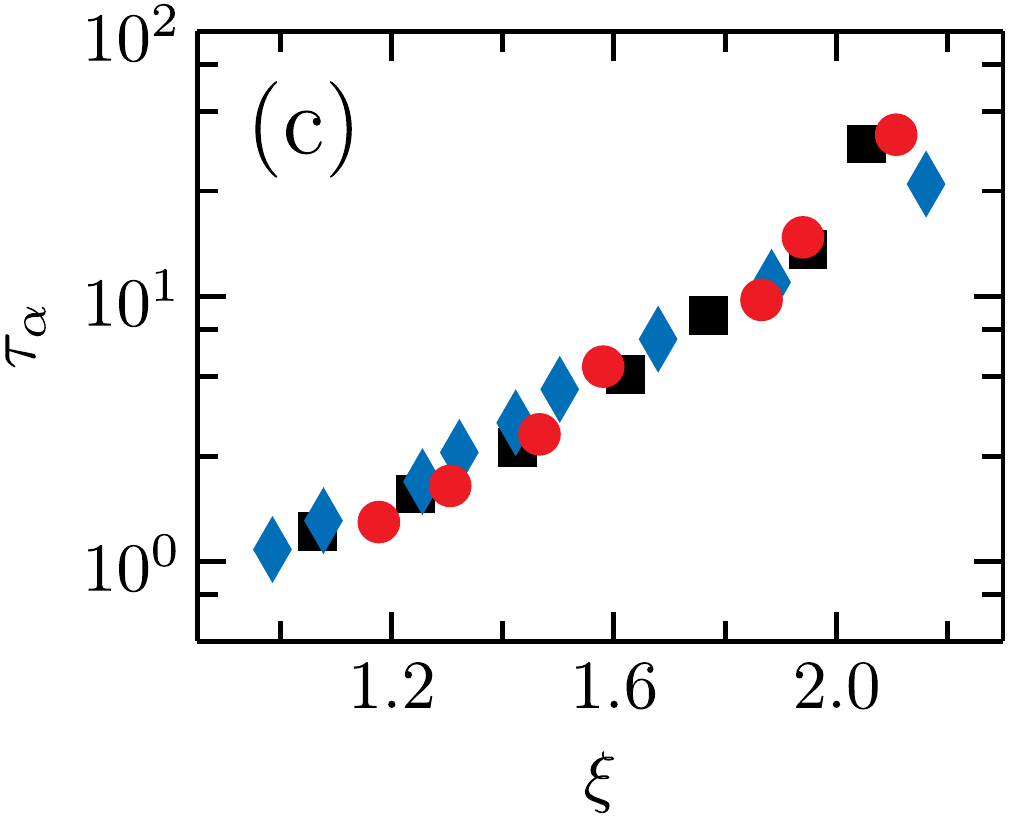}
	\caption{Correlation between the point-to-set correlation length $\xi$, and the relaxation time in different glass former models: LJ binary mixture (black squares), same mixture but truncated according to the Weeks-Chandler-Andersen model (blue diamonds), and inverse power law potential (red circles). Taken from \cite{hockyGrowingPointtoSetLength2012}.}
	\label{sfig:pts}
	\end{subfigure}\hfil
	\begin{subfigure}[t]{0.54\linewidth}
	\includegraphics[width=\textwidth, height=5cm]{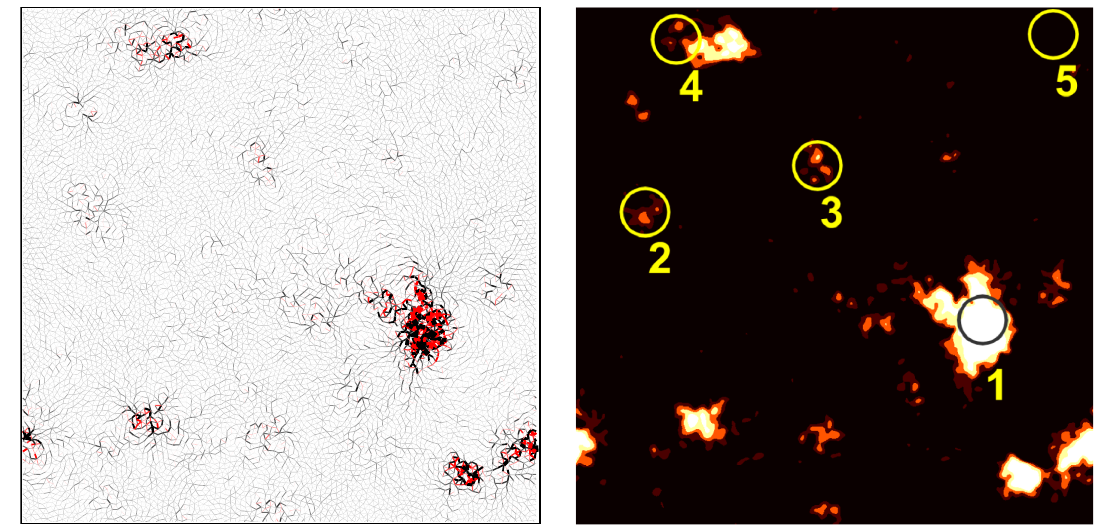}
	\caption{Left: Local thermal energy (LTE) in a 2D inverse power law glass former. Lines' thickness represent the magnitude of the LTE, with black (red) indicating a positive (negative) value. LTE incorporates anharmonic effects by construction. Right: Coarse-grained LTE field and its correlation with plastic rearrangements (identified by the numbered circles) caused by shear deformations.
	Taken from \cite{zylbergLocalThermalEnergy2017}.}
	\label{sfig:lte}
	\end{subfigure}\\
	\begin{subfigure}{\linewidth}
		\includegraphics[width=\textwidth]{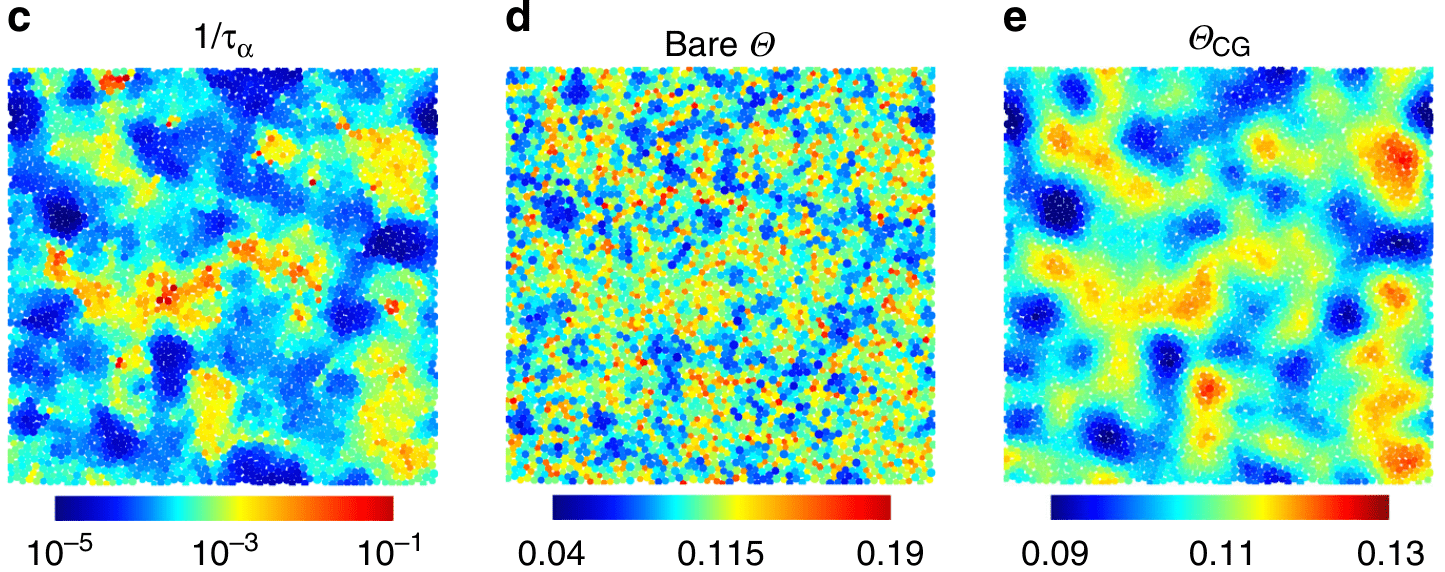}
		\caption{Left: Inverse relaxation time, at single particle level. Centre: Map of the ordered parameter obtained by measuring the deviation of the contact angles with respect to a locally favoured structure. Right: Map of the coarse-grained version of the same order parameter; notice the high correlation with the left-most panel. The system is a mixture of harmonic particles.
		Taken from \cite{tongStructuralOrderGenuine2019}.}
		\label{sfig:bond-orientation}
	\end{subfigure}
\caption{Different methods for connecting the local structure with dynamical features or plastic rearrangements in a variety of models.}
\label{fig:connection-structure-dynamics-other-methods}
\end{figure}

Ever since the discovery of dynamical heterogeneities people have been looking for connections with different structural variables\supercite{britoHeterogeneousDynamicsMarginal2007,berthier_biroli_theoretical_2011,berthierDynamicalHeterogeneitiesGlasses2011,candelierSpatiotemporalHierarchyRelaxation2010}.
Most of the works proceed in a similar fashion: the mobility of individual particles is measured by their squared displacement ($\abs{\delta \vb{r}_i(t)}^2 = \abs{\vb{r}_i(t) - \vb{r}_i(0)}^2$), and then interactions with their local environment are used to construct physical observables, expecting to find significant correlations between them. Of course, the crux of these studies consists in where such locality ends and which are the relevant physical quantities. In any case, with this approach, the influence of thermal fluctuations, shear deformations, etc. on particles rearrangements and plastic dislocations are studied.
The numerous set of structural quantities that have been proposed is a proof of how difficult the problem is. Relevant (and personally biased) examples include: point-to-set correlations\supercite{berthierStaticPointtosetCorrelations2012,hockyGrowingPointtoSetLength2012,charbonneauLinkingDynamicalHeterogeneity2016,karmakarLengthScalesGlassforming2016} (where a length scale over which particles are pinned by their neighbours is identified; see Fig.~\ref{sfig:pts}), local thermal energy\supercite{zylbergLocalThermalEnergy2017} (where the thermal energy of each interacting pair is used to predict plastic instabilities related to anharmonicity; see Fig.~\ref{sfig:lte}),  or plain geometrical parameters like bond orientations between nearby particles\supercite{tongRevealingHiddenStructural2018,tongRevealingInherentStructural2019,tongStructuralOrderGenuine2019} (Fig.~\ref{sfig:bond-orientation}). This list is far from exhaustive, but an insightful review can be found in \cite{royallReviewRoleLocalStructure2015}.
A fruitful approach uses the vibrational modes of the system\supercite{dynamic_criticality_jamming,henkesExtractingVibrationalModes2012,franzUniversalSpectrumNormal2015,rottlerPredictingPlasticitySoft2014}, and relates the density of states (DOS) with the collective dynamics of the system. This link is justified because the Fourier transform of the velocity autocorrelation function is closely related to the DOS\supercite{henkesExtractingVibrationalModes2012}. This relation, however, relies on the validity of an harmonic approximation for the interaction energy of the system, a scenario that in several cases might not be true. Furthermore, the vibrational spectra is also a relevant property of jammed and nearly jammed systems\supercite{silbertNormalModesModel2009,wyartEffectsCompressionVibrational2005,wyartGeometricOriginExcess2005,charbonneauUniversalNonDebyeScaling2016,arceriVibrationalPropertiesHard2020,wuResponseJammedPackings2017,franzImpactJammingCriticality2019}, so it will have an important role in this thesis; see Secs.~\ref{sec:network of contacts}, \ref{sec:normal modes} and Sec.~\ref{sec:comparison normal modes} in Chapter \ref{chp:inferring-dynamics}.

\begin{figure}[htb!]
	\centering
	\begin{subfigure}[t]{0.49\linewidth}
		\includegraphics[width=\textwidth]{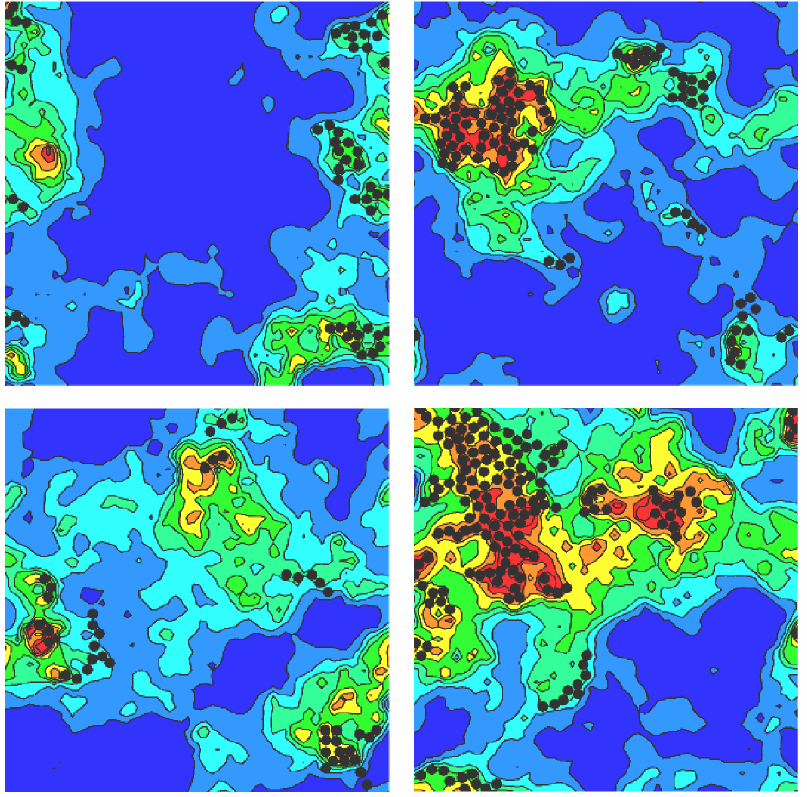}
		\caption{Colour map of the particles' mobility, showing in red (blue) the most (least) mobile particles, of four independent configurations. The filled circles are the particles with the highest Debye--Waller factor. Note however that this quantity is not exactly a structural one.
			Taken from	\cite{widmer-cooperPredictingLongTimeDynamic2006}.}
		\label{sfig:ice-dw-factor}
	\end{subfigure}
	\begin{subfigure}[t]{0.49\linewidth}
		\includegraphics[width=\textwidth]{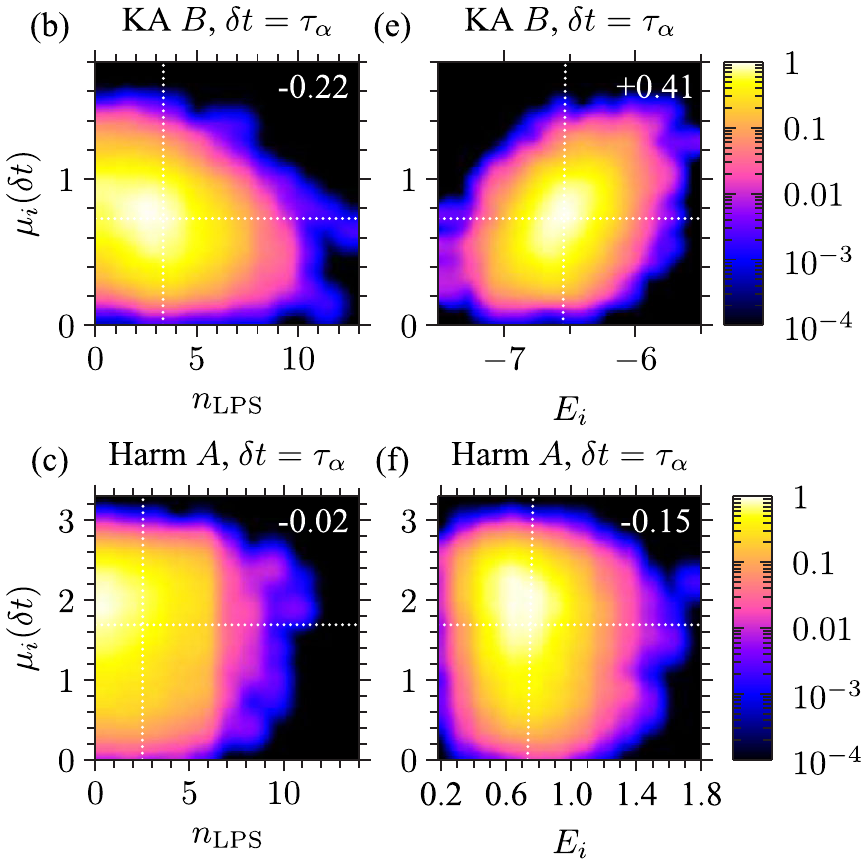}
		\caption{Histograms of the mobility of individual particles ($\mu_i$) as a function of the number of locally preferred structures (left column) and single particle energy (right). The upper row presents the results of a Kob--Andersen system, while the lower one is a binary mixture of harmonic spheres. On the top right of each panel the corresponding (Spearman) correlation is reported. Notice that its value is highly system dependent (see text). 
			Taken from \cite{hockyCorrelationLocalOrder2014}.}
		\label{sfig:ice-model-dependence}
	\end{subfigure}
\caption[Dynamics-structure correlation in the isoconfigurational ensemble.]{The usual approach in the isoconfigurational ensemble is to compute the particles' propensity, defined as the mobility averaged over several trajectories, and test whether it is correlated with physically sound (but arbitrary) structural variables.}
\label{fig:connection-structure-dynamics-ice}
\end{figure}

An alternative and rather more intuitive approach to the same problem can be adopted by using the so called “isoconfigurational ensemble”\supercite{widmer-cooperHowReproducibleAre2004,widmer-cooperRelationshipStructureDynamics2005,widmer-cooperStudyCollectiveDynamics2007,jackInformationTheoreticMeasurementsCoupling2014} (ICE) where the dynamics of a system is studied by simulating several trajectories, departing from the same initial configuration (hence the name). For instance, if molecular dynamics (MD) simulations are employed, the particles' initial position is fixed, while the velocities are randomly assigned at each run according to the Maxwell-Boltzmann distribution with a given temperature.
Clearly, this  method provides a way of sampling the space of all possible trajectories of the configuration and, in doing so, it allows to identify how the local environment of a particle influences its average mobility. 
Its only basic assumptions are that (i) the thermal noise is washed out when a large enough number of MD realizations are used; and (ii) that trajectories are self-averaging. The main advantage of the ICE is that regions where clusters of particles undergo major rearrangements during their dynamics can be easily identified. Moreover, such regions can be related to structural properties such as the Debye--Waller factor or the local energy density\supercite{widmer-cooperIrreversibleReorganizationSupercooled2008,widmer-cooperPredictingLongTimeDynamic2006}; see Fig.~\ref{fig:connection-structure-dynamics-ice}. Recently, information-theoretic methods have been used to assess more accurately the role of these and other structural quantities\supercite{jackInformationTheoreticMeasurementsCoupling2014}, as well as how reproducible the observed trajectories are. However, it has also been pointed out that providing a link between dynamical and static properties using exclusively the particles' mobility might be inadequate\supercite{berthierStructureDynamicsGlass2007}. Moreover, the main problem of the ICE approach is that even if a structural variable shows a high correlation with the dynamics in a given system, it may lead to poor predictions in a different model\supercite{hockyCorrelationLocalOrder2014}. In other words, sampling from the ICE leads to structure-based inferences whose quality might be very sensitive to type of glass former model employed.
These features will be discussed in more detail in Chp.~\ref{chp:inferring-dynamics}, where  I will show that the ICE can also be used to study the dynamics of amorphous solids near their jamming point and, notably, overcome several of the issues just mentioned.

\begin{figure}[htb!]
	\centering
	\begin{subfigure}[c]{0.43\linewidth}
		\includegraphics[width=\textwidth]{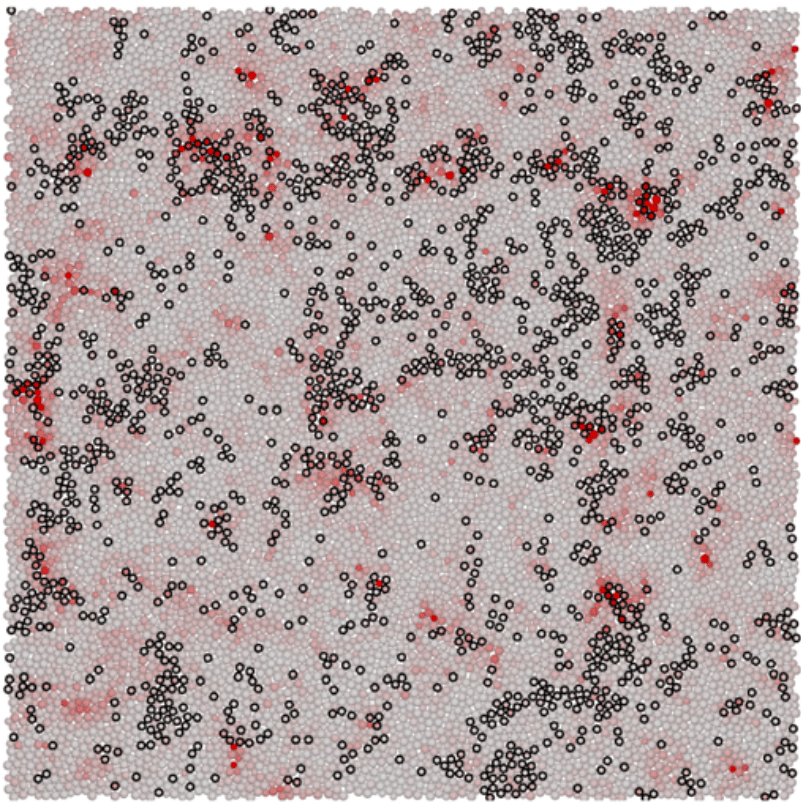}
		\caption{A LJ sheared system, where particles are coloured from grey to red according to their displacement, with the latter being the ones moving the most. The particles identified by the SVM algorithm as most prone to move are highlighted in black.
			Taken from \cite{cubukIdentifyingStructuralFlow2015}.}
		\label{sfig:svm-prediction}
	\end{subfigure}
	\begin{subfigure}[c]{0.56\linewidth}
		\includegraphics[width=\textwidth]{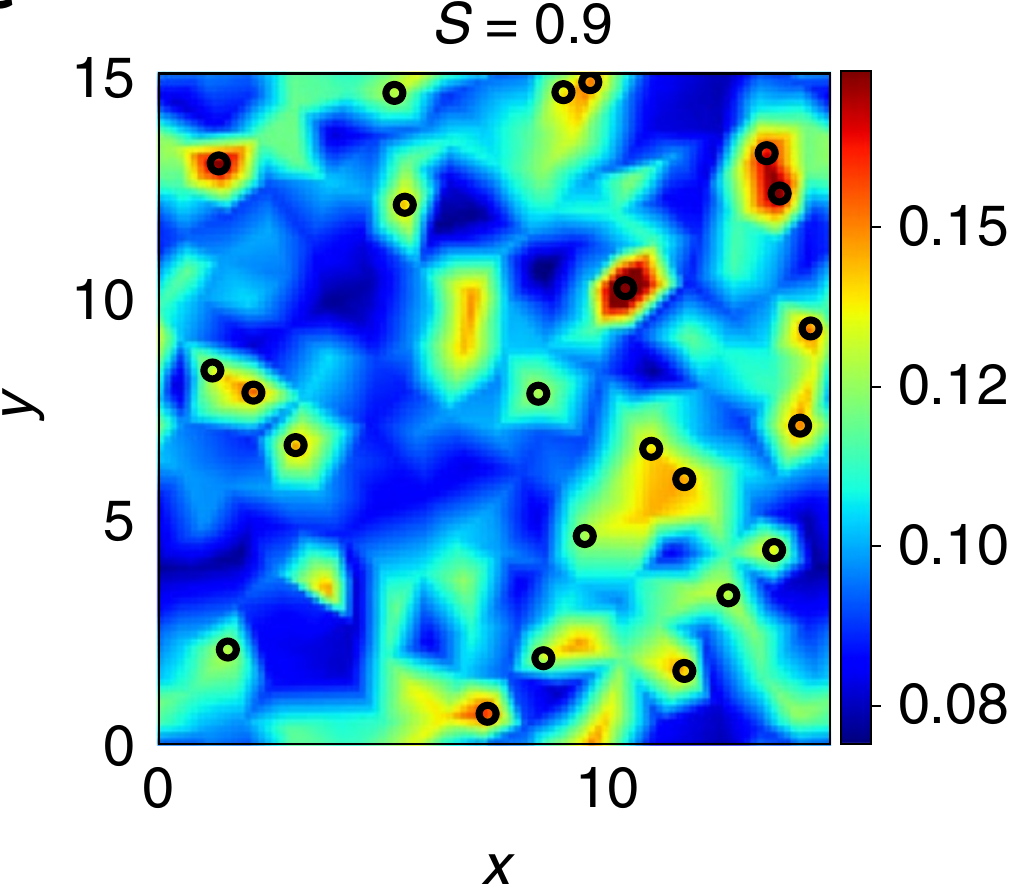}
		\caption{Colour map of the particles' mobility and its comparison with the top $10\%$ most mobile particles (black circles) as predicted by a graph neural network.
			Taken from \cite{bapstUnveilingPredictivePower2020}.}
		\label{sfig:gnn-prediction}
	\end{subfigure}
\caption[Machine Learning methods attain a high predictability of dynamical features, but use artificial representations of the particles' local environment.]{Machine Learning methods yield high quality predictions of particles rearrangements, but they rely on artificial representations of the particles' local environment.}
\end{figure}

On the other hand, a major step forward has been recently achieved by employing Machine Learning methods\supercite{cubukIdentifyingStructuralFlow2015,schoenholzCombiningMachineLearning2018}. In these works, the information of a particle's local structure is encoded in a new variable termed ``softness'' that is then related to the displacement of such particle in a given time interval. A Support Vector Machine is used to find a hyperplane (in features space) dividing the mostly movable particles from the mostly arrested ones. Then, the softness is computed as the signed distance of each particle's features to such an hyperplane, thus ``soft'' particles are prone to be displaced by a significant amount while ``hard'' ones will remain mostly fixed (see Fig.~\ref{sfig:svm-prediction}). It has been shown that softness is strongly correlated with physical quantities such as local energy and coordination number\supercite{cubukStructuralPropertiesDefects2016}, as well as being a useful variable for modelling the Arrhenius behaviour observed in supercooled liquids\supercite{schoenholzRelationshipLocalStructure2017,schoenholzCombiningMachineLearning2018,schoenholzStructuralApproachRelaxation2016}. Yet, it performs better at identifying mobile particles than these other structural variables. Furthermore, new results using Graph Neural Networks\supercite{bapstUnveilingPredictivePower2020} have shown to outperform several physically motivated variables in inferring (and predicting) dynamical properties of glassy systems (Fig.~\ref{sfig:gnn-prediction}). However, these methods are based on parametrizing a particle's local environment in terms of (artificial) feature functions, yielding a synthetic representation of the structure. Therefore, even though Machine or Deep Learning techniques yield high quality predictions about which particles are the likeliest to be highly mobile, their associated structural variables lack a clear physical meaning. In short, even if Machine Learning methods can be used to construct good predictors of particles' mobility, they fail to provide an answer about which are the real physical variables that determine such mobility.

As a final remark, note that for several of the methods described here, additionally to the structural variable considered, a coarse-graining procedure is needed to obtain better inferences; \textit{e.g.} Figs.~\ref{sfig:lte}, \ref{sfig:bond-orientation}, \ref{sfig:ice-dw-factor}, \ref{sfig:gnn-prediction}. Likewise, in the case of the SVM, even though predictions for individual particles can be obtained, the features functions are constructed essentially as weighted radial distribution functions. Hence, structural information of a small region is compressed into a single number, and can thus be consider as another instance of coarse-graining. In contrast, in Chp.~\ref{chp:inferring-dynamics} I will show that the method we developed for inferring the statistics of the particles' dynamics works using only information of individual and well identified particles.

To close this introductory section I should emphasise that there are many interesting topics that I have left completely uncover such as: ageing\supercite{biroliCrashCourseAgeing2005}, spin glasses\supercite{castellaniSpinglassTheoryPedestrians2005,zamponiMeanFieldTheory2010,mezardSpinGlassTheory1986}, complexity and configurational entropy\supercite{berthierConfigurationalEntropyGlassforming2019}, the ideal glass transition\supercite{royallRaceBottomApproaching2018}, kinetically constrained models\supercite{ritortGlassyDynamicsKinetically2003}, and a long etcetera.
%

%
%
%
%

\section{Hard sphere systems: From liquid to jamming with a glassy interlude} \label{sec:hs-fluids}

\begin{figure}[htb!]
	\centering
	\includegraphics[width=0.9\linewidth]{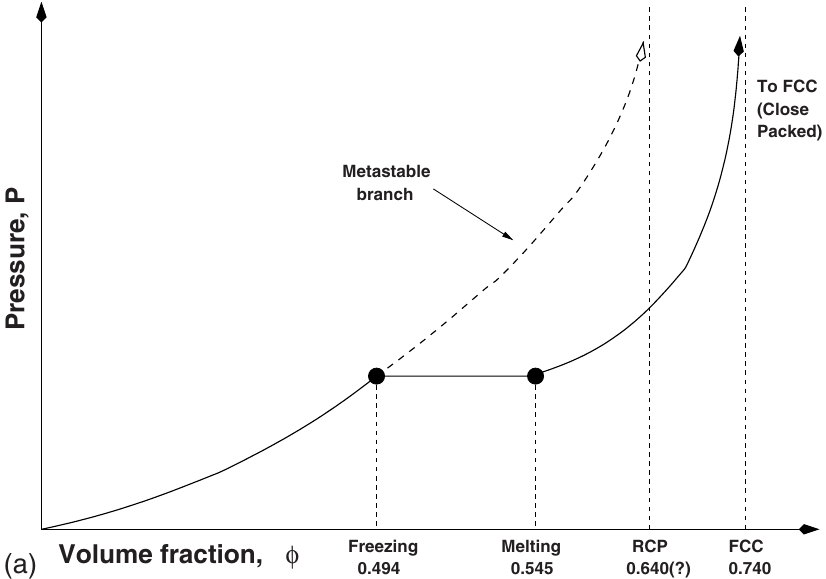}
	\caption[Phase diagram of hard sphere fluids: liquid and solid phase]{Phase diagram of monodisperse HS fluids. The solid black line is the equilibrium branch, while the dashed one indicates the metastable liquid (and glass) phase, beyond the freezing point (first dotted vertical line). The second vertical line marks the melting point, while the last two indicate the maximal possible density of the glass and solid phases, respectively. As discussed in the text, the former is associated to the so called \textit{random close packing}, and the latter to an FCC crystal.
		Taken from \cite{parisi_zamponi_2010}.}
	\label{fig:HS-sklog-wiki}
\end{figure}

In this section, I will quickly review some of the basic properties of simple hard sphere (HS) fluids, based mostly on Refs.~\cite{hansenTheorySimpleLiquids2013,santosStructuralThermodynamicProperties2020,puz_book}. That is, systems composed of frictionless and infinitely rigid spherical particles. Hence, no interaction is present between the spheres except when they come into contact with one another and an elastic collision occurs. Notwithstanding its simplicity, HS liquids display a non-trivial behaviour giving rise to a liquid-solid phase transition. This is summarized in Fig.~\ref{fig:HS-sklog-wiki} in the pressure-density plane for a monodisperse configuration. The solid line represents the stable equilibrium branch, showing that the liquid freezes at a density\footnote{Throughout this thesis, the terms density and packing fraction will be used interchangeably and will be denoted by $\vp$.} $\vp_f\approx 0.494$, while the solid melts at $\vp_m \approx 0.545$. In other words, for densities larger than $\vp_m$ an equilibrated HS system is in a crystalline phase, which ends when the spheres occupy the maximum possible volume.
As conjectured by Kepler and proved much later\supercite{halesProofKeplerConjecture2005}, this happens when the spheres are packed following a face-centred cubic (FCC) structure and each particle is in contact with its closest neighbours. The associated density is $\vp_{FCC}= \frac{\pi}{3\sqrt{2}} \approx 0.740 $, which thus acts as an upper limit for the density in monodisperse HS systems. Fig.~\ref{fig:HS-sklog-wiki} also shows (dashed curve) the metastable liquid branch, which has been found to finish at the so called “random close packing” density, $\vp_{RCP} \approx 0.64$. I will discuss at length what happens \emph{near} and \emph{at} this point in the following sections and chapters, but for the time being let me just mention that configurations with $\vp_{FCC}$ and $\vp_{RCP}$ are two examples of jammed packings. On the other hand, the presence of the metastable branch for $\vp>\vp_f$ suggests that HS liquids could also be used as glass-formers. And indeed a glass transition, taking place around\supercite{parisi_zamponi_2010,santosStructuralThermodynamicProperties2020} $0.58 \lesssim \vp_g \lesssim 0.62$, has been found. 
Here, two points are worth considering. First, note that for HS fluids the melting, glass, and other transitions are specified in terms of the density instead of the temperature, in contrast with the systems considered in the previous section. The reason is that in HS systems, the interaction energy is either zero or infinity, and therefore the temperature only acts as a scaling parameter of the different thermodynamic variables; see \textit{e.g.} Eqs.~\eqref{eq:P-rdf-hs}, \eqref{def:reduced P},  \eqref{eq:p virial series} below. 
I will argue in the rest of this section that HS liquids exhibit many of the complex phenomenology described above for usual supercooled liquids and glasses. But to make a proper comparison, it is useful to map the role of specific volume and temperature (in standard glass formers) to the pressure and density (in HS systems).

The second point worth mentioning is the rather unexpected crystallization at high densities in HS liquids. Indeed, how come there is a phase transition if no attractive force is present? Moreover, because the potential energy of HS fluids is always zero, the energetic contribution to their free energy is only due to a trivial kinetic term. This means that for large packing fractions the entropy of the crystal becomes \emph{larger} than the liquid's one. This very counterintuitive result brings us back to the Kauzmann paradox mentioned above. 
But, mainly, it shows that there is no reason \emph{a priori} to assume that the ordered, crystalline phase has a lower entropy than the inherently disordered liquid\supercite{berthier_biroli_theoretical_2011}. This sort of “entropic ordering” can be understood by considering that the entropy has a contribution associated with the (logarithm of the) accessible volume per particle\supercite{frenkelOrderEntropy2015}. The idea is that, while arranging the particles following a regular structure certainly reduces the entropy due to orientational degrees of freedom, it may also increase the part related to translational degrees of freedom. For high enough densities, the gain in the latter exceeds the loss caused by a reduced number of possible orientations and therefore (at least partial) crystallization occurs. In other words, the only way in which particles can benefit from a larger accessible volume, and thus allow the system to further increase its density, is by forming crystalline domains. 
Historically, the case of HS is very important because the first “empirical” signature of the freezing transition was obtained in the debut of molecular dynamics\supercite{alderPhaseTransitionHard1957} and Monte Carlo\supercite{woodPreliminaryResultsRecalculation1957} simulations as computational techniques to study the thermodynamics of many body systems. The case of hard disks followed soon afterwards\supercite{alderPhaseTransitionElastic1962}. As will be discussed in Chapter \ref{chp:lp-algorithm}, the simple potential of HS systems makes their dynamics trivial, rendering them specially amenable for the computational capabilities available in the late 1950's. The details of the story are fascinating --see \textit{e.g.} \cite{history_hs_MolDyn,history_hs_MonteCarlo,hooverHardSpheresCubes2020}-- starring an episode where several scientific personalities pondered about the freezing phase transition and “decided” \emph{by voting} whether it really occurred or not. I cannot help finding Sisyphean that, so many years later, we are still studying HS systems, with ever faster and more powerful computers and techniques, and yet marvelling by the many striking physics hiding underneath.

%

\subsection{Basic Statistical Mechanics of liquids}\label{sec:stat-mech-liquids}

Before proceeding in the analysis of HS systems, it is convenient to develop some general results of liquids theory, which is that I will do next. Later, in Sec.~\ref{sec:hs-stat-mech}, I will particularize such results to HS liquids and show that many simplifications are possible due to their peculiar potential, see Eq.~\eqref{def:hs-potential}. Additionally, this section is important because I will introduce some notation that will be used throughout this work. First of all, $N$ will denote the number of particles (or system's size), $\vb{r}_i$ will be used to indicate the centre's position of the $i$-th particle, while $\va{r} = \{\vb{r}_i\}_{i=1}^N$ will be the $3N$-dimensional vector that specifies the location of the system in configuration space. Analogously, let $m_i$ and $\vb{p}_i$ be the mass and momentum vector of the $i$-th particle and $\va{p}$ the $3N$-dimensional vector of momenta. Clearly, the same notation is applicable when dealing with systems in dimensions higher than $d=3$, and I will do so when needed.
Hence, the state of the system in phase space is specified in general by $\va{z}\equiv(\va{r},\va{p})$.

To continue, I will assume that particles interact via a (sufficiently well-behaved)\footnote{What is meant by “sufficiently well behaved”, of course depends on which properties we are interested in. For instance, it is clear that the HS interaction, Eq.~\eqref{def:hs-potential}, defines a potential that is \emph{not} continuous, yet it is suitable to model the properties of HS systems in the thermodynamic limit. For a more detailed discussion see \cite[Sec.~2.1.1]{puz_book} and references therein.}  potential function, $u(r)$, that only depends on the distance between pairs of particles, $r_{ij}\equiv \abs{\vb{r}_i-\vb{r}_j}$. We can thus define the Hamiltonian as
\begin{equation}\label{def:hamiltonian}
\H(\va{z}) = \sum_{i=1}^N \frac{\abs{\vb{p}_i}^2}{2m_i} + U(\va{r}) = \sum_{i=1}^N \frac{\abs{\vb{p}_i}^2}{2m_i} +
\frac12 \sum_{i\neq j}^{1,N} u(r_{ij}) \, .
\end{equation}
%
I will focus on deriving the equation of state (EOS) of the fluid in the canonical ensemble, at a fixed temperature $T=1/\beta$ and volume $V$. The partition function is given by
\begin{equation}\label{eq:partition-function}
Q_N(\beta, V) = \frac{1}{N!\ h^{3N}} \int \dd{\va{z}} e^{-\beta \H(\va{z})} = \frac{Z_N(\beta,V)}{N! \Lambda^{3N}} \qc
\end{equation}
where $h$ is the Planck's constant and $\Lambda = \sqrt{\beta h^2/m}$ is the De Broglie wavelength, that appears when the Gaussian integral of each component of $\va{p}$ is performed. On the other hand, $Z_N$ defines the configurational integral
\begin{equation}\label{def:config-integral}
Z_N(\beta, V) = \int \dd{\va{r}} \exp(-\beta U(\va{r})) \, .
\end{equation}
Note that the dependence of $Z_N$ on $V$ is implicitly contained in the integration limits of $\int \dd{\va{r}}$.

From the partition function \eqref{eq:partition-function}, we obtain the Helmholtz free energy:
\begin{equation}\label{eq:helmholtz F}
F(N,V,T) = -T \log Q_N(T,V) = 3T N \log \Lambda - T \log(\frac{Z_N(\beta, V)}{N!}) \, .
\end{equation}
Obviously, if for a given potential $u(r)$ we were able to compute the integral \eqref{def:config-integral} exactly and thus obtain a close form for $F$, all the thermodynamic variables would follow easily. For instance, the average energy and pressure are obtained simply by differentiating the free energy:
\begin{subequations}
	\begin{align}
	\avg{ E} & = \pdv{(\beta F)}{\beta}\qc \\
	P & = - \pdv{F}{V}\, . \label{seq:P-derivative-F}
	\end{align}
\end{subequations}
Unfortunately, no closed expression is known for $Z_N$ for the HS potential, let alone a general one as I am considering for the moment. Nevertheless, it is still possible to obtain some important results and thus gain some insight of the physical properties such as the variables just introduced. To do so, let us first note that in the thermodynamic limit, $N\to \infty$ and $V\to \infty$, the (intensive) free energy becomes a function of the temperature and the \emph{number density}, $\rho \equiv N/V$. Thus, letting
\[
\tf(\rho,T) \equiv \lim_{N,V \to \infty} \frac{F(N,V,T)}{N}\qc
\]
the free energy can be conveniently decomposed into two independent parts: (i) a term related to the energy of the system as if it was an ideal gas; and (ii) an “excess” contribution taking into account the interaction between particles. That is\supercite{hansenTheorySimpleLiquids2013,puz_book},
\begin{subequations}\label{eq:intensive free energy}
	\begin{align}
	\tf(\rho, T) & = \tf^{(\text{id})}(\rho, T) + \tf^{(\text{ex})}(\rho, T)\\
	\intertext{where}
	\tf^{(\text{id})}(\rho, T) & = 3 T \log \Lambda - T (1-\log \rho)\\
	\tf^{(\text{ex})}(\rho, T) & = -T \lim_{N,V \to \infty, \rho=N/V} \frac1N \log(\frac{Z_N}{^N}) \, .
	\end{align}
\end{subequations}
Note that this implies that in the thermodynamic limit the pressure is simply given by\supercite{puz_book}
%
\begin{equation}\label{eq:P-derivative-f}
P = \rho^2 \pdv{\tf(\rho,T)}{\rho} \, .
\end{equation}
This seems a rather unimpressive rewriting of Eq.~\eqref{seq:P-derivative-F}, but it makes clear that, for a constant $T$ value, the pressure can be written only in terms of $\rho$, for instance, as a power series:
\begin{equation}\label{eq:virial expansion}
\beta P = \rho + \sum_{n=2}^N B_n(T) \rho^n \, .
\end{equation}
This is the famous \emph{Virial Expansion}. The first term is easily obtained from the contribution of ${\tf}^{(\text{id})}$ to the total free energy. Therefore, terms of order 2 and higher in $\rho$ are due to “deviations” from the ideal gas caused by the interaction between particles.

As I will argue next, working in terms of densities brings several other advantages; see also Sec.~\ref{sec:MF dynamics and glass transition}. So let me spend some time working out some of the most important results. First of all, given the Hamiltonian of Eq.~\eqref{def:hamiltonian}, we can compute the probability of finding a given configuration $\va{r}$ as
\[
\varrho(\va{r}) = \frac1{N! h^{3N} Q_N} \int \dd{\va{p}} e^{-\beta \H(\va{r}, \va{p})} = \frac{1}{Z_N} e^{-\beta U(\va{r})} \, .
\]
From this distribution and the indistinguishability of particles it follows that the so called \emph{n-particle density}, is given by
\begin{equation}\label{eq:n-particle density}
\varrho_N^{(n)} \qty(\va{r}^{(n)}) = \frac{N!}{(N-n)! Z_N} \int \dd{\va{r}^{(N-n)}} \exp(-\beta U(\va{r}) )\qc
\end{equation}
where $\va{r}^{(n)}$ is the $3n$-dimensional vector obtained from considering the components of $\{\vb{r}_i \}_{i=1}^n$, while the integral of $\dd \va{r}^{(N-n)}$ is performed over the remaining coordinates, \textit{i.e.} $\{\vb{r}_i\}_{i=n+1}^N$. This type of functions determine the probability of finding $n$ particles in the volume element $\dd{\va{r}^{(n)} }$ independently of the positions of the rest of the particles and the configuration's momenta. From the definition of $Z_N$, Eq.~\eqref{def:config-integral}, it is easy to obtain that
\[
\int \dd{\va{r}^{(n)} } \varrho_N^{(n)}\qty(\va{r}^{(n)}) = \frac{N!}{(N-n)!} \qc
\]
and therefore, for $n=1$,
\[
\int \dd{\vb{r}} \varrho_N^{(1)} (\vb{r} ) = N\, .
\]
It then follows that, for a uniform system the single-particle density coincides with the number density, $\varrho_N^{(1)}(\vb{r}) = \rho$.

A closely related set of functions are the \emph{n-particle distributions}, $\tg_N^{(n)} \qty(\va{r}^{(n)} )$, defined in terms of the n-particle density as
\begin{equation}\label{eq:n-particle function}
\tg_N^{(n)} \qty(\va{r}^{(n)} ) = \dfrac{\varrho_N^{(n)} \qty(\va{r}^{(n)})}{\prod_{i=1}^n \varrho_N^{(1)}(\vb{r}_i)  } \, .
\end{equation}
I will only be interested in homogeneous systems and thus, this last equation reduces to $ \rho^n \tg_N^{(n)} \qty(\va{r}^{(n)} )  = \varrho_N^{(n)} \qty(\va{r}^{(n)})$. By far, the most important of these functions is the pair distribution function, $\tg_N^{(2)}(\vb{r}_1, \vb{r}_2)$. It is amply used to distinguish between gas, liquid and solid phases, but it provides a lot of information about the structure of a system as I will show next.
For isotropic systems, as I will consider here, $\tg_N^{(2)} (\vb{r}_1, \vb{r}_2) = \tg_N^{(2)} (\vb{r}_1 -\vb{r}_2)$, and it is called \emph{radial distribution function} (RDF). Furthermore, for central potentials the RDF only depends on the norm of its argument and will be simply denoted $\tg(r)$. From the identity,
\[
\avg{\delta(\vb{r}-\vb{r}_1)} = \frac1{Z_N} \int \dd{\va{r}} \delta(\vb{r}-\vb{r}_1) e^{-\beta U(\va{r})}
= \frac1{Z_N} \int \dd{\va{r}^{(N-1)}} \exp(-\beta U(\vb{r},\vb{r}_2, \dots, \vb{r}_N) )
\]
it follows easily\supercite{hansenTheorySimpleLiquids2013} that $\varrho_N^{(1)}(\vb{r}) =  \avg{\delta (\vb{r}-\vb{r}_i)}$. An analogous relation for $\varrho_N^{(2)}(\vb{r}_1, \vb{r}_2)$ leads to a convenient representation for $\tg(\vb{r})$ in homogeneous systems:
\begin{equation}\label{def:rdf}
\tg(\vb{r}) = \frac{1}{N \rho} \avg{ \sum_{i\neq j}^{1,N} \delta(\vb{r}-\vb{r}_j + \vb{r}_i)} \quad
\implies \tg(r) =
\frac{1}{4\pi N r^2 \rho} \avg{ \sum_{i\neq j}^{1,N} \delta(r-r_{ij})} \, ;
\end{equation}
where the rightmost equation follows if the system is isotropic.

The RDF allows to compute several structural properties of interest. For instance, defining the \emph{total correlation function}, $h(\vb{r})=\tg(\vb{r})-1$, the static structure factor can be obtained by computing its Fourier transform:
\begin{equation}\label{def:structure factor}
S(\vb{q}) \equiv  1 + \rho \int \dd{\vb{r}} h(\vb{r}) e^{-i \vb{q}\cdot \vb{r}} = 1 + \rho h(\vb{q}) \, .
\end{equation}
As mentioned in the previous section, $S(\vb{q})$ measures the density fluctuation in a scale of order $\sim 1/\abs{\vb{q}}$ and is a very relevant quantity since it can be measured accurately in experiments. Moreover, it can be directly linked to response functions, such as the isothermal compressibility, $\kappa \equiv \qty(-V \pdv{P}{V})^{-1}$\supercite{santosStructuralThermodynamicProperties2020}:
\begin{equation}\label{eq:compress-structure-factor}
\frac{\rho \kappa }{\beta} = 1 + 4\pi \rho \int \dd{r} r^2 h(r) = S(0).
\end{equation}
Even more importantly for our purposes, $\tg(r)$ is related to the average energy and pressure through the following equations:
\begin{subequations}\label{eqs:E-and-P-through-rdf}
	\begin{align}
	\avg{E} & = \frac32 N T + 2\pi N \rho \int \dd{r} u(r) \tg(r) \qc  \label{eq:E-by-rdf}\\
	\frac{\beta P}{\rho} & = 1 - \frac{2\pi \beta \rho}{3} \int \dd{r} r^3 u'(r) \tg(r) \, . \label{eq:P-by-rdf}
	\end{align}
\end{subequations}
Comparing expression \eqref{eq:P-by-rdf} with the virial expansion, Eq.~\eqref{eq:virial expansion}, suggests at a first sight that only the $\rho^2$ term is needed in order to compute the pressure of the liquid. Unfortunately, this is not the case because $\tg(r)$ itself depends on the density and on $\varrho_N(2)$. Thus, $\rho$ is implicitly contained in the integrand of Eq.~\eqref{eq:P-by-rdf}.

Finally, another important quantity is the \textit{direct correlation function}, $c(r)$, defined implicitly through the integral equation\supercite{hansenTheorySimpleLiquids2013,santosStructuralThermodynamicProperties2020}
\begin{equation}\label{def:dcf}
h(r) = c(r) + \rho \int \dd{\vb{r}'} c(\abs{\vb{r}-\vb{r}'}) h(r') \, ,
\end{equation}
called Ornstein--Zernike relation. Intuitively, it states that the total correlation between a pair of particles is made of two contributions: first, the direct correlation between both particles; and, in second place, an indirect contribution caused by correlations with the rest of the particles as intermediaries. Note however that, even though this equation defines $c(r)$, it is not possible to obtain a closed expression for it. Nevertheless, the Fourier transforms of $h(r)$ and $c(r)$ are related through simple, algebraic expressions:
\[
\hat{h}(q) = \frac{\hat{c}(q)}{1-\rho \hat{c}(q)}\, .
\]
Whence another expression for the compressibility is readily obtained,
\begin{equation}\label{eq:compress-dcf}
\frac{\rho \kappa }{\beta} = \frac{1}{1-\hat{c}(0)}\, .
\end{equation}

The set of equations \eqref{eq:compress-structure-factor}, \eqref{eqs:E-and-P-through-rdf}, \eqref{eq:compress-dcf}, together with the Ornstein--Zernike relation Eq.~\ref{def:dcf}, are important because they provide direct routes to compute the EOS, \emph{provided we knew the exact form of} $\tg(r)$ --or $h(r)$, $c(r)$ or even their Fourier transforms for that matter. Unfortunately, except for the ideal gas, this is never the case and several approximations are needed to make any progress. Nevertheless, in HS systems the situation is more tractable and many important results are readily available as discussed next.

\subsection{Equation of state of HS liquids}\label{sec:hs-stat-mech}

The HS interaction is modelled through a contact, pair-wise potential of the following form:
\begin{equation}\label{def:hs-potential}
u_{HS}(x; a) = \begin{dcases}
0 & x> a \qc \\
\infty & x \leq a \, .
\end{dcases}
\end{equation}
Naturally, if $\vec{\sigma}=\{\sigma_i\}_{i=1}^N$ is the set of diameters, $a=\frac{\sigma_i + \sigma_j}{2}$ is a parameter equal to the sum of a pair's radii. However, throughout this part I will assume that all particles are identical (thus $\sigma_i=\sigma$) and  henceforth omit reference to the parameter $a$ unless it is necessary. Next, given that a particle's volume is equal to $v=\pi \sigma^3/6$ and that spheres never overlap, the system's density or packing fraction is simply
\begin{equation}\label{def:packing fraction}
\vp = N v /V = \rho v \qc
\end{equation}
where $V$ is the system's volume (usually a cubic box of size $L$), while $\rho$ is the number density as above.

Next, notice that for $u_{HS}$ the integral of Eq.~\eqref{eq:E-by-rdf} identically vanishes, whence $\avg{E} = \frac32 N T$. This makes sense since HS liquids do not have any potential energy and therefore $\avg{E}$ is purely kinetic. Similarly, Eq.~\eqref{eq:P-by-rdf} can be considerably simplified. Introducing the \textit{cavity distribution function}\supercite{hansenTheorySimpleLiquids2013} $y(r)= \tg(r)e^{\beta u_{HS}(r)}$, and a function $w(r)=e^{-\beta u_{HS}(r)}$ the integrand can be written as
\[
u'_{HS}(r) \tg(r) = -\frac1\beta w'(r) y(r) = - \frac1\beta y(r) \delta(r-\sigma)\, .
\]
The last equation follows from the fact that $w(r)=1$ when $r>\sigma$ and vanishes otherwise; \textit{i.e.} it is the step function $w(r) = \Theta(r-\sigma)$. It is important to mention that $y(r)$ is a continuous function even if the potential is not\supercite{hansenTheorySimpleLiquids2013,santosStructuralThermodynamicProperties2020}, and therefore the integral is well defined. The $\delta$-function makes the integration trivial thus obtaining
\begin{equation}\label{eq:P-rdf-hs}
\frac{\beta P}{\rho} = 1 + 4 \vp \tg^{+}\, ; \qquad \tg^+ \equiv \lim_{r\to \sigma^+} \tg(r) \, .
\end{equation}
This means that the full equation of state for HS systems can be derived solely from the value of $\tg(r)$ at the contact distance. Although such value is not known exactly, many good approximations have been devised\supercite{santosStructuralThermodynamicProperties2020} with ever greater accuracy. Before continuing, it is worth recalling that in HS liquids temperature only acts as a scale parameter, fixing the average kinetic energy. For instance, consider the configurational integral:
\[
Z_N = \int \dd{\va{r}} e^{-\beta U_{HS}(\va{r})}
= \int \dd{\va{r}} \prod_{i<j}^{1,N} w(r_{ij})
= \int \dd{\va{r}} \prod_{i<j}^{1,N}  \Theta\qty(r_{ij}-\sigma)
\, .
\]
This last expression clearly shows that its value is independent of $T$. For convenience I will henceforth adopt a “temperature free” description for HS systems. Therefore, many of the quantities I will describe next are assumed to be scaled by $T$ and $\rho$ in order to make them dimensionless. The most relevant example is the \textit{reduced pressure},
\begin{equation}\label{def:reduced P}
p \equiv \frac{\beta P}{\rho} \qc
\end{equation}
which will appear often.

Now, as mentioned above, were we to know exactly either $\tg(r)$ or $c(r)$\footnote{The path I have been following in fact suggests that $c(r)$ is a secondary function, defined in terms of $h(r)=\tg(r)-1$, and therefore implying that the RDF is fundamental function to consider. However, as argued in \cite[Chp.~3]{hansenTheorySimpleLiquids2013} $c(r)$ is more properly defined as a functional derivative of $F$ with respect of $\varrho_N^{(1)}(\vb{r})$, and thus independent of $\tg(r)$. Similarly, it can also be shown that the RDF is obtained from the functional derivative of $F$ with respect to $v$.} the EOS would follow easily. But these two correlation functions are actually coupled through the Ornstein--Zernike relation. To overcome this additional obstacle and obtain a closed integral equation approximations are unavoidable. The idea is to write $c(r)$ as a functional of $h(r)$, $\mathcal{C}[h(r)]$. One of the most famous of such expressions is the Percus--Yevick (PY) closure relation\supercite{santosStructuralThermodynamicProperties2020},
\begin{equation}\label{eq:Percus-yevick dcr}
\mathcal{C}[h(r)] = \qty(1- h(r)) \qty(1 - e^{\beta v(r)})\, .
\end{equation}
Importantly, for HS systems the integral equation derived from the PY closure can be solved exactly and, for instance, obtained a simple expression for $\tg^+$:
\[
\tg^+ = \dfrac{1 + \frac12 \vp}{(1-\vp)^2}\, .
\]
Plugging it into Eq.~\eqref{eq:P-rdf-hs} we obtain a (long sought) EOS:
\begin{equation}\label{eq:eos PY virial}
p_{PY} = \frac{1+2\vp +3 \vp^3}{(1-\vp)^2} \, .
\end{equation}
Unfortunately, if we proceeded by first computing the compressibility, say, via Eq.~\eqref{eq:compress-structure-factor} or \eqref{eq:compress-dcf} , a \emph{different} EOS is obtained\supercite{santosStructuralThermodynamicProperties2020}:
\begin{equation}\label{eq:eos PY compress}
p_{PY,\kappa} = \frac{1+\vp + \vp^2}{(1-\vp)^3} \, .
\end{equation}
This rather disappointing result is termed \textit{thermodynamic consistency problem}. It is related to the different \emph{routes}\supercite{santosStructuralThermodynamicProperties2020} in which an EOS can be obtained. In general, Eqs.~\eqref{eqs:E-and-P-through-rdf} are called the energy and virial routes, while Eq.~\eqref{eq:compress-structure-factor} corresponds to the compressibility route. Even when there are other types, these are the most common ones. 
What causes the thermodynamic consistency problem is that once a given approximation for the RDF (or the other related quantities such as $S(q)$) is chosen, the free energy obtained when tracing back each of these routes is different. This is evinced, for instance, by noting that according to Eq.~\eqref{eq:P-derivative-f} we can recover $\tf$ by integrating with respect to $\vp$ the EOS for the pressure. But given that the available EOS's are different (cf. Eqs.~\eqref{eq:eos PY virial}, \eqref{eq:eos PY compress}, or \eqref{eq:eos CS} below), we would obtain a \emph{different} expression for the free energy depending on which expression for $p$ we choose.


On the other hand, let us consider the EOS resulting from the virial expansion, Eq.~\eqref{eq:virial expansion}. As with other quantities in HS systems, the virial coefficients $\{B_n\}_{n=2}^\infty$ are independent of the temperature. Moreover, values of $B_2, B_3,$ and $B_4$ are known exactly in $d=3$ and higher dimensions\supercite{puz_book}, while numerical estimates are known up to $B_{12}$ for spherical particles\supercite{santosStructuralThermodynamicProperties2020}. When the first few terms are included, the following series is obtained for the reduced pressure\supercite{hansenTheorySimpleLiquids2013}:
\begin{equation}\label{eq:p virial series}
p = 1 + 4\vp +10 \vp^2 + 18.365 \vp^3 + 28.224 \vp^4 + 39.82 \vp^5 + 53.34 \vp^6 \dots
\end{equation}
Surprisingly, a very good approximation is obtained if the virial coefficients are restricted to integer values. This was first proposed by Carnahan and Starling (CS) in \cite{carnahanEquationStateNonattracting1969}, by noting that the first 2 coefficients are integers, while the next few ones can be well approximated by $B_n = n^3 + 3n$. Plugging this expression into Eq.~\eqref{eq:virial expansion} results in a power series of $\vp$ that can be easily summed, resulting in a fairly simple formula:
\begin{equation}\label{eq:eos CS}
p_{CS} = \dfrac{1+\vp + \vp^2 - \vp^3}{(1-\vp)^3} \, .
\end{equation}
Although $p_{CS}$ can also be obtained from a combination of $p_{PY}$ and $p_{PY,\kappa}$, its merit relies on its notable accuracy (compared with numerical simulations) in a wide range of densities. Although it tends to underestimate the pressure, it does so by a small amount (about $3\%$)\supercite{hansenTheorySimpleLiquids2013}. And while some other rather simple approximations for the EOS exist\supercite{roblesNoteEquationState2014}, the CS expression is accurate enough for the topics that I will consider.

\begin{figure}[htb!]
	\centering
	\includegraphics[width=\linewidth]{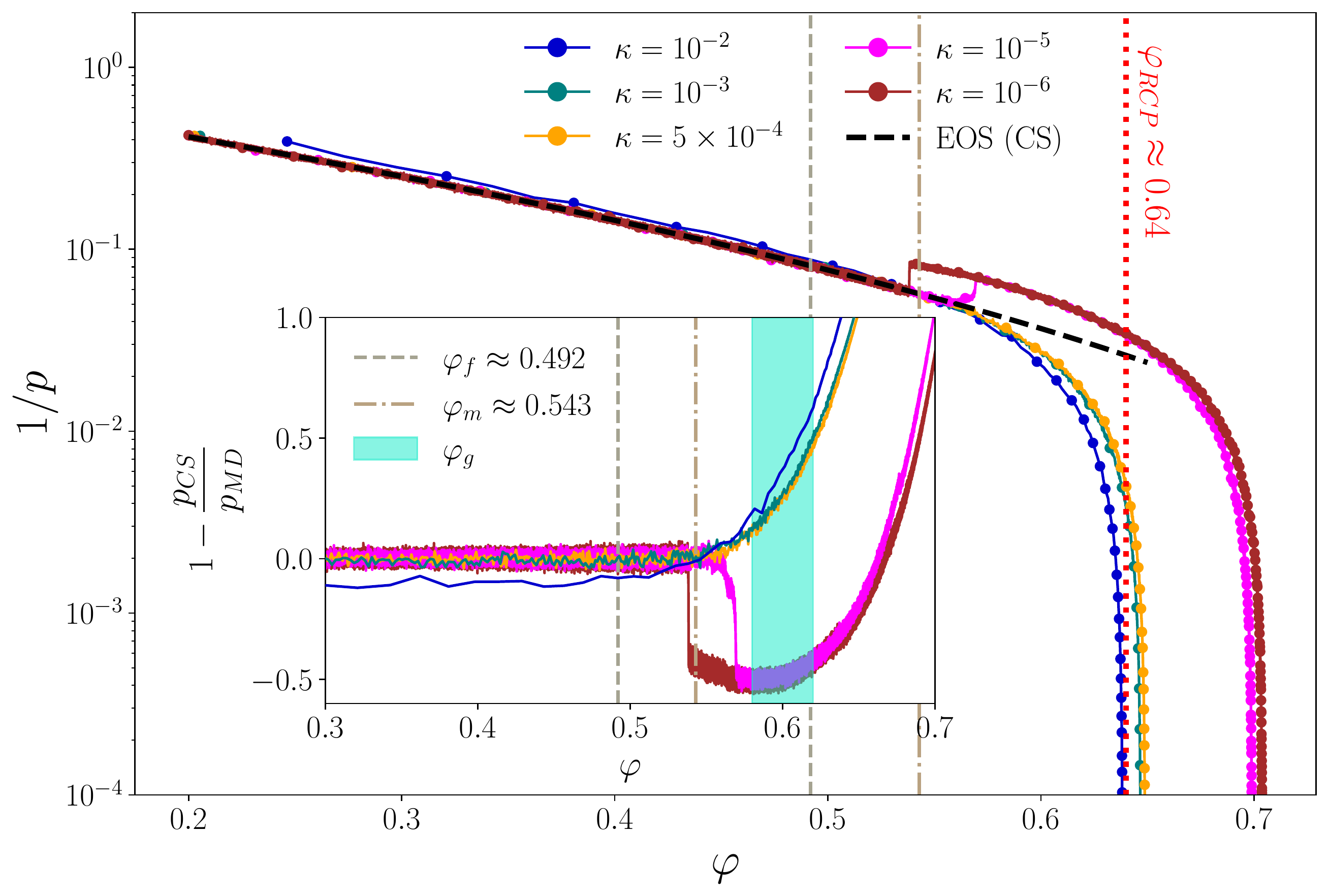}
	\caption[Liquid equation of state in HS systems, for different compression protocols]{Main: Inverse (reduced) pressure as a function the packing fraction. Markers correspond to estimations of $p$ through molecular dynamics simulations\supercite{md-code} with different compression rates ($\kappa$), as indicated by the labels. The system closely follows the liquid CS EOS (black, dashed line) even beyond the freezing point, $\vp_f$.
	Inset: Accuracy of the CS EOS. It also shows that the system is unable to remain in the liquid branch much beyond the melting density $\vp_m$. The region identified with the glass transition is shaded in light blue. See text for a more detailed discussion.
		}
	\label{fig:HS-EOS--diff-growth-rate}
\end{figure}

Once an EOS for HS systems is available, we can explore how the liquid branch continues beyond the freezing and melting points, thus forming a “super-compressed” liquid. In Fig.~\ref{fig:HS-EOS--diff-growth-rate} I present numerical results of such experiment, in the $(\vp, 1/p)$ plane. The simulations were done using an event-driven molecular dynamics algorithm\supercite{md-code} with a Lubachevsky--Stillinger compression protocol\footnote{This algorithm will be discussed in much more detail in the next chapter.}. Different curves correspond to different compression rates ($\kappa$) and the CS EOS is also included for comparison (black dashed line). First of all, note that the curve corresponding to the fastest compression is always off the EOS line, indicating that the system is never allowed to relax and remains out of equilibrium. However, by reducing the compression rate the configuration is able to remain in the liquid branch well beyond the freezing density, until at higher densities it eventually falls out of equilibrium. In turn, when $\kappa$ is further decreased the pressure follows very closely the EOS until, around $\vp_m$, the configuration suddenly crystallizes (partially). This is signalled by a sudden decrease in $p$ (seen as a “jump” in the pink and brown curves of Fig.~\ref{fig:HS-EOS--diff-growth-rate}), caused by the particles rearranging themselves in a regular structure, increasing their free volume as discussed above.

The most salient feature of these results is that they qualitatively reproduce the phenomenology discussed in Sec.~\ref{sec:glasses} for other glass-former models. For instance, compare Figs.~\ref{fig:glass-transition-phase-diagram} and \ref{fig:HS-EOS--diff-growth-rate}: it is clear that by rotating the former by $90^\circ $ the curves of both figures have the same qualitative behaviour. Moreover, notice that in all the cases where the system is able to remain in the liquid phase, all curves approach the $p\to \infty$ limit around the same packing fraction, namely, $\vp_{RCP}\simeq0.64$. Such value is evidently different from the one attained by systems with partial crystallization in the same high $p$ limit, as can be seen from the gap in the two groups of curves as $1/p \to 0$. This gap also mimics the analogous volume difference between a glass and a crystal in the cases mentioned in the previous section. 

Results of Fig.~\ref{fig:HS-EOS--diff-growth-rate} also show that monodisperse HS systems in $3d$ have a high propensity to crystallize. That is the reason why it is very hard to extend the liquid branch beyond $\vp_m$, even by small amount. Although the same happens with other types of algorithms\supercite{isobeHardsphereMeltingCrystallization2015}, in this case it is essentially due to the fact that I have employed a linear compression protocol. 
Consider, for instance, the inset of Fig.~\ref{fig:HS-EOS--diff-growth-rate}, which shows the difference between the pressures obtained according to the EOS from Eq.~\eqref{eq:eos CS} and the numerically from molecular dynamics. We can observer that even if $\kappa$ is chosen so that crystallization is avoided, the liquid surpasses coexistence region $[\vp_f, \vp_m]$ just slightly. This is, in fact, a general feature of protocols with constant compression (or cooling). As explained in detail in \cite[Sec.~3]{cavagna_supercooled_2009}, using constant cooling rates it is impossible to avoid both the formation of crystals and prolonging the metastable liquid phase until the glass transition is reached. Analogously for HS systems, by using a constant compression rate the liquid is bound either to form crystalline domains or fall out of equilibrium soon after $\vp_m$, and certainly before the glass transition density. 
To overcome this difficulty, a common technique is to use a polydisperse mixture (so crystallization is suppressed) and thus allow the liquid to continue for considerably higher densities\footnote{In higher dimensions such a trick is unnecessary because the crystalline state can be effectively avoided with particles of the same size, despite ordered configurations having a smaller free energy\supercite{puz_book}.}. 
As we will see next, polydispersity and a clever type of Monte Carlo algorithms can be used to explore what happens deep in the super-compressed and glassy phases. But in any case, all the considerations so far show that HS systems can be used as a minimal model to study the physics of glassy systems.

Let me close this subsection with a remark about the validity of the extrapolations of the liquid's EOS. Note that the EOS we considered here, Eqs.~\eqref{eq:eos PY virial}-\eqref{eq:eos CS}, does not contain any singularity when $\vp<1$. Such singularity would identify the end of the liquid branch, so this property is obviously unphysical, given that all the region $\vp>\vp_{FCC}\simeq 0.74$ is inaccessible. More realistic relations can be obtained from approximated EOS's that are constructed explicitly to match the observed behaviour in the high density regime. A myriad of proposals haven been put forward\supercite{muleroTheoryHSSimulations}, with different degrees of success, and with $p$ developing a singularity at some density $\vp \in [\vp_{RCP}, \vp_{FCC}]$. It is widely believed that such divergence is reached at either of the interval's extremes, but it has not been established if $p= \infty$ at the random close packing or FCC densities\supercite{santosStructuralThermodynamicProperties2020,muleroTheoryHSSimulations}. However, in a recent work\supercite{berthierEquilibriumSamplingHard2016} the authors showed that  polydisperse liquids can be thermalised even when $\vp \gtrsim \vp_{RCP}$. 

\subsection{Glass phase and Gardner transition in HS systems} \label{sec:phase-diagram-hs}

Let us now analyse the deeply compressed regime of HS systems. From now on, I will assume that no order is present in the configuration, and therefore the $p \to \infty$ divergence occurs at $\vp_{RCP} \approx 0.64$, or equivalently, that the accessible jammed states correspond to randomly arranged configurations. In other words, I will henceforth assume that the only possible jamming density is $\vp_J = \vp_{RCP}$.

To begin this part, let us consider the glass EOS\supercite{salsburgEquationStateClassical1962,md-code,berthierGrowingTimescalesLengthscales2016} in $d$ dimensions:
\begin{equation}\label{eq:eos glass hs}
p(\vp; \vp_J) = \frac{d}{1 - \vp/\vp_J} \, .
\end{equation}
This equation can be derived from the free-volume approximation of the partition function for HS, which becomes exact in the $N\to \infty$ limit, and the condition of forming a stable packing when particles are in contact\supercite{salsburgEquationStateClassical1962}. Importantly, this EOS depends explicitly on the jamming packing fraction. But this is natural given that $\vp_J$ marks the point where the free volume per particle vanishes. However, I want to stress that Eq.~\eqref{eq:eos glass hs} is \emph{not} a thermodynamic EOS, \textit{i.e.} it has no free energy associated with it, nor does it incorporate any reference to the equilibrium liquid state. Yet, it describes quite accurately the values of $p$ and $\vp$ found in numerical simulations as I will show here. I should also mention that no exact thermodynamic EOS is known for HS in any finite $d$, but in Sec.~\ref{sec:MF glasses thermo} I will describe how such relation is constructed within mean-field theory.

\begin{figure}[htb!]
	\centering
	\includegraphics[width=\linewidth]{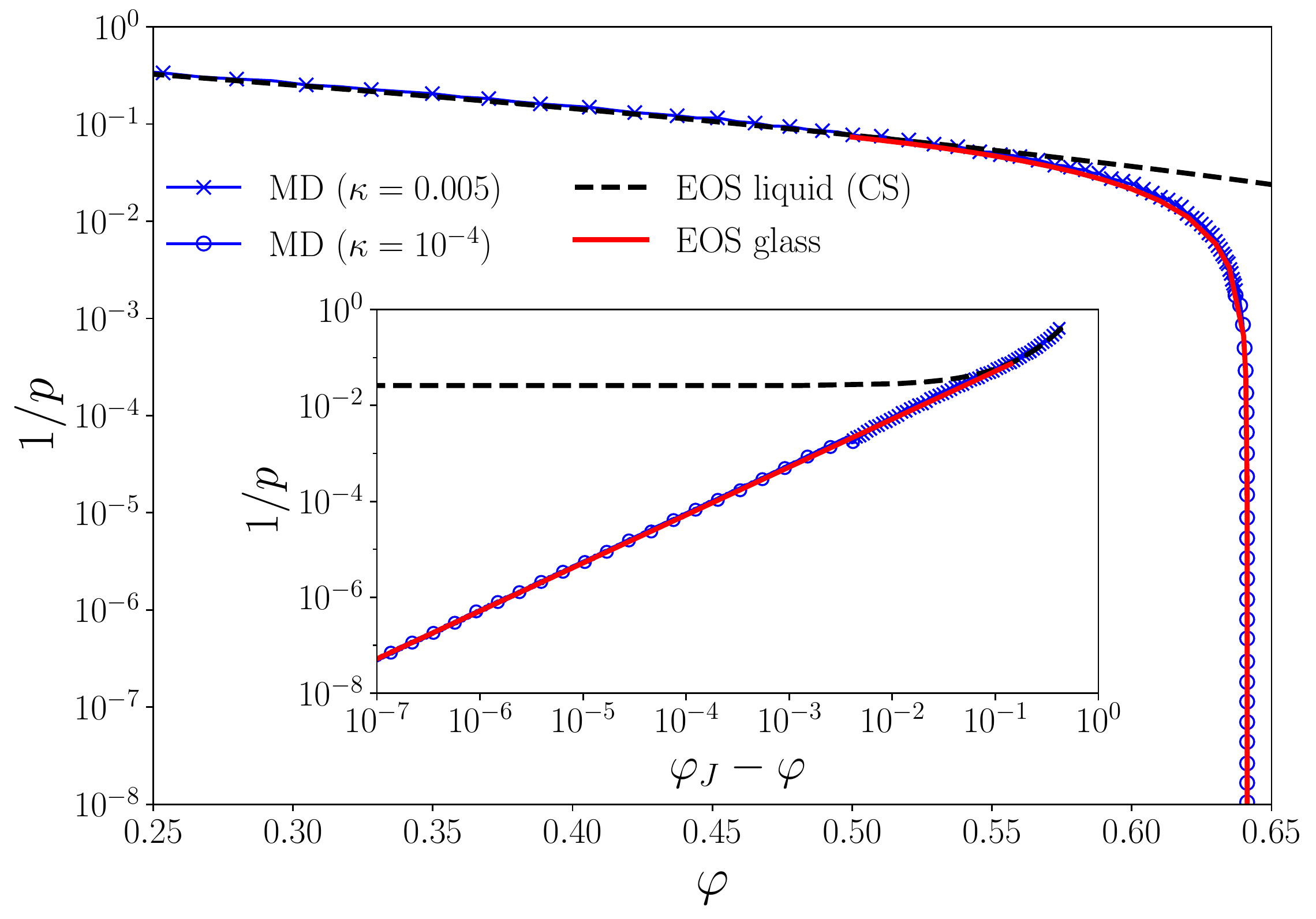}
	\caption[Liquid and glass equations of state in HS systems, comparing theory with numerical simulations.]{Main: Liquid (black, dashed) and glass (red) EOS in a HS system and its comparison with molecular dynamics simulations. Numerical results are obtained using two different compression protocols as indicated by the legend in the plot.
		Inset: Divergence of pressure, as expected from the glass EOS of Eq.~\ref{eq:eos glass hs}. Note however that this equation does not inform us how to obtain the density at the singularity, $\vp_J$.}
	\label{fig:HS-EOS-liquid-and-glass}
\end{figure}

In any case, a relevant property that is readily derived from Eq.~\eqref{eq:eos glass hs} is that $p\sim (\vp_J - \vp)^{-1}$; this is an important scaling that will show up again in Secs.~\ref{sec:jamming criticality} and \ref{sec:MD for jamming}. The accuracy of Eq.~\eqref{eq:eos glass hs} is put to test in Fig.~\ref{fig:HS-EOS-liquid-and-glass}, where the pressure is computed numerically using the same protocol\supercite{md-code} than in the previous figure. In order to closely follow the equilibrium branch for each value of $\vp$, I used two different compression rates. The system is initialized as a low density liquid and is compressed relatively fast up to $p=500$. In this way, a glass is formed by suppressing eventual relaxation to the crystalline phase. Then, to simulate a well thermalised glass, a small value of $\kappa$ is used to further compressed the system, approaching the $p\to \infty$ limit. The excellent agreement between numerical estimations of $p$ and the those from Eq.~\eqref{eq:eos glass hs} is outstanding, specially considering its simple form. Nonetheless, it is important to mention that the value of $\vp_J$ was obtained after bringing the configuration to its jamming point using the iterative Linear Programming (iLP) algorithm described in Chapter \ref{chp:lp-algorithm}. It should also be considered that although these MD simulations allow to attain very high pressures, extrapolations to  $1/p=0$ might yield a \emph{different} jammed state than the one obtained using iLP. This feature will also be analysed in Chapter \ref{chp:lp-algorithm}.

\begin{figure}[htb!]
	\centering
	\includegraphics[width=\linewidth]{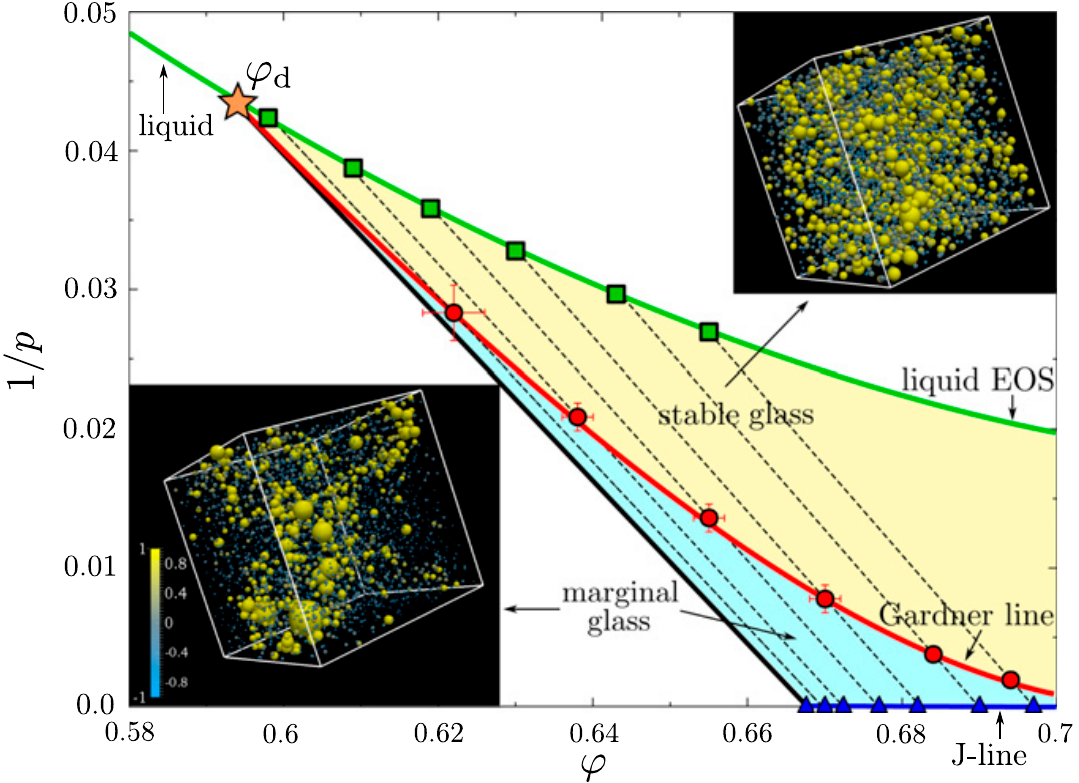}
	\caption[Phase diagram of $d=3$ HS polydisperse mixture.]{Phase diagram of $d=3$ HS polydisperse mixture. The liquid and \emph{equilibrated} glass EOS are shown in green, while the line where the Gardner Transition ensues is shown in red. Note that this line separates the glass phase in two regions: above the GT glasses are stable, but below they are only marginally so. This latter phase ends in the jamming line (in blue). Markers corresponds to values obtained through numerical simulations. See text for a more detailed discussion.
		Taken from Ref.~\cite{berthierGrowingTimescalesLengthscales2016}.}
	\label{fig:HS-phase-diagram}
\end{figure}

On the other hand, the simple aspect of Fig.~\ref{fig:HS-EOS-liquid-and-glass} can be misleading, given that a very rich phenomenology is “hidden” deep in the glass phase. However, to properly investigate this regime numerically, a special type of Monte Carlo (MC) algorithms\supercite{grigeraFastMonteCarlo2001,ninarelloModelsAlgorithmsNext2017,berthierEquilibriumSamplingHard2016} must be used that, rather unfortunately, only work in polydisperse systems. The idea is that simulations are performed using the standard MC dynamics but, after a fixed number of steps a swap between two particles is performed with a given probability. Impressively, this type of SWAP algorithms are able to equilibrate liquids \emph{well beyond} the glass transition of various systems. To explore the glass phase in HS configurations, once an equilibrated liquid has been generated at $\vp > \vp_g$\footnote{Note that the value of $\vp_g$ depends on the amount and type of polydispersity employed. Nevertheless, $\vp_g$ and $\vp_J$ are generally larger compared to those of monodisperse configurations.} an annealing compression is performed, \textit{e.g.} using the same MD protocol as above. This results in the phase diagram of Fig.~\ref{fig:HS-phase-diagram}, taken from Ref.~\cite{berthierGrowingTimescalesLengthscales2016}. A lot of information is contained in this picture, so let us analyse it carefully. First of all, the (estimated) glass transition density is indicated with the star at $\vp_d$, while the equilibrated liquid states are the green squares on top of the liquid EOS. Each of them is then compressed into the glass region until the jamming density is reached; notice that they follow the glass free-volume EOS, \textit{i.e.} the analogous of Eq.~\eqref{eq:eos glass hs} but for configurations with polydispersity. However, the most salient feature is that the glassy phase is divided in two by another transition (red line) that takes place \emph{within} such phase. This corresponds to a Gardner transition\supercite{berthierGardnerPhysicsAmorphous2019} --named after Elizabeth Gardner, who in \cite{gardnerSpinGlassesPspin1985} discovered the corresponding phenomenon in the $p$-spin-- that divides a region of stable glasses from one where stability is only marginal.

Many interesting features come about inside the marginal, Gardner phase, with the phase diagram further complexified by considering different types of potentials and the effects of shear\supercite{mft_review,jinStabilityreversibilityMapUnifies2018,urbaniShearYieldingShear2017,krzakalaPottsGlassRandom2008}. Once again however, space-time limitations dictate that I should omit giving a detailed description and thus I only refer to the very illuminating accounts \cite{berthierGardnerPhysicsAmorphous2019} and \cite{scallietMarginallyStablePhases2019}. Suffice it to say that a peculiar characteristic of glasses in their Gardner phase is the formation of hierarchical basins in their free energy landscapes. Off course, detecting such structure is no trivial task\supercite{charbonneauHoppingStokesEinstein2014,charbonneauNumericalDetectionGardner2015,fel_2014} and, as explained in Ref.~\cite{berthierGrowingTimescalesLengthscales2016}, the best signature of the Gardner transition is the appearance of a peak in the \emph{fluctuations of dynamical variables}. Specifically, the variance and skewness of overlap between the displacement of two clones of a configurations. The difference of such fluctuations is illustrated in the insets of Fig.~\ref{fig:HS-phase-diagram}, where at the centre of the $i$-th particle a sphere is drawn, with a radius proportional to $\abs{\vb{r}_i^{(A)} - \vb{r}_i^{(B)}}$, \textit{i.e.} the difference in displacement between two clones of the system. The snapshot that belongs to the stable glass phase does not show any peculiar structure. In contrast, beyond the Gardner line signatures of spatial heterogeneity are clearly present.

Importantly, jammed states are always located within the Gardner phase. This is not exclusive of HS system, but instead a general property of several models. Moreover, given that jammed configurations are identified with minima of the free energy landscape, understanding its structure is a fundamental step towards building a complete theory of jamming. In the $d\to \infty$ limit of mean-field, the situation has been carefully analysed\supercite{mft_exact_2,puz_book} and theory predicts a sharp transition in the Gardner line\supercite{fel_2014}. Yet, in $d\leq3$ systems the situation is less clear due to the possibility of a crossover or an avoided transition\supercite{hicksGardnerTransitionPhysical2018,zhangMarginallyJammedStates2020} and is further muddled by the \emph{absence} of the marginal phase in some exemplary glass-former models\supercite{scallietAbsenceMarginalStability2017}. Nevertheless, very near the jamming point numerical studies are promising\supercite{fel_2014,artiacoExploratoryStudyGlassy2020,dennisJammingEnergyLandscape2020} and suggest that the fractal, ultrametric structure of the energy landscape, predicted by mean-field theory, remains valid.

\section{Mean-field theory: The $d \to \infty$ limit} \label{sec:MF theory}

In this section, I will go through the main results of a recently developed mean-field (MF) theory for glasses and jammed systems. The theory itself is a \textit{tour de force} --see \textit{e.g.} Refs.~\cite{mft_exact_1,mft_exact_2,mft_exact_3,fel_2014,parisi_zamponi_2010} among others-- that provides an \emph{exact} description of the transition from a liquid to a glass, as well as the physics that ensues near and at the jamming point. In other words, it is an exact theory of the glass and jamming transitions. Furthermore, it also provides an accurate account of what happens as a system goes across such transitions, through its glassy phase. Naturally, the Gardner transition (GT), introduced in the end of the last section, is the most important feature that a glass stumbles upon in such path. Actually, the GT was first predicted and verified in MF models\supercite{mft_exact_2,charbonneauNumericalDetectionGardner2015} and later discovered in other glass formers. 

Yet, the cost paid for exactness is that systems are assumed to have very high dimensionality. It is undoubtedly intriguing that the $d\to\infty$ limit renders a given scheme more tractable and makes possible an analytical treatment. Hence, in this section I will try to give a very summarized account of why this is so. Happily for the curious reader, this MF theory has been recently put together in a coherent, self-contained, and pedagogical form in Ref.~\cite{puz_book}. A short and very nice review is also available in Ref.~\cite{mft_review}, which conveniently focuses on HS systems in large dimensions, and therefore I will follow it in the rest of this section (except in Sec.~\ref{sec:MF liquids} which is based on Ref.~\cite{puz_book}). Besides, adhering to its spirit (and further shortening the theory) I will not present any details of the calculations involved, but will mostly restrict myself to showing how many of the phenomena that have been described in this chapter are the “low dimensional reminiscence“ of the MF predictions. 

\subsection{Liquids in high dimensions}\label{sec:MF liquids}

In my opinion, the first thing that needs to be explained is why the high $d$ regime simplifies the analysis of liquids and glasses. However, this is not an exclusive property of this kind of systems. For instance, consider $N$ Ising spins, $\vec{S}=\{S_i\}_{i=1}^N$ with $S_i =\pm1$, whose Hamiltonian is given by
\begin{equation}\label{eq:H spins}
\H\qty(\vec{S}) = -\frac12 \sum_{i,j}^{1,N} J_{ij} S_i S_j - \sum_{i=1}^N B_i S_i \qc 
\end{equation}
where $B_i$ is a local field acting on the $i$-th spin. $J_{ij}=J_{ji}$ are the exchange couplings between spins $i$ and $j$, and naturally $J_{ii}=0$. Eq.~\eqref{eq:H spins} is very general, since it can describe many spins systems with pairwise interactions by the appropriate choice of couplings. For instance, fully connected models are such that $J_{ij}\neq 0$ for all $i\neq j$, while the a graph or lattice structure can be easily included by setting $J_{ij}\neq 0$ if an edge connects spins $i$ and $j$, and $J_{ij}= 0$ otherwise. An important case is the $d$-dimensional cubic lattice, where each spin interacts with its $2d$ nearest neighbours. It is a well known result from statistical mechanics that the free energy of the system, $F$, can be expressed, in principle, as a function of the configuration's local magnetizations, $\vec{m}=\{m_i\}_{i=1}^N$, with $m_i = \avg{S_i}$. This means that, thermodynamically, the configurations observed in such spin system are the ones that produce a given $\opt{\vec{m}}$ such that the free energy is minimized. That is, $\vec{m}$ plays the role of an order parameter and the configurations that dominate the statistics are the ones satisfying
\begin{equation}\label{eq:m opt deriv F}
\eval{\pdv{F(\vec{m})}{m_i}}_{\opt{\vec{m}}} = 0 \qc \forall i=1,\dots, N\, .
\end{equation}

Of course, finding a closed expression for $F(\vec{m})$ is extremely difficult, if not impossible, even in simple cases, making the equation above unusable. Resorting to approximations is thus unavoidable. Among the several possibilities, high temperature (small $\beta$) expansions are quite fruitful. For instance, considering terms up to order $\beta^3$ it can be shown\supercite{georgesHowExpandMeanfield1991} that
\begin{equation}\label{eq:F-spin high T}
\begin{aligned}
-\beta F(\vec{m}) & = - \sum_i \qty[\frac{1+m_i}2 \log \frac{1+m_i}2 +\frac{1-m_i}2 \log \frac{1-m_i}2 ] \\
 & + \beta \qty[\frac{1}{2} \sum_{i,j} J_{ij}m_i m_j +  \sum_i B_i m_i] + \frac{\beta^2}{4} \sum_{i,j} J_{ij}^2 (1-m_i^2)(1-m_j^2)\\
 & + \frac{\beta^3}{6} \qty[ 2 \sum_{ij} J_{ij}^3 m_i(1-m_i^2)(1-m_j^2) + \sum_{i,j,k} J_{i,j}J_{ik}J_{jk}(1-m_i^2)(1-m_j^2)(1-m_k^2) ] \, .
\end{aligned}
\end{equation} 
In the case a cubic lattice of dimension $d$, with an uniform external field, and setting\footnote{This choice guarantees that the interaction energy remains extensive in the thermodynamic limit, and of the same magnitude as the external potential.} $J_{ij}= \frac{1}{2d}$, the situation is considerably simplified: translational invariance indicates that $m_i=m\ \forall i$. The expression above then reads
\begin{equation}\label{eq:F-spin cubic lattics}
\begin{aligned}
\frac{F}{N} & = - \frac1\beta s_0(m) - \frac12m^2 -B m - \frac{\beta}{8d}(1-m^2)^2 - \frac{\beta^2}{12d^2}m^2(1-m^2)^2 \, .
\end{aligned}
\end{equation}
I have introduced $ s_0(m) = - \qty[ \frac{1+m}{2} \log(\frac{1+m}{2})  + \frac{1-m}{2} \log(\frac{1-m}{2}) ]$ to denote the entropy associated to a single spin.
Comparing Eqs.~\eqref{eq:F-spin high T} and \eqref{eq:F-spin cubic lattics} it is easy to conclude that the $d\to\infty$ limit is equivalent to the $\beta \to 0$ one. Moreover, the series in powers of $1/d$ maps precisely to the high temperature expansion. And given that for high $T$ the phenomenology is usually much simpler, this analogy suggests that by considering large values of $d$ theory might end up being more tractable.

In the case of liquids the advantages of a large $d$ are even clearer. To see why, let first us consider the virial expansion of the free energy. Combining Eqs.~\eqref{eq:intensive free energy} and \eqref{eq:virial expansion} we have that
\begin{equation}\label{eq:free energy virial}
-\beta \tf (\rho, T) = - \beta \tf^{(id)}(\rho ,T) - \sum_{n=2}^\infty \frac{B_n}{n-1} \rho^{n-1}\, .
\end{equation}
Of course, some clarification is in order to ease the (reasonable) doubts as to why we can use a set of $3d$ equations to describe large dimensional systems. First of all, we should consider that just as in the spin systems the free energy can be written in terms of $\vec{m}$, in liquids theory $F$ can be expressed as a \emph{functional} of the systems density, $F[\varrho(\va{r})]$. The formalism thus obtained, termed \textit{density functional theory}, is very powerful and actually easily generalizable to an arbitrary number of dimensions\supercite{hansenTheorySimpleLiquids2013,puz_book}. The analogies with spin systems are indeed deep, since the equilibrium thermodynamic state can also be found as the density profile $\opt{\varrho(\va{r})}$ that makes the (functional) derivative of $F$ vanish, \textit{i.e.} 
\[
\eval{\frac{\delta F[\varrho(\va{r})]}{\delta \varrho(\va{r})} }_{\opt{\varrho(\va{r})}} = 0 \, .
\]
Moreover, if we impose that solutions to the equation above should be uniform and isotropic, crystalline states can be readily discarded. And indeed it can be shown that $ F[\varrho(\va{r})] $ is minimized by $\varrho(\va{r}) = \rho$, that is, by a system with uniform density\supercite{puz_book}. Notice that this is just mimicking the homogenous solution $m_i=m\ \forall i$ discussed above by enforcing an uniform external magnetic field.

Hence, once that we know that $F$ can be expressed, for any $d$, as a function of the system's density, it is natural to write it as the series expansion of Eq.~\eqref{eq:free energy virial}. Clearly, the values of the virial coefficients $B_n$ depend, in general, on the dimensionality of the system and the interaction potential. But nevertheless, for sufficiently low densities and high temperatures, such virial expansion should be reasonably accurate.

The charm of the high dimensional limit is that it allows to truncate the series just after the first virial term. More precisely, if the density of the system does not scales exponentially with $d$, it can be shown that $B_2$ dominates the whole series. Therefore, under these conditions it is possible to only consider the ideal gas contribution and the first virial coefficient to obtain the free energy, and concomitantly the EOS. Besides, the value of $B_2$ is known exactly for HS systems in any dimensions:
\begin{equation}\label{eq:B2 virial}
B_2 = \frac12 2^d V_d \, ; \qquad V_d = \frac{\pi^{d/2} \sigma^d}{2^d\Gamma(1+d/2)} \, .
\end{equation}
In this last expression, $V_d$ is the volume of a $d$-dimensional hypersphere of diameter $\sigma$ and $\Gamma$ is the Gamma function. Introducing the rescaled packing fraction,
\begin{equation*}
\tilde{\vp} = 2^d \vp = 2^d \rho V_d \qc  
\end{equation*}
we have that, for any $d$, the free energy and (reduced) pressure are, to leading order,
\begin{subequations}\label{eqs:F-P leading virial}
\begin{align}
	- \beta \tf (\rho, T) & = -\beta \tf^{(id)}(\rho, T) - \frac{\tilde{\vp}}{2}\qc \\
	p(\rho) & = 1 + \frac{\tilde{\vp}}{2}\, . \label{eq:p leading virial}	
\end{align}
\end{subequations}

The proof\supercite{puz_book} of this result involves (i) identifying the different terms of the virial expansion as (irreducible) Mayer diagrams; (ii) showing that ring diagrams dominate at each order; and (iii) that their contribution to the $n$ order of the free energy is such that
\[
\abs{\tf_n} \sim \tilde{\vp}^{n-1} \qty(\frac{n^{n-2}}{(n-1)^{n-1}})^{d/2}\, .
\]
Now, because for $n\geq 3$ the term in parenthesis is smaller than $1$, $\abs{\tf_n}\to 0$ as $d\to \infty$. This means that Eqs.~\eqref{eqs:F-P leading virial} are actually \emph{exact} if conditions (ii) and (iii) can be met. Or, in other words, if the series \eqref{eq:free energy virial} is convergent. Currently, a lower bound of its convergence radius is known: $\tilde{\vp}^{(conv)} \geq 0.145$. Within the interval $ \tilde{\vp}^{(conv)} \leq \tilde{\vp} <1$ the series is conjectured to be convergent, while if $1<\tilde{\vp} < e^{\gamma d}$ with $\gamma<(1-\log 2)/2$ the series is divergent, but this would imply that the packing fractions grows exponentially with $d$. Nonetheless, even in this case resuming the virial expansion results in a series that to leading order also agrees with Eqs.~\eqref{eqs:F-P leading virial}. Importantly, very similar results hold for several other potentials, like LJ and SS whose interaction is a power-law of the overlap between them. See \cite[Sec.~2.3]{puz_book} for details.

\subsection{Dynamics and the glass transition} \label{sec:MF dynamics and glass transition}

We have seen that the thermodynamics of liquids in high $d$ can be solved exactly. But another important feature of the MF theory is that also the \emph{dynamics} is amenable to an exact treatment\supercite{maimbourgSolutionDynamicsLiquids2016}. As a major outcome, the glass transition comes about within this perspective as a \emph{sharp dynamical transition}. As we will see, in finite $d$ systems such transition is blurred due to the finite lifetime of metastable states. Nevertheless, many of the phenomena described in Sec.~\ref{sec:glasses} can be understood as the manifestation of the (avoided) large $d$ transition.

The solution begins by considering that particles follow a Langevin dynamics, with a friction term $\zeta$ and white noise $\vb*{\xi}$ with zero mean and variance $\avg{\xi_i^\mu(t) \xi_j^{\mu'}(t')} = 2 T \zeta \delta_{ij} \delta_{\mu \mu'} \delta(t-t')$. Whenever equilibrium can be reached an important result follows: the long time limit of the MSD ($\Delta$) can be used as the order parameter relevant to the dynamic transition. Let me briefly describe the argument.
First, the equilibrium assumption is important to guarantee that both time translational invariance and the fluctuation-dissipation theorem hold. If this is the case, it can be shown\supercite{maimbourgSolutionDynamicsLiquids2016,mft_review} that the dynamics of the MSD is given by
\begin{equation}\label{eq:MF langevin MSD}
\hat{m} \ddot{\Delta}(t) + \hat{\zeta}\dot{\Delta}(t) = T - \beta \int_{0}^{t} \dd{u} M(t-u) \dot{\Delta}(u) \, .
\end{equation}
Here $\hat{m}= \sigma^2 m/(2d^2)$ and $\hat{\zeta}=\sigma^2 \zeta / (2d^2)$ are the scaled mass and drag coefficients. Such rescaling with $d$ is important in order to keep the variables finite as $d\to \infty$. For later use, let me now introduce the scaled packing fraction
\begin{equation}\label{def:rescaled phi}
\hat{\vp} = 2^d \vp / d = \frac{\tilde{\vp}}{d}\, .
\end{equation}
On the other hand, $M(t)$ is a memory kernel that is determined self-consistently by solving the following equations: 
\begin{subequations}\label{eqs:MF kernel dynamics}
	\begin{align}
	\hat{m} \ddot{y}(t) + \hat{\zeta} \dot{y}(t) &= T - \tilde{v}'[y(t)] - \beta \int_{0}^{t} \dd{u} M(t-u) \dot{y}(u) + \Xi(t)\\
	M(t-t') & = \frac{\hat{\vp}}{2} \int \dd{y_0} \exp(y_0 - \beta \tilde{v}[y_0]) \avg{\tilde{v}'[y(t)] \tilde{v}'[y(t')]}_{\Xi} \, ;
	\end{align}
\end{subequations}
where $\tilde{v}\qty(d(x-\sigma))=v(x)$ is the rescaled potential and $\tilde{v}'$ its derivative. Additionally, the effective noise is correlated as $\avg{\Xi(t)\Xi(t')} = 2\hat{\zeta} T \delta(t-t') + M(t-t')$. Note that in this way, instead of dealing with $dN$ degrees of freedom, a single one is needed: $y(t)$. Its dynamics is driven by an effective potential $\tilde{v}[y(t)] - T y(t)$ and is influenced by the presence of coloured noise, whose memory kernel $M$ is determined by the force-force correlation\supercite{mft_review}.

From these equations a series of far-reaching predictions arise. First, there is a density $\hat{\vp}_d$ at which a \emph{dynamical glass transition} occurs. This is characterized by the fact that the $t\to\infty$ limit value of the MSD remains finite. Thus, let\footnote{The notation, now somewhat spread, honours Sam Edwards and Philip W. Anderson who in 1975 discovered an analogous order parameter in the theory of spin glasses; see \cite{edwardsTheorySpinGlasses1975}.} $\Delta_\text{EA} \equiv \lim_{t\to\infty} \Delta(t)  < \infty$. Importantly, MF theory is able to predict both $\hat{\vp}_d$ and $\Delta_\text{EA}$. In fact, it is found that $\hat{\vp}_d = 4.807$. Because $\hat{\vp}= \tilde{\vp}/d$, for sufficiently large $d$ this packing fraction is clearly within the interval where Eqs.~\eqref{eqs:F-P leading virial} are exact. Besides, the fact that the long time limit of the MSD remains finite implies that a given configuration can never escape the a region whence it departed, and thus remains out of equilibrium. The size of that region is measured by $\Delta_\text{EA}$.

Second, from the scaled diffusion constant $\hat{D}= \frac{2d^2}{\sigma^2} D$ and viscosity $\hat{\eta} = \frac{2^d V_d}{d^2}\eta$ an expression resembling the Stokes--Einstein relation can be obtained:
\begin{equation}\label{eq:MF stokes-einstein}
\left.
\begin{array}{c}
\hat{D} = \dfrac{T}{\hat{\zeta} + \beta \int_{0}^{\infty} \dd{u}M(u)}\\[6mm]
\hat{\eta} = \beta \hat{\vp} \int_{0}^{\infty} \dd{u}M(u)
\end{array}
\right\rbrace
 \qquad
\implies \quad \hat{D} = \frac{T}{\hat{\zeta} + \hat{\eta}/\hat{\vp}} \, .
\end{equation}
In the glass state, the configuration keeps memory of its initial state for arbitrarily long times. In other words, $M_\text{EA} \equiv \lim_{t\to\infty}M(t)  >0$. When this happens, the integral of the expressions above clearly diverges and therefore diffusion is suppressed and the viscosity becomes infinite. This is a sharpened analogy of the drastic increase in the viscosity discussed in Sec.~\ref{sec:glasses}. Interestingly, notice that even in this case, a sort of Stokes--Einstein relation remains valid, $\hat{D}\hat{\eta} \sim T \hat{\vp}$.

Another important outcome is that, for $\vp \lesssim \vp_d$, the way the MSD approaches and leaves its plateau value is different, namely,
\begin{subequations}\label{eqs:MSD two behaviours near plateau}
	\begin{align}
	\Delta(t) & \simeq \Delta_\text{EA} - A t^{-a}\qc \\
	\Delta(t) & \simeq \Delta_\text{EA} + B t^{b}\, .
	\end{align}
\end{subequations}
Note the very close resemblance with Eq.~\eqref{eq:Fq MCT} for the intermediate scattering function, $F(q,t)$, derived from MCT. Eq.~\eqref{eqs:MSD two behaviours near plateau} also captures the two steps relaxation discussed above, and illustrated in Fig.~\ref{fig:two-steps-msd}. Moreover, the two \textit{dynamical critical exponents} $a$ and $b$ are related through Eq.~\eqref{eq:critical dynamical exponents}, the same equation as in MCT. Importantly, for HS systems the ratio of Gamma functions defining the dynamical critical exponents can be compute exactly\supercite{mft_exact_2}. It equals  $0.707$, whence it follows that $a=0.324\dots$ and $b=0.629\dots$.

Another signature of the transition is obtained by analysing the behaviour of the four-point susceptibility
\[
\chi_4(t) = N \qty[ \overline{\Delta^2(t)} - \overline{\Delta(t)}^2  ] \qc 
\]
near $\vp_d$. Theory shows that $\chi_4$ diverges as $\vp_d$ is approached from either the glass phase (as $(\vp-\vp_d)^{-1/2}$) or from the liquid. In this latter scenario, however, it happens that $\chi_4$ is peaked at $\tau_\a$, which also diverges as $\vp\to \vp_d^-$. (Compare with Fig.~\ref{fig:4point-chi}.) The form of this divergence can also be deduced analytically, $\tau_\a \sim \hat{\eta} \sim (\hat{\vp}_d - \hat{\vp})^{-\gamma} $, with $\gamma = 1/(2a) + 1/(2b)$.

In finite $d$ however, \emph{no real transition comes about}. This is caused by the fact that activation processes  help to overcome dynamical arrest and, therefore, a configuration eventually escapes the basin whence it departed. That is, metastable states can only persist indefinitely in a large $d$ description, where energy barriers are always high enough to prevent relaxation towards equilibrium. Fortunately, many of the MF predictions remain \emph{qualitatively} valid in low dimensional systems, due to the fact that activation processes are exponentially suppressed when increasing $d$. Furthermore, several of them have already been put to test in finite $d$ systems since they coincide, or at least mimic, the predictions of MCT, which precedes MF theory.
Some of the most important findings of this latter approach are: (i) as $d$ increases, the range of validity of the power-law divergence of $\tau_\a$  is extended; (ii) estimations of $\lambda$ constantly approach the predicted value as systems increase their dimensionality; (iii) the product $\hat{D} \hat{\eta}$ seemingly converges to the modified Stokes-Einstein relation for higher $d$. See \cite[Fig.~1]{mft_review} for a summary in $d=3-8$.

\subsection{Thermodynamics of mean-field glasses}\label{sec:MF glasses thermo}

Let us now describe what happens beyond $\hat{\vp}_d$. Because diffusion is absent, there is no need to explicitly solve the dynamics and a statistical mechanical approach can be used instead. Once again, the relevant order parameter will be the long time limit of the MSD. As we will shortly see, the thermodynamic analysis is based on the fact that \emph{equilibrium} configurations can be used to define, and follow, glassy states. The emphasis should warn about the fact that this is a strong assumption because, by definition, configurations cannot be equilibrated in the glass phase. Indeed, we just discussed that beyond $\hat{\vp}_d$ the relaxation time diverges. Nevertheless, as a purely theoretical resource we can assume that equilibrated glasses exist (because they were prepare in such state since the beginning of time) and that we are fortunate enough to have some of them at hand for us to study. More realistically, as long as $d<\infty$, $\tau_\a$ will never be infinite and thus equilibration beyond $\hat{\vp}_d$ can be realized. As discussed in Sec.~\ref{sec:phase-diagram-hs}, one way of doing so is through MC-swap type of algorithms, which exploit an extended phase space to equilibrate systems at densities that would be otherwise unreachable. An even more important step in the “realism ladder” is accomplished via the ultra-stable glasses generated experimentally through vapour deposition\supercite{singhUltrastableGlassesSilico2013}. Finally, see Sec.~\ref{sec:compression-protocol} for an example of a MF model in which equilibration can be reached by a technique called \emph{planting}. Hence, we can say that using equilibrated glass configuration is a strong, but plausible assumption.

So, let $\va{s}$ be one of such equilibrated glasses, with a density $\hat{\vp}_g>\hat{\vp}_d$ and pressure $p_g$\footnote{Note that because it is an equilibrium state, once $\hat{\vp}_g$ is given, the corresponding pressure is automatically determined through the EOS. Thus, I will avoid making explicit reference to $p_g$.}. We want to argue that $\va{s}$ can be used to precisely identify metastable states and to study their thermodynamics. In a first scenario, it will be the initial condition for a trajectory, $\va{r}(t)$, with $\va{r}(0)=\va{s}$ and the same packing fraction, that evolves according to the Langevin dynamics of the liquid, as mentioned above. Given that beyond $\hat{\vp}_d$ the dynamics is arrested, the distance of $\va{r}$ from such reference state will be finite, even for long times; \textit{i.e.} $\displaystyle \lim_{t\to\infty} \abs{\va{r}(t)-\va{s}}^2 = \Delta_r < \infty$. In other words, this means that the glass is unable to escape the region of phase space around $\va{s}$. Now as a second scenario, let once again $\va{r}(0)=\va{s}$ but of a configuration with different density $\vp'\neq \vp_g$ and pressure $p'\neq p_g$. Then the long time limit of the distance between such compressed (or decompressed) configurations can be shown to be finite, $\displaystyle \Delta' \equiv \lim_{\tau \to \infty} \qty( \lim_{t\to\infty} \Delta[\va{r}(t+\tau), \va{r}(t)]  ) < \infty$, where $\Delta[\va{r}_1, \va{r}_2] = \frac{d^2}{N\sigma^2} \sum_{i=1}^N \abs{\vb{x}_{1,i} - \vb{x}_{2,i}}^2$ is the mean squared distance between configurations $\va{r}_1$ and $\va{r}_2$. That is, $\Delta_r$ measures the change in the configurations with respect to a reference one, while $\Delta'$ measures the change between configurations compressed up to $\hat{\vp}'$. Therefore, a glass state is determined by the values of $(\Delta', \Delta_r)$ because they specify the set of (isolated\footnote{They are isolated because glassy states belonging to difference reference states cannot approach each other. That is, $(\hat{\vp}_g, p_g)$ define a basin --since it is an equilibrium state-- isolated from basins with different densities and pressures.}) configurations whose mutual distance is $\Delta'$, and are $\Delta_r$ apart from the reference state.

Once we have seen how to identify glass states by specifying $\va{s}$ and $\hat{\vp}_g$, as well as the characteristic distances $\Delta_r$ and $\Delta'$, we can define a probability density constrained by the values of these variables. This density will thus determine how configurations $\va{r}$ with a density $\hat{\vp}'$, and a distance $\Delta_r$ away from $\va{s}$, are sampled. More precisely,
\begin{equation}\label{eq:metastable glass probability}
\begin{aligned} 
P(\va{r}, \hat{\vp}'| \va{s}, \hat{\vp}_g) & = \dfrac{e^{-\beta \H(\va{r}; \hat{\vp}')}}{Z\qty[\Delta_r, \hat{\vp}'| \va{s}, \hat{\vp}_g]} \ \delta(\Delta_r - \Delta[\va{r},\va{s}] ) \, ;\\
Z\qty[\Delta_r, \hat{\vp}'| \va{s}, \hat{\vp}_g]  &= \int \dd{\va{r}}\ e^{-\beta \H(\va{r}; \hat{\vp}')} \ \delta(\Delta_r - \Delta[\va{r},\va{s}] )
\end{aligned}
\end{equation}
%
We can then compute the free energy associated to the metastable state selected by $\va{s}$ from
\begin{equation}\label{eq:metastable glass free energy}
\tf [\Delta_r, \hat{\vp}'| \va{s}, \hat{\vp}_g] = - \frac1{N\beta} \log Z[\Delta_r, \hat{\vp}'| \va{s}, \hat{\vp}_g] \, .
\end{equation}

Certainly, there is nothing special about $\va{s}$ since there are many possible equilibrium configurations with the same density $\hat{\vp}_g$. Hence, the real thermodynamic free energy is given by averaging Eq.~\eqref{eq:metastable glass free energy} over all such possible initial configurations, conditioned by the values of $\hat{\vp}_g$ and $\Delta_r$. Performing such average we obtain the so called Franz--Parisi (FP) potential\supercite{puz_book,franzRecipesMetastableStates1995}:
\begin{equation}\label{def:Franz-Parisi potential}
V_\text{FP} (\Delta_r, \hat{\vp}'| \hat{\vp}_g) = \overline{\tf [\Delta_r, \hat{\vp}'| \va{s}, \hat{\vp}_g]}^{\va{s}} =  - \frac1{N\beta} \int \dd{\va{s}} \frac{ e^{-\beta \H(\va{s},\hat{\vp}_g)}}{Z[\hat{\vp}_g] } \log Z[\Delta_r, \hat{\vp}'| \va{s}, \hat{\vp}_g] \qc 
\end{equation}
where $Z[\hat{\vp}_g] = \int \dd{\va{s}} e^{-\beta \H(\va{s},\hat{\vp}_g)}$ is the equilibrium partition function. Despite its rather abstract construction, the FP potential can actually be computed explicitly through the replica method\supercite{mezardSpinGlassTheory1986} and thus employed to analyse the stability of the states selected by $\va{s}$. In fact, it is this latter step what introduces the dependence on $\Delta'$ as distances between replicas are considered. Once such computation has been carried out, the thermodynamics follows in the usual manner. That is, if $V_\text{FP}(\Delta_r, \hat{\vp}'|\hat{\vp}_g)$ has a minimum at $\Delta_r$, then it is very likely that some metastable states are found within a distance $\Delta_r$ from $\va{s}$. In contrast, if $V_\text{FP}$ has no minimum, then states are not trapped around $\va{s}$ for long times, \textit{i.e.} the system is a liquid.

\begin{figure}[htb!]
	\centering
	\includegraphics[width=\linewidth]{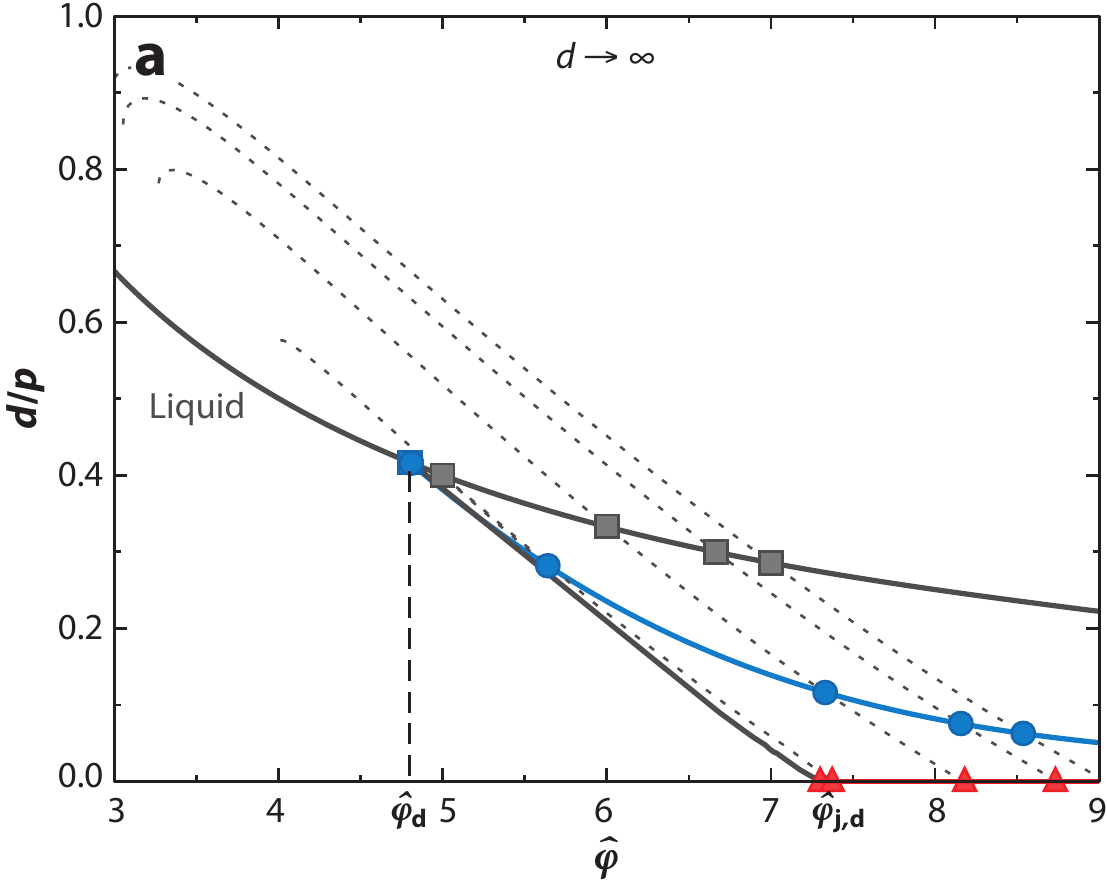}
	\caption[Phase diagram of HS systems in the $d\to\infty$ limit.]{
		Phase diagram of HS systems in the $d\to\infty$ limit. Liquid and equilibrated glassy states fall on top of the black solid line. Glasses in equilibrium can be compressed or decompressed \emph{beyond} the equilibrium line through state following (dashed lines). Deep in the glass phase, a Gardner Transition takes place (blue line), rendering the glasses marginally stable. Such marginal phase ends in the jamming line (red). See the close resemblance with Fig.~\ref{fig:HS-phase-diagram}.
		Taken from \cite{mft_review}; see also \cite{rainoneFollowingEvolutionGlassy2016}.}
	\label{fig:phase-diagram HS infinite d}
\end{figure}

Another advantage of the FP potential is that the EOS in the glass phase can be obtained via the relation $p = \hat{\vp}' \pdv{(\beta V_\text{FP})}{\hat{\vp}'}$. The first important instance is clearly $\hat{\vp'}=\hat{\vp}_g$, in which case the resulting EOS corresponds to the pressure of the \emph{equilibrated} glass, \textit{i.e.} a liquid in equilibrium \emph{beyond} $\hat{\vp}_d$. In other words, the pressure in this situation corresponds to the continuation of the liquid's EOS beyond $\hat{\vp}_d$. This means that, thermodynamically, the liquid is replaced by a set of glassy metastable states, all with density $\hat{\vp}_g$. The surprising result is that each of such metastable states has a pressure that equals the one the liquid would have at that density. 
The second relevant scenario is the behaviour of $p$ upon adiabatic compression, for a fixed reference state $\hat{\vp}_g$. That is, we want to explore what happens as $\hat{\vp}'$ changes infinitely slowly, from a fixed reference density $\hat{\vp}_g$. 
In contrast to the free-volume EOS of a introduced in Eq.~\eqref{eq:eos glass hs}, the MF approach allows to obtain the true thermodynamic EOS. Conceptually, this is evinced by the fact that the FP potential depends explicitly on equilibrium quantities (through $\hat{\vp}_g$), and on “how faraway from equilibrium” we have driven the system (through $\Delta_r$ and $\hat{\vp}'$). As mentioned above, no such information is contained in Eq.~\eqref{eq:eos glass hs}. Another relevant difference is that, while for finite $d$ systems the curves finish at the equilibrium EOS, the construction obtained through the FP potential shows that the metastable glassy states can be \emph{de}compressed beyond such equilibrium line. In other words, the system presents hysteresis. Each of these (out of equilibrium) glassy EOS ends at an spinodal point. The behaviour of the EOS is depicted in Fig.~\ref{fig:phase-diagram HS infinite d}. Note that the reference glass states (grey squares) fall on the continuation of the liquid EOS. On the other hand, the dashed lines, above (resp. below) the solid one, correspond to the glass EOS obtained by adiabatic compression (resp. decompression) or, analytically, through the FP potential. The glass phase is delimited on one side by the jamming line at $p=\infty$, but before this boundary is met, the Gardner transition ensues as we will discuss in the next part. Before concluding, note the very close resemblance of this figure with its $d=3$ counterpart, Fig.~\ref{fig:HS-phase-diagram}.

To recap, what we have seen in this part is that a glass prepared in equilibrium at a given density $\hat{\vp}_g> \hat{\vp}_d$ and pressure $p_g$, can be used to follow the evolution of glass states up to a target $(\hat{\vp}', p')$. These two states are connected through (i) the typical distance ($\Delta_r$) between the reference configurations at $(\hat{\vp}_g, p_g)$ and the target one; and (ii) the typical distance ($\Delta'$) between different states with the same target $(\hat{\vp}', p')$ . Such distance is necessarily finite, because diffusion is suppressed in the glass phase, and consequently states can never get too far from their departing point. In other words, metastable glasses remain stuck near the equilibrium state used as reference. This is important because the $t\to\infty$ limit of dynamical observables, as well as thermodynamical variables, can thus be computed by defining a probability density in phase space, conditioned on the values of $\hat{\vp}_g, p_g$, and $\Delta_r$. This procedure is what the Franz--Parisi construction is about\supercite{puz_book}.

\subsection{Gardner Transition and fractal free energy landscape}\label{sec:MF gardner transition}

I have just mentioned that the FP potential is computed through the replica method. More precisely, a 1-step Replica Symmetry Breaking (RSB) solution is found near the equilibrium line of glassy states. This solution always exists but, as the systems is further compressed to $\hat{\vp}_G>\hat{\vp}_g$, it becomes unstable and is replaced by a fullRSB one. This scenario is akin to the Gardner transition (GT) found in the mean-field $p$-spin model\supercite{gardnerSpinGlassesPspin1985,montanariNatureLowtemperaturePhase2003} and constraint satisfaction problems\supercite{montanariInstabilityOnestepReplicasymmetrybroken2004}. Very loosely speaking, a $k$-step RSB implies that there are $k$ order parameters involved in the computation of the free energy. In turn, in a fullRSB scheme there is a continuum set of order parameters. For the cases of glasses in particular, this feature applies to the MSD, which still acts as the order parameter. Importantly, this also determines how the states are organized within a basin, as I will explain next. 

\begin{figure}[htb!]
	\centering
	\includegraphics[width=\linewidth]{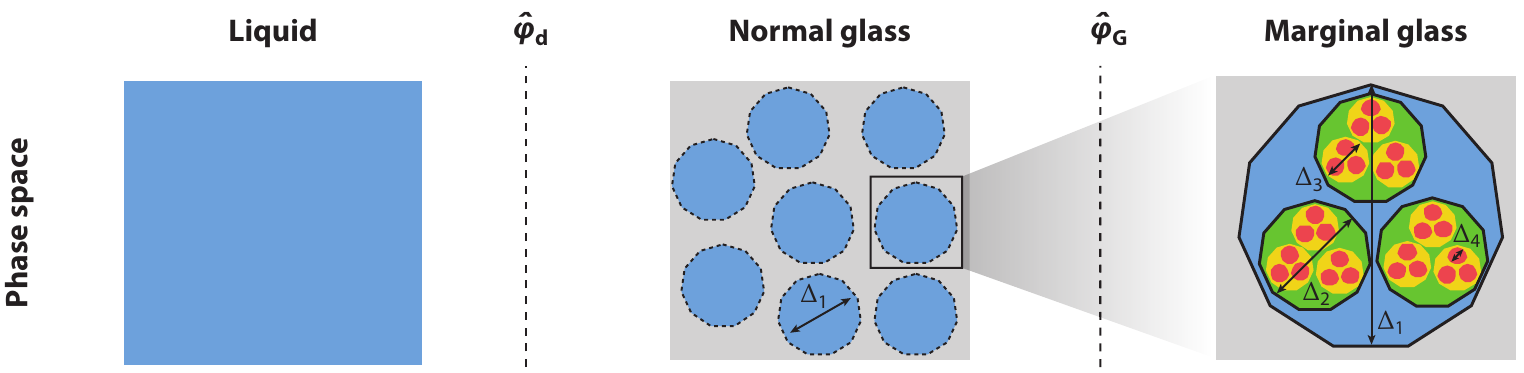}
	\caption[MF picture of the fragmentation of the phase space at different densities.]{MF picture of the fragmentation of the phase space at different densities. In the stable glass phase, \textit{i.e.} for $\hat{\vp}_d < \hat{\vp} < \hat{\vp}_G$, the phase space is clustered into disconnected basins, each of which defines a (metastable) glass. For $\hat{\vp}>\hat{\vp}_G$, \textit{i.e.} in the Gardner, marginal phase, each basin is decomposed into a meta-basin. This means that a hierarchy of sub-basins appears (only 4 levels shown), furnishing an ultrametric structure to the phase space. 
		Taken from \cite{mft_review}.}
	\label{fig:fragmentation phase-space gardner}
\end{figure}

Let me first consider the case of the FP potential when the 1RSB solution is stable. In such situation, an equilibrium configuration corresponds to a minimum of the energy landscape and thus can be used to identify a basin. Metastable glassy states associated to it are then recognized as the configurations within a typical distance of $\Delta_r = \Delta_1$ from the equilibrium state. Let me now assume that the transition is such that, beyond a certain density, the new stable solution corresponds to a $2$-step RSB. In this scenario, the equilibrium configuration would still define a basin, although one with richer structure, since smaller sub-basins would appear within it. Glassy states would now be determined by two order parameters --\textit{i.e.} values of the MSD--, namely $\Delta_{1,2}$ with $\Delta_2 < \Delta_1$. This means that if two metastable states are part of the same sub-basin, then their typical distance would be $\Delta_2$, while if they belong to different sub-basins they will be apart $\Delta_1$. This process can clearly be generalized to a $k$-step RSB solution, and is illustrated in Fig.~\ref{fig:fragmentation phase-space gardner} for the case of $k=4$. So, for a given level $k$ of RSB, basins are divided into a hierarchy of $k$ sub-basins and glassy states are specified according to the set of order parameters $\{\Delta_k, \Delta_{k-1},  \dots, \Delta_1\}$ as possible values for the MSD. Because these values are such that $\Delta_m < \Delta_{m-1}$, glassy states are said to be ordered hierarchically. These means that if two configurations, $\va{r}_1$ and $\va{r}_2$, belong to a common $n$-level sub-basin, then $\Delta[\va{r}_1, \va{r}_2] = \Delta_n < \Delta_{n-1}<\dots \Delta_1$. Hence, if another configuration $\va{r}_3$ is such that $\Delta[\va{r}_1, \va{r}_3] = \Delta_m$, \textit{i.e.} shares with $\va{r}_1$ a $m$-level sub-basin, with $m<n$, then it most also happen that $\Delta_m > \Delta_n$. Importantly, note that because of the hierarchical structure, it follows that $\Delta[\va{r}_2,\va{r}_3] = \Delta_m$. In general, we can therefore conclude that for any triplet of configurations, $\va{r}_1, \va{r}_2, \va{r}_3$, the following relation holds,
\begin{equation}\label{eq:ultrametricity}
\Delta[\va{r}_1, \va{r}_3] \leq \max \qty( \Delta[\va{r}_1, \va{r}_2], \Delta[\va{r}_2, \va{r}_3] )\, .
\end{equation}
In other words, \emph{the metastable glassy states have an ultrametric structure}.

A fullRSB scheme is the $k\to\infty$ generalization of the processes just described. Hence, basins that appeared featureless under the 1RSB solution are transformed, as the GT takes places, into meta-basins; \textit{i.e.} a hierarchy of sub-basins within sub-basins. Such meta-basins are not only hierarchical, but also ultrametric according to Eq.~\eqref{eq:ultrametricity}. Even more, the free energy landscape acquires a \emph{fractal structure}\supercite{fel_2014} as a consequence of the infinite sequence of decomposition of a single basin into ever smaller sub-basins.

Naturally, the precise value of $\hat{\vp}_G$ depends on the density $\hat{\vp}_g$ at which the equilibrated glass was driven away from equilibrium. This is clear from the (conditional) dependence of $V_\text{FP}$ on $\hat{\vp}_g$ in Eq.~\eqref{def:Franz-Parisi potential}. However, any equilibrium glass that is sufficiently compressed will undergo a GT sooner or later, independently of its initial value of $\hat{\vp}_g$ \footnote{At least in the case of HS systems, although the situation can be much more complicated for other potentials or in presence of shear; see Refs.~\cite{biroliBreakdownElasticityAmorphous2016,biroliLiuNagelPhaseDiagrams2018,rainoneFollowingEvolutionGlassy2016}.}. This is seen in Fig.~\ref{fig:phase-diagram HS infinite d} as the blue line running across the full glass phase. Now, as if all the intriguing features described in the previous paragraphs were not enough, there are a couple of extra properties that come about in the Gardner phase: dynamical heterogeneity and marginal stability. The first one is related to the fact that, beyond $\hat{\vp}_G$, particles vibrate around non-fixed positions, although their displacement is very slow. The amplitudes of such vibrations show correlation over large regions and these glasses are considerably more heterogeneous than when the density is such that $\hat{\vp}_g < \hat{\vp} < \hat{\vp}_G$. This is an important finding because the standard picture of a featureless basin is unable to explain the short timescales of inter-basin relaxation observed deep in the glass phase\supercite{fel_2014}. Consider also the insets of Fig.~\ref{fig:HS-phase-diagram} where the contrast in the mobility fluctuations below and above the Gardner line is shown.

On the other hand, the marginality of glasses in the Gardner phase is a direct consequence of the fact that at least one eigenvalue of the stability matrix, obtained by computing the replicated entropy, is zero\supercite{mft_exact_3}. Physically, this is revealed by the fact that $\chi_4=\infty$ throughout this phase: a result that, by the way, is also inconsistent with the simple basin picture\supercite{fel_2014}. A diverging four-point susceptibility thus signals the marginal stability of glasses because it implies that they respond abruptly even to infinitesimal perturbations. As we will see next, this have deep consequences for the jamming transition. The criticality of this last transition is characterized by three non-trivial exponents ($\theta$, $\gamma$, $\kappa$) and is caused by the fact that jammed states are always situated in the marginal phase. It has been a major success of MF theory to be able to predict all three exponents. The role of the first two will be thoroughly discussed in the next section. For the third one, suffice it to say that it equals $\kappa= 1.4157$ and that it controls the rate at which the innermost basin shrinks, as the system is compressed towards its jamming point\supercite{fel_2014,mft_review}, $\Delta \sim p^{-\kappa}$ \footnote{Note that this scaling contrasts with the ones expected for a normal glass, $\Delta \sim p^{-1}$, or a crystal, $\Delta \sim p^{-2}$.}. Its value also determines the fractal dimension of the free energy landscape, it is\supercite{fel_2014} $2/\kappa \approx 1.413$.

\section{The jamming transition} \label{sec:jamming-transition}

We thus finally arrive at the jamming transition. I would like to begin by stressing that the glass and jamming transitions are different. 
I hope that, by comparing the material from the previous sections to the current one, it becomes clear that the physics of both phenomena markedly differ. When discussing the glass transition above, much emphasis was put on dynamical variables, relaxation times, etc. \emph{At} jamming, such properties make no sense because there is no dynamics whatsoever\footnote{This is not strictly true if a finite temperature is present. However, I will only consider jammed packings at $T=0$ or as $T\to0$.}. Moreover, not even in the vicinity of the jamming transition should the same features arise, if only, because jammed states are contained in the marginal glass phase. That is, the Garner transition (see Secs.~\ref{sec:phase-diagram-hs} and \ref{sec:MF gardner transition}) separates glasses and jammed configurations. 
Naturally, also the variables of interest will be different. For instance, I will discuss about the critical scalings of the energy, pressure, bulk and shear moduli, etc. in Sec.~\ref{sec:jamming criticality}, while in Secs.~\ref{sec:forces-and-gaps} and \ref{sec:marginal stability} microscopic structural variables such as contact forces and interparticle gaps will be considered. 
As I will show in Sec.~\ref{sec:network of contacts}, these latter quantities are perfectly well defined at jamming and their value is obviously fixed. But how could the analogous structural variables, with the same precise definition, be defined in a glass? In any case, before beginning, I want to explicitly state that many of the critical properties I discuss next apply to what has been termed \emph{isostatic} jamming point.

\subsection{All the roads lead to jamming}\label{sec:jamming in many systems}

Throughout this chapter, the path I have followed is clearly the one of HS glass formers as they are compressed deep in their glass phase. Yet, the jamming transition would also be a “stop” if the journey had proceeded instead by analysing SS. But not only. As I will discuss next, the jamming transition is shared by a wide variety of systems and under a broad range of circumstances; recall Fig.~\ref{fig:jamming-examples}. What is more, several of its peculiar critical properties are common to all these systems as well. In other words, jamming is a critical point where physics from diverse systems and models coalesces. In this section, I will consider configurations of both SS and HS, favouring the first type because calculations are more transparent with such model. But the next chapter will be entirely devoted to HS configurations. This should not be taken as an example that jamming is only relevant for spheres. Quite the opposite. But systems made out of spherical particles are a very convenient \emph{minimal} model to explore the non-trivial phenomenology that comes about near and at the jamming transition. As expected, when more complicated shapes or scenarios are considered, the phenomenology is even richer.

As a first standpoint, jamming can be considered as a purely mathematical satisfiability problem\supercite{torquatoReviewJammedHardparticlePackings2010,cohnPackingCodingGround2016,conwaySpherePackingsLattices2013}, namely, that of finding possible packings of non-overlapping identical spheres. As simple as it may sounds, it encompasses very hard problems, such as, finding the densest possible packing (also called close packing) in an arbitrary dimension $d$. The archetypical example is the $d=3$ case, for which Kepler conjectured four centuries ago that the maximum possible density was, precisely, $\vp_{max,3d}=\vp_{FCC} = \pi/\sqrt{18} \approx 0.74$. Yet, a complete proof was made available less than 30 year ago by Hales\supercite{halesProofKeplerConjecture2005}. Even the “much simpler“ analogous problem for disks was completely solved not earlier than 1940\supercite{torquatoReviewJammedHardparticlePackings2010}. 
In this case, the largest packing fraction is attained by placing the disks in a triangular lattice; it yields $\vp_{max,2d} = \pi / \sqrt{12} \approx 0.907$. Besides these two results, and the trivial one $\vp_{max,1d}=1$, the value of $\vp_{max}$ is only known for $d=8$ and $d=24$\supercite{puz_book}. Importantly, in all these cases the optimal value of $\vp_{max}$ is realized by using Bravais lattices. In other dimensions the situation is identical: all the maximal density packings known correspond to periodic arrays; see \cite[Chp.~8]{puz_book} and references therein for a detailed account. With this in mind, the question “what is the maximum density that \emph{random} packings can attain?” seems more than challenging. Of course, such a question is not well posed until a proper definition of \textit{random} is given. And once again, this is no trivial task. See, \textit{e.g.}, \cite{torquatoReviewJammedHardparticlePackings2010} for an excellent review of the many intricacies of defining “disorder metrics”, as well as many other properties that are relevant for analysing jammed configurations, both of spherical and more complex shapes.

On the other hand, the mathematical approach pays little attention to how often a given possible packing is actually found empirically\footnote{I avoided making any reference to what is found \textit{in nature}, because it is clear that packings in $d\geq 4$ would be thus irrelevant. However, considering dimensions other than 2 and 3 is very relevant (and doable) in numerical experiments and in theoretical models. Thus, I will refer to both results coming from experiments and simulations as \textit{empirical}.}. And while it is true that regular structures and crystals are common in our daily lives (\textit{e.g.} salt), it is also true that the most common structures are in fact disordered. Therefore, from a physical point of view, analysing packings without any visible periodicity or long-range order is very relevant. Besides, random packings thus produced can also be thought of as instances of a satisfiability problem, in the mathematical sense. It is nonetheless astonishing that many of the features of such \emph{random} close packings (RCP) are very robust and easily reproducible. Here, an intuitive notion of randomness is actually enough. For instance, consider spheres placed via a Poisson process in a fixed volume and at a density large enough that many of them initially overlap. The position of each sphere is independent of the others, so obviously no correlation is expected in their locations. If an interaction energy is introduced between pairs of overlapping spheres, it is clear that by finding the ground state\footnote{In practice, the situation is more complicated because besides an energy minimization, the particles' size should be simultaneously decreased. Nevertheless, several algorithms exist that can accomplish such task in configurations of tens or even hundreds of thousands of spheres.} a valid jammed packing is obtained. And because the initial condition was random, so will be the final configuration with very high probability. Experimentally, the situation is, \emph{in principle}, straightforwardly feasible. Consider placing the spherical beads in a large container and tapping or vibrating it until no rearrangements take place. Clearly, no order is expected in the particles positions after such process. And yet, experiments done with hundreds to tens of thousands particles\supercite{bernalPackingSpheresCoordination1960,scottDensityRandomClose1969}, as well as numerical simulations in several dimensions and with different protocols, show that, on average, the properties of the packings thus produced are surprisingly similar. The most important examples are the packing fraction and the average coordination number $\overline{z}$. In $d=3$ for example, all the instances have approximately the same value, $\vp_{RCP} \approx 0.64$ --recall Fig.~\ref{fig:HS-EOS--diff-growth-rate} or the experimental results reported in \cite{scottDensityRandomClose1969}-- and $\overline{z} \approx 6$. These findings, as well as several others\supercite{parisi_zamponi_2010,torquatoReviewJammedHardparticlePackings2010}, motivate an \emph{ensemble approach} when considering jamming of RCP configurations. That is, the study of jammed packings of configurations that are most likely to be observed empirically. I will adopt such approach in this work since it is particularly useful for the type of systems I will consider; see however Refs.~\cite{donevCommentJammingZero2004,ohernReplyCommentJamming2004} for a more detailed discussion. Furthermore, given that only disordered configurations will be considered, I will implicitly assume that any jammed packing is an RCP instance of the ensemble and, henceforth, use $\vp_J$ instead of $\vp_{RCP}$.

Additionally, the ensemble approach is well suited for studying athermal systems, such as grains, beads, sand, etc. The reason is that it allows to apply, in theory, the full methodology of statistical mechanics. This approach was pioneered by Sam F. Edwards\footnote{Yes, the same Sam F. Edwards of the Edwards--Anderson order parameter of Sec.~\ref{sec:MF dynamics and glass transition}.}, and is nicely reviewed in Ref.~\cite{bauleReviewEdwardsStatisticalMechanics2018}. The basic idea is that, within such framework,  all the relevant observables can be computed as ensemble averages of an appropriately defined probability measure. As a result, it is possible to define metastability of jammed states (by considering the changes in packing fraction resulting from particle displacements); compute the configurational entropy of packings; as well as to identify a hierarchy of basins connected through their degree of constraint.
Interestingly, this method shows a very close correspondence with classifications relying mainly on mathematical aspects, such as the so called \textit{local}, \textit{collective}, and \textit{strictly} categories of jamming\supercite{torquatoReviewJammedHardparticlePackings2010}. Even more, it can be shown that the ground states of jammed granular materials are in an analogous fullRSB phase, just like glass formers (see Sec.~\ref{sec:MF gardner transition}), thus demonstrating that what connects jamming of these two type of systems is not just a mere analogy, but their underlying thermodynamical properties. 
Similarly, several of the scalings present at the jamming transition (see Sec.~\ref{sec:jamming criticality}) are also found in soft materials such as foams, colloids, and emulsions\supercite{vanheckeReviewJammingSoftParticles2010}.

\begin{figure}[htb!]
	\centering
	\includegraphics[width=0.8\linewidth]{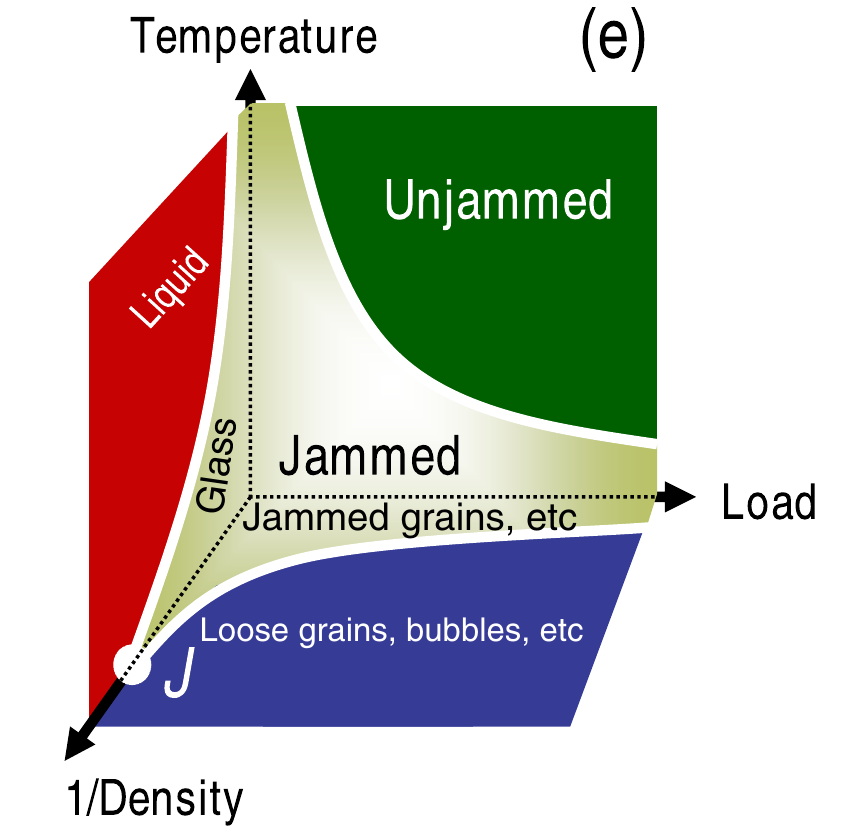}
	\caption[Liu--Nagel phase diagram]{Liu--Nagel phase diagram. It sketches that the jamming transition can be identified with a critical point --common for several systems and reachable by tuning different physical variables-- the so called “J-point”. Note however that the distinction between glasses and jammed states is not clear from this particular plot.
		Taken from \cite{vanheckeReviewJammingSoftParticles2010}, after the proposal in \cite{liuNonlinearDynamicsJamming1998,ohernJammingZeroTemperature2003}.}
	\label{fig:liu-nagel-plot}
\end{figure}

Jamming thus seems to be rather pervasive, on the one hand, while it consistently exhibits the same features, on the other. This latter point will be discussed in the following sections, but before doing so, it is worth asking why the jamming phenomenology is so persistent in so many diverse systems. 
In their seminal works\supercite{liuNonlinearDynamicsJamming1998,ohernJammingZeroTemperature2003,liuJammingTransitionMarginally2010}, Liu, Nagel and co-workers provided an answer of far-reaching consequences: jamming is ubiquitous because it is a common critical point (also termed \textit{J-point}) of all this kind of systems.
This behaviour can be illustrated using the so called \textit{Liu--Nagel plots}, as the one shown in Fig.~\ref{fig:liu-nagel-plot}. From this perspective it becomes clear that, although different systems reach jamming by tuning different physical variables, several properties near and at the onset of jamming are shared by all of them. It is also worth mentioning that equivalent results can also be derived from MF theory, as in, \textit{e.g}, \cite{biroliBreakdownElasticityAmorphous2016,biroliLiuNagelPhaseDiagrams2018}.

To finish this part, let me mention that jamming also occurs outside the physics realm. Two prominent examples are the perceptron\supercite{franzSimplestModelJamming2016,franzCriticalJammedPhase2019a,kallusScalingCollapseJamming2016} and the SAT-UNSAT transition of continuous constraint satisfaction problems\supercite{franzUniversalitySATUNSATJamming2017,krzakalaLandscapeAnalysisConstraint2007}. These problems exhibit all the critical properties expected from a MF perspective and are thus important theoretical models to analyse. Likewise, the Mari--Kurchan (MK) model\supercite{mariDynamicalTransitionGlasses2011} is another MF-type of model that is very useful as a reference for comparison. It consists in particles interacting through a randomly shifted distance, $D(\vb{r}_i, \vb{r}_j) = \abs{\vb{r}_i - \vb{r}_j + \zeta \vb{A}_{ij}}$, where $\vb{A}_{ij}$ is a fixed random vector, different for each pair of particles. $\zeta$ is a parameter that allows to interpolate between fully MF $(\zeta \to \infty)$ and finite $d$ ($\zeta=0$) behaviours. The MK model in $d=2$ and $d=3$ has been used to validate several of the predictions of MG theory\supercite{charbonneauHoppingStokesEinstein2014}, in particular the GT\supercite{charbonneauNumericalDetectionGardner2015}. Some other aspects about the liquid and glass phases of this model will be given in Sec.~\ref{sec:MD for jamming}, while in Chapter \ref{chp:fss} I analyse in detail its microstructure at jamming, so for the moment I just mention that it has also been used to test some of the jamming criticality properties\supercite{hexnerCanLargePacking2019}. Finally, jamming is also present in other MF models\supercite{mariJammingGlassTransitions2009}, neural networks\supercite{spiglerJammingTransitionOverparametrization2019,geigerJammingTransitionParadigm2019,franzJammingMultilayerSupervised2019}, and statistical inference\supercite{antenucciGlassyNatureHard2019}. This shows the relevance of jamming as a current and relevant research topic.


\subsection{Criticality approaching the jamming transition}\label{sec:jamming criticality}

To analyse the critical behaviour of (isostatic) jammed configurations, I will first consider the case of soft spheres that interact only when there is an overlap between them. It is convenient to define the dimensionless gaps,
\begin{equation}\label{def:gaps}
h_{ij}(\va{r}_i, \va{r}_j) = \dfrac{\abs{\vb{r}_i - \vb{r}_j}}{\sigma_{ij}} -1 \qc 
\end{equation}
where $\sigma_{ij} = \frac12(\sigma_i +\sigma _j)$ is the sum of the two particles' radii. That is, for the time being, I will consider the more general case of polydisperse configurations for reasons that will become clear shortly. In this way, an overlap corresponds to a negative gap. The contact potential modelling the interaction between particles can then simply be expressed as a function of the gaps. Following the literature\supercite{puz_book,charbonneauJammingCriticalityRevealed2015,liuJammingTransitionMarginally2010} I will consider a contact potential of the form
\begin{equation}\label{eq:contact potential soft spheres}
v(h) = \frac{\epsilon}{\a} \abs{h}^\a \ \Theta(-h) \qc 
\end{equation}
where $\Theta$ is the Heaviside step function, $\epsilon$ is a constant that fixes the energy scale, while $\a$ is a parameter that can be used to vary the “softness“ of the spheres.

Once the potential has been defined, a valid jammed packing can be obtained with a suitable energy minimization procedure. Recently, the FIRE algorithm\supercite{FIRE} has been extensively used for this purpose, although several options are possible. 
Note however that, independently of the algorithm employed, jammed configurations correspond to an energy minimum with $T=0$. That is, in scenarios where thermal noise is included to enhance the exploration of the configuration space and thus find lower minima, it is important to have in mind that a proper critical jammed configuration will only be achieved in the $T\to 0$ limit\footnote{At least in the sense that I will discuss here. But see Ref.~\cite{degiuliTheoryJammingTransition2015} for a study with $T>0$.}.

At this point, I want to bring attention to a peculiar feature of the jamming transition. So far, finding jammed packings seems only a minimization problem, similar to others found in statistical physics. Consider for the sake of simplicity the Ising model with no external field and equal couplings, $J_{ij}=J$. Finding its ground state in the $T\to 0$ limit, will always result in well defined physical variables, such as $\avg{\abs{m}}= 1$, and analogously for its energy, heat capacity, etc. In contrast, for jammed packings the situation is very different, although this has been only implicit in the discussion up to this point. To make the matters clearer, consider the phase diagrams of Figs.~\ref{fig:HS-phase-diagram} and \ref{fig:phase-diagram HS infinite d}. Note that in those figures, jammed states are not concentrated around a point, but actually lie along a \emph{line}. This means that the jamming density can take any value within an interval\footnote{Clearly, there is a lower bound for such interval, since at very low densities not even a glass can be formed. In contrast, deriving an upper bound is not an easy task. However, it can be shown that, the maximal density an amorphous packing can have is of order $\sim d \log (d 2^{-d})$; see \cite[Chp.~8]{puz_book}.}.
In fact, the value of $\vp_J$ depends on both the protocol used and the configuration itself. In the case of monodisperse systems, this effect is considerably suppressed, given that particles are identical and consequently the J-\emph{line} is shrunk to a narrow interval. Nonetheless, the dependence of the protocol remains. This is evinced by the results of Fig.~\ref{fig:HS-EOS--diff-growth-rate}, in which all the MD simulations departed from the same initial condition and only the compression rate ($\kappa$) was varied. It is evident that although the final packing fractions are similar, using a slower compression consistently produces denser packings. Similarly, in the case of the polydisperse configurations, letting the system remain longer in equilibrium allows to reach higher values of $\vp_J$. Fortunately, these observations also provide the answer to understand the strange dispersion in the values of $\vp_J$:
that jamming is an \emph{out of equilibrium} transition\supercite{charbonneauUniversalMicrostructureMechanical2012}. In other words, the jammed state reached by whatever methods depends fundamentally on both (i)  the state at which the system abandons equilibrium and (ii) how it evolved afterwards.
%
Consequently, also some of the observables at jamming will show such dependence. Now, in the thermodynamic limit and with an infinitesimally slow compression or decompression protocol, $\vp_J$ has a well defined value, that nonetheless depends on the equilibrated glass state whence it departed (see previous section), \textit{i.e.} $\vp_J = \vp_J(\vp_g)$. When analysing jamming criticality, these considerations are important because we are interested in how thermodynamical variables scale as a function of $(\vp-\vp_J)$. 
So, in principle, we should use the thermodynamic limit value of the jamming density,  $\vp_J(\vp_g)$ , to properly observed the power-law scalings. In practice, however, one should keep in mind that the value of $\vp_J$ will be unique of each configuration.

Now, jamming criticality can be intuitively justified as follows. If SS interact through a potential like Eq.~\eqref{eq:contact potential soft spheres}, then it is clear that for $\vp< \vp_J$ the energy is identically zero, while for densities larger than $\vp_J$ it is strictly positive. Thus, the energy must be singular at $\vp_J$. The scaling of the energy and other thermodynamical variables are summarized in the following expressions\supercite{ohernJammingZeroTemperature2003,vanheckeReviewJammingSoftParticles2010,liuJammingTransitionMarginally2010}:
\begin{subequations}\label{eqs:criticality overjammed}
	\begin{align}
	e(\vp) & \sim (\vp-\vp_J)^{\alpha} \qc \label{eq:energy near jamming}\\
	P(\vp) & = \rho^2 \dv{e}{\rho} \propto \vp^2 \dv{e}{\vp} \sim \qty(\vp - \vp_J)^{\a -1} \qc	\label{eq:P near jamming}\\
	G (\vp) & \sim \qty(\vp -\vp_J)^{\alpha - 3/2} \qc \label{eq:G near jamming}  \\
	B (\vp) - B_0(\a) & \sim \qty(\vp - \vp_J)^{\a -2}  \label{eq:B near jamming} \qc \\
	\overline{z}(\vp)-z_c & \sim (\vp - \vp_J)^{1/2} \, . \label{eq:coordination near jamming}
	\end{align}
\end{subequations}
In the last equation, $z_c$ is the average number of contacts observed for a particular configuration with $\vp_J$. Finite size effects causes this variable to exhibit small sample to sample fluctuations. But in the thermodynamic limit its value is perfectly well defined and independently of the jamming protocol; it is $z_c=2d$. In Sec.~\ref{sec:network of contacts} I will show that this value is a consequence of the mechanical stability of the packings. On the other hand, the term $B_0(\a)$ in Eq.~\eqref{eq:B near jamming} is such that $B_0(\a) >0$ if $\a \leq 2$ and zero otherwise. Let me consider briefly the harmonic potential, $\a=2$. Such potential is, quite reasonably, an important case and it is useful in modelling emulsions, colloids, etc.\supercite{vanheckeReviewJammingSoftParticles2010} The scalings of Eqs.~\eqref{eq:B near jamming} and \eqref{eq:G near jamming} imply that a jammed packing of harmonic SS is unable to withstand shear forces ($G\to0$), and yet remains rigid ($B\to B_0>0$, \textit{i.e.} it supports compression). In fact, the difference of the critical exponents of the shear and bulk moduli is a consequence of $G$ being more susceptible to non-affine effects. In contrast, $B$ is determined through the effective stiffness ($k$) of the spheres, which is usually determined by affine deformations; see \cite{vanheckeReviewJammingSoftParticles2010} for a detailed discussion. Moreover, if the rigidity of a configuration is measured by normalizing with respect to $k$, then the behaviour at $\vp_J$ resembles a first order phase transition, given that $B/k$ remains finite for any $\vp>\vp_J$ but vanishes for smaller densities. The same jump is observed in the coordination number since $\avg{z}=0$, for $\vp < \vp_J$.

At any rate, the power-law scalings summarized in Eqs.~\eqref{eqs:criticality overjammed} have been extensively verified in several works\supercite{ohernJammingZeroTemperature2003,liuJammingTransitionMarginally2010,goodrichFiniteSizeScalingJamming2012,goodrichJammingFiniteSystems2014,vanheckeReviewJammingSoftParticles2010} and, surprisingly, they are unaffected by dimensionality or polydispersity of the configurations. Additionally, the finite size effects expected from these relations have been confirmed\supercite{goodrichFiniteSizeScalingJamming2012} and a Widom-like scaling ansatz has been derived for all of them\supercite{goodrichScalingAnsatzJamming2016}. Furthermore, various studies have identified correlation lengths associated to the characteristic length scales of vibrational response to perturbations\supercite{goodrichStabilityJammedPackings2013,schoenholzStabilityJammedPackings2013}, the fluctuations in the number of contacts\supercite{hexnerTwoDivergingLength2018,hexnerCanLargePacking2019}, and the fluctuations of particle mobility\supercite{dynamic_criticality_jamming}, all of which diverge at the jamming point.

In the case of HS similar relations hold. However, because the interaction potential is always zero, the energy identically vanish for any $\vp < \vp_J$. Likewise, before the jamming point is reached, the configurations are not rigid\supercite{liarteJammingMulticriticalPoint2019} nor do they sustain shears, hence both bulk and shear moduli vanish. Therefore, the relevant scaling relations only concern the reduced pressure and the intensive entropy\supercite{parisi_zamponi_2010}:
\begin{subequations} \label{eqs:criticality underjammed}
	\begin{align}
	p(\vp) & \sim \qty(\vp_J - \vp)^{-1} \qc  \label{eq:p near jamming} \\
	s(\vp) & \sim \log(\vp_J - \vp) \, . \label{eq:s near jamming}
	\end{align}
\end{subequations}
Indeed, the second of these expressions can easily be obtained from the first one given that for HS (i) $p=\beta \rho \pdv{\tf}{\rho}$ and (ii) $\tf = - T s$. It is then straightforward to obtain $p=- \vp \pdv{s}{\vp}$ whence \eqref{eq:s near jamming} follows by integration. It is also worth noting that Eq.~\eqref{eq:p near jamming} is the same as the free-volume of a HS glass, cf. Eq.~\eqref{eq:eos glass hs} and Fig.~\eqref{fig:HS-EOS-liquid-and-glass}. This implies that the algebraic dependence of $p$ on $\vp_J - \vp$ is valid throughout the glass phase and not only near jamming.

Another important result is that effects of finite temperature can also be included near jamming\supercite{degiuliTheoryJammingTransition2015,dynamic_criticality_jamming,jacquinMicroscopicMeanFieldTheory2011} and thus join the behaviours on both sides of $\vp_J$. For instance, Eqs.~\eqref{eq:P near jamming} and \eqref{eq:p near jamming} can be combined in a single scaling relation, that depends on both $\vp$ and $T$:
\begin{equation}\label{eq:p scaling near jamming}
p(\vp, T) = T^{-1/\a} \ \mathcal{P} \qty( T^{-1/\a} (\vp - \vp_J) ) \qc \quad 
\mathcal{P}(x) \sim \begin{dcases}
\abs{x}^{-1} & \text{for } x\to - \infty\qc \\
x^{\a-1} &  \text{for } x \to \infty \, .
\end{dcases}
\end{equation} 
Scalings for other variables are listed in Ref.~\cite{degiuliTheoryJammingTransition2015}. All of these results thus provide some of the strongest evidence in support of the critical nature of the jamming transition.

The picture that unfolds by putting  together the exact MF description (Sec.~\ref{sec:MF theory}), the scalings of Eqs.~\eqref{eqs:criticality overjammed}-\eqref{eq:p scaling near jamming}, and the robustness of numerical experiments for several dimensions and different protocols\supercite{parisi_zamponi_2010,charbonneauUniversalMicrostructureMechanical2012,md-code,liuJammingTransitionMarginally2010,ohernJammingZeroTemperature2003} suggests that the jamming transition of spherical particles defines a universality class. Even more, it is a rather broad class that encompasses all of the different systems mentioned at the end of Sec.~\ref{sec:jamming in many systems}. Nevertheless, we also know that if inherent order is enforced -- \textit{e.g.} by using minimally polydisperse crystals\supercite{charbonneauGlassyGardnerlikePhenomenology2019,tsekenisJammingCriticalityNearCrystals2020,ikedaJammingReplicaSymmetry2020}-- or particles are not spherical\supercite{ikedaInfinitesimalAsphericityChanges2020,britoUniversalityJammingNonspherical2018b} universality is broken. Properly demarcating such universality class is therefore an open problem that falls outside the scope of this work. Nevertheless, I will briefly come back to the subject in Chp.~\ref{chp:fss}, which deals with a systematic assessment of the jamming criticality of the \emph{microscopic variables} introduced in Sec.~\ref{sec:forces-and-gaps}.


\subsection{Network of contacts and isostaticity}\label{sec:network of contacts}

In this section, I will analyse the properties of the network of contacts (NC) formed \emph{exactly at a jammed state}. This is a fundamental section for several of the results that will be presented in the following chapters. From the analysis of the NC the most important properties that I will show are:
\begin{enumerate}
	\item A jammed state reached with a given interaction potential is an equally valid jammed configuration with any other potential.
	\item In the case of spherical particles, the minimal requirement for a jammed packing to be stable is that it has a \emph{single state of self stress}. This means that the number of contacts ($N_c$) is equal to the number of degrees of freedom \emph{plus one}; \textit{i.e.} $N_c = N_{dof}+1$.
	\item Under such conditions, the forces exerted by pairs of particles in contact are determined solely by the particles positions.
	\item The structure of the NC determines the Hessian of the system at jamming, and vice-versa.
\end{enumerate}
Several of these results have been developed in a series of important works, \textit{e.g.} \cite{wyartEffectsCompressionVibrational2005,wyartGeometricOriginExcess2005,lernerLowenergyNonlinearExcitations2013,lernerBreakdownContinuumElasticity2014,degiuliForceDistributionAffects2014}.  While in \cite{charbonneauJammingCriticalityRevealed2015} they have been put together and generalized. Thus, I will mostly follow the discussion presented there, as well as in \cite[Sec.~9.2]{puz_book}.

To show the properties mentioned above, let me begin by analysing the Hessian, $H$ of a configuration \emph{above} jamming, \textit{i.e.} I will study a configuration of SS with $\vp>\vp_J$. To do so, I will use the following notation: $\ctc{ij}$ with $i<j$ will denote the contact between particles $i$ and $j$ and it will also be used as a contact index, \textit{i.e.} $\ctc{ij}\in \{1,\dots. N_c\}$; $\vb{n}_{ij} = \dfrac{\vb{r}_i - \vb{r}_j}{\abs{\vb{r}_i - \vb{r}_j}}$ is a unity vector pointing from particle $j$ towards particle $i$ (clearly $\vb{n}_{ij}=- \vb{n}_{ji}$); $f_{ij}=f_{ji}$ is the magnitude of the force of contact $\ctc{ij}$. Additionally, let $\mathcal{C}=\{\ctc{ij}_c\}_{c=1}^{N_c}$ be the set of contacts. With this notation, the total potential energy can be expressed as
\[
V(\va{r}) = \sum_{c\in \mathcal{C}} v(h_{c}(\va{r})) \qc
\]
where, for the time being, I am assuming that the gap function can depend on the full configurational vector $\va{r}=\{\vb{r}_i\}_{i=1}^N$ and not just on the pair of particles involved in the contact $c$. Assuming a general power-law contact potential like the one of Eq.~\eqref{eq:contact potential soft spheres} we have that
\begin{equation}\label{eq:derivative potential at jamming}
\pdv{V(\va{r})}{r_i^{\mu}} = \sum_{c\in \mathcal{C}}  v'(h_c) \pdv{h_c}{r_i^\mu}
= - \epsilon \sum_{c=1}^{N_c} \abs{h_c}^{\a-1} \pdv{h_c}{r_i^\mu}
\, ; \qquad \mu=1,\dots, d \, .
\end{equation}
The minus sign in the rightmost term comes from the assumption $h_c<0$, given that $\vp>\vp_J$, so the $\Theta$ function is consequently omitted. Physically, it represents the fact that a force should increase in magnitude as the corresponding gap becomes more negative. Similarly, the Hessian is a $dN\times dN$ symmetric matrix $\pdv[2]{V(\va{r})}{\delta \va{r}}$ whose entries are  given by
\begin{equation}\label{eq:Hessian at jamming}
H_{i,\mu}^{j,\nu}  \equiv  \epsilon \sum_{c=1}^{N_c} \qty[ (\a-1) \abs{h_c}^{\a-2}\  \pdv{h_c}{r_i^\mu} \pdv{h_c}{r_j^\nu} - \abs{h_c}^{\a-1} \pdv{h_c}{r_i^\mu}{r_j^\nu}    ] \, .
\end{equation}

Now, let $\va{r}_0$ be an energy minimum, although not necessarily a jammed configuration. In that case, Eq.~\eqref{eq:derivative potential at jamming} vanishes and the energy cost of a small perturbation $\delta \va{r}$ around $\va{r}_0$ is well approximated by
\begin{equation}\label{eq:perturbation jamming}
\begin{aligned}
\delta V(\va{r}) & = V(\va{r}_0 + \delta \va{r}) - V(\va{r}_0)   \approx \frac12 \delta \va{r} \cdot H \cdot \delta \va{r} \\
& = \frac{\epsilon}{2} \sum_{c=1}^{N_c} 
\qty[ (\a-1) \abs{h_c}^{\a-2}  \underbrace{ \qty(  \delta \va{r} \cdot \pdv{h_c}{\delta \va{r}})^2}_{\text{harmonic}} 
 \  \underbrace{-\ \delta \va{r}\cdot \pdv[2]{h_c}{\delta \va{r}} \cdot \delta \va{r} }_{\text{prestress}} \abs{h_c}^{\a-1} ] \, .
\end{aligned}
\end{equation}
Note that the harmonic term is a stabilizing contribution to the perturbation, since each contact suppresses perturbations along the direction of its gradient. Thus, of the possible $dN$ directions in which a perturbation could move away from $\vb{r}_0$, $N_c$ are blocked by the contacts. To proceed with this argument, I will restrict to the case where the system is subject to periodic boundary conditions, which reduces in $d$ the number of degrees of freedom, $N_{dof}$ (see below a detailed argument), \textit{i.e.} $N_{dof}=dN-d$. This means that $N_{dof}-N_c$ directions remain along which the system could be perturbed at zero energy cost. In other words, there are $\max \{N_{dof}-N_c, 0\}$ non-trivial soft or zero modes, and $d$ trivial ones, related to uniform displacements of the configuration.

Because I assumed that $\va{r}_0$ is a minimum $H$ must be a positive semi-definite matrix and $\delta V>0$. To ensure stability, some constraints most hold for the prestress term. There are two possibilities, either $ \pdv[2]{h_c}{\delta \va{r}}$ is a positive-definite matrix, or its not. In the first case, it will have a \emph{de}stabilising effect in $H$. In particular, the prestress term is negative\supercite{puz_book} along the $N_{dof}-N_c$ (possible) zero modes of the harmonic contribution. Hence, it must be the case that $N_c \geq N_{dof}$ in order to guarantee stability. Otherwise $\va{r}_0$ would not be a minimum, in contradiction with the initial assumption. The second possibility in which $ \pdv[2]{h_c}{\delta \va{r}}$ is not positive-definite is more complicated because it can happen that the prestress term can actually stabilize some of the zero modes from the harmonic part. In this case, the situation $N_c < N_{dof}$ is in principle allowed, although it is system dependent. When this happens, the system is said to be \textit{hypostatic}, the condition $N_c = N_{dof}$ corresponds to an \textit{isostatic} configuration, while if $N_c>N_{dof}$ the system is called \textit{hyperstatic}.

I will show now that for particles whose gaps are given by Eq.~\eqref{def:gaps} the energy minima can never by hypostatic. In particular, this is true for SS, because symmetry constrains pairwise gaps to be only a function of the centres' distance. Hence, letting $h_\ctc{kl}$ be of form of Eq.~\eqref{def:gaps}, it is straightforward to derive that
\[
\begin{aligned}
\pdv{h_\ctc{kl}}{r_i^{\mu}} & = \frac1\sigma \frac{r_k^\mu - r_l^\mu}{\abs{\vb{r}_k - \vb{r}_l}} (\delta_{ik}- \delta_{il}) \qc \\
\pdv{h_\ctc{kl}}{r_j^\nu}{r_i^\mu} & = \frac{(\delta_{ik}- \delta_{il})(\delta_{jk}-\delta_{jl})}{\sigma} 
\qty[
\frac{\delta_{\mu\nu}  }{\abs{\vb{r}_k - \vb{r}_l}} - \frac{ (r_k^\mu - r_l^\mu)(r_k^\nu - r_l^\nu) }{\abs{\vb{r}_k - \vb{r}_l}^3} 
] \, .
\end{aligned}
\]
And thus, plugging these expressions into Eq.~\eqref{eq:perturbation jamming} and performing the dot products gives, 
%
\begin{equation}\label{eq:perturbation jamming soft spheres}
\begin{aligned}
\delta V(\va{r}) & \approx \frac{\epsilon}{2\sigma} \sum_{\ctc{ij} \in \mathcal{C}}
\qty[
\frac{(\a-1)}{\sigma} \abs{h_\ctc{ij}}^{\a-2} \qty( \delta \vb{r}_{ij} \cdot \vb{n}_{ij})^2 -
\abs{h_\ctc{ij}}^{\a-1} \frac{ \abs{\delta \vb{r}_{ij}}^2 - \qty( \delta \vb{r}_{ij} \cdot \vb{n}_{ij})^2}{r_{ij}} 
]\\
& =\frac{\epsilon}{2\sigma} \sum_{\ctc{ij} \in \mathcal{C}}
\qty[
\frac{(\a-1)}{\sigma} \abs{h_\ctc{ij}}^{\a-2} \qty( \delta \vb{r}_{ij} \cdot \vb{n}_{ij})^2 -
\abs{h_\ctc{ij}}^{\a-1} \frac{ \abs{\delta \vb{r}_{ij}^{\perp}}^2 }{r_{ij}} 
] \qc 
\end{aligned}
\end{equation}
where $\delta \vb{r}_{ij}= \delta\vb{r}_i - \delta \vb{r}_j$, $r_{ij} = \abs{\vb{r}_{i,0} - \vb{r}_{j,0}}$, and $\delta \vb{r}_{ij}^{\perp}$ is the component of $\delta \vb{r}_{ij}$ orthogonal to $\vb{n}_{ij}$. Because the second term inside the square brackets is the sum of non-positive terms the prestress matrix is thus negative semi-definite. And according to the discussion above, this implies that $N_c \geq N_{dof}$. Therefore, energy minima of spherical particles are at least isostatic.

Note that so far these results are valid for any type of interaction potential of the form \eqref{eq:contact potential soft spheres} and for any state that defines an energy minimum. If, besides being an equilibrium point, $\va{r}_0=\va{r}_J$ is also a jammed state, further important results can be derived. First of all, note that at jamming the overlaps go to zero, and as long as $\a >1$\footnote{The case $\a = 1$ is very interesting, since it corresponds to a singular potential. It can be shown however, that all the results derived so far also apply. See Refs.~\cite{franzCriticalEnergyLandscape2020a,franzCriticalJammedPhase2019a}.}, the prestress term vanishes, proportionally to the pressure, and quicker than the harmonic one. We can thus define the contact forces at jamming as,
\begin{equation}\label{def:contact forces jamming}
f_{ij} = f_{ji} = \lim_{h_\ctc{ij} \to 0^-} - \abs{ \pdv{v'(h_\ctc{ij})}{\vb{r}_{ij}}} =  \lim_{h_\ctc{ij} \to 0^-} \frac{\epsilon}{\sigma} \abs{h_{ij}}^{\a-1} > 0 \, .
\end{equation}
The last inequality follows from the fact that at the jamming transitions configurations are \emph{rigid}, which means that there is a non-zero force between particles in contact. This a subtle issue, so let me analyse it more carefully. Recall that in the previous section I mentioned that the value of the bulk modulus at jamming is, in general, finite. This was a consequence of the spheres' stiffness, $k$, remaining finite at jamming. Now, $k$ is proportional to the curvature of the energy minimum at $\va{r}_J$, which must also be positive due to stability. Explicitly, $k = \frac{\epsilon}{\sigma} \abs{h}^{\a-2}>0$, or equivalently, $f_{ij} = k_{ij}^{\frac{\a-1}{\a-2}}>0$. 
Moreover, the finiteness of the contact forces can be deduced from mechanical stability considerations as I will now show. A very important outcome of the proof is that contact forces can indeed be obtained only from the particles' position $\va{r}$, without making any reference to the form of the interaction potential.

By definition, the jammed state corresponds to a state of mechanical equilibrium, and therefore the total force acting on particle $i$ is given by
\[
\vb{F}_i + \sum_{j \in \partial i} \vb{n}_{ij} f_{ij} = 0 
\]
where $\partial i$ is the set of particles in contact with $i$ and $\vb{F}$ is any possible external force. If we denote by $\uv{f}= \{f_\ctc{ij}\}_{\ctc{ij} \in \mathcal{C}} $ the $N_c$-dimensional vector of contact forces, the equilibrium condition can be expressed for the complete configuration as $ \va{F} + \S^{T} \uv{f} = 0 $, where $\S$ is the $N_c\times dN$ \textit{contact matrix}, whose entries read
\begin{equation}\label{def:S matrix}
 \S_\ctc{jk}^{i,\mu} = (\delta_{ij}- \delta_{ik}) n_{jk}^{\mu} 
 \qc \quad \text{ or } \quad \S_\ctc{jk}^i = (\delta_{ij}- \delta_{ik}) \vb{n}_{jk}
 \, .
\end{equation}
Note that the same expression, except for a proportionality faction of $\frac{\epsilon}{\sigma}$, would follow by plugging in the expression given for $\pdv{h_{jk}}{r_i^\mu}$ into Eq.~\eqref{eq:derivative potential at jamming}, with the definition of $f_{ij}$ according to Eq.~\eqref{def:contact forces jamming}. 

From now on, I will focus on systems with periodic boundaries and with no external forces acting on them, but see Ref.~\cite{charbonneauJammingCriticalityRevealed2015,lernerLowenergyNonlinearExcitations2013} for the general analysis. The second of these conditions makes the mechanical equilibrium to be succinctly expressed as
\begin{equation}\label{eq:mechanical equilibrium jamming}
\S^{T} \uv{f} = 0 \, \quad \iff \quad \sum_{\ctc{jk}} S_\ctc{jk}^{i,\mu} f_{jk} = 0 \qc \forall i=1,\dots, N \text{ and } \forall \mu = 1, \dots d \, .
\end{equation}
The last expression corresponds to a system of $dN$ homogeneous linear equations in $N_c$ unknowns. But not all of them are independent because the total force in the system must be zero, due to absence of external forces. Because this condition holds independently for each spatial dimension we have that,
\[
\sum_{i=1}^{N} \ \sum_{\ctc{jk}} S_\ctc{jk}^{i,\mu} f_{jk} = 0 \qc \forall \mu = 1\dots d \, .
\]
Hence, there are only $d(N-1)$ independent equations. But this is precisely the number of (relevant) degrees of freedom! To see why, we just need to consider that the periodic boundary conditions make the system invariant under translations, so $d$ degrees of freedom should be subtracted from the configurational dimension, thus leading to
\begin{equation}\label{eq:Ndof}
N_{dof} = d(N-1) \, .
\end{equation}
Equivalently, the same conclusion is obtained by noting that a uniform translation $\delta \va{r} = \va{a}$ implies that $\delta \vb{r}_{ij}=0$, in which case Eq.~\eqref{eq:perturbation jamming soft spheres} vanishes identically. This means that uniform translations are zero modes of the Hessian, thus effectively reducing the degrees of freedom as in Eq.~\eqref{eq:Ndof}. On the other hand, we already know that a condition valid for all energy minima is that $N_c \geq N_{dof}$. In particular, for a jammed state the mechanical equilibrium condition, Eq.~\eqref{eq:mechanical equilibrium jamming}, defines a set of homogeneous equations that therefore has $\max \{N_c - d(N-1), 0\}$ non-zero solutions. But experience teaches us that jammed packings do exist, so there is least one non-trivial solution to  Eq.~\eqref{eq:mechanical equilibrium jamming}. Therefore the number of contacts must be, at least,
\begin{equation}\label{def:N single self stress}
N_{1SS} \equiv  N_{dof}+1 = d(N-1) +1 \, .
\end{equation}
This condition corresponds to the so called \textit{single state of self-stress} (1SS), although in literature it is more commonly termed \emph{isostaticity}. Nevertheless, I will use the former name and reserve the latter for the case $N_c = N_{dof}$. This shows that, whenever $N_c \geq N_{1SS}$ it necessarily happens that $f_{ij}>0$ $\forall \ctc{ij}\in \mathcal{C}$. 

Before moving forward, I would like to mention that a precise enumeration of the degrees of freedom is of utmost important to correctly analyse jamming critical properties, because several of them manifest when $N_c=N_{1SS}$. In fact, at the beginning of this section I mentioned that the results discussed here were valid for isostatic configurations, but there I implicitly assumed that $N=\infty$. However, the correct statement is that the number of constraints should be 1 above isostaticity. (Of course, in the thermodynamic limit such distinction is irrelevant.) At any rate, to properly count the degrees of freedom it is mandatory to consider the symmetries as extra constraints of the system, which begs the question: should central potentials not further reduce $N_{dof}$? The answer is: it depends on the boundary conditions. In fact, for closed systems, $N_{dof}$ is further reduced by $\frac{d(d-1)}{2}$. However, in systems with periodic boundary conditions angular momentum is \emph{not} conserved\supercite{kuzkinAngularMomentumBalance2015}, and such extra symmetries need not to be included. I am of the mind that these considerations are commonly overlooked, or simply stated without further thought, sometimes leading to apparently different criteria for $N_{1ss}$, \textit{e.g.} \cite{wyartEffectsCompressionVibrational2005,donevPairCorrelationFunction2005,hopkinsDisorderedStrictlyJammed2013,goodrichFiniteSizeScalingJamming2012,haghBroaderViewJamming2019}. (However, once the inherent symmetries of each case are accounted for, the criteria are consistent.) So I wanted to spend some time addressing this issue, at least for the case of periodic boundaries, for which Eq.~\eqref{def:N single self stress} provides the correct value. A careful discussion in other scenarios can be found in Refs.~\cite{goodrichFiniteSizeScalingJamming2012,donevPairCorrelationFunction2005}.

Now, despite its simplicity, Eq.~\eqref{def:N single self stress} is an important result. So let me discuss it further. First, because each contact involves two particles, the average number of contacts is
\begin{equation}\label{eq:avg z isostaticity}
\overline{z}_J = \frac{2N_{1SS}}{N} = 2d + \frac{1-d}{N}  \quad  \underset{N\to\infty}{\longrightarrow}  \quad 2d = z_c \, .
\end{equation}
The critical value of the average coordination number in Eq.~\eqref{eq:coordination near jamming} is thus recovered. As discussed above, this is the smallest number of contacts required for a configuration to be stable. This type of counting argument was first derived by Maxwell, and applies to the stability analysis of a large class of problems, like networks of harmonic springs\supercite{haghBroaderViewJamming2019}. Second, mechanical stability also dictates that each particle must have $d+1$ contacts, otherwise at least one degree of freedom remains unconstrained.
Particles that do not fulfil this bound are called \textit{rattlers} and their contribution to the NC should be removed since they do not contribute to the rigidity of the packing. Nevertheless, they are scarce in the vast majority of the configurations, and their quantity decreases with dimensionality. 
Along these lines, it is an interesting fact that the coordination number is a self-averaging quantity, which means that at jamming, each particle very likely has $2d$ contacts. Moreover, the fluctuations around this value are suppressed rapidly as $d$ increases\supercite{charbonneauJammingCriticalityRevealed2015}.
As a final remark, I should mention that these last expressions make clear that in the thermodynamic limit there is no distinction between isostaticity and 1SS. Nevertheless, such difference will be relevant for the results of Chp.~\ref{chp:fss}, in which the effects of finite size in jammed packings are studied. 
At any rate, the role of the additional contact of 1SS configurations, with respect to isostaticity, cannot be overstated because it implies that $\uv{f}$ is the unique zero mode of $\S$, which only depends on $\va{r}_J$. Consequently,  \emph{the forces magnitudes are entirely determined by the particles' position}. 

Notice that in deriving these results I have not made any assumption about the potential between spheres, thus showing that once a jammed configuration is reached, it is a valid jammed state for any potential. Nevertheless, a direct connection with the energetic approach is possible by noticing that the Hessian can be expressed in terms of a rescaled contact matrix. For instance, in the most usual case of an harmonic interaction, from Eq.~\eqref{eq:Hessian at jamming} and the expression for $\pdv{h_\ctc{kl}}{r_j^\nu}{r_i^\mu}$, it readily follows that
\[
H_{i,\mu}^{j,\nu} = \delta_{ij} \sum_{k\in \partial i} n_{ik}^{\mu} n_{ik}^{\nu} - n_{ij}^{\mu} n_{ij}^{\mu} \delta(\ctc{ij}) = (\S^T \S )_{i,\mu}^{j,\nu}
\]
In this last expression, I have used $\delta(\ctc{ij})$ to denote a function that is 1 if particles $i$ and $j$ are in contact, and 0 otherwise. Even more, for a jammed configuration $\va{r}_J$ obtained with a general potential, we can construct $\S$ and then solve Eq.~\eqref{eq:mechanical equilibrium jamming} for $\uv{f}$. Concomitantly, we can compute the effective stiffness $k_{ij}= \frac{(\a-1)\epsilon}{\sigma} f_{ij}^{\frac{\a-2}{\a-1}}$ and thus define
\begin{equation}\label{eq:scaled S matrix}
\widetilde{\S}_\ctc{jk}^{i,\mu} = k_{jk}^{1/2} (\delta_{ij}-\delta_{ik}) n_{jk}^{\mu} =  k_{jk}^{1/2}  \S_\ctc{jk}^{i,\mu} \, .
\end{equation}
Then, the relation 
\begin{equation}\label{eq:Hessian as contact matrix}
H = \widetilde{\S}^{T} \widetilde{\S}
\end{equation}
for expressing the Hessian in terms of the (rescaled) contact matrix is still fulfilled. Notice that once the Hessian is obtained, we can obtain all the normal modes and the spectrum of the system; see Sec.~\ref{sec:normal modes}. This result will be used in Chp.~\ref{chp:inferring-dynamics}.

As a final comment, it should be mentioned that the contact matrix, the Hessian, and the closely related symmetric matrix
\begin{equation}\label{def:N matrix}
\mathcal{N} = \S \S^{T} \qc
\end{equation}
are very useful to analyse the mechanical properties of jammed configurations. It is easy to show that $H$ and $\mathcal{N}$ have the same spectra. Besides, because $\mathcal{N}$ and $\S$ have the same zero modes, another way to calculate $\uv{f}$ is to extract the eigenvector associated with the zero eigenvalue of $ \mathcal{N} $. Numerically, this is often more convenient\supercite{charbonneauJammingCriticalityRevealed2015}. Moreover, all these matrices provide information about the floppy modes, response to a dipolar force opening a contact, etc. A very complete discussion can be found in \cite{charbonneauJammingCriticalityRevealed2015, lernerLowenergyNonlinearExcitations2013,lernerBreakdownContinuumElasticity2014,degiuliForceDistributionAffects2014}.

\subsection{Criticality of the microscopic structure \emph{at} the jamming point}\label{sec:forces-and-gaps}

Jamming exhibits another, rather peculiar, type of criticality. It involves the contact forces of the NC and the gaps between near contacts. Both of these variables also exhibit a power law scaling once the configuration \emph{has reached} its jamming point. Importantly, MF theory predicts the exponents of such scalings to be universal.
Note that this is different from the criticality discussed in Sec.~\ref{sec:jamming criticality}, because in that case, the scaling relations \eqref{eqs:criticality overjammed} and \eqref{eqs:criticality underjammed} were derived as the jammed state is \emph{approached}. In other words, the usual type of scalings studied in critical phenomena\supercite{MC_book,goldenfeldLecturesPhaseTransitions2018}. In contrast, the results of this part concerns the situation where $\vp=\vp_J$ and a jammed configuration $\va{r}_J$ has already been realized. In this situation, it has been found that the distributions of interparticle gaps and contact forces are power-laws with non-trivial critical exponents.

I begin with the gaps, defined according to Eq.~\eqref{def:gaps}. At variance with the previous section, only positive gaps will be considered because the aforementioned scaling pertains particles that are almost touching. Now, gaps values are randomly distributed because packings are disordered, but theoretical predictions\supercite{fel_2014} state that the distribution of small enough gaps should scale as 
\begin{equation}\label{eq:pdf-gaps}
g(h)\sim h^{-\gamma}, \qquad \text{with } \gamma = 0.41269\dots
\end{equation}
(The distribution $g$ just introduced should not be confused with the radial distribution function, $\tg(r)$, defined in Eq.~\eqref{def:rdf}, although they are closely related; see Eq.~\eqref{eq:rdf at jamming} in the next chapter.) Current numerical evidence supports the universality hypothesis of $\gamma$, since the probability density function (pdf) of Eq.~\eqref{eq:pdf-gaps} has been verified in different dimensions and using different protocols\supercite{charbonneauUniversalMicrostructureMechanical2012,md-code,fel_2014}.

Similarly, the distribution of small contact forces is predicted to scale as, $p(f) \sim f^{\theta}$, but a strong dependence of $\theta$ on dimensionality and jamming protocol was initially reported, in apparent contradiction with the theoretical expectation. This paradox was resolved by recognizing that two different types of forces contribute in this regime\supercite{charbonneauJammingCriticalityRevealed2015,degiuliForceDistributionAffects2014}. This is a consequence of the two distinct responses that can occur when opening the contact between a pair of particles: (i) a localized rearrangement of neighbouring particles; or (ii) a displacement field that extends over the whole configuration, without decaying with distance; see Fig.~\ref{fig:floppy-modes}. The former is associated with a buckling motion, and hence remains localized. The latter is associated with a correlation length of the same order as the system size, and hence is a clear example of the criticality of jammed packings. Considering these two types of forces separately yields two power laws with different exponents, 
\begin{subequations}\label{eq:pdf-forces}
	\begin{align}
	p_\ell(f) & \sim f^{\theta_\ell} , & \text{with } & \theta_\ell\simeq 0.17\qc  \label{eq:pdf-f-loc}  \\ 
	p_e(f) & \sim f^{\theta_e} , & \text{with } &  \theta_e = 0.42311\dots \, ; \label{eq:pdf-f-ext}
	\end{align}
\end{subequations}
for localized and extended excitations, respectively.

The ability of MF theory\supercite{puz_book,fel_2014,mft_review} to predict the non-trivial values of $\gamma$ and $\theta_e$ is a major success. No MF prediction, however, exists for $\theta_\ell$, because bucklers are an intrinsically low-dimensional feature\supercite{charbonneauJammingCriticalityRevealed2015}, and are therefore subdominant in the $d\to\infty$ description. In any case, the MF prediction for $\th_e$ and $\gamma$ is based on the  solutions of the fullRSB differential equations, in the limit of $\Delta \to 0$. The calculations are far from trivial, but when an scaling ansatz is proposed for the form of the solutions, it can be shown that conditions leading to $g$ and $p_e$ are indeed universal. The details of the calculations are presented in \cite{mft_exact_3,franzUniversalitySATUNSATJamming2017,rainoneFollowingEvolutionGlassy2016} and in abbreviated manner in \cite[Secs.~9.2-9.3]{puz_book}. 
On the other hand, besides the theoretical predictions several works provide numerical results supporting the relations of Eqs.~\eqref{eq:pdf-gaps} and \eqref{eq:pdf-forces} in several dimensions and in different scenarios; see  \cite{md-code, charbonneauUniversalMicrostructureMechanical2012,charbonneauJammingCriticalityRevealed2015,degiuliForceDistributionAffects2014,lernerLowenergyNonlinearExcitations2013,franzCriticalEnergyLandscape2020a,franzCriticalJammedPhase2019a,kallusScalingCollapseJamming2016,mft_exact_3,franzSimplestModelJamming2016,franzUniversalitySATUNSATJamming2017}.

\subsection{Marginal stability}\label{sec:marginal stability}

The last aspect I would like to discuss concerns the marginal stability of jammed packings. The mathematical derivation follows mostly Ref.~\cite{lernerBreakdownContinuumElasticity2014}, which offers a more general result than the one presented in a previous, very nice work\supercite{wyartMarginalStabilityConstrains2012}. Additionally, the role of marginal stability in jammed packings and in other common disordered systems has been reviewed recently in \cite{mullerMarginalStabilityStructural2015}.

The argument goes as follows. Let us focus on a given contact $\lambda \equiv \ctc{kl}$ and open it an amount $s$. We are interested in analysing the response of the system to such contact opening, while leaving the rest of them unchanged. The rearrangement of particles caused by opening $\lambda$ is called a \textit{floppy mode}. Notice that if the floppy mode associated to $\lambda$ leads to a denser packing, then the original configuration would be unstable. 
Denoting the displacement experienced by particle $i$ after opening $\lambda$ as $\delta \vb{r}_i^{(\lambda)}$, and $\delta \vb{r}_{ij}^{(\lambda)} = \delta \vb{r}_i^{(\lambda)} - \delta \vb{r}_j^{(\lambda)}$ being the relative displacements, it can be shown that the equation such displacements must obey, up to the leading nonlinear contribution, reads:
\begin{equation}\label{eq:displacements contact opening}
\delta \vb{r}_{ij}^{(\lambda)} \cdot \vb{n}_{ij} + \frac{\qty(\delta \vb{r}_{ij}^{(\lambda)} \cdot \vb{n}_{ij}^{\perp})^2}{2 r_{ij}} + \order{s^3} = s \delta_{\ctc{ij},\lambda} \qc
\end{equation}
where $\vb{n}_{ij}^{\perp}$ is a vector orthogonal to $\vb{n}_{ij}$. Multiplying this last expression by $f_{ij}$ and summing over all the contacts we obtain the virtual work theorem:
\begin{equation}\label{eq:virtual work theorem jamming}
p\  \delta V^{(\lambda)} \approx  s f_\lambda - N \frac{c_\lambda \avg{f}}{\sigma} s^2  \, .
\end{equation}
In this equation $ \delta V^{(\lambda)}$ is the volume change caused by the force unbalance when the contact $\lambda$ was opened, while $c_\lambda$ is a dimensionless number, containing the total contribution of the $ \qty(\delta \vb{r}_{ij}^{(\lambda)} \cdot \vb{n}_{ij}^{\perp})^2$ terms, in the limit $s\to 0$. Physically, it is related to the magnitude of the displacement field averaged over the whole system and, as argued in Ref.~\cite{wyartMarginalStabilityConstrains2012}, its value is determined by average force in the system, and is of order $s^2$.

Notice that the nonlinear term is important, otherwise Eq.~\eqref{eq:virtual work theorem jamming} would imply that any floppy mode would increase the volume and therefore packings would be inherently stable. In contrast, when such term is taken into account it is easy to see that there exists $s^*$ such that for
\begin{equation}\label{eq:def s-star}
s< s^{*} \equiv \frac{\sigma f_\lambda}{N c_\lambda \avg{f}} \qc 
\end{equation}
no packing with smaller volume can be formed. To prevent such situation, a new contact must be formed elsewhere in the configuration before $\lambda$ is opened as much as $s^*$. Denoting by $s^\dagger$ the distance at which such new contact is formed, a configuration will be stable as long as $s^\dagger < s^*$. Otherwise, a denser packing could be formed before another contact is closed, signalling that the configuration was initially unstable. The conditions for stability are thus derived by (i) estimating $s^\dagger$ using the gaps distribution, and  (ii) computing $s^*$ from the distributions of the contact forces. It is convenient to consider a \textit{stability index}
\begin{equation}\label{def:stability index}
\Sigma_\lambda \equiv s^* / s^\dagger = \frac{\sigma f_\lambda}{N s^\dagger c_\lambda \avg{f}} \qc 
\end{equation}
so that $\Sigma_\lambda$ greater (smaller) than 1 corresponds to a stable (unstable) packing. It is then reasonable that we centre our analysis on the smallest forces, since they are likeliest to produce destabilising floppy modes. 

\begin{figure}[htb!]
	\centering
	\includegraphics[width=\linewidth]{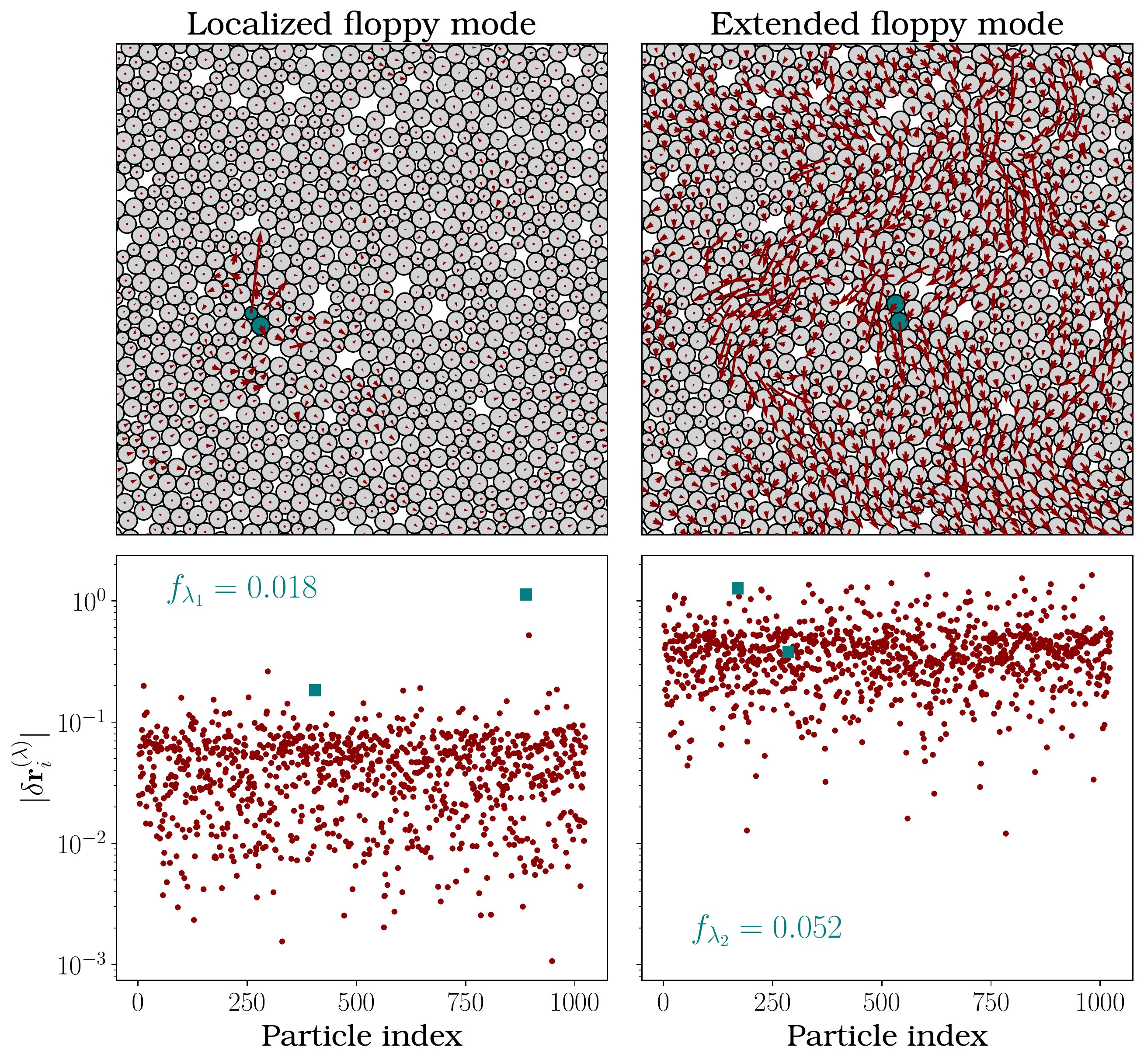}
	\caption[Floppy modes of localized and extended excitations.]{Upper panels: Localized (left) and extended (right) floppy modes obtained by opening the contact involving highlighted particles. The displacement field is obtained following the methods developed in Ref.~\cite{charbonneauJammingCriticalityRevealed2015}.
	Lower panels: Norm	of each particle's displacement in the floppy mode. The data of the particles involved in the contact openings are similarly highlighted. Notice that both forces have a similar magnitude, but their corresponding floppy modes are clearly different.
}
	\label{fig:floppy-modes}
\end{figure}

As argued in Ref.~\cite{lernerLowenergyNonlinearExcitations2013}, there are two (approximately independent) possibilities for having a small force; whence let us write $f_\lambda \sim b_\lambda W_\lambda$, where the meaning of these two factors will be explain next. The first possibility is that a contact force between a pair of particles is weakly coupled to the rest of the forces  exerted on them by the rest of their neighbours. For instance, if all but one contacts of a particle are nearly coplanar, in which case the remaining force should be orthogonal to that plane and very small. This would lead to a bucking motion, whose displacement field thus remains mostly localized, as mentioned in the previous part; see also the analysis of Ref.~\cite{charbonneauJammingCriticalityRevealed2015}. If we let $b_\lambda$ measure the median of the displacements, then it is clear that for these forces $b_\lambda \ll 1$. More precisely, it can be shown\supercite{lernerLowenergyNonlinearExcitations2013} that a localized floppy mode is such that $b_\lambda \ll 1/\sqrt{N}$. I should emphasize that localized modes produce displacements of considerable magnitude, but only near the originating contact. It is farther away that the displacement field is negligible. This justifies using the median and not the mean for properly identifying them, given that it would be dominated by the large displacements near the opened contact. In contrast, it could also happen that a contact force is small because it is weakly coupled to the \emph{bulk} --not just forces of other neighbour contacts-- and thus produces a very small, but \emph{uniform and non-decaying}, floppy mode. $W_\lambda$ is thus a measure of the bulk coupling to contact $\lambda$. Calculating it is not straightforward, and its definition even depends on the boundary conditions of the system. But in any case, it can be estimated through the response of the full configuration to an extended floppy mode; the details are given in \cite{lernerLowenergyNonlinearExcitations2013}. The two type of floppy modes are illustrated in Fig.~\ref{fig:floppy-modes}, for a configuration of $1024$ disks with periodic boundary conditions. Note that both values of $f_\lambda$ (lower panels) are similar, but the displacement fields they produce are very different.

Another important assumption for $\{b_\lambda\}_{\lambda\in \mathcal{C}}$ and $\{W_\lambda\}_{\lambda\in \mathcal{C}}$ is that, besides being statistically independent, they are power-law distributed as
\begin{equation}\label{eq:power-laws p-b p-W}
\begin{aligned}
p(b) &\sim b^{\th_\ell} \qc \\
p(W) & \sim W^{\th_e} \,.
\end{aligned}
\end{equation}
So the same distributions as the bucklers and extended forces in Eq.~\eqref{eq:pdf-forces}. In fact, if MF theory is not considered, the distributions of contact forces can only be found empirically and the factorization $f_\lambda \sim b_\lambda W_\lambda$ is thus justified \textit{a posteriori}. Besides, given that MF calculations\supercite{mft_exact_3} make no prediction for localized modes, in practice the effects of $\{b_\lambda\}_{\lambda\in \mathcal{C}}$ should be checked after the full NC is known. In any case, the factorization assumed for the forces also allows to split $c$ into two contributions as
\[
c_\lambda \sim b_\lambda^2 + \frac1N \qc 
\]
in such a way that for localized modes $b\ll 1/\sqrt{N}$ we have $c\sim 1/N$, while for extended modes $c_\lambda \sim b_\lambda^2$.

Let us now estimate the maximal displacement $s^\dagger$ from the distribution of gaps. If $h_{min}$ is the smallest gap in the configuration and gaps are distributed according to Eq.~\eqref{eq:pdf-gaps}, then
\[
\frac{1}{\sigma^{1-\gamma}} \int_0^{h_{min}} \dd{h} h^{-\gamma} = \qty(\frac{h_{min}}{\sigma})^{1-\gamma} \sim \frac1N \quad 
\implies \quad  h_{min} \sim \sigma N^{-\frac1{1-\gamma}} \, .
\]
On the other hand, from the arguments above we have that $s^\dagger \sim h_{min}/b_\lambda \sim 
\frac{\sigma}{b_\lambda} N^{-\frac1{1-\gamma}}  $. The last step needed to estimate the stability index, defined in Eq.~\eqref{def:stability index}, is to compute $f_\lambda/\avg{f}$. To do so, let me first consider that opening $f_\lambda$ causes an extended response in the system, so (1) the smallness of $f_\lambda$ is caused by a small value of $W_\lambda$ and (2) $c_\lambda \sim b_\lambda^2$
Assuming the extreme case of which $f_\lambda$ is of the order of the smallest force in the system we have that
\[
\int_0^{f_{min}}  \frac{ \dd{f} f^{\th_e}}{\avg{f}^{1+\th_e}} \sim
\int_0^{W_{min}}  \frac{ \dd{W} W^{\th_e}}{\avg{W}^{1+\th_e}} \sim 1/N \quad \implies \frac{W_{min}}{\avg{W}} \sim  \frac{f_\lambda}{\avg{f}} \sim N^{-\frac{1}{1+\th_e}} \qc 
\]
where the $\avg{\bullet}$ terms are introduced for normalization purposes.
Plugging these results into Eq.~\eqref{def:stability index} we obtain
\[
\Sigma_\lambda \sim \frac{f_{min}}{N h_{min}} \sim \frac{N^{\frac{\gamma}{1-\gamma}}}{N^{\frac{1}{1+\th_e}}} \, .
\]
Analogously for a localized mode, if $f_\lambda$ is small due to a very small value of $b_\lambda$, a worst case scenario analysis leads in this case to
\[
\Sigma_\lambda \sim \frac{f_{min}^2}{\frac1N h_{min}} \sim \frac{N^{\frac{1}{1-\gamma}}}{N^{\frac{2}{1+\th_\ell}}} \, .
\]
From these two last expressions, it is easy to verify that if we require $\Sigma_\lambda>1$ to guarantee stability, then the following pair of inequalities must hold:
\begin{subequations}\label{eq:marginality-bounds}
	\begin{align}
	\gamma & \geq \frac{1}{2+\theta_e} \qc  \label{eq:gamma_theta_ext}\\
	\gamma & \geq \frac{1-\theta_\ell}{2}\ . \label{eq:gamma_theta_loc}
	\end{align}
\end{subequations}

Values derived from MF theory, reported in Eqs.~\eqref{eq:pdf-gaps} and \eqref{eq:pdf-f-ext}, as well as estimations from simulations suggest that Eq.~\eqref{eq:gamma_theta_ext} is actually an equality. The same is valid for \eqref{eq:gamma_theta_loc}, although in this case the value of $\th_\ell \approx 0.17$ is obtained directly from numerical results. Therefore, given that both inequalities are saturated, we conclude that jammed packings are \emph{marginally} stable; see \cite{mullerMarginalStabilityStructural2015} for a detailed discussion. Importantly, this result is consistent with the MF picture mentioned above, which always locate the jamming line within the Gardner phase. Recall that, as discussed in Sec.~\ref{sec:MF gardner transition}, the meta-basin structure acquired by the free energy landscape is a consequence of the (fullRSB) decomposition of a basin into an infinite hierarchy of smaller sub-basins, where each minimum identifies a marginal state. Marginality is observed, for example, in the divergence of $\chi_4$ throughout the Gardner phase. Jammed states thus inherits the marginality property, but it is not longer visible through a correlation of dynamical variables, since no dynamics takes place whatsoever. In turn, it is a feature exhibited by the microscopic structure of the network of contacts, as I have just argued. More precisely, a mathematical analysis of the fullRSB equations that fix the values of the exponents $(\th_e, \gamma, \kappa)$ show that they are equivalent to the marginal stability condition; see \cite{fel_2014,mft_exact_3}.

\subsubsection{Density of states at jamming}\label{sec:normal modes}

As a final topic, I will briefly consider the vibrational density of states (DOS), $D(\omega)$. In amorphous solids, the DOS has been extensively studied (see Refs.~\cite{wyartGeometricOriginExcess2005,silbertVibrationsDivergingLength2005,wyartEffectsCompressionVibrational2005,silbertNormalModesModel2009,lernerStatisticsPropertiesLowFrequency2016,franzUniversalSpectrumNormal2015,charbonneauUniversalNonDebyeScaling2016,ikedaUniversalNonmeanfieldScaling2019,mizunoContinuumLimitVibrational2017,degiuliEffectsCoordinationPressure2014,manningVibrationalModesIdentify2011,arceriVibrationalPropertiesHard2020} for a \emph{partial} selection) because the normal modes and their spectrum control several physical quantities of interest, such as specific heat, transport properties, response to shear on compression. For a well defined Hamiltonian, the normal modes are obtained easily by diagonalizing the corresponding Hessian. The spectrum (\textit{i.e.} the set of frequencies $\{\omega\}_{i=1}^{dN}$) is simply related to the corresponding eigenvalues ($\{\lambda_i\}_{i=1}^{dN}$) of the Hessian by $\omega_i = \sqrt{\lambda_i}$. Whenever $\omega_i=0$, the corresponding mode is termed \textit{soft mode} to emphasize that particles can be perturbed at no energy cost.

It has been known for some time now that in glasses and amorphous solids $D(\omega)$ differs from the expected behaviour of a crystal, namely, the Debye scaling at low frequencies, $D_\text{crystal} \sim \omega^{d-1}$, with $d$ the dimensionality of the system. However, the actual from of $D(\omega)$ has been amply debated. For instance, MF analytical calculations\supercite{franzUniversalSpectrumNormal2015} (in the perceptron) and effective medium theory\supercite{degiuliEffectsCoordinationPressure2014} predict that the DOS present different behaviours parametrized by three characteristic frequencies, ($\omega_0,\ \omega_{\max}\,\ \opt{\omega}$), and has approximately the following form\supercite{puz_book}:
\begin{equation}\label{eq:dos}
D(\omega) = \begin{dcases}
0, & \omega \notin [\omega_0, \omega_{\max}] \qc \\
\qty(\frac{\omega}{\opt{\omega}})^2, & \omega_0 \ll \omega \ll \opt{\omega} \qc \\
\text{constant}, & \opt{\omega} \ll \omega \ll \omega_{max} \, .
\end{dcases}
\end{equation}
The role of $\omega_0$ and $\omega_{\max}$ is thus to define the domain of the DOS, while $\opt{\omega}$ determines a characteristic frequency at which the distribution detaches from the constant value plateau. Now, a fullRSB calculation shows that $\omega_0=0$, implying a \emph{gapless} spectrum. Moreover, numerical studies\supercite{silbertVibrationsDivergingLength2005, wyartEffectsCompressionVibrational2005,charbonneauUniversalNonDebyeScaling2016} in both the over- and under-jammed\supercite{arceriVibrationalPropertiesHard2020} phases have shown that $\opt{\omega} \sim (\vp - \vp_J)^{1/2}$. Therefore, \emph{at} jamming, $D(\omega)$ does \emph{not} decay to zero, even as $\omega\to 0$. 

Such behaviour is actually another manifestation of the marginal stability of jammed packings, because a non-vanishing DOS as $\omega \to 0$ implies that excitations of arbitrary low frequencies are always present --apart from the trivial $d$ zero modes from uniform translations. Hence, jammed configurations have an extensive response to external perturbations. See Ref.~\cite{franzUniversalSpectrumNormal2015} for a detailed calculation of $D(\omega)$ and how marginal \emph{mechanical} stability --derived from the 1SS criterion-- and marginal \emph{landscape} stability --characterized by the soft modes of $D(\omega)$-- are related.

For completeness, I should also mention that another regime, proportional to $\qty(\frac{\omega}{\opt{\omega}})^{4}$ has been identified far away from jamming, and related to localized modes. However, the extent of this additional regime has not been fully established due to its dependence on $d$\supercite{arceriVibrationalPropertiesHard2020,ikedaUniversalNonmeanfieldScaling2019}. On the other hand, no theoretical argument has been found to account for such behaviour, given that it mainly is a low dimensional feature, and thus out of the scope of MF theory.

\chapter{Linear Programming algorithm to generate jammed configurations}\label{chp:lp-algorithm}

In this chapter I will describe in detail a method we developed to reach the jamming point of a system through an iteration of Linear Programming (LP) optimizations. LP-based algorithms have been used before to generate jammed packings (see for instance \cite{donevLinearProgrammingAlgorithm2004,krabbenhoftGranularContactDynamics2012}, or \cite{torquatoRobustAlgorithmGenerate2010} for an iterative approach similar to ours). Nonetheless, our implementation provides a simple and robust mechanism to reach the jamming point, as well as accessing the critical structural variables, \textit{i.e.} the network of contact forces and of interparticle gaps. It has been described briefly in Refs.~\cite{paper-dynamics,artiacoExploratoryStudyGlassy2020}, but here I will present a more careful analysis, putting special attention in demonstrating that it produces stable packings.
The code with the algorithmic implementation will be made publicly available shortly.

I should anticipate that the applicability of LP algorithms is limited by the system's size, given that each particle introduces several constraints to be considered for the optimization. Nonetheless, we were able to generate configurations of more than $10^4$ particles with a generic optimization software.
Despite such disadvantage, LP jamming algorithms are a reliable way of producing packings of hard spheres (HS) because the non-overlapping constraints are explicitly introduced in the optimization problem. Carefully exploring HS systems as jamming ensues is important because, as discussed in the previous chapter, most of the numerical studies have focused on soft sphere systems. Therefore, developing algorithms capable of precisely probing the jamming criticality as approached from the unjammed phase is still relevant.

This chapter is structured as follows. First, in Sec.~\ref{sec:LP algorithm} I discuss the connections of jamming with (linear) optimization theory and I show that jammed packings can be produced by iteratively maximising the density of a configuration, which acts as the objective function. To continue, I enumerate few results from Convex optimization theory (Sec.~\ref{sec:convex optimization}) to show afterwards that, upon convergence, our iterative LP algorithm (iLP) generates well defined and mechanically stable jammed packings (\ref{sec:LP network of contacts}). With the \hyperref[alg:LP algorithm]{explicit algorithmic description} of our method, this section contains the main theoretical results of this chapter.
As exemplified by Figs.~\ref{fig:jammed-config-2d} and \ref{fig:jammed configuration}, our algorithm is able to inflate a configuration from a low density phase all the way to its jamming point. 
However, using a configuration with a very high pressure --and thus closer to jamming-- as initial condition drastically enhances its performance. Sec.~\ref{sec:MD for jamming} is devoted to describing an efficient molecular dynamics (MD) compression protocol that allows to produce such high $p$ configurations. In this section I also present some results of the glassy phase of the Mari--Kurchan model, for which I developed the corresponding MD code to implement the compression. By complementing our iLP algorithm with the MD compression, we are able to obtain, rather quickly, \emph{typical} jammed packings; that is, without any signatures of partial ordering.
Later, in Sec.~\ref{sec:characterization MD-LP} I present a recent characterization of our iLP algorithm. I first consider the influence on the jammed state of the final pressure and compression rate of the MD part (Sec.~\ref{sec:LP dependence initial confs}) and show that the results agree with the fractal energy landscape framework\supercite{fel_2014} mentioned in Sec.~\ref{sec:MF gardner transition}; then I describe the phenomenology of the MD+iLP route to jamming (Sec.~\ref{sec:MD-and-LP}) ; and finish by analysing the computational performance of our algorithm as a function of the system's size (Sec.~\ref{sec:LP scaling with size}). Sec.~\ref{sec:conclusions-lp} closes the chapter presenting some conclusions and topics to study in future works.

\section{Jamming as a Linear Programming problem}\label{sec:LP algorithm}

The methodology I will present in this section has been previously described in \cite{tesi-claudia,artiacoExploratoryStudyGlassy2020,paper-dynamics}. Nevertheless, here I will present a complete account of how the iLP algorithm works and show that, upon convergence, it reproduces the conditions for mechanical stability discussed in Sec.~\ref{sec:network of contacts}; specifically, see Eq.~\eqref{eq:mechanical equilibrium jamming}. As mentioned in that section, jammed packings must have a single state of self-stress (1SS), or, in other words, have an extra contact with respect to isostaticity. Naturally, this property is also present in the packings produced with our iLP algorithm.
The idea behind it is fairly simple. Suppose that we begin with a given configuration of $N$ infinitely hard (hyper) spheres (HS) in $d$ dimensions, whose positions\footnote{I will follow the notation introduced in Sec.~\ref{sec:stat-mech-liquids}.} $\va{r}$ and diameters $\va{\sigma}$ are known. Because HS can never overlap, the configuration's packing fraction necessarily satisfies $\vp \leq \vp_J$, where $\vp_J$ is the jamming density that will be eventually reached. Without loss of generality I will assume the strict inequality, otherwise the jammed state has already been realized.

One way to approach the jamming point is to inflate all the particles by a factor $\sqrt{\lpf}>1$. (The reason for including the square root will be evident soon.) Because particles cannot overlap, $\lpf$ is limited by the distance between the closest pair of particles, $\sqrt{\lpf} = \min_{ij} \frac{1}{\sigma_{ij}} \abs{\vb{r}_i - \vb{r}_j}$. But such process would leave the vast majority of particles without neighbours in contact. A more clever strategy would be to let particles move in such a way that a larger value of $\sqrt{\lpf}$ was possible. Even more, we could look for the optimal displacement $\va{s}$ such that, when particles are rearranged as $\va{r}\to \va{r}+\va{s}$, $\lpf$ is maximal. (Because the square root is a monotonic increasing function, if $\lpf$ reaches its maximum, so does $\sqrt{\lpf}$.) Mathematically, this is equivalent to the following optimization problem (OP):
\begin{subequations}\label{def:Jamming as optimization problem}
\begin{align}
\max & \ \lpf\\
\abs{ \vb{r}_i + \vb{s}_i - (\vb{r}_j + \vb{s}_j)}^2 & \geq \lpf \sigma_{ij}^2 \, \quad \forall i\neq j = 1,\dots, N\, . \label {seq:exact non-overlap constraints}
\end{align}
\end{subequations}
In this OP the \textit{design variables}, \textit{i.e.} the variables on which the objective function and constraints depend, are $\va{s}$ and $\lpf$. Additionally, any vector $\va{s}$ that satisfies all the constraints is called \textit{feasible}. Note that by including the square root in the inflating factor, this OP depends on $\lpf$ only linearly. At this point, this is no big advantage, but it will prove useful later.
Expectedly, in such general setup this problem is incredibly hard to solve. Moreover, even if we consider a simplified version where all particles have the same size, we know from Sec.~\ref{sec:jamming in many systems} that the optimal solution is, very likely, a crystal. But this is a configuration we want to avoid in the jamming configurations we are looking for\footnote{This might not be the case in general. For instance, analysing jammed packings with some inherent ordering is interesting; see \cite{jiaoNonuniversalityDensityDisorder2011,torquatoRobustAlgorithmGenerate2010}. Nevertheless, for this thesis only amorphous systems will be relevant.}. To overcome both difficulties and be able to obtain useful jammed packings from this OP some manipulations are needed.

\begin{figure}[!htb]
	\centering
	\begin{subfigure}{0.49\linewidth}
	\includegraphics[width=\textwidth]{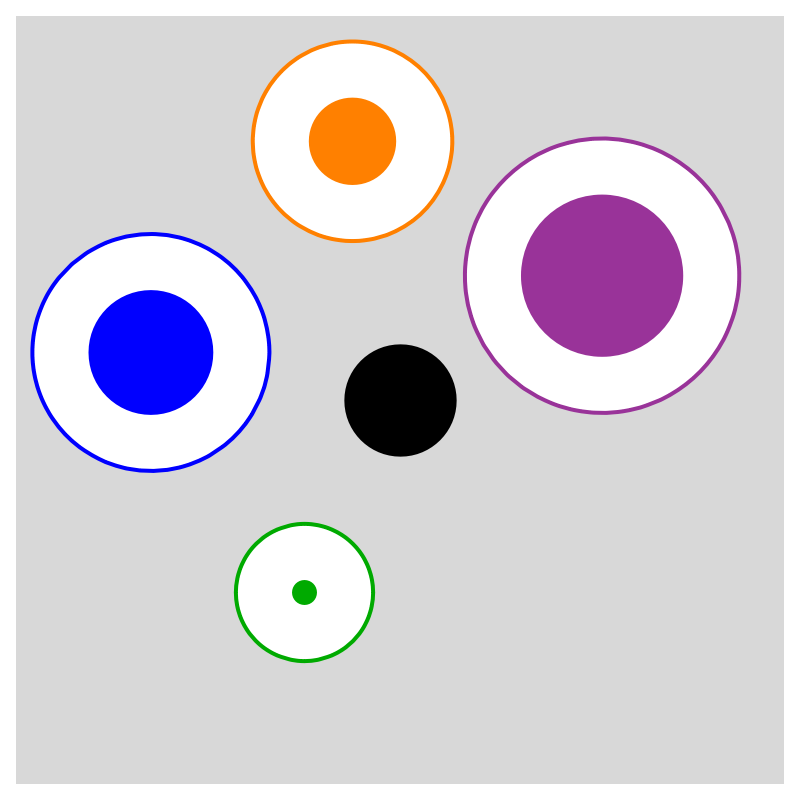}
	\caption{Exact constraints of the OP \eqref{def:Jamming as optimization problem}.\\
		The feasible region is \emph{non}-convex.}
	\label{fig:exact-constraints}
	\end{subfigure}
	\centering
	\begin{subfigure}{0.49\linewidth}
	\includegraphics[width=\textwidth]{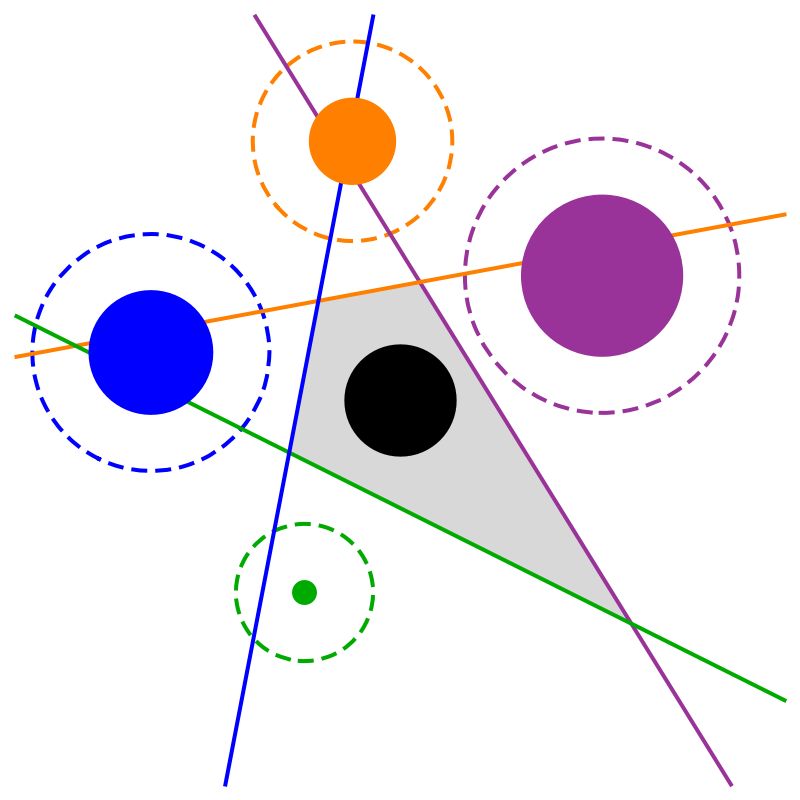}
	\caption{Linearised (solid) and exact (dashed) constraints.}
	\label{fig:linear-constraints}
	\end{subfigure}
	\caption[Exact and linearised non-overlapping constraints of the optimization problem of HS jamming.]{Exact (left) and linearised (right) non-overlapping constraints according to Eqs.~\eqref{def:Jamming as optimization problem} and \eqref{def:Jamming as LP}, respectively. Each particle induces on the black disk's centre a constraint identified by its colour, with the resulting feasible region shaded in grey. Only in the second case it is convex. Coloured particles are assumed to be fixed, so only the constraints affecting the black one are shown.}
	\label{fig:diagram constraints}
\end{figure}

The first thing to mention is that \eqref{def:Jamming as optimization problem} effectively poses the generation of a jammed packing as an instance of an OP. Its objective function, $\lpf$, is trivial. So what makes the problem hard to solve are the non-overlapping constraints, Eq.~\eqref{seq:exact non-overlap constraints}, because they are \emph{non}-convex inequalities. To see why, let $\vb{r}_{ij}=\vb{r}_i - \vb{r}_j$ be the vector pointing from $\vb{r}_j$ towards $\vb{r}_i$, with an analogous definition for $\vb{s}_{ij}$ and $\sigma_{ij}=\frac{\sigma_i+\sigma_j }{2}$ the sum of the radii.
For a given pair of particles, inequality \eqref{seq:exact non-overlap constraints} thus reads
\begin{equation}\label{eq:exact constraints}
	\abs{\vb{s}_{ij}}^2 + 2 \vb{r}_{ij}\cdot \vb{s}_{ij} - \lpf \sigma_{ij}^2 = 
	\va{s} \cdot Q \cdot \va{s} + 2 \va{b} \cdot \va{s} - \lpf \sigma_{ij}^2 \geq - \abs{\vb{r}_{ij}}^2 \, .
\end{equation}
Where $\va{b} = (\vb{0}, \dots, \vb{r}_{ij}, \vb{0}, \dots, -\vb{r}_{ij}, \vb{0}, \dots ,\vb{0})$ is a $dN$ dimensional vector whose only non-zero entries are the $d$ components associated to particles $i$ and $j$. Similarly, $Q$ is a $dN\times dN$ symmetric matrix with entries
\[
Q_{k,\mu}^{l,\nu} = \begin{dcases}
\delta_{\mu,\nu} & \text{if } i=k \text{ or } j=l \qc \\
- \delta_{\mu,\nu} & \text{if } j=k \text{ or } i=l \qc \\
0 & \text{ otherwise.}
\end{dcases}
\]
In other words, it is a matrix of $N\times N$ blocks, each of dimension $d\times d$. The $i$th and $j$th ones along the diagonal are $\mathbb{I}_d$ (the identity matrix in $d$ dimensions), while the ones in the $i$th and $j$th rows and column are $-\mathbb{I}_d$. All the remaining entries are zero.
%
Now, linear terms are always convex, so what determines the convexity of the constraints is the quadratic term, $\va{s} \cdot Q \cdot \va{s}$. But $Q$ is a positive semi-definite matrix, since for any vector $\va{x}\in \mathbb{R}^{dN}$, $\va{x}\cdot Q \cdot \va{x} = \abs{\vb{x}_i - \vb{x}_j}^2 \geq 0$. Therefore the right hand of Eq.~\eqref{eq:exact constraints} \emph{is} a convex function of the design variables. It is thus anticlimactic that the \emph{non}-convexity is rooted in the inequality sign. More precisely, the \textit{feasible domain} of the OP \eqref{def:Jamming as optimization problem} is a non-convex set, as a consequence of the inequality sign of its constraints. Fig.~\ref{fig:exact-constraints} illustrates this feature in the case of 4 fixed (coloured) disks constraining the available location of another  movable (black) particle, all of different sizes. In this diagram, the set of constraints \eqref{eq:exact constraints} corresponds to the white space inside the four circles, and the feasible region --\textit{i.e.} the space where the centre of the black disk is allowed to be-- is shaded in grey. It is clearly non-convex, since there are many line segments that pass through the space excluded by the constraints. 

%

Fortunately, we can sidestep the difficulty of solving a non-convex OP by assuming that the given configuration $\va{r}$ and its diameters $\va{\sigma}$ are sufficiently close to the jamming point (a method to achieve this will be discussed in Sec.~\ref{sec:MD for jamming} below). In this situation, the quadratic term can be neglected and thus the original OP can be transformed into a Linear Programming (LP) problem,
\begin{subequations}\label{def:Jamming as LP}
\begin{align}
\max &\ \lpf \\
 2 \vb{r}_{ij}\cdot \vb{s}_{ij} - \lpf \sigma_{ij}^2 &  \geq - \abs{\vb{r}_{ij}}^2 , \quad \forall i\neq j = 1,\dots, N\, . \label {seq:linear non-overlap constraints}
\end{align}
\end{subequations}
As mentioned few lines above, OP's involving only linear functions of the design variables are always convex, for which many efficient solving algorithms exist; see, \textit{e.g.} \cite{boydConvexOptimization2004,vanderbeiLinearProgrammingFoundations2014,luenbergerLinearNonlinearProgramming2016,nocedalNumericalOptimization2006}. Moreover, because we have neglected a non-negative term, any feasible solution of \eqref{def:Jamming as LP} also satisfies the constraints of the original OP \eqref{def:Jamming as optimization problem}. 
Geometrically, each of the linear non-overlapping constraints defines the half-space of an hyper-plane perpendicular to the segment joining the centres of the corresponding pair of particles. Hence, the complete set of constraints forms a polytope as sketched in Fig.~\ref{fig:linear-constraints}. Given that each bisection of space by a hyper-plane is itself convex, and that the intersection of convex sets is also convex\supercite{boydConvexOptimization2004}, the resulting polytope is guaranteed to be convex. 

On the other hand, neglecting the quadratic term causes any solution of the LP problem to be sub-optimal with respect to the original OP. In other words, an optimal solution ($\opt{\va{s}}, \opt{\lpf}$) of \eqref{def:Jamming as LP} is (very likely) not a jammed configuration. Yet, the new configuration thus produced $\va{r}'=\va{r}+\va{s}$, with diameters $\va{\sigma}' = \sqrt{\opt{\lpf}} \va{\sigma}$, is definitely closer to the jamming point. Then, if this new configuration is used to define another instance of the LP \eqref{def:Jamming as LP}, the second optimal solution will be even closer to jamming. If this process is repeated, and it converges, then the resulting configuration will correspond to a proper jammed state. This is, essentially, the method we developed to generate jammed packings.
I mention in passing that if the initial condition is random, then this algorithm is also able to overcome, in most of the cases, the issue of possibly ending up with an ordered packing. This is so because the linear constraints impose much tighter bounds to the possible displacements of each particle, even at relatively low densities; cf. the feasible regions of both figures in Fig.~\ref{fig:diagram constraints}. This means that at each LP step, only local changes are performed, which are consequently unable to rearrange a disordered configuration into a regular one. For this to happen, major collective rearrangements are needed, which are significantly disfavoured in our algorithm. Put it simply, it means that by linearising the non-overlapping constraints, our algorithm suppresses jumps across (entropic) barriers. The only exception we have found so far are configurations of monodisperse disks, because they are particularly prone to partially crystallize. But it is widely known that crystallization in monodisperse disks is very hard to avoid, so it is not an issue specific to our algorithm. In contrast, in $3\leq d\leq5$ we have verified that with this method crystallization does not occur if a random initial condition is used. 

In summary, our algorithm works as follows (see however Alg.~\ref{alg:LP algorithm} for more detailed description):
\begin{enumerate}
	\item Choose an initial random configuration $\va{r}$, and its set of diameters $\va{\sigma}$.
	\item Use $(\va{r},\va{\sigma})$ to construct an instance of the LP problem \eqref{def:Jamming as LP}.
	\item Obtain the optimal solution of the LP problem \eqref{def:Jamming as LP}, corresponding to displacements $\opt{\va{s}}$ and growing factor $\opt{\lpf}$.
	\item \texttt{If} $\opt{\va{s}}= \va{0}$ and $\opt{\lpf}=1$ \texttt{break} (convergence was reached).
	\item \texttt{else}
		\begin{enumerate}
			\item Update the particles position according to $\va{r} \to \va{r} + \opt{\va{s}}$ and $\va{\sigma} \to \sqrt{\opt{\lpf}} \va{\sigma}$.
			\item Repeat steps 2-5.
		\end{enumerate}
\end{enumerate}
However, the considerations of the previous paragraph hint that the solution thus obtained is not globally optimal, since only a “local search” has been performed. Nevertheless, from the discussion of the previous chapter, we know that jammed states correspond to local minima of the free energy landscape (FEL), because they are obtained from the metastable glass states. Even more, recall that because of their abundance, local minima actually dominate the probability measure. So the local optimality obtained through our method is an expected property from any other technique and, moreover, it is a desired feature for sampling correctly the FEL of HS.
Note also that in our algorithm convergence is reached when the only solution to \eqref{def:Jamming as LP} is $\opt{\va{s}}=0$ and $\opt{\lpf}=1$, and this implies that particles cannot be further inflated because all the degrees of freedom are blocked. Heuristically, this is precisely what defines a jammed state according to the discussion of Sec.~\ref{sec:jamming-transition}. 
Furthermore, upon convergence, the final configuration obtained from a sequence of LP optimizations of \eqref{def:Jamming as LP} coincides with a (possibly local) optimal solution of \eqref{def:Jamming as optimization problem}. 

All of these arguments hint that our algorithm is capable of producing jammed packings, although so far I have not shown that such packings are stable, nor that they fulfil the mechanical equilibrium condition. To do so, I will introduce some mathematical results of convex optimization theory. Then in Sec.~\ref{sec:LP network of contacts} I will show that from such results the full network of contacts can be obtained and the stability of packings can be established (in the sense of the results of Sec.~\ref{sec:network of contacts}). Once again, linear constraints will play a crucial role because their associated dual variables correspond to the contact forces in the jammed configuration.

\subsection{Some results of Convex Optimization} \label{sec:convex optimization}

Here I will mention (without proving) some mathematical results needed to analyse the solution obtain from the iLP algorithm described above. I will follow Ref.~\cite{boydConvexOptimization2004}, a very good reference on convex optimization problems, both from a theoretical and practical point of view.

Let me first introduce some basic terminology. Consider the following general OP:
\begin{subequations}\label{def:OP}
\begin{align}
\min & \ F_0(\va{x})  \label{eq:obj OP}  \\
\text{subject to } F_a(\va{x}) & \leq 0 \qc \quad a=1,\dots, m \,;   \label{eq:ineqs OP} \\
\qquad H_a(\va{x}) & = 0\qc \quad a=1,\dots, q\, ; \label{eq:eqs OP}
\end{align}
\end{subequations}
where $\va{x}\in \mathbb{R}^n$. The function $F_0$ is called the \textit{objective function}, while $F_a$ and $H_a$ are the constraint equations. The domain of the OP \eqref{def:OP} is $\mathcal{D} = \cap_{a=0}^m \text{dom } F_a \cap \cap_{a=1}^p \text{dom }H_a \subseteq \mathbb{R}^n  $.  A point $\va{x}\in \mathcal{D}$ is said to be \textit{feasible} if it satisfies the equality and inequality constraints, while the OP is called feasible if there is at lease one feasible point. The \textit{feasible set} or \textit{feasible domain} is the collection of all possible feasible points. Additionally, the optimal value of the OP \eqref{def:OP}, denoted $\opt{p}$, is defined as
\begin{equation}\label{def:CO optimal}
\opt{p} = \inf \qty{ F_0(\va{x}) \ | \ F_a(\va{x}) \leq 0,\ a=1,\dots,m \ ; \ H_a(\va{x})=0,\ a =1,\dots, q } .
\end{equation}
It is customary to assume that $\opt{p}$ is within the \emph{closed} interval $\opt{p} \in [-\infty, \infty]$, because $\opt{p}=-\infty$ means that $F_0$ is an unbounded objective function, while $\opt{p}=\infty$ for an infeasible OP. Analogously, a point $\opt{\va{x}}$ is termed \textit{optimal point} if (i) $\opt{\va{x}}$ is a feasible point; and (ii) $F_0(\opt{\va{x}}) = \opt{p}$. Finally, given a feasible point $\va{x}$, each of the inequality constraints $\{F_a\}_{a=1}^m$ can be classified as: \textit{active} at $\va{x}$ if $F_a(\va{x})=0$, or \textit{inactive} at $\va{x}$ if $F_a(\va{x}) < 0$. A constraint is \textit{redundant} if the feasible set is unchanged when it is deleted.

All of this applies to any OP, but I will be interested only in \emph{Convex} optimization problems (COP). But first, some extra terminology. A set $C$ is convex if for any $\va{x}_1, \va{x}_2 \in C$ and $0 \leq \th \leq 1$, then $\th \va{x}_1 + (1-\th) \va{x}_2 \in C$. Similarly, a function $F: \mathbb{R}^n \to \mathbb{R}$ is said to be \textit{convex} if its domain, $\text{dom }F$ is convex and, if for all $\va{x}$, $\va{y} \in \text{dom } F$ and $0 \leq \th \leq 1$ we have that
\begin{equation}\label{def:convexity}
F(\th \va{x} + (1-\th)\va{y}) \leq \th F(\va{X}) + (1-\th)F(\va{x}) \, .
\end{equation}
If only the inequality holds in this last expression $F$ is \textit{strictly convex}. On the other hand, $F$ is called (strictly) \textit{concave} if $-F$ is (strictly) convex.
Now, COP's have the same form as Eq.~\eqref{def:OP} but the functions $\{F_a\}_{a=0}^m$ are required to be convex, while the function of the equality constraints are required to be \emph{affine}. That is, $H_a(\va{x}) = \va{k}_a\cdot \va{x} - \va{c}_a$, for constant vectors $\va{k}_a$ and $\va{c}_a$. So, a COP in \textit{standard form} is stated as
\begin{subequations}\label{def:COP}
	\begin{align}
	\min & \ F_0(\va{x})  \label{eq:obj COP}  \\
	\text{subject to } F_a(\va{x}) & \leq 0 \qc \quad a=1,\dots, m \,;   \label{eq:ineqs COP} \\
	\qquad \va{k}_a\cdot \va{x} & = \va{c}_a\qc \quad a=1,\dots, q\, . \label{eq:eqs COP}
	\end{align}
\end{subequations}
Because the intersection of convex sets is convex, the domain of this COP is also convex. This means that we are minimising a convex function over a convex set. Note that this type of problem also includes trivially the case of maximising a \emph{concave} function $\tilde{F}_0$, because it suffices to consider the objective function $-\tilde{F}_0$ and minimise it.
 One of the most important properties of COP's is that \emph{any locally optimal point is also \emph{globally} optimal.} Moreover, if $F_0(\va{x})$ is strictly convex, then the optimal set of \eqref{def:COP} contains at most one point.

LP problems are a particular kind of COP, for which some special results hold. Yet, I will not deal here with such particular cases because the most important result I will use is valid for COP's in general. (But Refs.~\cite{luenbergerLinearNonlinearProgramming2016,vanderbeiLinearProgrammingFoundations2014} contain a detailed account of LP problems and special methods for their solution.) So, to continue, I will mention some results of (Lagrangian) \textit{Duality theory}, where for the sake of generality I first assume a general OP as \eqref{def:OP}. The idea of Lagrangian duality is to incorporate the inequality and equality constraints into a new objective function called the \emph{Lagrangian}, $\mathcal{L}:\mathbb{R}^n \times \mathbb{R}^m \times \mathbb{R}^q \to \mathbb{R}$ and defined as
\begin{equation}\label{def:lagrangian}
\mathcal{L}(\va{x}, \vb*{\lambda}, \vb*{\nu}) = F_0(\va{x}) + \sum_{a=1}^m \lambda_a F_a(\va{x}) + \sum_{a=1}^q \nu_a H_a(\va{x}) \, .
\end{equation}
$\vb*{\lambda}$ and $\vb*{\nu}$ are called \textit{dual variables}. A closely related function is the (Lagrange) \textit{dual function} $G:\mathbb{R}^m \times \mathbb{R}^q \to \mathbb{R}$ defined as the infimum of $\mathcal{L}$ over $\va{x}$, \textit{i.e.}
\begin{equation}\label{def:dual}
G(\vb*{\lambda}, \vb*{\nu}) = \inf_{\va{x}} \qty{  F_0(\va{x}) + \sum_{a=1}^m \lambda_a F_a(\va{x}) + \sum_{a=1}^q \nu_a H_a(\va{x}) } \, .
\end{equation}
It is very important to note that $G$ is defined as a pointwise infimum of affine functions of the dual variables. This implies that $G$ is concave even if the OP \eqref{def:OP} is \emph{not} convex.
Additionally, the dual, $G$, is closely related to the original OP because it satisfies that
\[
G(\vb*{\lambda}, \vb*{\nu}) \leq \opt{p} \, ; \qquad \forall \vb*{\lambda} \succeq \vb{0} \qc \forall \vb*{\nu}.
\]
(I will use $\succeq$ and $\preceq$ to denote the element wise inequalities $\geq$ and $\leq$, respectively.) Hence, $G(\vb*{\lambda}, \vb*{\nu}) = -\infty$ if $\mathcal{L}$ is unbounded from below.

From the concavity of $G$ and the upper bound just given, it follows that by maximising $G$ we can “get close to” or obtain --in the best case-- the optimal value $\opt{p}$ of the original OP. So we can define the \textit{dual} OP of \eqref{def:OP} as
\begin{subequations}\label{def:dual OP}
	\begin{align}
	\max & \ G(\vb*{\lambda}, \vb*{\nu}) \\
	\text{subject to } \vb*{\lambda} & \succeq \vb{0} \, .
	\end{align}
\end{subequations}
In analogy with the \textit{primal} OP -- i.e. \eqref{def:OP}--, $(\vb*{\lambda}, \vb*{\nu})$ are dual feasible if (i) $ \vb*{\lambda}  \succeq \vb{0}$; (ii) $(\vb*{\lambda}, \vb*{\nu}) \in \text{dom } G$; and (iii) $G(\vb*{\lambda}, \vb*{\nu}) > -\infty$. Additionally, let $\opt{d}$ and $(\opt{\vb*{\lambda}}, \opt{\vb*{\nu}})$ denote the optimal value and optimal solution, respectively, of \eqref{def:dual OP}. Whenever $\opt{d}=\opt{p}$ we say that we have \textit{strong duality}.

Strong duality may occur even for non-convex problems, but in the case of COP \emph{and when the inequality constraints $\{F_a\}_{a=1}^m$ are affine}, it is possible to show that strong duality holds. This is a corollary of Slater's Theorem; see \cite[Secs.~5.2-5.3]{boydConvexOptimization2004}. But, in general strong duality yields a very useful relation between constraints and dual variables:
\[
\begin{aligned}
\opt{p} & = F_0(\opt{\va{x}}) = G(\opt{\vb*{\lambda}}, \opt{\vb*{\nu}}) = \opt{d}\\
& = \inf_{\va{x}} \qty{  F_0(\va{x}) + \sum_{a=1}^m \opt{\lambda}_a F_a(\va{x}) + \sum_{a=1}^q \opt{\nu}_a H_a(\va{x}) }\\
& \leq F_0(\opt{\va{x}}) + \sum_{a=1}^m \opt{\lambda}_a F_a(\opt{\va{x}}) + \sum_{a=1}^q \opt{\nu}_a H_a(\opt{\va{x}})\\
& \leq F_0(\opt{\va{x}}) = \opt{p} \qc 
\end{aligned}
\]
where the last line follows from the fact that $H_a(\va{x})=0$, while $\opt{\vb*{\lambda}}\succeq \vb{0}$ and $F_a(\va{x})\leq 0$, making $ \sum_{a=1}^m \opt{\lambda}_a F_a(\opt{\va{x}})$ a sum of non-positive terms. But from the first and last lines, it is obvious that this sum equals zero. That is,
\begin{equation}\label{eq:complementary slackness}
\sum_{a=1}^m \opt{\lambda}_a F_a(\opt{\va{x}}) = 0 \quad \implies \quad  \opt{\lambda}_a F_a(\opt{\va{x}}) = 0 \quad \implies
 \ \begin{cases}
\opt{\lambda}_a > 0 & \implies F_a(\opt{\va{x}})=0 \\
F_a(\opt{\va{x}}) < 0 & \implies \opt{\lambda}_a =0
\end{cases} \, .
\end{equation}
This property is called \textit{complementary slackness} and it is essential for relating the dual variables with the contact forces in the LP problem \eqref{def:Jamming as LP}.

A final results that I will make use of are the so called Karush--Kuhn--Tucker (KKT) optimality conditions. From them, the stability analysis of the packings generated through our iLP algorithm follows easily. They are stated in the following
\begin{theorem}[KKT optimality conditions]\label{theo:kkt}
Assume $\{F_a\}_{a=0}^m$ and $\{H_a\}_{a=1}^q$ are differentiable, and let $\opt{\va{x}}$ and  $(\opt{\vb*{\lambda}}, \opt{\vb*{\nu}})$ be any primal optimal and dual optimal points, respectively, for which strong duality holds. We know that since $\opt{\va{x}}$ minimizes $\mathcal{L}(\va{x}, \opt{\vb*{\lambda}}, \opt{\vb*{\nu}} )$ its gradient vanishes
\begin{equation}\label{eq:kkt gradient}
\nabla \eval{\mathcal{L}(\va{x}, \opt{\vb*{\lambda}}, \opt{\vb*{\nu}} )}_{\opt{\va{x}}} = \nabla F_0(\opt{\va{x}}) + \sum_{a=1}^m \opt{\lambda}_a \nabla F_a(\opt{\va{x}}) + \sum_{a} \opt{\nu}_a \nabla H_a(\opt{\va{x}}) = 0 \, .
\end{equation}
Whence the KKT conditions follow
\begin{enumerate}
	\item \label{it:kkt compl slack} $F_a(\opt{\va{x}}) \opt{\lambda}_a = 0$, so either $F_a(\opt{\va{x}}) \leq 0$ or  $\opt{\lambda}_a \geq 0$;  for all $a=1,\dots,m$.
	\item $H_a(\opt{\va{x}})= 0$ for all $a=1,\dots, q$.
	\item \label{it:kkt zero grad} $ \nabla F_0(\opt{\va{x}}) + \sum_{a=1}^m \opt{\lambda}_a \nabla F_a(\opt{\va{x}}) + \sum_{a} \opt{\nu}_a \nabla H_a(\opt{\va{x}}) = 0$ .
\end{enumerate}
\end{theorem}

The KKT conditions are necessary conditions to be fulfilled by $\opt{\va{x}}$  and $(\opt{\vb*{\lambda}}, \opt{\vb*{\nu}})$ in any OP. However, in the case of a COP where the objective and inequality constraint functions are differentiable, such conditions are also \emph{sufficient}. These relations are important, because in several cases, it is easier to solve the KKT conditions for $\va{x}$ and the dual variables than the original COP. Notice also that condition (\ref{theo:kkt}-\ref{it:kkt compl slack}) just restates the complementary slackness condition derived above. That property as well as (\ref{theo:kkt}-\ref{it:kkt zero grad}) lie at the basis of the proof of the stability of jammed packings obtained through LP, as I will show next. However, their relevance for physical applications is not limited to the problem discussed in this thesis, but actually is of great generality. Indeed, if $F_0$ is a potential function, while $\{F_a\}_{a=0}^m$ and $\{H_a\}_{a=1}^q$ are physical constraints, condition  (\ref{theo:kkt}-\ref{it:kkt zero grad}) is nothing but the force balance equation for each degree of freedom and the dual variables play precisely the role of contact forces. See \cite[Sec.~5.5]{boydConvexOptimization2004} for a more detailed discussion of the KKT conditions as well as other applications to physical and other type of problems.

\subsection{Stability and network of contacts from iLP}\label{sec:LP network of contacts}

Using the KKT theorem \eqref{theo:kkt}, it is now easy to analyse the stability of packings obtained upon convergence of the LP algorithm described above. In principle, this means that we need to apply the KKT conditions (\ref{theo:kkt}-\ref{it:kkt compl slack}) and (\ref{theo:kkt}-\ref{it:kkt zero grad}) in the limit $\opt{\va{s}\to 0}$, $\opt{\lpf} \to 1$. (Because the LP problem \eqref{def:Jamming as LP} does not contain equality constraints, there is no need to consider the second condition of Theorem \eqref{theo:kkt}.) Let me begin with the latter of these conditions, involving the gradient with respect of the design variables, \textit{i.e.}
\[
\nabla = \qty(\pdv{\vb{s}_1}, \ \pdv{\vb{s}_2}, \ \dots, \pdv{\vb{s}_N},\ \pdv{\lpf}) \, .
\]
Notice that in order to convert the jamming LP problem \eqref{def:Jamming as LP} into a COP in standard form, like \eqref{def:COP}, we need to minimise the objective function $F_0= - \lpf$.
Hence,
\begin{equation} \label{eq:grad objective}
\nabla F_0 = \qty(0, \dots, 0, -1) \, .
\end{equation}
Next, letting 
\[
F_{(i,j)}(\va{s}, \lpf) = F_{(j,i)}(\va{s}, \lpf)= -2 \vb{r}_{ij} \cdot \vb{s}_{ij} + \lpf \sigma_{ij}^2 - \abs{\vb{r}_{ij}}^2 \leq 0 \qc \text{ with } i\neq j
\]
be the linear constraints associated to \eqref{seq:linear non-overlap constraints} in standard form, we have that
\begin{equation}\label{eq:grad constraints}
\nabla F_{(ij)} = \qty(0, \dots, 0, \underbrace{-2 \vb{r}_{ij}}_{i\text{th entry}}, 0 \dots, \underbrace{2\vb{r}_{ij}}_{j\text{th entry}}, 0 ,\dots, \sigma_{ij}^2) \, .
\end{equation}

Plugging in Eqs.~\eqref{eq:grad objective} and \eqref{eq:grad constraints} into (\ref{theo:kkt}-\ref{it:kkt zero grad}) and considering the $i$th component of the gradient we have that\footnote{As in Sec.~\ref{sec:network of contacts}, $\partial i$ denotes the set of particles in contact with $i$, while  $\ctc{ij}$, with $i<j$ is the index of the contact between particles $i$ and $j$. The full network of contacts is denoted as $\mathcal{C}$ and $N_c = \abs{\mathcal{C}}$ is the number of contacts in a jammed configuration.}
\[
0 = \sum_{j\neq i}^{1,N} -2 \opt{\lambda}_{(i,j)} \vb{r}_{ij} = 
-2 \sum_{ \substack{j \in \partial i, \\ \ctc{ij}\in \mathcal{C} }} \opt{\lambda}_{(i,j)} \vb{r}_{ij} \ \qc 
\]
where the last equality follows from the complementary slackness condition, \eqref{eq:complementary slackness}. In this case, such condition means that the sum over all active constraints --\textit{i.e.} the constraints whose dual variables are positive-- is equivalent to a sum over all \emph{linear} contacts. Linear, because they saturate the linear non-overlapping constraints, $F_{(i,j)}$ defined above. Nevertheless, at convergence $\opt{\va{s}}=0$ and $\opt{\lpf}$=1, so $F_{(i,j)}(\opt{\va{s}},\opt{\lpf}) = 0$ clearly implies $\abs{\vb{r}_{ij}} = \sigma_{ij}$. In other words, in the final configuration of the iLP algorithm, linear contacts are equivalent to real physical contacts. But the correspondence is not only geometrical, but also mechanical. Rescaling the dual variables as  $\opt{\lambda}_{(i,j)} = \frac{f_{ij}}{\abs{\vb{r}_{ij}}}$, where $f_{ij}$ are yet unknown (although they clearly correspond to the contact forces), the last expression reads
\begin{equation}\label{eq:force balance LP}
0 = \sum_{j \in \partial i} \vb{n}_{ij} f_{ij} \, .
\end{equation}
And this is nothing but the mechanical equilibrium condition in the absence of external forces. Moreover, when all the spatial components of $ \sum_\ctc{ij}^{1,N_c} \opt{\lambda}_{(i,j)} \nabla F_{(i,j)} (\va{0}, 1)=0$ are considered, the resulting expressions correspond precisely to the force balance equations of the whole jammed configuration, \textit{i.e.} Eq.~\eqref{eq:mechanical equilibrium jamming}! This equation was derived in Sec.~\ref{sec:network of contacts}, where I also showed that jammed packings are mechanically stable provided that it has a non-zero solution. To verify that this is the case in the output of our algorithm notice that the component of (\ref{theo:kkt}-\ref{it:kkt zero grad}) corresponding to $\pdv{\lpf}$ fixes the scale of the contact forces:
\begin{equation}\label{eq:grad lpf}
\sum_{\ctc{ij} \in \mathcal{C}} \frac{\sigma_{ij}^2 f_{ij}}{\abs{\vb{r}_{ij}}} - 1 = 0 \quad 
\implies \quad \sum_{\ctc{ij} \in \mathcal{C}} f_{ij} \sigma_{ij} = 1 \, \qc
\end{equation}
with the last equation being valid at jamming. But given that all dual variables are non-negative, this last expression implies that at least one of them is positive. Hence, the homogeneous solution to the force balance equation is never a solution of the LP problem. From the arguments of Sec.~\ref{sec:network of contacts} we can directly conclude that jammed packings thus produced are mechanically stable because their number of contacts satisfies $N_c\geq N_{dof}+1= d(N-1)+1$, where I assumed periodic boundary conditions for the last relation. 

Here, two remarks are in order. First, although it is possible that more contacts are present, in practice we found that in the vast majority of cases this last inequality is saturated, just as with other algorithms. Hence, most of the configurations have a single state of self-stress (1SS). Second, as a consequence of this feature the packings thus obtained are, actually, predominantly isostatic, \emph{if the whole set of design variables are considered},  \textit{i.e.} $(\va{s},\lpf)$. The reason is that $\lpf$ acts just as another degree of freedom upon which the optimization is carried out. Physically, this is equivalent to using the density as an extra variable of the configuration, whence the need of the extra contact can be justified. Or, from another perspective, the extra constraint comes from the requirement of jammed states to be rigid.

Incidentally, for monodisperse systems, or whenever $\sigma_{ij} \sim \overline{\sigma}$ --\textit{i.e.} the distribution of radii is peaked around its mean-- Eq.~\eqref{eq:grad lpf} also determines the typical scale of contact forces. Because for a given configuration contained in a fixed volume its packing fraction scales as $\vp \sim N \overline{\sigma}^3$, Eq.~\eqref{eq:grad lpf} implies that
\begin{equation}\label{eq:scaling avg f}
N \overline{f} \sim \frac{1}{\overline{\sigma}} \quad \implies \overline{f} \sim N^{-2/3}  \, .
\end{equation}

Therefore, the results of this part show that configurations obtained with our method are well defined jammed states (\textit{i.e.} determined by the same mechanical equilibrium conditions introduced in Sec.~\ref{sec:network of contacts} of the previous chapter),  whose stability is guaranteed because they have, at least, 1SS. Moreover, the \emph{active} dual variables corresponding to the non-overlapping constraints provide direct access to the contact forces between particles, whence the full network of contacts can be easily constructed; see Fig.~\ref{fig:jammed-config-2d}.

\subsection{Details of the algorithm and some examples}\label{sec:details algorithm}

\begin{figure}[!htb]
	\centering
	\begin{subfigure}[t]{0.49\linewidth}
		\includegraphics[width=0.95\linewidth]{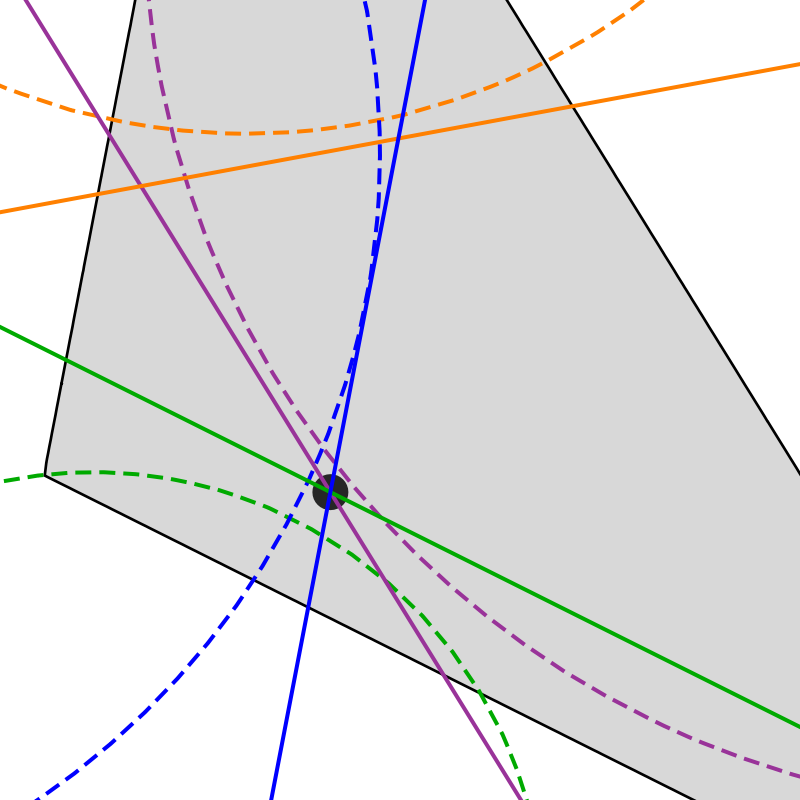}
		\caption{Optimal displacement after 1 LP step. The grey shaded area is the original feasible region, while the solid straight lines (dashed curves) show the linearised (exact) constraints using $\opt{\lpf}$. The position of the disk's centre after one optimization is indicated by the black dot. Clearly, the optimal solution saturates (some of) the \emph{linear} constraints. See main text for a more detailed discussion.}
		\label{fig:LP-solution}
	\end{subfigure} \hfil
	\begin{subfigure}[t]{0.49\linewidth}
		\includegraphics[width=0.95\linewidth]{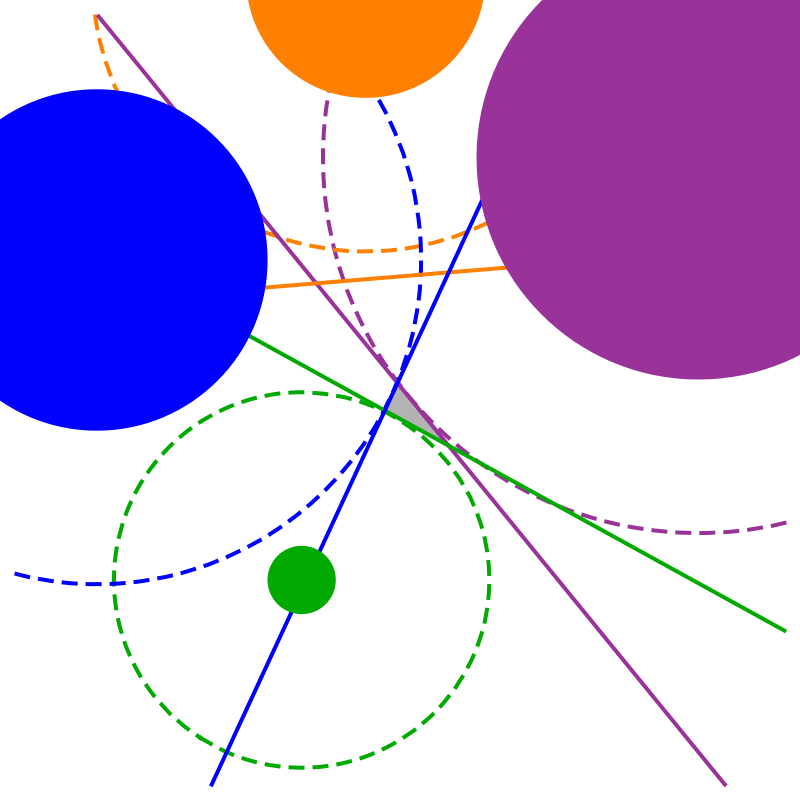}
		\caption{New instance of the LP problem after updating the black particle's position (not shown) and all the radii using the optimal values of displacement and inflation factor. Notice the considerable reduction of the feasible region (cf. Fig.~\ref{fig:linear-constraints}) and that most of the linear constraints (solid) are tangent to the exact ones (dashed). The constraint induced by the orange disk is an example of a redundant constraint.}
		\label{fig:LP-constraints-after-1-step}
	\end{subfigure}
	\caption{Optimal solution and new constraints after the first LP optimization of the linear problem in Fig.~\ref{fig:linear-constraints}.}
	\label{fig:LP-solution-after-1-step}\vspace*{-5mm}
\end{figure}

Some additional remarks are in order to better understand the properties of our algorithm. Adopting first  a mathematical point of view, note that the LP problem \eqref{def:Jamming as LP} satisfies the Slater's condition given that its constraints are linear and thus affine, so strong duality always holds.
This is important because the solution of the primal problem can be obtained simply by maximising its dual, which amounts to solving the set of linear equations obtained from (\ref{theo:kkt}-\ref{it:kkt zero grad}).
Second, such equations hold at \emph{each} step of LP optimization, and not only at convergence. This is due to the fact that neither $\nabla F_0$ nor $\nabla F_{(ij)}$ depend on the design variables, and therefore Eq.~\eqref{eq:force balance LP} is fulfilled even if $\opt{\va{s}}\neq \va{0}$ and $\opt{\lpf}>1$. Thus, at each step, configurations are “mechanically stable” with respect to the linear constraints. Geometrically, this feature represents the fact that linear constraints are always saturated, even if no physical contact occurs between particles. In Fig.~\ref{fig:LP-solution} this situation is illustrated after one optimisation step of the LP problem of Fig.~\ref{fig:linear-constraints}. The initial feasible set (defined by constraints with the initial positions and sizes) appears in grey; the straight lines are the same linear constrains but using the optimal value of the inflating factor, $\opt{\lpf}$; and the dashed curves are initial the exact constraints, but also evaluated with $\opt{\lpf}$ . The centre of the black disk has been translated according to the optimal displacement vector obtained from the solution and is indicated by the black dot. Notice that it saturates three of the linear constraints but \emph{none} of the exact ones, \textit{i.e.} there is not a single real contact, but three “linear” ones. The solution, nonetheless, satisfies “force balance” with respect to this latter type.

This last consideration evinces the importance of performing several steps of LP, until linear contacts are equivalent to the physical ones. For this to happen, it is not enough for the linear constraints to have “zero gap” with respect of the exact ones, as the blue constraint of Fig.~\eqref{fig:LP-solution} shows. Here, both type of constraints coincide at a point, but the optimal solution is not on it, so neither in this case a physical contact is realized. It is only when the optimal position coincides with the tangency points of both type of constraints that real contacts take place. And for this to happen it is necessary that $\opt{\va{s}}=0$ and $\opt{\lpf}=1$. 
On the other hand, Fig.~\ref{fig:LP-constraints-after-1-step} shows that, if particles initially have a lot of free volume, the initial inflation is so large that the feasible region reduces rather quickly. Consequently, linear and exact constraints are expected to match after few steps. However, one should also kept in mind that these figures only consider one movable particle. In the real scenario where all of them are allowed to move more steps are required to ensure real contacts. 
Fig.~\ref{fig:LP-solution-after-1-step} also reveals that constraints that were initially relevant, might become redundant closer to the jamming point (compare orange constraints in Figs.~\ref{fig:linear-constraints} and \ref{fig:LP-constraints-after-1-step}). Even more, the set of inequalities \eqref{seq:linear non-overlap constraints} of the jamming LP problem establishes a non-overlapping constraint for \emph{each pair} of particles. But whenever particles are far apart such constraints are obviously redundant and can be omitted so the performance of the algorithm is improved.
It is therefore convenient to define a radius of influence for each particle, $\ell_i(\vp)$, whose value (possibly) depends on the system's packing fraction. Such dependence is justified because for low densities, most of the particles will be displaced by a large amount, and therefore even if a given pair is not relatively close, it might overlap after such large displacement. Conversely, for densities close to $\vp_J$, only the closest particles need to be considered because very small rearrangements are performed after each LP solution. Note that, in principle, $\ell_i$ could be different for each particle, hence the subscript. This might be useful, for instance, if a given particle is small enough to escape the cage formed only by its closes neighbours. However, such situation never occurred in the configurations considered here, so I will assume $\ell_i = \ell\ \forall i$. With this in mind, in what follows the constraints \eqref{seq:linear non-overlap constraints} are to be included only if $\abs{\vb{r}_{ij}}< \ell (\vp)$. This is the so called “neighbours list” approach, in which a given particle $i$ only interacts with the set of neighbouring particles $\tilde{\partial} i \equiv \{j \text{ such that } \abs{\vb{r}_{ij}} \leq \ell(\vp)\}$. A final remark: for our periodic systems we used the  \textit{nearest image convention}\supercite{donevNeighborListCollisiondriven2005} to compute distances and thus identify the relevant constraints. After all these considerations, our iLP method can be described algorithmically as follows:

\begin{algorithm}[H]\label{alg:LP algorithm}
	\SetAlgoLined
	\KwIn{Particles' position $\va{r}$ and diameters $\va{\sigma} \succ 0$}
	\KwResult{Jammed configuration ($\va{r}_J$, $\va{\sigma}_J$) and network of contacts forces $\mathcal{C} = \{ (f_{ij}, \vb{n}_{ij})  \}_{\ctc{ij}=1}^{N_c}$.}
	\BlankLine
	\BlankLine
	Compute initial density, $\vp$, and radius of influence, $\ell(\vp)$\;
	Define convergence tolerance $(tol_{\vb{s}}, tol_\lpf)$\;
	Initialize displacements $\va{s}$ and inflation factor $\lpf$ so they are outside tolerance range\;
	\BlankLine
	\While{$\max_{i} \abs{\vb{s}_i}> tol_{\vb{s}}$ \textbf{or} $\lpf-1 > tol_\lpf$ }{
		\tcc{Find relevant constraints according to $\ell(\vp)$}
		\nlset{Constrs.}\For{$i=1,\dots, N$}{\label{loop:add-constraints}
		Construct $\tilde{\partial} i$, defined as the set particles $j$ such that $\vb{r}_{ij}<\ell(\vp)$ and $j>i$
		\tcc*[r]{So each potential contact is counted once.}
			\For{$j$ in $\tilde{\partial} i$}{
				Include the corresponding linearised non-overlapping constraint, Eq.~\eqref{seq:linear non-overlap constraints}\;
				}
			}
		\BlankLine
		\nlset{LP step}Solve LP problem \eqref{def:Jamming as LP} with the constraints defined in \ref{loop:add-constraints} and obtain optimal displacements $\opt{\va{s}}$ and inflation factor $\opt{\lpf}$ \label{step:LP} \;
		Store the active dual variables $\opt{\vb*{\lambda}}$\;
		$\va{s} \leftarrow \opt{\va{s}}$ and $\lpf \leftarrow \opt{\lpf}$\;
		Update particles' position $\va{r} \ \leftarrow\ \va{r} + \va{s}$ and diameters $\va{\sigma}\ \leftarrow\ \sqrt{\lpf} \va{\sigma}$\; 
		Update system's packing fraction using the new diameters\;
		Recompute $\ell(\vp)$\;	
		}
	\BlankLine
	\tcc{Construct output variables from data at convergence}
	$(\va{r}_J,\ \va{\sigma}_J)  \leftarrow (\va{r},\ \va{\sigma})$\;
	\For{$\ctc{ij}$ in $\{\opt{\lambda}_{(i,j)} \text{ such that } \opt{\lambda}_{(i,j)} >0 \}$ }{
	$\vb{n}_{ij} \leftarrow \frac{\vb{r}_{ij}}{r_{ij}}$\;
	$f_{ij} \leftarrow \frac{\opt{\lambda}_{ij}}{\sigma_{ij}}$
	}

	\caption{iterative LP  (iLP) algorithm to generate jammed packings}
\end{algorithm}

\begin{figure}[!htb]
	\centering
	\includegraphics[width=\linewidth]{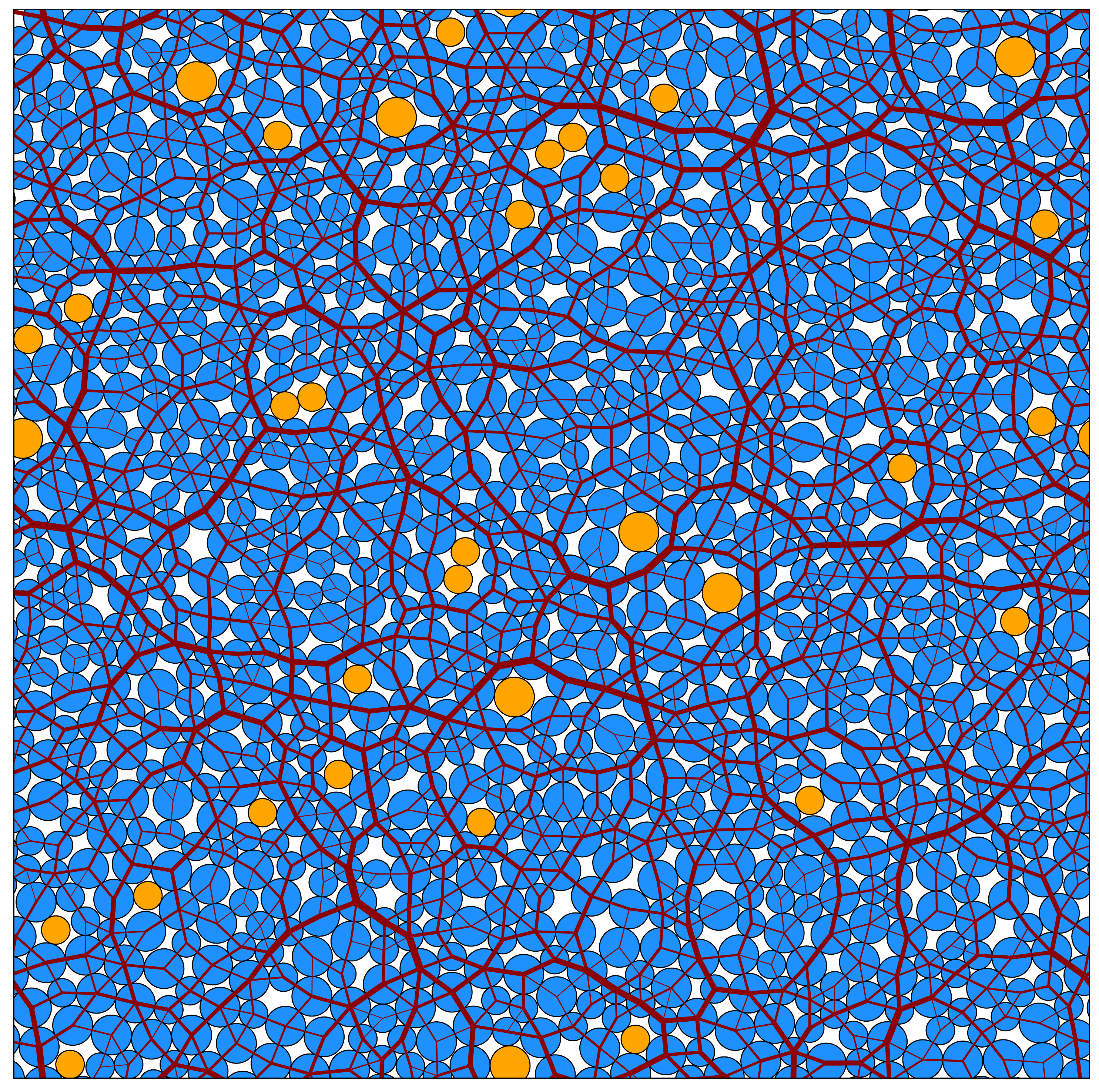}
	\caption[Jammed configuration and network of contact forces of $N=1024$ bidisperse disks and periodic boundary conditions.]{Jammed configuration and network of contact forces of $N=1024$ bidisperse disks and periodic boundary conditions. Rattlers, \textit{i.e.} particles with less than $d+1=3$ contacts, are coloured in orange. The edges of the network are the contact forces, whose thickness indicates the magnitude of the force. The configuration was produced using the Algorithm \ref{alg:LP algorithm} with particles' initial position drawn uniformly from the box area, initial density $\vp_0 = 0.4$, and keeping a big/small radii ratio of 1.4.}
	\label{fig:jammed-config-2d}
\end{figure}

In this way, packings such as the one depicted in Figs.~\ref{fig:jammed-config-2d} and \ref{fig:jammed configuration} can be generated. On the other hand, notice that Alg.~\ref{alg:LP algorithm} does not mention anything about the initial condition. I did it on purpose to stress that our algorithm is robust enough to produce jammed packings even if the initial condition is not very close to the jamming point. In fact, the configuration of Fig.~\ref{fig:jammed-config-2d} was produced with an initial density of $\vp_0 \approx 0.4$ (cf. $\vp_{J,2d} \approx 0.84$), while in the one of monodisperse spheres of Fig.~\ref{fig:jammed configuration} the initial packing fraction was $\vp_0 \approx 0.2$ ($\vp_{J,3d} \approx 0.64$). Naturally, the closer the initial condition is to the jamming point the faster the iLP algorithm converges. 
But besides being a matter of speed, we should also keep in mind the physics simulated during the implementation of our method. Considering that all particles' displacements come from the series of LP solutions and not from any thermal energy, our algorithm acts as an immediate quench to $T=0$ and a very fast compression of the input configuration. In other words, it is a crunching algorithm that, accordingly, does not necessarily follows the glass EOS. And even though the final configuration is a valid jammed state, if only the iLP method is used it may happen that it is rather untypical. For instance, it could be one of the highest minima of the fractal free energy landscape\supercite{artiacoExploratoryStudyGlassy2020}. Therefore, if we want to explore the typical properties of jammed packings it is convenient to use a glass configuration as thermalised as possible. So I will next describe a method to obtain such configurations.

\section{Approaching jamming using molecular dynamics} \label{sec:MD for jamming}

In order to generate a sufficiently well thermalised glass HS configuration we used the Molecular Dynamics (MD) algorithm described in Ref.~\cite{md-code}. Such algorithm works in an event-driven fashion, taking advantage of the fact that particles only interact when they collide with each other and the corresponding collision times can be easily computed. I will first describe why event-driven MD methods are very efficient when using the HS potential, Eq.~\eqref{def:hs-potential}, and then discuss how a compression protocol can be implemented. Besides the standard HS configurations, I will also present results of MD simulations in the Mari--Kurchan (MK) model\supercite{mariDynamicalTransitionGlasses2011} of HS with infinite range random shifts. In this latter case, the simulations were performed from my own implementation of the event-driven MD algorithm. Producing numerous MK glasses near their jamming point was essential for the results of Chp.~\ref{chp:fss}. Expanding on what I mentioned in Sec.~\ref{sec:jamming in many systems}, in the MK model particles interact according to a randomly shifted distance,
\begin{equation}\label{eq:MK distance}
D_\zeta(\vb{r}_i, \vb{r}_j) = \abs{\vb{r}_i - \vb{r}_j + \zeta \vb{A}_{ij}} \qc 
\end{equation}
where $\vb{A}_{ij}$ is a quenched, uniformly distributed random vector and $\zeta$ is a parameter that tunes the effect of the shift. Setting $\zeta=0$ the usual Euclidean distance is recovered, while $\zeta \to \infty$ corresponds to the Mean-field (MF) limit. In this latter case, an exact solution to its thermodynamics is available, whence a simple closed-form of its equation of state (EOS) in the liquid phase can be derived\supercite{mariDynamicalTransitionGlasses2011}. This is possible because by introducing the random shifts three body interactions (leading to short range correlations) are avoided. Being explicit, if $\sigma$ denotes the particles diameter\footnote{For simplicity, I will henceforth deal with monodisperse configurations. But the results can be easily generalized to systems with polydispersity.}, in MK configurations even when $D_\infty(\vb{r}_i, \vb{r}_j) \sim \sigma$ and $D_\infty(\vb{r}_i, \vb{r}_k) \sim \sigma$, it is almost certain that $D_\infty(\vb{r}_j, \vb{r}_k) \gg \sigma$. So neighbours of a given particle are very unlikely neighbours between themselves. MK configurations are very useful because they allow to test many of the MF predictions in finite dimensional systems, while being amenable for numerical simulations. I will henceforth assume that $\zeta=\infty$ whenever referring to MK systems. As mentioned earlier, it has been shown that MK configurations of HS display all the rich phenomenology of real glass formers\supercite{charbonneauHoppingStokesEinstein2014} as well as having a Gardner\supercite{charbonneauNumericalDetectionGardner2015} and jamming transitions\supercite{hexnerCanLargePacking2019}. Along this section, I will discuss few other properties of the MK model, as they become relevant to the topics of the MD simulations.

\subsection{Basics of event-driven MD} \label{sec:event-driven MD}

\begin{figure}[!htb]
	\centering
	\includegraphics[width=\linewidth]{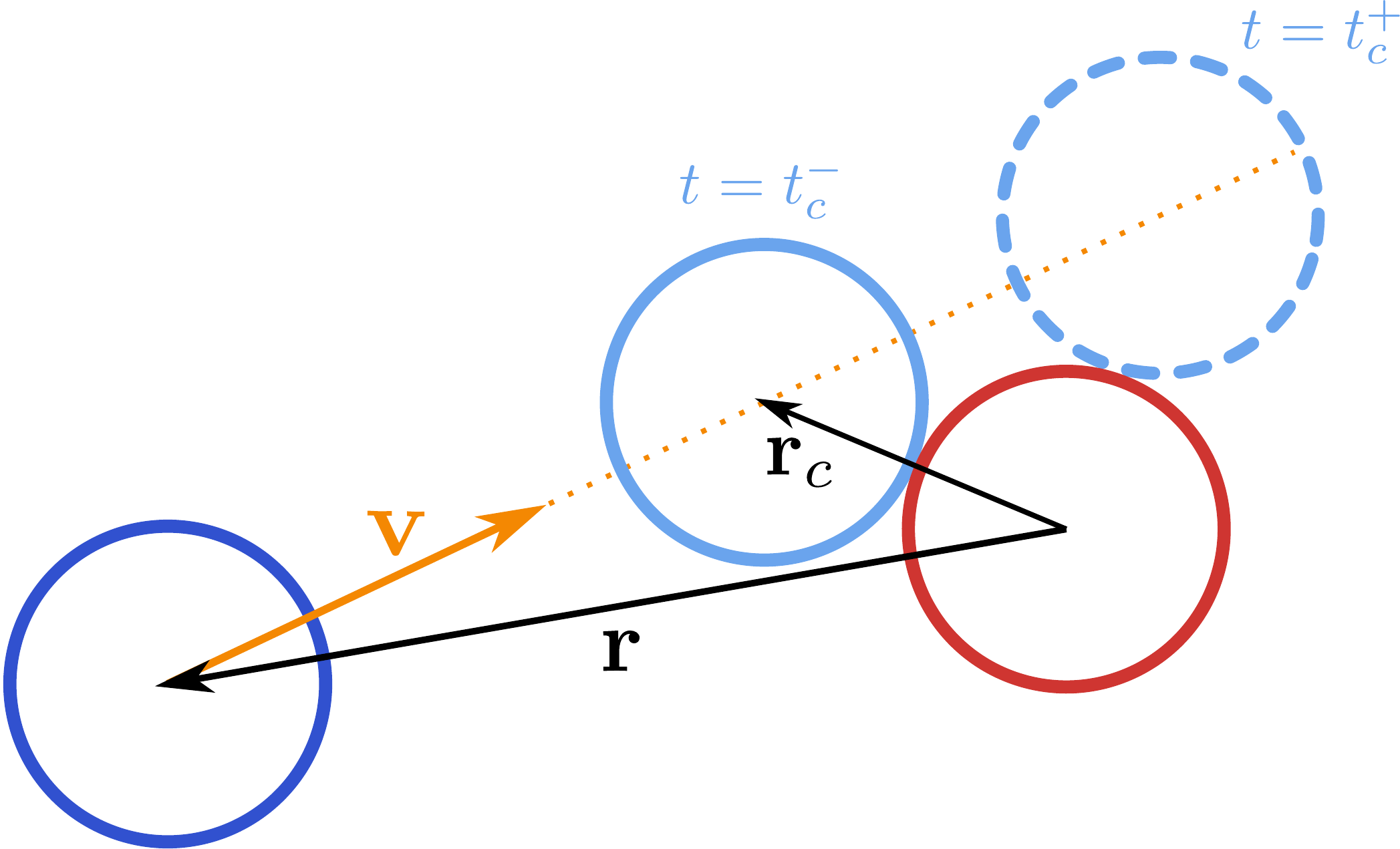}
	\caption[Diagram to determine the collision time of HS in MD simulations.]{A (blue) particle collides with another one (red). To determine the collision time $t_c$, only their relative position $\vb{r}$ and velocity $\vb{v}$ are needed. The collision (lighter blue) occurs at the smaller of the solutions of Eq.~\eqref{eq:collision-particles}, $t_c^-$. Although the second solution $t_c^+$ is valid mathematically, it lacks physical meaning (dashed).}
	\label{fig:diagram-collision}
\end{figure}

MD simulations are particularly well suited for studying the thermodynamics of configurations of HS because their potential renders the dynamics trivial. More precisely, given that particles only interact through elastic collisions, each particle moves ballistically until it bounces off another particle. Compared with most of other types of interaction potentials --for which integrating the equations of motion is mandatory-- in HS systems only the computation of collision times is required. Even more, the collision time between a pair of particles is readily obtained by solving a quadratic equation. Such equation is easy to derive considering the diagram of Fig.~\ref{fig:diagram-collision}, which depicts a (blue) particle colliding with another one (in red), that can be used as reference. That is, let $\vb{r}$ be the vector from the reference particle to the incident one, and $\vb{v}$ their relative velocity. I will also assume that both particles have the same mass and diameter, $\sigma$.

The two particles will only collide (lighter blue) if the equation
\begin{equation}\label{eq:collision-particles}
\abs{ \vb{r} + \vb{v}t}^2 = \sigma^2 \qquad \Longleftrightarrow \qquad \abs{\vb{r}}^2 + 2 \vb{r}\cdot \vb{v} t + \abs{\vb{v}}^2 t^2 - \sigma^2 =0
\end{equation}
admits at least one real, positive solution for $t$. Using the general formula for quadratic equations it is straightforward to show that this will only happen if: (i) the discriminant is positive (otherwise they do not cross with each other); and (ii) $\vb{r}\cdot \vb{v}<0$, which means that they are approaching each other. If only (ii) does not hold, a collision “would have occurred in the past”, \textit{i.e.} for $t<0$. Because Eq.~\eqref{eq:collision-particles} is a quadratic equation, two solutions exist, $t_c^\pm$, of which only the smaller (positive) one, $t_c^-$, is physically relevant. As illustrated with the dashed particle, the other solution also provides a \emph{geometrically} valid collision, but physically unfeasible. 

In this way, the collision times between each pair of particles are easily obtained, and by selecting the smallest of such times, the next collision is identified. Hence, the state of the full configuration can be advanced up to that instant and the collision simulated by exchanging the component of the particles' momenta parallel to $\vb{r}_c$ at that time. With the updated velocities, the next collision time is obtained. The dynamics is thus evolved following successive collision events. Moreover, because momentum and energy are conserved at each collision, and the number of particles and system's volume are fixed, the dynamics corresponds to trajectories sampling the microcanonical ensemble (also called $NVE$ ensemble in the MD jargon).

This is the basic idea of \emph{even-driven} MD simulations. It is a very efficient algorithm when dealing with HS systems and its performance can be improved using the following two techniques. First, only potential collisions between particles nearby need to be taken into account. This can be done using the “neighbours list” approach described above for the LP jamming algorithm. However, a faster program is obtained if the simulation volume is divided into cells of fixed size and hence collisions can only occur between particles inside the same cell or the ones next to it. The gain in efficiency is mainly due to the updates of the neighbours lists in the first approach, which requires, every so often, recomputing the distances between all the particle pairs. In turn, in the grid approach the cells remain fixed, so there is only need of bookkeeping the cell to which each particle belongs to. Transfers between cells can also be easily computed and handled as a second type of event.
The MD algorithm of Ref.~\cite{md-code} that I used for HS systems belongs to the second class. However, the results of the MK configurations to be considered here and in Chp.~\ref{chp:fss} were obtained with an implementation of event-driven MD using the neighbours list method, following the algorithm described in \cite{donevNeighborListCollisiondriven2005}. Indeed, in this model the grid method leads to a poorer performance because each particle would need to have an independent grid, given that random shifts are different for each pair. Moreover, a cell transfer for a given grid does not necessarily correspond to a change of cell in the grid of another particle. Hence, every time a particle moves one must check which of the remaining $N-1$ cell assignments are still valid and which ones need to be updated.

The second way to accelerate event-driven MD is to use asynchronous dynamics (Refs.~\cite{lubachevskyHowSimulateBilliards1991,donevNeighborListCollisiondriven2005} provide detailed and very clear descriptions). Actually, this is by far the major performance improvement and the reason why event-driven approaches are so useful.  Asynchronous dynamics works by computing the impending event for each particle independently and storing the respective times. Because a collision involves only a pair of particles, each event can be processed by updating the state of only those two particles, while leaving the rest of the configuration unaltered. Note that after a collision has taken place, only the new events of the particles involved need to be computed, so most of the remaining collision times previously found remain valid. After the new times have been calculated, the next event on the list is processed and the dynamics continues. Besides, storing and extraction of collision times can be done very efficiently employing a \textit{binary heap}\footnote{A heap is a data structure that allows to store, extract, and update ordered data more efficiently than by using a sorted list. To compare, given a set of $n$ data, implementations of both quick sorting and a binary heap scale, on average, as $\order{n\log n}$. However, once the objects are constructed the time complexity of inserting and deleting elements in a sorted array scales as $\order{n}$, while using a heap structure it is reduced to $\order{\log n}$.}. In this work, all the MD simulations were carried out in this way.

In general, MD simulations can be used to compute many variables of interest by averaging over time and/or different realizations. As discussed in Sec.~\ref{sec:hs-fluids}, for HS systems the relevant variables are the (reduced) pressure $p$, Eq.~\eqref{def:reduced P}, and packing fraction $\vp$, because the temperature is just a scale parameter. In fact, its role is restricted to assigning the initial velocities, which are sampled from a Maxwell--Boltzmann distribution before the MD takes places. Because the kinetic energy is\footnote{Recall that throught the text I am using $k_B=1$.} $K=\frac{d}{2} N T = \frac12 N m \abs{\va{v}}^2$, for a fixed volume, the temperature of a HS system simply determines the temporal scale according to $t\sim1/\sqrt{T}$.
Now, computing the packing fraction is straightforward, but the pressure requires some care when the system is subject to periodic boundary conditions as I will assume throughout this work. Under such conditions no walls are containing the system, so $p$ cannot be measured as the force (per unit area) exerted on them. Nevertheless, a kinetic pressure, associated to the momentum exchange during collisions, can be computed. In fact, from the Virial Theorem\supercite{hansenTheorySimpleLiquids2013} we have that
\begin{equation}\label{eq:p in MD}
p = 1 + \frac{\beta}{dN} \avg{ \sum_{i<j}^{1,N} \vb{r}_{ij} \cdot \vb{F}_{ij}} = 1 + \frac{1}{2K} \avg{\sum_c \frac{\vb{r}_c \cdot \Delta \vb{p}_c}{\Delta t_c}}_t \, .
\end{equation}
To transform the Virial relation into the last expression, the sum over interaction forces has been converted into a sum of the momentum exchanged during each collision, $\Delta \vb{p}_c$, and divided by the time passed between successive the collisions, $\Delta t_c$. The vector $\vb{r}_c$ is the contact vector at the collision time, illustrated in Fig.~\ref{fig:diagram-collision}. Note also that the average in the rightmost equation is taken over long times (or, equivalently considering a large amount of collisions), while the one in the virial equation is the usual ensemble average.

\begin{figure}[!htb]
	\centering
	\includegraphics[width=\linewidth]{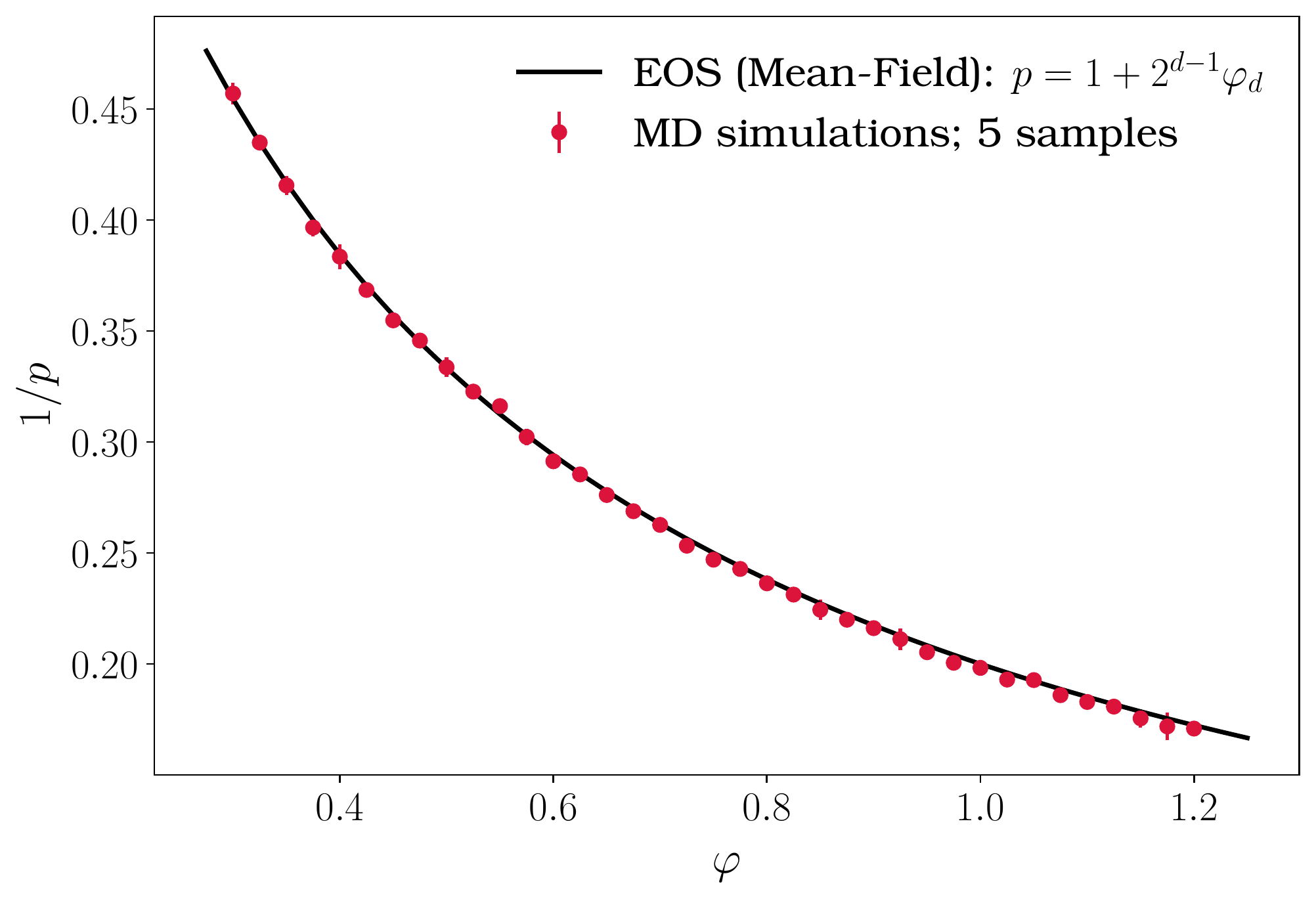}
	\caption[Comparison of event-driven MD simulations and the exact liquid EOS in the MK model]{Comparison of event-driven MD simulations (markers) and the EOS, Eq.~\ref{eq:p leading virial} (black curve). MD data corresponds to the average and standard deviation over 5 independent estimations of $p$ for each $\vp$. Note that because of the MF nature of the MK configurations, truncating the virial series after the second term leads to the \emph{exact} EOS. Additionally, because of the random shifts, MK configurations can reach densities that would be unphysical in standard HS systems, \textit{i.e.} $\vp>\vp_{FCC}$ in $3d$ or even $\vp>1$. All of this while remaining in the liquid phase.}
	\label{fig:eos-mk}
\end{figure}

In Figs.~\ref{fig:HS-EOS--diff-growth-rate} and \ref{fig:HS-EOS-liquid-and-glass} I showed that the MD algorithm of Ref.~\cite{md-code} correctly reproduces the liquid and glass EOS, estimating the pressure according to Eq.~\eqref{eq:p in MD}, and using as reference the Carnahan--Starling formula \eqref{eq:eos CS} for the liquid and Eq.~\eqref{eq:eos glass hs} for the glass\footnote{Although these figures were obtained from MD simulations with a compression protocol (as described in the next subsection), the compression was very slow so a good estimate of $p$ was available for each density.}.
Similarly, with the analogous algorithm for the MK systems the liquid phase can be accurately simulated as illustrated in Fig.~\ref{fig:eos-mk}. In this case, the respective EOS of an MK liquid is given, exactly, by the MF equation \eqref{eq:p leading virial} in the $d\to \infty$ limit. The excellent agreement between the simulations and the closed expression confirms the MF nature of the model. In other words, random shifts effectively suppress correlations between three or more particles, so only the leading term of the virial expansion, associated to pairwise correlations, is non-zero. For the same reasons. although in a different situation, within the MF theory this is the only non-vanishing correction to the ideal gas pressure; see Sec.~\ref{sec:MF liquids}

Fig.~\ref{fig:eos-mk} also shows that, due to random shifts, MK configurations can attain densities that are prohibitively high in usual models, \textit{e.g.} $\vp >1$. Once again, this is caused by the fact that neighbours of any given particle are not mutual neighbours between themselves, so they can be overlapping from the point of view of such particle. Consequently, a particle has much more neighbours than if random shifts were absent. From an algorithmic point of view, this feature hinders the MD simulations because neighbours lists are considerably larger than for usual HS systems, so more potential collisions are to be considered when obtaining the set of potential $t_c$.
I should also mention that for $\vp>0.7$, the initial configuration of the MD simulations was generated via \textit{planting}. That is, for a given value of $\vp$, the set of particles' initial positions $\va{r}_0$ is drawn from a random distribution, uniform over all the volume. And only \emph{afterwards} the shifts $\vb{A}_{ij}$ are drawn and assigned, but with the condition that no overlap occurs for any pair of particles, so $\vb{A}_{ij}$ is repeatedly drawn until such condition is met.
With this trick, liquid MK configurations of arbitrarily high densities\supercite{charbonneauHoppingStokesEinstein2014} can be produced without altering the equation of state. This method will be important for simulating correctly the glass phase as I discuss next.

\subsection{Compression protocol} \label{sec:compression-protocol}

\begin{figure}[!htb]
	\centering
	\includegraphics[width=\linewidth]{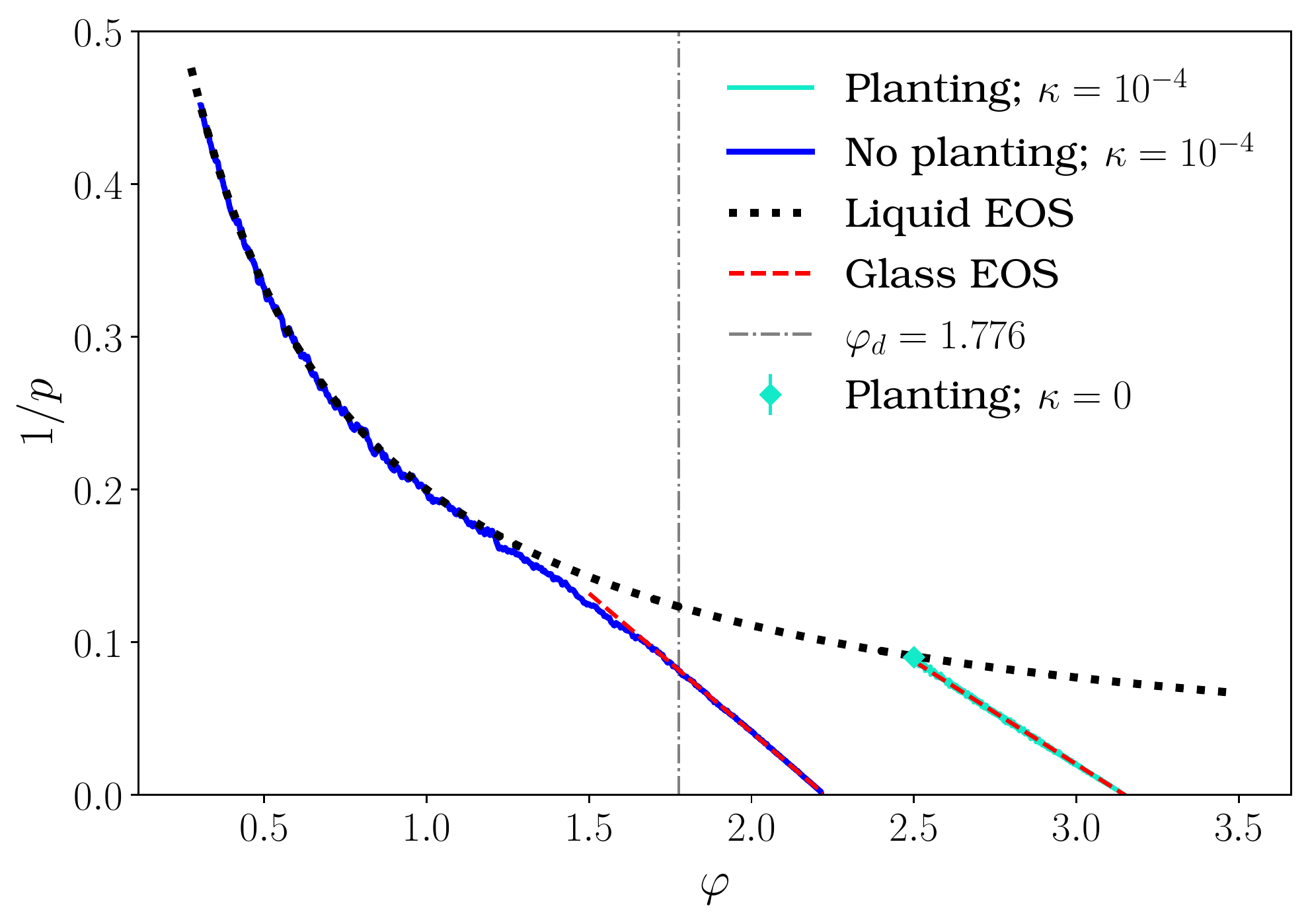}
	\caption[Liquid and glass branches in the MK model, both with and without planting.]{Liquid (dotted, black) and glass (dashed, red) branches in the MK model. The dynamical glass transition of this model, characterized by a divergent relaxation time, occurs at $\phi_d$ (dash-dotted). If no planting is used, the liquid is unable to equilibrate beyond $\vp_d$ using any compression protocol, including the LS one. However, planting allows to obtain an equilibrated liquid (cyan diamond) at an arbitrarily high density, which can then be used to probe the glass phase when compression is “switched on” (cyan curve). The glass EOS curves were obtain from a fit of the numerical data to Eq.~\eqref{eq:eos glass hs}.}
	\label{fig:MK-liquid-and-glass}
\end{figure}

For compressing the liquid into a glass we used a Lubachevsky--Stillinger (LS) compression protocol\supercite{lubachevskyGeometricPropertiesRandom1990}. It consists in increasing the particles' diameter at a given rate $\dot{\sigma}(t)$ and it is therefore easy to implement in periodic systems as we consider here. As shown from the results of Ref.~\cite{md-code}, in HS systems a glass can be readily obtained from a liquid using the simplest compression protocol, \textit{i.e.} one with a constant growing rate $\dot{\sigma} = \kappa$. In that work, MD simulations with the LS protocol were tested in monodisperse systems in several dimensions, and latter, in \cite{berthierGrowingTimescalesLengthscales2016}, the same methodology was used for $3d$ polydisperse configurations. 
Analogously, such LS compression method has been used to study MK configurations in their glass phase; see \cite{charbonneauHoppingStokesEinstein2014,charbonneauNumericalDetectionGardner2015}. Similarly, the implementation I developed of the LS protocol is able to reproduce the expected EOS of MK liquids and glasses, as shown in Fig.~\ref{fig:MK-liquid-and-glass}. 
An important feature of the MK model is that it exhibits a dynamical glass transition, just as the one described by MF theory in Sec.~\ref{sec:MF dynamics and glass transition}, at $\vp_d \approx 1.776$. Notice that a slow compression, say  $\kappa=10^{-4}$ (blue line in Fig.~\ref{fig:MK-liquid-and-glass}), yields an out of equilibrium liquid at densities well below $\vp_d$, and decreasing the growth rate by an order of magnitude barely improves the situation\supercite{charbonneauHoppingStokesEinstein2014}. This is a manifestation of the true dynamical transition predicted by MF theory, and therefore no compression protocol is able to produce an equilibrated state for $\vp>\vp_d$.

On the other hand, these results also show the importance of using planted initial conditions in order to access the full glassy phase at high densities. When such method is used with no compression (\textit{i.e.} $\kappa=0$), MD simulations correctly estimates the pressure of the \emph{equilibrated} glass (cyan diamond) for densities considerably above $\vp_d$ (vertical dash-dotted line). Yet, independently of whether planting is used or not, the LS protocol generates a dynamics that correctly reproduces the metastable glass behaviour, as indicated by the excellent agreement with the red dashed lines. These curves were obtained from a fit to Eq.~\eqref{eq:eos glass hs}, which describes the glass EOS in HS systems. However, when fitting the data I treated $d$ in this equation as a free parameter, given the lack of inherent dimensionality in MK systems.

Before continuing, some clarification of the glass EOS is in order. In Figs.~\ref{fig:HS-EOS-liquid-and-glass} and \ref{fig:MK-liquid-and-glass}, I have used Eq.~\eqref{eq:eos glass hs} as the expected theoretical behaviour of a glass, and showed that it is nicely verified by numerical simulations. However, it is important to stress that this equation is \emph{not} derived from thermodynamic considerations, but only from free volume (\textit{i.e.} geometric) arguments; see Sec.~\ref{sec:hs-fluids} and \cite{salsburgEquationStateClassical1962}. (Calling it “Equation of State” is actually an abuse of terminology.) As discussed in Sec.~\ref{sec:MF dynamics and glass transition} and \ref{sec:MF gardner transition}, the true EOS is obtained, at least for MF models, by the technique of state following, and requires using the Franz--Parisi potential in a 1RSB and fullRSB schemes. Restricting the discussion to the MK model, for which a full MF treatment is possible\supercite{charbonneauHoppingStokesEinstein2014,charbonneauNumericalDetectionGardner2015,rainoneFollowingEvolutionHard2015}, notice that the true thermodynamic EOS of a glass depends on the values $(p,\vp)$ at equilibrium --for instance, the cyan diamond in Fig.~\ref{fig:MK-liquid-and-glass}. In contrast, the glass pressure using the free volume EOS depends, quite naturally, on the density at jamming. Consequently, it does not provide information about any equilibrium properties, not even when crossing the liquid EOS line\footnote{Beyond $\vp_d$, the “liquid” branch actually corresponds to an equilibrated glass. See Sec.~\ref{sec:MF theory} for a more detailed discussion.}. As expected, no trace of the Gardner transition is present in this latter EOS, while it is a salient feature of the thermodynamic construction based on state-following; see \cite{charbonneauNumericalDetectionGardner2015,rainoneFollowingEvolutionHard2015,rainoneMetastableGlassyStates2017}. Even more, no other thermodynamic variables or free energies can be obtained from Eq.~\eqref{eq:eos glass hs}. As a side note, I take the opportunity to mention that planting is another type of method that allows to equilibrate \emph{some} systems beyond the dynamical glass transition; see the discussion at the beginning of Sec.~\ref{sec:MF glasses thermo}. 
Unfortunately, this is not true in general since the free energy of a planted system usually differs from the real thermodynamic one. But in the case of the MK model, planting is a very useful trick because the annealed (\textit{i.e.} planted) and quenched free energies \emph{of the liquid} coincide\supercite{charbonneauHoppingStokesEinstein2014}.

Let me now briefly discuss the main implications of using a LS compression protocol in the event-driven MD algorithm discussed above. Given that the particles' diameter is also a function of time, the most obvious consequence is that Eq.~\eqref{eq:collision-particles} cannot be solve analytically for a general growth function $\dot{\sigma}(t)$. However, when a constant compression is used, as in all the cases studied here, the diameter is simply $\sigma(t) = \sigma_{t_0} + \kappa t$, so Eq.~\eqref{eq:collision-particles} remains quadratic in time.
In contrast, the effect of particles' growth on the momentum transfer due to collisions is not so obvious. If two particles $i$ and $j$ collide, the new velocity of particle $i$ (resp. $j$) has an extra contribution\supercite{lubachevskyGeometricPropertiesRandom1990,md-code} equal to $\vu{r}_{ij} \kappa$ (resp. $\vu{r}_{ji} \kappa$), where $\vu{r}_{ij} = \frac{\vb{r}_i - \vb{r}_j}{\abs{\vb{r}_i - \vb{r}_j}}$ is a unit vector connecting the particles' during the collision. This extra contribution can be understood by considering the case of two static particles, where clearly no collision is expected. However, if their size is being increased, sooner or later they will come into contact. When this happens, the only source of “impulse” they have is due to the speed at which their radius is increasing, and the “collision” takes place along the line connecting their centres. The same argument applies when particles are not static, thus the extra contribution to their momenta is justified. Because $\vu{r}_{ij}=-\vu{r}_{ji}$, momentum is still conserved, but the energy is not\footnote{Thus, strictly speaking the LS compression protocol modifies the sampling ensemble of the MD simulations.}.
In fact, the difference in kinetic energies before and after the collision scales as $\kappa^2$. A simple solution is to rescale, after a fixed number of events, all the particles velocities so that their kinetic energy matches $K=\frac{d}2 N T$. On the other hand, $2\kappa$ should be subtracted from $\Delta \vb{p}$ in each collision in order to avoid the extra velocity contribution to affect the estimation of $p$ in Eq.~\eqref{eq:p in MD}.

As a final verification that the compression algorithm described here is able to produced well thermalised configurations for very high pressures, I will compare the gaps distributions obtained from MD simulations with the radial distribution function (RDF) at jamming\supercite{md-code}: 
\begin{equation}\label{eq:rdf at jamming}
\tg(h) = \frac{\overline{z} \sigma }{\rho s(\sigma)} \delta(h) + c g(h) \, .
\end{equation}
In this equation, $\overline{z}=2d$ is the average coordination number, $\rho=N/V$ is the number density, $s(\sigma)$ is the surface of a hypersphere of radius $\sigma$, $c$ is a geometric constant and $g(h)$ is the critical gaps distribution defined in Eq.~\eqref{eq:pdf-gaps}. (Notice that this latter quantity is written in italic, while the RDF is in regular typeface.) Additionally, instead of considering the RDF as a function of $r$, as I did when it was introduced in Sec.~\ref{sec:hs-fluids}, I am using $h=\frac{r}{\sigma}-1$, which amounts to rescaling the distances by $\sigma$ and translating the plot of $\tg(r)$ by the same value. Notice that this is just the gap definition given in Eq.~\eqref{def:gaps} and used to study jammed packings in Sec.~\ref{sec:jamming-transition}, but particularized for monodisperse systems. Now, Eq.~\eqref{eq:rdf at jamming} tells us that, at jamming, the RDF is made of a $\delta$ singularity due to particles in contact ($r=\sigma$) that decays following a non-trivial power-law: $g(h)\sim h^{-\gamma}$, with $\gamma\approx0.41$ as discussed in Sec.~\ref{sec:forces-and-gaps}. This is nothing but a consequence of $\tg(r)$ being proportional the probability of finding a pair of particles a distance $r$ apart. Indeed, as I will show next, when $p$ is high enough, these characteristic properties of $\tg$ are reflected by the probability distribution function (pdf) of $h$.

\begin{figure}[!htb]
	\centering
	\includegraphics[width=\linewidth]{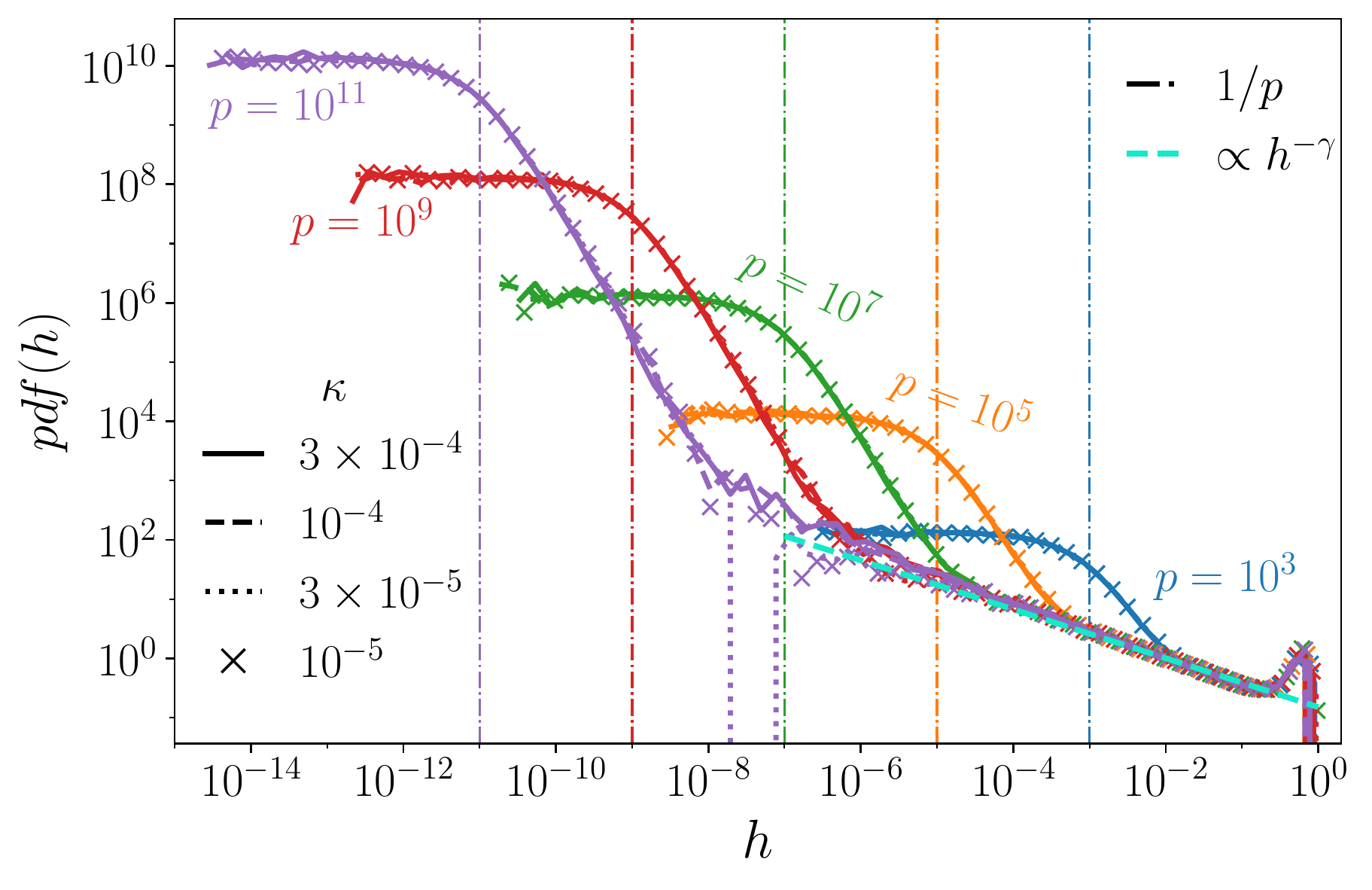}
	\caption[Probability distribution function of interparticle gaps from MD simulations after reaching different values of $p\gg 1$]{Probability distribution function of interparticle gaps $h=\frac{r}{\sigma}-1$ obtained from the MD based compression. The different values of the final pressure are indicated by different colours, while the distinct line styles identify the values of the particles' growth rate. A comparison (in cyan) with the critical gaps distribution expected at jamming is also included, while the predicted onset of the plateaus at $h=1/p$ is indicated by the dash-dotted vertical lines. See main text for a discussion of why these two features are important.}
	\label{fig:gaps after MD}
\end{figure}

To estimate the pdf of $h$, I obtained the histogram of interparticle gaps using the compression method just described in 20 systems of $N=1024$ HS, at different final pressures and with different values of $\kappa$. The results are reported in Fig.~\ref{fig:gaps after MD}, where a comparison with the power-law of $g(h)$ for $10^{-7} \leq h \leq 1$ is also included. The numerical results are in excellent agreement with the theoretical prediction over a range that, naturally, extends as $p$ increases. On the other hand, the effect of changing $\kappa$, even by an order of magnitude is barely noticeable. This is an important point to which I will come back later. 
A second relevant attribute is the formation of plateaus whose height also increments with $p$. To understand this behaviour, recall that Eq.~\eqref{eq:P-rdf-hs} relates the pressure of a HS system with the value of the RDF at $r=\sigma$, $\tg^+$. This expression was obtained for a liquid, but the only underlying assumption is that the pressure can be obtained via the Virial theorem and, as argued few paragraphs above, this is the case in MD simulations, provided that $\kappa$ is small enough so that averaging $p$ over a large number of collisions matches its thermal average. Besides, note that when the final pressure of the MD simulations is increased by, say, two orders of magnitude the respective plateau is raised by the same amount. We can then conclude that, except for geometric factors, the height of the left part of the histograms equals $\tg^+ \sim p$.
The plateaus are thus caused by a nascent singularity that gives rise to the $\delta$ function in Eq.~\eqref{eq:rdf at jamming}. This means that as $p\to \infty$ the length of such plateaus should vanish, whence it is expected that their onset is reduced with $p$. In fact, when substituting $\overline{z}$ and $s(\sigma)$ by their value in $d=3$, some simple algebra leads to
\[
\lim_{\epsilon\to 0 } \ \int_{0}^{ \epsilon} \tg(h) \dd{h} = \frac{1}{4\vp} \, .
\]
In turn, from the histograms obtained numerically, we have that the same integral is approximately $\epsilon \tg^+$. But from Eq.~\eqref{eq:P-rdf-hs} we know that for $p\gg1$, $\tg^+ = \frac{p}{4\vp}$, whence we obtain that $\epsilon \sim 1/p$. This result is confirmed by the dashed-dotted, vertical lines of Fig.~\ref{fig:gaps after MD}, all of which are drawn at $1/p$ and that are in excellent agreement with the rightmost part of the plateaus.

To make an explicit connection with the inners of the MD simulations, let me give an additional argument of why it is expected that the plateaus begin when $h\sim 1/p$. It goes as follows: according to Eq.~\eqref{eq:p in MD}, the value of $p$ depends on how often collisions occur in the system. Given that velocities follow a Maxwell--Boltzmann distribution their variance is finite, so typical temporal intervals between collisions are of order $\Delta t \sim \frac{h}{v} \sim \frac{h}{\sqrt{T/m}}$. Because $p \sim 1/\Delta t$, it then follows that gaps which contribute the most to the kinetic estimation of $p$ are, at most $\order{1/p}$. These considerations show that before jamming is reached, the $\delta$ singularity of Eq.~\eqref{eq:rdf at jamming} is smoothed out as a plateau of increasing height but decreasing width, caused by particles colliding with their nearest neighbours. These interactions define (some of) the contacts that will form at $\vp_J$. Additionally, there are other set of very small gaps (although several orders of magnitude larger than the former), that begin to delineate the critical gaps distribution; see Eq.~\eqref{eq:pdf-gaps} in the previous chapter. Fig.~\ref{fig:gaps after MD} shows that if jamming has not been reached, there is a crossover between these two sets, where it is impossible to know if a given gap should be classified as a contact or as a true gap. Moreover, this intermediate interval is not affected by using different compression rates, so it is not caused by a lack of thermalisation of the configurations.

Interestingly, this same argument also shows that the MD-based compression identifies the $p=\infty$ point with the situation where $\Delta t=0$. But notice that this situation does \emph{not} necessarily correspond to a jammed state. For instance, there could be a percolating chain of contacts, leading to a set of particles blocked by the event-driven dynamics, and yet be an unjammed configuration. The case of the contact forces is harder to solve: even if contacts could be perfectly identified by the $h\sim1/p$ criterion, the problem of defining the forces magnitude would remain.
Therefore, despite the similarities of the pdf of $h$ at $p\gg1$ but finite, and the expected behaviour at jamming, the configurations obtained using MD are detectably different from real jammed packings. Because all the results of this thesis are based on a precise identification of gaps and forces, we needed to complement the MD method with our iLP algorithm to guarantee that systems are \emph{at} jamming.
On the other hand, as I have argued along this part, the compression protocol considered here is a good approximation of the state-following scheme. Therefore, using the configurations thus produced as initial conditions of the iLP algorithm is a way to ensure that typical jammed states are obtained. 
Combining both algorithms yields a robust and relatively fast method, as I will now describe.

\section{Characterization of MD+iLP algorithm} \label{sec:characterization MD-LP}

As mentioned in the last part of the previous section, configurations compressed employing the MD algorithm will be used as initial conditions of the iLP algorithm described in Sec.~\ref{sec:LP algorithm}. Figs.~\ref{fig:HS-EOS-liquid-and-glass}, \ref{fig:gaps after MD} (for HS) and \ref{fig:MK-liquid-and-glass} (in the MK model) show that when a small value of $\kappa$ is used, configurations can be compressed very close to their jamming point, identified by the $1/p =0$ line.
Hence, such initial conditions will be parametrized by the target pressure \ptar at which the LS compression protocol stops, and by the diameter growth rate $\kappa$ used in such protocol. We decided to use the pressure instead of $\vp$ because the packing fraction at jamming is system dependent, and might even be distributed within an unbounded interval (as in the MK configurations). In turn, jamming of rigid particles always ensues when $1/p=0$. Additionally, the required fine tuning in $\vp$ is much harder to achieve, since $p$ can increase by 4 or more orders of magnitude, while $\vp$ changes its value by less than $1\%$, see Fig.~\ref{fig:HS-EOS-liquid-and-glass}. Hence, fixing a threshold value of the packing fraction would yield a sample with very heterogeneous pressures.

The results that I will present in this section are a detailed characterization of the properties and performance of the MD and iLP algorithms combined. I will only consider HS configurations, but the results should be qualitatively analogous for the MK model. As I have mentioned before, in $d=3$ monodisperse HS configurations tend to crystallize if a very slow compression is utilized. To avoid it, all the configurations I will consider were produced after an initial and fast ($\kappa=5\times 10^{-3}$) compression up to $\ptar^{(0)}=500$. From the results of Figs.~\ref{fig:HS-EOS--diff-growth-rate} and \ref{fig:HS-EOS-liquid-and-glass}, it is clear that this value of $\kappa$ is enough to suppress the formation of crystalline domains, while $\ptar^{(0)}$ is well into the glass phase. All the MD compressions were performed using the implementation of Ref.~\cite{md-code}, while each LP optimization step of our method was carried out using Gurobi\supercite{gurobi}. 

\subsection{Influence of the initial condition's parameters}\label{sec:LP dependence initial confs}

\begin{figure}[!htb]
	\centering
	\includegraphics[width=\linewidth]{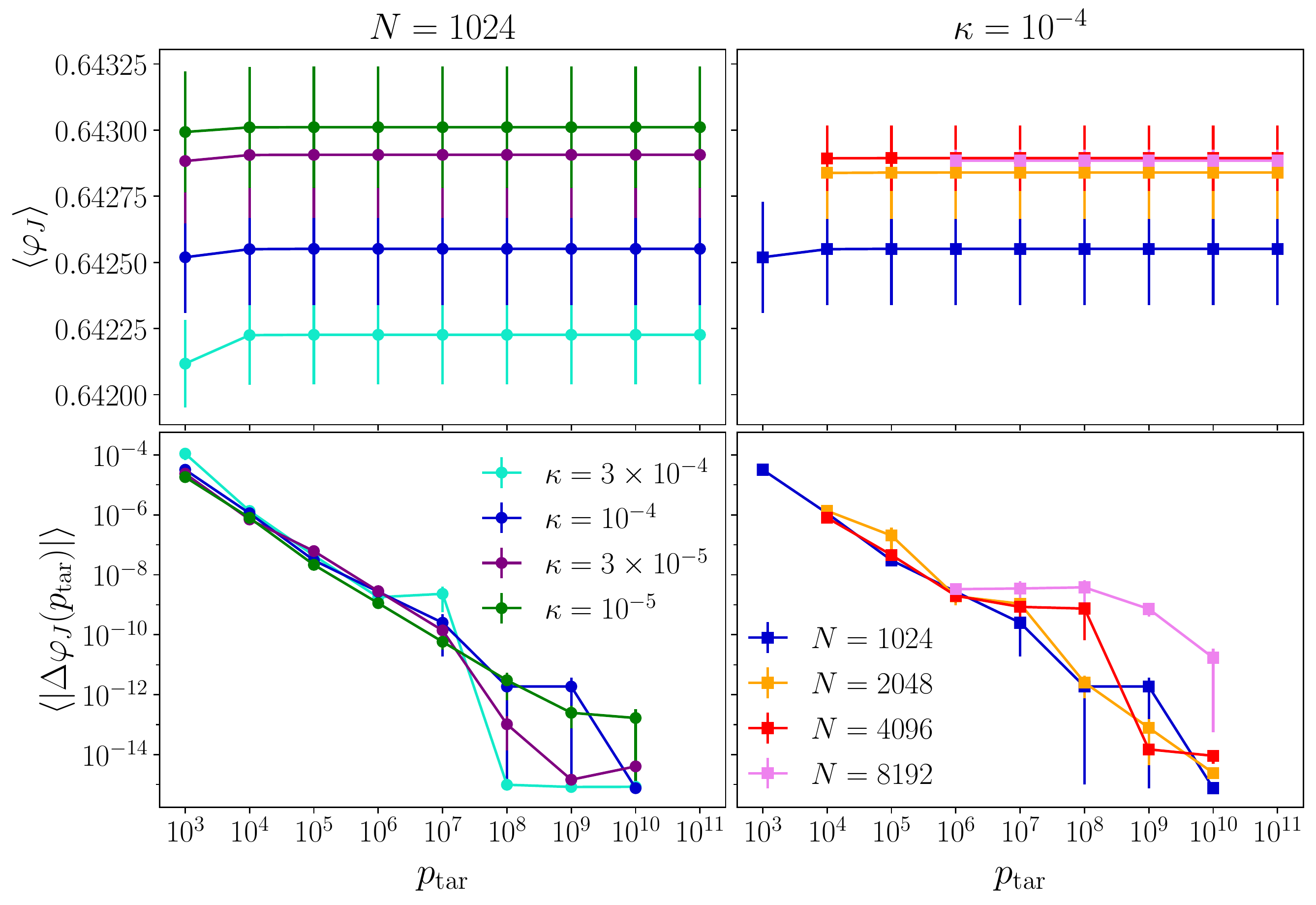}
	\caption[Jamming packing fraction as a function of $\ptar$, $\kappa$, and $N$]{Upper panels: Dependence of the jamming density $\vp_J$ on the target pressure from the compression protocol, $\ptar$. 
	Lower panels: Statistics of the difference in packing fractions, $\Delta \vp_J(\ptar) \equiv \vp_J^{(p_{\max})} - \vp_J^{(\ptar)}$; because for some samples such difference is not monotonic, the absolute value of the difference is considered. 
	Several growth rates with $N=1024$ are considered (left) and different system sizes with fixed $\kappa=10^{-4}$ are analysed (right). Markers denote the sample average, while the error bars are the standard error.}
	\label{fig:phi-vs-p-and-kappa}
\end{figure}

Possibly the first thing to consider is how the jamming packing fraction $\vp_J$ depends on the parameters of the initial condition, $(\ptar, \kappa)$. The results obtained for $\ptar \in [10^3, 10^{11}]$ and for several values $\kappa$ are reported in the upper panels of Fig.~\ref{fig:phi-vs-p-and-kappa}. Each point corresponds to the average value of $\vp_J$ (and the standard error) over $20$ (resp. 10) samples for $N=1024$ (resp. $N\geq 2048$); hence throughout this section $\avg{\cdot}$ denotes the average over samples. As anticipated, all the values are close to $0.64$, and changing the target pressure of the compression protocol has little or negligible influence for $\ptar \geq 10^6$. In contrast, the effect of changing the growth rate (left panel) and system size are notorious (right plot), albeit similarly small. As expected, lowering $\kappa$ yields a higher value of $\vp_J$ because it amounts to a better thermalised compression, so it can reach lower minima of the free energy landscape (FEL). Likewise, the jamming density attains higher values as $N$ augments. This result is also unsurprising given that the larger the system, the less it is constrained by border or periodic effects; hence it can better rearrange its particles to achieve a higher $\vp_J$. Of course, $\vp_J$ cannot grow indefinitely as $N\to \infty$, and indeed the data suggest a rather quick convergence to its thermodynamic limit value. I should emphasize that no crystallization occurs in these configurations (see Fig.~\ref{fig:rdf-vs-p}), so for a fixed $N$, $\vp_J$ is narrowly distributed around a well defined value\supercite{ohernJammingZeroTemperature2003,donevCommentJammingZero2004,ohernReplyCommentJamming2004}. If this was not the case, a much broader distribution of values could be obtained; see \cite{torquatoRobustAlgorithmGenerate2010} for an example were also an LP-based jamming algorithm is used.

Let me discuss in greater detail the dependence of the final configurations on $\ptar$. Such analysis is relevant since it puts to test the hypothesis of the fractal structure of the FEL. As explained in Ref.~\cite{fel_2014} a careful investigation is required to explore such scenario, and while the results presented here are not intended for those purposes, I will argue that they nevertheless support such hypothesis. To do so, I will consider as a reference for comparison the configurations obtained from the largest target pressure $p_{\max} \equiv 10^{11}$. First I will focus on the difference $\Delta \vp_J(\ptar) \equiv \vp_J(p_{\max}) - \vp_J(\ptar)$ as a measure of how much the final configuration changes if the iLP crunching begins at a different $\ptar$. The lower panels of Fig.~\ref{fig:phi-vs-p-and-kappa} depict the dependence of $\avg{\abs{\Delta \vp_J(\ptar)}}$ of the same samples considered before. The absolute value is necessary because, unexpectedly, the difference is not monotonic. That is, some samples attain a larger density when $\ptar < p_{\max} $, so in order to avoid deviations to larger values cancelling the ones to smaller values the absolute value is used. In any case, when $N=1024$ and the value of $\kappa$ is varied, our results show (bottom left panel) that the difference of  jamming packing fractions decreases almost steadily as $\ptar$ grows. This behaviour is precisely the expected one from a fractal FEL, because it implies that the minima --attained from different $\ptar$ and that define the jammed states-- are similar between themselves (because they all belong to the same meta-basin), but their specific features are realized gradually (because the configuration navigates the different sub-basins as $\ptar$ increases). Conversely, if the FEL possessed a simple basin structure, once $\ptar$ was high enough all the configurations would end up in the same minimum, so the final packings would be identical.
Second, for $\ptar \leq 10^6$ results are rather insensitive to the compression rate. However, while $\ptar \geq 10^8$ is enough to suppress fluctuations if $\kappa=3\times 10^{-4}$, detectable differences are present even if $\ptar =10^{10}$ when the slowest compression is used. Our data thus suggest that the hierarchical structure of the FEL can only be fully resolved with an infinitesimally slow compression. Similarly, when $\kappa$ is fixed and different system sizes are considered (bottom right panel), $\Delta \vp_J(\ptar)$ is monotonically decreasing up to $\ptar \sim 10^7$. For higher pressures, the difference with respect to $p_{\max}$ seem to increase with $N$. This is, very likely, once again a manifestation of the finite value of $\kappa$, which determines the \emph{radius} growth rate. Because the volume of all the systems is fixed, configurations made of more particles have smaller spheres and, consequently, using the same value of $\kappa$ for all of them produces different compression rates. Additionally, the discrepancy found for different $N$ is also possibly caused by finite size effects, given that the marginal fractal phase is strictly valid only in the thermodynamic limit. 

\begin{figure}[!htb]
	\centering
	\includegraphics[width=\linewidth]{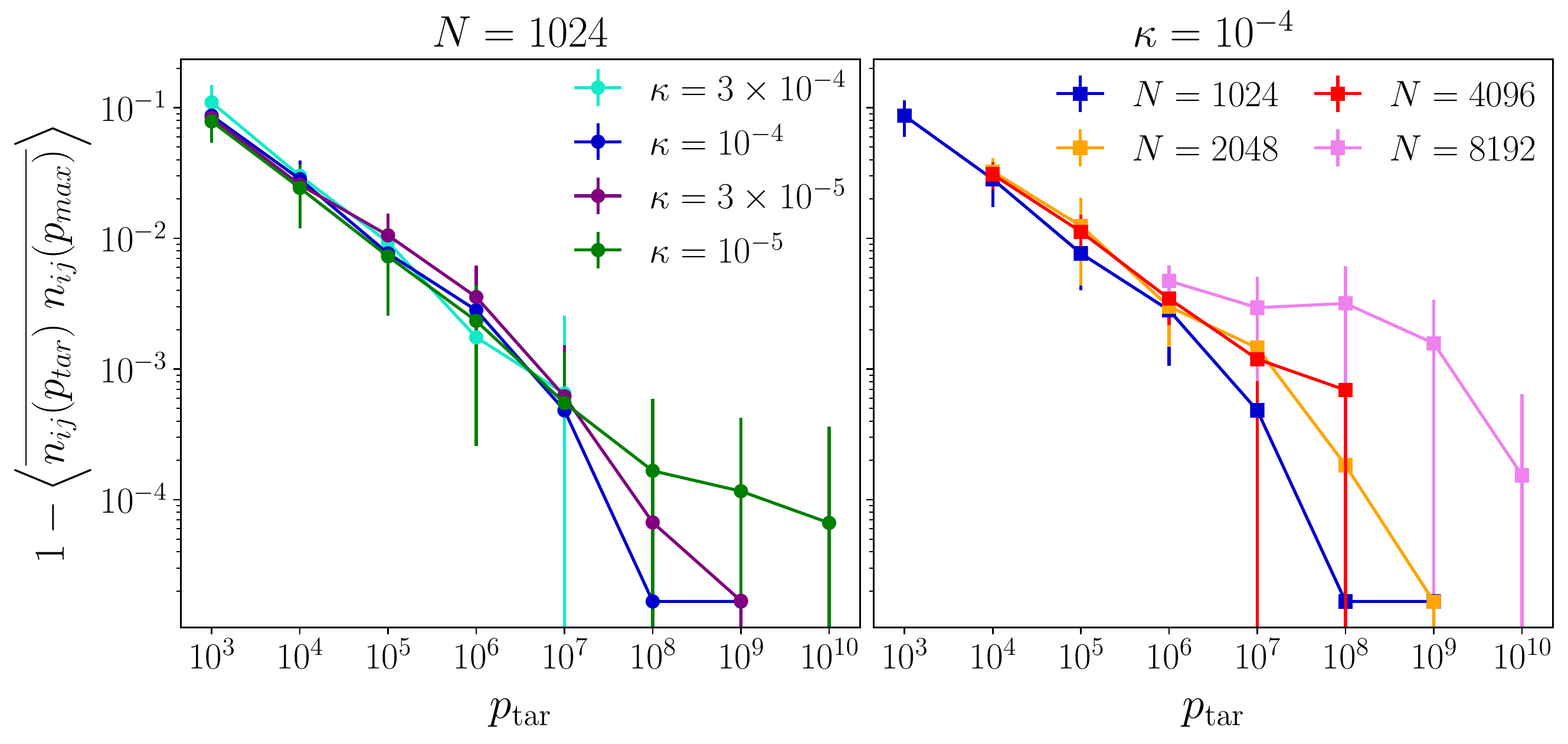}
	\caption[Similarity in the network of contacts as a function of $\ptar$, $\kappa$, and $N$.]{Overlap (defined in the main text) of the networks of contacts using as reference the one corresponding to $p_{\max}$. The overlaps increase \emph{gradually} as the configurations go deeper in the FEL, suggesting that it has a fractal structure. The data reported here correspond to the same configurations of Fig.~\ref{fig:phi-vs-p-and-kappa}.}
	\label{fig:similarity-contact-network}
\end{figure}

As a second measure of the similarity between jammed configurations I will consider the overlap of the network of contacts. Thus, let $n_{ij}(\ptar) = 1$ if particles $i$ and $j$ are in contact in the jammed configuration obtained from an initial condition with $\ptar$, and $0$ if they are not in contact. Taking once again $p_{\max}$ as reference, the overlap between two networks of contacts is  then $\overline{n_{ij}(\ptar) n_{ij}(p_{\max}) }$, where the overline denotes an average over all the contacts of the reference network. Notice that this variable is more sensitive to variations in the jammed state, given that the full microstructure of the packings is considered. In fact, this is the same quantity analysed in Ref.~\cite{fel_2014} to investigate the fractal structure of the FEL. However, the packings produced there were not precisely at jamming --but at a very high $p$-- and the authors were concerned with the overlap of \emph{different} configurations at the \emph{same} pressure. Here I am instead considering the jamming point of the same system, reached from different pressures. Fig.~\ref{fig:similarity-contact-network} shows the results of the similarity of the contact networks obtained from the configurations considered above. Also in this case the data support the fractal FEL hypothesis, because as $\ptar$ increases the network of contacts are more alike. 
Once again, this can be explained assuming that the broad structure of the contacts network is shared by packings within a given meta-basin, but the precise particle contacts are determined \emph{progressively} as a configuration goes down in the hierarchical structure of minima.
It is worth emphasizing that the discrepancies found when varying $\kappa$ and $N$ are in tune with the explanation given above, namely: the need to use an adiabatic compression and be in the thermodynamic limit in order to probe the fractal structure of the FEL at all scales and for every value of $p$. 
In fact,  when this is not the case, our results show that for some values of $\kappa$ and $N$, and at sufficiently high pressure, the networks of contacts are \emph{identical in all the samples}. 
At any rate, I should warn that even if these results are very promising and complimentary to the ones of Ref.~\cite{fel_2014}, they are still preliminary given the few samples considered. Additionally, a more systematic analysis of the influence of the compression protocol and system size is needed to explain satisfactorily the dispersion of values found for $\ptar \geq 10^8$. 

\begin{figure}[!htb]
	\centering
	\includegraphics[width=\linewidth]{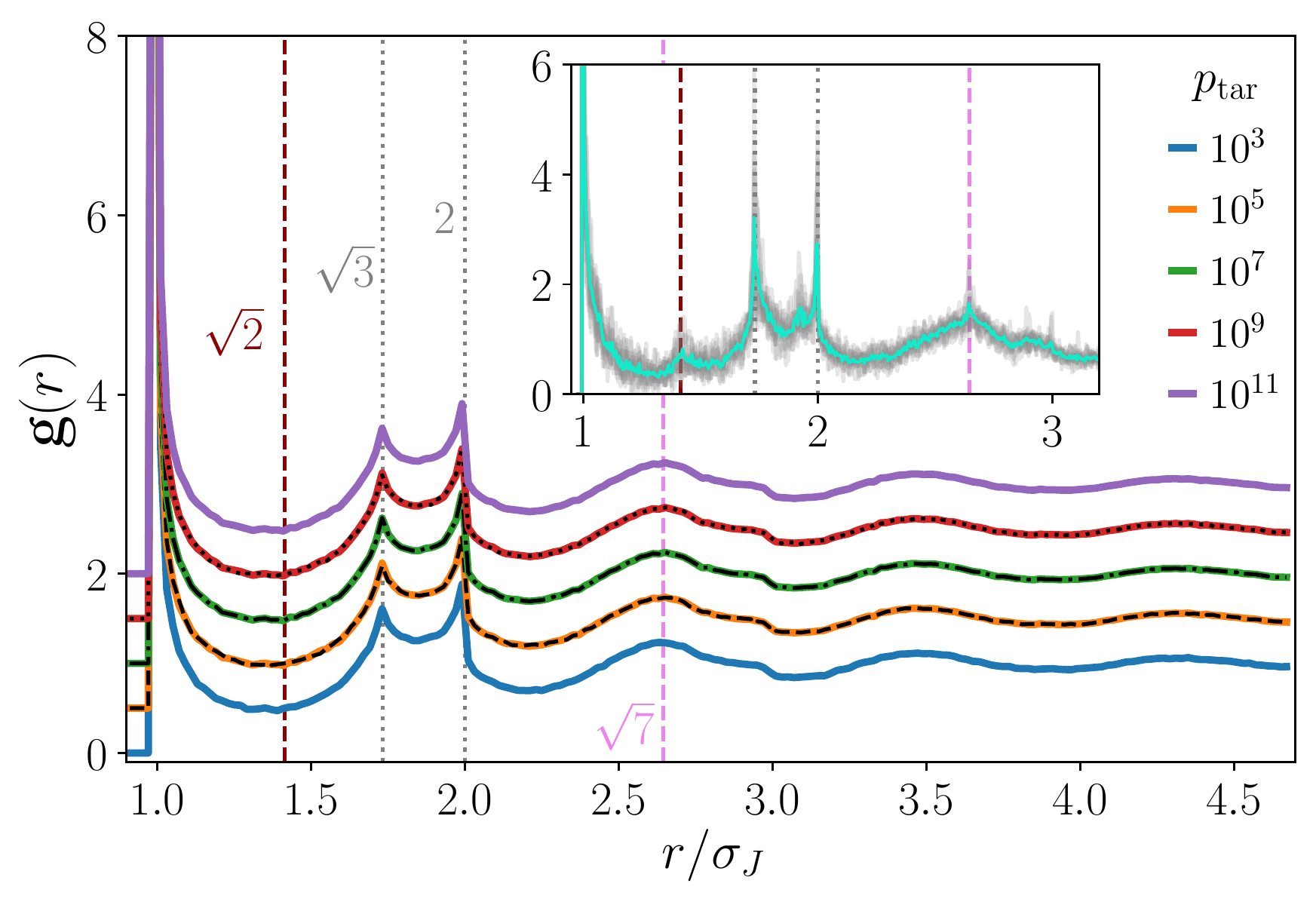}
	\caption[Radial distribution function of jammed packings obtained from different $\ptar$.]{Radial distribution function (with the distance scaled by the diameter at jamming, $\sigma_J$) and its dependence on the target pressure of the initial configuration. Each curve is obtained by averaging $20$ samples of $N=1024$ particles. The curves have been displaced vertically for clarity, because they overlap almost perfectly. All of them were obtained using the largest value of $\kappa$ in the MD compression, but changing the compression rate leaves $\tg(r)$ unaltered as shown by the dashed ($\kappa=10^{-4}$), dash-dotted ($\kappa=3\times 10^{-5}$), and dotted ($\kappa=10^{-5}$) lines.
	Vertical lines at $\sqrt{2}$ and $\sqrt{7}$ identify the position of expected peaks if crystallization occurs, while the grey dotted lines are present also in amorphous systems; see text.
	Inset: Sample of 10 configurations (in grey) and their average (cyan) where partial crystallization occurs. The presence of the peak at $\sqrt{2}$ is a clear signature of partial ordering, while the one at $\sqrt{7}$ is only smoothed out.
		}
	\label{fig:rdf-vs-p}
\end{figure}

As a final feature, I will show that no crystalline domains are present in the jammed packings produced by our MD+iLP method. To do so, I will resort to the RDF computed using Eq.~\eqref{def:rdf} and averaging over the 20 jammed packings of $N=1024$ particles. 
Fig.~\ref{fig:rdf-vs-p} contains several plots of $\tg(r)$ obtained using different values of $\ptar$ to generate the jammed packings, and fixing $\kappa=3\times 10^{-4}$. However, changing the compression rate leaves the RDF unaltered, as evidenced by the dashed, dash-dotted, and dotted black lines (corresponding to $\kappa=10^{-4},\ 3\times10^{-5},\ 10^{-5}$, respectively) superposed to the curves of the respective pressure. If crystallization did occur, a peak would appear exactly at $\sqrt{2}$, due to the distance between the pair of particles in the diagonal of a square\supercite{tesi-forze,rissoneLongrangeAnomalousDecay2020}. Given that $\tg(r)$ has a minimum very close to $r=\sqrt{2}\sigma_J$ (red dashed line), such peak would be easily noticeable. For instance, the inset shows the RDF of 10 configurations with a small degree of crystallization, $\vp \in (0.65, 0.675)$, all of which show an evident peak at the predicted position. Another signature of crystallization is the peak at $\sqrt{7}$ (pink dashed, line), although its presence is less pronounced and it is only smoothed out in amorphous packings. In any case, these configurations were not included in any of the analyses I present here.
On the other hand, the peaks at $\sqrt{3}$ and $ 2$ are expected in disordered configurations\supercite{donevPairCorrelationFunction2005,md-code,tesi-forze,rissoneLongrangeAnomalousDecay2020}. At $r=3 \sigma_J$ a discontinuity is expected, although it is hardly visible at this scale. A detailed analysis of the behaviour of the RDF in jammed packings can be found in \cite{donevPairCorrelationFunction2005,tesi-forze,rissoneLongrangeAnomalousDecay2020}, while a nice example of how the peaks disappear as a configuration moves away from the crystal is presented in Ref.~\cite{tsekenisJammingCriticalityNearCrystals2020}, albeit using polydispersity as the source of disorder. At any rate, the behaviour of $\tg(r)$ obtained from our method agrees with previous results far from contact -- that is, for $r/\sigma_J-1 \gg 1$-- confirming that the MD+iLP algorithm fully suppresses crystallization. 

\subsection{The MD+iLP route to jamming}\label{sec:MD-and-LP}

In this subsection I continue analysing some features of the jammed packings obtained by combining the MD and iLP algorithms. But now, instead of considering the properties of the final jammed configuration, I will focus on how such configurations are reached. In other words, I will describe some of the physics underlying the MD+iLP protocol. For this purpose, I begin by arguing that, \emph{based solely on a $(p,\vp)$ parametrization} of a jammed state, the packings obtained through the two algorithms combined cannot be distinguished from the ones corresponding to the  $p\to\infty $ limit using only the MD compression. 
To perform such comparison, note first that the density value extrapolated to this limit, $\vp^{(MD)}_J$, can be readily estimated by fitting the values of $\frac1p$ and $\vp$ obtained from MD simulations with a given $\kappa$ to Eq.~\eqref{eq:eos glass hs}. The results that I present next were obtained from the 20 configurations of $N=1024$ particles and the data of $p\in[10^5, 10^{11}]$ for each $\kappa \in \{3\times 10^{-4}, 10^{-4}, 3 \times 10^{-5}, 10^{-5}\}$. Instead of fixing $d=3$, I considered $d$ as a free parameter for the fit in order to account for the deviation with respect to the EOS and thus better estimate of $\vp_J^{(MD)}$. Nonetheless, in all the fits I found  $1  \leq d/3 \leq 1.07$ and $0.999 \leq R^2 $, indicating that they are reliable enough. $\vp_J^{(MD)}$ is then equal to $d$ divided by the estimated slope. 
The average densities resulting from such fits are reported in Table \ref{tab:phiJ-MD}, as well as the analogous packing fractions with target pressures $\ptar =10^3$ and $\ptar=10^5$. Results with larger target pressures are the same, considering the standard error.

\begin{table}[!htb]
	\centering
	\begin{tabular}{|c|c| c | c |}
		\hline
		$\kappa$ & $\displaystyle \avg{\vp_J^{(MD)}}$ & $\avg{\vp_J(\ptar=10^{3})}$ & $\avg{\vp_J(\ptar=10^{5})}$  \\[2mm]
		\hline
		$ 3\times 10^{-4} $	&	0.64223(19) & 0.64212(17) &  0.64223(19)   \\
		$ 10 ^{-4} $ &	0.64255(21) & 0.64252(21) & 0.64255(21) \\
		$ 3\times 10^{-5}$ &  0.64291(24) & 0.64288(24) & 0.64291(24)  \\
		$ 10^{-5} $ & 0.64301(23) & 0.64299(23) & 0.64301(23) \\
		\hline 
	\end{tabular}
	\caption{Average values of the jamming density of the MD compression protocol alone, and complimented by iLP. Each row corresponds to a different growth rate. Packing fractions in the $p\to\infty$ limit are estimated through a fit, as described in the text. The uncertainties reported are the standard errors, since these are larger than the ones obtained from the error propagation of the fits estimates.} \vspace*{-5mm}
	\label{tab:phiJ-MD}
\end{table}

\begin{figure}[!htb]
	\centering
	\includegraphics[width=\linewidth, height=10cm]{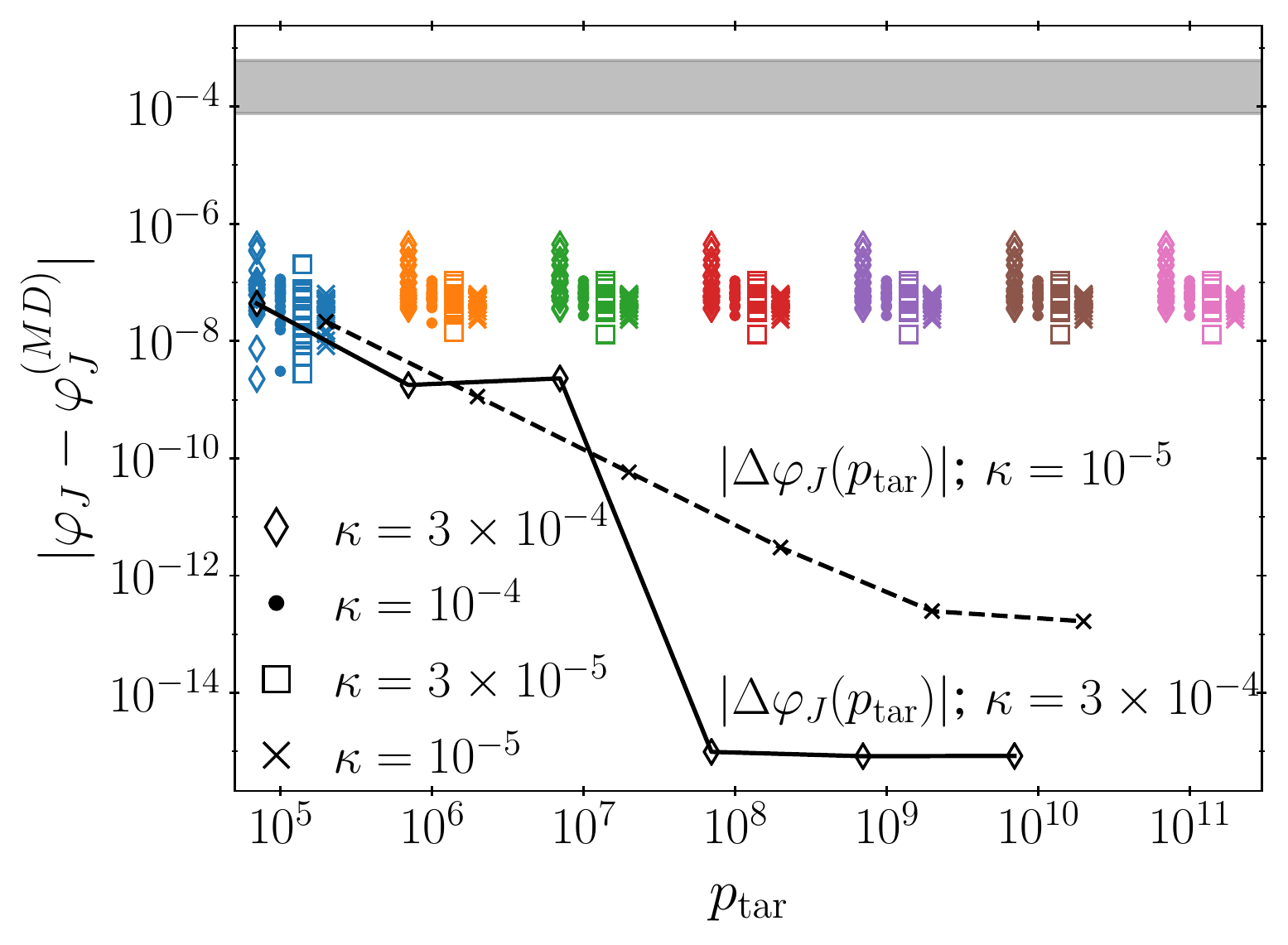}
	\caption[Difference in jamming densities obtained from iLP, $\vp_J$, and fitting data of MD simulations, $\vp_J^{(MD)}$.]{Difference in jamming densities obtained from iLP, $\vp_J$, and fitting data of MD simulations, $\vp_J^{(MD)}$. Markers are results obtained using 20 samples of $N=1024$.  Each marker style identifies a value of the growth rate, with the corresponding data horizontally displaced for clarity. To compare with the intrinsic variance of $\vp_J$ data of $\Delta \vp_J(\ptar)$ is also included. See text for more details.}
	\label{fig:diff phiJ LP and MD}
\end{figure}

It is clear that when the standard error associated to the sample is considered, jamming densities obtained from MD cannot be distinguished from the ones of  MD+iLP, even for the smallest $\ptar$ we considered here. Yet, because of sample to sample fluctuations of the values of $\vp_J$ and $\vp_J^{(MD)}$ (although always within a small range), it might be useless to compare their ensemble average. An alternative is  to consider the difference $\vp_J(\ptar) - \vp_J^{(MD)}$ from individual samples. However, the errors associated to the fitted parameters, whence $\vp_J^{(MD)}$ is calculated, lead to an uncertainty several times bigger than such difference. These findings are depicted in Fig.~\ref{fig:diff phiJ LP and MD}, where $\abs{ \vp_J(\ptar) - \vp_J^{(MD)}}$ is plotted for each sample, all values of $\kappa$ used so far, and $\ptar \geq 10^5$. For comparison, the range where uncertainty values of $\vp_J^{(MD)}$ lie is indicated by the grey shaded region, showing that it is at least two orders of magnitude larger than\footnote{I should mention that I have introduced  the absolute value only because for $\ptar \leq10^5$ some samples have $\vp_J < \vp_J^{(MD)}$, while for greater pressures the opposite is true in the vast majority of cases. Hence, removing the absolute value makes no difference in our conclusions.} $\abs{ \vp_J(\ptar) - \vp_J^{(MD)}}$. This implies that the two jammed states cannot be told apart \emph{macroscopically}.

On the other hand, note that the average values of $\Delta \vp_J$ are several orders of magnitude smaller than $\abs{ \vp_J(\ptar) - \vp_J^{(MD)}}$; cf. with the black curves included in Fig.~\ref{fig:diff phiJ LP and MD}. This means that even when there is a measurable difference in the iLP packing fraction, it is impossible to discern it only through a comparison with the values corresponding to MD. This would call for a \emph{microscopic} comparison between the two types of jammed states, if only the MD jammed packings were realizable. Unfortunately, this is clearly not the case. But even if a $p=\infty$ configuration could be produced through MD compression, there is no guarantee that it will be strictly jammed. As I argued in Sec.~\ref{sec:compression-protocol}, whenever $\kappa>0$, a $p=\infty$ \emph{could} correspond to a configuration in which only few particles are in contact, leaving several degrees of freedom unconstrained, and thus far from the 1SS requirement for jamming criticality (see Sec.~\ref{sec:jamming criticality}). In other words, although our simulations do not show any inconsistency with the hypothesis that $\vp_J$ and $\vp_{J}^{(MD)}$ identify the same jammed state, proving that this is the case is far from easy. 

%
%
%

Let me then analyse in detail the jammed states produced with iLP, for which micro- and macroscopic data are readily accessible. Recall that results of the previous subsection reveal that, when using the two smallest values of $\kappa$ in the $N=1024$ systems, an increase in $\ptar$ is accompanied by a measurable change in the jammed state reached through iLP. Such change is detectable both by the values of $\vp_J$ and of the similarity of contact networks (see Figs.~\ref{fig:phi-vs-p-and-kappa} and \ref{fig:similarity-contact-network}). Nevertheless, when larger values of $\kappa$ were used, these differences persist up to a certain threshold pressure, $\ptar^{(\rm{th})}$, above which all jammed states are virtually identical. That is, for $p\geq \ptar^{(\rm{th})}$ it happens that $\phi_J(p) =\phi_J(p_{\max})$ and $n_{ij}(p)=n_{ij}(p_{\max})$ for all configurations, at least within numerical precision. It is natural to assume that $\ptar^{(\rm{th})} = \ptar^{(\rm{th})}(\kappa, N)$, with a larger threshold pressure associated to a smaller growth rate and larger $N$ because the hierarchical structure of the FEL should  be realized in the thermodynamic limit with an infinitely slow compression. For instance, our simulations indicate that for $N=1024$ systems $\ptar^{(\rm{th})}(\kappa=3\times10^{-4}) \simeq 10^8$, $\ptar^{(\rm{th})}(\kappa=10^{-4})\simeq 10^{10}$, and $\ptar^{(\rm{th})} > p_{\max}=10^{11}$ for slower compressions. We can then conjecture that for any finite $\kappa$, there should be a threshold pressure above which iLP jammed states become identical, even using a \emph{microscopic} characterization. This latter claim is further supported by the results presented soon below in Figs.~\ref{fig:displacements-vs-ptar-and-kappa} and \ref{fig:msd-during-LP}.



%
%

\begin{figure}[!htb]
	\centering
	\includegraphics[width=\linewidth]{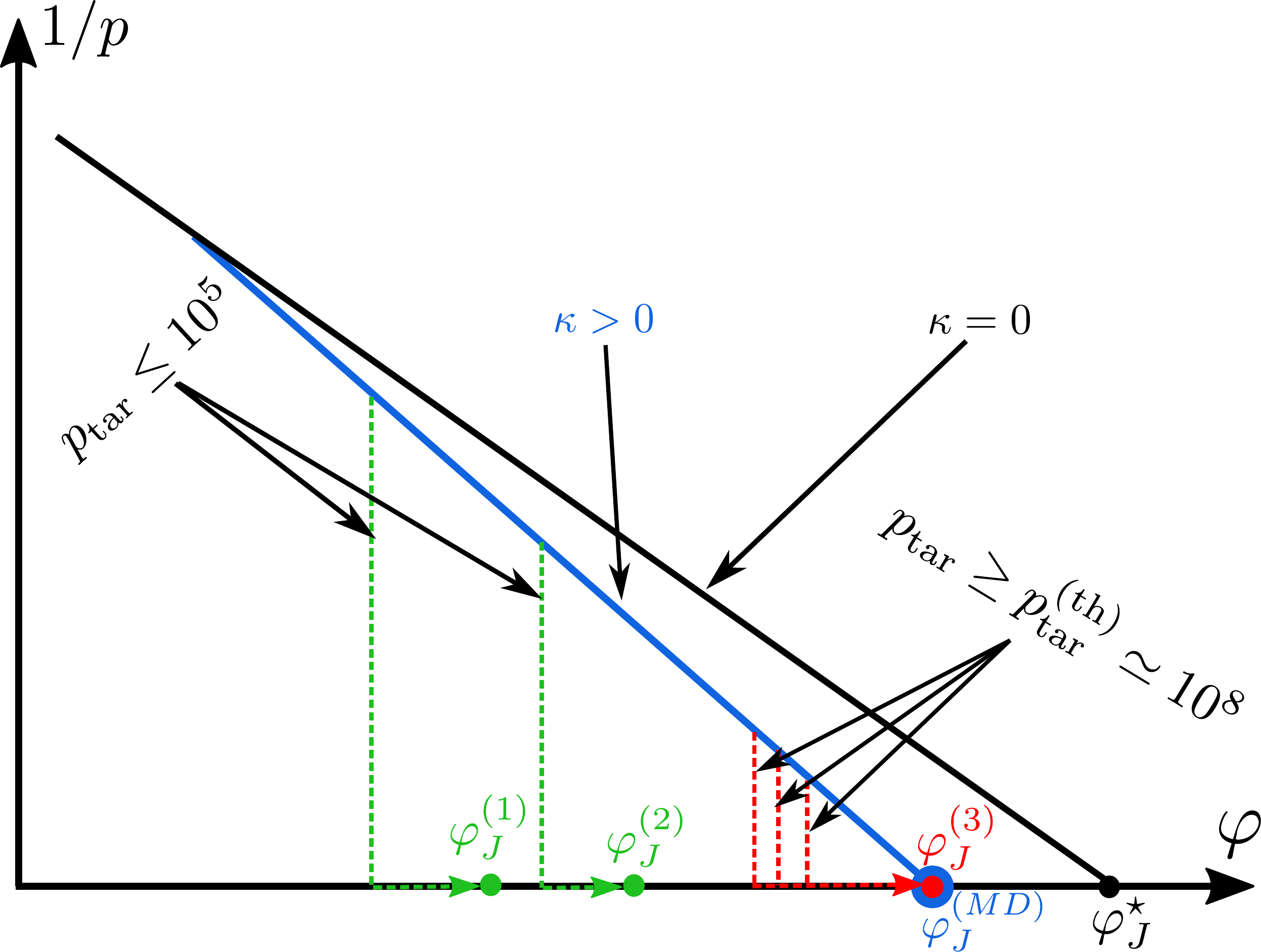}
	\caption[Diagram of the MD+iLP route to jamming]{Sketch of iLP and MD paths to jammed states, identified by their density. The equilibrium line obtained by adiabatic compression (solid black line) is followed closely by the MD compression protocol (blue), but due to the finite growth rate, it eventually detaches from the equilibrium line. Its $p=\infty$ extrapolated value is identified by $\vp_J^{(MD)}$ (although it is possibly not a strictly jammed state; see text), using a big circle to highlight the uncertainty in its estimation.
	When $\ptar$ is high, but not too much, using the configurations from MD as seeds of the iLP algorithm leads to different packings (green) with different packing fractions $\vp_J^{(1)}< \vp_J^{(2)}$. However when $\ptar$ is very large, different initial conditions lead to the same jammed state (red), $\vp_J^{(3)}$. Yet, an accurate comparison of $\vp_{J}^{(MD)}$ and $\vp_{J}^{(3)}$ with the available data is impossible given the uncertainty of the former. In any case, they are likely smaller than the jammed density estimated of an adiabatic protocol, $\opt{\vp}_J$. The values of $\ptar$ included are representative of the samples I consider here with $\kappa>10^{-4}$, and therefore are not universal.
}
	\label{fig:diagram iLP and MD states}
\end{figure}

The scenario that emerges from the discussion so far is that jammed states potentially reached by pure MD compression as well as those generated using iLP are different from the “thermodynamic” jammed state ($\opt{\vp}_J$) obtained via, say, adiabatic compression. The full picture is outlined in Fig.~\ref{fig:diagram iLP and MD states} in the $(1/p, \vp)$ plane close to jamming, \textit{e.g.} $p\gg10^3$.
The thermodynamic EOS is shown in black, while the path followed by MD compression is blue coloured. 
$\kappa$ controls the relative slope between these two curves, which end at $\opt{\vp}_J$ and $\vp_J^{(MD)}$, respectively, with $\opt{\vp}_J > \vp^{(MD)}_J$. Now, we have already seen that if $\ptar$ is deep in the glass phase but not extremely high (say $\ptar \leq 10^5$), different initial conditions in the iLP algorithm lead to different packings; this is illustrated by the green paths, ending in $\vp_{J,1}$ and $\vp_{J,2}$. However, when the initial pressure is very large ($\ptar \geq  \ptar^{(\rm{th})}$), different values of $\ptar$ lead to the same final jammed state $\vp_{J,3}$ (red), even in its microscopic structure. Of course, the value of such $\ptar^{(\rm{th})}$ depends on the value of $\kappa$; cf. the cyan and green curves in Figs.~\ref{fig:phi-vs-p-and-kappa} and \ref{fig:similarity-contact-network}. 
In most of the cases, $\vp_{J,1} < \vp_{J,2} < \vp_{J,3}$, while the uncertainty in the estimation of $\vp_J^{(MD)}$ does not allow to derive a similar relation with respect to $\vp_{J,3}$. Nevertheless, it is likely that $\vp_{J,3}=\vp_J^{(MD)}$. Besides, given that the iLP algorithm works by crunching a given configuration, and in doing so it necessarily takes it out of equilibrium, it is also expected that $\vp_{J,3}< \opt{\vp}_J$. 
That is, because iLP functions by applying an immediate quench to $T=0$ and then bringing the configuration to a local minimum, essentially without going over barriers in the FEL (see Sec.~\ref{sec:details algorithm}), it is reasonable that a smaller density is achieved, but given that no precise calculation of $\opt{\vp}_J$ is available for monodisperse HS, I could not verify this explicitly.
In any case, note that this picture mimics the very recent findings of \emph{algorithmic} memory formation in jammed packings reported in \cite{charbonneauMemoryFormationJammed2020}.

\begin{figure}[!htb]
	\centering
	\includegraphics[width=\linewidth]{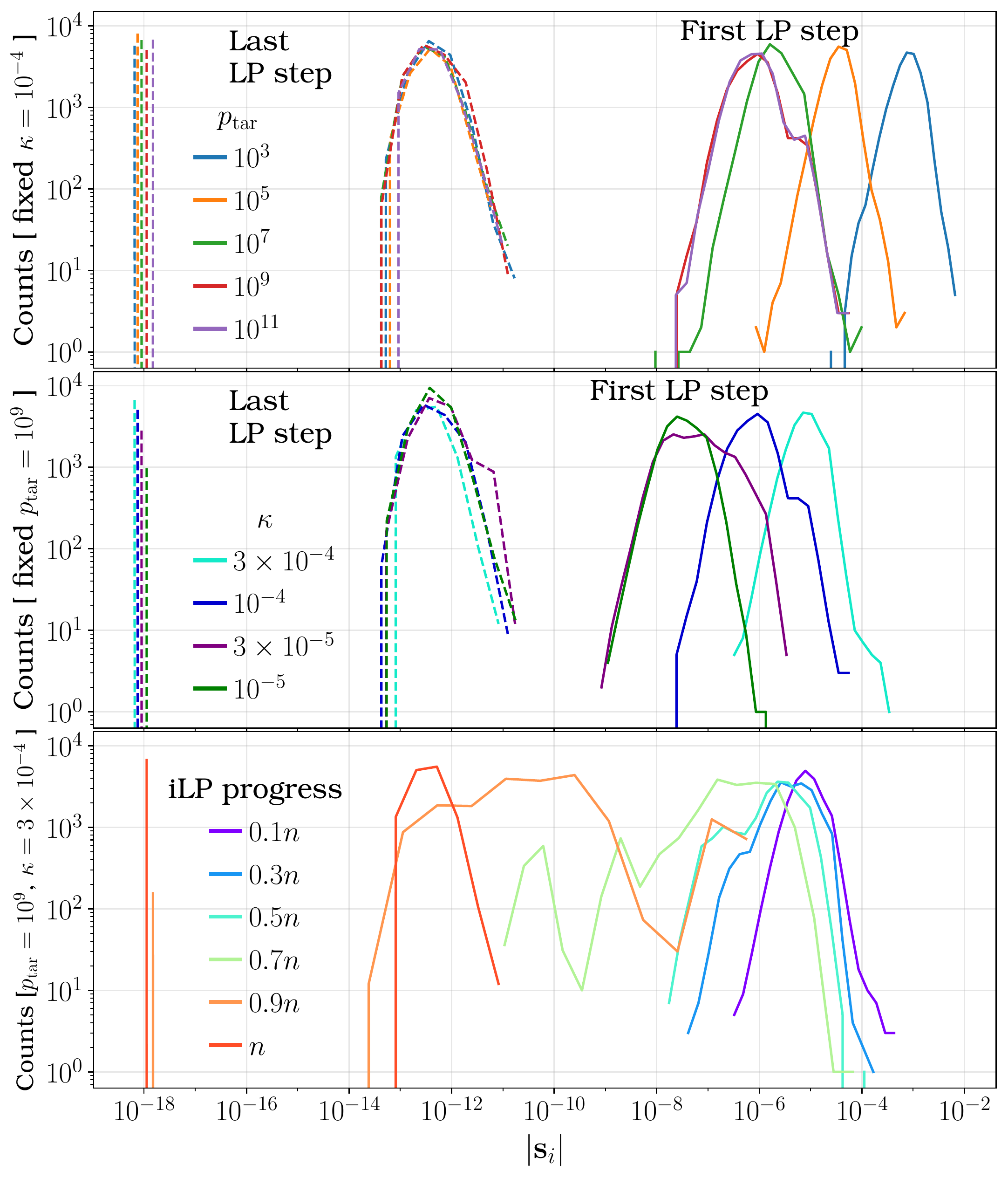}
	\caption[Histograms of displacements during iLP crunching]{Histograms of individual particle displacements, $\{\abs{\vb{s}_i}\}_{i=1}^{N'}$ (\textit{i.e.} excluding rattlers), at different steps of our iLP algorithm. Results changing the target pressure (resp. growth rate) are shown in the top (resp. middle) panel, for the first (solid lines) and last (dashed) LP steps. In order to also include the case of $\abs{\vb{s}_i}=0$, such null displacements are drawn approximately at $10^{-18}$. The bottom panel instead shows the change in the distributions of $\abs{\vb{s}_i}$ at different steps of the iLP crunching (with $\kappa$ and $\ptar$ as indicated on the left). These results show that our LP algorithm brings a configuration to its closest minimum as discussed in the text.}
	\label{fig:displacements-vs-ptar-and-kappa}
\end{figure}

To proceed with the microscopic characterization of the iLP algorithm, it is worth to analyse the particles rearrangements performed by the iLP algorithm as the jamming point is reached. If the scenario of Fig.~\ref{fig:diagram iLP and MD states} is true, for sufficiently high $\ptar$ we should observe that the particles displacements $\{\vb{s}_i\}_{i=1}^N$ become (roughly) independent of the target pressure of the initial condition. To test such hypothesis, I will consider the distribution of LP displacements, $\{\abs{\vb{s}_i}\}_{i=1}^{N'}$ (\textit{i.e.} without including rattlers) mainly during the first and last iterations. For later use, I will denote as $n$ the number of LP optimizations carried out to reach the jamming point of a configuration. (The dependence of $n$ on $\ptar$ and $\kappa$ is postponed to Fig.~\ref{fig:iters-vs-p-and-kappa} in Sec.~\ref{sec:LP scaling with size}.) Given that $\avg{n} \approx 3$ for $\ptar \geq 10^7$ and $\kappa\leq 10^{-4}$, in practice only the first and last LP steps can be consistently compared in all the samples.
Now, we need to distinguish between these two steps because, as argued in Sec.~\ref{sec:LP algorithm}, our iLP algorithm quickly reduces the free volume (\textit{i.e.} the feasible region) of the configuration. It is hence expected that the largest displacements take place during the initial steps. In contrast, just before convergence iLP performs displacements that are, very likely, employed to “fine tune” the contacts (for instance, by matching linear and real contacts, identified by tangency of constraints; see Sec.~\ref{sec:details algorithm}), but not on relatively large rearrangements. Furthermore, from the examples given in Sec.~\ref{sec:details algorithm} about how the feasible region is monotonically reduced, we can reasonably assume that during the iLP process, configurations do not really jump over free energy (or actually, entropic) barriers. With all this in mind, if initial displacements are similar for very large $\ptar$, while the last ones are minor “adjustments” (for all values of $\ptar$), we can conjecture that iLP works by crunching the configuration to the nearest minimum. Happily, the histograms of $\abs{\vb{s}_i}$ in the central and uppermost panels of Fig.~\ref{fig:displacements-vs-ptar-and-kappa} agree precisely with this picture. Specifically, the effect of $\kappa$ (central panel) is only visible for the first LP step, while if $\ptar \geq 10^{7}$ a small to negligible effect is observed even in such initial part (top panel). 
On the other hand, for the last LP iteration, there is essentially no difference when $\kappa$ or $\ptar$ are changed. The only visible effect is that the amount of null displacements, which for convenience are drawn around $10^{-18}$, increases monotonically with $\kappa$. This means that when the initial condition of iLP is better equilibrated --\textit{i.e.} when a smaller $\kappa$ is used-- more contacts can be fine tuned simultaneously. In contrast, if a faster compression is employed a larger portion of the configuration remains fixed, while the final missing contacts are determined.

To further explore the mechanisms of the iLP algorithm during its convergence process, we can consider the configurations obtained with $\kappa=3\times 10^{-4}$, given that in such case $\avg{n} \approx 10$ for the range of pressures $\ptar \geq10^7$. The results are presented in the bottom panel of Fig.~\ref{fig:displacements-vs-ptar-and-kappa}, whence we conclude that during the first half of iterations the displacements are practically of the same size, while it is only in the second half that a significant fraction of particles remain rather fixed. Besides, in the intermediate steps the distribution of $\abs{\vb{s}_i}$ broadens across several decades, while during the initial and final steps it is mostly peaked. This provides evidence that the iLP algorithm works by first converging to a small region of a meta-basin (by relatively large displacements) and then realizing progressively the network of contacts (by small but broadly distributed displacements).

\begin{figure}[!htb]
	\includegraphics[width=\linewidth]{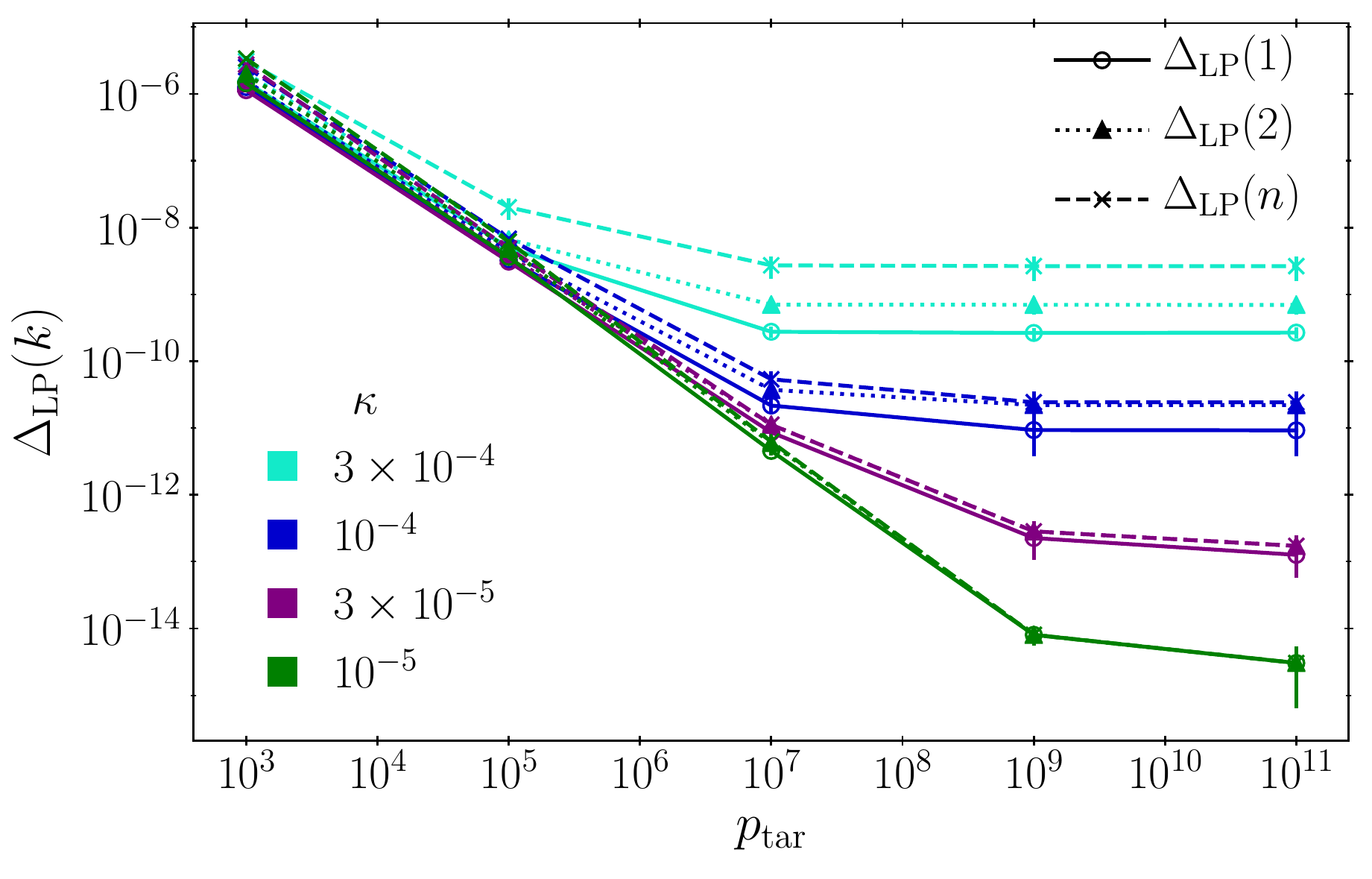}
	\caption[MSD during at different steps of iLP]{MSD for the iLP algorithm (defined in the text), for different growth rates and computed after the first, second, and last LP optimization step.}
	\label{fig:msd-during-LP}
\end{figure}

Finally, it is also interesting to keep track of the full configuration motion during the iLP process, so let me now analyse the MSD at each iteration, that is $\Delta_{\text{LP}} (k) := \frac1{N'} \sum_{i=1}^{N'} \abs{\vb{r}_i(k) - \vb{r}_i(\ptar)}^2$, where $k=1,\dots, n$ and $\va{r}(\ptar)$ is the configuration obtained from MD at a given target pressure. The values of this quantity for the different parameters considered here are reported in Fig.~\ref{fig:msd-during-LP} for the first, second and last iterations. The data confirm the discussion of the previous paragraphs about the fact that for a given $\kappa$, there exists a target pressure above which the rearrangements of the configurations are essentially unchanged. Moreover, even for very small $\kappa$ and very high $\ptar$, the first and last iLP steps have different roles: the former accommodates the particles, and the latter fixes the contacts. Put together, these two features suggest that for any finite growth rate and even in the $\ptar \to \infty$ limit of MD, some rearrangements are needed to obtain a proper jammed 1SS state.
To recap: in combination with the results above, these findings further support the claim that iLP states reached from $p\geq \ptar^{(\rm{th})}$ are identical, both from a macroscopic point of view (\textit{i.e.} using $\vp_J$) or from a microscopic one (characterized through $\{\abs{\vb{s}_i}\}_{i=1}^{N'}$ or $\overline{n_{ij}(\ptar) n_{ij}(p_{\max}) }$). It is also worth recalling that $\vp_J^{(MD)}$ apparently converges to the same limit value than iLP for sufficiently large pressure, which strongly suggests that iLP crunching is an efficient way to reach the infinite pressure limit; see Fig.~\ref{fig:diagram iLP and MD states}. 
However, two sources of error should be kept in mind: (1) fitting the glass EOS to MD data yields an estimation of $\vp_J^{(MD)}$ with a relatively large uncertainty; and (2) the configuration of the MD compression protocol in $p\to\infty$ limit might not be strictly jammed. 
Therefore, although we cannot ensure that both jammed states are equal, our findings suggest that they should be. In any case,  a more extensive study with even slower compressions and higher target pressures should be useful to firmly confirm this picture. 



As a last remark in this section, I would like to mention how iLP could be used in a future work to carefully explore the landscape structure. First, we could analyse if $\Delta_{\text{LP}}$ bears any relation with the hierarchy of meta-basins of the fractal FEL, itself characterized by the distribution of possible values of the MSD. A way to do this is to apply the MD compression up to a determined $\ptar$ and then let the dynamics of the system continue its evolution but \emph{without} compression. After a fixed number of collisions, the configurations are saved and crunched using iLP. By comparing the MSD of the actual dynamics, with the “distance” between jammed states, and also with $\Delta_{\text{LP}}$ we could further explore how iLP navigates the FEL towards a minimum. It should be interesting to find out if any of these quantities captures the jamming critical exponents. For instance, in Sec.~\ref{sec:MF gardner transition} I mentioned that the plateau value of the MSD scales as $\Delta \sim p^{-\kappa}$. A similar relation could hold between $\Delta_{\text{LP}}$ and $\lpf$. Nevertheless, note that the plateaus of $\Delta_{\text{LP}}$ that form for a given value of $\kappa$ might indicate that configurations cannot probe the finer structure of a meta-basin, even as they go down (\textit{i.e.} as $\ptar$ increases) the landscape. So this proposal might not be straightforward to implement.

%

\subsection{Complexity of iLP: Scaling with size}\label{sec:LP scaling with size}

A final property I will briefly analyse is the size scaling of the iLP algorithm. I will mostly omit the MD compression part mainly because we used a well known protocol and did not implement any new features\footnote{Except possibly for the MD simulations of the MK model, but we will consider its full characterization in a future work}. But also because the LS compression and the iLP crunching benefit from two different and excluding computational approaches. In the former case,  due to the deterministic dynamics and the serial nature of asynchronous event-driven algorithms\footnote{Although see \cite{bannermanDynamOFreeCal2011} and references therein for clever parallelised implementations, some allowing scaling of $\order{N}$.}, the MD compression of a single sample is not accelerated if more processors are available. (Clearly, if several systems are required, the LS compression can be trivially hastened by executing independent MD simulations in separate threads.) 
In contrast, the opposite is true for the iLP part, at least using the Gurobi solver\supercite{gurobi} as we did here\footnote{\label{fnt:gurobi}We decided to use Gurobi because a free academic license is available, which allows to solve large-scale optimization problems (other free licensed software limits the number of variables+constraints to about $3000$) and is definitely faster than other freely available libraries, like GLPK or HiGHS; some benchmarks are available \href{http://plato.asu.edu/ftp/lpsimp.html}{here}. The obvious drawback is that, being a proprietary software, it works as a “black-box” with few tunable parameters. That is, the user can specify the desired accuracy for the solution (within a given range), the solving method, etc. but cannot access the real algorithmic implementation.}.
The reason being that the optimization of each \ref{step:LP} in Algorithm \ref{alg:LP algorithm} is done using a \emph{non-deterministic} algorithm that runs multiple instances of the interior-point  method\supercite{boydConvexOptimization2004,vanderbeiLinearProgrammingFoundations2014,nocedalNumericalOptimization2006} with different random initial conditions. The solution reported is the first that converges within a user defined accuracy. Hence, the larger the amount of different initial conditions the faster the algorithm, so the solver works by implementing as many as available processors. (Unfortunately, however, the speed gain in increasing the number of initial conditions is not substantial: doubling the threads is far from halving the required computational time.)
On the other hand, the alternative of using a “single-thread” solution technique, such as the simplex method\footnote{\label{fnt:simplex-method}Most accessible implementations of the simplex method --both in its primal or dual variants-- are serial, although some parallelised versions have been recently developed\supercite{huangfuParallelizingDualRevised2018}. However, the most performant ones are proprietary and non-free; see footnote \ref{fnt:gurobi}.}, in order to maintain the serial property in the full MD+iLP method, severely slows down the LP optimization. We therefore decided to focus mostly on the iLP part (which is the real novelty of our approach) and just consider briefly the execution times of the \emph{full} MD+iLP procedure as I now discuss.

\begin{figure}[!htb]
	\centering
	\includegraphics[width=\linewidth]{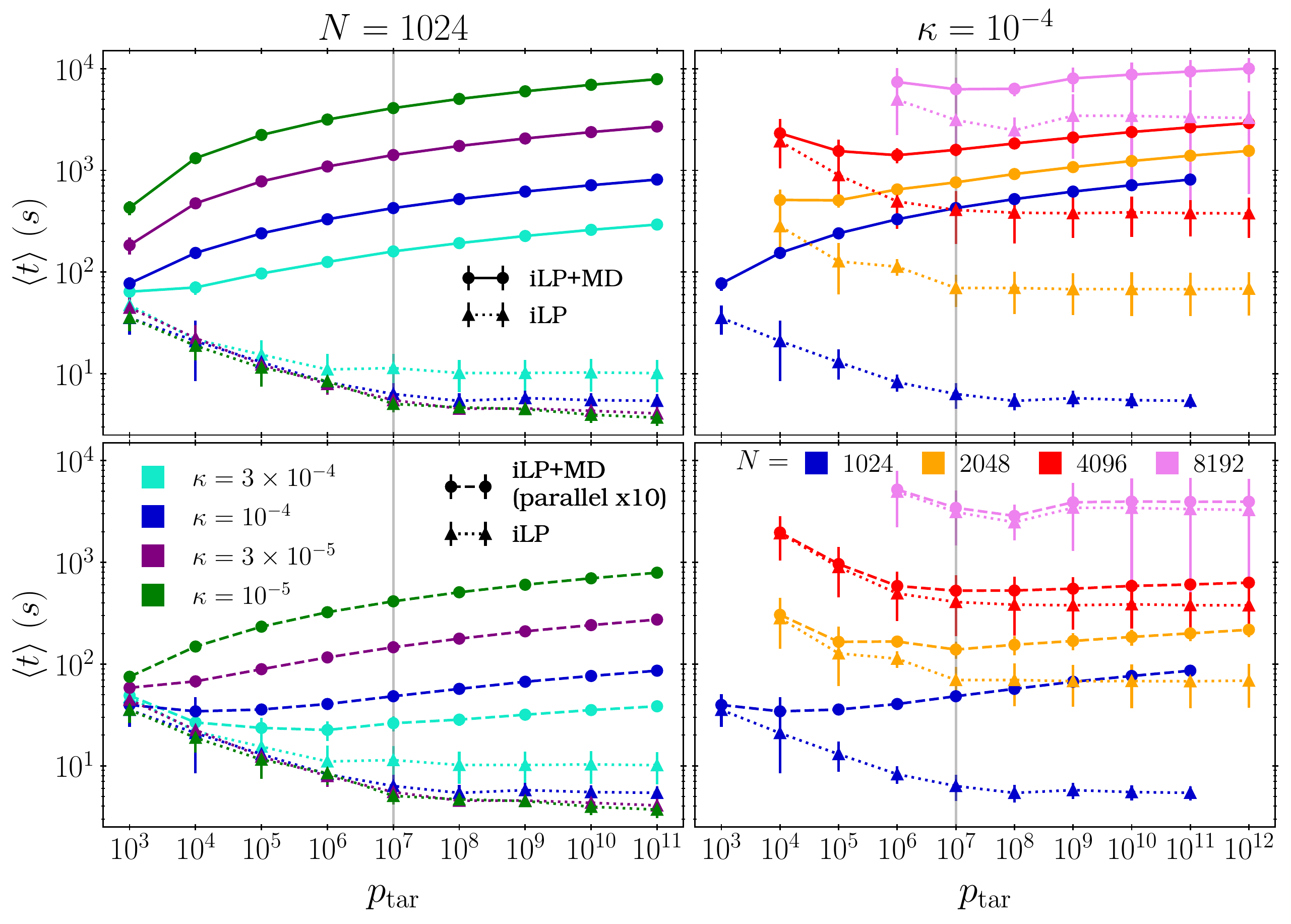}
	\caption[Execution times of the MD+iLP algorithm, for different values of $\kappa$, $\ptar$, and $N$.]{Execution times of the full MD+iLP process as a function of $\ptar$ using $N=1024$ and different growth rates (left panels) and for fixed $\kappa=10^{-4}$ and varying system sizes (right). Upper panels show the total time of MD+iLP (solid lines), while the lower plots show the same quantity, but assuming that 10 MD samples can be generated simultaneously. In all cases, the iLP times have also been included (dotted, cross markers) for comparison. The vertical grey lines indicate the value of $\ptar$ employed for most of the results presented in this section. For relatively small systems and $\ptar\geq 10^6$, it is clear that the most time consuming part is the MD compression. For larger systems instead, iLP always takes a considerable portion of the time. See text for more details.}
	\label{fig:total-times-MD-iLP}
\end{figure}

In tune with the results of the two previous parts, I will also explore the dependence of the running time ($t$) as a function of $\ptar$, for different values of $\kappa$ and $N$. All the results that follow were obtained running the simulations in a 6 cores computer, with processor Intel Core i7-8700 at 3.2 GHz. In the iLP part, each LP optimization was solved using the barrier method as implemented in Gurobi version 8.1, with at most one initial condition per core.
The statistics of $t$ are summarized in Fig.~\ref{fig:total-times-MD-iLP}, where mean running times of the full MD+iLP are shown (circular markers) and compared with the average running time of the iLP part only (crosses, dotted lines). The values reported correspond to the average over the same samples as above, and the error bars are their standard deviation. Left panels show the influence of the growth rate for fixed system size, $N=1024$;  while the graphs on the right explore the effects of the system size for fixed $\kappa=10^{-4}$. Additionally, the bottom panels plot the same quantities, but considering that several MD processes can be executed simultaneously in a single computer. In particular, for these plots I assumed that 10 configurations are produced in parallel, which corresponds to the most common situation when these data were generated. 

Let me begin by analysing the effects of the growth rate in the small systems. As expected, changing $\kappa$ mostly affects the duration of MD, while its impact on the iLP times is very small, especially for $\kappa < 10^{-4}$. This feature is also reflected in the number of LP iterations ($n$) performed by our crunching algorithm as reported in Fig.~\ref{fig:iters-vs-p-and-kappa} below. Yet, the most notable effect is that the time needed by iLP is roughly constant for $ \ptar\geq 10^7 $, as expected from the fact that $n$ remains essentially unchanged for such large pressures, and confirms the features summarized above when discussing Fig.~\ref{fig:diagram iLP and MD states}. Now, given that decreasing $\kappa$ obviously lengthens the MD simulations, a trade-off between a large (but not huge) value of $\ptar$ and a small growth rate is expected. This would be clearly signalled by a convex curve of the MD+iLP time, with the minimal value of $t$ being rather unaltered by changing $\kappa$. However, this is not the scenario observed for the $N=1024$ --where $t$ increases monotonically with both $\ptar$ and $\kappa$. This is easily explained because the iLP algorithm is very fast at crunching small systems, so extending the MD compression in order to reach a larger $\ptar$ is in general a bad strategy. Of course, this conclusion only applies to the target pressures considered here: if iLP is used from a liquid configuration (\textit{e.g.} using $\ptar \simeq 10$) this would not be true and the value of $t$ would be significantly larger. Interestingly, for small systems iLP is fast enough that even when considering the simultaneous MD compression of samples, only a small dip in the curves of $t$ can be appreciated. And it occurs at a relatively small target pressure, $\ptar\simeq 10^5$, and large values $\kappa$. Thus, moderate target pressures and fast compressions are optimal when jamming small systems. In turn, when dealing with large systems increasing $\ptar$ is more beneficial the higher $N$ is. Once again, however, the gain in speed of the iLP algorithm is capped for very large pressures, $\ptar \geq 10^9$, at least for the sizes considered here. Moreover, in the scenario of big configurations, it is clear that a significant amount of the computation time is caused by iLP. This is specially important when considering the possibility of simultaneous MD runs, as shown by the lower right panel of Fig.~\ref{fig:total-times-MD-iLP}. This plot evinces that for $N\geq 4096$ the major contribution to $t$ is precisely the iLP crunching. Note also that, for fixed $N$, varying $\ptar$ across 8 orders of magnitude rarely increases $t$ by more than a decade. However, changing the system's size can lead to an increase in $t$ (resp. $t_\text{LP}$) of almost two (resp. more than two) orders of magnitude, which means that reducing $t$ of large samples should be favoured over small ones. Put it simply, it is unfeasible to choose values of $(\kappa,\ \ptar)$ that are optimal for \emph{any} given system size. This is the reason why in our analysis of the complexity of the iLP algorithm as a function of $N$ we used a fixed value of $\ptar=10^7$ (grey vertical lines) and just tred two different MD compression protocols as described below. This value of $\ptar$, although suboptimal for small samples, provides a good trade-off when using several values of $N$.


I will henceforth restrict the analysis of $t$, iterations, etc. to the iLP algorithm exclusively. From the considerations of the first paragraph of this subsection, it would seem that the characterization I will present here heavily depends on the specific solver we used. Fortunately, the situation I described there is rather general, \textit{i.e.} interior point methods are usually faster than simplex-based ones (see, however, footnote \ref{fnt:simplex-method}). Additionally, the trend we found suggests that the complexity of the iLP jamming algorithm has a well defined dependence with the system size. 
A final technicality to consider when interpreting the results is that, when using an interior point method, the convergence rate to a solution within a given accuracy also depends on the system size\supercite{nocedalNumericalOptimization2006,boydConvexOptimization2004}. Here, we set the (absolute) tolerance for optimality and feasibility to $10^{-9}$, the most accurate option available with Gurobi. It is also worth noticing that our algorithm is able to generate jammed packings with such relatively low accuracy, given that other methods require quad-precision to avoid over-shooting the jamming point\supercite{FIRE,charbonneauJammingCriticalityRevealed2015}. All in all, this means that the convergence times reported next and related quantities are good proxies of the real algorithmic complexity of iLP, but the precise values should be taken with a grain of salt.

\begin{figure}[!htb]
	\centering
	\includegraphics[width=\linewidth]{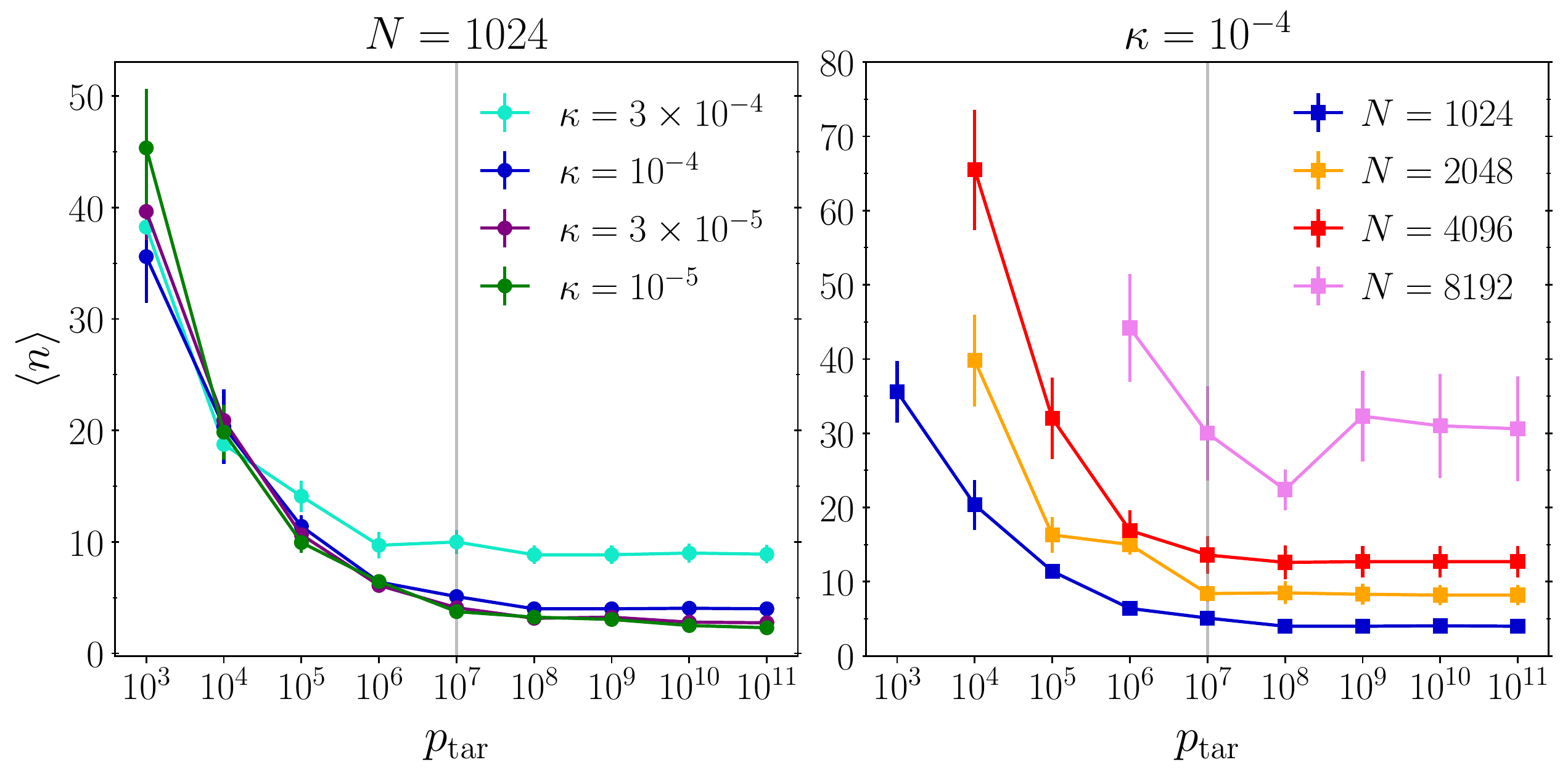}
	\caption[Number of LP optimizations required for convergence as a function of $\ptar$ and $\kappa$.]{Number of LP optimization steps required by the Algorithm \ref{alg:LP algorithm} as a function of the target pressure, varying the compression rate (left panel; 20 samples per point) and the system size (right panel; 10 samples for $N\geq 2048$). Data obtained from the same configurations of Sec.~\ref{sec:LP dependence initial confs}.
	}
	\label{fig:iters-vs-p-and-kappa}
\end{figure}

The first feature to analyse is the number of LP steps ($n$) needed to reach the jamming point and its dependence on $\ptar$ and $\kappa$. Analysing $n$ complements the results discussed above regarding Fig.~\ref{fig:total-times-MD-iLP}, as well as those of Secs.~\ref{sec:LP dependence initial confs}-\ref{sec:MD-and-LP}.
The statistics of $n$ are reported in Fig.~\ref{fig:iters-vs-p-and-kappa}, using data from the same configurations considered above. In agreement with the results presented previously, there is a negligible variation for $\ptar \geq 10^7$. In contrast, for fixed $N=1024$, $\avg{n}$ barely changes for $\kappa \leq 10^{-4}$, indicating that the  growth rate has a smaller influence on $n$ than on other quantities. Instead, when different values of $N$ are used, the results confirm the expected outcome that $\avg{n}$ increases monotonically with the system size. Note also that for sufficiently large $\ptar$, the number of LP steps remains practically constant and is never smaller than 3.
When considered together with the results of Figs.~\ref{fig:displacements-vs-ptar-and-kappa} and \ref{fig:msd-during-LP}, these data suggest that, for any $\ptar < \infty$ and finite $\kappa$, the iLP method requires at least two steps in order to reach the jamming point. In the first one, particles undergo relatively large displacements that define an inherent structure, while in the second step they are minimally displaced in order to turn linear (\textit{i.e.} non tangent) contacts into physical ones (Sec.~\ref{sec:details algorithm}).

The next feature to consider is the size dependence of $n$ and the convergence time $\tau$. The results that follow were obtained fixing $\ptar = 10^7$ and testing two different compression strategies. In the first one, a constant growth rate $\kappa = 10^{-5}$ was used for all values of $N= \{2^8, 2^9, \dots, 2^{14}\}$. Notice however that such procedure keeps the \emph{radius}' growth rate constant, but not the density one. For the latter, we must keep in mind that $\dot{\vp} \sim N \sigma^2 \dot{\sigma} \sim (N \vp^2)^{1/3} \kappa$. Hence, if for a given density $\vp$ we want configurations of different size to have the same compression rate, we should scale $\kappa_N \sim N^{-1/3}$. Thus, in the second strategy we used as reference $N_{\max}=2^{14}$ and then scaled $\kappa_N = (N_{\max}/N)^{1/3}\ 10^{-5}$ for smaller sizes. To perform an accurate analysis, we generated $M=100$ samples for each value of $N$ and for each compression strategy. 

\begin{figure}[!htb]
	\centering
	\includegraphics[width=\linewidth]{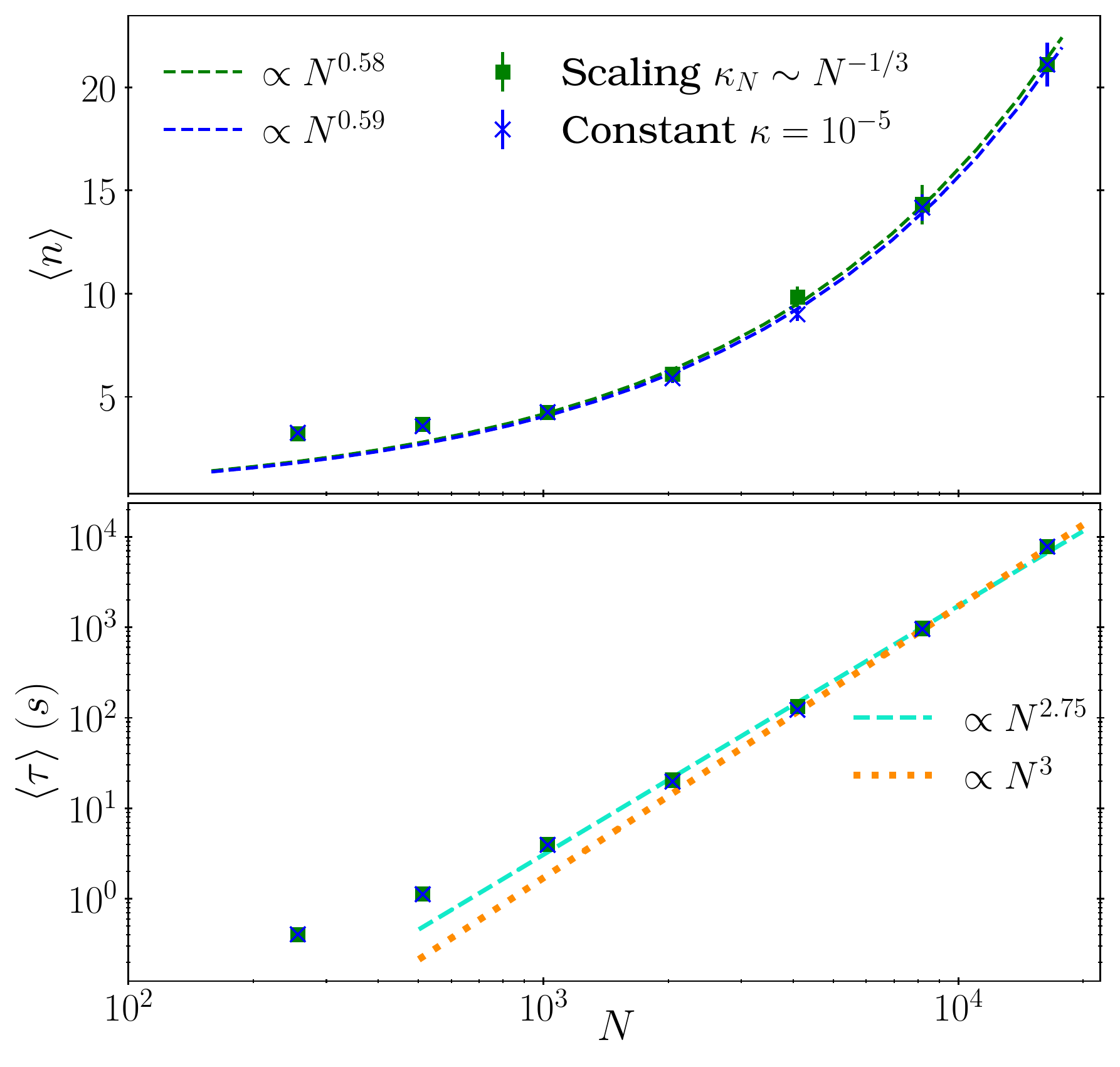}
	\caption[Iterations and convergence time as a function of the system's size]{Average number of LP optimizations (upper panel) and convergence time (lower) as a function of the system size $N$. Results from a constant (resp. scaled) compression rate are indicated by the blue (resp. green) markers, while dashed lines are least-squares fits obtained from data with $N\geq 1024$. Each marker corresponds to the average over 100 samples, with the error bars being the associated standard error. The orange dotted line suggests that if even larger systems were considered, a cubic scaling with $N$ is likely to result. For small $N$, the large proportion of constraints with respect to the number of variables explains the deviation from the large $N$ trend.}
	\label{fig:conv-time-vs-N}
\end{figure}

The average values of $n$ (upper panel) and $\tau$ (bottom panel) are reported in Fig.~\ref{fig:conv-time-vs-N} for all the system sizes considered and both compression procedures. Error bars correspond to the standard error, although for $\tau$ the logarithmic scale renders them invisible. 
The first thing to notice is that both compression strategies produce very similar results, although less iterations are obtained in general when $\kappa$ is fixed for all $N$, as expected. Such similarity is easy to understand because given the values of $N$ considered here, we have that $\kappa_{N_{\min}} / \kappa_{N_{\max}} = (2^{14}/2^{8})^{1/3} = 4$. In other words, the compression rate changes at most by a factor $4$, which is not enough to compensate the influence of changing the system size; recall results from Fig.~\ref{fig:iters-vs-p-and-kappa}.
Importantly, both series of results exhibit a well defined scaling with $N$, with $\avg{n}$ scaling as $N^{0.59\pm 0.02}$ ($N^{0.58\pm0.01}$) for the fixed (scaled) compression rate, while $\tau \sim N^{2.75 \pm 0.08}$ in both cases. These exponents were found by least-squares fits, which are also included in Fig.~\ref{fig:conv-time-vs-N} (dashed lines). Such fits were obtained only from data of $N\geq 1024$, because finite size effects are likely to influence the results for smaller $N$. The reason is that, given the particles' diameter for $N=256$ and $512$, $\ell(\vp)$ is about $1/3$ or more of the system's length, and thus the ratio of number of constraints/number of variables is considerably higher than for larger $N$. Furthermore, current data are not enough to rule out a faster scaling, such as $N^3$ (dotted, orange line) for $N\geq 2048$. Nevertheless, it is expected that the real exponent should be close to $3$, although larger system sizes are needed to confirm this conjecture.


\begin{figure}[!htb]
	\centering
	\includegraphics[width=\linewidth]{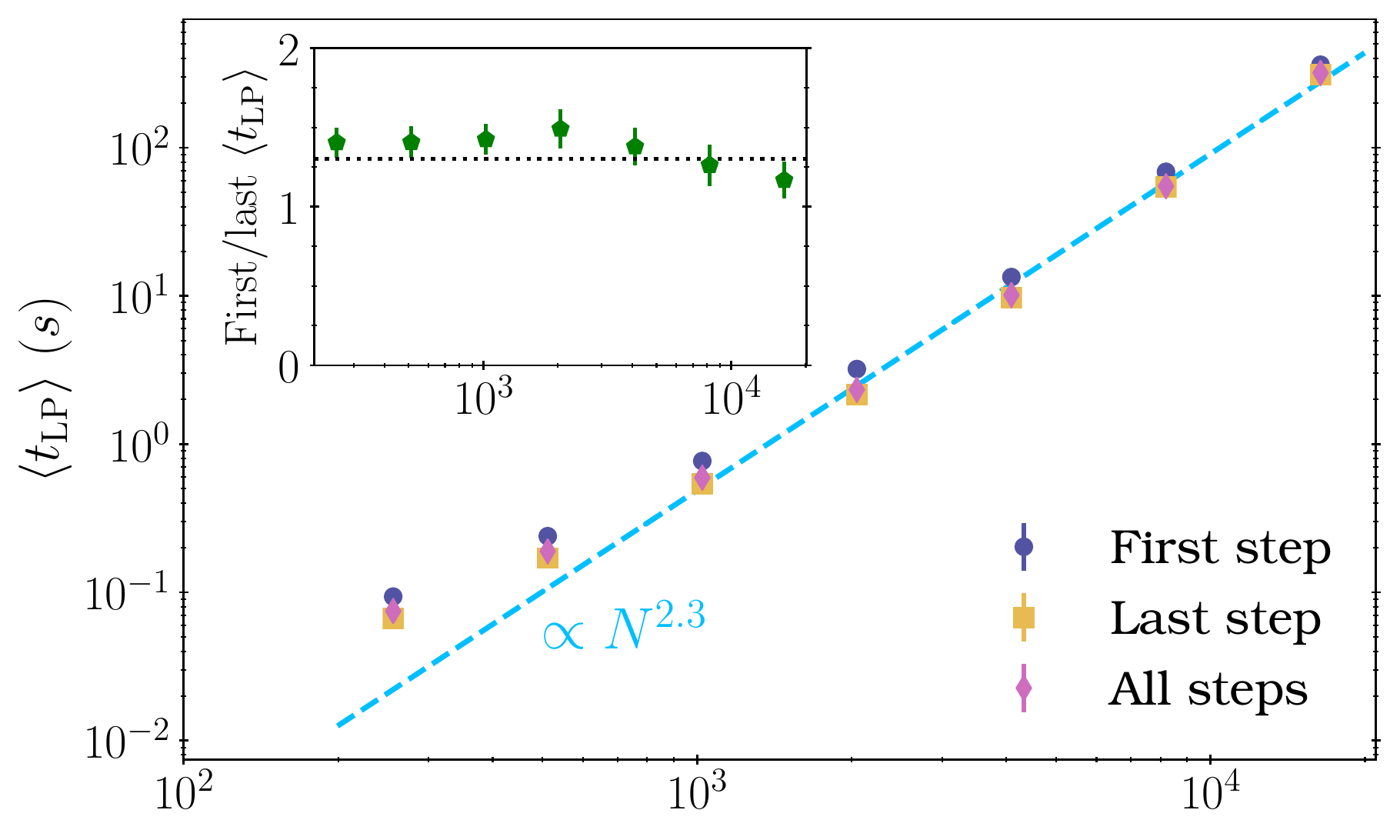}
	\caption[Time per LP iteration and its dependence on $N$.]{Main: Average time of LP optimization steps. The first and last steps are considered separately, since the corresponding displacements are very different; see Fig.~\ref{fig:displacements-vs-ptar-and-kappa}. The behaviour of intermediate steps is also considered by including all the LP steps to estimate $\avg{t_{\text{LP}}}$. The dashed, blue line corresponds to a least-squares fit and shows that $t_{\text{LP}}$ of the three sets scales as $N^{2.3\pm 0.1}$. Inset: average ratio of $t_{\text{LP}}$ between the first and last optimization. The dotted line signals that such ratio is roughly 1.5 for all system sizes, confirming that essentially the same scaling with $N$ holds throughout the iLP algorithm
	}
	\label{fig:time-per-iteration-vs-N}
\end{figure}

As a final feature, I will consider only the time spent in solving each \ref{step:LP} of the iLP algorithm, $t_{\text{LP}}$. This is an important variable because it measures how intrinsically complex the LP optimizations are, without taking into account the computational costs of identifying the relevant constraints, incorporating periodic effects, computing the force network, etc.
Fig.~\ref{fig:time-per-iteration-vs-N} shows the results of $\avg{t_{\text{LP}}}$ obtained using a constant $\kappa$, and distinguishing between $t_{\text{LP}}$ of the first and last iterations, (circular and square markers, respectively). Recall that in Fig.~\ref{fig:displacements-vs-ptar-and-kappa} we have already seen that the first LP steps are considerably different from the last ones, therefore there is no reason \textit{a priori} to assume that they scale equally with $N$ and considering them separately is thus important. Nevertheless, our simulations demonstrate (inset) that their ratio is approximately constant for all sizes, implying that they scale with the same power of $N$. This is further confirmed by the values of exponents found by fitting the data of $N\geq 1024$ as above. The analysis yields that the time of the first (resp. last) scale as $N^{2.22\pm 0.05}$ (resp. $N^{2.30\pm 0.07}$). Moreover, if all the iterations are considered, it is found that $\avg{t_{\text{LP}}} \sim N^{2.27\pm0.07}$, which closely follows the other two cases. Hence, by simply assuming $N^{2.3}$ all the data for $N\geq 1024$ can be nicely captured, as evinced by the blue, dashed line.

Putting together these results, the first important thing that emerges is that the size scaling of $\tau$ is \emph{different} from the one naively obtained by considering $n\times t_{\text{LP}}$, although barely within the margin of error: cf. $N^{2.9\pm0.1}$ and $N^{2.75\pm0.08}$. This implies that although the most costly part is the optimization itself, the additional operations --like computing distances between particle pairs, assigning constraints, and computing the structural variables-- cannot be simply neglected. However, it is likely that as bigger systems are considered, the dominant contribution will be due to $t_{\text{LP}}$, increasing the size scaling of $\tau$ slightly. 
On the other hand, I want to emphasize the empirical character of the analysis I have presented here. That is, the scalings reported were obtained only through curve-fitting (although showing very good agreement) and not from analysing the algorithmic complexity of each part of our iLP method. Even performing a more detailed analysis of the optimization part is not trivial, mainly, due to the proprietary code we used. But even the “turning the knobs of the black-box” approach is challenging because of the several secondary steps carried our by the solver, such as presolving, Cholesky factorization, barrier cross-over\footnote{Each of these subroutines is almost independent from the others, so analysing their complexity is beyond the scope of this thesis, but an accessible exposition of each of them can be found in \cite{nocedalNumericalOptimization2006}. See also \href{https://www.gurobi.com/resource/parallelism-linear-mixed-integer-programming/}{this tutorial} for a simple assessment of their computational costs specific to the Gurobi solver.}. It is thus unsurprising that when combined, the size dependence of our iLP algorithm is not necessarily a rational power-law. A final remark is that the role of dimensionality has been completely neglected in this characterization, and although we tested our algorithm in up to 5 dimensions, we leave the corresponding complexity analysis as a topic for future research. Nevertheless, some important properties can be anticipated. First, the number of contacts increase linearly with $d$, as stated in Eq.~\eqref{def:N single self stress} and discussed in Sec.~\ref{sec:LP network of contacts}, which means that complexity of LP optimizations is at least linear in $d$. However, not only touching particles are considered when introducing in the \ref{loop:add-constraints} step of Algorithm \ref{alg:LP algorithm}, but any potential contact within $\ell(\vp)$ must be included. Unfortunately, the contribution of near contacts scales much faster with $d$, and results in $d=4-6$ suggest that their amount is close to the kissing number\supercite{torquatoReviewJammedHardparticlePackings2010,md-code}, which grows exponentially with dimensionality. It is thus likely that LP-based methods become very time consuming in higher dimensions.

\section{Conclusions and future work} \label{sec:conclusions-lp}

In this chapter I have presented in detail (Sec.~\ref{sec:LP algorithm}) the iterative Linear Programming algorithm we developed to produce jammed packings of hard spheres. In contrast with other algorithms were an effective potential is introduced, or the impenetrability of particles is somewhat relaxed, our method does not use any of those techniques while maintaining the non-overlapping constraint between particle pairs at all times. I showed that upon convergence, the method always produces stable packings (Sec.~\ref{sec:LP network of contacts}), and it is empirically found that the vast majority of configurations have a single state of self-stress. It is a robust algorithm capable of producing packings of polydisperse particles in different dimensions without any major change.  Besides, models with a modified distance function (like the MK one, Eq.~\eqref{eq:MK distance}) can be trivially incorporated.

Furthermore in Sec.~\ref{sec:characterization MD-LP} I showed that our iLP method can be readily complimented by other compression protocols (such as the Lubachevsky--Stillinger protocol) that are faster for obtaining a very compressed glass, but fail in producing a proper jammed state. Under this scenario, I studied the scale with size of our algorithm (Sec.~\ref{sec:LP scaling with size}) and obtained that the time needed to generate a packing of size $N$ scales as $\order{N^3}$ at most.

On the other hand, the results I have presented in this chapter motivate some interesting topics to investigate further. In the first place, the discussion of Sec.~\ref{sec:LP dependence initial confs} about the fractal structure of the free energy landscape should be carried out in a more systematic way, for instance, using more samples and with a more careful evaluation of the effects of the compression rate. This analysis would be complimentary to the one carried out in Ref.~\cite{fel_2014}, with the benefit that our method allows us to precisely define the full network of contacts. Recall that the peculiar structure of the landscape is one of the most important predictions of Mean-field theory (Sec.~\ref{sec:MF gardner transition}), and recent results\supercite{dennisJammingEnergyLandscape2020,artiacoExploratoryStudyGlassy2020} support the same picture for finite dimensional models. Hence, a thorough assessment of the energy landscape's structure, as the one I am proposing here, will serve to dispel any remaining doubts about the range of validity of Mean-field theory.

Second, in Ref.~\cite{charbonneauMemoryFormationJammed2020} it was recently suggested the existence of a Gardner \emph{algorithmic} transition when the jamming point was approached. Similarly to the case studied here, the authors also considered hard spheres models, but in several dimensions. However, their algorithm relied on an effective potential and the associated energy minimization. It would thus be interesting to explore if the same features can be reproduced with our method, since this would imply that such transition is an inherent property of hard spheres systems and not of the specific algorithm used.

Finally, using our algorithm we could also investigate the density of states (Sec.~\ref{sec:normal modes}) of glass formers in the \emph{under}-compressed phase. As discussed above, the spectrum of disordered solids is an important quantity because it is connected with several properties of interest, and it is known to display a critical behaviour as the jamming point is approached. However, current studies are restricted to the over-compressed regime because they usually rely on computing the Hessian, which in turn depends on the interaction potential. Because of the non-singular interaction of hard spheres, these systems have not been studied with as much detail, except possibly in \cite{arceriVibrationalPropertiesHard2020}, albeit also by introducing an effective potential. With our method both difficulties can be overcome by considering the set of \emph{linear} contacts at each iteration of our algorithm. More precisely, recall that at each LP step we can obtain the active dual variables, and that they always satisfy the “mechanical equilibrium” condition (Sec.~\ref{sec:LP network of contacts}), implying that they act as “contact forces”. Strictly speaking, only at jamming these variables correspond to the real forces, however for slightly lower densities they can be used to construct a good approximation to the contact matrix, $\mathcal{S}$, whence an approximated Hessian can be obtain according to Eq.~\eqref{eq:Hessian as contact matrix}. Moreover, because of the equilibrium condition, we can guarantee that such Hessian corresponds to a minimum, and therefore its normal modes should be similar to the real ones. Additionally, it could also be checked that such contact forces follow the predicted critical distribution, Eq.~\eqref{eq:pdf-forces}. In this way, we could extend the analysis of the density of states to real hard spheres systems.

\chapter{Inferring the particle-wise dynamics near the jamming point}\label{chp:inferring-dynamics}

In this chapter I will address the issue of using the structural information contained in the network of contact forces at jamming to analyse the dynamics that takes place near such point. Most of the results have already appeared in  Ref.~\cite{paper-dynamics}. As discussed in Sec.~\ref{sec:dynamics-and-local-structure-glasses}, finding a structure-dynamics connection in glassy systems is still an unsolved problem despite being a very active field of research. In fact, the intricacy of the problem is well exemplified by the ample variety of proposals of methods and structural variables, often producing results that are not consistent between one and other\supercite{hockyCorrelationLocalOrder2014,berthierStructureDynamicsGlass2007}.
In an attempt to get rid of several difficulties addressed in Sec.~\ref{sec:dynamics-and-local-structure-glasses}, we opted for two complimentary strategies. First, we make use of a simple glass former: a monodisperse system composed of frictionless, spherical particles. Second, and more important, we restrict our analysis to the short-time dynamics occurring very near the jamming point. It is worth mentioning that this is a regime that has received little attention from a theoretical and numerical point of view, (Refs.~\cite{dynamic_criticality_jamming,hentschelDiffusionAgitatedFrictional2019} are notable exceptions, albeit in a somewhat different scenario from the one we will consider here). This fact is even more surprising considering that experimental techniques allow to precisely probe this type of dynamics\supercite{dauchotDynamicalHeterogeneityClose2005,lechenaultCriticalScalingHeterogeneous2008,britoElementaryExcitationModes2010,deseigneCollectiveMotionVibrated2010,deseigneVibratedPolarDisks2012,coulaisHowIdealJamming2014,seguinExperimentalEvidenceGardner2016}.

As discussed at length in the first chapter, at the jamming transition the dynamics is completely frozen due to geometric frustration. Unsurprisingly, it remains very sluggish near such critical point. Nevertheless, a jammed state is uniquely determined by the particles position\footnote{This is true, at least, for frictionless spheres which is the case I will focus on.}, so we expect that any possible connection between structural properties and dynamics should be more noticeable in this scenario. Hence our choice to analyse such a constrained regime. As I will show, this approach allows us to obtain a particle-wise description, both of the relevant structural quantities and of the statistics of the dynamics.
Our method is rooted on the network of contact forces that is formed at jamming, whereby we construct structural variables that allow us to characterize, individually, the displacement of particles. Our description identifies the most mobile particles but also if they exhibit any preferential direction of motion, a point that has been rarely addressed. 
Note that by using the network of contacts, we dispel many of the issues mentioned in Sec.~\ref{sec:dynamics-and-local-structure-glasses} because it is a well defined physical quantity. Moreover, the simplicity of the structural variables we consider --see Eqs.~\eqref{def:tot CV} and \eqref{def:tot dot CVs} below-- makes our method versatile and robust enough to infer the particles motion in detail. For instance, we are able to identify any inherent anisotropy in the particles trajectory rather independently of the magnitude of their displacement.


Before concluding this introduction, I mention that our assumption that the statics has some bearing with the dynamics is motivated by the picture of the fractal free energy landscape (FEL) of structural glasses\supercite{fel_2014} mentioned in Sec.~\ref{sec:MF gardner transition}. Recall that within this framework a jammed system can be thought of as being in one of the many possible minima of the FEL, while the dynamics takes place as the system explores the associated meta-basin and possibly the neighbouring ones. Hence, it is reasonable to expect that if one of these minima is used as initial condition for the dynamics, the trajectory of the configuration as it moves in phase space should be influenced by the specific jammed stated from where it departed, at least for short times.

For reading ease I summarise here the main results: the starting point is the jammed configuration of spheres shown in Fig.~\ref{fig:jammed configuration} and the resulting network of contact forces. With this information we construct, at the single particle level, two quantities that are physically well defined --the sum of contact vectors ($\vb{C}$, see Eq.~\eqref{def:tot CV}) and sum of all pairs of dot products between them ($S$, see Eq.~\eqref{def:tot dot CVs})-- and that will be shown to be related to the dynamics. Some of their properties are analysed in Sec.~\ref{sec:structural variables} and later in Sec.~\ref{sec:vectors vs forces}.
I then continue by analysing the dynamical variables introduced in Sec.~\ref{sec:dynamical variables} and studying the statistical properties of particles trajectories. I emphasize that this dynamical regime has been little studied so far, specially from a particle-wise perspective (see however \cite{coulaisHowIdealJamming2014}), which justifies the need of first finding a robust characterization of the particles motion. 
From the results of our numerical simulations, we gather that the statistical distribution of trajectories of each particle can be succinctly described by considering the first moment of its displacement (as a vectorial quantity indicating some anisotropy in the particles motion) and its norm squared (as a measure of its mobility). A sample of these distributions is depicted in Fig.~\ref{fig:pdf displacement}, whence we deduce, on the one hand, that some particles have preferential directions of motion, while on the other hand, there is a broad distribution of mobility values, signalling that even close to jamming particles are constrained by their local environment in a heterogeneous way.
I then turn to show that there is a strong link between these dynamical features and the aforementioned structural variables computed in terms of the contact vectors; see Fig.~\ref{fig:MD-displacements-vs-totCV} (resp. \ref{fig:MC-displacements-vs-totCV}) for the relation with the preferential directions in the Molecular Dynamics (resp. Monte Carlo) simulations, and Fig.~\ref{fig:MD-mobility-vs-dotCV} (resp. \ref{fig:MC-mobility-vs-dotCV}) for the connection with mobilities. To further test our approach, we verified that we can also make statistical predictions of single particle trajectories in a wide variety of systems by simply ranking the particles according to their value of $\vb{C}$ and $S$ (see Fig.~\ref{fig:ranking-forces-and-displacements}). Interestingly, in Figs.~\ref{fig:decorrelation-MD} and \ref{fig:decorrelation-MC} I provide evidence that the structural information contained in the jammed configuration gets lost rather slowly and \emph{independently} of the dynamical protocols used for the simulations and the different parameters modelling their interactions. Some remarks about the physics behind this set of results are given in Sec.~\ref{sec:vectors vs forces}, with special attention to the fact that ignoring the forces magnitude produces more informative structural variables. I anticipate that our guiding model will be the fractal FEL picture introduced in Sec.~\ref{sec:MF gardner transition} and briefly studied numerically in Sec.~\ref{sec:LP dependence initial confs}. 
As a further benchmark, in Sec.~\ref{sec:comparison normal modes} we use the vibrational modes obtained from the Hessian (at jamming) as an alternative scheme to analyse the short time dynamics, obtaining negligible correlations between them. This indicates that, in contrast with our method, the normal modes description fails to capture the statistics of the single-particle trajectories in the dynamical regime we consider here. These results show that the formalism we developed, based solely on exploiting the details of the network of contacts at jamming, provides a more effective statistical inferential technique than previous works.
Finally, Sec.~\ref{sec:conclusions-inferring-dynamics} includes our conclusions and gives some perspectives for future research.

\section{Statics: Physical quantities at the jamming point}\label{sec:statics}

In Sec.\ref{sec:jamming-transition} of the first chapter I explained that isostatic jammed packings are characterized customarily by their packing fraction, $\vp_J$, given that such value shows a very weak dependence on the (possibly random) initial condition, changes of the system's parameters, and even the specific algorithm used\supercite{ohernJammingZeroTemperature2003,parisi_zamponi_2010,md-code}. Hence, the most common situation is that random packings exhibit very similar densities\supercite{ohernJammingZeroTemperature2003,ohernReplyCommentJamming2004}, although a broad range of values of $\vp_J$ can be accessed\supercite{jiaoNonuniversalityDensityDisorder2011,hopkinsDisorderedStrictlyJammed2013,torquatoRobustAlgorithmGenerate2010,donevCommentJammingZero2004} by tuning properties such as the average coordination number or by seeding some regularity in the particles position\supercite{jiaoNonuniversalityDensityDisorder2011, tsekenisJammingCriticalityNearCrystals2020, charbonneauGlassyGardnerlikePhenomenology2019}. In any case, in the unbiased scenario the value of $\vp_J$ is mainly determined by the dimensionality of the system\supercite{puz_book,md-code,torquatoRobustAlgorithmGenerate2010} and the distribution of the particles sizes. When $d=3$ (the case I will be concerned with in this chapter), one can safely state that typical monodisperse isostatic jammed packings have $\vp_J \approx 0.64$.

On the other hand, from a microscopic point of view, a jammed state is fully determined by the particles positions and radii\supercite{charbonneauJammingCriticalityRevealed2015}, $(\va{r}^{(J)}, \vec{R}_J) := \qty(\{\vb{r}_i^{(J)},\ R_{i,J} \}_{i=1}^N)$. (Note that I am following the convention used so far where $\vec{X}$ denotes the set of $N$ values of quantity $X$ and boldface variables are $d$ dimensional vectors. But, in contrast with such convention, I am parametrizing the particles' size in terms of the radius instead of the diameter because (i) the radius defines the units of time in the molecular dynamics simulations; and (ii) $\sigma$ will be used in this chapter to denote the standard deviation of dynamical variables.) As explained in Sec.~\ref{sec:network of contacts}, in mechanically stable configurations these quantities suffice to determine the $N_c$ contact forces, $\{ \vb{f}_\ctc{ij} \}_{\ctc{ij}=1}^{N_c}$, where $\ctc{ij}$ with $i<j$ denotes an ordered pair and is used as a contact's index. In that same section I also mentioned that a requirement for mechanical stability is that the number of contacts is greater than the number of degrees of freedom\footnote{This requirement in the number of contacts is enough to guarantee that particles are in mechanical equilibrium, \textit{i.e.} $\sum_{j \in \partial i} \vb{f}_{ij}=0$, where $\partial i$ is the set of all the neighbours of particle $i$. $\vb{f}_{ij}=\vb{f}_{\ctc{ij}}$ if $i<j$, or $\vb{f}_{ij}=-\vb{f}_\ctc{ij}$ otherwise. See Secs.~\ref{sec:network of contacts} and \ref{sec:LP network of contacts} for more details.}, $N_{dof}$. To calculate $N_{dof}$ (Sec.~\ref{sec:network of contacts}) we should consider that we are using a system with periodic boundary conditions, and that a small fraction of rattlers --particles that do not contribute to the rigidity of the network of contacts-- are present. In $d$ dimensions, this latter type of particles are characterised by having less that $d+1$ contacts. Hence, if $N'\lesssim N$ is the number of non-rattlers, we have that $N_{dof}=d(N'-1)$. In order to generate a  stable jammed packing we used the iterative Linear Programming (iLP) algorithm described in Chp.~\ref{chp:lp-algorithm}, but without the Lubachevsky--Stillinger compression protocol\footnote{The reason for not including the molecular dynamics part described in Sec.~\ref{sec:MD for jamming} is that at the moment the characterization of the MD+iLP algorithms was not available.}. As discussed in Sec.~\ref{sec:LP algorithm}, our iLP algorithm allows us to easily access the full network of contacts and always produces a stable jammed state. Moreover, the vast majority of times, such state has a single self-stress (1SS), implying that exactly $N_{1SS}:= N_{dof}+1$ contacts are present. Consequently, the configurations thus produced are marginally stable\supercite{mft_review,berthierGardnerPhysicsAmorphous2019,mullerMarginalStabilityStructural2015}. (Strictly speaking, this means that we are dealing with an \emph{hyper}static configuration, since there is an extra contact with respect to isostaticity; see Sec.~\ref{sec:network of contacts} for a detailed discussion.)

\begin{figure}[h!]
	\centering
	\begin{subfigure}{0.6\textwidth}
		\includegraphics[width=\textwidth]{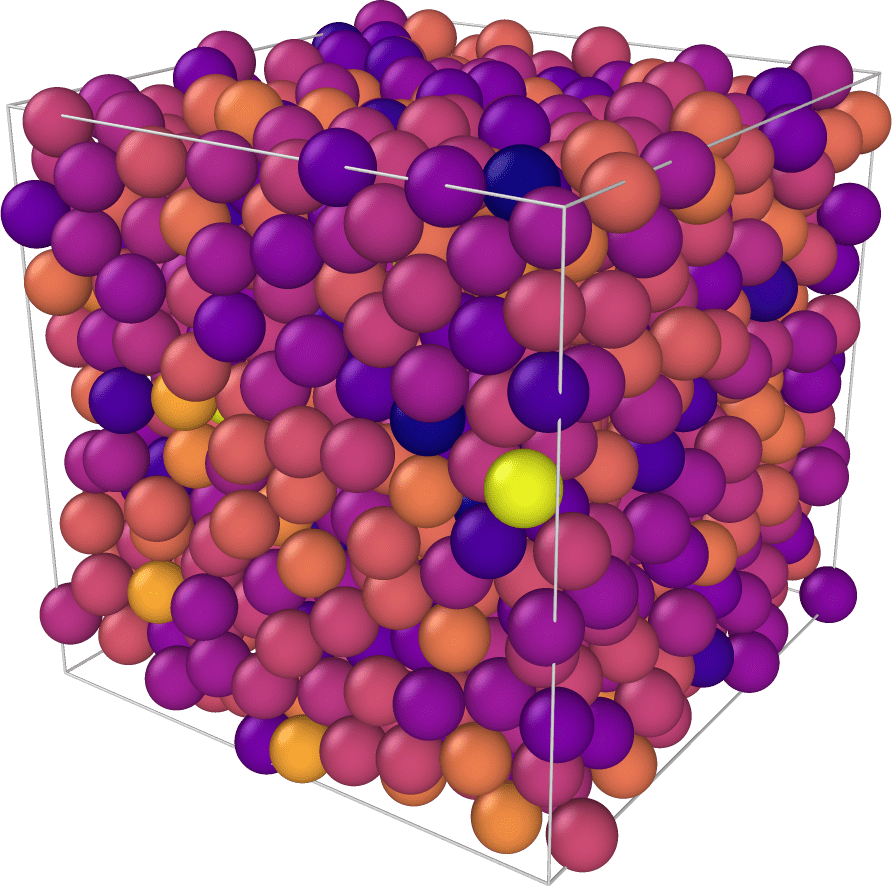}
	\end{subfigure} \\
	\begin{subfigure}{0.6\textwidth}
		\centering
		\includegraphics[width=\textwidth]{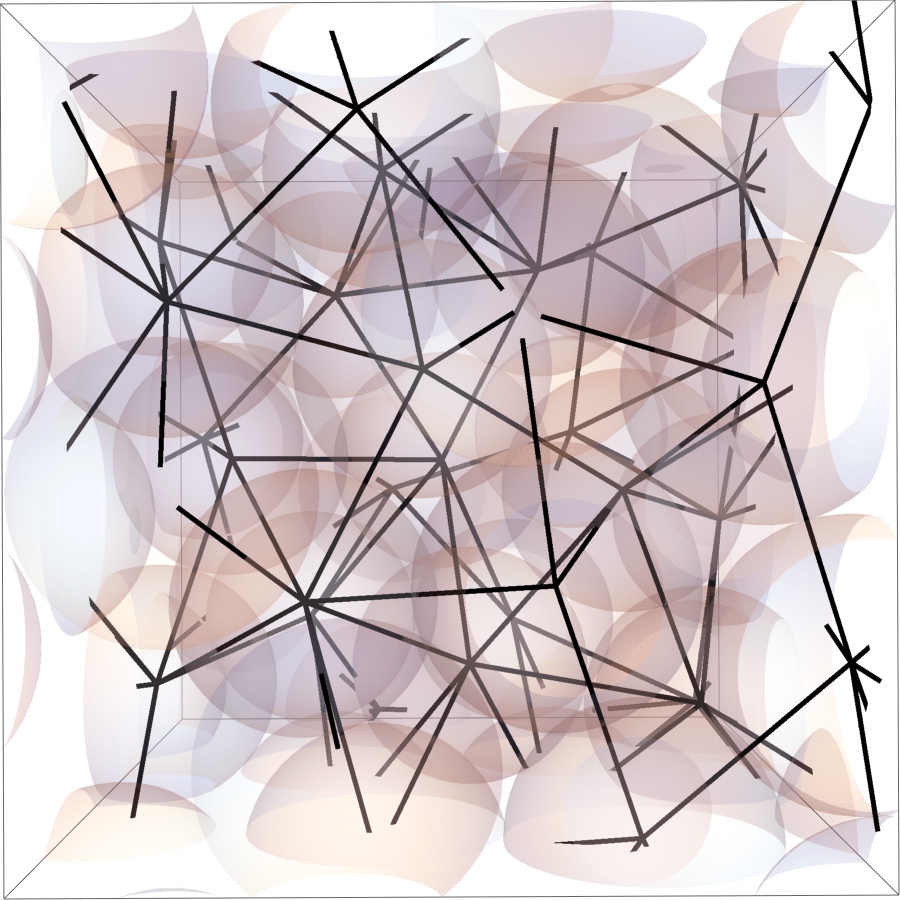}
	\end{subfigure}
	\caption[Jammed configuration of $N=1024$ spheres used to perform dynamical simulations.]{Top: Jammed configuration of $N=1024$ spheres used to perform the dynamics simulations. Its packing fraction is $\vp_J=0.635$ and each sphere is coloured according to its value of $S_i$, defined in Eq.~\eqref{def:tot dot CVs}, from light yellow (big values) to dark blue (small, negative values); see also the colour scale in Fig.~\ref{fig:pdf displacement}b. Bottom: Zoom in a region of the same configuration (without colouring) showing the contact vectors obtained according to the method described in the previous chapter.}
	\label{fig:jammed configuration}
\end{figure}

We thus produced several monodisperse configurations of $N=1024$ spheres. Fig.~\ref{fig:jammed configuration} illustrates\supercite{ovito} a typical realization that we employed afterwards in the dynamical simulations. In the left panel, the spheres have been coloured according to their value of $S_i$, defined soon below in Eq.~\eqref{def:tot dot CVs}, while the right panel depicts the resulting contact vectors between neighbouring spheres in a small region of the same configuration. Importantly, our choice of using particles of the same size is grounded on their convenience for the statistical inference we aimed to achieve. Anticipating the features to be discussed in Secs.~\ref{sec:dynamical variables}-\ref{sec:results Monte Carlo simulations}, notice that were we to use spheres of different radii, the smallest ones would be, presumably, the most mobile ones because they would experience less collisions with their neighbours. Yet, their relatively unimpeded motion would be a consequence exclusively of their size, and therefore \emph{independent any structural property of their vicinity}.
In order to avoid this sort of trivial inference and to really isolate the contribution coming solely from the structure, we restricted our analysis to monodisperse systems, for which I will henceforth use $R_J$ to denote the particles' radius at jamming.
The configuration used here and depicted in Fig.~\ref{fig:jammed configuration} has a packing fraction of $\vp_J=\frac{4\pi N R_J^3}{3L^3}=0.635$, with $R_J=0.0529$ being the spheres' radius once the jamming point is reached. Only $1.3\%$ of the $N=1024$ particles are rattlers and none have any contact forces acting on them. As we will see later, this lack of constraints causes that most rattlers are able to move more freely than the rest of the particles.

\subsection{Critical properties of the jammed configurations} \label{sec:critical properties jamming}

As a security check, we verified that the packings obtained with our iLP algorithm exhibit all the properties encountered before using other algorithms\supercite{wyartMarginalStabilityConstrains2012,lernerLowenergyNonlinearExcitations2013,degiuliForceDistributionAffects2014,haghBroaderViewJamming2019} and predicted by the best current available theory based on mean field (MF)\supercite{puz_book,charbonneauJammingCriticalityRevealed2015,mft_review}: the critical distributions of (i) forces and (ii) gaps; and (iii) a constant density of states (DOS) for $\omega \ll 1$. The importance of these properties has been introduced in Secs.~\ref{sec:forces-and-gaps} and \ref{sec:normal modes}. Therefore, here I will only mention that MF theory predicts, in the $d\to\infty$ limit, that small contact forces and interparticle gaps are distributed as $f^{\theta_e}$ and $h^{-\gamma}$, respectively. The respective exponents are highly non-trivial: $\theta_e=0.42311\dots$ and $\gamma=0.41269\dots$. Similarly, marginal stability considerations and numerical results indicate that $D(\omega)\to $ constant as $\omega \to 0$. 

\begin{figure}[!htb]
	\centering
	\includegraphics[width=\linewidth]{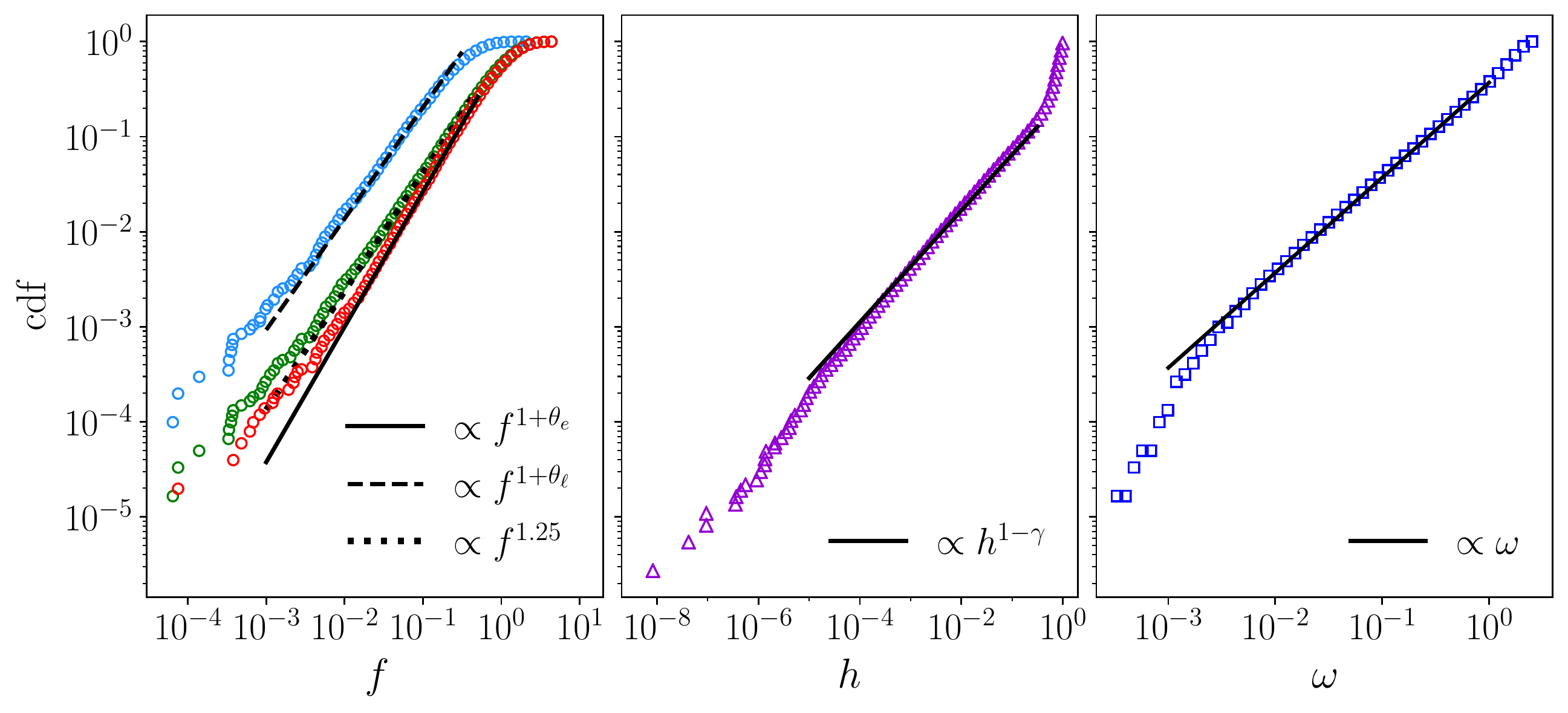}
	\caption[Cumulative distributions of contact forces, gaps, and frequencies.]{Cumulative distribution functions (cdf) of contact forces (left), interparticle gaps (central), and frequencies (right). In all cases, the black lines are comparisons with the power laws predicted theoretically or found in previous studies. For the data of panel (a), I have split the forces in the ones associated to localized (blue) and extended modes (red), using the smallest force per particle for the former type and the criterion of Ref.~\cite{charbonneauJammingCriticalityRevealed2015} for the latter one. The green markers are all the forces put together and the exponent of the associated fit was taken from the same reference. 
	In panel (b) I show the cdf of the dimensionless interparticle gaps (defined in Eq.~\eqref{def:gaps}) and that it closely follows the expected scaling, Eq.~\eqref{eq:pdf-gaps}. 
	Finally, in the last panel I used the fact that the density of states is predicted to be constant for small values of $\omega$ and hence its cdf (the integral of $D(\omega)$) should be proportional to the frequency. The spectrum, $\{\omega_i\}_{i=1}^{dN'}$, was obtained by diagonalizing the Hessian \emph{at} jamming, following Eq.~\eqref{eq:Hessian as contact matrix} with the harmonic contact matrix.
	Deviations in all of these cases are due to finite size effects as studied in the next chapter.
	}
	\label{fig:properties-jamming}
\end{figure}

Fig.~\ref{fig:properties-jamming} shows that our configurations indeed fulfil these properties, by comparing the empirical cumulative distributions (markers) obtained from 20 independent packings\footnote{The configurations employed here, either to perform the simulations reported in the main text or the ones reported in the Appendix \ref{sec:second-configuration}, are two independent systems selected at random from this set.
} (to improve the sampling), with the theoretical predictions (solid lines in each panel). 
In the case of the forces, however, finite dimensional systems exhibit two separate contributions, due to extended and buckling excitations; see Eq.~\ref{eq:pdf-forces}. Only the former follows the MF distribution, while no such prediction exists for the latter. As explained in Secs.~\ref{sec:forces-and-gaps}-\ref{sec:marginal stability}, the two types of excitations can be differentiated by the number of contacts of their associated particles. Thus, bucklers are, with very high probability, associated with particles having $q_i=d+1=4$ kissing neighbours, \textit{i.e.} the minimum number of contacts to guarantee stability\footnote{I mention in passing that although not all particles with $q_i=d+1$ contacts are strictly bucklers, with very high probability any buckler will have $d+1$ contacts; see Ref.~\cite{charbonneauJammingCriticalityRevealed2015}. I will thus abuse on terminology and probability by simply referring to any particle with 4 contacts as a buckler.}. Nevertheless, the associated distribution in the leftmost panel of Fig.~\ref{fig:properties-jamming} (identified by the blue circles) was obtained computing instead the smallest contact force per particle. This distinction is not important --because bucklers are, in any case, usually associated with small forces-- but helped in our case to obtain a cleaner separation of the two contributions\footnote{In the next chapter, where these critical distributions are studied in great detail I have used the criterion in terms of the coordination number to identify the bucklers and show that they do follow the previously known power-law. See Figs.~\ref{fig:forces-3d} and \ref{fig:forces-loc-3d}.}. At any rate, the agreement between theory and numerical results is remarkable, specially considering that our systems are rather far from the limit of infinite dimensions. Moreover, several of the observed deviations from the predicted power-laws are due to finite size corrections as analysed in detail in Chapter \ref{chp:fss}.

\subsection{Structural variables of interest} \label{sec:structural variables}

I now address one of the main goals of this chapter: how to use the network of contacts formed at jamming to construct well defined physical quantities that can be used as predictors of the dynamics that takes place close to such point. Yet, before answering this question, I want to remark why this approach differs from previous ones. Most importantly, note that by using a jammed sate as the initial configuration for the dynamics our knowledge is augmented in comparison with the scenario where the initial state is only \emph{close} to a jammed configuration. In this latter case we would be missing data about the true contacts between particles, and thus our description would be limited to the usage of ``coarse-grained'' variables, for instance, the local density or a pair correlation function defined within some small vicinity of each particle, not unlike methods introduced in Sec.~\ref{sec:dynamics-and-local-structure-glasses}. 
In contrast, here I will be considering a scenario where the dynamics departs from a configuration in which we can identify the actual neighbours of each particle and, consistently, only include such relevant particles in our description. This distinction is crucial for uncovering preferential directions in the particles' displacements, as we investigate in the next sections. 
Moreover, because in our dynamical simulations we only consider contact potentials between the spheres, a ``coarse-grained'' description would presumably fail to provide a realistic picture of the main interactions driving the particles trajectories. In other words, given that we explore the particles' motion in the vicinity of a jamming point, it is to be expected that the correlation between structure and dynamics should be stronger than in other scenarios. Hence, if we have at hand more accurate information about the structure, we will be able to achieve a better and longer lasting statistical inference of the dynamics.

With this in mind, I will use the network of contacts to introduce some quantities that will be used in the rest of this chapter to describe the statistics of the particles' trajectories. Following the notation from Sec.~\ref{sec:network of contacts}, $\vb{n}_{ij} = \frac{ \vb{r}_i^{(J)} - \vb{r}_j^{(J)}}{\abs{\vb{r}_i^{(J)} - \vb{r}_j^{(J)}} } = \frac{ \vb{r}_i^{(J)} - \vb{r}_j^{(J)}}{2R_J}$ will denote the unit contact vector, pointing from particle $j$ towards $i$ (assumed to be in contact). Clearly, the set of such vectors defines the edges of the network depicted in the lower panel of Fig.~\ref{fig:jammed configuration}. Using the contact vectors, we can easily construct the following two quantities: (i) the vectorial sum of all the contact vectors acting on the $i$-th particle,
\begin{equation}\label{def:tot CV}
\vb{C}_i : = \sum_{j \in \partial i} \vb{n}_{ij} \qc
\end{equation}
and, (ii) the sum of all pairs of scalar products of the contacts acting on the same particle,
\begin{equation}\label{def:tot dot CVs}
S_i := \sum_{j<k \in \partial i} \vb{n}_{ij}\vdot \vb{n}_{ik} = \frac12 \left( \abs{\vb{C}_i}^2 - q_i \right) \qc
\end{equation}
where $q_i$ is the coordination number\footnote{In other parts of this thesis I have used $z_i$ for the same purpose, but to avoid confusion with the $z$ component of a particle's trajectory, in this chapter I will use $q_i$.} at jamming of particle $i$. The colours used in the upper panel of Fig.~\ref{fig:jammed configuration} indicate the value of $S_i$ for each particle of the system, with a bigger (smaller) value corresponding to a colour on the yellow (dark blue) part of the scale reported in Fig.~\ref{fig:pdf displacement}. The complex distribution of colours displayed by the configuration resembles the ones found using other structural variables and order parameters, which in turn have been linked with the heterogeneous dynamics observed in glassy systems\supercite{schoenholzStructuralApproachRelaxation2016,zylbergLocalThermalEnergy2017,tongRevealingHiddenStructural2018,tongRevealingInherentStructural2019,berthierStaticPointtosetCorrelations2012,charbonneauLinkingDynamicalHeterogeneity2016,hentschelDiffusionAgitatedFrictional2019,tongStructuralOrderGenuine2019}.

\begin{figure}[!htb]
	\centering
	\includegraphics[width=\linewidth]{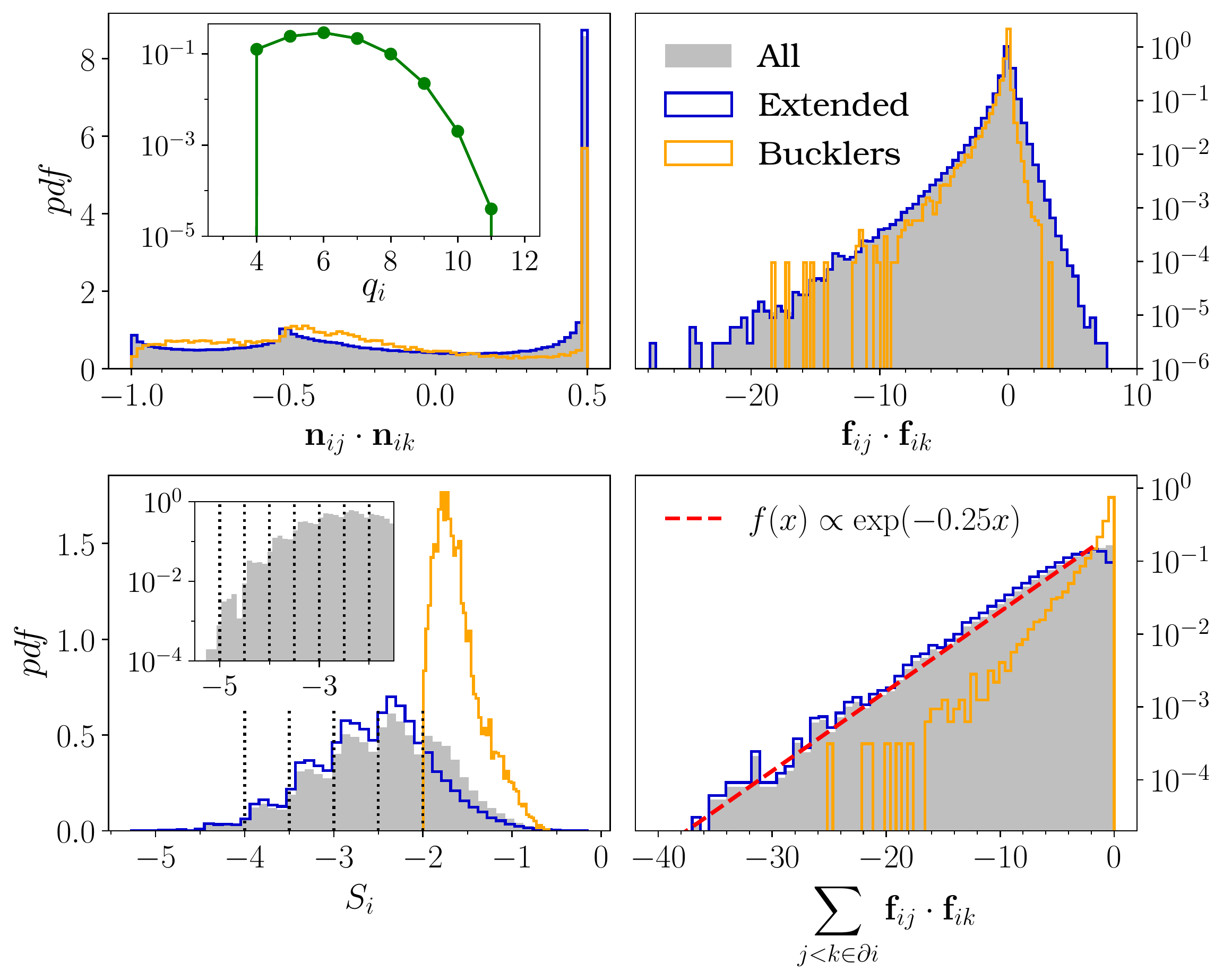}
	\caption[Distributions of dot products of contact vectors and contact forces.]{Distribution of dot products of all contact vectors (upper left panel) and contacts forces (upper right) acting on the particles and their respective sum per particle (lower panels). Rattlers have been excluded, and the separate contributions of bucklers (orange) and the rest of the particles (blue) have been also included. The inset in the pdf of $S_i$ --defined in Eq.~\eqref{def:tot dot CVs}-- depicts the same distribution in a logarithmic scale to show that the influence of the bound $S_i\geq q_i/2$ (dotted lines) is persistent for all possible values of $q_i$; see text for more details.
		The pdf of  $q_i$ is plotted in the inset of the upper left panel.}
	\label{fig:dist-contacts}
\end{figure}

As I will argue in the next sections, it is the contact \emph{vectors} and not the \emph{forces} ($\vb{f}_{ij}= f_{ij}\vb{n}_{ij}$) what actually convey more information about the dynamics of the configuration near its jamming point. But before exploring such connection, it is convenient to analyse their differences in the packings we produced. First, note that the analogous of Eq.~\eqref{def:tot CV} for contact forces trivially vanishes -- \textit{i.e.} $\sum_{j\in \partial i} \vb{f}_{ij}=0$-- given the mechanical equilibrium property of jammed states. That is, it cannot provide information about the dynamics because it has the same value for all particles. 
The case of the scalar structural variable is different and a more careful analysis should be carried out. In this part I will show that, despite being closely related quantities, the dot products between pairs of contact forces and contact vectors have strikingly different probability distributions (pdf). In later sections I will compare their performance as predictors of the dynamics. The claim about their pdf's is summarized by Fig.~\ref{fig:dist-contacts}, where the data of all 20 configurations is included. Left panels depict distributions of quantities involving only contact vectors, while the right ones show the distributions of the analogous quantities using contact vectors. 
On the other hand, the upper panels show the distribution of all dot products between pairs of particles in contact, and the lower ones their sum per particle (thus the lower left panel is the distribution of $S_i$). Additionally, the contributions of bucklers and particles associated to an extended response (see Secs.~\ref{sec:forces-and-gaps}-\ref{sec:marginal stability}) have been plotted separately. 
Let me first consider the pdf's in the upper panels, At a first glance, the distribution of $\vb{f}_{ij}\cdot \vb{f}_{jk}$ suggests that the vast majority of contact forces are nearly orthogonal (notice the log-scale in the vertical axis). But when compared with pdf of $\vb{n}_{ij}\cdot \vb{n}_{jk}$ we see that the likeliest arrangement is the one consisting in three spheres touching each other (hence the peak at $\cos \pi/3=0.5$), while any other arrangement is roughly equally likely. This in turn implies that the noticeable peak at zero in the first case is due to the existence of many small forces whose effect is to blur any intrinsic geometrical feature of the configuration's structure.
Given the few amount of bucklers (see inset in the upper left panel for the pdf of $q_i$), the distributions of all the particles and the ones associated to extended excitations are virtually identical. Considering the distribution of $\vb{n}_{ij}\cdot \vb{n}_{jk}$ restricted to bucklers we see a relative deficit of three mutually touching spheres, and a slight excess of triplets spanning an angle \emph{close to} $\frac{2\pi}{3}=\arccos(-\frac12)$. In combination with the low fraction of particles almost perfectly aligned ($\vb{n}_{ij}\cdot \vb{n}_{jk} \simeq -1$), these results support the hypothesis that bucklers are very likely particles with $d$ almost coplanar contacts. However, if this simplified picture is entirely valid, there is a peak “missing” at $\vb{n}_{ij}\cdot \vb{n}_{jk} \simeq 0$ corresponding to the almost orthogonal force-balancing contact. The absence of such peak is likely due to the fact that the remaining contact does not always bear a small force, and therefore is not necessarily orthogonal. See more details in \cite[SI, Sec.~II]{charbonneauJammingCriticalityRevealed2015}.

Now, the sums of these dot products, \textit{i.e.} $S_i$ and $\displaystyle \sum_{j<k \in \partial i} \vb{f}_{ij}\cdot \vb{f}_{ik}$, are also distributed in a very different fashion, as shown in the lower panels of Fig.~\ref{fig:dist-contacts}. For instance, the pdf of $S_i$ displays an irregular shape because this structural variable is lower bounded by $-\frac{q_i}{2}$, as can be easily derived from its definition. This causes its value to change rather abruptly in regular intervals of $1/2$ for $S_i\leq -2$ (because the smallest number of neighbours is 4), as shown by the dotted lines. As the inset shows, this feature is preserved all the way down to $q_i=11$, the largest connectivity observed. Importantly, in Ref.~\cite{paper-dynamics} this distribution was incorrectly compared to a Gaussian, because the lack of resolution (due to the small number of samples considered) prevented the discontinuities at $-\frac{q_i}{2}$ to be observed.
In any case, the analogous sum using contact forces clearly follows a different distribution, if only by its monotonic form. This latter case shows that, not only there is an excess of very small forces, but there are particles whose full set of contact forces is given by forces of minute magnitude. Even more, the probability of having a different value for this variable decreases exponentially, $p(x) \sim e^{0.25 x}$, as indicated by the red dashed line. Finally, notice that the distribution being upper bounded at zero is a straightforward consequence of the identity 
\[
\sum_{j<k \in \partial i} \vb{f}_{ij}\cdot \vb{f}_{ik} = \frac12 \qty( \sum_{j\in \partial i} \vb{f}_{ij} \cdot \sum_{k \in \partial i} \vb{f}_{ik} - \sum_{j\in \partial i} \abs{\vb{f}_{ij}}^2) = -\frac12 \sum_{ j \in \partial i} \abs{\vb{f}_{ij}}^2 \, .
\]


To close this section and to put into context the main results of this chapter, I emphasize that both $\vb{C}_i$ and $S_i$ --introduced in Eqs.~\eqref{def:tot CV} and \eqref{def:tot dot CVs}-- are well defined physical variables for each particle and, consequently, our approach will provide a description of the system's dynamics at the single particle level. We thus extend other techniques where the characterization of the particles' displacement was done in terms of clusters or mesoscopic regions within the system, as exemplified in Sec.~\ref{sec:dynamics-and-local-structure-glasses}.

\section{Dynamical variables of interest}\label{sec:dynamical variables}

Even though the jamming regime is by itself interesting and complicated enough, many of the most salient physical properties of amorphous solids are linked with the dynamical slowing down that takes place when, say, their density increases. So I now turn to the main part of this chapter in which we first characterise the dynamics of our configuration of spheres as it moves away from its jamming point, \textit{e.g.}, by reducing the system's packing fraction by a small amount and providing the particles with momenta. 
Then, in a second stage, we establish a connection between the sluggish motion of the configuration and its static properties computed at the jamming point. 

At first sight, it might seem that making such two-step division of the analysis is somewhat redundant, since we could proceed instead by directly trying to construct some quantity using the network of contact forces and then relate it with certain dynamical features. This is the standard methodology in studies exploring the link between the local environment of particles and their square displacement, where this last quantity is always taken as a measure of the ``mobility'' of a particle.
As mentioned in Sec.~\ref{sec:dynamics-and-local-structure-glasses}, this technique has been used to relate structural features of several model systems with their dynamics.  Nevertheless, given that here we are considering the dynamics in a different and rather unexplored regime, we opted for first trying to answer the question of how particles move near the jamming point. In particular, do they have any preferential direction of motion? And if they do, how much do they move along it? To show that these attributes are not necessarily determined by the mobility of a particle, consider the case in which its closest neighbours are distributed uniformly around it forming a relatively large cage. It is clear that the particle will be more mobile than if the cage was small, rendering its motion very constrained. And yet, in neither of these two cases the particle's displacement would exhibit a preferred direction. Conversely, if the arrangement of neighbours is quite irregular, the motion of the particle should be facilitated towards the direction where fewer obstacles are present, however, this favoured path will only be discernible if the particle is also allowed to move enough, because otherwise the preferred direction will be washed out, for example, by thermal noise. These elementary examples illustrate that even a very simple study of the dynamics near jamming should take into account these two properties in order to yield a sufficiently comprehensive picture. Therefore, our approach is based on analysing the particles' displacement both as a vectorial quantity (that is, component-wise) and as a scalar one (using its norm squared, henceforth also termed ``mobility''):
\begin{subequations} \label{eq:displacement}
	\begin{align}
	\delta \vb{r}_i(t) & = \vb{r}_i(t) - \vb{r}_i(0) = (\delta x_i(t), \delta y_i(t), \delta z_i(t) ) \label{seq:displacement}\\
	\abs{\delta \vb{r}_i(t)}^2 & = \delta x_i^2(t) + \delta y_i^2(t) + \delta z_i^2(t) ; \label{seq:square displacement}
	\end{align}
\end{subequations}
where $\vb{r}_i(t)$ is the position of the $i$-th particle at time $t$. Expectedly, another variable of interest will be the mean squared displacement (MSD), introduced previously but whose expression I reproduce here for convenience:
\begin{equation}\label{eq:MSD}
\Delta (t) = \avg{ \frac1N \sum_{i=1}^N \abs{\vb{r}_i(t) - \vb{r}_i(0)}^2 }\, .
\end{equation}
When it was considered in Sec.~\ref{sec:glasses}, the meaning of $\avg{\bullet}$ was a thermal average. In this chapter however, it will denote an average over the different trajectories generated, as explained next.

We want to probe how the configuration evolves as it explores the phase space close to the FEL minimum identified by the jammed state. 
It is then natural to use the jammed configuration as initial condition, \textit{i.e.} $\va{r}(0) = \va{r}^{(J)}$, and then generate many independent trajectories. This approach is equivalent to sampling from the isoconfigurational ensemble\supercite{widmer-cooperHowReproducibleAre2004,widmer-cooperRelationshipStructureDynamics2005,widmer-cooperPredictingLongTimeDynamic2006,widmer-cooperStudyCollectiveDynamics2007,widmer-cooperIrreversibleReorganizationSupercooled2008,jackInformationTheoreticMeasurementsCoupling2014} (ICE), since the initial positions are kept fixed while the configuration evolves according to some prescribed dynamical protocol, following a different trajectory each time. Sampling from the ICE thus allows to uncover the contribution of the network of contacts to the dynamics because any stochastic contribution to the dynamics is expected to cancel out if the number of trajectories is large enough. With this in mind, the meaning of $\avg{\bullet}$ is simply the ICE average, whose value we estimate through several independent simulations of the dynamics.

In the next two sections I present the statistics of the particles trajectories in the ICE, sampled using two different dynamical schemes, namely, Molecular Dynamics (MD) and Monte Carlo (MC) simulations. In the former type, the infinitely hard sphere (HS) model is kept and the system's density is reduced with respect to $\vp_J$. For the MC simulations we considered instead soft spheres (SS) with different interaction potentials, but with keeping the system's packing fraction constant at $\vp_J$. The reason for using these two types of simulations is that we want to investigate the applicability of our method as we change the model system and its parameters. This is specially important because other studies\supercite{hockyCorrelationLocalOrder2014} using the ICE have reported that the level of correlation between local structure and dynamics is very sensitive to the glass former model and the type of interaction.

\section{Results with MD simulations} \label{sec:results MD simulations}

I begin by showing the results obtained using MD simulations, for which the radius of all spheres was reduced by the same factor in order to reach a new density $0.9 \leq \vp / \vp_J \leq 0.999$, that was kept constant afterwards. For each value of $\vp$ we performed $M_{MD}=5000$ independent simulations of event-driven MD as described in Sec.~\ref{sec:MD for jamming} and using the implementation of Ref.~\cite{md-code}. As mentioned above, the initial conditions were set to $\va{r}^{(J)}$ in all cases, while the initial velocities were assigned randomly according to a Maxwell--Boltzmann distribution at inverse temperature $\beta=10$ (in the reduced units of the algorithm). As discussed in the previous chapter, this MD code uses an event-driven scheme, which implies that the ballistic regime will be mostly absent from all of our data. Additionally, in event-driven type of algorithms the natural units for the time step of the dynamics are the number of collisions per particle that have occurred. In order to make a clear distinction between this time unit and the ``physical time'' $t$ in Eqs.~\eqref{eq:displacement} and \eqref{eq:MSD} above, I will denote as $\tau_{MD}$ the number of events per particle. However, Fig.~\ref{fig:collisions-vs-t} in Sec.~\ref{sec:supplementary MD} shows that once we have fixed $\vp$, there is a well defined relation between $\tau_{MD}$ and $t$. This incoming part is devoted to further details of the MD simulations that are not essential for the rest of the chapter, and can be skipped. Then, in Sec.~\ref{sec:statistics MD trajectories} I show the statistics of the particles trajectories and how they can be captured by their moments. Finally, in Sec.~\ref{sec:MD correlation with structure} I show that these statistics are highly correlated with the structural variables introduced in \ref{sec:structural variables}.


\subsection{Further details of the MD simulations} \label{sec:supplementary MD}

\begin{figure}[!htb]
	\centering
	\includegraphics[width=0.99\linewidth]{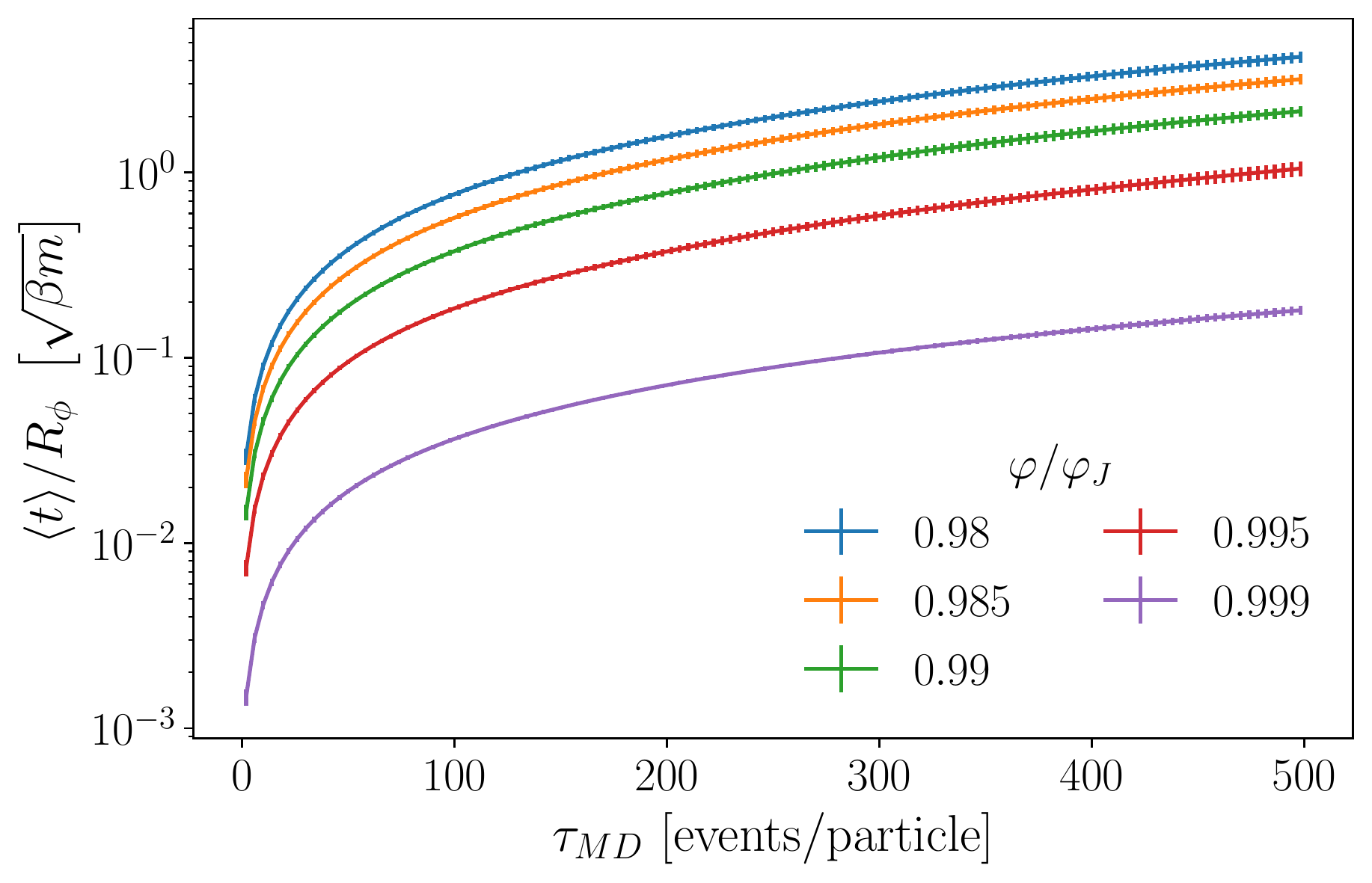}
	\caption[Relation of the physical time and collision events in MD simulations.]{Relation of the (average) global time of the MD simulations as a function of the number of collision events, for different packing fractions (different colours). Each marker correspond to an instant at which the particles positions where stored and the solid lines are just a guide to the eye. Errorbars indicate the value of the standard deviation of the values.}
	\label{fig:collisions-vs-t}
\end{figure}

In Sec.~\ref{sec:MD for jamming} I mentioned that event-driven MD algorithms are well suited for simulating systems of HS given their trivial dynamics. With this method, the state of the system is advanced until a collision takes place, where the particles exchange momenta, and the next collision is predicted with the new velocities. In such a way, the evolution of the system is updated each time a collision event occurs and the global time $t$ of the simulation is just the sum of the elapsed times between events.  Another consequence of using the HS potential is that the only characteristic length of the system is the particles size, which in turn fixes the time units as $\sqrt{\beta m R^2}$, where $m$ and $R$ are the spheres mass and radius, respectively. Without loss of generality, we set $m=1$, $\beta=10$, while $R$ is determined by the value of $\vp$ we chose when running the simulations. In Fig.~\ref{fig:collisions-vs-t} I show that for a fixed value of the packing fraction, there is well defined relation between the $t$, averaged over all the trajectories, and the value of $\tau_{MD}$, which measures the number of collisions that have taken place. The very small error bars suggest that the distribution of times is considerably peaked around $\avg{t}$, so I will henceforth use $t$ to the denote its sample average.

%
%

\begin{figure}[!htb]
	\centering
	\includegraphics[width=0.99\linewidth]{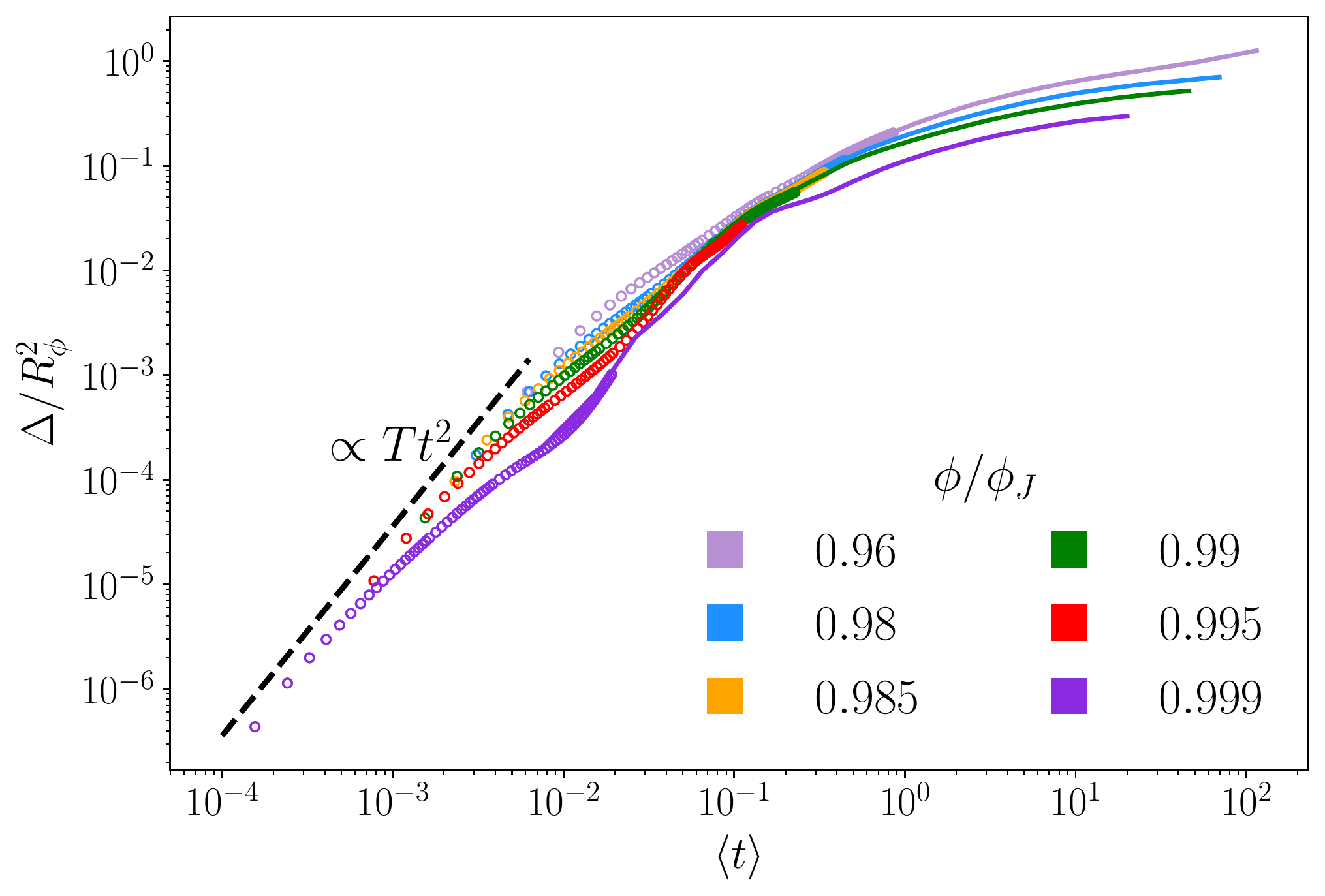}
	\caption[Temporal evolution of MSD for different packing fractions.]{Temporal evolution of the MSD for different packing fractions (different colours). The circular markers show the values of MSD obtained from the main simulations, while the solid lines correspond to simulations 100 times as longer, but for a reduced number of values of $\vp$. The black dashed line is included as a reference for comparison with a ballistic behaviour. Note that all the results presented in the main text correspond to the evolution of the configuration much before the MSD reaches its plateau value, which is the regime that so far has received most attention.}
	\label{fig:MD-msd-vs-t}
\end{figure}

To better characterize the dynamical regime studied here, we performed additional MD simulations (with respect to the ones used for our main results) letting the configuration evolve for $100$ times as many collision events, using configurations of packing fractions $\vp/\vp_J = \{0.96, 0.99,0.98, 0.999\}$ and a smaller number of trajectories, namely $M_{long}=1000$. In Fig.~\ref{fig:MD-msd-vs-t} we present the results of the $\Delta$ as a function of time, for several packing fractions using the data of the main simulations (circular markers) and from the longer ones (solid lines). 
For reference, I included the expected trend from a purely ballistic growth (black dashed line). 
As explained in Chapter \ref{chp:intro}, most of the unique features of glassy systems occur after the configuration has abandoned such regime and its MSD has reached its characteristic plateau. But as this figure shows, we are exploring the dynamical evolution well before the MSD attain its asymptotic value. 
As a final remark, I want to stress that the ballistic regime is, expectedly, almost entirely lost due to the event-driven nature of our simulations.
%

\subsection{Particle trajectories and their statistics} \label{sec:statistics MD trajectories}

\begin{figure}[!htb]
	\centering
	\includegraphics[width=1.0\textwidth]{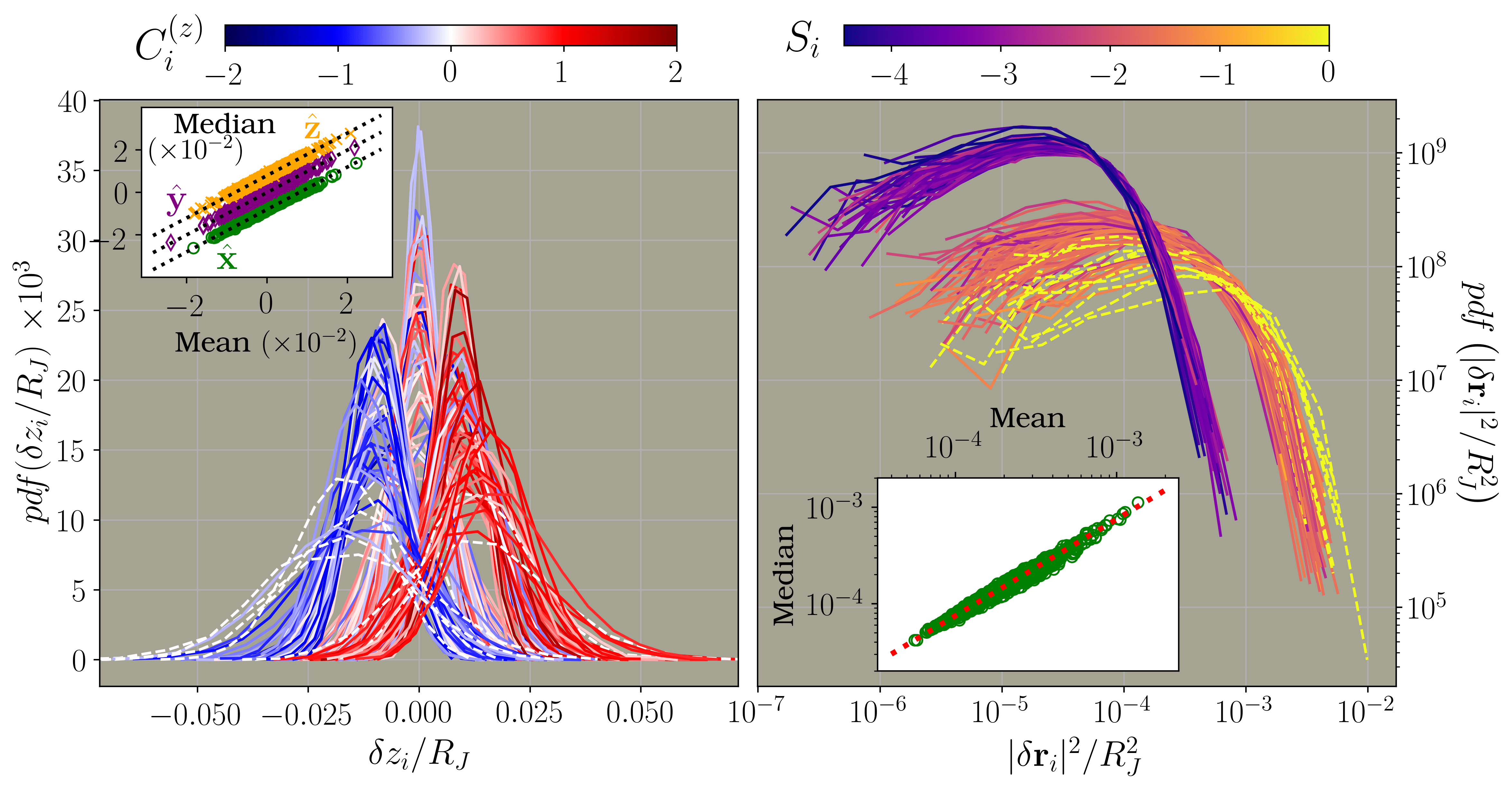}
	\caption[Probability distributions of single particle displacements and mobilities.]{
		Main figures: Probability distributions (pdf) of the variables of Eqs.~\eqref{eq:displacement} at a fixed time $\tau_{MD}=20$ and packing fraction $\vp/\vp_J=0.995$, obtained from $M_{MD}=5000$ independent trajectories sampled from the ICE. Each curve corresponds to the pdf of a single particle, but only $15\%$ of the configuration is shown as described in the main text. Left panel: pdf of particles' displacement along the $\vu{z}$ direction (similar results hold for the other directions). Right panel: pdf of their square displacement. Curves colours correspond to the particles value of $C_i^{(z)}$ (left) or $S_i$ (right) as indicated in the colour scales on top, while dashed lines are used to identify if the curve is associated with a rattler. Insets: Correlation between mean and median of the distributions. In the left one, the points of each set have been displaced vertically by a fixed amount for clarity reasons and the identity lines (also displaced by the same amount) are shown in black dotted, while in the right panel's inset the red dotted line corresponds to $f(x)=0.75x$.
	}
	\label{fig:pdf displacement}
\end{figure}

After generating $M_{MD}\gg 1$ trajectories in the ICE for a fixed packing fraction, we can access a statistical characterization of the particles' motion. For instance, the left panel of Fig.~\ref{fig:pdf displacement} shows the pdf of the $\vu{z}$ component of the particles displacement (the distributions of the other components are very similar) at $\tau_{MD}=20$ and using $\vp=0.995\vp_J$.
Keeping in mind that each curve is the pdf of a single particle, it is clear that all the particles have a well defined mean displacement, many of which are \emph{different from zero}. This is the first of our main results, because it indicates that some spheres indeed have a preferred direction of motion and that it can be identified despite the statistical fluctuations coming from the thermal noise and sample-to-sample variations. More importantly, we can infer such direction using the sum of contact vectors defined in Eq.~\eqref{def:tot CV}. This can be seen from the evident division of colours to the left (right) corresponding to a negative (positive) value $C_i^{(z)}:= \vb{C}_i \cdot \vu{z}$, as indicated in the scale on top. Only $15\%$ of the particles were used to generate the main figure, and for clarity reasons they where chosen as the $10\%$ ($5\%$) with the largest (smallest) absolute value of $\avg{\delta z_i}$, but no structural information nor other statistical property whatsoever was considered. 
Now, since all the distributions shown are unimodal and rather narrow, we can expect the value of $\avg{\delta z_i}$ to be descriptive enough. To test this idea, we compared the mean value and the median of each particle in the configuration and obtained the results included in the inset of the same figure, where data points of different colour correspond to the different spatial components as indicated. The fact that all of them lie very close to the identity line (black dotted curves) confirms that $ \avg{\delta \va{r}}$ can be used to discern the existence of preferential directions in the possible trajectories of the configuration. I note in passing that the presence of anisotropy in the particles' motion has only been recently studied in few works, \textit{e.g.} \cite{rottlerPredictingPlasticitySoft2014,patinetConnectingLocalYield2016,xuPredictingShearTransformation2018,barbotLocalYieldStress2018,schwartzman_nowikAnisotropicStructuralPredictor2019}.

Analogous results for the particles' square displacement are shown in the right panel of Fig.~\ref{fig:pdf displacement}, also selecting $15\%$ of the particles according to their mobility using the same criterion as before. Correspondingly, the colour scale indicates the value of $S_i$ defined in Eq.~\eqref{def:tot dot CVs}.
Importantly, in this case the shape of the distributions is notoriously different with respect to the ones of $\delta \va{r}$. This is exemplified by the fact that the mean and median of the mobility do not coincide in this case (see the inset) because the mean is a statistic more sensible to extreme events, such as trajectories where the mobility can be about 10 times larger than the most typical ones. Nevertheless, there is a simple linear relation between both statistics, as illustrated in the inset where the red dotted line indicates the curve $f(x)= \frac{3}{4}x$.
In summary, the distributions of $\{\delta \vb{r}_i\}_{i=1}^N$ and $\{|\delta \vb{r}_i|^2\}_{i=1}^N$ have clearly a different shape, with unimodal, well peaked curves for the former, and broader, asymmetric probability densities in the latter case. Such difference evinces that these two variables have different behaviours and should therefore be studied separately as we do here. I also remark that, by the same token, it is not obvious that the ICE mean of a particle's mobility, \textit{i.e.} $\avg{|\delta \vb{r}_i|^2}$, would be a good descriptor of its full distribution.  Therefore, to justify the usage of such average to characterize the statistics of $|\delta \vb{r}_i|^2$, I provide the following reasons:
i) the support of the distributions shown in the figure differs, roughly, by an order of magnitude, which is enough to tell apart the least mobile particle from the most mobile ones; ii) the inset shows that there is still a linear relation between the median and the mean, implying that a higher value of the latter is accompanied by a proportional translation of the full distribution also to higher values; iii) other studies using the ICE have found that several dynamical features worth studying in glassy systems are indeed captured by $\avg{|\delta \vb{r}_i|^2}$.
Additionally, a word of caution is in place to avoid possible confusions regarding how to interpret the mobility distributions. First of all, the first moment of $|\delta \vb{r}_i|^2$ in general can be different to its typical values, as has already been reported in \cite{berthierStructureDynamicsGlass2007}. Additionally, the grouping of the curves depicted in the right panel of Fig.~\ref{fig:pdf displacement} in two different sets should not be understood as if there were two limiting distributions of the mobilities, but instead as a consequence of the fact that we selected only $15\%$ of the particles trajectories; that is, the remaining $85\%$ of the curves fill the gap between the apparently two different sets.

\begin{figure}[!htb]
	\centering
	\includegraphics[width=\linewidth]{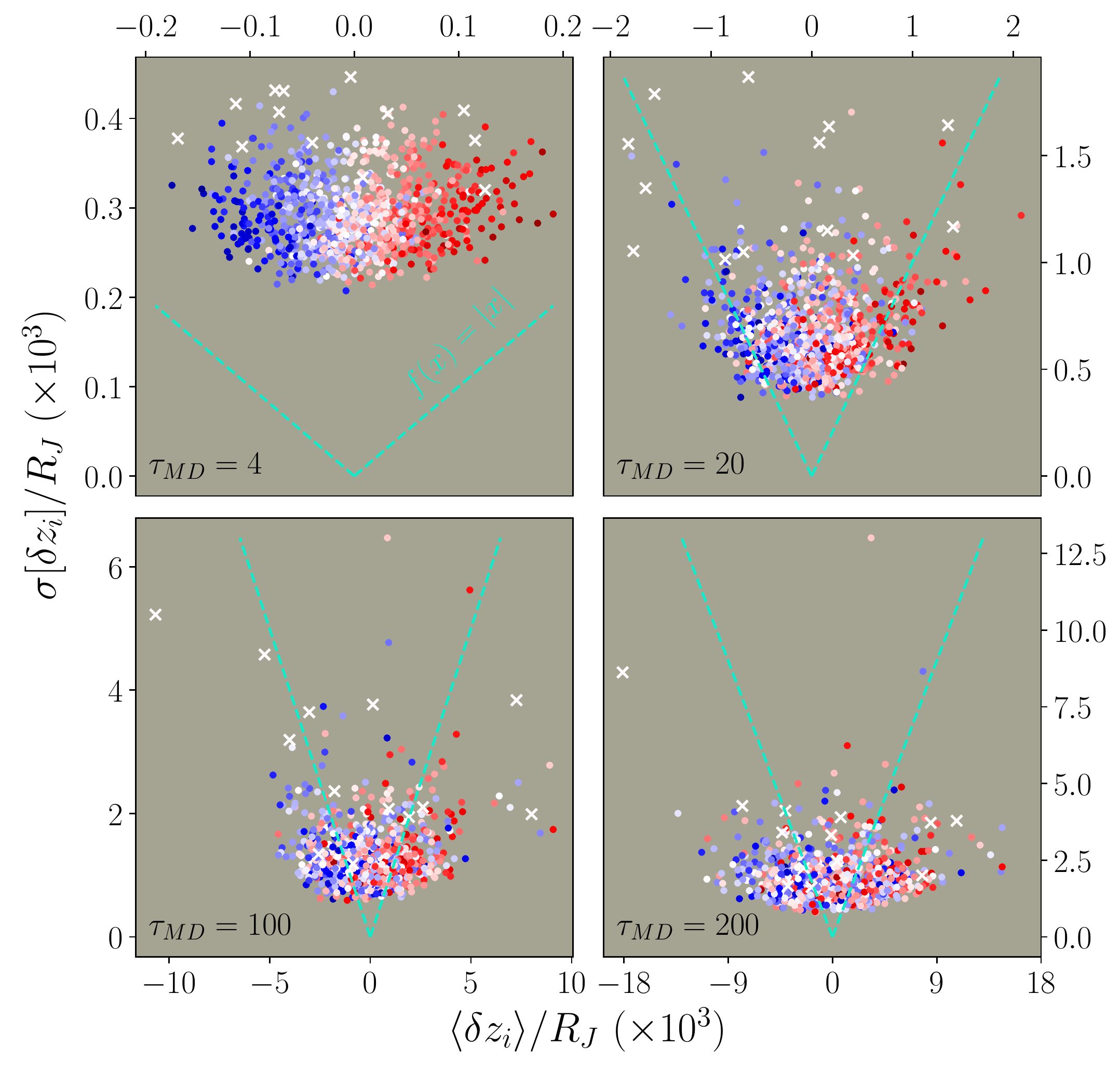}
	\caption[Comparison of mean and standard deviation of the single-particle displacements.]{Comparing the average displacement and its standard deviation along the $\vu{z}$ component of the full configuration, at four different times as indicated by the label on the bottom of each panel. The colour of each point corresponds once again to the scale of $C_i^{(z)}$ as in the previous figure, while rattlers are identified by crosses. In cyan I have included the $\pm$ identity lines for a better comparison, so that whenever a point lies beneath them it means that the mean dominates over sample-to-sample fluctuations. These results also demonstrate the clear correlation between $\avg{\avg{\delta \va{r}}}$ and $\va{C}$, but the lack of it for $\sigma[\delta \va{r}]$. This is analysed in detail in later sections.}
	\label{fig:mean-vs-std--MD-displacements}
\end{figure}

It is worth analysing more closely the variation around the mean displacements with the purpose of verifying that preferential directions can be identified despite sample to sample fluctuations. To this end, in Fig.~\ref{fig:mean-vs-std--MD-displacements} I compare the expected value of the displacements, $\avg{\delta \va{r}}$, to their standard deviation, $\sigma[\delta \va{r}]$, at different times (as indicated by the label of each panel). Without loss of generality, I once again restrict the analysis to the $\vu{z}$ component. 
These plots include the statistics of the full configuration, with the rattlers identified by crosses. Anticipating the results of the next section, the colour of each point is associated to its value of $C_i^{(z)}$, using the same scale as in Fig.~\ref{fig:pdf displacement}. The first thing to notice is the very high correlation between $\va{C}$ and $\avg{\delta \va{r}}$ at short times (upper panels). However, no such correlation is present for $\sigma[\delta \va{r}]$, as will be further analysed in the next part.
More importantly for the purposes of this part, except for the cases for which $\avg{\delta \vb{r}_i}\approx 0$, the mean displacement is of the same order (and sometimes even larger) than the corresponding $\sigma[\delta \vb{r}_i]$; \textit{cf.} with the $\pm$ identity lines included in all the panels. This implies that if particles move, on average, predominantly along a given direction, different realizations of the dynamics will exhibit a bias for similar trajectories. These results also show that for longer times (lower panels) preferential directions are better defined, as evinced by the larger fraction of points below the $\pm$ identity lines. Unfortunately, this temporal regime coincides with the one where the dynamics has lost most of its initial structural information. This is exemplified by the mixture of colours, but below I provide a more quantitative characterization. 

The results of this part constitute our first main finding. Namely, that near the jamming point, the trajectories of individual particles show preferential directions of motion. During the very initial times, such preferred directions might be hard to spot due sample to sample fluctuation as well as the small value of $\avg{\delta \va{r}}$. But as time passes by, an increasing number of particles have a mean displacement such that $\abs{\avg{\delta \va{r}_i}} > \sigma[\delta \vb{r}_i]$, implying that the mean value becomes the dominant one and most of the trajectories are similar to the preferred one. 
Similarly, statistics of the particles' mobility are well described by their mean and, actually, in this case it is customary that $\avg{\abs{\delta \vb{r}_i}^2} > \sigma[\abs{\delta \vb{r}_i}^2]$ even for short times. Hence, the characteristic mobility of particles is a persistent feature across different samples, despite the fact that its average might not coincide precisely with a typical value.

\subsection{Correlation with structure} \label{sec:MD correlation with structure}

\begin{figure}[!htb]
	\centering
	\includegraphics[width=0.99\textwidth]{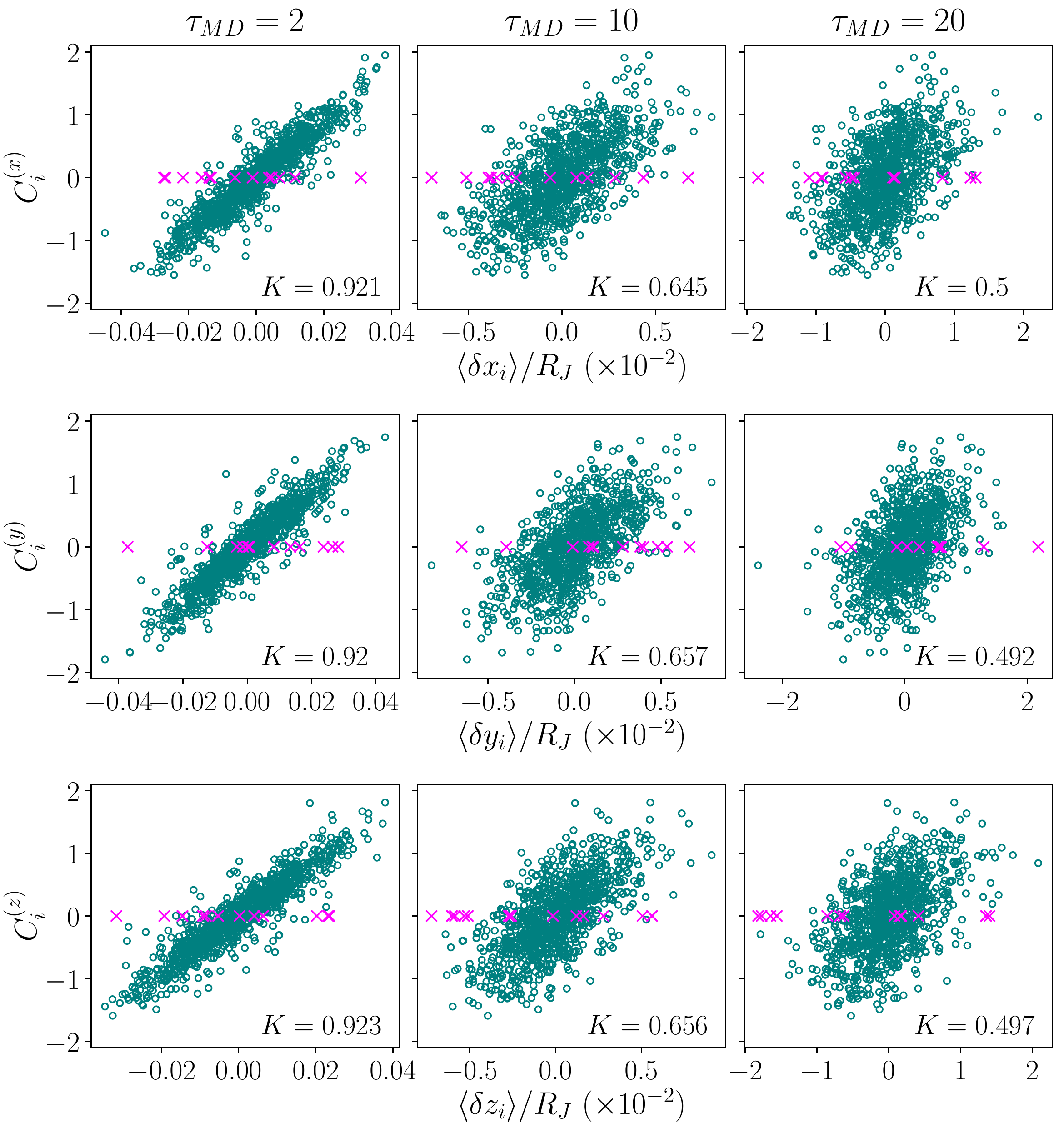}
	\caption[Scatter plots of $\avg{\delta \vb{r}_i}$ (from MD simulations) and $\vb{C}_i$.]{Scatter plots of the particles' mean displacement (scaled by the jamming radius $R_J$) and the sum of contact vectors, $\vb{C}_i$, defined in Eq.~\eqref{def:tot CV}. The reported values are the ICE averages with $\vp=0.995 \vp_J$. Each row corresponds to one spatial dimension, while each column depicts the values of the mean displacement at different times (measured in collision events, $\tau_{MD}$). Rattlers are identified by pink crosses and the value of the Spearman's rank correlation coefficient is indicated in the lower right part of each panel.
	}
	\label{fig:MD-displacements-vs-totCV}
\end{figure}

I now turn to the question of finding how much information the jamming point contains about the dynamics close to it. This means that the correspondence “similar local environment leads to similar dynamics”, illustrated in Fig.~\ref{fig:pdf displacement} for few particles (by the clustering of distributions of similar colours), must now be extended to the full configuration. 
A convenient way to achieve this is to use just few statistics of the distributions, instead of the complete pdf. As argued above, the first moment comes across as a quantity descriptive enough, and the results of Fig.~\ref{fig:mean-vs-std--MD-displacements} support this approach. However, to perform a more quantitative analysis and assess whether there is an underlying connection between $\avg{\delta \va{r}}$ and $\va{C}$, Fig.~\ref{fig:MD-displacements-vs-totCV} shows the scatter plots of these two quantities for the $N$ particles. Each of the three rows in the figure corresponds to a single direction, while the different columns refer to the different times of the dynamics at which the mean displacement was calculated. Once again, I mention that the distribution of values of the mean displacement is broad enough to conclude that many of the particles have a preferential motion direction, \textit{i.e} $\avg{\delta \va{r}} \neq \va{0}$.
It is worth emphasizing that, since we used a large set of trajectories, this latter fact cannot be attributed to statistical noise. Another relevant feature that validates our hypothesis is that the rattlers (shown with pink crosses in Fig.~\ref{fig:MD-displacements-vs-totCV} and whose pdf are drawn with dashed lines in Fig.~\ref{fig:pdf displacement}) generically exhibit a larger value of mean displacement, which is consistent with the expectation that their trajectories are less constrained because they are not part of the network of contacts; cf. Fig.~\ref{fig:jammed-config-2d}. Similarly, it is also normal that initially the dynamics of each particle is highly correlated with its value of $\vb{C}$, but as more collisions occur this relation eventually gets lost.

These results confirm that there is indeed a correlation between $\avg{\delta \va{r}}$ and the sum of unit contact vectors, which means that we can use our knowledge about $\va{C}$ to identify preferential directions in the motion of individual particles. To measure the quality of such inference we used the Spearman's rank correlation coefficient, $K$, whose value is reported in the lower-right part of each panel. The advantages of using $K$ for studying the correlation between local structure and dynamics in disordered systems are twofold: for one part, this coefficient has proven to be very sensitive to changes of the different parameters of a glass former model\supercite{hockyCorrelationLocalOrder2014,tongStructuralOrderGenuine2019}; and on the other, it naturally reflects our inferential approach based on the hypothesis that the particles ranking by a suitably chosen static variable (in this case $\vb{C}_i$) should have a direct connection with the ranking obtained via a dynamical variable (here $\avg{\delta \vb{r}_i}$).

Intuitively, the influence of a particle's set of contact vectors in establishing its preferred direction of motion can be understood because $\va{C}$ provide a good description of where the particles' neighbours are located and, therefore, which are the directions that will be favoured by possible collisions. Thus, a large value of, say, $C_i^{(x)}$ indicates that the particle's nearest neighbours are arranged in such a way that their net effect is to predominantly push it in that direction. In contrast, a mostly uniform arrangement will lead to a very small value of $\vb{C}_i$ and thus the particle would mainly remain in a reduced vicinity.

\begin{figure*}[!htb]
	\centering
	\includegraphics[width=0.95\textwidth]{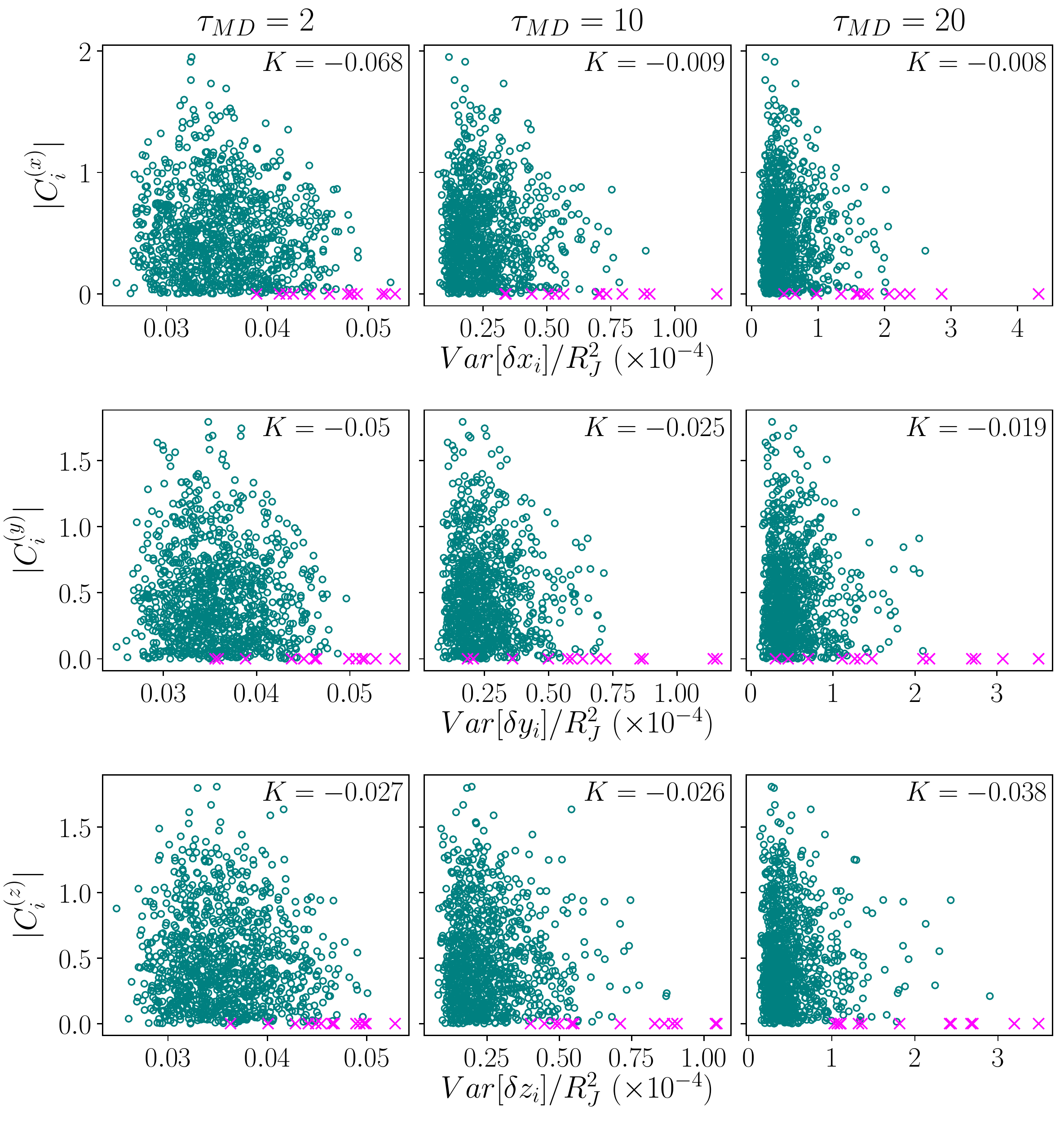}
	\caption[Scatter plot of ${Var \qty[\delta \vb{r}_i ]}$ (from MD simulations) and the absolute value of $\vb{C}_i$.]{Scatter plot of the components (one in each row) of $Var[\delta \vb{r}_i]$ and the absolute value of the components of $\vb{C}_i$ in the MD simulations at different times (different columns). Rattlers are identified by pink crosses as before.}
	\label{fig:MD-variance-vs-totCV}
\end{figure*}

On the other hand, we do not expect $\vb{C}_i$ to appropriately describe the fluctuations of a particle's motion around its mean displacement mainly because the information about the directionality is lost when computing the second moment of its displacement. Fig.~\ref{fig:MD-variance-vs-totCV} corroborates this prediction by a component-wise comparison of $Var[\delta \va{r}]$ with the absolute value of the sum of contact vectors. (The data were obtained with the same parameters as in Fig.~\ref{fig:MD-displacements-vs-totCV}, and rattlers are also identified by pink crosses). As anticipated from the results of the last part, there are no correlations between these two quantities and the values of $K$ found in this case are notably small, reflecting the fact that contact vectors only provide information about preferential directions in the particle's motion, but not about the their corresponding deviations.

\begin{figure}[!htb]
	\centering
	\includegraphics[width=0.99\textwidth]{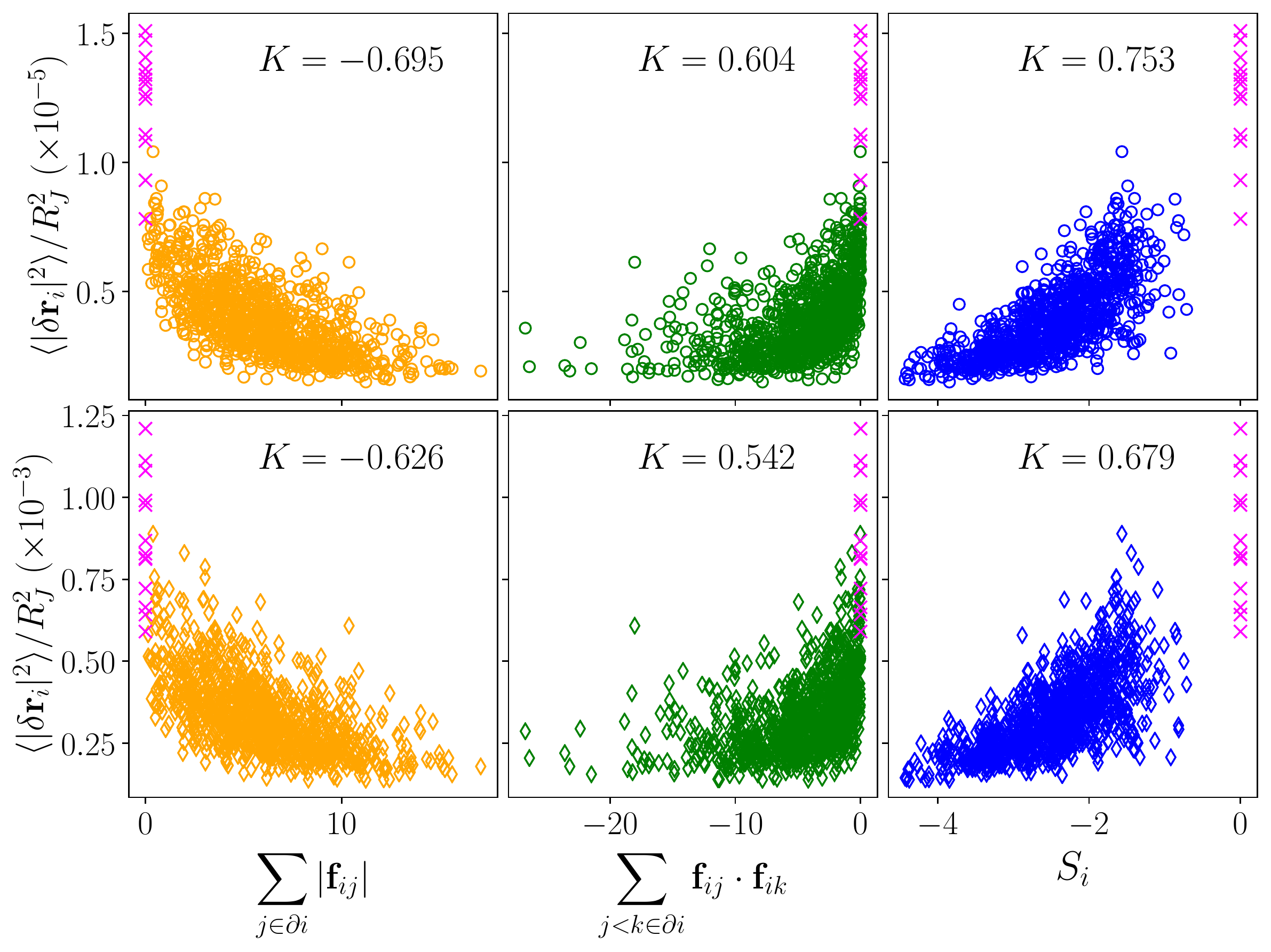}
	\caption[Scatter plots of the particles' average mobility (from MD simulations) and three scalar quantities.]{
		Scatter plots of the particles' average mobility and three possible scalar quantities that can be constructed using the network of contact forces at the jamming point: the sum of forces magnitudes (first column), the sum of dot product between pairs of contact \emph{forces} (second column), or using the contact \emph{vectors}, $S_i$, (third column) as defined in Eq.~\eqref{def:tot dot CVs}. In the upper panels we report the results from simulations with density $\vp/\vp_J=0.999$, while the lower ones show the analogous results using $\vp/\vp_J=0.99$. As in the previous figure, rattlers are also identified with pink crosses and the Spearman's rank correlation coefficient is also included. Note that only $S_i$ cleanly separates this type of particles.
	}
	\label{fig:MD-mobility-vs-dotCV}
\end{figure}

To continue, I will study the correlation with the particles' square displacement. $ \avg{|\delta \vb{r}_i|^2} $ is an important quantity to take into account in the high density regime because it provides an estimate of the ``cage'' size in which the particle is moving.
As anticipated in Fig.~\ref{fig:pdf displacement}, there is a close correlation between $\{\avg{|\delta \vb{r}_i|^2} \}_{i=1}^N$ and $\vec{S}$. However, here there is no reason \emph{a priori} why this variable should be used instead of other observables that can be obtained from the network of contacts. In particular, the role of the forces magnitudes has been so far ignored. Given that $\sum_{j\in \partial i}\vb{f}_{ij}=0$, their role as analogous variables of $\va{C}$ can be immediately discarded. But this is not the case when scalar combinations are considered.
Therefore, I will now show that it is more convenient to include only the contacts direction and not their magnitude in order to obtain a better predictor of the particles' mobility. To do so, I will consider here two other scalar quantities that are, both, physically well defined (in terms of the network of contacts at jamming), and intuitively related to the statistics of $\abs{\delta \va{r}}^2$. The first one is the sum of all the contact forces magnitudes, $\displaystyle \sum_{j \in \partial i} \abs{\vb{f}_{ij}}$; while the second one consists in the sum of all the pair of dot products between contact \emph{forces}\footnote{Note that because their are many very small forces and only few large ones, despite the discussion at the end of Sec.~\ref{sec:structural variables}, these two variables will have different distributions.}, $ \displaystyle \sum_{j<k \in \partial i} \vb{f}_{ij}\cdot \vb{f}_{ik}$. Fig.~\ref{fig:MD-mobility-vs-dotCV} compares the correlation between $\{\avg{|\delta \vb{r}_i|^2} \}_{i=1}^N$ and these two observables (first two columns), as well as $\vec{S}$ (right-most column). (The values reported correspond to $\tau_{MD}=10$.) Upper (resp. lower) panels show the results from the MD trajectories with a density of $\vp/\vp_J=0.999$ (resp. $\vp/\vp_J=0.99$). As before, data associated with rattlers are indicated with pink crosses and the corresponding values of $K$ are also included for comparison. Note that rattlers are distinguished much more clearly through $S_i$. As expected, in all cases the correlations attained in the system of higher density are larger because the local structure resembles more the one of the jammed state. The dynamics is therefore considerably influenced by the collisions with the closest neighbours at such state. Yet, at first sight it seems paradoxical that including more information, namely, the magnitudes of the forces, effectively reduces the predictive power. However, before explaining why this happens, I will present the analogous results from the MC simulations, where we obtained very similar findings. Hence, a common explanation applies to both types of simulations as discussed later in Sec.~\ref{sec:vectors vs forces}.
To close this section, I point out that we tested this same methodology on another independent jammed configuration, as described in the Appendix \ref{sec:second-configuration}, obtaining identical results; see Figs.~\ref{fig:MD-displacements-vs-totCV-2nd-config} and \ref{fig:MD-mobility-vs-dotCV-2nd-config}.

\section{Results with MC simulations} \label{sec:results Monte Carlo simulations}

\begin{figure}[!htb]
	\centering
	\includegraphics[width=0.99\linewidth]{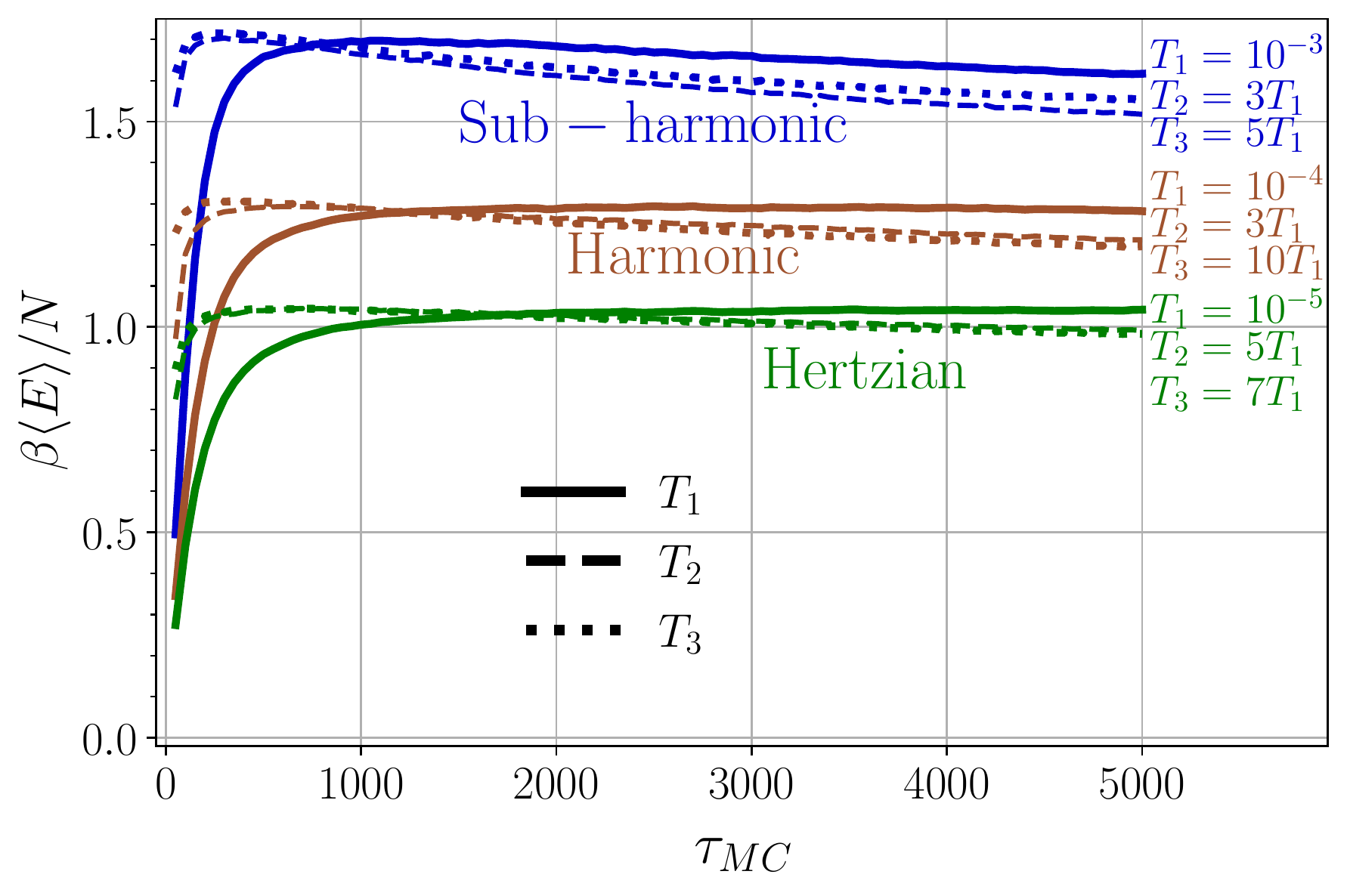}
	\caption[Average energy per particle as a function of the number of MC steps using three different potentials.]{Average energy per particle as a function of the number of MC steps ($\tau_{MC}$) for the three potentials used: $\a=3/2$ or sub-harmonic (blue), $\a=2$ or harmonic (brown), and $\a=5/2$ or Hertzian (green). Different line styles correspond to different temperatures ($T_1<T_2<T_3$), as indicated in the right legends.}
	\label{fig:MC-energy-vs-t}
\end{figure}

In contrast with the MD simulations, for the MC ones we fixed the spheres' radius at $R_J$ and soften the interaction between them by introducing a contact pair potential as the one introduced in Sec.~\ref{sec:jamming criticality}, but without scaling by the exponent, \textit{i.e.}
\begin{equation}\label{def:potential soft sphere for dynamics}
U(\vb{r}_i, \vb{r}_j) = \abs{2R_J - \abs{\vb{r}_{ij}}}^\a \ \Theta\qty(2R_J - \abs{\vb{r}_{ij}}) \, ;
\end{equation}
where $\vb{r}_{ij}=\vb{r}_i - \vb{r}_j$ and $\Theta$ is the Heaviside step function. $\a$ plays the role of a ``stiffness'' parameter that we varied to explore the effect of different types of interactions. Because we generated trajectories using different temperatures, introducing the proportionality constant of $1/\a$ simply rescales the temperature, $T$. Analogously to the MD simulations, we used the jammed configuration as initial condition to generate a trajectory using the Metropolis-Hastings algorithm at a fixed $T$. Hence, for this type of dynamics, the natural time unit is the number of MC steps performed, which I will denote as $\tau_{MC}$ to distinguish it from the time scale of the MD simulations and the physical time. We tried different interaction potentials by setting $\a=\{3/2,2,5/2\}$ (henceforth referred as sub-harmonic, harmonic, and Hertzian interaction, respectively), performing $M_{MC}=1000$ MC simulations for each type of potential and temperature. The different values of $T$ were selected in such a way that the samplings were done with a similar acceptance rate. This is shown in Fig.~\ref{fig:MC-energy-vs-t} by plotting the temporal evolution of the average energy per particle scaled by the inverse temperature, $ \beta \avg{E}/N$. The figure includes the results of the three interactions potentials and the different temperatures used for each one. For a fixed value of $\a$, the energies of the different simulations are comparable between them, once the scaling with $\beta$ is considered, indicating that a similar sampling was carried out. We can then conclude that the trajectories generated, using different values of $T$ and $\a$, belong to the same dynamical regime given that all the values of $\beta \avg{E}$ lie  within a small range of each other.


\begin{figure*}[!htb]
	\centering
		\includegraphics[width=0.95\textwidth]{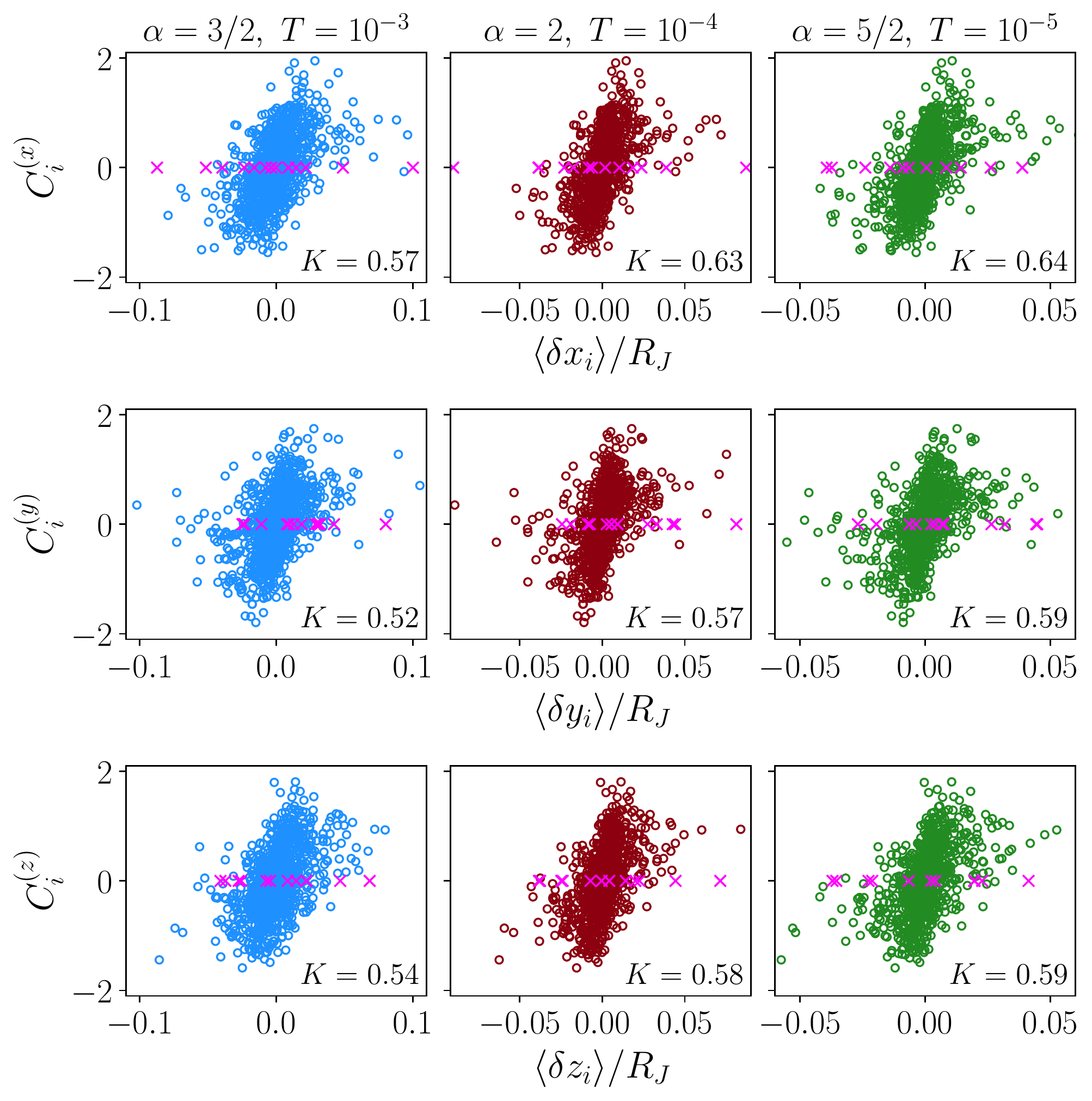}
	\caption[Scatter plots of $\avg{\delta \vb{r}_i}$ (from MC dynamics) and $\va{C}_i$.]{
		Scatter plots of $\avg{\delta \va{r}}$ and $\va{C}$ (each row corresponds to one direction), and for the three types of interaction considered: sub-harmonic (first column), harmonic (second), and Hertzian (third); different colours are included just for clarity sake. The temperature used to generate the MC trajectories for each potential is indicated at the top, and the values of $\avg{\delta \va{r}}$ reported correspond to $\tau_{MC}=250$ MC steps. For each value of $\a$, the corresponding values of temperature were chosen so that the average energy of the system divided by $T$ remained within a small range; see Fig.~\ref{fig:MC-energy-vs-t}. As in the analogous plots of MD simulations, the quantities associated with rattlers are identified by pink crosses and the value of $K$ is also included.
	}
\label{fig:MC-displacements-vs-totCV}
\end{figure*}

\begin{figure*}[!htb]
	\centering
	\includegraphics[width=0.95\textwidth]{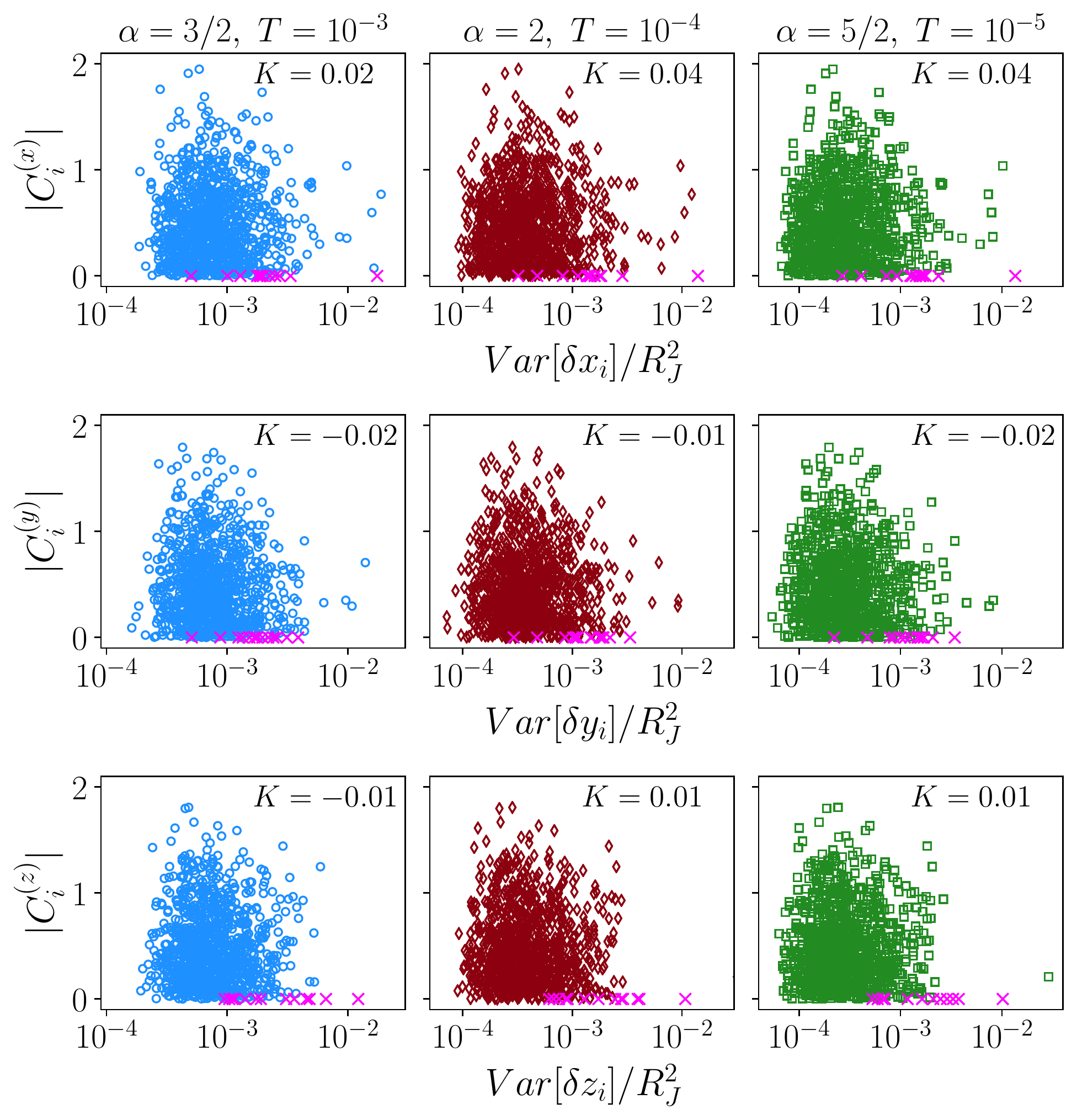}
	\caption[Scatter plot of neighbouring particles ${Var\qty[\delta \vb{r}_i ]} $ (from MC simulations) and the absolute value of $\vb{C}_i$.]{Scatter plot of the components (one in each row) of $Var[\delta \va{r}]$ and the absolute value of the components of $\va{C}$ in the MC simulations for the three different potentials considered in this work (different columns). All the data reported here correspond to $\tau_{MC}=250$ and the value of $T$ used throughout the simulations of each potential are reported above the corresponding column. Different colours are used for the sake of clarity, except for the rattlers data, which as usual are depicted with the pink crosses. The very small values of $K$ obtained are also included and confirm that these two variables are uncorrelated.}
	\label{fig:MC-variance-vs-totCV}
\end{figure*}

\begin{figure*}[!htb]
	\centering
	\includegraphics[width=0.95\textwidth]{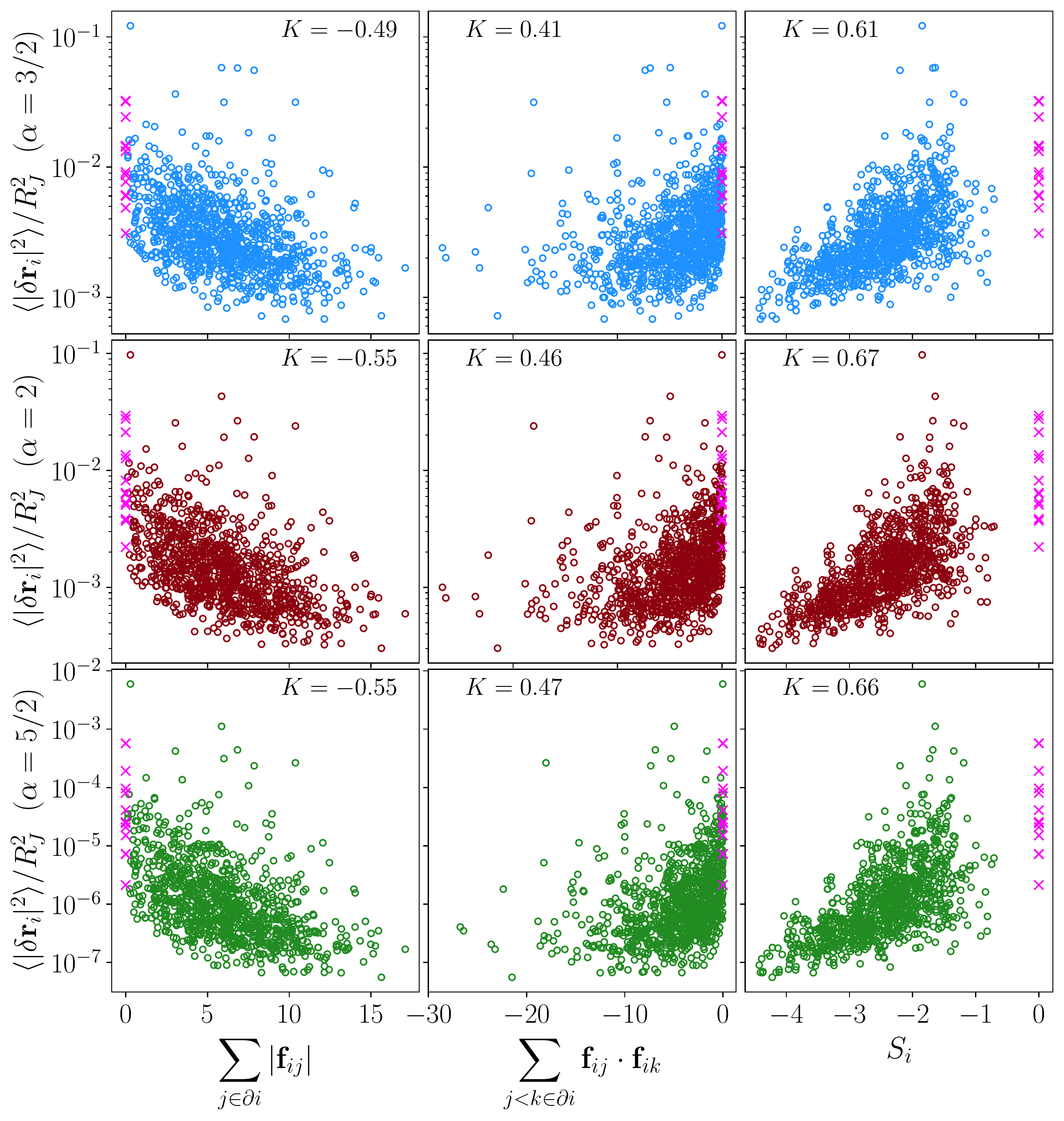}
	\caption[Scatter plots of the particles mobility (from MC dynamics) and three scalar quantities.]{Scatter plots of each particle's average mobility at $\tau_{MC}=250$ with the same interaction parameters and temperatures of Fig.~\ref{fig:MC-displacements-vs-totCV}. Each row consists on the same values of $\avg{|\delta \vb{r}_i|^2}$ as function of different scalar quantities that can be constructed using the network of contact forces at the jamming point, like the sum of forces magnitudes (first column), the sum of dot product between pairs of contact forces (second column), or using the contact \emph{vectors} (third column) as defined in Eq.~\eqref{def:tot dot CVs}. As in the previous figures, points associated with rattlers are identified with pink crosses and the Spearman's rank correlation coefficient is also included.
	}
	\label{fig:MC-mobility-vs-dotCV}
\end{figure*}

I now show that with this different dynamical protocol we obtain results in excellent agreement with those of MD simulations. In Fig.~\ref{fig:MC-displacements-vs-totCV} I compare the correlations between $\avg{\delta \va{r}}$ and $\va{C}$ for the three types of interactions (one in each column), and with the temperatures indicated on top. As mentioned above, the value of $T$ was selected in such a way that a similar acceptance rate was obtained for all potentials. Each row corresponds to a component of $\avg{\delta \vb{r}_i}$, computed at $\tau_{MC}=250$ in all cases, and with a different colour for each value of $\a$. Additionally, rattlers are identified in all panels rattlers by pink crosses as in previous figures. The values of $K$ are comparable for all the interaction potentials tested and also with the ones obtained in the MD simulations, thus signalling that our approach is robust enough to be applied to both types of dynamics. It is also worth mentioning that there is a systematic increase of the correlation with $\a$, \textit{i.e.} the smallest (largest) value of $K$ was obtained with the sub-harmonic (Hertzian) interaction, independently on the components. However, as I discuss later, this last result is related to the fact that different values of $\a$ produce different decorrelation rates, if measured by $\tau_{MC}$.
%
In Sec.~\ref{sec:decorrelation} I show that if the MSD is used instead as a measure of the evolution of the configuration, the decrease of $K$ follows a universal behaviour. On  the other hand, in Fig.~\ref{fig:MC-variance-vs-totCV} I present the analogous results of the correlation between $Var[\delta \va{r}]$ and the absolute value of $\va{C}$, using the same format and parameters than in the previous figure. In agreement with the MD results, these data demonstrates that the sum of contact vectors is a very poor indicator of the single-particle displacement's variance, even when the interaction potential has been softened. This claim is quantified by the almost vanishing values of $K$ reported in each panel.

More importantly, when considering the mean mobility of particles and its correlation with the scalar observables constructed from the network of contacts, we also found that $\vec{S}$ is the most informative variable in comparison with the ones that include the magnitudes of the contact forces. This finding holds independently of the type of interaction as shown in Fig.~\ref{fig:MC-mobility-vs-dotCV}, where each column compares the correlation between $\avg{|\delta \vb{r}_i|^2}$ and the three scalar quantities considered before, while different rows (from top to bottom) correspond to the different interaction types (from $\a=3/2$ to $\a=5/2$). Once again, we obtained values of $K$ comparable to the ones of the MD simulations, although this time the highest value of $K$ was obtained with the harmonic interaction between particles. But this is due to the fact that there is a different decorrelation rate between different statistics, as I argue in Sec.~\ref{sec:decorrelation}.

To close this section I want to stress that the structural variables considered here, namely $\va{C}$ and $\vec{S}$, are computed exclusively from the properties of the configuration at jamming. Consequently their value remains fixed and is independent of the type of dynamical protocol employed. Yet, our results demonstrate that they can be utilized as proxies of the statistics of single-particle trajectories in a rather broad range of circumstances.

\section{Contact vectors \textit{vs} contact forces}\label{sec:vectors vs forces}

Let me now go back to the question of why the contact \emph{vectors} ($\vb{n}_{ij}$) are variables more informative about the dynamics near jamming than the contact \emph{forces} ($\vb{f}_{ij}$). 
In other words, $\va{C}=\{\vb{C}_i\}_{i=1}^N$ and $\vec{S}=\{S_i\}_{i=1}^N$, defined respectively in Eqs.~\eqref{def:tot CV} and \eqref{def:tot dot CVs}, are predictive enough to allow us to identify a preferential direction of motion for each sphere, as well as to spot the most mobile particles. However, the value of these observables does not depend on the magnitude of the contact forces but only on their direction. And the results of Secs.~\ref{sec:results MD simulations} and \ref{sec:results Monte Carlo simulations} show that when the forces magnitude is included the quality of  the prediction \emph{worsens}. Moreover, this is the case independently of the model, \textit{i.e.} either HS or SS, and even the interaction potential in the latter case. It is thus worthwhile to analyse more closely this counterintuitive finding.

Let me begin by considering the same problem from a purely static point of view. As derived in detail in Sec.~\ref{sec:network of contacts}, in configurations having 1SS --as we know ours do-- the forces magnitude can be obtained as the unique zero mode of the \textit{contact matrix}, $\S$ defined in Eq.~\eqref{def:S matrix}. Therefore, magnitudes do not contain any extra information than what is already present in the spheres position.
Second, considering for the time being only the MD dynamical protocol, it is clear that as soon as the spheres are no longer touching each other, the magnitude of a contact force loses its physical meaning, while the true force is related to the momenta exchange during the collisions. In contrast, the corresponding contact vector still contains information about which directions constrain a particle's motion the most.
Similarly, in the case of SS the value of the contact force at jamming is not the relevant physical quantity either, because the real forces driving the dynamics in the MC protocol depend explicitly on both, their mutual overlap and the type of interaction potential, which we have parametrized here through $\a$. Yet, independently of the choice of $\a$, and hence the value of the force magnitude, the contact vectors can be used as a proxy of the force field experienced by each particle, at least for the initial part of the dynamics. 
So, in either case, the set of contact vectors provide an estimation of the optimal direction in which a particle should move in order to minimize the free energy of the system.
Additionally, as can be noted from the rightmost equality of Eq.~\eqref{def:tot dot CVs}, $\vec{S}$ also incorporate the influence of the number of neighbours of each particle, $q_i$. This feature intuitively explains that regardless of the vectorial contribution of the contacts, particles with more (less) neighbours are expected to be more (less) constrained. Besides, it also explains why a significant correlation was found when considering $\sum_{j \in \partial i} \abs{\vb{f}_{ij}}$ and $\sum_{j<k \in \partial i} \vb{f}_{ij}\cdot \vb{f}_{ik}$ as a predictors, the reason being that they are essentially a weighted sum of the number of contacts. Another fact to consider is that the distribution of dot products between contact forces and the analogous one between contact vectors are significantly different, as discussed in Sec.~\ref{sec:structural variables}; that is, Fig.~\ref{fig:dist-contacts} provides further evidence that these two quantities convey different information.

Even more importantly, another physical mechanism is suggested by the fractal FEL picture; see Secs.~\ref{sec:MF gardner transition} and \ref{sec:LP dependence initial confs}. The argument goes as follows: in Ref.~\cite{fel_2014} the authors measured numerically the overlap of contact \emph{vectors} between pairs of jammed states (\textit{i.e.}  pairs of minima in the landscape), defined by a target pressure $p^\star$, as a function of a common initial pressure of the configurations, $p_0$. What they found is that as $p_0$ was closer to $p^\star$, the overlap increased \emph{gradually}. This implies that, although jammed configurations belonging to a same (fractal) meta-basin share a similar structure, the specific neighbours with which a particle will eventually be in contact, once in the jammed state, are not defined unequivocally by its local environment above jamming. Instead, they are determined progressively as such state is approached. By the same token it also indicates that as a configuration goes up in the landscape, the jammed state it departed from will still encode useful structural information, since a significant amount of the network of contacts is shared by the nearby minima. Now, note that we can think of the short time dynamics studied here as several realizations of trajectories followed by a configuration in its way to explore a small neighbourhood above its initial minimum. Under these assumptions, the original meta-basin should contain all the phase space available as this brief exploration takes place, which in turn implies that many of the interactions between nearest neighbours influencing the dynamics are going to remain roughly unchanged, and therefore we can expect major correlations with the actual network of contact vectors.

\section{Further statistical properties}

\subsection{Predicting the mean displacement and mobility} \label{sec:prediction}

In Sec.~\ref{sec:dynamics-and-local-structure-glasses} in the first chapter it was mentioned that there have been many proposals to tackle the problem of the connection between local structure of disordered solids and their dynamics. However, one of the most pressing issues is not only to find such correlations between statical and dynamical properties, but instead using the former to \emph{predict} the latter. To the best of our knowledge, just a handful of works have dealt in detail with
this problem using well established physical variables\supercite{widmer-cooperPredictingLongTimeDynamic2006,candelierSpatiotemporalHierarchyRelaxation2010,coulaisDynamicsContactsReveals2012}, although more recently in\cite{schoenholzCombiningMachineLearning2018,cubukIdentifyingStructuralFlow2015,cubukStructuralPropertiesDefects2016, schoenholzStructuralApproachRelaxation2016,schoenholzRelationshipLocalStructure2017,cubukStructuralPropertiesDefects2016,maHeterogeneousActivationLocal2019,bapstUnveilingPredictivePower2020} researchers addressed similar questions employing machine learning methods. As I have commented before, this latter techniques have an outstanding prediction power, but they rely on an artificial representation of the particles' environment at the \emph{mesoscopic} scale, which is an important difference with the approach we have adopted here. At any rate, as a further benchmark of our method we now address the predictability issue.

First, I should clarify that this is the inverse scenario of the one explored so far. Consider in particular 
Figs.~\ref{fig:pdf displacement} and \ref{fig:mean-vs-std--MD-displacements}, where we began by distinguishing the particles with some prior information about the dynamics (\textit{e.g.} the ranking based on the first moments of $\delta \vb{r}_i$ and $\abs{\delta \vb{r}_i}^2$) and showed that such distinction carried over to the structural features we here study (as exemplified by the clustering of data with similar colours).
In contrast, I now focus on the reciprocal problem where our starting point is the static information (namely, the contact vectors) which will then be used to infer a property about the dynamics of the configuration. I will show that within our approach, prediction is indeed possible insomuch as we can use $\va{C}$ to reveal preferential direction in the particles' motion, while $S_i$ can be utilized to identify the most mobile ones, \emph{without any prior information about the dynamics}. As expected, the validity of such predictions is highest at the initial part of the trajectories and later it decays as time goes by and the system evolves. 

To be more precise about how the initial correlation is lost, it is important to find a way to consistently compare the evolution of the configuration in both types of dynamical protocols and for different values of the parameters. A natural candidate is the (ICE averaged) MSD, defined in Eq.~\eqref{eq:MSD}. Intuitively, it is a measure of how much the configuration has changed from its initial state, $\va{r}^{(J)}$.
Because we focused on the short-time dynamics, the values attained by the MSD during our simulations are well below the one of its characteristic long time plateau --identified with the onset of the Debye behaviour in the density of states\supercite{dynamic_criticality_jamming}-- as I have shown in Sec.~\ref{sec:supplementary MD}. Furthermore, Fig.~\ref{fig:MD-msd-vs-t} there shows that all our simulations were done in the same dynamical regime, where the MSD have a similar behaviour for the different packing fractions we used.
Thus, I will henceforth report the evolution of the quantities we analyse as a function of $\avg{\Delta}$ instead of as a function of time, $\tau_{MD}$, or $\tau_{MC}$.

\begin{figure*}[!htb]
	\centering
	\includegraphics[width=0.99\textwidth]{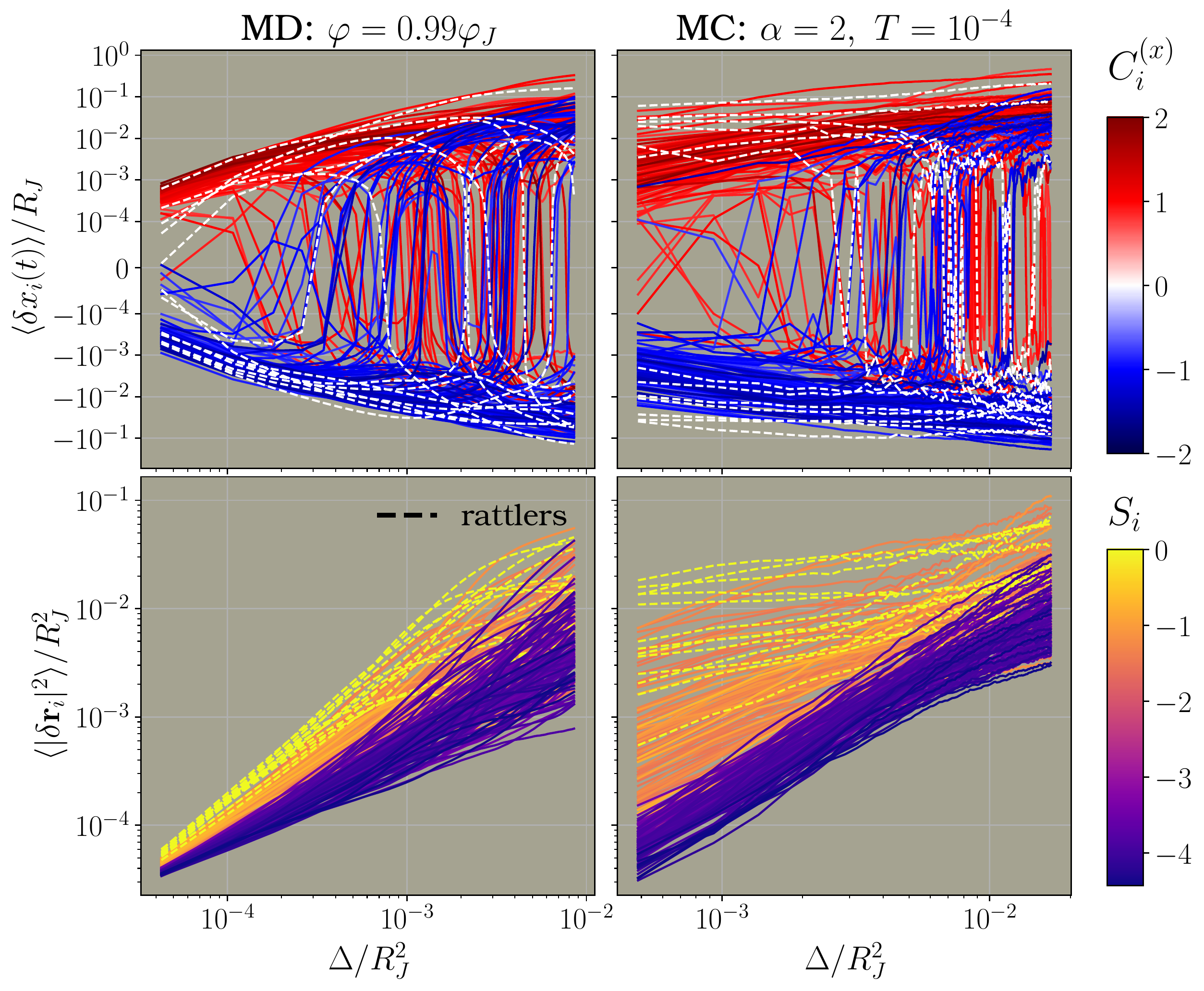}
	\caption[Predicting preferential directions and mobility from the ranking of structural variables.]{
		Evolution of the $\vu{x}$ component of the mean displacement (upper panels) and the average squared displacement (lower panels) of $20\%$ of the particles for the two type of dynamics utilized here (MD on left panels and MC on the right ones), with the parameters specified on top. The trajectories on the upper (resp. lower) panes were selected by choosing the top and bottom $10\%$ of the particles ranked according to their value of $\vb{C}_i$ (resp. $S_i$); rattlers were also included in the first case. Trajectories are coloured according to the scale shown in the right. Thus, the division of colours demonstrates that we can identify both: which particles are the most mobile ones and if they have a preferred direction of motion. We used the mean squared displacement, $\Delta$, to measure the dynamical evolution for the reasons explained in the main text.
	}
	\label{fig:ranking-forces-and-displacements}
\end{figure*}


We can pose the problem of predicting the preferred directions in configuration space as investigating whether particles for which the observable $\vb{C}_i$ has a large absolute value in one of its components also exhibit trajectories that move predominantly along that same component. We can answer in the affirmative this question following a straightforward method: we rank the particles according to their value of $\vb{C}_i^{(x)}$, then select the top and bottom $10\%$, and plot the evolution of their trajectories. This results in the plots in the upper panels of Fig.~\ref{fig:ranking-forces-and-displacements} for the $\vu{x}$ component of the trajectories generated via the MD simulations with $\vp/\vp_J=0.99$ (left panel) or the MC algorithm with harmonic interaction and $T=10^{-4}$ (right panel). In both cases, the lines colour indicates the value of the sum of contact vectors according to the scale on the right. These figures show that most particles which share a close value of $\vb{C}_i$, and thus whose respective curves are shown in a similar colour, move in the same direction. Trajectories of rattlers are also included (as dashed lines) and even though we can not predict any preferred direction in their motion, we can nevertheless identify them as highly mobile particles  and thus can be used as an estimate of bounds delimiting the maximum displacement of individual particles.
Analogously, if we rank the spheres according to their value of $S_i$ we have enough information to identify the most mobile particles, as demonstrated in the lower panels of Fig.~\ref{fig:ranking-forces-and-displacements}. In this case, the effect of the different protocols for the simulations is more evident for the initial part, since the MD simulations show a much narrower range of values of the squared displacement, while the MC ones yield values of $\avg{\abs{\delta \vb{r}_i}^2}$ that differ by more two orders of magnitude. However, the grouping of trajectories with similar colours is still preserved, confirming that $S_i$ can be used to distinguish between mostly mobile and mostly blocked particles. Note that in this case, rattlers also provide a good estimate for an upper bound on the value of $\avg{\abs{\delta \vb{r}_i}^2}$ and that only for these particles $S_i$ is identically zero, while for the others $S_i \simeq -1$ at most. As anticipated, the clear division of colours, reflecting an almost perfect correspondence between contact vectors and the statistical properties of the dynamics, is subsequently lost as the configuration moves away from its initial state, or, in other words, as $\Delta$ gets larger.

\subsection{“Universal” decorrelation rate} \label{sec:decorrelation}

\begin{figure}[!htb]
	\centering
	\includegraphics[width=\linewidth]{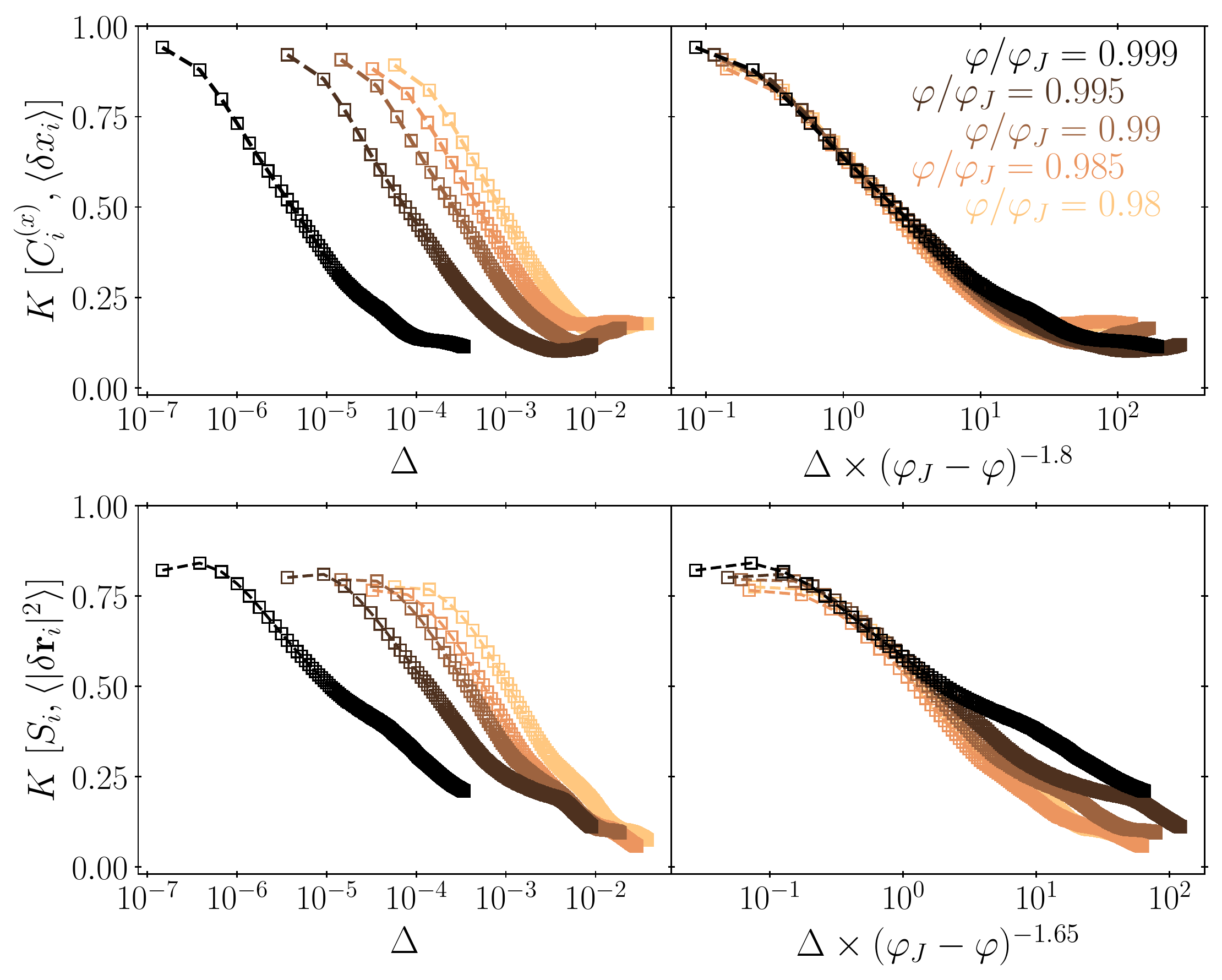}
	\caption[Evolution of the Spearman correlation between preferential directions and total contact vector and between mobility and the sum of dot products of contact vectors. Results from MD simulations]{Evolution of the Spearman correlation between preferential directions and total contact vector (upper panels) and between mobility and the sum of dot products of contact vectors pairs (lower panels), for several values of $\vp$ (each one in a different line colour). The right panels show that if the MSD is rescaled by a factor that measures the distance from the jamming point, simulations with different parameters follow the same behaviour. See main text for a detailed discussion. }
	\label{fig:decorrelation-MD}
\end{figure}

\begin{figure}[!htb]
	\centering
	\includegraphics[width=\linewidth]{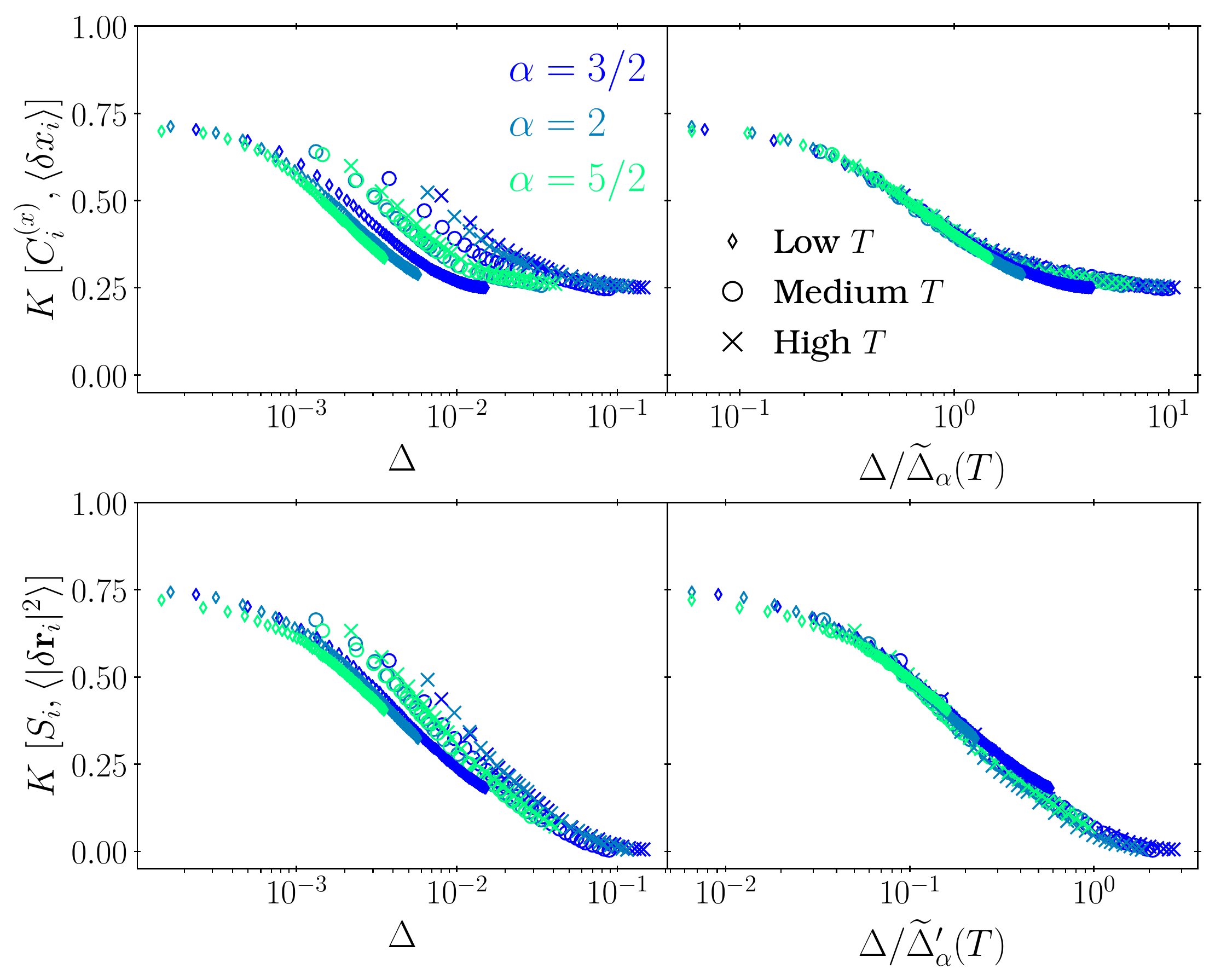}
	\caption[Evolution of the Spearman correlation between preferential directions and total contact vector and between mobility and the sum of dot products of contact vectors. Results from MC dynamics.]{Analogous to Fig.~\ref{fig:decorrelation-MD} but for MC dynamics, modelling SS with different  contact potentials (different colours) and at different temperatures (different markers shape). Note that once again, if the MSD is rescaled by a function of the distance to the jamming point (in this case determined by $T$), the different curves can be collapsed into a single one (right panels).}
	\label{fig:decorrelation-MC}
\end{figure}

Results so far show that significant correlations exist between the structural properties at jamming and the dynamics near the such point. Nonetheless, it is clear that as time evolves such initial correlation should eventually disappear. Therefore a crucial question is how fast and in which way this initial information is lost. To answer it, we computed the value of $K[\vb{C}_i, \avg{\delta \vb{r}_i}]$ and $K\qty[S_i, \avg{\abs{\delta \vb{r}_i}^2}]$ at different times. (For brevity, I introduced the notation $K[X,Y]$ to denote the Spearman's rank correlation coefficient between variables $X$ and $Y$.)
Expectedly, different systems and different parameters yield diverse decorrelation rates, using $\Delta$ to measure the temporal evolution as above. 
The results obtained with the different parameters using the MD and MC simulations are reported in the left panels of Figs.~\ref{fig:decorrelation-MD} and \ref{fig:decorrelation-MC}, respectively, confirming this behaviour. 
However, a surprising feature is revealed when $\Delta$ is rescaled in terms of the “distance to the jamming point”: all the curves can collapsed on top of each other for a significant fraction of the dynamics. This finding signals the existence of a common process by which the configuration decorrelates from its initial state. For instance, considering first the correlation between the total contact vector and the mean displacement in the MD simulations, it is natural to use $\vp$ to parametrize the distance from jamming. In fact, we found the appropriate rescaling factor to be $(\vp_J-\vp)^{1.8}$ and, as the upper right panel of Fig.~\ref{fig:decorrelation-MD} shows, when the MSD is divided by such factor the different curves fall on top of each other almost two orders of magnitude. 
Very similar findings hold for $K[S_i, \avg{\abs{\delta \vb{r}_i}^2}]$ as reported in the lower right panel of Fig.~\ref{fig:decorrelation-MD}, except that in this case the scaling factor turns out to be $(\vp_J-\vp)^{1.65}$.

When considering the dynamics of SS, it is reassuring that we found completely analogous results using the MC dynamics. More precisely, each value of $\a$ fixes an intrinsic softness to the spheres, while the temperature determines how much they are allowed to overlap on average. Hence, once the interaction parameter $\a$ is set, the rescaling of the MSD should only depend on $T$. With this in mind, we can define a “thermal scale” for the MSD as $\widetilde{\Delta}_\a(T) = c_\a T^{\nu_\a}$, where $c_\a$ and $\nu_\a$ are constants that only depend on the interaction type; we report the latter in the first row of Table \ref{tab:values-nus}. By measuring $\Delta$ in terms of the corresponding thermal scale, all the different curves of $K[\vb{C}_i, \avg{\delta \vb{r}_i}]$ can be collapsed; see upper right panel of Fig.~\ref{fig:decorrelation-MC}. For the case of the mobility, the same type of rescaling applies, albeit with different constants, \textit{i.e.} $\widetilde{\Delta}'_\a(T) = c'_\a T^{\nu'_\a}$; the values of these exponents are reported in the last row of Table \ref{tab:values-nus}. 
\begin{table}[!htb]
	\centering
	\begin{tabular}{|c|c|c|c|}
		\hline
		& $\a=3/2$ & $\a=2$ & $\a=5/2$\\[1mm]
		\hline
		$\nu_\a$ & $0.84$ & $0.64$ & $0.51$ \\[2mm]
		$\nu_\a'$ & 0.45 & 0.40 & 0.35\\
		\hline
	\end{tabular}
	\caption{Values of the exponents used for the rescaling of the MSD with the temperature in the MC simulations, producing the curves shown in Fig.~\ref{fig:decorrelation-MC}. Values in the first row correspond to the results of $K\qty[\vb{C}_i, \avg{\delta \vb{r}_i}]$ while in the second row we report the associated values for $K\qty[S_i, \avg{\abs{\delta \vb{r}_i}^2}]$. }
	\label{tab:values-nus}
\end{table}

To better comprehend these findings, some remarks are useful. First, it is clear that we obtained considerably high values of $K$ for both types of structural variables and dynamical protocols. Additionally, the fact that during the intervals considered here the value of $K$ decreases by roughly an order of magnitude, while the (rescaled) MSD increases by at least two decades, reveals that the decorrelations are actually rather slow. 
It is also worth emphasizing that, in all cases, the rescaling factors essentially measure how far away from the jamming point the dynamics takes place: either by reducing its packing fraction (in the HS scenario) or by providing some thermal energy (in the case of SS with a contact potential). Unfortunately, no theory exists for predicting the form of these scaling variables, and here we had to determine them by inspection. Nevertheless, if we resort to the fractal FEL picture, the curves collapse suggests that, independently of the type of dynamical protocol or even the interaction potential, the same mechanism drives the loss of correlation as the system explores its meta-basin. It is tempting to rationalize this feature in terms of the jamming universality (see Sec.~\ref{sec:jamming in many systems}), but I consider that a careful, dedicated study, with a more systematic characterization of the dynamics and other physical variables, is needed for such purpose. Finally, we can understand that the correlations with $\va{C}$ and $\vec{S}$ yield different scalings exponents because these two observables provides information about different features of the dynamics, namely, the directions in which the particles' displacements are facilitated and how much they are expected to move, respectively. 
I finish this section mentioning that we also verified that the same results hold for the other configuration we tested, as shown in Fig.~\ref{fig:correlations-vs-t-2nd-config} in the Appendix \ref{sec:second-configuration}.


\section{Comparison with normal modes and other works} \label{sec:comparison normal modes}

In the Sec.~\ref{sec:dynamics-and-local-structure-glasses} it was mentioned that the dynamical properties of systems near jamming have been investigated predominantly in experiments, and for the scope of this chapter it is relevant to consider Ref.~\cite{coulaisHowIdealJamming2014}. In this study, the authors found a clear connection between the dynamics of their samples, which consisted in a bidisperse mixture of photoelastic soft disks, and the network of contacts formed among them. They studied both the under- and over-compressed parts of the dynamics, although an important difference with our work is that in their case the dynamics was driven by an external vibrating apparatus rather than by thermal fluctuations as we considered here. At any rate, our results should be considered as complimentary to each other because while we focused on the statistics of the trajectories for a given configuration, their results are based on several realizations on the network of contacts, \textit{i.e.} over several jammed states. The fact that during our simulations we found displacements of the same magnitude as the ones reported experimentally is reassuring.

On the other hand, this reference is also important because their experiments managed to probe several of the dynamical features analysed theoretically by Ikeda \textit{et al.} in Ref.~\cite{dynamic_criticality_jamming}. In this latter work, the authors identify a “critical regime” characterized by strong anharmonicities that make a vibrational description unavailable. As I will argue below, our simulations fall precisely within this same critical region. To provide the necessary context, I briefly summarize the results obtained by Ikeda \textit{et al.} that are most important for this thesis. Their main finding is that the changes in the behaviour of the MSD can be associated with corresponding changes in the DOS, $D(\omega)$, where this latter quantity is calculated via the Fourier transform of the velocity autocorrelation function of the particles, $d(t)= \frac{1}{3N'T} \sum_{i=1}^{N'} \avg{\vb{v}_i(t) \vb{v}_i(0)}$ (as above, $N'$ is the number of non-rattlers). For instance, they showed that the time at which the MSD deviates from the ballistic regime can be associated with a frequency for which the rapid decay of the DOS at large $\omega$ occurs. Similarly, for the time at which the MSD reaches its plateau value at high densities, there is a corresponding frequency which indicates the onset of the Debye behaviour for small frequencies, $D(\omega)\sim \omega^{d-1}$. And although this identification is blurred out as the systems move away from their jamming point, say, by raising the temperature or changing the packing fraction, the authors were able to define a region where the harmonic description is always valid. Indeed, they showed that there is a temperature, $T^\star \sim \abs{\vp -\vp_J}^\a$, such that whenever $T<T^\star$, then the collective properties of the dynamics are well described by the vibrational modes. In turn, if for a given $\vp$ the temperature of the system is large in comparison with $T^\star$, then the dynamics is dominated by anharmonic effects and no vibrational description is available and the system is in the so called critical regime.

Importantly, the instances considered in our work lie precisely in this latter regime as I will now show. To begin with, recall that the MC simulations were performed exactly at $\vp=\vp_J$, meaning that for any finite temperature, the dynamics are not be purely vibrational. In fact, Ikeda \textit{et al.} state that in this scenario, the jamming DOS cannot be used to infer properties of the vibrational dynamics for any $T>0$. Additionally, the temperature range that we considered for the harmonic interaction --the only case analysed numerically by the authors-- is several orders of magnitude above $T^\star$. So even if the harmonic description was appropriate at the packing fraction we utilized, all its effects in our simulations would have been washed out due to thermal noise.

The case of the MD simulations using HS is more subtle, but in Ref.~\cite{coulaisHowIdealJamming2014} it was pointed out as well that the experimental configurations analysed in \cite{lechenaultCriticalScalingHeterogeneous2008,lechenaultSuperdiffusionRigidityTransition2010,coulaisDynamicsContactsReveals2012}, which consisted of hard brass disks, also belong to the anharmonic regime. Clearly, these are empirical realizations that closely resemble the infinitely rigid spheres of our simulations, and suggest that our findings with $\vp<\vp_J$ should lay beyond an interpretation in terms of vibrational modes.
Nevertheless, we deem indispensable to stringently test that normal modes are unable to described the single-particle dynamics in the regime considered here, so I will present some results obtained following such method.
I should emphasize that our procedure cannot mimic exactly the one of Ref.~\cite{dynamic_criticality_jamming} because, as mentioned above, the authors used the velocity autocorrelation function in order to compute $D(\omega)$. However, note that $d(t)$ is intrinsically a dynamical property, while our approach relies on exclusively using knowledge about the structure of the configuration. This limitation is, fortunately, only apparent because using the formalism developed in Refs.~\cite{charbonneauJammingCriticalityRevealed2015,lernerLowenergyNonlinearExcitations2013,degiuliForceDistributionAffects2014,lernerBreakdownContinuumElasticity2014} and reproduced in detail in Sec.~\ref{sec:network of contacts}, we can compute the Hessian ($H$) at the jamming point and then obtain the \emph{exact} DOS by diagonalizing it.
I think it is worth remarking that, formally, this is the procedure that should be followed to compute $D(\omega)$, although under certain conditions it has been shown\supercite{henkesExtractingVibrationalModes2012} to coincide with the DOS obtained from the Fourier transform of $d(t)$, which is the technique most commonly utilized.
%

In the case of the harmonic potential, the Hessian is given by $H=\S^T \S$, where $\S$ is the $N_c\times d N'$ contact matrix mentioned above an whose entries are spelled out in Eq.~\eqref{def:S matrix}. As a side note, I bring the attention to the fact that the vibrational formulation is unable to take the rattlers into account, since they are omitted from $\mathcal{S}$ by construction; in contrast, our method incorporates them readily. Now, when dealing with a different type of interaction, universality ensures that a jammed state obtained with a given potential is also a valid jammed configuration for any other potential (see Secs.~\ref{sec:jamming criticality}, \ref{sec:network of contacts} and Refs.~\cite{charbonneauJammingCriticalityRevealed2015,mft_review}). In the case of HS, the entries of the contact matrix should be rescaled by the corresponding force magnitude (see Eq.~\eqref{eq:scaled S matrix}), whence a similar expression for $H$ is recovered, as stated in Eq.~\eqref{eq:Hessian as contact matrix}. Explicitly, its entries are
\begin{equation}\label{eq:H-hard-spheres}
H_{i\alpha}^{j\alpha'}  =
\delta_{ij} \sum_{k\in \partial i} f_{ki}^2 n_{ki}^\alpha n_{ki}^{\alpha'}
- \delta \qty(\ctc{ij}) f_{ij}^2 n_{ij}^\alpha n_{ij}^{\alpha'}
\end{equation}
where $\delta\qty(\ctc{ij})$ is a function that is one only if particles $i$ and $j$ are in contact and zero otherwise.

To obtain the normal modes of the system, we need to obtain the eigenvalues of the Hessian, which in turn determine the normal frequencies of the system and, thus, the DOS: if $\{\lambda_n\}_{n=1}^{3N'}$ are the eigenvalues of $H$, then the associated frequency is $\omega_n=\sqrt{\lambda_n}$. The $3N'$ eigenvectors\footnote{I will also use the notation $\va{\bullet}$ to denote the $3N'$ dimensional vectors of the configuration space, in order to highlight the fact that any displacement $\delta \va{r}$ can be expressed as a linear combination of the eigenvector. Of course, such decomposition assumes that \emph{rattlers are excluded} from the components of $\delta \va{r}$.}, $\{\va{v}_n\}_{n=1}^{3N'}$, correspond precisely to the vibrational modes, each of which describe a unique oscillatory mode of the configuration around the jamming energy minimum. For further use, I introduce here the Inverse Participation Ratio (IPR), which can be calculated in terms of the eigenvectors as\supercite{charbonneauUniversalNonDebyeScaling2016},
\begin{equation}\label{eq:IPR}
	Y(\omega) = \frac{\sum_{i=1}^{N'} \abs{\vb{v}_i(\omega) }^4  }{\qty(\sum_{i=1}^{N'} \abs{\vb{v}_i(\omega)}^2 )^2}
\end{equation}
where $\vb{v}_i(\omega)$ is a $d=3$ dimensional vector obtained from the eigenvector $\va{v}$ with corresponding frequency $\omega$ by taking the components associated with particle $i$. Intuitively, $Y(\omega)$ quantifies the localization of the mode with frequency $\omega$. That is, for very small values ($Y\sim 1/N'$) of the IPR, the associated mode is essentially extended, meaning that many particles are collectively displaced, all of them by roughly the same amount. In contrast, for large values ($Y\sim 1$) few particles participate in the displacement excitation, so the mode is localized.

\begin{figure}[!htb]
	\includegraphics[width=\linewidth]{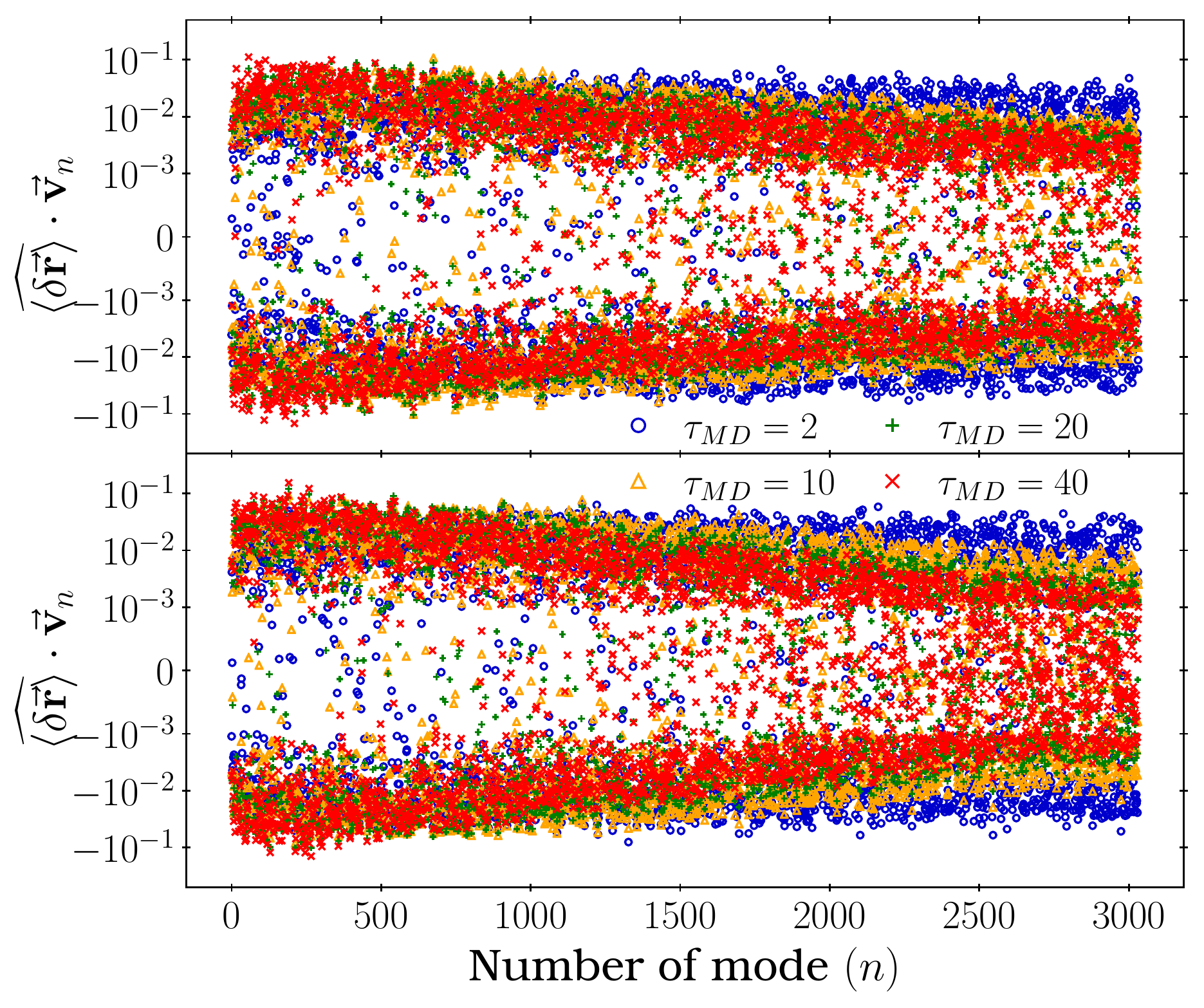}
	\caption[Projection of the single particle mean displacement along each normal mode.]{Value of the projection of $\widehat{ \avg{\delta \va{r}}}(t)$ along each of the normal modes, $\{\va{v}_q\}_{q=1}^{N'}$, obtained from the Hessian using the harmonic (upper panel) or hard-sphere interaction potential (lower panel). The different times at which the dot products were computed are identified by different colours and markers, according to the legend. Values of $\widehat{ \avg{\delta \va{r}}}$ were obtained from the simulations with $\vp/\vp_J=0.995$. It is only for later times that more weight is assigned to few of the low frequency modes, but there is no way of knowing, \textit{a priori}, how to select them.}
	\label{fig:projections-mean-disp-per-mode}
\end{figure}

Now, if the particles were \emph{only vibrating} around an energy minimum their displacements would be easily captured by the normal modes formalism. Moreover, if this was the case, we could expect the lowest energy modes to carry most of the weight of the configuration's displacement. To put to test this assumption, we measured the projections of the configurational vector of mean displacement along the different modes, \textit{i.e.} $\widehat{ \avg{\delta \va{r}}} \cdot \va{v}_n$, where only non-rattlers components are included and $\widehat{\avg{ \delta \va{r}}}$ indicates that the average displacement vector has been normalized. Such normalization is performed  at each time, so the different values of the projections remain comparable to each other and a fair comparison is possible as time evolves.
The resulting values from the MD simulations with $\vp/\vp_J=0.995$ are reported in Fig.~\ref{fig:projections-mean-disp-per-mode}, where the Hessian was obtained with both the harmonic (upper panel) and hard sphere (lower panel) interaction potentials, while different colours correspond to different times at which $\avg{ \delta \va{r}}$ was calculated. The scatter of the points hints that, for either type of potential, the collective particles' displacement is not concentrated in few modes, specially during the initial times. For longer times, more weight is allocated in the first modes, but still there are many other low frequency modes with negligible weight. We consider that such lack of criterion to select the most relevant modes exhibits one important drawback of the normal mode description at the single-particle level, and, unfortunately, it carries over to the IPR as I discuss next.

\begin{figure}[!htb]
	\includegraphics[width=\linewidth]{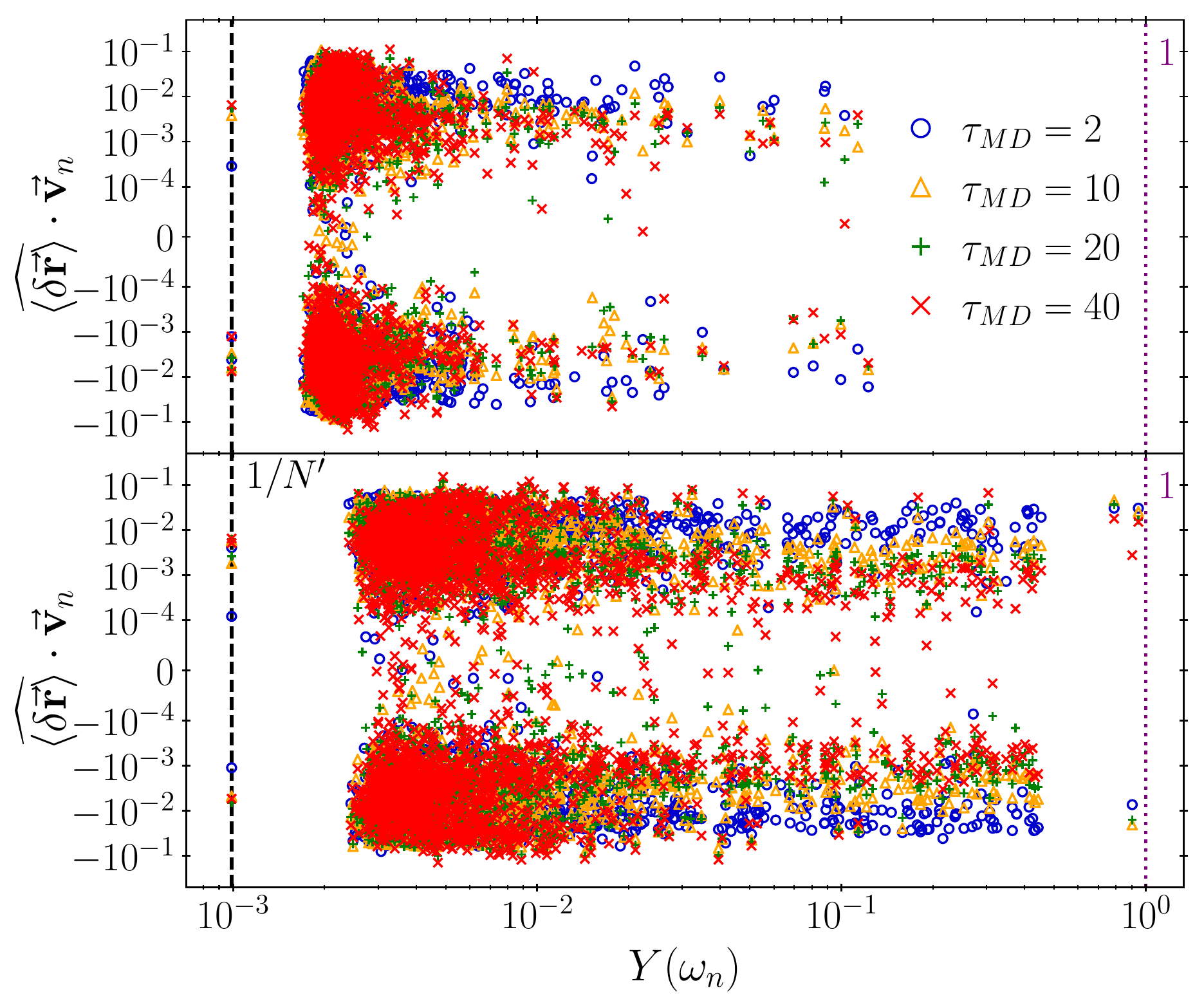}
	\caption[Projection of the single particle mean displacement along each normal mode, as a function of the participation ratio.]{Same projections as in Fig.~\ref{fig:projections-mean-disp-per-mode} but plotted as a function the IPR defined in Eq.~\eqref{eq:IPR}, using the harmonic and hard-sphere potentials, panels (a) and (b) respectively. The black vertical dashed line at $1/N'$ is the lowest value that can be attained by $Y(\omega)$ and correspond to maximally extended modes, while the dotted line at $1$ identifies the most localized ones. The few points with $Y(\omega)\approx 1/N'$ are associated with the $d=3$ zero modes caused by the translational invariance of our systems and their respective weight can be used as a lower threshold for determining when a mode can be considered relevant (see main  text).}
	\label{fig:IPR-vs-projections}
\end{figure}

Considering the IPR is important because, independently of their frequency, we expect that not all the modes are equally important, due to the fact that the trajectories we studied here consist in displacements of all the particles of the system. Hence, a more reasonable property to study is whether the extended modes posses a higher weight than the localized ones. Were this to be the case, then choosing the modes with the lowest value of $Y(\omega)$ could be taken as a criterion for selecting the vibrational modes that influence the dynamics the most. Yet, as Fig.~\ref{fig:IPR-vs-projections} demonstrates, even using the IPR we are unable to identify the correct normal modes to account for the configuration's displacement statistics. That is, even when the IPR is very small, \textit{i.e.} $Y(\omega) \in (2\times 10^{-3}, 10^{-2})$, the corresponding values of the projections lie within a range of 2-3 orders of magnitude. Such a broad distribution exemplifies that the collective dynamics cannot be described as if it consisted of few extended excitations. Note that, on the one hand, these results are independent of the type of potential, while on the other, allowing the dynamics to evolve for longer times only affects the localized modes, the vast majority of which loses most of its weight.
A more quantitative comparison is at hand by considering the three most extended modes (the ones lying on the vertical line $1/N'$) corresponding to the zero modes of $H$, which in turn are associated with the translational invariance of our system induced by its periodic boundary conditions. These modes determine directions along which the system could be collectively displaced without any energy cost, but since we know that during the dynamics the systems is \emph{not} being uniformly translated, we can use their weight as a lower threshold to distinguish relevant modes for the dynamics. What we observe from Fig.~\ref{fig:IPR-vs-projections} is that there are several modes (also considerably extended) that have an even smaller weight, while we still lack a guideline to find the ones more aligned with $\widehat{ \avg{\delta \va{r}}}$. Furthermore, note that the sign of the projection is particularly important if we really want to be able to identify preferential directions. But still, the vibrational scheme fails to provide this information because it is assumed that the motion mainly consists in oscillations around a given minimum, so the direction of motion becomes irrelevant. Nonetheless, our results of the previous sections prove that this is not the case for the dynamical regime we are studying. For completeness, I mention that we verified that our results are unmodified if use $K\qty[\widehat{ \avg{\delta \va{r}} }, \va{v}_n]$ instead of the projection for measuring the correlation between the configurational displacement and the normal modes.

\begin{figure}[!htb]
	\includegraphics[width=\linewidth]{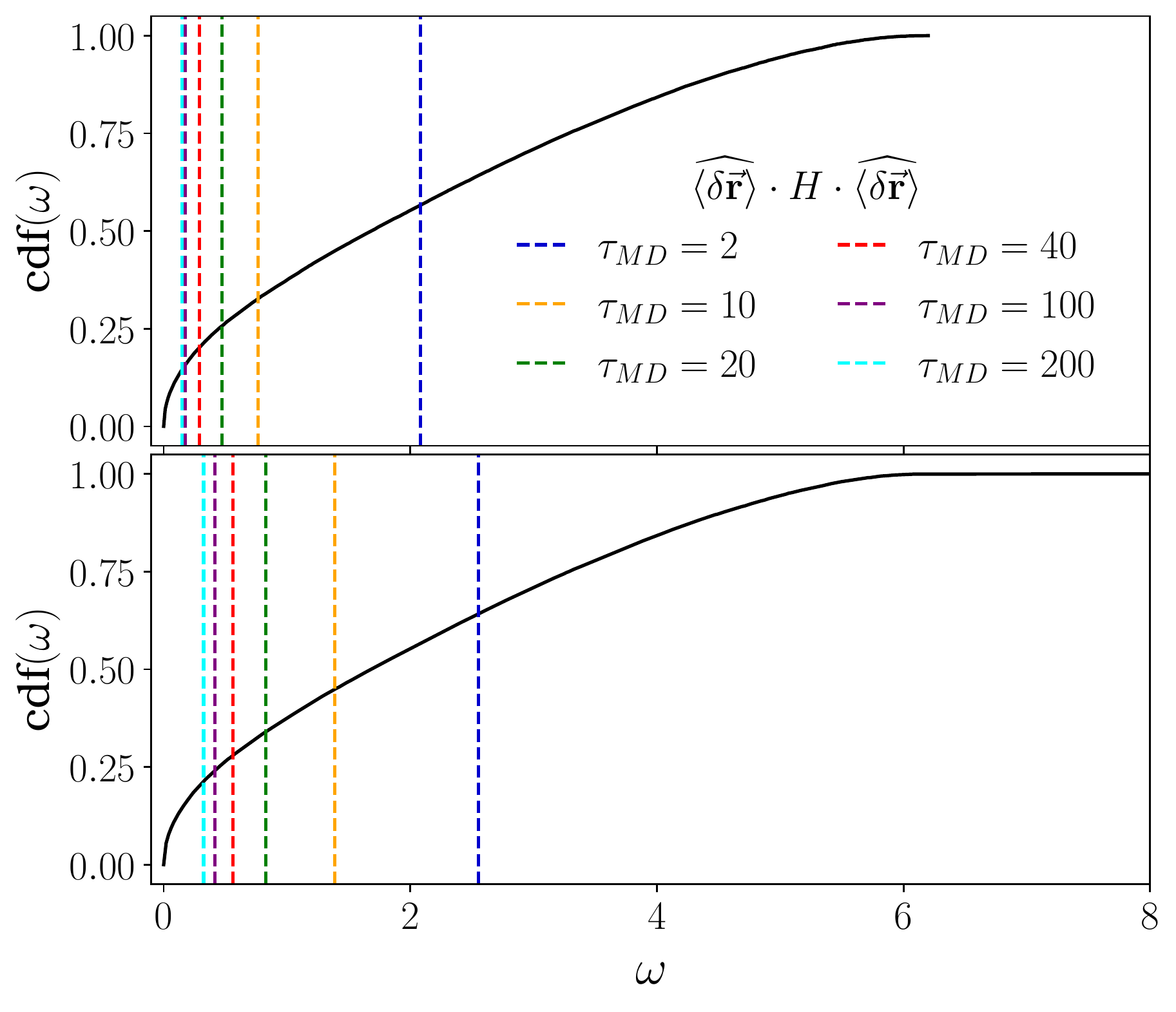}
	\caption[Curvature induced by the Hessian and the comparison with the density of states.]{Comparison of the ``curvature'', measured by $\widehat{\avg{\delta \va{r}}}\cdot H \cdot \widehat{\avg{\delta \va{r}}}$, with the spectrum of $H$ computed using the harmonic (top) and hard-sphere (bottom) interaction. Values of this curvature at different times are indicated by the vertical dashed lines, and their intersection with with cdf$(\omega)$ (black curve) equals the fraction of frequencies smaller than $\omega$. The fact that the fraction of frequencies smaller than a given value of curvature is always extensive, even for long times, reveals that the dynamics probed during our simulations are not captured by the normal modes approach.
	}
	\label{fig:curvature-hessian-cdf-omega}
\end{figure}

As I argued before, the lack of connection between the mean configurational displacement and the distinct vibrational modes can be explained by the fact that these modes are not suitable to examine the trajectories generated with our simulations, given that they belong to the critical region where the harmonic approximation is not valid. And even though the results presented so far support this hypothesis, we have not yet verified that the dynamics is mainly driven by anharmonic effects. To do so, we computed the value of $\widehat{\avg{\delta \va{r}}}\cdot H \cdot \widehat{\avg{\delta \va{r}}}$, which should provide an estimate of the ``curvature''induced by the dynamics at the FEL's minimum. The idea is that if the harmonic description was valid, this curvature should be small given that particles move predominantly along the lower-energy (\textit{i.e.} flatter) directions. Conversely, a large value signals that the forces driving the dynamics are not captured by the linear assumption implicit in the harmonic approximation. To test these hypotheses, in the upper (resp. lower) panel of Fig.~\ref{fig:curvature-hessian-cdf-omega} we plot the cdf of the frequencies for the harmonic (resp. hard-sphere) potential and we indicate the values of $\widehat{\avg{\delta \va{r}}}\cdot H \cdot \widehat{\avg{\delta \va{r}}}$ with vertical dashed lines, using different colours to distinguish different times. Recall we have normalised the average displacement at each time, so the intersection of each vertical line with the curve of cdf$(\omega)$ determines the fraction of frequencies smaller than the corresponding value of the curvature.
Clearly, during the initial times the value of the curvature is quite large, \textit{i.e.} greater than $40-60\%$ of the frequencies, specially considering the hard-sphere potential. But also for longer times this fraction remains around $20\%$ (\textit{i.e.} an extensive number of modes), corroborating that the preferential directions of motion, inferred via the vibrational modes of low energy, do not correspond to the real trajectories followed by the particles. This finding establishes that our systems were in the anharmonic region, according to the classification of Ref.~\cite{dynamic_criticality_jamming}.

\section{Conclusions}\label{sec:conclusions-inferring-dynamics}


I presented a simple and robust method to connect the structural information present at the jamming point of a configuration of $N$ spherical particles and the dynamics that happens close to such point. I first analysed the statistics of the spheres' displacement after producing several trajectories originating from the positions at the jammed state. The resulting distributions show, on the one hand, the existence of preferential directions of motion, and on the other, the presence of heterogeneity in the particles' mobility, reflected by the fact that some spheres remain mostly fixed, while others are more mobile by several orders of magnitude, see Fig.~\ref{fig:pdf displacement}. I also argued that these features are captured by their respective first moments, \textit{i.e.} $\avg{\delta \vb{r}_i}$ and $\avg{\abs{\delta \vb{r}_i}^2}$ for $i=1,\dots,N$, and thus we used these statistics to succinctly represent the full distributions.

The main result is that we found a straightforward and significant connection, at the single particle level, between these statistics and the network of contact vectors that is formed at the jammed state. In particular, we found that the sum of contact vectors acting on a given particle, $\vb{C}_i$ (defined in Eq.~\eqref{def:tot CV}), is a good predictor of its preferred displacement direction, if any; while the sum of dot products between all pairs of contact vectors,  $S_i$ (defined in Eq.~\eqref{def:tot dot CVs}), is highly correlated with the particle's mobility. Our approach was tested using two different types of dynamical protocols, namely Molecular Dynamics and Monte Carlo simulations, and verified that it remains valid in both cases and for the different parameters we studied; cf. Figs.~\ref{fig:MD-displacements-vs-totCV}-\ref{fig:ranking-forces-and-displacements}.
The statistical correlation, measured by the Spearman coefficient, $K$, decays rather slowly as the dynamics evolves, thus showing that the jammed state used as the initial condition of the trajectories has a persistent influence on the particles displacements. More precisely, using the MSD as a measure of how far the configuration is from its initial state, we found that while $K$ decreases its value by an order of magnitude or less, the MSD has increased by at least two decades. Because we are not probing the ballistic regime, the change in the MSD is only due to interaction between particles, and thus implies that $\va{C}$ and $\vec{S}$ are relevant after a small, but definitely measurable, rearrangement of the whole configuration has taken place. Furthermore, in Figs.~\ref{fig:decorrelation-MD} and \ref{fig:decorrelation-MC} I showed that when $\Delta$ is rescaled by a factor that depends on how far from the jamming point the dynamics occurs, the curves of the loss of correlation follow a single master curve. This suggests that for a given dynamical protocol the decorrelation rate is, surprisingly, independent of the distinct parameters used in our simulations. We therefore conclude that the network of contact vectors formed at a jamming point contains significant information to infer the short-time features of the particles' motion near such point.

Importantly, our results show that in the dynamical regime we studied, and for the timescales of our simulations, the particles' motion does not consist of simple vibrations around a reference state, which in our case is naturally identified with the initial jammed configuration. This novel finding is supported by the fact that the vibrational modes fail to capture the real mean displacement of the configuration, as argued in the previous section. The negligible correlations obtained in this case are reported in Figs.~\ref{fig:projections-mean-disp-per-mode} and \ref{fig:IPR-vs-projections}. 
I should emphasize that the scope of our work is restricted to finding a connection between dynamical and structural (and hence, static) variables, therefore we only used data about the contact forces as input for our statistical inference. These data can be used to compute the Hessian at the jamming point, whence the normal modes can be easily obtained. This shows that, utilising the same structural information as input, our approach yields a better description of the statistics of the single-particle dynamics than the normal modes approach. Moreover, the level of statistical correlations we obtained is comparable to the ones derived from more sophisticated, state of the art methods.

Additionally, I want to mention that I am optimistic that our method could be applied, given that it is rather simple and straightforward, to other type of systems having jamming as a critical point (see Sec.~\ref{sec:jamming-transition}). In other words, it is likely that the structural variables we identified here, namely $\va{C}$ and $\vec{S}$, also convey information about the particles trajectories when the dynamics is driven by other physical variables, \textit{e.g.}, the application of a load or a stress. 
A related feature is that because jamming criticality is noteworthy robust to changes in dimensionality and interaction types, our approach presumably has the same range of validity. Besides, recall that the jamming universality class is composed not only by amorphous solids and glass formers but also by several constraint satisfaction problems and learning algorithms. Thus, once the analogous structural variables are identified in those cases, our method could also be used to extract new insights about the \emph{algorithmic} dynamics of that sort of problems.
 
Now, some remarks about the regime of validity of our results are also in order. For this purpose, I will once again follow the picture of the fractal FEL, in particular the results of Ref.~\cite{fel_2014} and Sec.~\ref{sec:LP dependence initial confs}.
There it was shown that a jammed state is realized gradually as the configuration goes down towards a minimum and the contact forces are concomitantly determined. Therefore, similar jammed states exhibit similar network of contacts and, more importantly, different realizations of jammed states, originated from a single initial configuration, would have a significant fraction of common contacts. Of course, the closer this initial configuration is to an energy minimum the higher the fraction of common contacts in the final jammed packings. Now, assuming that the short time dynamics essentially explores a small vicinity around the reference jammed sate, these findings set the Gardner phase as a rough limit beyond which we do not expect our method to remain valid. That is as long as the dynamics departs from a state (in parameter space) within this marginal phase, it is likely that we find non-negligible correlations between our structural variables and the single-particle displacement statistics.
To see why this is the case, let us imagine that we use a jammed configuration as the initial condition of a given dynamical protocol. After a short time we stop the evolution, then bring the configuration to a new jammed state and, finally, we restart the dynamics using this second state as initial condition. Since the two jammed states are similar, we expect that the second trajectory to be correlated, at least to some extent, with the structure of the first jammed state. Yet, note that for this scenario to be true it is crucial that the dynamics occurs inside the Gardner phase because only in this way we can guarantee that the structural variables from the original configuration are going to be similar to the ones of the nearby sub-basins explored during the dynamics. Nevertheless, this argument does not provide a criterion for determining the rate and sharpness of the loss of correlation as the stable glass phase is approached. This and other open questions are described in more detail next.

\subsection{Future work}

I finish by pointing out several routes of improvement that can be considered. The first one is to construct other variables that incorporate information of the ``second nearest neighbours''. In other words, as it is now, our method only considers the role of a particle's first neighbours, but it is expected that a more complete description can be attained when a larger vicinity is considered. An independent possibility to explore is that, with enough data, we could aim to infer the full distribution of the particles displacements and not only their moments as we did here. This would provide a complete description of the trajectories followed by amorphous solids near their jamming point. Clearly, both approaches can be combined, where the degree of statistical agreement of the full distribution expectedly increases as more complicated models are taken into account, which in turn would include observables constructed from an even larger set of nearby particles and more parameters. This opens the way to the application of Bayesian model selection, which has proven to be very fruitful in a surprisingly broad range of topics.

On the other hand, as I stressed through this chapter, we have used a particle-wise approach in which the most relevant structural features are captured by the position of a particle's nearest neighbours. But one could also think on implementing a more ``coarse-grained'' description in which the dynamics of a group of particles (say their mean mobility) is characterized by a structural quantity of the corresponding region (for instance the average value of $S_i$ in the cluster). This clusterised approach would be similar to the more common methods that explore the influence of the local environment on the dynamics of glassy system. Hence we expect that the more we adopt a cluster description, the longer our inferences should remain valid. In this way, one could extend the duration of high correlations by including the information of more neighbours and analysing their statistical properties.

The possibilities I just mentioned are concerned with developing techniques to improve the quality and duration of the statistical inference, but our results also hint at some properties that deserve to be explored, from a physical point of view, with greater detail. First of all, the spatial correlations of $\va{C}$ and $\vec{S}$. That is, based on the fact that particles are dynamically correlated at the large scale, while we have shown that dynamics and local structure are connected, we can anticipated that $\va{C}$ and $\vec{S}$ (or other similar structural variables) should also exhibit signatures of correlations on longer length scales.
Unfortunately, due to high fluctuations in these structural variables, a very large number of independent jammed configurations should be used in order to extract a meaningful signal and thus identify the presence of truly extensive structural correlations. Nonetheless, we deem that putting this hypothesis to test is particularly relevant since it would provide stronger evidence of the influence of the local structure on the dynamics.

Additionally, a promising path for extension, also related to long range correlations, is how the presence of system-wide order or regularity would influence our results. Candidates for such studies could be, for instance, either low density systems --such as tunnelled crystals\supercite{jiaoNonuniversalityDensityDisorder2011}-- or, on the contrary, high density ones --formed by slightly polydisperse crystals\supercite{charbonneauGlassyGardnerlikePhenomenology2019,tsekenisJammingCriticalityNearCrystals2020}. Besides the usual marginal stability of common jammed packings, which leads to structural correlations lengths comparable to the system size (see Chp.~\ref{chp:fss}), these systems are characterised by periodic ordering of their particles. Therefore, two different types of structural correlations coexist across the whole configurations. It is clear that our method can be directly applied to such systems because the network of contacts is equally well defined in all of them, but the extent to which their inherent order increases the relevance of the second nearest neighbours in the dynamics, and thus the level of correlation our approach would yield, is something that needs to be explored.

Another independent feature to investigate is how the dynamics of the configuration in its original basin affects the loss of correlation in time. In other words, it would be interesting to study if there is a connection between this decorrelation and the distance from the original minimum to nearest one at a given state of the configuration.
This would provide evidence that the decorrelation is strongly influenced by the overlap between the many jammed sates in a basin of the FEL and, moreover, it could suggest how to use the structure of these other free energy minima to enhance the quality and the duration of the statistical inferences we presented here or other similar ones. Additionally, if a stronger link between such statistical correlations and the overlap of jammed sates within a meta-basin is found, it could be used to set a more accurate limit for the range of applicability of the sort of inference tools we developed here.

\begin{subappendices}

\section{Reduced analysis on an independent configuration} \label{sec:second-configuration}

\begin{figure*}[!htb]
	\centering
	\includegraphics[width=0.85\linewidth]{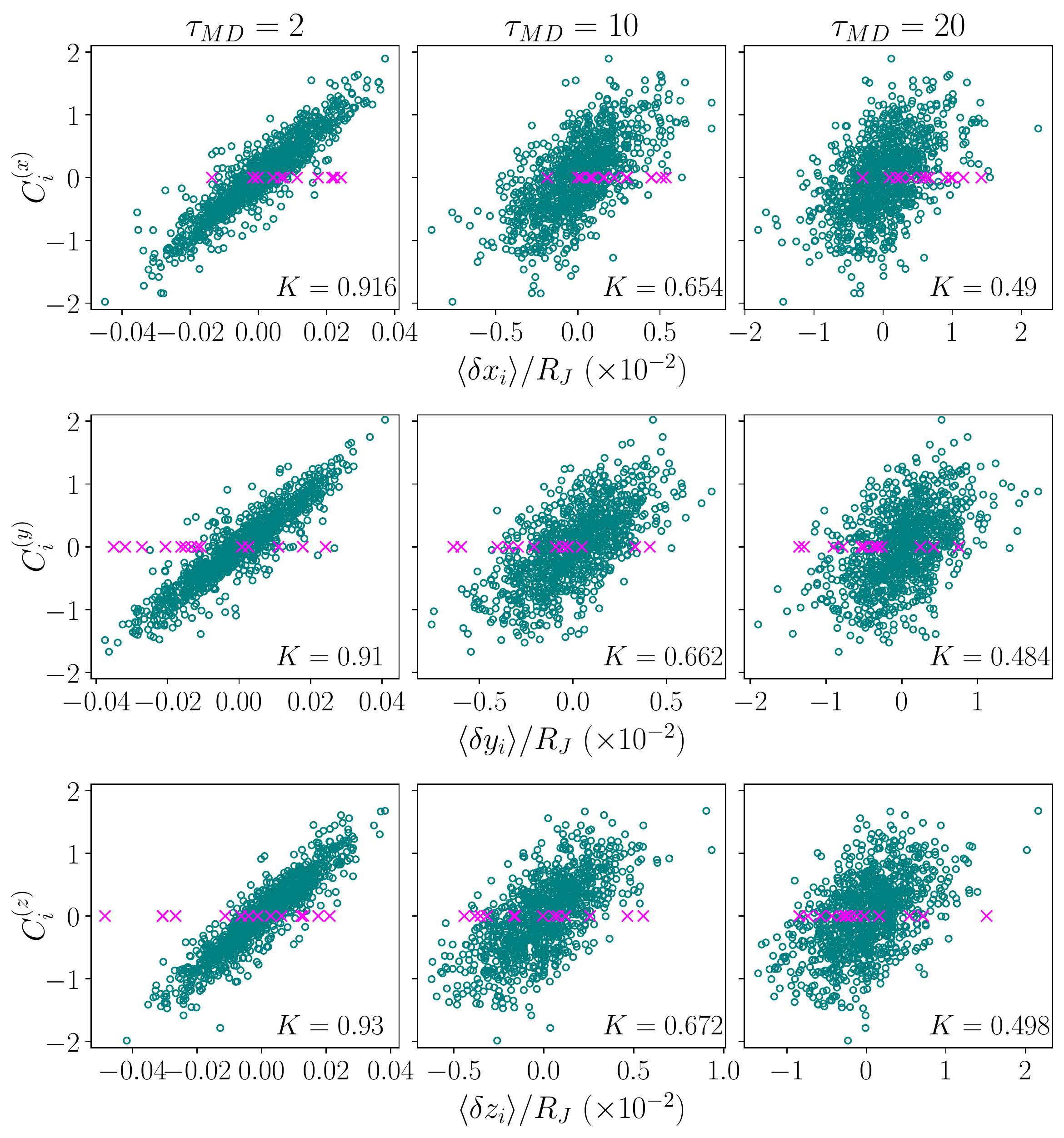}
	\caption[Analogous of Fig.~\ref{fig:MD-displacements-vs-totCV} but for a second, independent configuration.]{Correlations of mean displacements and $\vb{C}_i$ for a second independent configuration. Note the close resemblance to the results reported in Fig.~\ref{fig:MD-displacements-vs-totCV}. The packing fraction used is $\vp/\vp_J=0.995$.}
	\label{fig:MD-displacements-vs-totCV-2nd-config}
\end{figure*}

To verify that our methodology is system independent, we performed the same analysis on another configuration selected at random from the set of configurations analysed in Sec.~\ref{sec:structural variables}. It also consists of $N=1024$, with $1.4\%$ of rattlers, and packing fraction $\vp_J=0.6350$. We used the same MD algorithm to explore its dynamics, although in a reduced set of packing fractions, obtaining very similar results and values of correlations with the two structural variables of interest. For instance, the scatter plots of Fig.~\ref{fig:MD-displacements-vs-totCV-2nd-config} show the correlations between $\avg{\delta \va{r}}$ and $\va{C}$, analogous to Fig.~\ref{fig:MD-displacements-vs-totCV}.
Likewise, we tested the three scalar variables considered in Figs.~\ref{fig:MD-mobility-vs-dotCV} and \ref{fig:MC-mobility-vs-dotCV} and once again found that the value of $\vec{S}$ is a better predictor of the particles mobility. We thus confirm that, compared to contact forces, contact vectors are more useful to infer dynamical statistical properties.

\begin{figure}[!htb]
	\centering
	\includegraphics[width=0.99\linewidth]{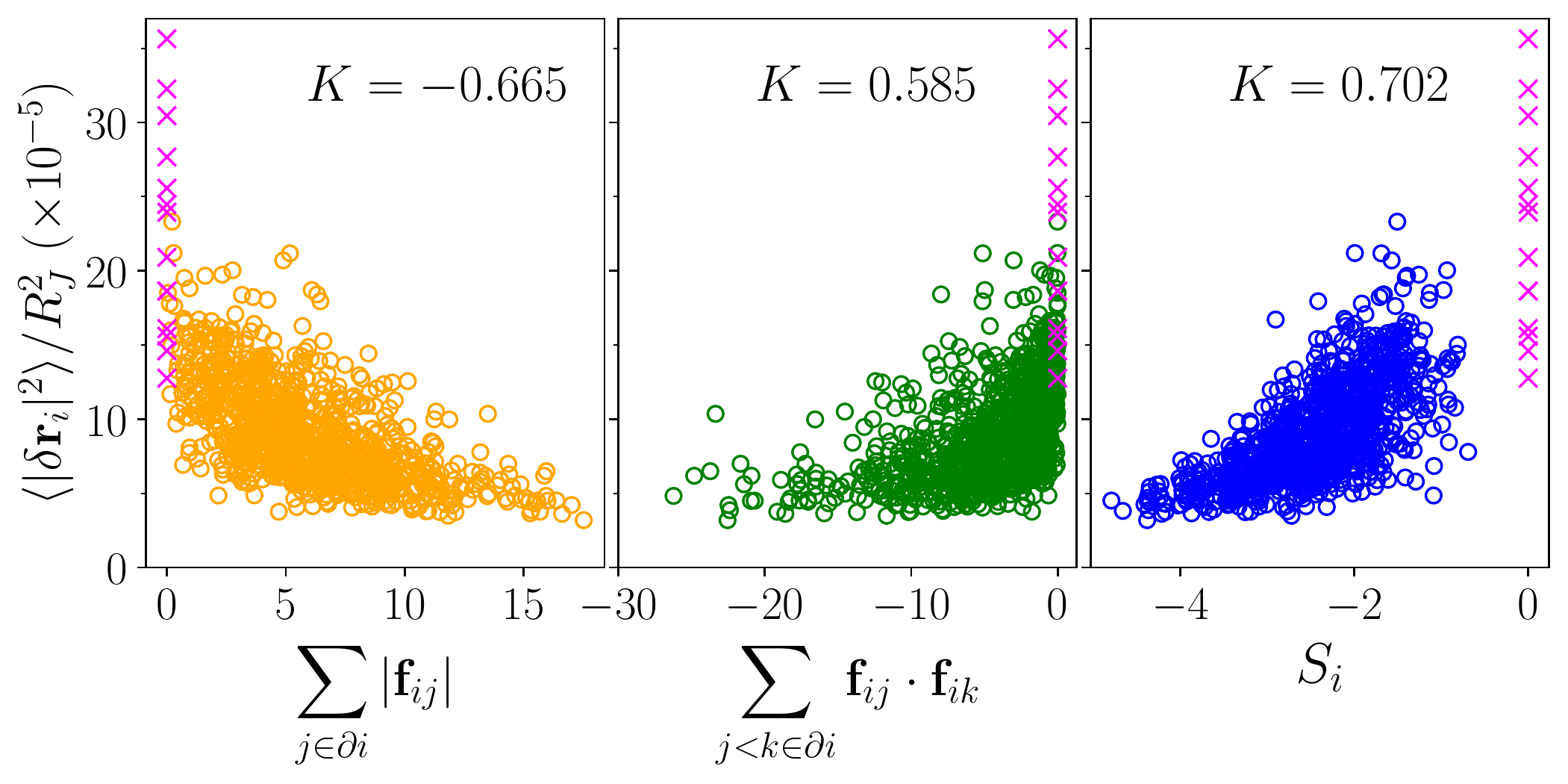}
	\caption[Analogous of Fig.~\ref{fig:MD-mobility-vs-dotCV} but for a second, independent configuration.]{Correlations of particles' mobility and three different structural variables obtained from the network of contact forces. Note that these results reproduce the ones reported in Fig.~\ref{fig:MD-mobility-vs-dotCV}. The packing fraction used is $\vp/\vp_J=0.995$.}
	\label{fig:MD-mobility-vs-dotCV-2nd-config}
\end{figure}

\begin{figure}[!htb]
	\centering
	\includegraphics[width=1.02\linewidth]{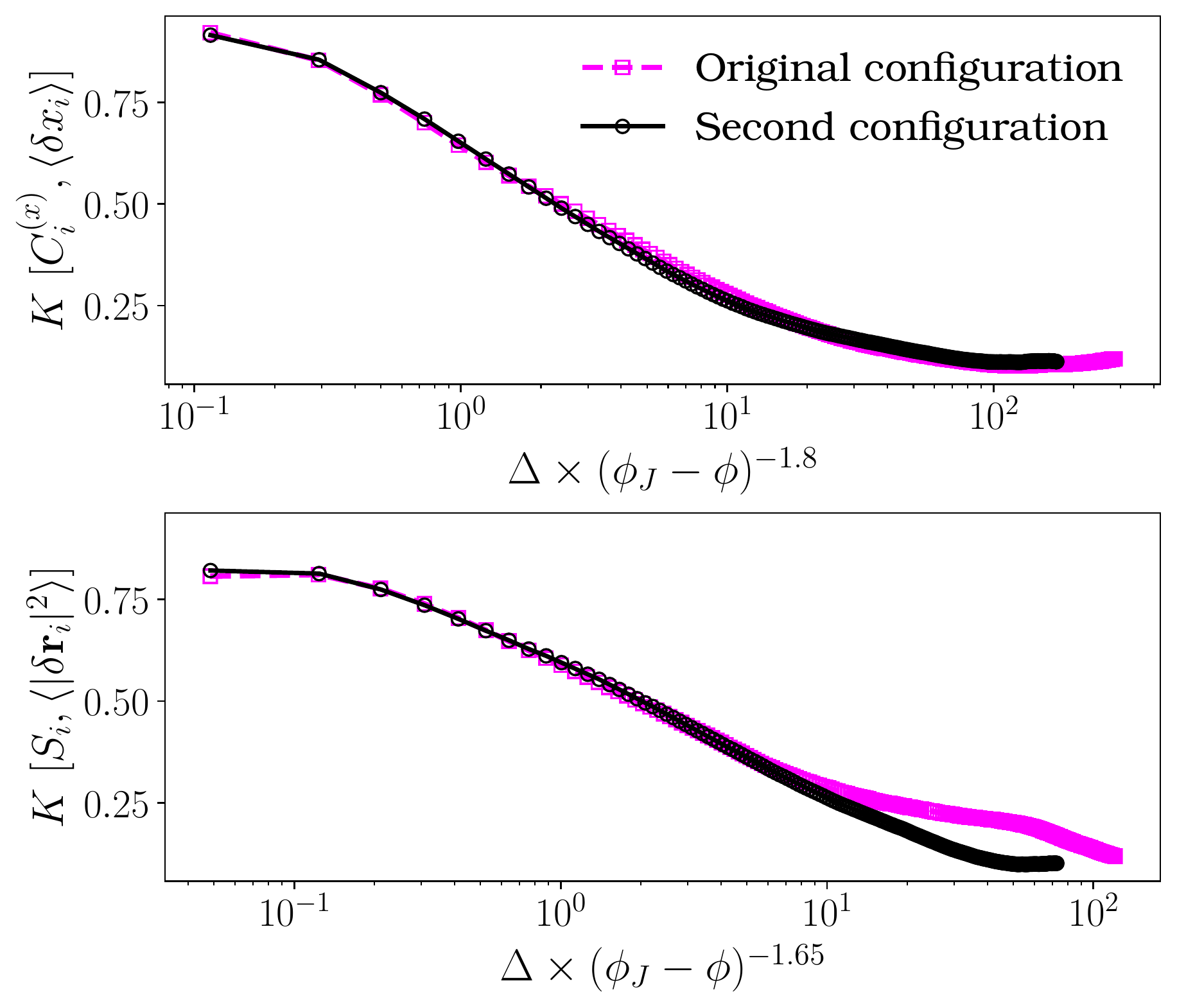}
	\caption{Comparison of the temporal evolution of the statistical correlations between structure and dynamics in two independent configurations.}
	\label{fig:correlations-vs-t-2nd-config}
\end{figure}

Finally, we also verified that the decorrelation from the initial configuration follows the same universal behaviour as the ones reported in Fig.~\ref{fig:decorrelation-MD}. To do so, we computed the values of $K[\avg{\delta \vb{r}_i}, \vb{C}_i]$ and $K[\avg{|\delta \vb{r}_i|^2}, S_i]$ as they evolved in time, and compared them with the ones of the original configuration at the same packing fraction. The comparison is shown in Fig.~\ref{fig:correlations-vs-t-2nd-config}. The fact that the two curves resemble each other very closely shows that our approach is robust enough to be applied to any jammed configuration of spherical particles.

\end{subappendices}

\chapter{Finite size scaling of microscopic structural variables}\label{chp:fss}

\section{Introduction}\label{sec:itroduction-fss}

This chapter contains a detail study of the finite size effects in the critical distributions of contact forces and interparticle gaps; see Secs.~\ref{sec:forces-and-gaps}-\ref{sec:marginal stability} and Eqs.~\eqref{eq:pdf-gaps}-\eqref{eq:pdf-forces} in particular. For convenience, I reproduce here such equations:
\begin{subequations}
	\begin{align*}
	g(h) & \sim h^{-\gamma}, & \text{with } & \gamma = 0.41269\dots \ ;   \tag{\ref{eq:pdf-gaps}}\\ 
	p_\ell(f_\ell) & \sim f^{\theta_\ell} , & \text{with } & \theta_\ell\simeq 0.17\qc  \tag{\ref{eq:pdf-f-loc}}  \\ 
	p_e(f_e) & \sim f_e^{\theta_e} , & \text{with } &  \theta_e = 0.42311\dots \, ; \tag{\ref{eq:pdf-f-ext}}
	\end{align*}
\end{subequations}

Many of the results of this chapter have already been presented in Ref.~\cite{paper-fss}, in collaboration with several colleagues to whom I am deeply grateful. There, we studied said distributions in four models: monodisperse spheres, the Mari--Kurchan (MK) model (see Secs.~\ref{sec:jamming in many systems} and \ref{sec:MD for jamming}) in $3d$, polydisperse disks, and minimally polydisperse FCC crystals. However, here I will only present results from the first two, because those are the ones to which I contributed more actively. Nevertheless, in the case of the monodisperse spheres, I will also make use of the numerical data of our colleagues to exemplify the very nice match obtained when the jamming point is approach from either side of the transition using hard spheres (HS) or soft spheres (SS); see Figs.~\ref{fig:forces-3d} and \ref{fig:gaps-3d} below.
Actually, this last point is but a reason for the need of further testing the non-trivial power law distributions of forces and gaps, despite what has been said in Sec.~\ref{sec:jamming-transition} about the universality and robust criticality of the jamming transition. To explain why, I will briefly come back to some of the points already discussed in that section to give some perspective of how they affect the topics of this chapter. Specifically, a precise determination of the exponents of the distribution of forces, $\{\theta_e, \theta_\ell \}$, and gaps, $\gamma$.

Recall that in Sec.~\ref{sec:jamming criticality} I mentioned that several thermodynamic variables in configurations of spheres exhibit scalings in the vicinity of the jamming point but that, importantly, these relations differ if jamming is approached from the \emph{over-compressed} (OC) phase --see Eqs.~\eqref{eqs:criticality overjammed}-- or from the \emph{under-compressed} (UC) one\footnote{Throughout this chapter, I will only use physics-based terminology. But it is important to keep in mind that many of the arguments considered here also apply from the constraint satisfiability perspective of jamming (see Secs.~\ref{sec:jamming in many systems} and \ref{sec:LP algorithm}). For instance, the UC regime (or when jamming is approached from below) corresponds to the SAT phase, while the OC one (or approaching from above), is equivalent to the UNSAT phase, where at least one constraint is not satisfied.} --Eqs.~\eqref{eqs:criticality underjammed}. The most salient example is the (reduced) pressure that, assuming an harmonic interaction in the OC regime, scales as $p \sim \abs{\vp - \vp_J}^{\pm 1}$, as $\vp \to \vp_J^\pm$, where once again $\vp_J$ is the system's density at jamming. Such power laws behaviours of thermodynamic variables have been amply validated both by numerical results\supercite{ohernJammingZeroTemperature2003,liuJammingTransitionMarginally2010,dynamic_criticality_jamming} and by detailed scaling analyses\supercite{goodrichFiniteSizeScalingJamming2012,goodrichJammingFiniteSystems2014,goodrichScalingAnsatzJamming2016}. Additionally, we now know that these features are essentially unmodified by changes in dimension, polydispersity, etc. And even though they are strictly valid only in the $T\to 0$ limit, the role of temperature has been partially incorporated in some studies\supercite{degiuliTheoryJammingTransition2015,dynamic_criticality_jamming}. This behaviour is not exclusive of thermodynamic variables, but is also observed in quantities such as the coordination number, $z$, for which numerical simulations show that its mean value, $\overline{z} = \frac1N \sum_{i=1}^N z_i$, exhibits a discontinuity exactly as $\phi\to\phi_J^-$, and then grows as $\overline{z}(\phi)-\overline{z}(\phi_J)\sim(\phi-\phi_J)^{1/2}$ for $\phi>\phi_J$\supercite{ohernJammingZeroTemperature2003,charbonneauUniversalMicrostructureMechanical2012}.

Hence, we should not immediately take for granted that the exponents of the critical distributions have the same value if jamming is approached from the high or the low density phase. In fact, to the best of my knowledge, the only work analysing this question in detail is Ref.~\cite{charbonneauUniversalMicrostructureMechanical2012} and they obtained \emph{different} exponents of the forces distributions in the OC regime ($\theta_{SS} \approx 0.42$) and in the UC one ($\theta_{HS} \approx 0.28$). The situation was further complicated because shortly afterwards, the authors of Ref.~\cite{degiuliForceDistributionAffects2014} estimated $\theta_{SS} \approx 0.17$ after improving on the statistics (although they also reported $\theta'_{HS} \approx 0.23$ in $3d$ monodisperse packings). Besides, some dependence on the dimensionality has also been observed\supercite{lernerLowenergyNonlinearExcitations2013,degiuliForceDistributionAffects2014}.  All these works however reported a value of $\gamma \approx 0.4$.
To add even more confusion, the mean-field (MF) theory --developed almost simultaneously\supercite{mft_exact_1,mft_exact_2,mft_exact_3,fel_2014,parisi_zamponi_2010}-- predicted \emph{exactly}, in the limit of very high dimensions, the values $\gamma=0.41269$ and $\theta=0.42311$, as stated in Eqs.~\eqref{eq:pdf-gaps} and \eqref{eq:pdf-forces}. So how to reconcile all the numerical results between themselves and explain the occasional agreement with MF theory?  Fortunately, an elegant way out of this dilemma was soon devised\supercite{degiuliForceDistributionAffects2014,charbonneauJammingCriticalityRevealed2015}: the MF predictions are correct also for low dimensional systems, and the apparent contradictions observed in the value of $\theta$ are caused by the presence of two different types of contact forces. The difference is that opening a contact associated to one or other type gives rise to either an extended or to a localized excitation; see Fig.~\ref{fig:floppy-modes}. (For brevity, I will henceforth use “localized” and “extended” to distinguish the corresponding contact \emph{forces}, although they obviously are only present at the contact point.) When the two contributions are considered separately, it is \emph{consistently} found that localized (extended) forces are distributed according to $\theta_\ell \approx 0.17$ ($\theta_e\approx 0.42$) for any dimension $d\geq 2$. However, the relative fraction of localized and extended modes strongly depends on the dimensionality and, possibly, also on the preparation protocol (and maybe even on polydispersity in $d=3$). This is caused by the fact that localized forces are predominantly associated with buckling excitations which occur, with very high probability, in particles with $z_\ell \equiv d+1$ contacts. The reason is that buckling forces occur mainly because $d$ contacts are approximately coplanar and thus, the remaining force should be almost orthogonal and of small magnitude. Note that as $z_i$ increases it is ever more unlikely to achieve a similar nearly coplanar arrangement, and therefore the vast majority of bucklers have $z_\ell$ contacts; see details in \cite[SI. II]{charbonneauJammingCriticalityRevealed2015}.
Importantly, the amount of buckling particles decreases very rapidly with dimension, because fluctuations away from the isostatic value  $z_c=2d$ are exponentially suppressed\supercite{charbonneauJammingCriticalityRevealed2015}. Putting all of these arguments together we can satisfactorily explain, on the one hand, why MF is unable to provide a description of localized modes (namely, because they are an exclusive low $d$ phenomenon) and, on the other, the discrepancy of values of $\theta$. In particular, the results of Ref.~\cite{charbonneauJammingCriticalityRevealed2015} reveal that in $d=2$ localized forces are dominant, an hence if the two forces are considered together it is found that $\theta\approx 0.19$; in $d=3$, the same analysis yields $\theta\approx 0.25$ (cf. $\theta_{HS}$ and $\theta_{HS}'$ above), and $\theta\approx 0.30$ in $d=4$. But these dissimilar values are actually simply caused by the dependence of the amount of bucklers with the dimension. Even more, when localized and extended forces are considered separately, in all these cases their probability function (pdf) follow very closely  $p(f_\ell)\sim f_\ell^{\theta_\ell}$ and $p(f_e)\sim f_e^{\theta_e}$, where $\theta_\ell=0.17462$ and $\theta_e=0.42311$, in agreement with the MF prediction for the latter.


Unfortunately, once this matter was solved, no further test was made to carefully verify that the three values $(\gamma, \theta_e, \theta_\ell)$ match as $\vp \to \vp_J^\pm$. That is, the available numerical evidence strongly supports that this is the case but, as with many other phase transitions, there is no reason \textit{a priori} to suppose that the jamming criticality is unaffected by the direction of the limit. Moreover, a precise estimation of their value is needed because the three exponents are related by mechanical stability bounds as derived in Sec.~\ref{sec:marginal stability}. MF predictions of $\theta_e$ and $\gamma$ saturate the corresponding bound, Eq.~\eqref{eq:gamma_theta_ext}, and current estimates of $\theta_\ell$ indicate that the same is true for the other relation, Eq.~\eqref{eq:gamma_theta_loc}. Hence, it is fundamental  to perform a careful analysis of the values of $(\gamma, \theta_e, \theta_\ell)$ inferred from numerical simulations in order to assess the stability, if only marginal, of jammed configurations.
Conducting such an analysis is especially important considering that
packings of slightly polydisperse crystals are reported to exhibit a microstructure characterized by exponents that significantly differ from those of Eqs.~\eqref{eq:pdf-gaps} and \eqref{eq:pdf-forces}\supercite{charbonneauGlassyGardnerlikePhenomenology2019,tsekenisJammingCriticalityNearCrystals2020}. Additionally, recent works have shown that many of the salient features of spherical packings depend sensitively on particle shape. For instance, introducing even an infinitesimal amount of asphericity changes the universality class\supercite{ikedaInfinitesimalAsphericityChanges2020,britoUniversalityJammingNonspherical2018b}, in which the isostatic condition no longer holds. The relevance of investigating the extent of the jamming universality class and thoroughly testing its many theoretical predictions is therefore needed\supercite{berthierGardnerPhysicsAmorphous2019}, specially considering that jamming criticality is, up to date, the most precise prediction reaped from the replica method applied to realistic materials models.

With this in mind, in this chapter I will present a detailed analysis of the finite-size scaling of the distributions of interparticle gaps and contact forces. These distributions are one of the fundamental consequences of the presumed non-trivial critical behaviour of jammed packings, hence their testing is a key step toward rigorously validating a whole set of critical properties. Although a similar analysis has been carried out for the perceptron\supercite{kallusScalingCollapseJamming2016} and for the gaps distribution of a two-dimensional binary mixture\supercite{ikedaInfinitesimalAsphericityChanges2020}, no systematic result exists for jammed packings of spherical particles, nor for amorphous packings with other sources of disorder. As I mentioned above, in Ref.~\cite{paper-fss} we considered several models, both in $d=2$ and $d=3$, with the aim to define precisely which are the most robust features of jamming criticality, and thus better demarcate its physical universality. For instance, to properly account for size effects in $2d$ systems logarithmic corrections\supercite{goodrichJammingFiniteSystems2014,wangMonteCarloStudy1993,ruiz-lorenzoLogarithmicCorrectionsSpin1998,kennaFiniteSizeScaling2004} are important, thus confirming that $d=2$ is the upper critical dimension\supercite{charbonneauJammingCriticalityRevealed2015,lernerLowenergyNonlinearExcitations2013,goodrichJammingFiniteSystems2014}. (Further evidence comes from recent studies\supercite{ikedaJammingUpperCritical2020,zhangMarginallyJammedStates2020} where a different criticality is observed in quasi-one-dimensional systems.)
Similarly, we also found that crystalline ordering breaks the jamming universality and renders jammed configurations unstable according to the criterion of the usual bounds, Eqs.~\eqref{eq:marginality-bounds}.
Nevertheless, I will not consider such systems here an instead only focus in $3d$ systems of monodisperse spheres, and the analogous MK model.

This chapter is organized as follows, in Sec.~\ref{sec:methods} I provide some details about the generation of jammed packings and in Sec.~\ref{sec:fss-theory} I explain how to study the size effects we are interested id. Then in Sec.~\ref{sec:spheres-fss} I present a meticulous analysis of the finite size effects in the critical distributions of packings of spherical particles. An unexpected outcome is the striking contrast of such effects on the distributions $f_e$ and $h$. Then, in Sec.~\ref{sec:MK-fss} I carry out a similar study for the MK configurations, which shows that size effects are much pronounced in this model and that localized forces exhibit a different distribution. A tentative explanation is advanced for both features in terms of the particular properties of the MK model. Finally, a discussion and brief conclusion are given in Sec.~\ref{sec:discussion}.


\section{Methods for generating jammed packings}\label{sec:methods}

I will first consider the case of three-dimensional configurations of $N$ spheres of equal diameter $\sigma$ in a cubic box under periodic boundary conditions. In a certain sense, this choice is the minimal model for producing jammed packings. Lower-dimensionality systems inevitably crystallize unless polydisperse mixtures are used, but ordering can be avoided for monodisperse spheres in $d\geq3$. Sphere positions then serve as the only source of disorder. For configurations initially in the UC regime, a hard-sphere potential is used and a combination of molecular dynamics (MD) and linear optimization algorithms are employed to reach $\phi_J$. 
More precisely, we start from a low-density configuration ($\vp = 0.2$) with particles' position assigned randomly and uniformly in the box volume. Then, we use event-driven MD with a Lubachevsky--Stillinger growth protocol\supercite{md-code} (see Sec.~\ref{sec:MD for jamming} for details) to increase the (reduced) pressure up to $p=500$. This first step is performed with a fast compression rate ($\dot{\sigma}=10^{-2}$) in order to avoid any partial crystallization and is then followed by a second, much slower ($\dot{\sigma}\leq 10^{-3}$), growth protocol until $p \gtrsim 10^7$. (Details about the influence of $\dot{\sigma}$ and the final $p$ on the jammed configuration thus produced are reported in Sec.~\ref{sec:LP dependence initial confs}.) At this point the configuration is used as input for the iterative Linear Programming (iLP) algorithm described in Sec.~\ref{sec:LP algorithm}. As discussed there, at each step, the LP algorithm finds the optimal rearrangement of particles that  allows to maximize their radius, considering a linearised version of the non-overlapping constraint between all pairs of particles. Upon convergence, this algorithm produces a jammed configuration, because neither particle displacements nor size increases are possible. With this we can easily reconstruct the full network of contacts at jamming, because genuine contact forces can be extracted from the active dual variables associated to the non-overlapping constraints. As I proved in Sec.~\ref{sec:LP network of contacts}, our iLP algorithm always produces a mechanically stable configuration and, the majority of times, it has a single state of self-stress (1SS). As discussed in Sec.~\ref{sec:network of contacts}, such a state is characterized for having one contact above isostaticity, \textit{i.e.} when the total number of constraints in a system ($N_c$) matches its number of degrees of freedom ($N_{dof}$), and determines where the jamming criticality comes about. The extra contact with respect to isostaticity is required in order to achieve a finite bulk modulus \supercite{haghBroaderViewJamming2019,goodrichFiniteSizeScalingJamming2012}. Put differently, the system density is an additional variable that needs to be fixed, and thus requires one additional contact above isostaticity\supercite{donevPairCorrelationFunction2005} (Sec.~\ref{sec:LP network of contacts}).

On the other hand, when in the OC phase\footnote{All the simulations in the OC regime were performed by our collaborators, so I just describe them briefly.} and given the position vectors, $ \{\vb{r}_i\}_{i=1}^N $, we considered an harmonic contact potential,
\begin{equation}
U\qty( \{\vb{r}_i\}_{i=1}^N ) = \frac{\epsilon}{2} \sum_{i,j}  (\sigma - \abs{\vb{r}_i - \vb{r}_j})^2 \ \Theta \qty( \sigma - \abs{\vb{r}_i - \vb{r}_j}  ) \qc
\end{equation}
where $\epsilon$ is a constant that defines the energy scale, and $\Theta$ is the Heaviside step function. Hence, a pair of particles only interacts if there is an overlap between them. Starting with $\vp>\vp_J$ and a Poisson random distribution of spheres, a series of energy minimization steps and packing fraction reduction steps are performed until the system has 1SS. More precisely, at a given density the FIRE algorithm\supercite{FIRE} is used to achieve force balance in the configuration. The energy of the configuration is then calculated and the known scaling relation\supercite{charbonneauJammingCriticalityRevealed2015}, $U\propto \left(\phi-\phi_J\right)^2$, is used to determine by how much the sphere radii should be uniformly decreased to reduce the system energy by a fixed fraction. Thus, after several iterations of this procedure, the packing has precisely $N_c=N_{dof}+1\equiv N_{1SS}$ contacts. 
After removing rattlers (see below), the dynamical matrix\supercite{vanheckeReviewJammingSoftParticles2010} is used to ensure that the packing is jammed. This algorithm is implemented in the pyCudaPacking software using general purpose graphical processing units and quad-precision calculations, see details in Refs.~\cite{morseGeometricSignaturesJamming2014, charbonneauUniversalNonDebyeScaling2016, morseEchoesGlassTransition2017}.

Once the contact forces have been computed for a given configuration, they are normalized so that their mean value equals 1. We then proceeded to classify them as localized or extended. The criterion we used is based on the coordination number of each particle, $z_i$. Thus, if a particle has $z_\ell=4$ contacts, it is automatically considered a buckler and all its contact forces are classified as localized. In turn, if a particle has $z_i>z_\ell$ all the related forces, except the ones possibly associated with bucklers, are classified as extended. Clearly, this criterion is far from perfect, because (i) not all particles with $z_\ell$ neighbours necessarily have a buckling contact and (ii) not all bucklers have $z_\ell$ contacts. However, the fact that a buckling force is associated with a specific geometric configuration as discussed above and that relatively few particles have less than $z_c=6$ contacts justifies that by doing so, we are capturing most of the relevant localized modes. Of course, more accurate methods have been devised\supercite{lernerLowenergyNonlinearExcitations2013,degiuliForceDistributionAffects2014,charbonneauGlassyGardnerlikePhenomenology2019}, but as shown in Ref.~\cite{charbonneauJammingCriticalityRevealed2015} this criterion often suffices. Besides, another way to extract the distribution of localized and extended forces is to consider the smallest and median force \emph{per particle}. When this is done, it is found that $p(f_{small}) \sim f_{small}^{\theta_\ell}$ and $p(f_{med})\sim f_{med}^{\theta_e}$. We verified that our results hold with either methodology, so I will only present the data obtained from the connectivity criterion.

On the other hand, computing the dimensionless gaps, $h$ (defined in Eq.~\eqref{def:gaps}), in each configuration is straightforward once the jamming point has been reached: it is enough to compute the distance between \emph{non}-touching particles, rescale it by $\sigma_J$ and subtract 1. However, given that the jamming power-law only applies for small gaps, it is convenient to introduce a cut-off to the distances considered. Here, we set it to $2\sigma_J$, which amounts to $h_{max}=1$, a value well beyond the critical regime (see Figs.~\ref{fig:gaps-3d} and \ref{fig:gaps-MK}). As a side comment, note that when the iLP algorithm is employed, another advantage of using active dual variables to define contacts between particle pairs is that such pairs can be directly excluded from the list of distances to be considered.

Finally, for the MK model we restricted our analysis to configurations of HS, and hence, approaching jamming from below. As described in previous chapters, this model is characterized by particles interacting through a randomly shifted distance, as in Eq.~\eqref{eq:MK distance}. For our simulations, we assumed the random shifts to be quenched vectors uniformly distributed in space, corresponding to the MF limit. To generate the packings we followed essentially the same procedure relying on the MD+iLP algorithm. With respect to HS packings, there were only a couple of differences in the MD compression part of MK (the actual protocol is illustrated in Fig.~\ref{fig:MK-liquid-and-glass}). First, that a single value of $10^{-3} \leq \dot{\sigma} \leq 2\times 10^{-3}$ was used, because there is no risk of crystallization in this model, so there is no need to begin with a fast compression. Second, the initial configuration for the dynamics was a planted configuration with density $\vp=2.5$. In this way, we drastically reduced the time of the MD simulations and guaranteed to go beyond the dynamical transition point\supercite{charbonneauNumericalDetectionGardner2015,charbonneauHoppingStokesEinstein2014} (see Fig.~\ref{fig:MK-liquid-and-glass}). However, I should mention that a disadvantage of doing so is that the fraction of rattler increases considerably to about $15\%-20\%$. Nevertheless, we always obtained a configuration having 1SS with the help our iLP algorithm. As a final remark I mention that generating a jammed packing of $N$ particles following this procedure takes considerably longer than for a usual HS configuration of the same size. This is due to the fact that any given particle is surrounded by many more particles than in the case with Euclidean distance. Besides, the neighbours-list approach\supercite{donevNeighborListCollisiondriven2005} for the MD compression is also significantly slower than the cell-based one\supercite{md-code}; more details are given in Sec.~\ref{sec:MD for jamming}.

\section{Methods for studying finite size effects}\label{sec:fss-theory}

We used the methodology just described to produce several jammed configurations for each system size, parametrized by $N$. For systems of spheres with the usual Euclidean (resp. MK) distance, we tried $N\in \{2^8, 2^{14}\}$ (resp. $N\in \{2^8, 2^{12}\}$). Moreover, to ensure that we sampled all the systems with the same accuracy, for a given value of $N$, $M_N$ independent packings were produced in such a way that  $N\times M_N \simeq 2.2\times 10^6$ ($5.5\times 10^6$) for standard HS (SS) configurations, while $N\times M_N \simeq 10^6$ for the MK ones. The reason of the smaller sizes and reduced dataset in this latter model is that both the MD and iLP algorithms perform much slower than in the common case as mentioned above.
These data provide enough accuracy to study the forces and gaps across many orders of magnitude, and finite-size corrections can be systematically identified. 

It is known that using a logarithmic binning when testing for power-laws in a pdf often leads to poor comparisons as a consequence of the loss of resolution when grouping enough data in a single bin\supercite{newmanPowerLawsPareto2005}. To tackle this difficulty, we instead utilized the cumulative distribution function (cdf). Note that if a random variable, $x$, is distributed according to a pdf of the form $\rho(x) \sim x^\alpha$ for $\alpha > -1$, then its cdf follows $c(x) \sim x^{1+\alpha}$. The cdf allows the accuracy of the data acquired to be preserved and, in the case of independent random variables, maximum likelihood methods can be used to obtain a more precise estimation of $\alpha$ than by least-squares fit\supercite{newmanPowerLawsPareto2005}. However, for the cases I consider here, a size scaling analysis leads to better results (see discussion in Sec.~\ref{sec:fss gaps MK}).

To empirically estimate the cumulative distributions, we opted to put together the data of all the $M_N$ samples for each system size. Doing so improves the statistics of the smallest values of the structural variables we investigate here, so we are able to probe with some detail the left tail of all the distributions. Such analysis is an important part of our results as I describe below. However, given the $M_N$ independent samples for each system size, this is not the only way in which one could estimate the corresponding cdf. Nevertheless in Appendix \ref{sec:equivalence-cdfs} I show that this approach produces essentially the same results than when the cdf of single samples are considered.

Now, when fitting a distribution to empirical data it should be considered that, even if $x$ strictly follows such a distribution all the way down to $x= 0$, finite sampling inevitably leads to deviations. Here, the situation is further complicated by our consideration of marginals of \emph{correlated variables}. Gaps and forces distributions of finite $N$ configurations are indeed prone to exhibit deviations from their expected form due to both finite sampling and system-wide correlations. As I will argue in the remaining part of this chapter, introducing a scaling function (as is usually done in the study of critical phenomena\supercite{amitFieldTheoryRenormalization2005,MC_book}) can account for both effects, allowing the $N$ dependence of the cdf to be carefully analysed.
To derive the size scaling of the distributions of $x$, we first note that in a sample of size $N \gg 1$, we can estimate the order of the smallest value observed in the data, $x_\text{min}$ from the probability mass assigned to the extremes of the distribution:
\begin{equation}\label{eq:scaling-xmin}
\int_0^{x_\text{min}} \rho(x) \dd{x} \sim x_\text{min}^{1+\alpha} \sim \frac{1}{N}.
\end{equation}
It then follows that $x_\text{min} \sim N^{-1/(1+\alpha)}$. Note that strictly speaking in this last equation $N$ should be replaced by $N_c$ when analysing, for instance, the distribution of contact forces. However, given that $N_c \sim d N$ and that we are mostly concerned with the scaling exponent, we can safely neglect the associated proportionality constants. The behaviour of the gaps distribution is expected to be similar, in the sense that particles almost in contact should be self-averaging. Following the traditional path for analysing size scalings, we can write the pdf as
\begin{equation}\label{eq:scaling-p}
\rho(x) \sim N^\beta \tilde{\rho}\qty(x N^{\frac1{1+\alpha}} )
\end{equation}
where $\tilde{\rho}$ is the scaling function of the pdf such that $\tilde{\rho}(x) \sim x^\alpha$ for $x \gtrsim 1$. The exponent $\beta$ can be easily determined by requiring that $\rho(x)$ exhibit no $N$ dependence. We thus get that $\beta = -\frac{\alpha}{1+\alpha}$, whence the expressions used for the scalings studied in Ref.~\cite{kallusScalingCollapseJamming2016} are recovered. For the cumulative distributions, repeating the above analysis for $c(x) \sim N^{-\delta} \tilde{c}\qty(x N^{\frac1{1+\alpha}} )$ gives $\tilde{c}(x) \sim x^{1+\alpha}$, and it immediately follows that $\delta=1$, whence the relevant scaling relation is
\begin{equation}\label{eq:scaling-cdf}
c(x) \sim N^{-1}\  \tilde{c} \qty(x N^{\frac{1}{1+\alpha}} )  .
\end{equation}
Using the correct $\alpha$ should remove any dependence on $N$ in this last equation. That is, data for different system sizes can be rescaled in such a way that they follow a common curve, $\tilde{c}$. Finding a good collapse of the curves for different $N$ thus indicates that deviations from the expected algebraic behaviour are caused by our systems necessarily being off the thermodynamic limit, but not because variables follow a different power-law scaling. Additionally, showing that the system size influences the cdf of a given variable strongly evinces that such a variable is correlated across the whole system. Hence, an upper bound to the correlation length can then be estimated.

Notice that Eq.~\eqref{eq:scaling-xmin} suggests another way to estimate the value of $\alpha$ from the datasets obtained from different system sizes. And in fact, this was the method employed in Ref.~\cite{kallusScalingCollapseJamming2016} for a similar study in the perceptron. The idea is the following. For a fixed $N$, instead of computing the cdf of a given variable by putting together the data of the $M_N$ samples, the typical $x_{min}$ is approximated by its sample average, that is,
\begin{equation}\label{eq:scaling-avg-xmin}
\overline{x_{\min}} \equiv \frac{1}{M} \sum_{m=1}^M x_{\min}^{(m)} \sim N^{-\frac1{1+\a}} \qc
\end{equation}
where $x_{min}^{(m)}$ is the smallest value of the random variable $x$ in the $m$-th sample. In practice, this expression can be tested by showing that $N^{\frac1{1+\a}}\ \overline{x_{\min}} \approx $ constant for different values of $N$. Unfortunately, this method cannot be used in the cases we study here due to corrections to Eq.~\eqref{eq:scaling-p} for very small values, as I now describe. 

Given that our jammed configurations have one extra contact than $N_{dof}$ (see Secs.~\ref{sec:network of contacts}, \ref{sec:LP network of contacts}), theoretical predictions\supercite{ikedaMeanFieldTheory2019,franzUniversalitySATUNSATJamming2017,ikedaInfinitesimalAsphericityChanges2020} establish that the microstructural critical variables should exhibit a different scaling in their left tails. For instance, MF theory  predicts that interparticle gaps are distributed as $h^{-\gamma}$ only for values larger than a cut-off $h^\star \sim \delta z^\frac{1}{1-\gamma}$, where $\delta z$ is the excess of contacts in a system. In our case, $\delta z \sim 1/N$, so instead of Eq.~\eqref{eq:pdf-gaps} the pdf describing the distribution of $h$ reads,
\begin{equation}\label{eq:pdf-gaps-v2}
    g(h) \sim \begin{dcases}
    N^{\frac{\gamma}{1-\gamma}}\ g_0\qty(h N^{\frac{1}{1-\gamma}}), & h N^{\frac{1}{1-\gamma}} \ll 1 \\
    h^{-\gamma}, & h N^{\frac{1}{1-\gamma}} \gtrsim 1
     \end{dcases} \, ;
\end{equation}
where $g_0(x)\sim 1$ for $x\ll 1$\supercite{ikedaInfinitesimalAsphericityChanges2020}. Analogously, for extended forces Eq.~\eqref{eq:pdf-f-ext} should be replaced by
\begin{equation}\label{eq:pdf-f-ext-v2}
    p(f) \sim \begin{dcases}
    N^{\frac{-\theta_e}{1+\theta_e}}\ p_0\qty(f N^{\frac{1}{1+\theta_e}}), & f N^{\frac{1}{1+\theta_e}} \ll 1 \\
    f^{\theta_e}, & f N^{\frac{1}{1+\theta_e}} \gtrsim 1
     \end{dcases} \, ;
\end{equation}
where $p_0(x) \sim 1$ for very small values is to be expected.
Equations~\eqref{eq:pdf-gaps-v2} and \eqref{eq:pdf-f-ext-v2} are indeed consistent with Eq.~\eqref{eq:scaling-p},
and, repeating the same arguments as above, it is straightforward to derive that both regimes can be captured by Eq.~\eqref{eq:scaling-cdf} using a single scaling function, such that
\begin{equation}\label{eq:general scaling cdf}
	\tilde{c}(x) \sim \begin{dcases}
		x\, , & x \ll 1 \\
		x^{1+\alpha} \, , & x \gg 1   
	\end{dcases} \, .
\end{equation}	
That is, using the correct $\a$ in Eq.~\eqref{eq:scaling-cdf} accounts for size effects that give rise to deviations from the main power-law scaling as well as the appearance of the linear regime in the left tails. By plotting $Nc$ as a function of $N^{\frac1{1+\alpha}} x$ both corrections can thus be tested from a single scaling collapse.

Importantly, from this last equation and the preceding arguments, it is easy to conclude that the left-tail corrections become important in the same regime than the finite $N$ effects, thus what we will see in a realistic scenario is a crossover from the MF algebraic distributions to a flat one, given the form of $g_0$ and $p_0$. This is the reason why Eq.~\eqref{eq:scaling-avg-xmin} yields a very poor estimation of the critical exponents.

Before concluding this section I want to emphasize that for microscopic variables of jammed configurations the situation is conceptually different from that of standard critical phenomena, because the systems are already \emph{at} the critical point. That is, all the results that I will present here do not deal with the question of how the distributions of contact forces and gaps converge to the expected ones as we move away from $\vp_J$. Instead, I am going to analyse how the system size affects the range over which power-law scalings are followed. Equation~\eqref{eq:scaling-cdf} can nevertheless be used to estimate the scaling functions of the cdf of gaps and forces, obtained by integrating Eqs.~\eqref{eq:pdf-gaps} and \eqref{eq:pdf-f-ext}, respectively. I should mention as well that most of our results are concerned with extended forces and gaps since theory and previous numerical studies suggest that these variables are critically correlated across the whole system. In contrast, the distribution of buckling forces is expected to be independent of the system size, given the localized nature of their corresponding floppy modes. As I show in the next sections, our findings verify this picture.

\begin{figure}[!htb]
	\centering
	\includegraphics[width=0.8\textwidth]{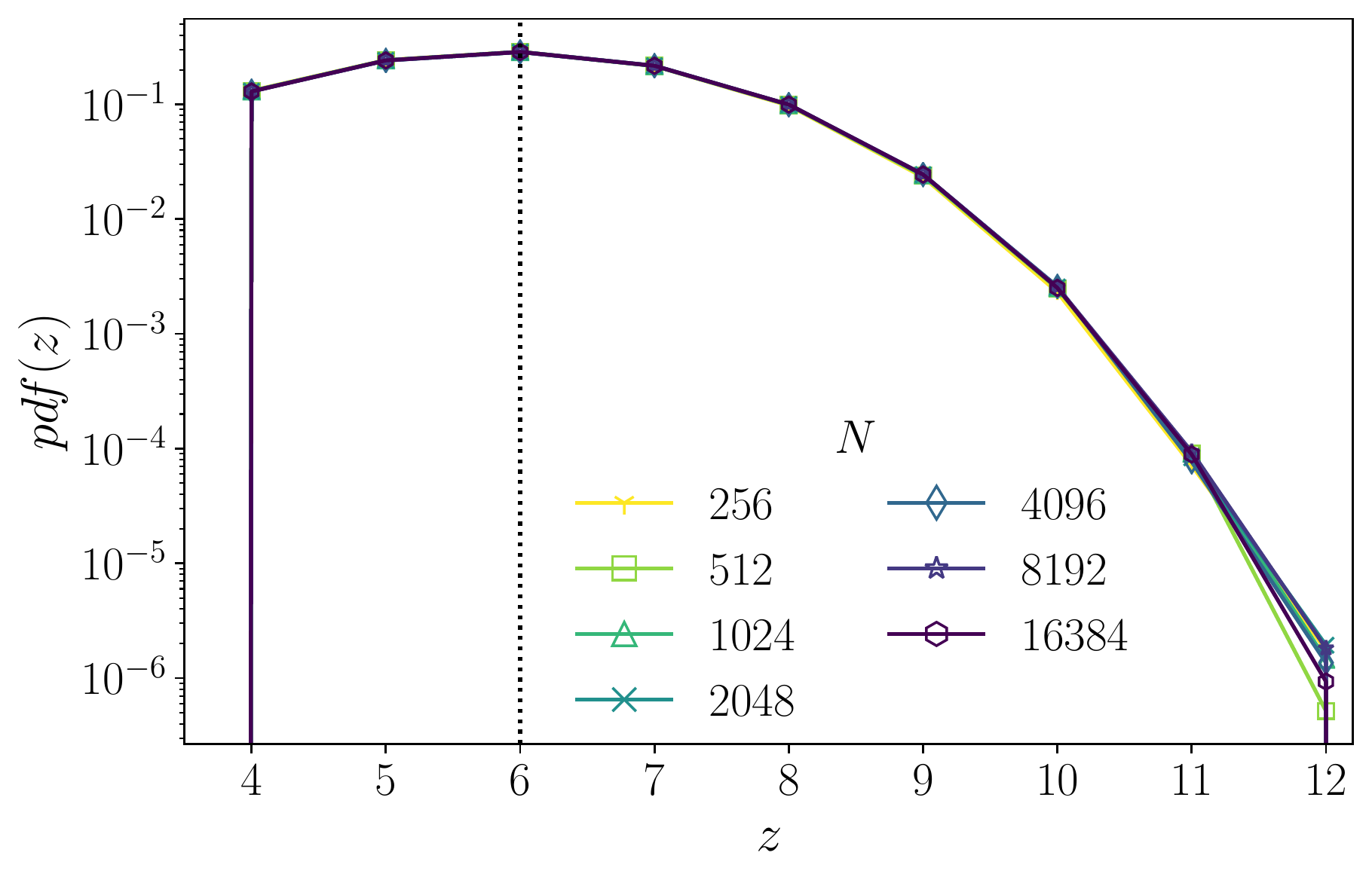}
	\caption[Probability density function of the coordination number for different system sizes.]{Probability density function of the coordination number for different system sizes. The mode is located at $z_c=2d$ as expected from the marginal stability of our jammed packings. No visible size effects are found.}
	\label{fig:pdf z}		
\end{figure}

\section{Finite size effects in spheres systems}\label{sec:spheres-fss}

\subsection{Some structural properties}\label{sec:fss structure spheres}

In this section, I briefly analyse two quantities of the HS jammed packings, namely, the distribution of coordination number, $z$, and the radial distribution function (RDF), $\tg(r)$, both for all the values of $N$ tested. The idea is to verify that the system's size does not affect the expected properties of configurations with at the jamming point. Hence, the results here are not concerned with the size scaling, but are more of a “security check” for the scalings presented in the next part. The pdf of $z$ is shown in Fig.~\ref{fig:pdf z} for all values of $N$. Clearly, no signatures of size effect are present, and in all the cases it is found that $z_c=2d=6$ is the likeliest value. This last feature is expected on the basis of the constraints counting argument at the jamming transition, introduced in Sec.~\ref{sec:network of contacts}. The very small amount of particles with $10\leq z \leq12$ (the latter being the $3d$ kissing number) suggest that no ordering is present in our configurations.

\begin{figure}[!htb]
	\centering
	\includegraphics[width=0.8\textwidth]{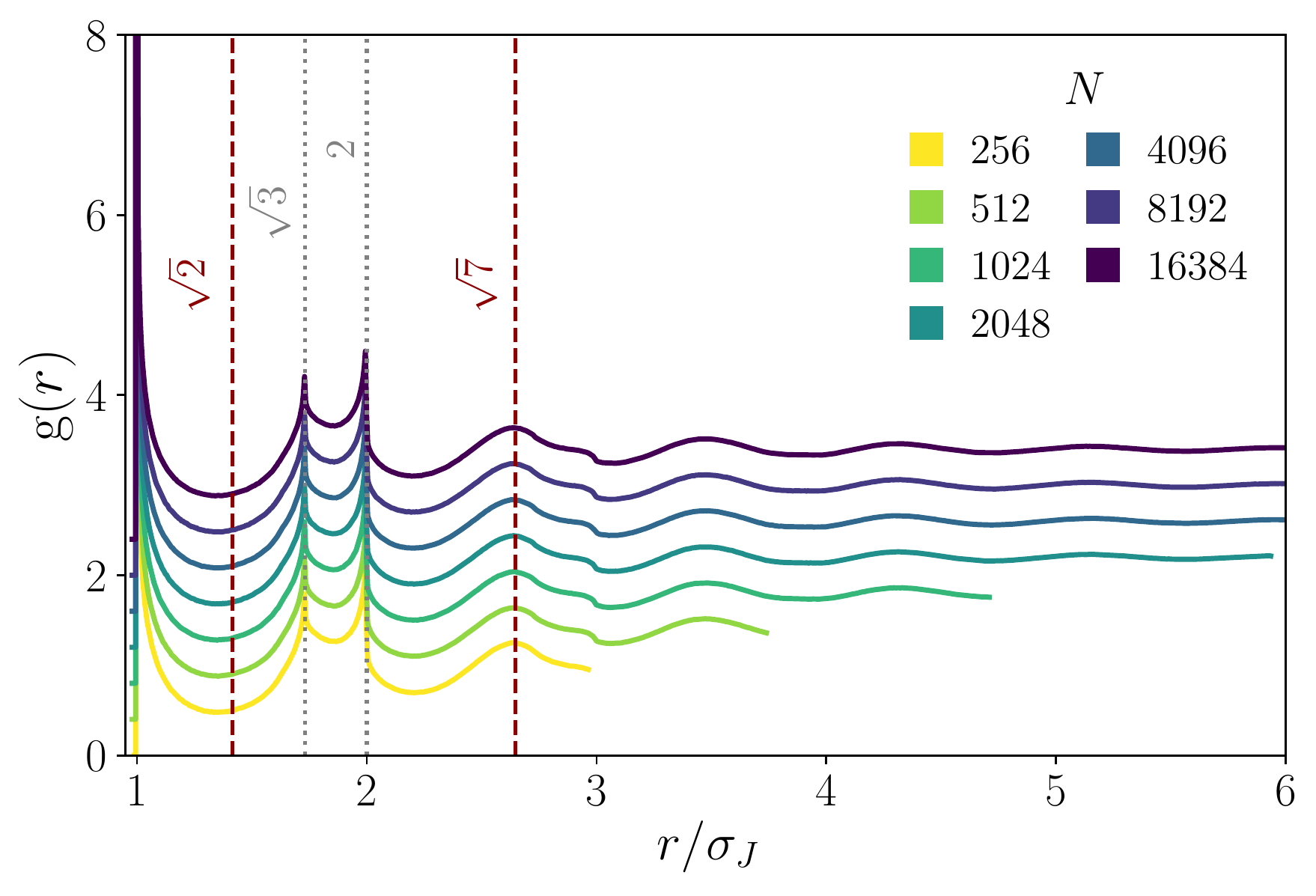}
	\caption[Radial distribution function of HS jammed packings.]{Radial distribution function of HS jammed packings (data average over at least 100 samples for each system size). Curves of different values of $N$ have been displaced vertically for clarity, while the domain of each curve is limited by $r<L/2$ due to the periodic boundaries, with $L$ being the size of the box containing the system. No influence of the system size is found.
		Grey dotted lines indicate the expected position of peaks or discontinuities in $\tg(r)$ present in \emph{amorphous} packings, while red dashed lines are the analogous positions for \emph{ordered} structures; cf.~Fig.~\ref{fig:rdf-vs-p} and see text for details.}
	\label{fig:rdf-vs-N}		
\end{figure}

This last feature can be tested more stringently by looking at the peaks of the RDF's, computed according to Eq.~\eqref{def:rdf} and presented in Fig.~\ref{fig:rdf-vs-N}. (In this case curves associated to a different value of $N$ have been displaced vertically for clarity.) Notice first that no size dependence is found in this case either, except for the trivial fact that periodic boundary conditions constrain $r$ to be restricted to less than half of the box. That is, given that $\sigma \sim N^{-1/3}$, the smaller $N$ the faster periodic effects are important. But this does not affect the behaviour of the RDF. Now, the positions of the peaks commonly found in \emph{disordered} packings are indicated by grey dotted lines, showing an excellent agreement. In contrast, the red dashed lines correspond to the positions of peaks associated with the formation of crystalline domains (see Fig.~\ref{fig:rdf-vs-p} and the related discussion). The absence of peaks at such locations confirms that no partial crystallization is present in our samples, for all values of $N$.

\subsection{Finite size scaling of the critical distributions} \label{sec:fss distributions spheres}

\subsubsection{Distribution of extended forces}\label{sec:fss fext spheres}

\begin{figure}[!htb]
	\includegraphics[width=0.99\linewidth]{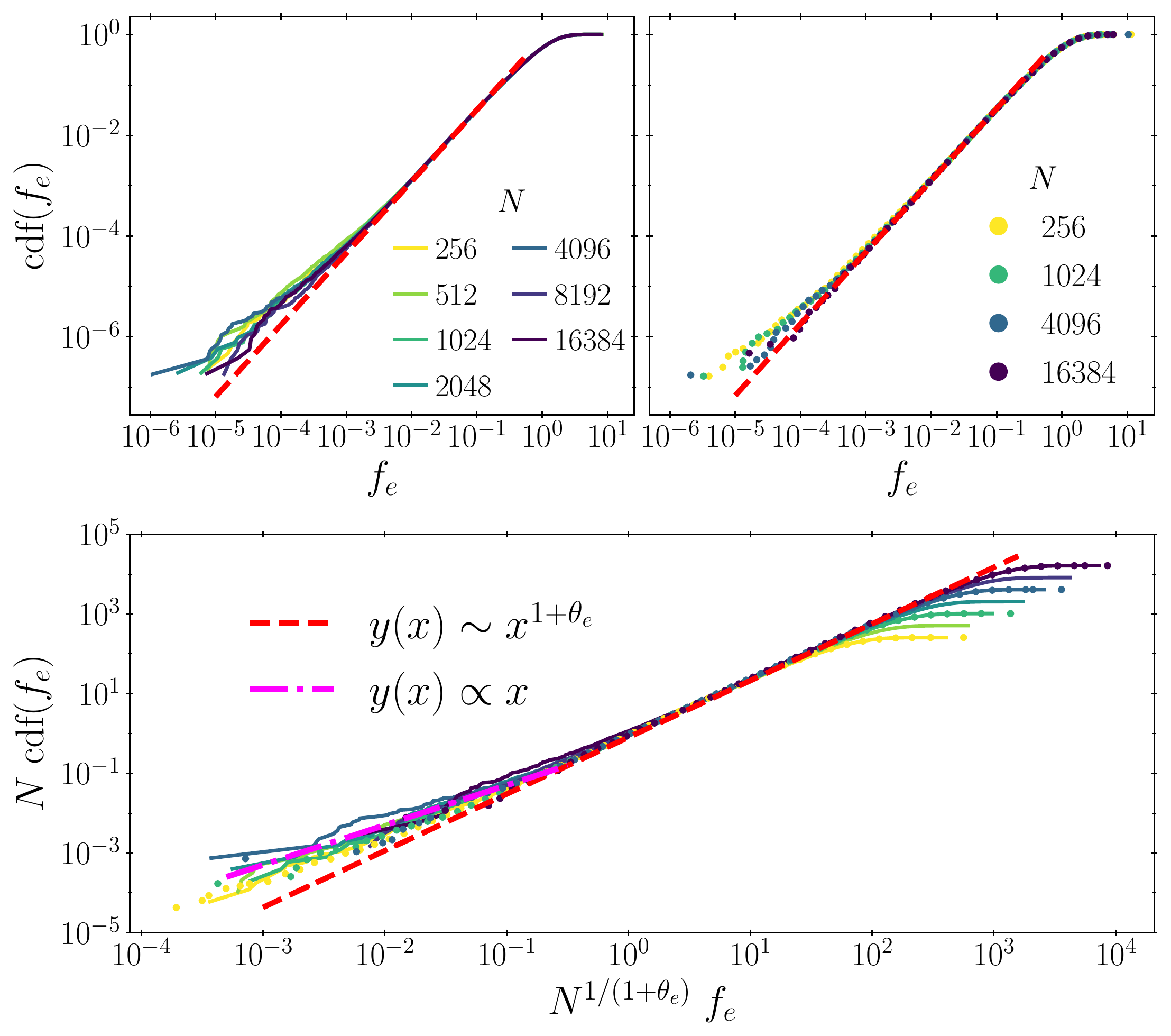}
	\caption[Size scaling of the cdf of $\{f_e\}$ in HS and SS configurations.]{Cumulative distributions of contact forces associated with extensive excitations in monodisperse configurations of frictionless spheres for different system sizes $N$. The upper left panel (resp. upper right panel) show the data as the jamming point is reached from below or UC regime (resp. above or OC regime). To better distinguish between the two different regimes, results belonging to the UC (OC) phase are identified by solid lines (circular markers). Bottom panel: Rescaling data from the upper panels according to Eq.~\eqref{eq:scaling-cdf} clearly collapses the data. The red dashed line corresponds to the power-law scaling of Eq.~\eqref{eq:gamma_theta_ext}, and shows an excellent agreement between the MF predictions and our numerical results.  The coincidence of results from the UC phase and OC phase for various $N$ confirms that $\theta_e$ is the same when jamming is reached from either direction. In the left tail of the distributions of  the bottom panel I also include a comparison with the linear scaling expected for very small values, following Eq.~\eqref{eq:pdf-f-ext-v2}.}
	\label{fig:forces-3d}
\end{figure}

Once we know that our packings do not present signatures of crystallization, let me begin by analysing the size effects in the cdf of the extended forces, $\{f_e\}$.
Figure~\ref{fig:forces-3d} shows the distributions of $f_e$ obtained coming from below (UC, upper left panel) and from above (OC, upper right panel). Comparing the results with the theoretically predicted power-law (dashed red) reveals an outstanding agreement over at least three decades. More importantly, no visible signature of finite-size corrections can be detected over the range of $N$ considered. To verify more stringently the absence of finite-size effects, we attempted to collapse the different curves by rescaling the extended forces and their cdf following Eq.~\eqref{eq:scaling-cdf}, obtaining the curves reported in the bottom panel. 
This latter figure contains our first main result: strong evidence that the same critical distribution of forces is found independently of whether the jamming point is generated from the UC or OC regimes.

\begin{figure}[!htb]
	\includegraphics[width=\linewidth, height=17.5cm]{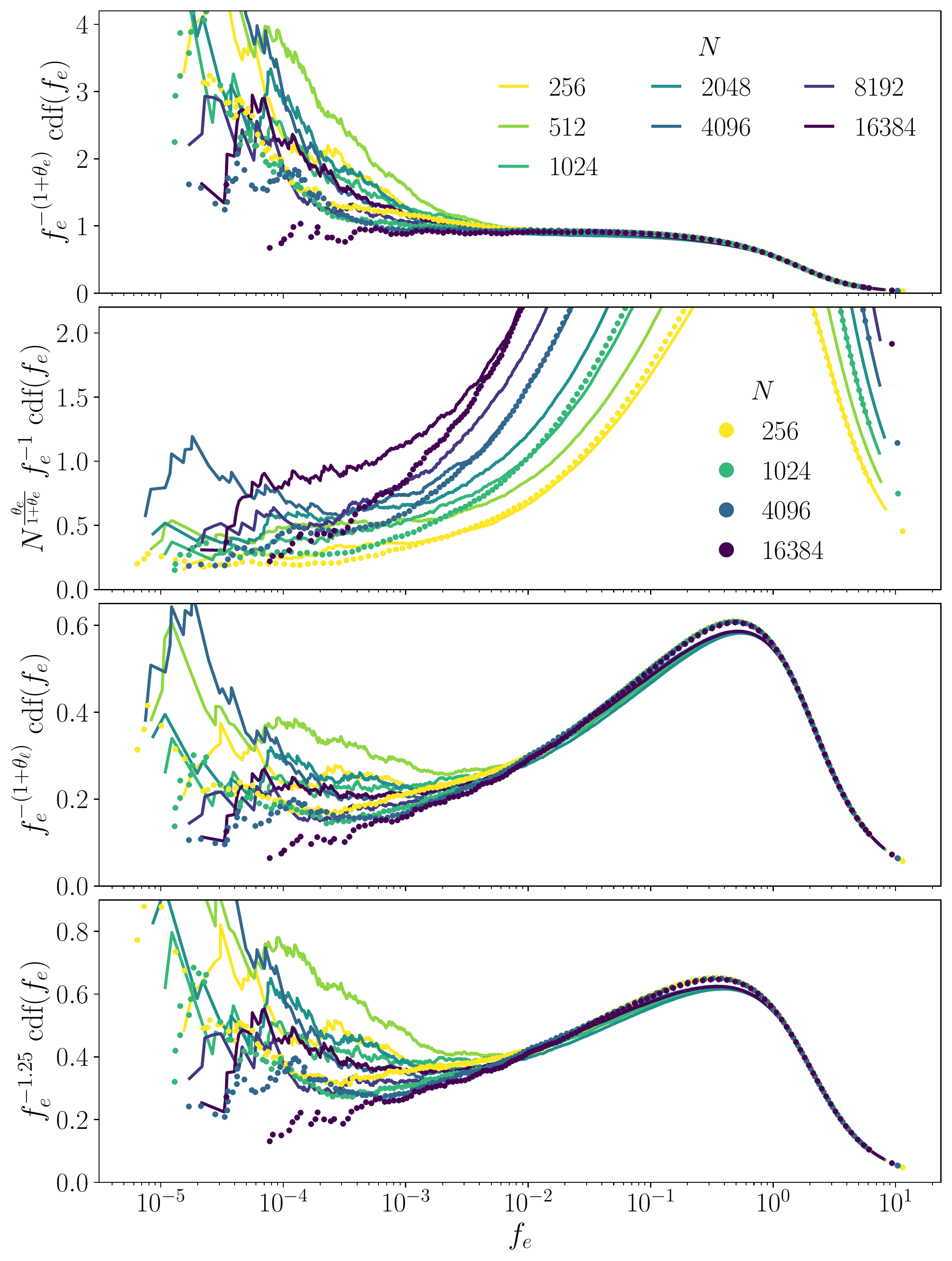}
	\caption[Rescaled cdf of $\{f_e\}$ in spheres packings to study the two different power-laws.]{Same distributions of Fig.~\ref{fig:forces-3d} but rescaled by $f_e^\a$ to examine more carefully the different scalings present in the cdf. The value of $\a$ in the different panels, from top to bottom, is $1+\theta_e$, $1$, $1+\theta_\ell$, $1.25$, for the reasons explained in the main text Notice that the cdf values in the second panel (from top to bottom) have also been multiplied by $N^\frac{\theta_e}{1+\theta_e}$, following Eq.~\eqref{eq:pdf-f-ext-v2}; this serves to verify that once rescaled, the behaviour of the left tail should be a constant of order 1. The data from the UC phase (lines) seem to be consistent with a linear scaling in the left tail, while OC data suggest that localized modes might also be present, inducing small deviations from the MF behaviour.}
	\label{fig:rescaled cdf fext}
\end{figure}

\begin{figure}[!htb]
	\includegraphics[width=\linewidth]{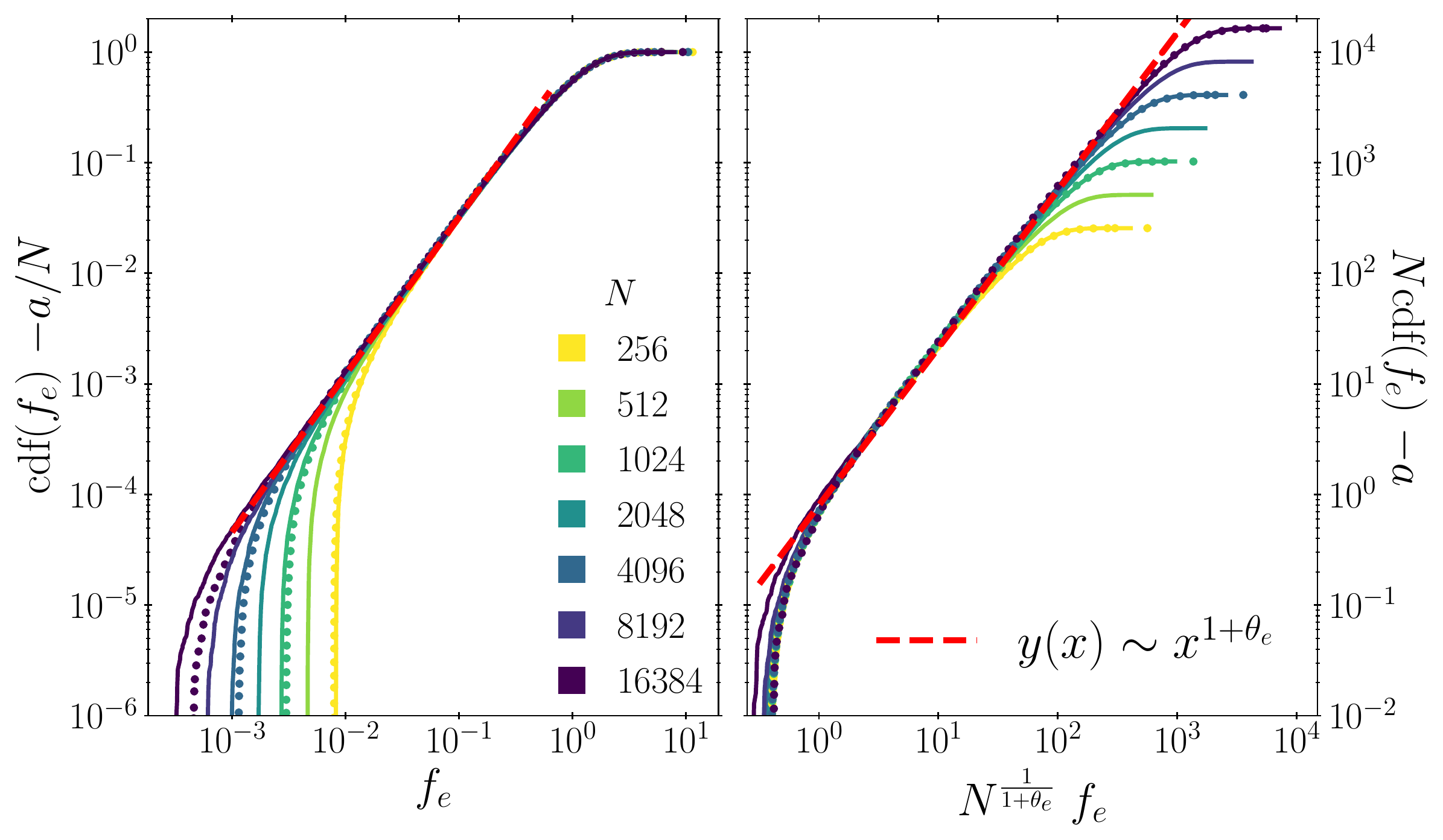}
	\caption[Cdf of $f_e$ in spheres packings with contribution of extremes removed.]{Same distributions of Fig.~\ref{fig:forces-3d} but subtracting $1/N$ to the cdf. This is a simple strategy for isolating the main scaling domain, by removing the contribution of the extremes. This reduced cdf clearly shows (left panel) that finite $N$ effects can be identified in the distribution of extended forces and, more importantly (right panel), accurately accounted for by the scaling function  of Eq.~\eqref{eq:scaling-cdf}, using the MF value of $\theta_e$.}
	\label{fig:forces-3d-minus-1N}
\end{figure}

Yet, it is clear that our packings exhibit an excess of very small forces (an effect more noticeable when jamming is reached from below; cf. upper panels of Fig.~\ref{fig:forces-3d}), echoing earlier observations\supercite{charbonneauJammingCriticalityRevealed2015,lernerLowenergyNonlinearExcitations2013,charbonneauGlassyGardnerlikePhenomenology2019,kallusScalingCollapseJamming2016}. Note that the scaling of Eq.~\eqref{eq:scaling-cdf} does not remove these deviations from the predicted distribution, and that they occur, roughly, at the same scaled force: $N^\frac{1}{1+\theta_e} f_e \lesssim 1$. It is therefore likely that forces are subject to size effects caused by the onset of a different behaviour, $p(f)\sim 1$ (see Eq.~\eqref{eq:pdf-f-ext-v2}). 
One way to analyse more carefully these two regimes, and in particular the linear left tail of the cdf, is to divide the cdf by the corresponding power law. That is, assuming $c(x)\sim x^{1+\a}$, plotting $x^{-1-\a} c(x)$ should result in a rather constant function. Performing such rescaling in the cdf of $f_e$ yields the curves shown in Fig.~\ref{fig:rescaled cdf fext}, where the exponent for the rescaling factor used in each panel is, from top to bottom, $(1+\theta_e, 1, 1+\theta_\ell, 1.25)$. The reason for including $1+\theta_\ell$ and $1.25$ is to check whether the left tail is due only to localized modes, or, to the fact that the “$z_i=z_\ell \implies \text{localized}$” criterion fails for the smallest forces, respectively\footnote{Recall that in the introduction I mentioned that in $d=3$ the distribution of the two types of forces combined has an exponent of $0.25$. This was shown in Ref.~\cite{charbonneauJammingCriticalityRevealed2015} and we verified that the same is true for our data.}.   
Thus, the topmost figure confirms, in the right part of the distribution, the MF prediction for the extended forces, while the panel below it tests the linear behaviour mentioned for the left tail. Notice that in this latter case the cdf has also been scaled by $N^\frac{\theta_e}{1+\theta_e}$, as predicted by Eq.~\eqref{eq:pdf-f-ext-v2}, in order to verify that the $f_e\to 0$ tail leads to a constant of order 1\footnote{Clearly, this “extra” factor has the effect of splitting vertically the curves of different values of $N$. However, I verified that this effect can be removed by rescaling the \emph{horizontal} axis by $N^\frac{1}{1+\theta_e}$, as expected.}. Data from the UC phase (lines) show a reasonable agreement, \textit{i.e.} a slight plateau, for $f_e \in [10^{-4}, 10^{-3}]$ approximately. Nevertheless, the same comparison with data from the OC regime (circles) is slightly worse. Moreover, the second to last and bottom panels of the same figure show that, arguably, the cdf corresponding to SS data is better rescaled by assuming some influence of the localized modes. In contrast, this does not seem to be the case for HS data. Unfortunately, with the available data it is impossible to prove (or even to disprove) the linear form of the left tail, which calls for a more assiduous future study of the true behaviour of the extremes of $p(f_e)$.


As a final “trick” to isolate the main finite-$N$ effects in the cdf notice that, according to Eq.~\eqref{eq:scaling-xmin}, by subtracting $c(x)-a/N$ the influence of the left tails can be teased out, with $a$ a constant of order 1 to be determined. In other words, given that the weight of the extremes is of order $1/N$, by removing their contribution to the cdf only the main, MF power-law should remain, albeit in a reduced domain. Hence, by applying the same scaling arguments in such main region a more rigorous analysis is obtained.
In Fig.~\ref{fig:forces-3d-minus-1N} I show the outcome of this technique when applied to the usual cdf (left panel), as well as to the distribution rescaled according to Eq.~\eqref{eq:scaling-cdf}  with the MF value of $\theta_e$ (right panel). Notice that in the first case, the size effects are now evident, but the second figure shows a remarkable collapse onto the scaling function, confirming the MF prediction for $\theta_e$. I assumed $a=0.25$, given the value of the approximate plateaus in the second panel of Fig.~\ref{fig:rescaled cdf fext}, but the results are rather insensitive to this value.

\subsubsection{Distribution of gaps}\label{sec:fss gaps spheres}

\begin{figure}[!htb]
	\includegraphics[width=0.99\linewidth]{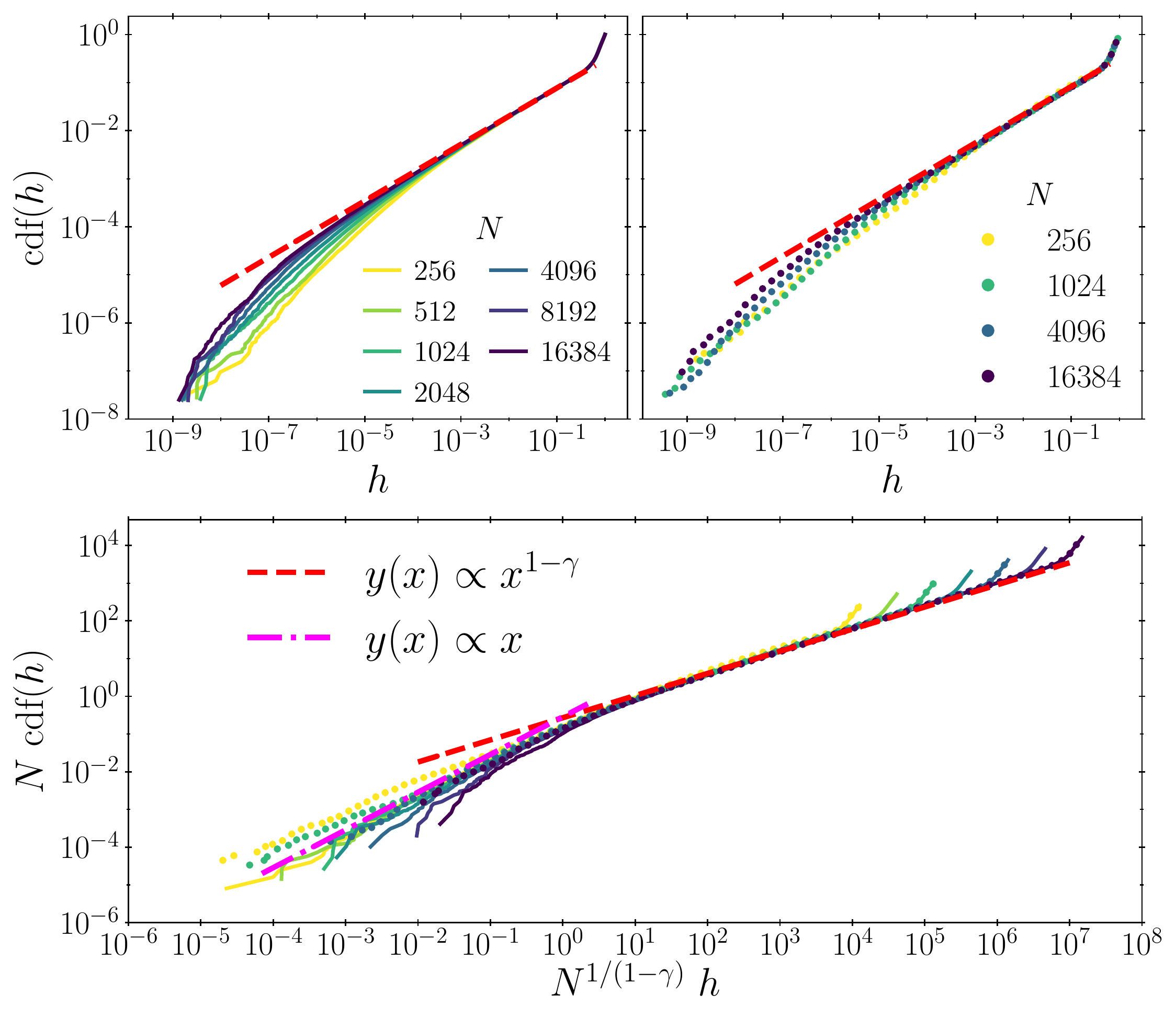}
	\caption[Size scaling of the cdf of $\{h\}$ in HS and SS configurations.]{Cumulative distributions of interparticle gaps for the same configurations of Fig.~\ref{fig:forces-3d}, as their jamming point is reached from below or UC phase (upper left) and from above or OC phase (upper right). Lower panel: rescaling both datasets according to Eq.~\eqref{eq:scaling-cdf} shows that finite-size corrections can be accounted for in all cases. For comparison, the power-law scaling derived from MF theory, Eq.~\eqref{eq:pdf-gaps}, is also shown (red dashed line). Once again, the fact that datasets from both phases, \textit{i.e.} UC (markers) and OC (lines), neatly superimpose confirms that the exponents at the jamming point are the same, independently of how  $\phi_J$ is approached. Additionally, the secondary scaling regime $g(h)\sim 1$ of Eq.~\eqref{eq:pdf-gaps-v2}, also predicted by MF theory, can be observed for very small values. Its associated linear cdf is shown (magenta dash-dotted line).
	}
	\label{fig:gaps-3d}
\end{figure}

I will now analyse the distribution of interparticle gaps, $\{h\}$, and their associated size effects. Fig.~\ref{fig:gaps-3d} shows the corresponding cdf's obtained from the UC (upper left panel) and OC (upper right) from the same configurations as above. Once again, the data are in very good agreement with the predicted distribution, Eq.~\eqref{eq:pdf-gaps}, independently of the direction in which jamming is approached. More interesting however, is the fact that the cdf$(h)$ exhibits strong signatures of finite $N$ effects, in contrast to $p(f_e)$. Note that the scaling correction given in Eq.~\eqref{eq:scaling-cdf} using the MF value of $\gamma$ precisely corrects for such effects over almost \emph{seven} orders of magnitude (lower panel of Fig.~\ref{fig:gaps-3d}). We thus verified that the MF value of $\gamma$ holds on both sides of the jamming transition as well.

\begin{figure}[!htb]
	\includegraphics[width=\linewidth]{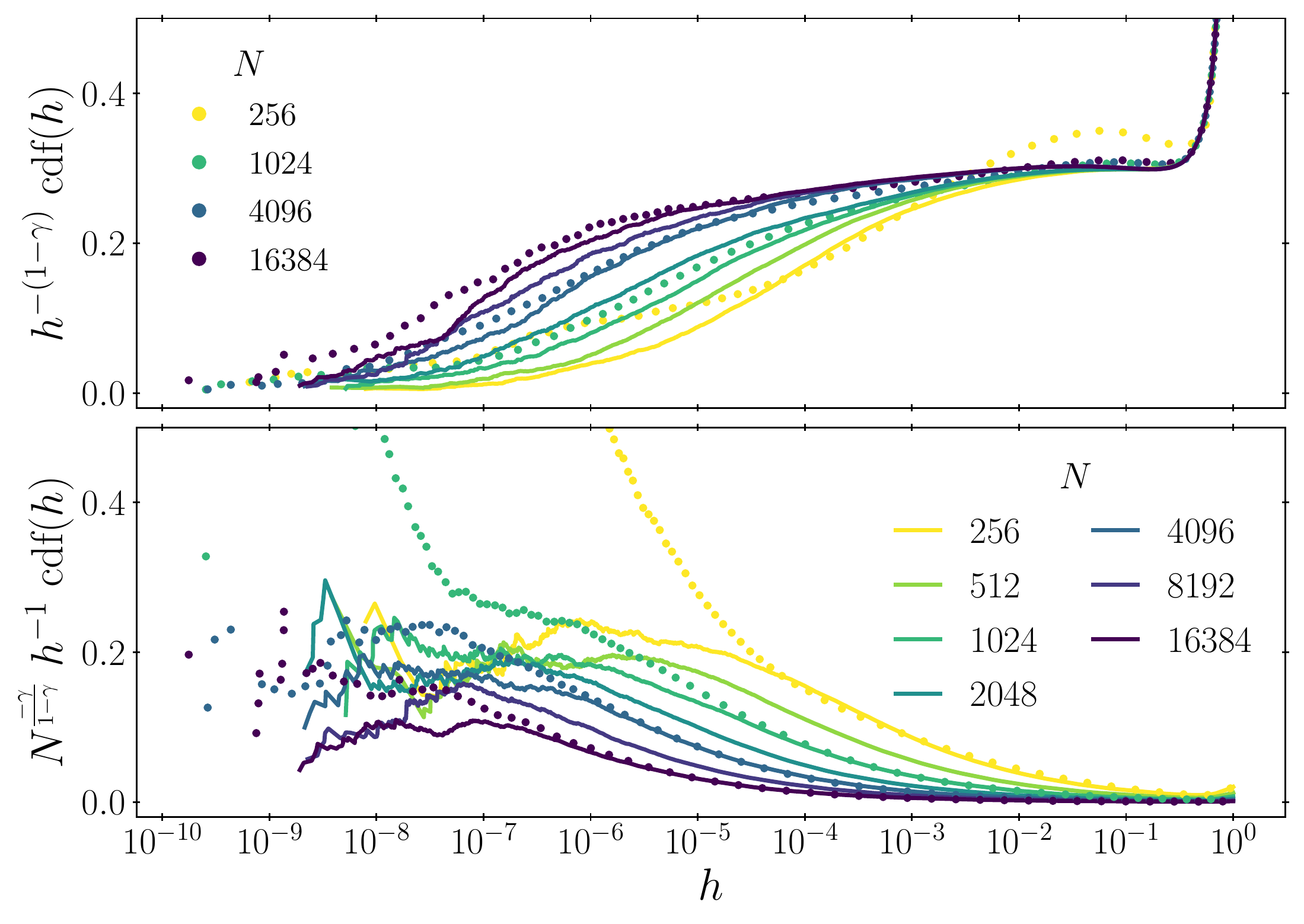}
	\caption[Rescaled cdf of $\{h\}$ in SS and HS configurations to study the two different power-laws.]{Same distributions of Fig.~\ref{fig:gaps-3d} but rescaled by $ h^{-(1-\gamma)}$ (top plot) and $N^\frac{-\gamma}{1-\gamma}\ h^{-1} $ (bottom one) to examine more carefully the different scalings present in the cdf's. The $N$ dependence of the rescaling factor in the second case follows from Eq.~\eqref{eq:pdf-gaps-v2}; see main text.
	As with the case of extended forces, configurations of HS (lines) show a better agreement with the linear dependence assumed, while the smallest packings generated from the OC phase using SS (circles) exhibit significant deviations in this extremal regime.
		}
	\label{fig:rescaled cdf h}
\end{figure}

On the other hand, the growing deficit of very small gaps as the system size decreases is, very likely, another manifestation of the cut-off in the main power law of $g(h)$. It leads to a secondary linear regime, as stated in Eq.~\eqref{eq:pdf-gaps-v2}, that is in reasonable agreement with the numerical results shown in the lower panel of Fig.~\ref{fig:gaps-3d}. Yet, to assess the presence of such regime more thoroughly we can once again resort to the method of dividing the cdf by $h$ raised to the corresponding power of the regime we want to investigate. When doing so, we obtain the plots shown in Fig.~\ref{fig:rescaled cdf h}, where the rescaling factor is $h^{-(1-\gamma)}$ in the upper panel (to verify the theoretical exponent) and $N^\frac{-\gamma}{1-\gamma}\ h^{-1}$ in the lower one (to test for the presence of the linear left tail, including the prefactor spelled out in Eq.~\eqref{eq:pdf-gaps-v2}). Once again, by including such correction the $h\to0$ behaviour of the rescaled tail approaches a constant value of order 1, for almost all values of $N$. In fact, the only discrepancies observed are for the two smallest sizes of the systems obtained from the OC regime. But larger systems of SS as well as all the cases with HS reproduce the expected plateau. I deem important to emphasize that to remove as much as possible any size dependence of this secondary regime the $ N^\frac{-\gamma}{1-\gamma} $ term is needed. That is, if the cdf are rescaled only by $h^{-1}$ an increase in the left tails is observed, \textit{i.e.} an \emph{spurious} sub-linear behaviour.

At any rate, the different role of the system's size in the distributions of forces and gaps is, nonetheless, striking. Our numerical results indicate that distances between nearby spheres are significantly modified in finite $N$ configurations and, consequently, so is the distribution of gaps. This is an intriguing finding that I will attempt to rationalize heuristically relying on the response to contact openings in jammed packings. The argument goes as follows. In the thermodynamic limit, a system has always enough space to relax any perturbation caused by a contact opening, and hence it is always able to re-accommodate enough particles --even if this requires bringing many of them infinitesimally close to each other-- in order to guarantee stability.
In a finite system, by contrast, no such unconstrained relaxation can take place. Rearranging an extensive fraction of particles necessarily influences the pair of spheres involved in the contact just opened. Therefore, there is a certain scale below which the occurrence of small gaps is disfavoured. If the system was further relaxed, then at least one extra contact would form. This is consistent with the uniform distribution hypothesis, since such distribution decreases the weight of small enough gaps in comparison with the MF one, that actually predicts a divergence as $h\to 0$.

At this point, I wish to stress that our results demonstrate the existence of two different types of finite-size corrections to the distributions of extended forces and gaps. The first is a consequence of large scale correlations and can thus be readily taken into account by the scaling of the cdf given in Eq.~\eqref{eq:scaling-cdf}. Although this correction is practically absent in the forces distribution, for $g(h)$ it is the main source of deviation from the theoretical prediction. The second is a consequence of the critical scalings of Eqs.~\eqref{eq:pdf-gaps} and \eqref{eq:pdf-f-ext} being cut off at very small values. This effect, which is likely related to the extra contact with respect to $N_{dof}$ (see Sec.~\ref{sec:fss-theory}), affects both microstructural variables.
I get back to this point in Sec.~\ref{sec:MK-fss}, after having considered its signature in the MK model.

\subsubsection{Distribution of localized forces}\label{sec:fss floc spheres}

\begin{figure}[!htb]
	\includegraphics[width=\linewidth]{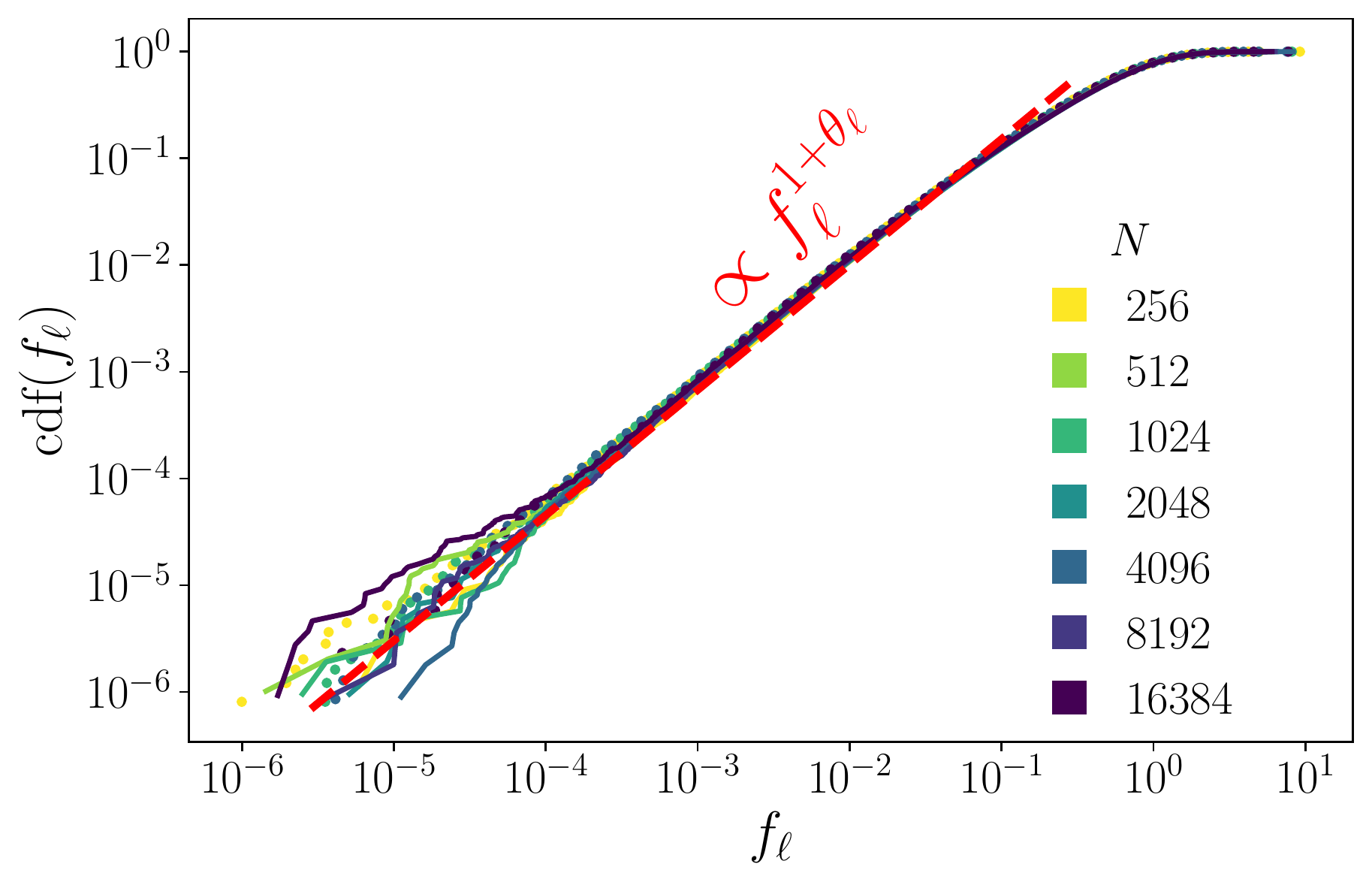}
	\caption[Cumulative distribution function of localized forces in spheres packings]{Cumulative distribution function of localized forces, $\{f_\ell \}$, in configurations obtained from the UC (lines) and OC (circles) phases and for different system sizes, as indicated by the colours in the legend. No signatures of finite $N$ dependence are observed.}
	\label{fig:forces-loc-3d}
\end{figure}

The only critical distribution that remains to be analysed is the one associated to the set of localized forces, $\{f_\ell\}$. As I have anticipated above, no size effects are expected for this variable and this claim is nicely confirmed by our numerical results reported in Fig.~\ref{fig:forces-loc-3d}. Note that in this case the estimation of cdf$(f_\ell)$ is very close to the predicted power-law for forces as small as $10^{-4}$ (cf. the case of extended forces in Fig.~\ref{fig:forces-3d}). Besides, for even smaller values the curves corresponding to different $N$ are disperse both above and below the theoretical line, which indicates that no systematic size dependence is present.

Before concluding this section, it is worth emphasizing that our numerical results are in excellent agreement with the MF, $d\to\infty$ predictions for the power-law scaling of the distributions of both the extended forces and the interparticle gaps, as well as analogous previous estimates concerning the localized forces.  These results confirm that the jamming criticality of these microstructural variables is also valid for low dimensional systems all the way down to $d=3$, in agreement with earlier albeit less accurate studies\supercite{charbonneauUniversalMicrostructureMechanical2012,lernerLowenergyNonlinearExcitations2013,degiuliForceDistributionAffects2014}. Because results from both OC and UC phases superimpose onto each other, we can further conclude that the critical behaviour is controlled by the same exponents on both sides of the jamming point.

\section{Finite size effects in the MK model}\label{sec:MK-fss}

I will now present the analogous results obtained from the jammed configurations in the MK model. Here however, packings were produced only from the UC phase, so only data for HS will be considered. In Sec.~\ref{sec:contact vectors MK} I begin by analysing some properties of contact vectors which are distinctive of the MK configurations. Then, in Sec.~\ref{sec:fss distributions MK} I present the main results of this section, concerning the size effects in the critical distributions.

\subsection{Distribution of contact vectors}\label{sec:contact vectors MK}

\begin{figure}[!htb]
	\centering
	\includegraphics[width=0.8\linewidth]{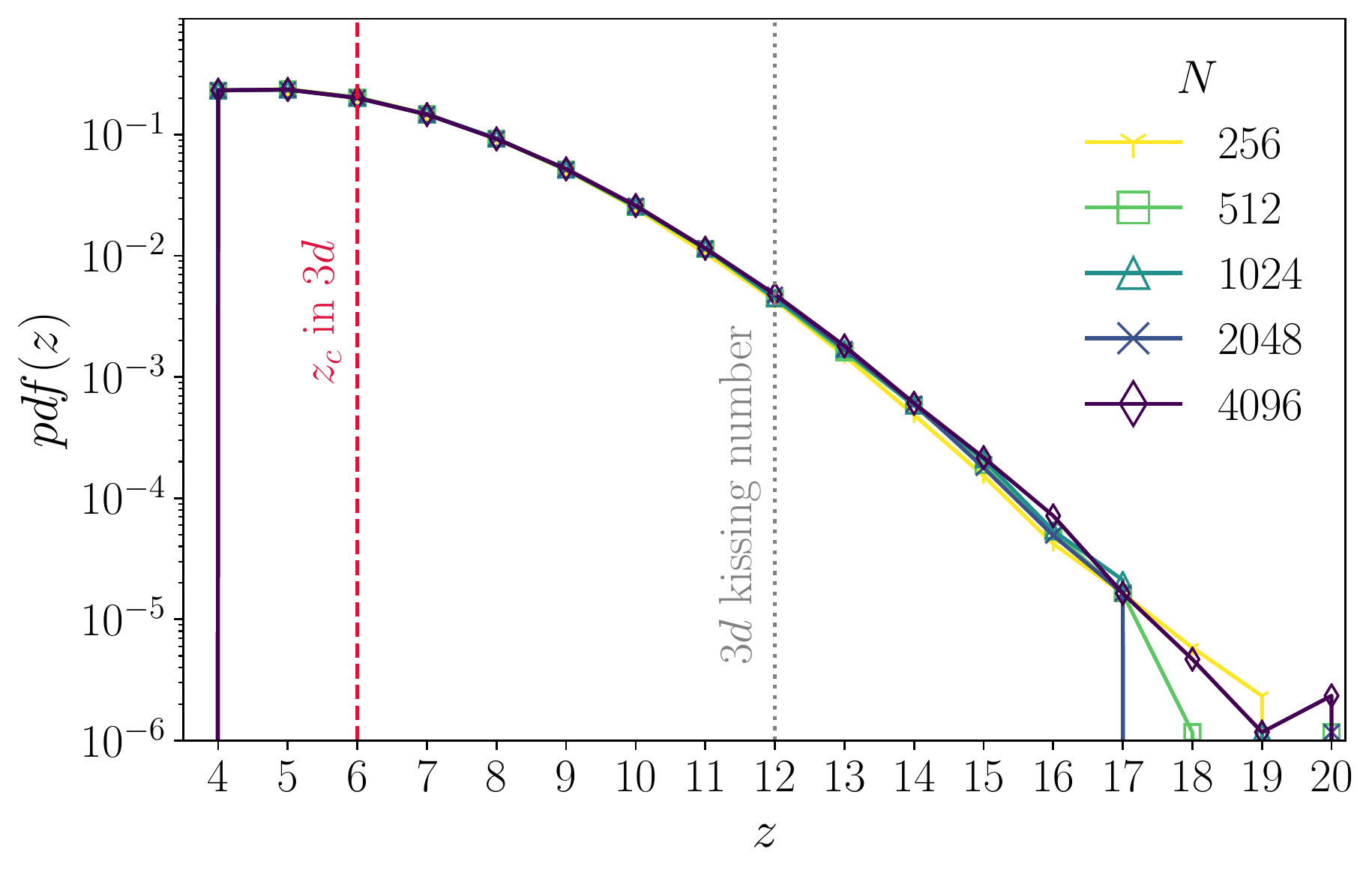}
	\caption[pdf of the coordination number in MK packings.]{Probability distribution of the coordination number per particle in MK configurations of different size, identified according to the legend. The expected value from isostaticity arguments ($z_c$) in $3d$, as well as the corresponding kissing number are highlighted. Cf. with Fig.~\ref{fig:pdf z} for the analogous results in HS.}
	\label{fig:pdf z MK}
\end{figure}

The first notable property of the MK packings is their connectivity. First, because the possible number of contacts per particle is no longer constrained by the respective kissing number. But also because $z_c=2d$ is no longer the most common value of the coordination number. Instead, our results show that $z_\ell=4$ and $z=5$ are the likeliest values, at least in our $3d$ configurations. Both features are illustrated in Fig.~\ref{fig:pdf z MK} where it is also shown that these properties are independent of the systems' size. Note that the distribution of $z$ is monotonically decreasing, in contrast with HS systems (see Fig.~\ref{fig:pdf z}). Along the same lines, given that $\avg{z}=z_c + \order{1/N}$,  the broader domain of the distribution shifts the mode to lower values than $z_c$, differing from previous results\supercite{charbonneauJammingCriticalityRevealed2015} in $d=2,\dots,8$. This is an unexpected result given that all our packings have 1SS, and therefore their average connectivity is indeed $z_c$ with corrections of order $1/N$.

\begin{figure}[!htb]
	\centering
	\includegraphics[width=0.8\linewidth]{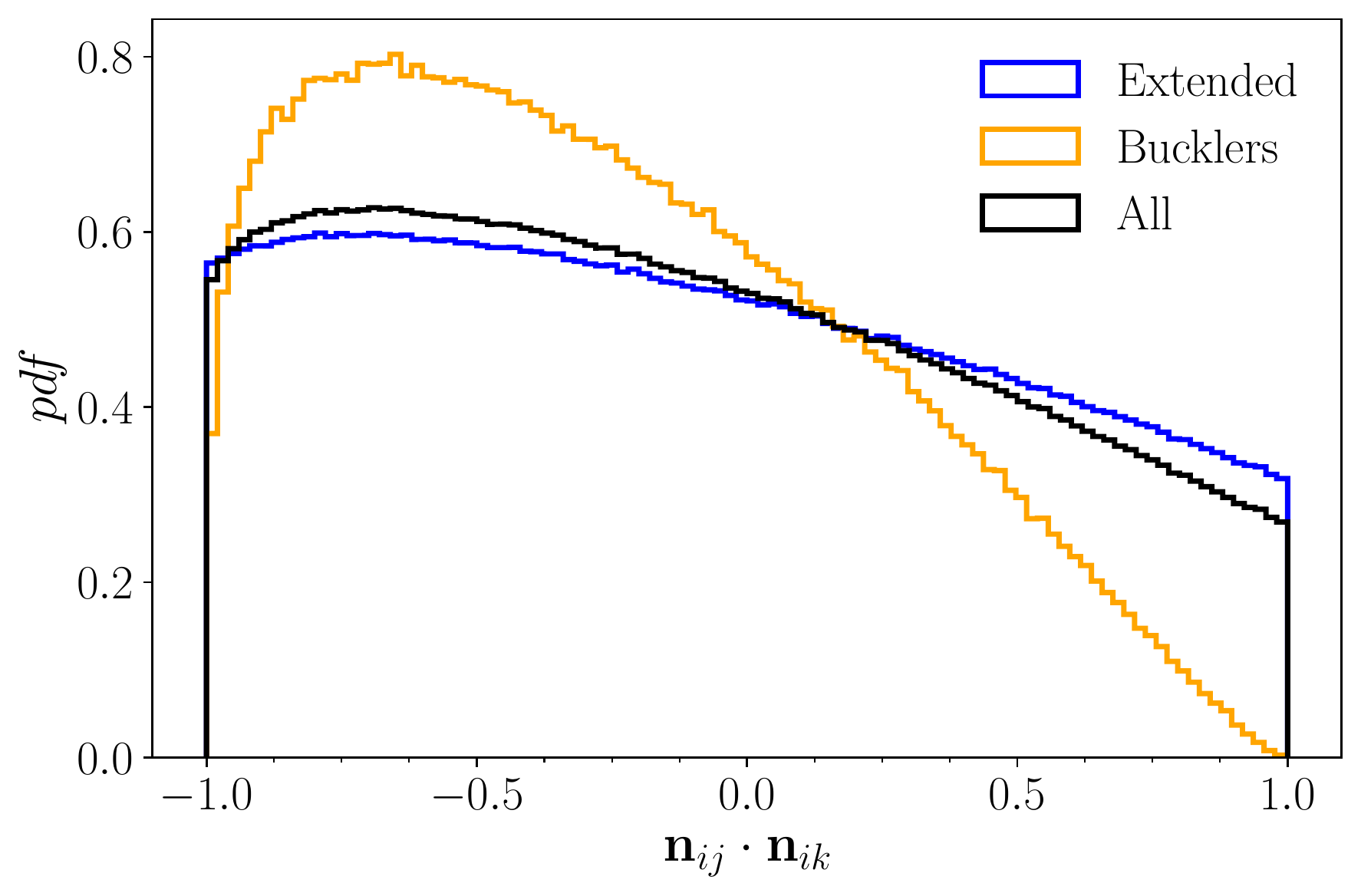}
	\caption[pdf of dot products between contact vectors in MK configurations]{Probability distribution of dot products between contact vectors in MK configurations of $N=2048$ particles. The contributions of bucklers (orange) and extended (blue) modes are also plotted separately. Cf. with Fig.~\ref{fig:dist-contacts} for analogous results in HS.}
	\label{fig:pdf dot products MK}
\end{figure}

For future use, I will now study the geometric arrangement of contacts, for which I will follow the same methodology than in Sec.~\ref{sec:structural variables}. Hence, I will first analyse the distribution of dot product between pairs of contact vectors, $\vb{n}_{ij}\cdot \vb{n}_{ik}$, for $j,k \in \partial i$. The pdf obtained from the data of the $N=2048$ configurations is shown in Fig.~\ref{fig:pdf dot products MK}, where I have also included the corresponding distributions when localized and extended modes are considered separately. Comparing with Fig.~\ref{fig:dist-contacts}, where the corresponding results in HS are presented, it is clear that random shifts in the MK configurations allow particles to form rather different structures. The most salient feature is that no peak is observed at $0.5$, and actually the distribution continues all the way down to $1$. This is expected, because many neighbours of a given particle can overlap (from the point of view of such particle), and these results show that they can even be concentric. As in the HS case, the distributions of the full set of particles and the ones associated to extended forces are very similar to each other, while the one of bucklers is in this case notoriously different. Note that for this latter type of particles more weight is given to contact vectors forming angles greater than $\pi/2$, which supports the near coplanarity assumption of bucklers. However, an accompanying peak around $0$ is clearly missing, signalling that other configurations that fulfil the mechanical equilibrium condition are possible.

\begin{figure}[!htb]
	\includegraphics[width=\linewidth]{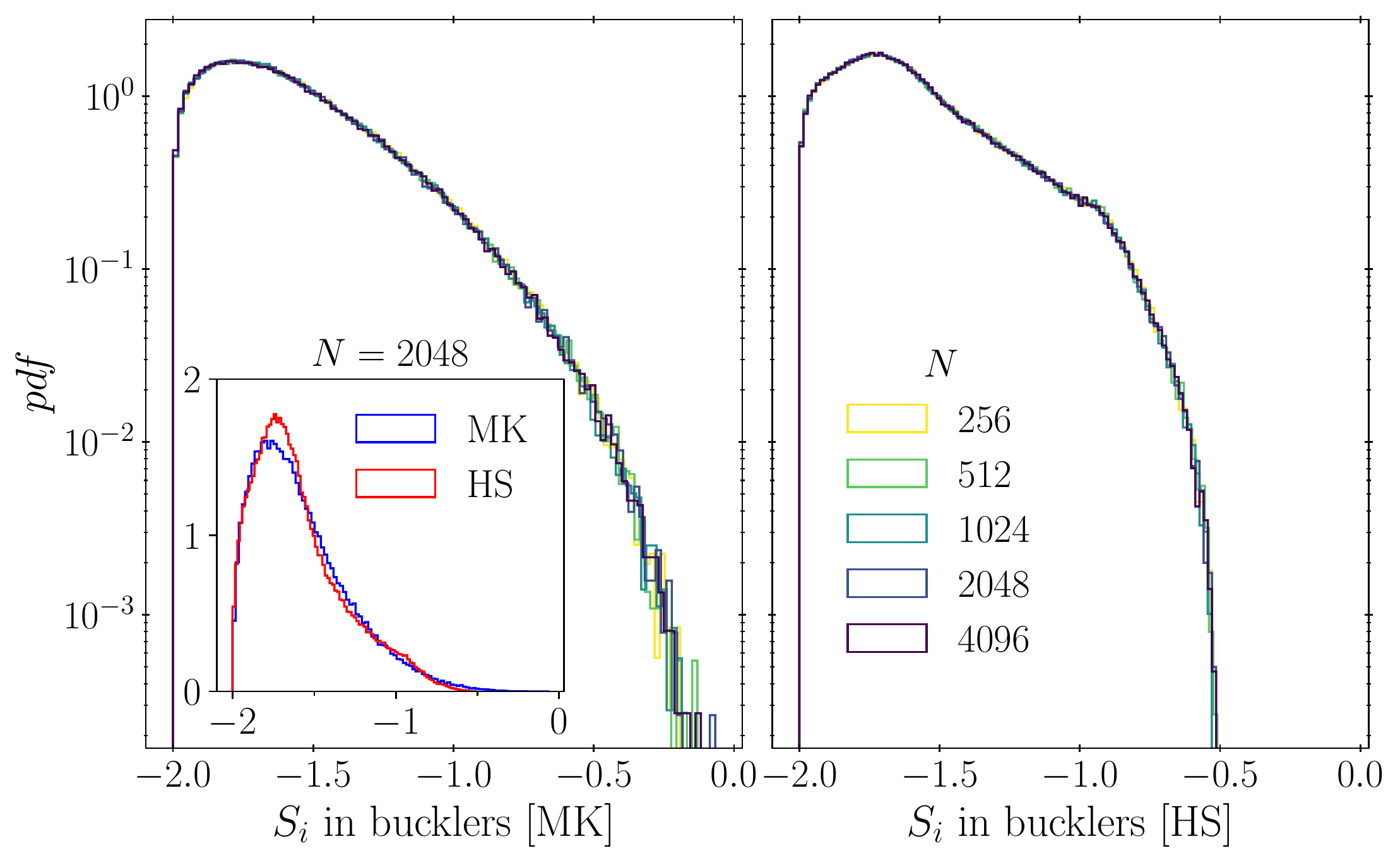}
	\caption[pdf of $S_i$ of bucklers in MK and HS of different size]{Pdf of $S_i$ of bucklers in MK (left) and HS (right) jammed packings of different sizes. Inset: comparison of the same distributions in both models (using a linear scale); data from configurations of $N=2048$ particles.}
	\label{fig:pdf Si bucklers}
\end{figure}

Finally, I consider the sum of such dot product per particle, \textit{i.e.} the variable $S_i$ defined in Eq.~\eqref{def:tot dot CVs}. Fig.~\ref{fig:pdf Si bucklers} shows the resulting distributions in both MK and HS configurations, and considering different system sizes. 
These results show, once again, that size effects do not influence the structural variables considered in the previous chapter. More importantly, we can observe that even if $S_i$ can take values greater than $-0.5$ in MK configurations, the effect on the distribution is rather small. In fact, bucklers in both models seem to have a very similar pdf, which is rather unexpected from the corresponding distributions shown in Figs.~\ref{fig:dist-contacts} and \ref{fig:pdf dot products MK}. All these results will be important to understand some of the findings presented in the following part, but the comparison included in this last figure will have a special role in the discussion of the distribution of $\{f_\ell\}$ (see Sec.~\ref{sec:fss fext MK}).

\subsection{Finite size scaling of the critical distributions}\label{sec:fss distributions MK}

\subsubsection{Distribution of extended forces}\label{sec:fss fext MK}

\begin{figure}[!htb]
	\includegraphics[width=\linewidth]{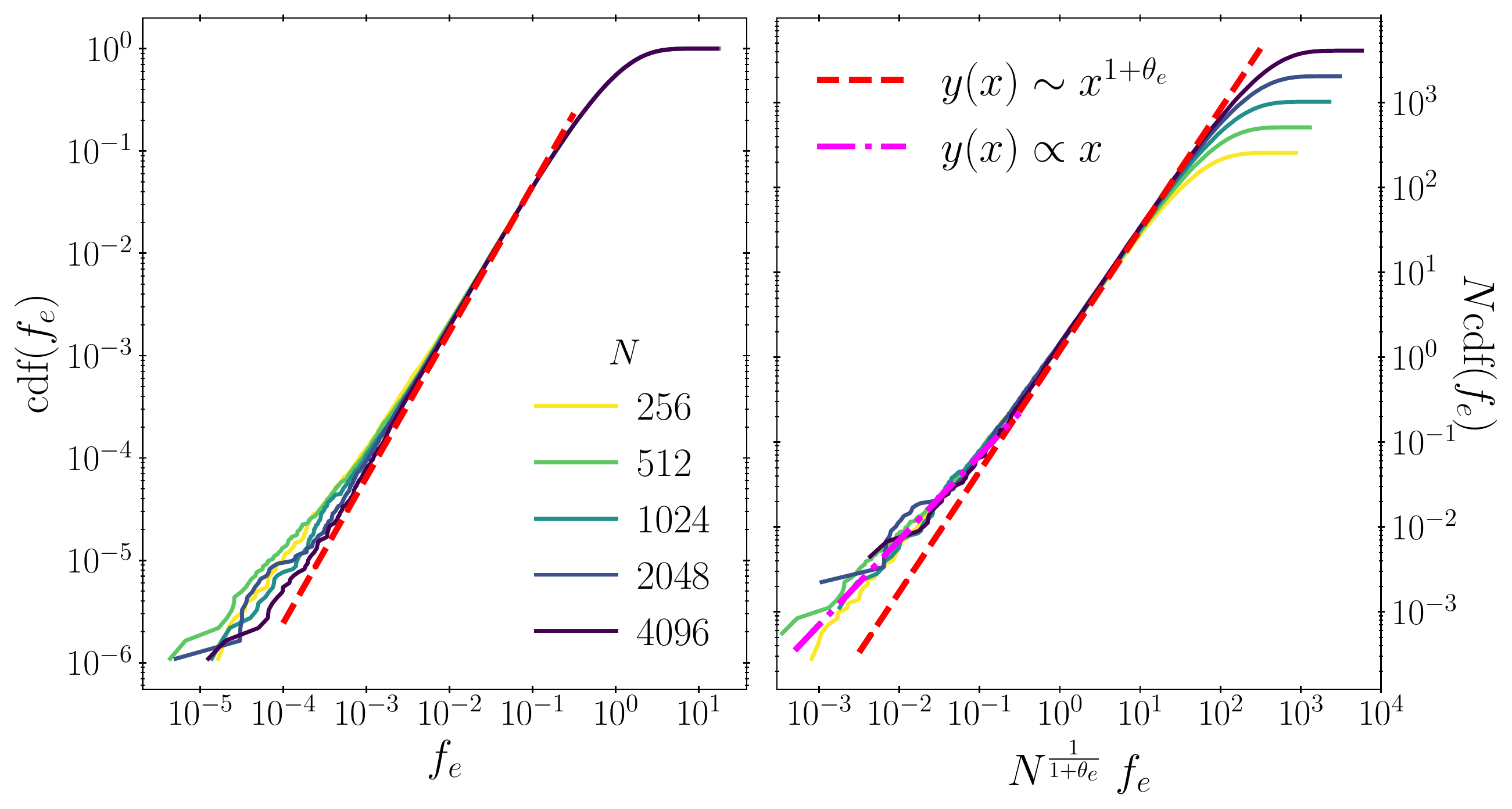}
	\caption[Size scaling of the cdf of $\{f_e\}$ in MK configurations]{Cdf of the set of extended forces in MK configurations of different size (left) and the resulting scaling (right) following Eq.~\eqref{eq:scaling-cdf} with the MF value of $\theta_e$. The red dashed line is the theoretical power-law, while the pink, dash-dotted one corresponds to the linear regime mentioned in Eq.~\eqref{eq:pdf-f-ext-v2}.}
	\label{fig:forces-MK}
\end{figure}

I will now proceed to analyse the size effects in the critical distributions, following the same methodology than for HS (Sec.~\ref{sec:fss distributions spheres}). Hence, I begin by the considering the distributions of extended forces for $N=256, \dots, 4096$, shown in the left panel of Fig.~\ref{fig:forces-MK}. Notice that in contrast with the HS results, here the size effects are noticeable, even if small. Interestingly, this seems to be a feature exclusive of the MK packings, since the other models studied in Ref.~\cite{paper-fss} do not show such dependence on $N$. The influence of the size effects in cdf$(f_e)$ is clearer when the size scaling is carried out, following Eq.~\eqref{eq:scaling-cdf}. The result is shown in the right panel of the same figure, where the collapse obtained using the MF value of $\theta_e$ is again outstanding. The reasons explaining the pronounced finite $N$ effects in the MK systems however is postponed to Sec.~\eqref{sec:fss gaps MK}, because similar findings hold for $g(h)$.

\begin{figure}[!htb]
	\includegraphics[width=\linewidth]{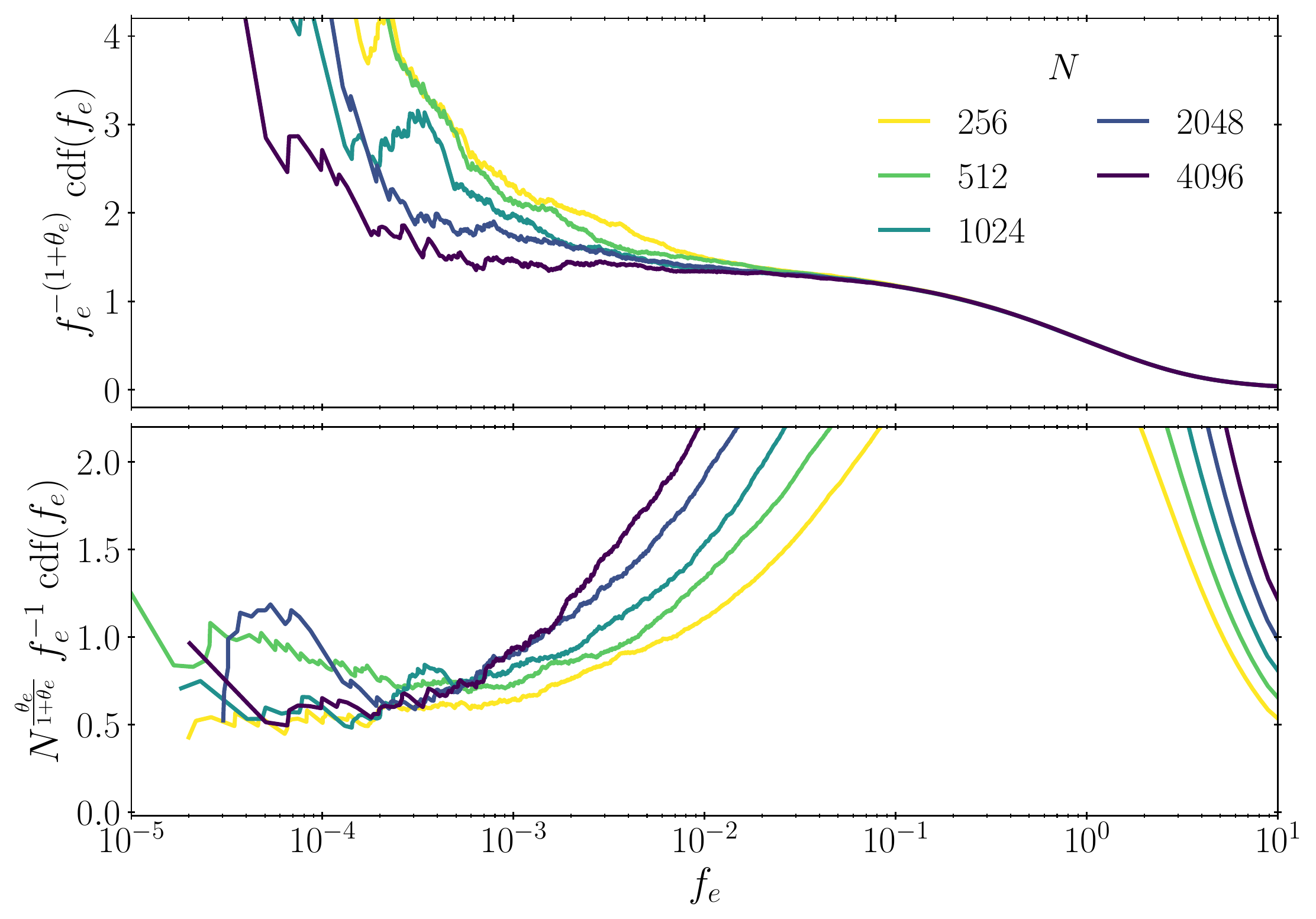}
	\caption[Rescaled cdf of $\{f_e\}$ in MK configurations to study the two different power-laws.]{Same distributions of Fig.~\ref{fig:forces-MK} but rescaled by $f_e^{1+\theta_e}$ (upper panel) and $ N^\frac{\theta_e}{1+\theta_e}\ f^{-1}$ (bottom one) to examine more carefully the different scalings present. Also in this case the left tail of cdf$(f_e)$ agrees reasonably well with a linear behaviour. Note that, in analogy to the analysis of the previous section, the size dependence of the left tail has also been accounted for, thus leading to a rather constant value of order 1.}
	\label{fig:rescaled cdf fext MK}
\end{figure}

On the other hand, an excess of very small forces is easy to observe, similar to the one found for HS (and the other systems of Ref.~\cite{paper-fss}). To analyse this left tail more carefully, I will make use of the same methodology than from HS, namely, rescale cdf$(f_e)$ by a power of $f_e$ in an attempt to find a constant function. The outcome of trying $1+\theta_e$ (to verify the main scaling) and $1$ (to test the hypothesis of a linear left tail) is presented in the top and bottom panels, respectively, of Fig.~\ref{fig:rescaled cdf fext MK}. Note that, once more, these two exponents describe the respective two regimes within an acceptable degree, but more data is needed to confirm the power-law of the left tail beyond doubts. I mention in passing that in this case there is no need to try $\theta_\ell$ or another power because for MK packings the localized forces actually follow a \emph{different} power-law; see Sec.~\ref{sec:fss floc MK}.

\subsubsection{Distribution of gaps}\label{sec:fss gaps MK}

\begin{figure}[!htb]
	\includegraphics[width=\linewidth]{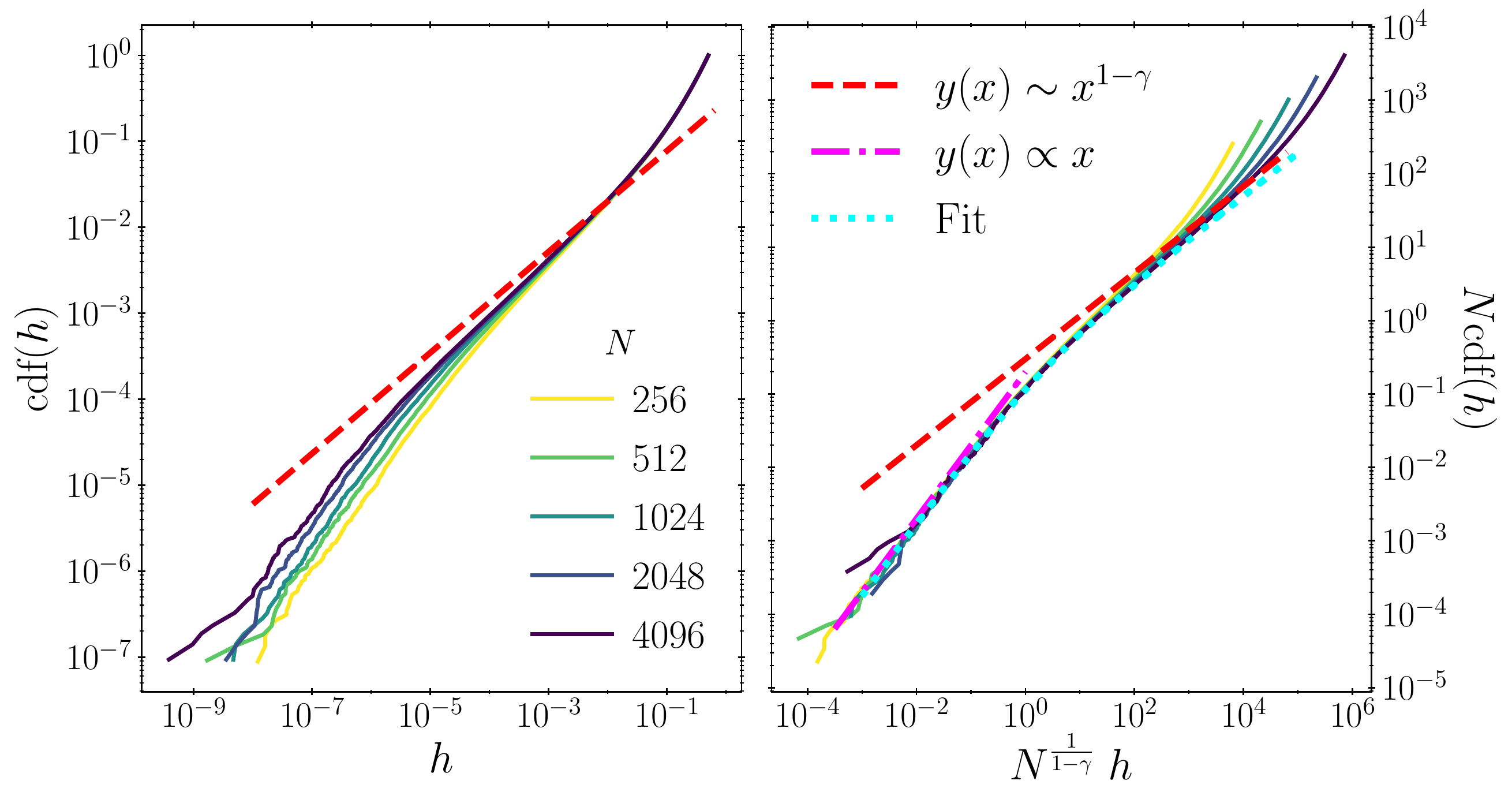}
	\caption[Size scaling of the cdf of $\{h\}$ in MK configurations.]{Cdf of the set of interparticle gaps in MK configurations of different size (left) and the resulting scaling (right) following Eq.~\eqref{eq:scaling-cdf} with the MF value of $\gamma$. The red dashed line is the theoretical power-law, while the pink, dash-dotted one corresponds to the linear regime mentioned in Eq.~\eqref{eq:pdf-gaps-v2}. Additionally, the dotted, cyan line is the result of fitting Eq.~\eqref{eq:fit-gaps} to the rescaled data of $N=4096$. This fit assumes the MF value of $\gamma$, but weighs the two different regimes of \eqref{eq:general scaling cdf} depending on the values of the scaling variable; see discussion in the main text.
	Interestingly, in this case, the curves do not seem to follow the predicted power-law due to strong finite size effects, but the very good collapsed obtained verifies that the exponent predicted by MF is the correct one. See text for details.}
	\label{fig:gaps-MK}
\end{figure}

To continue, I now analyse the cdf of the interparticle gaps. The distributions obtained from the same configurations as above, together with the size scaling of Eq.~\eqref{eq:scaling-cdf} using the MF value of $\gamma$, are shown in the left and right panels, respectively, of Fig.~\ref{fig:gaps-MK}. The analysis of these results is more interesting, given that in comparison with the analogous results in HS, here the curves do not really seem to follow the power-law behaviour predicted by MF. In fact, we can note that individual distributions of $h$ suggest that a smaller exponent would better fit the curves in the left panel of Fig.~\ref{fig:gaps-MK}. Yet, using the theoretical value of $\gamma$ produces an excellent collapse as shown in the right panel of the same figure, and trying different values significantly worsens the quality of such collapse.
This is, in fact, a typical situation of many critical scalings in finite $N$ systems\supercite{amitFieldTheoryRenormalization2005,MC_book}. In these cases, the most reliable way to determine true critical exponents is from the finite-size scaling analysis, as we have done here.

\begin{figure}[!htb]
	\includegraphics[width=\linewidth]{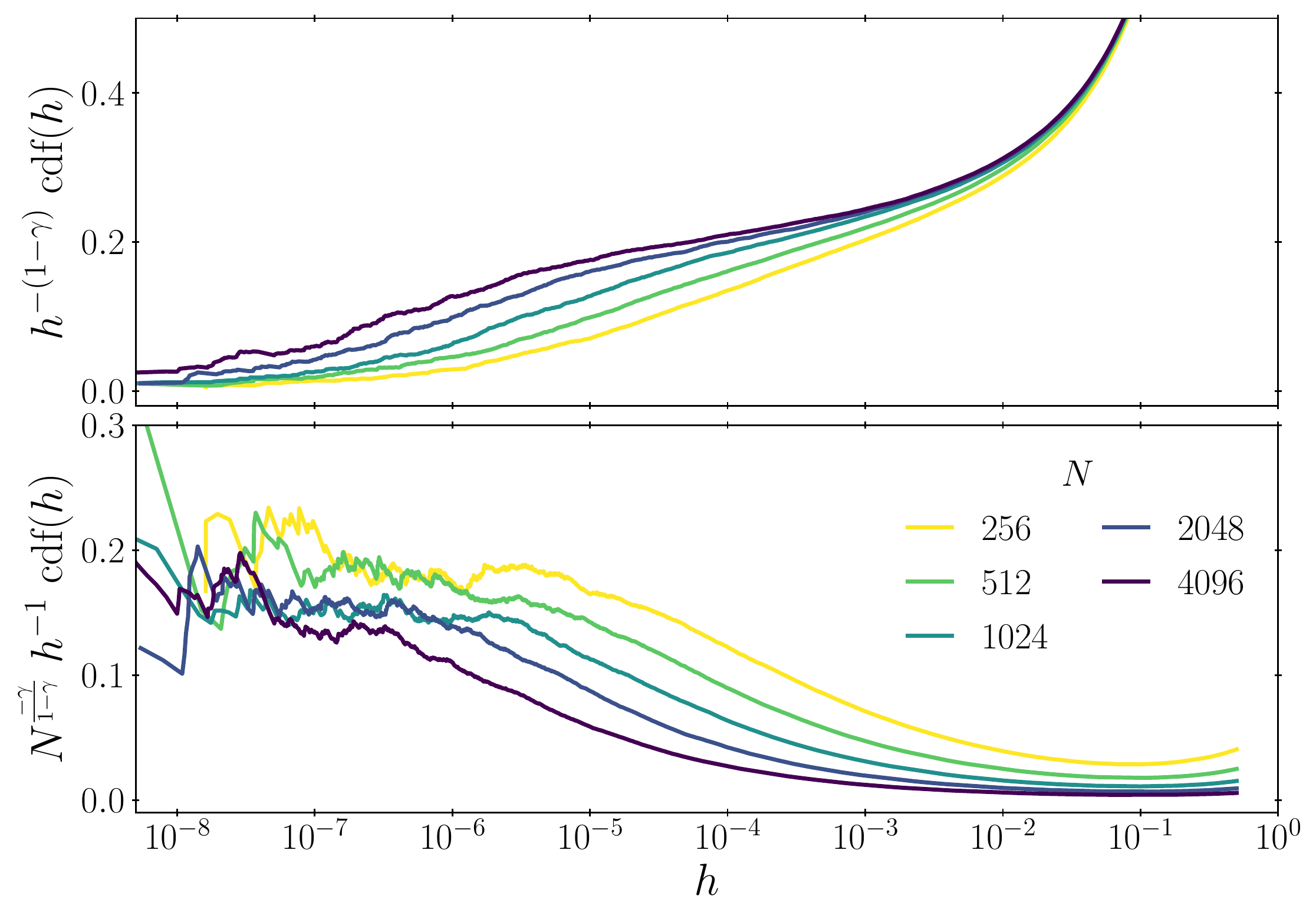}
	\caption[Rescaled cdf of $\{h\}$ in MK configurations to study the two different power-laws.]{Same distributions of Fig.~\ref{fig:gaps-MK} but rescaled by $h^{1-\gamma}$ (upper panel) and $ N^{\frac{-\gamma}{1-\gamma}}\ h^{-1}$ (bottom one) to examine more carefully the different scalings present. Here, the near plateau formed after rescaling the cdf with a linear factor is much clearer, and by including the $N$ dependence of the prefactor of Eq.~\eqref{eq:pdf-gaps-v2} the different curves tend to roughly the same value as $h\to 0$.}
	\label{fig:rescaled cdf gaps MK}
\end{figure}

Before further analysing the prominent size effects observed in this model, I test once again the hypothesis of a linear behaviour for very small values of gaps. Applying the technique used so far, in Fig.~\ref{fig:rescaled cdf gaps MK} I present the cdf rescaled by $h^{-(1-\gamma)}$ (upper panel) and $N^{\frac{-\gamma}{1-\gamma}}\ h^{-1}$ (bottom one). Notice that in MK packings, the curves do not really have a constant value the region associated to the main power law, \textit{i.e.} $h\in [10^{-5}, 10^{-2}]$, in contrast to the analogous results in HS and SS systems (cf. Fig.~\ref{fig:rescaled cdf h}). This  simply reproduces what I pointed out in the previous paragraph that a seemingly smaller value of $\gamma$ (greater slope) would better fit the cdf, specially for the smallest systems. Nevertheless, it is remarkable that also in these configuration the linear behaviour of the smallest values of $h$ remains unaltered, for at least two orders of magnitude. In fact, for this model the agreement with Eq.~\eqref{eq:pdf-gaps-v2} of such secondary regime is better than in other cases. This is likely caused by the MF power-law being realized only partially, leading to a quick crossover to the extremal statistics behaviour. Note also that when the $N$ dependence in the prefactor of the scaling function is included the plateau of all curves has a similar value.

\begin{figure}[!htb]
	\centering
	\includegraphics[width=0.9\linewidth]{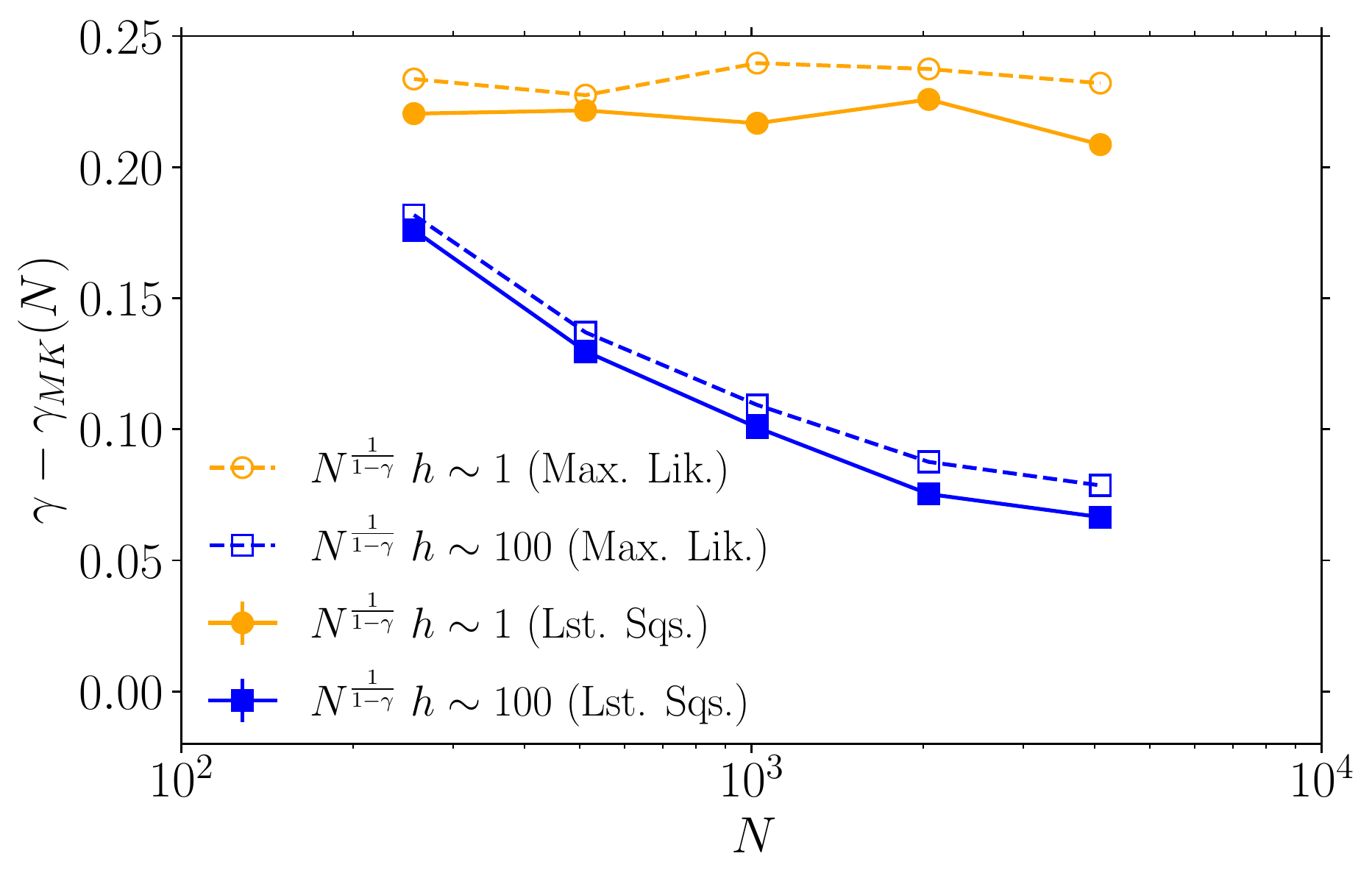}
	\caption[Difference of the gaps exponent in the MK model and its MF value]{Difference of the gaps exponent in the MK model and its MF value. The values of $\gamma_{MK}$ were determined using a least squares fit (solid lines, filled markers) or a maximum likelihood estimation (dashed lines, empty markers) at two different values of the scaling variable, $\tilde{h}=1$ (orange) and $\tilde{h}=100$ (blue). Note that even in this latter case the MF value is not fully recovered, indicating that larger systems, possibly by orders of magnitude, would be needed to observe the theoretical value of $\gamma$.}
	\label{fig:fits-gamma-MK}
\end{figure}

I now come back to the marked size effects found in this model, evinced by the size dependence of  $cdf(f_e)$ and the (apparently) different exponent of the gaps distribution. Both of these features contrast with the results of the previous section and the other models studied in Ref.~\cite{paper-fss}. For concreteness, I will restrict the analysis to the gaps cdf's, where finite size effects are most pronounced. We should first consider that, somehow surprisingly, the individual gaps distributions in the MK model, which by construction should closely resemble the MF value, do not display the right gap exponent. Indeed, from the right panel of Fig.~\ref{fig:gaps-MK} we observe that the scaling variable using the MF value of $\gamma$ is the correct one (curves do collapse when plotted as a function of $\tilde{h} = N^\frac{1}{1-\gamma} h$), but the slope of the different curves in the range covered in our simulations ($10^{-3}< \tilde{h} < 10^3$) differs from the MF prediction. An important concern is thus whether this deviation is due to finite-size corrections or whether it indicates a failure of the MK model. In order to resolve the matter, we used the expected form of the scaling function, Eq.~\eqref{eq:general scaling cdf}, to construct a fitting function, $F(\tilde{h})$, that \emph{assumes} the correct behaviour of the scaling function for large values of $\tilde{h}$. Explicitly:
\begin{equation}\label{eq:fit-gaps}
	F(\tilde{h}) = \qty[ (a\tilde{h})^d + (b \tilde{h}^{1-\gamma})^d ]^{1/d}
\end{equation}
This fitting function hence only depends on three parameters and fulfils the condition that $F(\tilde{h})\propto \tilde{h}$ for $\tilde{h}\ll 1$, while the MF form, $\tilde{h}^{1-\gamma}$, is recovered for large values of the scaling variable. 
Fitting $F(\tilde{h})$ to the largest system size results in the dotted, cyan line in the bottom panel of Fig.~\ref{fig:gaps-MK}, which clearly interpolates nicely between both regimes. (The values of the parameters obtained from a least squares fit read: $ a=0.179008\qc b= 0.223543\qc d=-1.25569 $.)  Therefore, the hypothesis that gaps in larger MK systems would eventually follow the MF power-law is very plausible or, at least, cannot be rejected. The convergence of the slope of the scaling function to the predicted value is nevertheless extremely slow in comparison with both SS and HS (and also considering the cases studied in Ref.~\cite{paper-fss}). This implies that very large values of $\tilde{h}$ are needed to measure the right slope.
Quantitatively, in Fig.~\ref{fig:fits-gamma-MK} I report the difference of $\gamma$ and our estimation of \emph{local} $\gamma_{MK}$ from the local slope estimated around $\tilde{h}=1$ and $\tilde{h}=100$ for the various system sizes. The value of $\gamma_{MK}$ was determined using a maximum likelihood estimation\supercite{newmanPowerLawsPareto2005} (empty markers, dashed lines) and a least squares fit (filled markers, solid line). Notice that both methods yield very similar results, although maximum likelihood values are systematically smaller than the ones of least squares. This is likely due to the fact that by truncating the interval used for the fit, the simple\footnote{There is also a generalization for truncated power-laws as explained, for instance, in \cite{delucaFittingGoodnessoffitTest2013}. However, assessing the goodness of fit is much more involved in this scenario and, for the case we are interested in, the results do not improve significantly.} maximum likelihood method is no longer exact and the resulting exponent is underestimated.
In any case, around $\tilde{h}=1$, the slope clearly differs from the MF prediction, but even when $\tilde{h}\sim 10^2$ very large system sizes are needed for it to approach the theoretical exponent. This deviation results in an \emph{apparent} size dependence of the global exponent, \textit{i.e.} $\gamma_{MK} = \gamma_{MK}(N)$, that is substantially more pronounced than for other models at similar $N$.

It is likely that this discrepancy is caused, at least in part, by the MK model being a fully connected one. In contrast with their sparse counterparts, fully connected models indeed require much larger system sizes for thermodynamic power-law scalings to be visible\supercite{lucibelloAnomalousFiniteSize2014,lucibelloFinitesizeCorrectionsDisordered2014,ferrariFinitesizeCorrectionsDisordered2013}.
In the MK configurations in particular this feature can be physically understood by recalling that the introduction of random shifts results in neighbours of a given particle (very likely) not being neighbours themselves. A particle can thus have many more contacts than normally allowed in Euclidean space, as reported in Fig.~\ref{fig:pdf z MK}. Data in that figure shows that it is not uncommon ($\sim1\%$) for particles at jamming to have as many as 12 contacts (the $d=3$ kissing number) or more. In general, particles are thus surrounded by many more particles --in actual and near contact-- than usual hard spheres.
Additionally, jamming densities in this model are much higher than in standard HS systems. Using the method described in Sec.~\ref{sec:methods} --\textit{i.e.} our MD+iLP algorithm with planted configurations\supercite{charbonneauNumericalDetectionGardner2015} to speed up the growing process-- produces packings with densities $\phi_{J,MK} \gtrsim 3.1 $ (cf. $\phi_{J,3d}\simeq 0.64$). Considering that $\phi\sim \sigma^{1/d}$, such values in the packing fraction imply that our MK configurations are made out of particles nearly twice as big as those of standard HS. When these two effects are combined, the outcome is that particles in MK packings are surrounded by an abundant cluster of relatively large neighbours. Therefore, with the effective size of the system being thus drastically reduced, the finite-size corrections are correspondingly more pronounced. We thus conclude that gaps in the MK model will probably follow the MF power-law scaling, as expected, but only at system sizes orders of magnitude larger than those considered here. In practice, the finite size effects are so important in the distribution of gaps in the MK model that its MF nature is, perhaps paradoxically, a strong limitation to study its MF behaviour in the thermodynamic limit. These same features are probably also responsible for the more pronounced size effects found in the distribution of extended contact forces.

\subsubsection{Distribution of localized forces}\label{sec:fss floc MK}

\begin{figure}[!htb]
	\includegraphics[width=\linewidth]{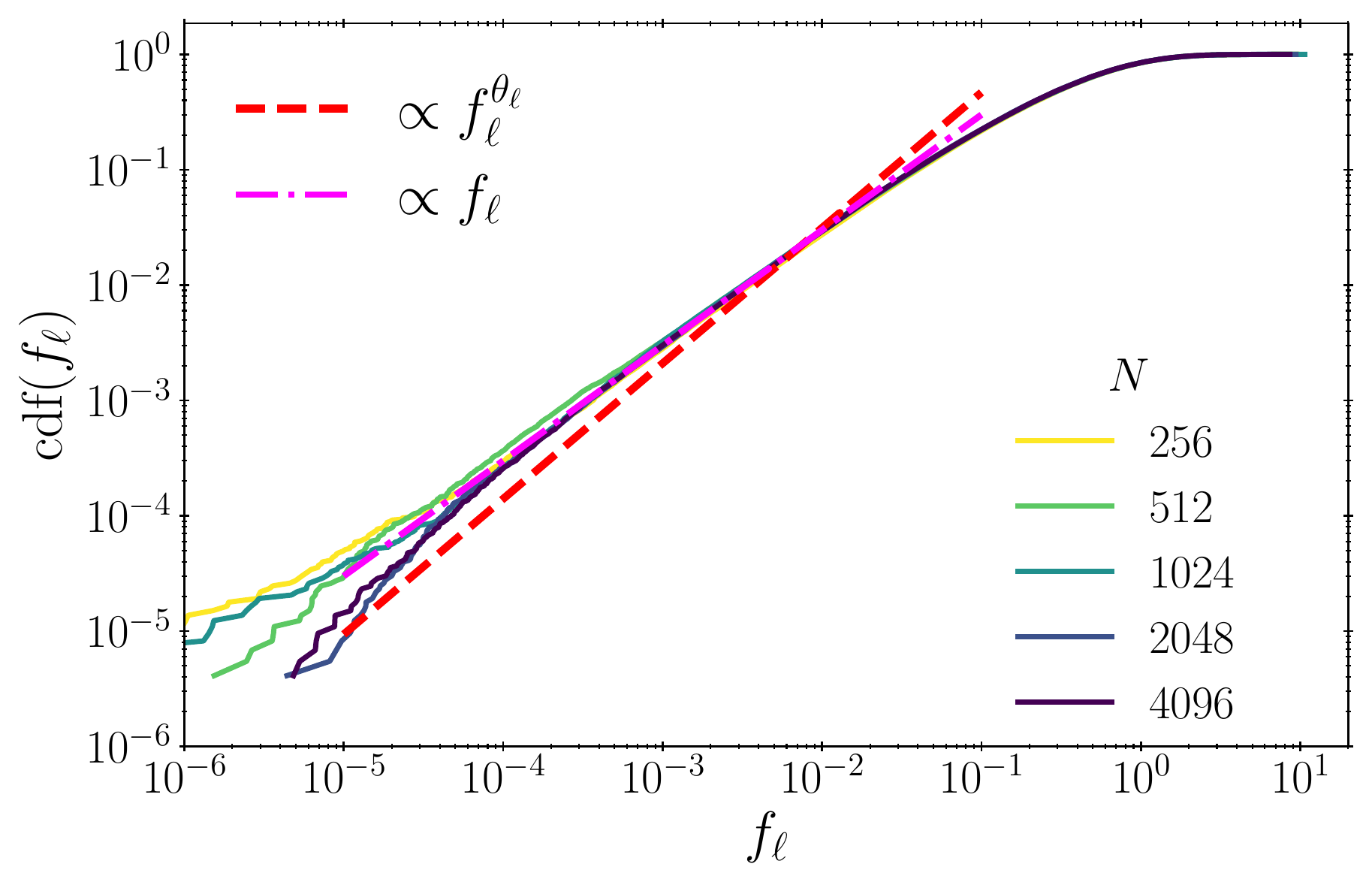}
	\caption[Cdf of localized forces in jammed MK configurations]{Cumulative distribution function of $\{f_\ell\}$ in MK packings of different sizes (as indicated by the legend). Note that, as anticipated, no signature of finite size effects are present. However, an unexpected finding is that $f_\ell$ is distributed linearly (pink dash-dotted line), in contradiction the behaviour expected from previous studies (red, dashed line), \textit{i.e.} with an exponent $1+\theta_\ell$.}
	\label{fig:forces-loc-MK}
\end{figure}

Lastly, I consider the cdf of localized forces whose cdf is presented in Fig.~\ref{fig:forces-loc-MK}, where no clear signatures of finite $N$ effects are visible, although some dispersion can be observed in the extremal values. Putting together these results with the analogous ones in standard configurations of spheres (Sec.~\ref{sec:fss floc spheres}) and considering that the same absence of size effects were observed in the other two models of Ref.~\cite{paper-fss}, we conclude that the expected behaviour is fully confirmed. Recall that this finding was anticipated previously in this chapter on the basis that the set $\{f_\ell\}$ corresponds to contact forces acting on bucklers, for which opening a contact mostly results in localized displacement field as derived in Refs.~\cite{charbonneauJammingCriticalityRevealed2015,lernerLowenergyNonlinearExcitations2013} and Sec.~\ref{sec:marginal stability}. In other words, given that opening any of the contacts associated with a buckler only has a non-negligible effect over few particle layers away from its origin (see Fig.~\ref{fig:floppy-modes}), it is reasonable to assume that their properties should be insensitive to $N$, or to any border or periodic effects, as we have here verified.

On the other hand, a surprising finding is that $f_\ell$ is distributed \emph{uniformly}, \textit{i.e.} $\theta_{\ell,MK}=0$, at variance with previous studies reporting $\theta_\ell \approx 0.17$.
A careful analysis suggests that this unexpected distribution is in tune with the spatial properties of MK packings. To explain why, first note that even though buckling forces follow a different pdf, selecting particles with $z_\ell$ contacts is still a valid selection criterion. (If their contribution had not been isolated, the remaining forces would not follow the MF power-law scaling given in Eq.~\eqref{eq:pdf-f-ext}, as it does in Figs.~\ref{fig:forces-MK}-\ref{fig:rescaled cdf fext MK}, whereas if both kinds of forces are considered together, their joint pdf scales with an exponent $\approx 1.1$, which differs from the analogous quantity for conventional spheres\supercite{charbonneauJammingCriticalityRevealed2015}.) Second, analysing the distribution of dot products between contact vectors as in Sec.~\ref{sec:contact vectors MK} reveals that particles with $z_\ell$ contacts in MK packings have a very similar distribution as those in standard hard sphere packings (cf. Fig.~\ref{fig:pdf Si bucklers}). Bucklers thus mainly give rise to a localized response thanks to them having three nearly coplanar contacts and one nearly orthogonal force. In order to understand why localized forces are uniformly distributed, I follow Ref.~\cite{lernerLowenergyNonlinearExcitations2013}, which showed that the two types of contact forces are related to two different types of floppy modes: (i) extended forces are related to floppy modes that can couple strongly to external perturbations, and hence their response is bulk dominated; and (ii) buckling forces are associated to floppy modes of a rapidly decaying displacement field. (See also the discussion of Sec.~\ref{sec:marginal stability}.) With this in mind, it is important to consider that the value of $\theta_\ell \approx 0.17$ was estimated from the statistics of \emph{displacements} in the latter type of contacts. There is therefore a strong connection between the distribution of forces in bucklers and the particle displacements their floppy modes produce. 
Now, let us assume that in an MK packing we open a buckling contact, $\ctc{ij}$, between particles $i$ and $j$, in order to describe the associated displacement field. In particular, let us focus on the remaining contacts of, say, particle $i$. Because of the random shifts, the other particles touching $i$ are (very likely) not constrained by each other nor by the other particles near $i$. Instead, the displacement of each neighbour of $i$ is limited by its own contacts, which are not neighbours themselves, and are typically far apart. By the same token, the effect on the rest of particles in contact with $j$ is determined by secondary contacts that --with high probability-- are distant from each other and from $\ctc{ij}$. As a result, opening a buckling contact produces a small series of uncorrelated displacements. No particular length scale is hence favoured over any other. Because of the close relation between localized forces and displacements just mentioned, it is natural for $f_\ell$ to be uniformly distributed.

At any rate, I would like to stress that our finding of a different value of $\theta_\ell$ is important not merely as a matter of scrupulous curve fitting, but mainly because it violates the stability bound related to local excitations, Eq.~\eqref{eq:gamma_theta_loc}. Interestingly, a similar result was obtained for the FCC packings studied in \cite{paper-fss}, suggesting that more general criteria are needed for assessing the stability of such excitations in other classes of disordered models. I comment further about this point below.

\section{Conclusions}\label{sec:discussion}

The results I presented here systematically corroborate the non-trivial power-laws of the distributions of forces and gaps in two disordered systems at jamming, fully supporting the description derived from the exact MF theory. I emphasize that for the archetypical minimal model of monodisperse spheres, such validation included changing the direction of approach to the jamming point. Our results obtained with the MK model, a mandatory reference for comparison given its MF-like behaviour, similarly reinforces the $d\to \infty$ predictions.

Yet, our main finding is the contrasting system-size dependence of these distributions. In the two models studied here, size effects in $p(f_e)$ are practically non-existent, while $g(h)$ exhibits clear and systematic signatures of finite-$N$ deviations from the expected power-law scaling. The same is true in the rest of the models considered in Ref.~\cite{paper-fss} and analysed with the same methodology.
I emphasize that testing for such size scalings not only rigorously assesses the critical exponents\supercite{amitFieldTheoryRenormalization2005,MC_book}, but also provides key insight into the length scale of their correlations. Hence, it can be concluded that the MF exponents for all gaps and extended forces distributions are correct. 
Moreover, a second and more informative conclusion is that the distribution $p(f_e)$ reaches its thermodynamic limit behaviour at smaller values of $N$ than $g(h)$. Two different correlation lengths, $\xi_{f_e}$ and $\xi_h$, therefore characterise the relevant length scales of correlations of contact forces and gaps, respectively. For the former, it must hold that $N^{1/d} \gg \xi_{f_e}$, and thus finite size corrections are negligible, while the analogous condition for gaps reads $N^{1/d} \lesssim \xi_h$. In other words, gaps are correlated over significantly larger distances than forces, \textit{i.e.} $\xi_h \gg \xi_{f_e}$. Such disparity in the correlation lengths is an unexpected consequence of our results, considering that both quantities are usually treated on an equal footing from the perspective of the SAT-UNSAT transition in the perceptron\supercite{franzSimplestModelJamming2016,franzCriticalJammedPhase2019a}, constraint satisfaction problems\supercite{franzUniversalitySATUNSATJamming2017,ikedaMeanFieldTheory2019}, and neural networks\supercite{franzJammingMultilayerSupervised2019} as well as from the point of view of marginal stability in amorphous solids\supercite{wyartMarginalStabilityConstrains2012,mullerMarginalStabilityStructural2015}.
MK results back this hypothesis if we consider that their very high densities and connectivity reduce the effective system size, as discussed in Sec.~\ref{sec:fss distributions MK}.
Observing the scaling of Eq.~\eqref{eq:scaling-cdf} for the cdf of $f_e$ is thus a manifestation of the smaller effective volume (for a similar $N$), which confirms that finite-size effects at jamming do occur for $p(f_e)$, but disappear for relatively small system sizes; see also the discussion leading to Fig.~\ref{fig:forces-3d-minus-1N}. The significantly more pronounced $N$ dependence of the distributions of $h$ (Figs.~\ref{fig:forces-MK} and \ref{fig:gaps-MK}) thus supports our finding that $\xi_h \gg \xi_{f_e}$.

On the other hand, the linear growth of cdf$(f_\ell)$ in the MK model is at odds with the stability condition of Eq.~\eqref{eq:gamma_theta_loc}. This finding is more surprising because in these systems there is no long-range order, as there is in near-crystalline packings where similar violations have been reported\supercite{charbonneauGlassyGardnerlikePhenomenology2019,tsekenisJammingCriticalityNearCrystals2020,paper-fss}. At the end of Sec.~\ref{sec:fss floc MK} I used the peculiar geometry of the MK packings to suggest a physical explanation for the uniform distribution of $f_\ell$, but this reasoning does not explain why the stability condition between $\gamma$ and $\theta_\ell$ is apparently violated. Given the drastic difference in the inherent structures of the near-crystals and MK packings, they highlight the need to better understand how the spatial correlation caused by external perturbations are influenced by the type of disorder in jammed packings.

Finally, the most persistent observation was that all cumulative distributions, of both gaps and extended forces, behave in a seemingly linear fashion at very small arguments (specially when jamming is reached from the UC phase), in agreement with the MF predictions, $p_0$ and $g_0$ in Eqs.~\eqref{eq:pdf-gaps-v2} and \eqref{eq:pdf-f-ext-v2}, respectively. We argued that such a cut-off in the main power-law is due to the extra contact with respect to isostatic configurations (see Sec.~\ref{sec:fss-theory}) and has been previously reported for the gaps distributions of disks packings\supercite{ikedaInfinitesimalAsphericityChanges2020}, but we are not aware of analogous findings in any other model or for the $\{f_e\}$ distributions. The universality of this secondary scaling has been previously predicted\supercite{franzUniversalitySATUNSATJamming2017} for all models that can be mapped to jamming of spherical particles, and it has been shown to occur even for non-spherical particles\supercite{ikedaInfinitesimalAsphericityChanges2020}, provided that their jammed states remain sufficiently close to isostaticity. Such robustness can be understood in part by considering that isostaticity is a global property of a system --where the amount of constraints matches the degrees of freedom-- and not to the specific distributions of its microstructural variables.
Because we have restricted our analysis to packings having 1SS, that is with exactly $N_c = N_{dof}+1$ contacts, the ubiquity of the linear growth in the extremal part of the cdf's supports the hypothesis that left tails of the critical distributions is determined by the 1SS property alone, and not by the inherent structure.
However, undersampling effects are very hard to avoid when dealing with extremal statistics, so a more stringent analysis must be carried out to verify that $p_0(x) \sim g_0(x) \sim 1$ when $x\ll1$, specially for the former pdf.
For completeness, I should mention that a previous work on the perceptron\supercite{kallusScalingCollapseJamming2016} also reported a similar transition to a uniform distribution of contact forces  that, unexpectedly, depended on the type of algorithm used to reach the jamming point. This is an important consideration because the MD+iLP and FIRE algorithms, used in the UC and OC phases, respectively, indeed yielded distributions with rather different left tails. Given the prominent role of the smallest forces and gaps in determining the global stability of the packings (Sec.~\ref{sec:marginal stability}), understanding their algorithmic dependence, if any, is undoubtedly an interesting topic for future research. 

\newpage

\begin{subappendices}
	
\section{Equivalence of cumulative distributions} \label{sec:equivalence-cdfs}

Here I briefly compare the cumulative distributions presented before in the chapter with the estimations using two other methods. Given $M$ independent samples, the first of such methods consists in computing the cdf of each individual sample, and then compare such distributions across the $M$ datasets. Because in general computing the cdf of a set of values involves ranking them, when different cdf's are compared, we can obtain an estimate of the typical value of the first, second, etc. element in such set. Moreover, the sample variance of the cdf thus estimated can also be provides useful information about how big are the fluctuations of the ranked elements.

The third way we tried also involves computing the cdf of each sample individually. But then, such function is evaluated at a fixed set of points, and this is repeated for all the systems. Clearly, the set of points at which all the cdf's are evaluated will not coincide in general with the sample values of any given system, but nonetheless the \emph{empirical} cdf can be evaluated in all of them. The reason is that the empirical cdf of a given system is computed as a stepwise function, whose increments occur at the values sampled on such system, but is constant in the intervals formed by such values. Thus, the value of the cdf is well defined essentially in any point (compatible with its domain). In any case, the advantage of this procedure is that by considering the cumulative distributions of the different samples we can obtain the statistics of the probability mass assigned to a fixed set of points. 

\begin{figure*}[!htb]
	\centering
	\includegraphics[width=0.99\linewidth]{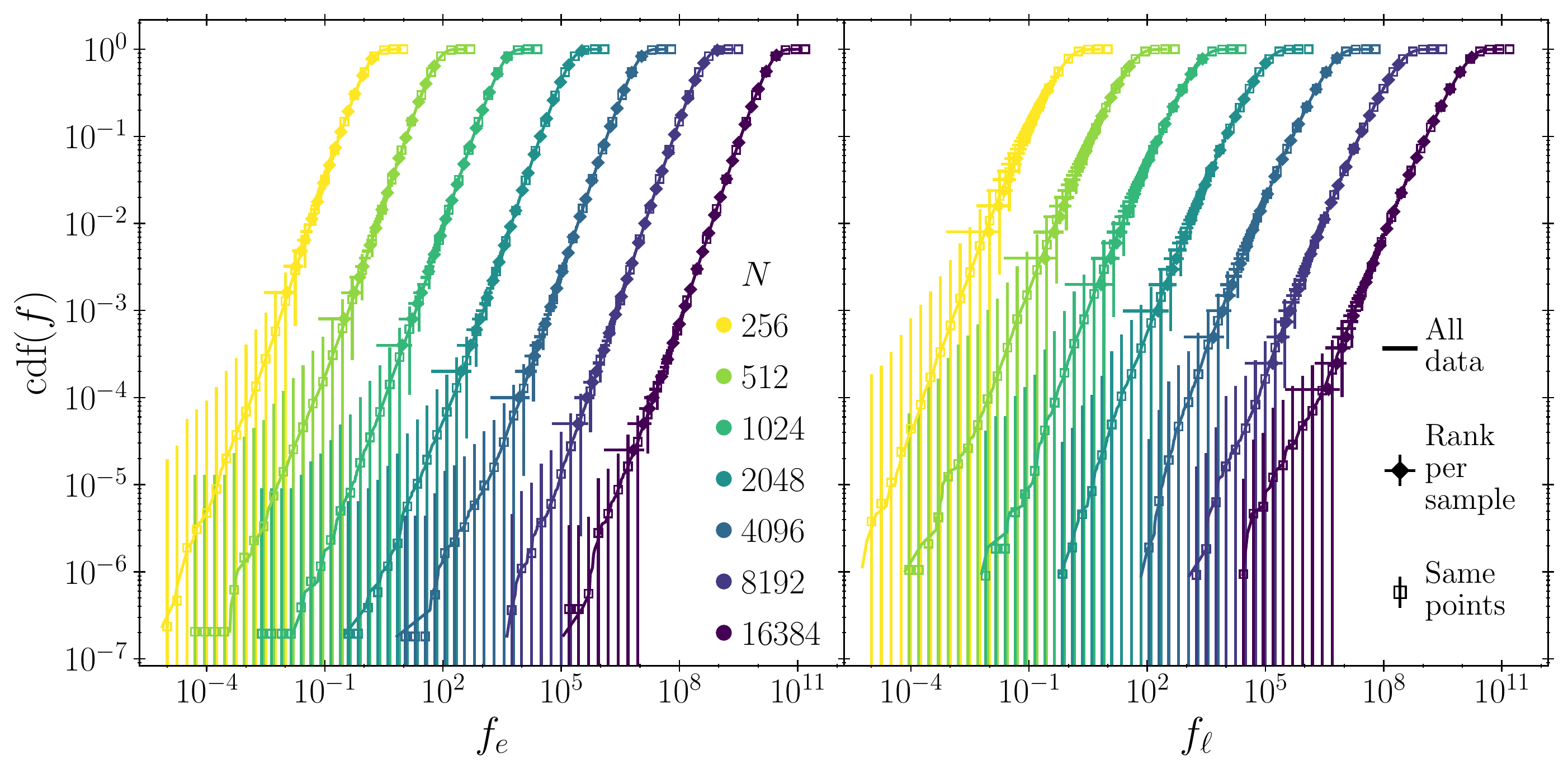}
	\caption[Three different (and compatible) ways of computing the cdf of forces in HS.]{Cumulative distribution functions of extended (left panel) and localized (right) forces in HS computed using three different methods and for all the system sizes considered in this chapter. The method used in the main text (lines) is clearly compatible with (i) the cdf obtained by ranking the forces in each sample (filled diamonds) and (ii) with evaluating each sample's cdf on the same set of points (empty squares). Curves of different $N$ have been displaced horizontally for clarity. Error bars correspond to the standard deviation in the $M_N$ samples considered for each size; their different meaning in the two latter cases is explained in the text.}
	\label{fig:equivalence-cdf-forces-spheres}
\end{figure*}

Naturally, with enough data available both methods should be compatible with each other, as well as with the estimation of the cdf used in the main text, \textit{i.e.} using a single, large dataset obtained from all the samples. However, the two methods described in this appendix are useful to obtain uncertainties of the cdf and are complimentary in this aspect: the method based on each sample's ranking mainly informs on the variance of the random variable itself, while evaluating different cdf's in the same points provides the fluctuations of the probability mass.

\begin{figure}[!htb]
	\centering
	\includegraphics[width=0.99\linewidth]{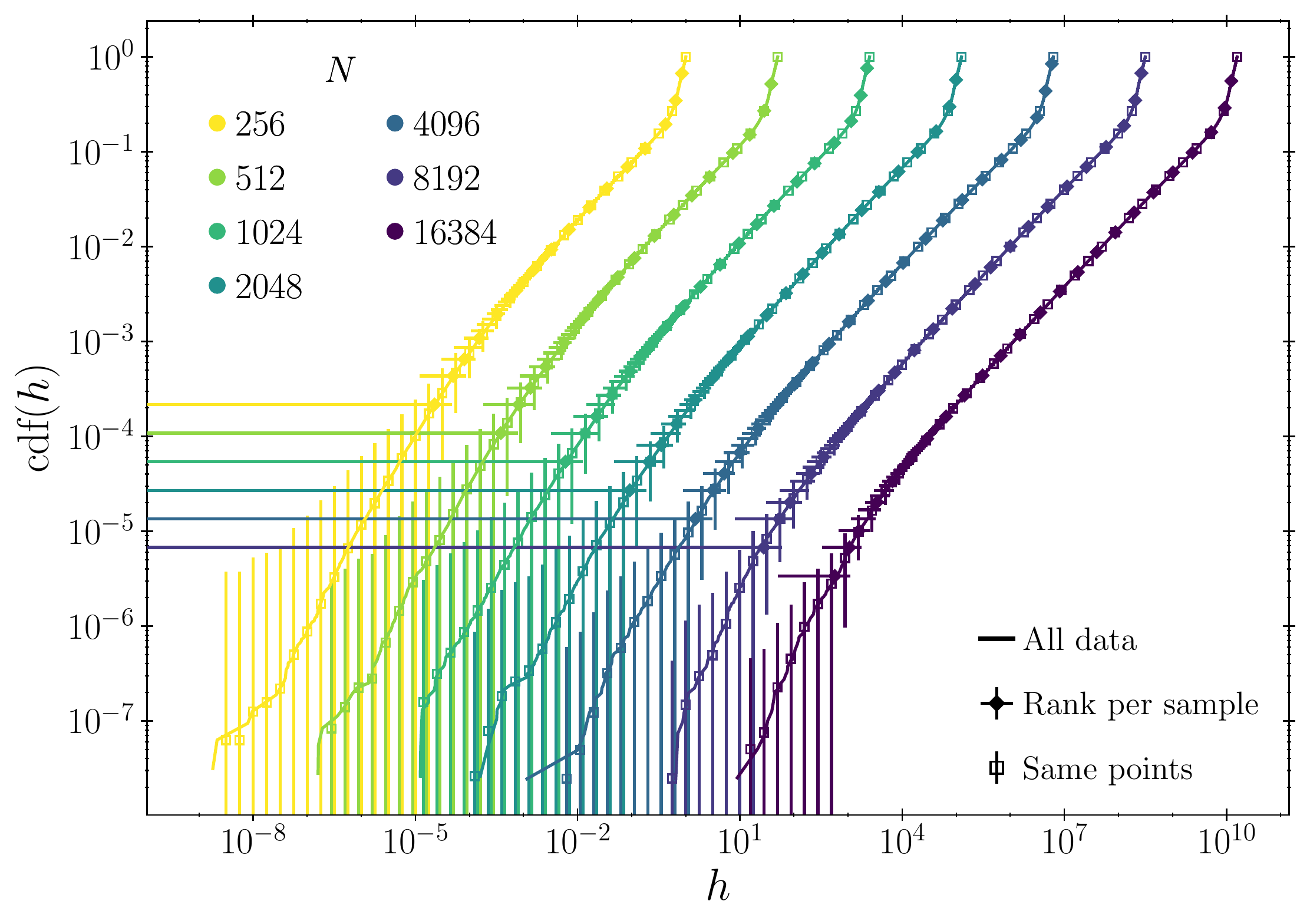}
	\caption[Three different (and compatible) ways of computing the cdf of gaps in HS.]{Cumulative distribution functions of interparticle gaps in HS computed following three different procedures and for all the system sizes considered in this chapter. Once again, combining all the data in a single set as done in the main text (lines) produces compatible results with (i) the cdf obtained from each sample's ranking (filled diamonds) and (ii) with evaluating each sample's cdf on the same set of points (empty squares). Curves of different $N$ have been displaced horizontally for clarity. Error bars correspond to the standard deviation in the $M_N$ samples considered for each size. Note that the smallest gaps show much broader sample-to-sample fluctuations than the smallest forces.}
	\label{fig:equivalence-cdf-gaps-spheres}
\end{figure}

In Figs.~\ref{fig:equivalence-cdf-forces-spheres} and \ref{fig:equivalence-cdf-gaps-spheres} I compare the cumulative distributions of forces and gaps, respectively, computed using each the three methods in all our HS systems. The agreement between them is excellent for all system sizes and the three of structural variables considered here. Additionally, the error bars of each data series confirms the interpretation given above for the corresponding method.

\end{subappendices}

\backmatter
\cleardoublepage
\phantomsection 

\printbibliography[heading=bibintoc]

\listoffigures

\end{document}